\documentclass[a4paper,12pt]{book}

\usepackage[english]{babel}
\usepackage[sectionbib]{natbib} % allows bibtex, \citet{}, \citep{} referencing
\usepackage[table]{xcolor}
\usepackage{graphicx,epic}
\usepackage{tikz} % must be done after xcolor.
\usepackage{float}
\usepackage{amsmath,amssymb}
\usepackage{enumerate}
\usepackage{multirow}
\usepackage{indentfirst} % indent the first line of a section
\usepackage{ifthen}
\usepackage{calc}
\usepackage[implicit=false, colorlinks=true, linkcolor=black, urlcolor=mycolorlink]{hyperref}

\newcounter{hours}
\newcounter{minutes}
\newcommand{\now}
	{
	\setcounter{hours}{\time/60}
	\setcounter{minutes}{\time-\value{hours}*60}
	\textcolor{red}{version: \today~(\thehours:\ifthenelse{\value{minutes}<10}{0}{}\theminutes)}
	}

\def\printerversion{0}

\definecolor{mycolordeco}{rgb}{0.941176,0.941176,0.941176}
\definecolor{mycolorbox}{rgb}{0.93,0.93,0.93}
\definecolor{mycoloracr}{rgb}{0.0,0.0,0.6}
\definecolor{mycoloracrbox}{rgb}{0.94,0.94,0.94}
\definecolor{mycoloracrline}{rgb}{0.85,0.85,0.85}

\ifthenelse{\equal{\printerversion}{1}}{
	\definecolor{mycoloracrlink}{rgb}{0.0,0.0,0.0}
	\definecolor{mycolorlink}{rgb}{0.0,0.0,0.0}
	\definecolor{mycolorcite}{rgb}{0.0,0.0,0.0}
	\definecolor{mycolorbiblio}{rgb}{0.0,0.0,0.0}
	}{
	\definecolor{mycoloracrlink}{rgb}{0.0,0.0,0.0}
	\definecolor{mycolorlink}{rgb}{0.0,0.0,0.6}
	\definecolor{mycolorcite}{rgb}{0.6,0.0,0.0}
	\definecolor{mycolorbiblio}{rgb}{0.0,0.0,0.6}
	}

\newcommand{\cdp}{\clearpage{\pagestyle{plain}\cleardoublepage}}  % insert even page with plain style.

\newcommand{\linkup}[1]{%
	\protect\begin{picture}(0,0)%
	\protect\put(0,12){\protect\hypertarget{#1}{}}%
	\protect\end{picture}%
	}

\newcommand{\myparskip}{10pt}
\newcommand{\myparskiptoc}{2pt}

\newcommand{\localseclabel}{sec:}
\newcommand{\localsubseclabel}{sec:}
\newlength{\templength}
\newcounter{saveenumi} % save counter for lists of publications
\newcounter{autotoc}
\newcounter{autolof}
\newcounter{autolot}

\def\thechapter{\Roman{chapter}}

\makeatletter

\def\@chapter[#1]#2#3{%
		\ifnum \c@secnumdepth >\m@ne
                       \if@mainmatter
				\refstepcounter{chapter}%
				\typeout{\@chapapp\space\thechapter.}%
				\addcontentsline{toc}{chapter}{\protect\numberline{\thechapter}#1}%
			\else
				\addcontentsline{toc}{chapter}{#1}%
			\fi
		\else
			\addcontentsline{toc}{chapter}{#1}%
		\fi
		\chaptermark{#1}%
		\addtocontents{lof}{\protect\addvspace{10\p@}}%
		\addtocontents{lot}{\protect\addvspace{10\p@}}%
		\pagestyle{normal}
		\thispagestyle{plain}
		\@makechapterhead{#2}{#3}%
		\@afterheading}

\def\@makechapterhead#1#2{%
	\vspace*{0\p@}%
	{\parindent \z@ \raggedleft \normalfont
	\ifnum \c@secnumdepth >\m@ne
		\if@mainmatter
			\begin{picture}(0,0)\put(0,25){\hypertarget{cha:#2}{}\hypertarget{autotoc:\theautotoc}{}\addtocounter{autotoc}{1}}\end{picture}%
			\label{cha:#2}%
			\large\textsc{\@chapapp} %
			\begin{picture}(40,0)
				\begin{tikzpicture}[overlay, mycolordeco]
					\fill (0pt,4pt) rectangle (40pt, 200pt);
					\fill (8pt,4pt) circle (8pt);
					\fill (40pt-8pt,4pt) circle (8pt);
					\fill (8pt,-4pt) rectangle (40pt-8pt,10pt);
				\end{tikzpicture}
			\put(0,0){\makebox[40pt][c]{\Huge\thechapter}}
			\end{picture}\\
			\vskip 7\p@
		\fi
	\fi
	\Huge \bfseries #1\par\nobreak
	\vskip 45\p@
	}}

\def\@schapter#1{%
	\if@twocolumn
		\@topnewpage[\@makeschapterhead{#1}]%
	\else
		\@makeschapterhead{#1}%
		\@afterheading
	\fi}

\def\@makeschapterhead#1{%
	\vspace*{29\p@}%
	{\parindent \z@ \raggedleft
	\normalfont
	\Huge \bfseries\begin{picture}(0,0)\put(0,25){\hypertarget{autotoc:\theautotoc}{}\addtocounter{autotoc}{1}}\end{picture} #1\\
	\vskip 45\p@}}

\def\@startsection#1#2#3#4#5#6{%  need to be redefined to send the argument #2 to \@ssect
	\if@noskipsec \leavevmode \fi
	\par
	\@tempskipa #4\relax
	\@afterindenttrue
	\ifdim \@tempskipa <\z@
		\@tempskipa -\@tempskipa \@afterindentfalse
	\fi
	\if@nobreak
		\everypar{}%
	\else
		\addpenalty\@secpenalty\addvspace\@tempskipa
	\fi
	\@ifstar
		{\@ssect{#2}{#3}{#4}{#5}{#6}}%
		{\@dblarg{\@sect{#1}{#2}{#3}{#4}{#5}{#6}}}%
	} 

\def\@sect#1#2#3#4#5#6[#7]#8{% called by \@starsection    WARNING: works for section and subsection!
	\ifnum #2>\c@secnumdepth
		\let\@svsec\@empty
	\else
		\refstepcounter{#1}%
		\protected@edef\@svsec{\@seccntformat{#1}\relax}%
	\fi
	\@tempskipa #5\relax
	\ifdim \@tempskipa>\z@
		\begingroup
			#6%
			\ifnum #2<2% Sections only
				\settowidth{\mysecwidth}{\@hangfrom{\hskip #3\relax\@svsec}\Large\bfseries#8}%
				\begin{tikzpicture}[overlay, mycolordeco]%
					\fill (10pt,-4.5pt) -- ++(\mysecwidth,0pt) -- ++(0pt,20pt) -- ++(-\mysecwidth,0pt) -- cycle;
					\fill (\mysecwidth+10pt, 5.5pt) circle (10pt);
					\fill (10pt, 5.5pt) circle (10pt);
					\fill[white] (10pt, 5.5pt) circle (8pt);
				\end{tikzpicture}%
				\hspace{5pt}%
			\fi%
			{\@hangfrom{\hskip #3\relax\@svsec}\interlinepenalty \@M #8\@@par}%
		\endgroup
		\csname #1mark\endcsname{#7}%
		\addcontentsline{toc}{#1}{%
		\ifnum #2>\c@secnumdepth \else
		\protect\numberline{\csname the#1\endcsname}%
		\fi
		#7}%
	\else
		\def\@svsechd{%
		#6{\hskip #3\relax
		\@svsec #8}%
		\csname #1mark\endcsname{#7}%
		\addcontentsline{toc}{#1}{%
		\ifnum #2>\c@secnumdepth \else
			\protect\numberline{\csname the#1\endcsname}%
		\fi
		#7}}%
	\fi
	\@xsect{#5}}

\def\@ssect#1#2#3#4#5#6{% called by \@starsection    WARNING: works for section and subsection!
	\@tempskipa #4\relax
	\ifdim \@tempskipa>\z@
		\begingroup
			#5%
			\ifnum #1<2% Sections only
				\settowidth{\mysecwidth}{\@hangfrom{\hskip #2}\Large\bfseries#6}%
				\begin{tikzpicture}[overlay, mycolordeco]%
					\fill (10pt,-4.5pt) -- ++(\mysecwidth,0pt) -- ++(0pt,20pt) -- ++(-\mysecwidth,0pt) -- cycle;
					\fill (\mysecwidth+10pt, 5.5pt) circle (10pt);
					\fill (10pt, 5.5pt) circle (10pt);
				\end{tikzpicture}%
			\fi%
			{\@hangfrom{\hskip #2}\interlinepenalty \@M \hspace{10pt}#6\@@par}%
		\endgroup
	\else
		\def\@svsechd{#5{\hskip #2\relax #6}}%
	\fi
	\@xsect{#4}}

\renewcommand\section{\@startsection {section}{1}{\z@}%
	{-3.5ex \@plus -1ex \@minus -.2ex}%
	{2.3ex \@plus.2ex}%
	{\noindent\begin{picture}(0,0)\put(0,25){\hypertarget{\localseclabel}{}\hypertarget{autotoc:\theautotoc}{}\addtocounter{autotoc}{1}}\end{picture}\normalfont\Large\bfseries}}
	
\renewcommand\subsection{\@startsection{subsection}{2}{\z@}%
	{-3.25ex\@plus -1ex \@minus -.2ex}%
	{1.5ex \@plus .2ex}%
	{\noindent\begin{picture}(0,0)\put(0,20){\hypertarget{\localsubseclabel}{}\hypertarget{autotoc:\theautotoc}{}\addtocounter{autotoc}{1}}\end{picture}\normalfont\large\bfseries}}

\makeatother

\newcommand{\chanonumber}[1]{% {title}   implementing other args may be difficult because toc, lof, lot and bibliography uses chapter* with only one arg.
	\cdp
	\chapter*{#1}
	\refstepcounter{chapter}% reset all dependent counters
	\setcounter{chapter}{0}% needed to get the Figure 12 format instead of II.12
	\pagestyle{nochapnumber}
	\thispagestyle{plain}
	\markboth{#1}{}
	\addcontentsline{toc}{chapter}{#1}
	\addtocontents{lof}{\protect\addvspace{10pt}}
	\addtocontents{lot}{\protect\addvspace{10pt}}
	}

\newlength{\mysecwidth}
\newcommand{\cha}[2]{\cdp\chapter{#2}{#1}} % {label}{title}
\newcommand{\sect}[2][]{\def\localseclabel{sec:#1}\section{#2}\label{sec:#1}}% [label]{title}   use the variable localseclabel to be called for the target in the \section
\newcommand{\subsect}[2][]{\def\localsubseclabel{sec:#1}\subsection{#2}\label{sec:#1}} % [label]{title}   use the variable localsubseclabel to be called for the target in the \section

\newcommand{\refcha}[1]{\hyperlink{cha:#1}{\textcolor{mycolorlink}{Chapter~\ref{cha:#1}}}}
\newcommand{\refapp}[1]{\hyperlink{cha:#1}{\textcolor{mycolorlink}{Appendix~\ref{cha:#1}}}}

\newcommand{\refsec}[2][]{\hyperlink{sec:#2}{\textcolor{mycolorlink}{Section~\ifthenelse{\equal{#1}{}}{}{\ref{cha:#1}.}\ref{sec:#2}}}}
\newcommand{\refsect}[2][]{\hyperlink{sec:#2}{\textcolor{mycolorlink}{Section~\ifthenelse{\equal{#1}{}}{}{\ref{cha:#1}.}\ref{sec:#2}}}, page~\pageref{sec:#2}}
\newcommand{\refsecp}[2][]{\hyperlink{sec:#2}{\textcolor{mycolorlink}{Section~\ifthenelse{\equal{#1}{}}{}{\ref{cha:#1}.}\ref{sec:#2}}} (page~\pageref{sec:#2})}

\newcommand{\fig}[3]{% file_name (no /figs nor .eps), short_caption, normal caption
	\begin{figure}
	\hypertarget{fig:#1}{}\hypertarget{autolof:\theautolof}{}\addtocounter{autolof}{1}
	\begin{center}
	\includegraphics{figs/#1.eps}
	\caption[#2]{#3}
	\label{fig:#1}
	\end{center}
	\end{figure}}

\newcommand{\note}{\textsc{Note:} }

\newcommand{\tab}[6]{% width, label, caption, format, content, note
	\settowidth{\templength}{#6}
	\begin{table}
	\hypertarget{tab:#2}{}\hypertarget{autolot:\theautolot}{}\addtocounter{autolot}{1}
	\begin{center}
	\begin{minipage}{#1}
	\caption{#3}
	\label{tab:#2}
	\vspace{0.2cm}
	\begin{tabular*}{\linewidth}{#4}
	\hline\hline
	#5
	\hline
	\ifthenelse{\lengthtest{\templength=0pt}}
		{\end{tabular*}}
		{
		\vspace{-0.3cm}
		\end{tabular*}
		{\footnotesize #6}
		}
	\end{minipage}
	\end{center}
	\end{table}}

\newcommand{\eqn}[2][]{% [label]{content}
	\begin{picture}(0,0)\put(0,0){\hypertarget{eqn:#1}{}}\end{picture}
	\begin{align}
	#2
	\label{eqn:#1}
	\end{align}
	}

\newcounter{myboxctr}[chapter]
\makeatletter \renewcommand \themyboxctr {\ifnum \c@chapter>\z@ \thechapter.\fi \@arabic\c@myboxctr} \makeatother % don't put the chapter number in case of chapter*
\newcommand{\follow}{\hfill$\cdots$}  % insert a symbol at the end of a page in a box

\newcommand{\mybox}[2][]{% [label]{content} (for a second page boxe, use label=nocnt)
	\ignorespaces
	\begin{center}
	\begin{minipage}{15cm}
	\ifthenelse{\equal{#1}{nocnt}}
		{}
		{
		\refstepcounter{myboxctr} % increase counter
		$\blacksquare$\,Box \themyboxctr\linkup{box:#1}\\
		\vspace{-0.8cm}
		}
	\begin{center}
	\setbox0=\hbox\bgroup
	\begin{minipage}{13.9cm}
	\setlength{\parindent}{15pt}
	\setlength\parskip{10pt}
	\ifthenelse{\equal{#1}{nocnt}}
		{\begin{flushleft}\vspace{-5pt}$\cdots$\vspace{-5pt}\end{flushleft}}
		{}
	#2
	\ifthenelse{\equal{#1}{nocnt}}{}{\begin{picture}(0,0)\put(0,0){\label{box:#1}}\end{picture}}
	\end{minipage}\egroup
	\fcolorbox{black}{mycolorbox}{\box0}\end{center}\end{minipage}
	\end{center}
	}

\makeatletter
\newcommand{\testacr}[3]{%
	\@ifundefined{s@#1}{\global\@namedef{s@#1}{1}#2\linkup{acr:#1}}{\hyperlink{acr:#1}{\textcolor{mycoloracrlink}{#1#3}}}%
}
\makeatother

\newcommand{\acr}[4][]{% nortarget(notarg), short, long     (use notarg as first arg to avoid the margin thing in captions, footnotes, boxes ...).   use optional arg to mark the plural form of the short version
	\ifthenelse{\equal{#2}{notarg}}% skip definition
		{\hyperlink{acr:#3}{\textcolor{mycoloracrlink}{#3#1}}}%
		{%
		\testacr{#3}% do we need definition? if yes, print the following. 
			{%
			\label{tempacr:#3}%
			\ifthenelse{\isodd{\pageref{tempacr:#3}}}% is this an odd page?   Note that \thepage maybe wrong in some (rare) configurations.
				{\begin{picture}(0,0)\put(0,-4){\textcolor{mycoloracrbox}{\rule{\widthof{#4}}{13pt}}}\color{mycoloracrline}\put(0,-4){\line(1,0){1000}}\end{picture}\hbox{#4}}% odd page
				{\begin{picture}(0,0)\put(0,-4){\textcolor{mycoloracrbox}{\rule{\widthof{#4}}{13pt}}}\end{picture}\hbox{#4}\begin{picture}(0,0)\color{mycoloracrline}\put(0,-4){\line(-1,0){1000}}\end{picture}}% even page
			\marginpar
				[\raggedleft\textcolor{mycoloracr}{\textsf{#3}}\space\begin{picture}(12,12)\put(-80,-6){\textcolor{white}{\rule{80pt}{3pt}}}\color{mycoloracr}\put(0,-4){\line(0,1){14}}\end{picture}]%
				{\begin{picture}(12,12)\color{mycoloracr}\put(12,-4){\line(0,1){14}}\put(12.2,-6){\textcolor{white}{\rule{80pt}{3pt}}}\end{picture}\raggedright\space\textcolor{mycoloracr}{\textsf{#3}}}%
			}{#1}%
		}%
	}
\newcommand{\reffig}[2][]{\hyperlink{fig:#2}{\textcolor{mycolorlink}{Figure~\ref{fig:#2}\ifthenelse{\equal{#1}{}}{}{.#1}}}}
\newcommand{\reffigt}[2][]{\hyperlink{fig:#2}{\textcolor{mycolorlink}{Figure~\ref{fig:#2}\ifthenelse{\equal{#1}{}}{}{.#1}}}, page~\pageref{fig:#2}}
\newcommand{\reffigp}[2][]{\hyperlink{fig:#2}{\textcolor{mycolorlink}{Figure~\ref{fig:#2}\ifthenelse{\equal{#1}{}}{}{.#1}}} (page~\pageref{fig:#2})}

\newcommand{\reftab}[1]{\hyperlink{tab:#1}{\textcolor{mycolorlink}{Table~\ref{tab:#1}}}}

\newcommand{\reftabp}[1]{\hyperlink{tab:#1}{\textcolor{mycolorlink}{Table~\ref{tab:#1}}} (page~\pageref{tab:#1})}

\newcommand{\refeqn}[1]{\hyperlink{eqn:#1}{\textcolor{mycolorlink}{Equation~\ref{eqn:#1}}}}
\newcommand{\refeqnt}[1]{\hyperlink{eqn:#1}{\textcolor{mycolorlink}{Equation~\ref{eqn:#1}}}, page~\pageref{eqn:#1}}
\newcommand{\refeqnp}[1]{\hyperlink{eqn:#1}{\textcolor{mycolorlink}{Equation~\ref{eqn:#1}}} (page~\pageref{eqn:#1})}

\newcommand{\refbox}[1]{\hyperlink{box:#1}{\textcolor{mycolorlink}{Box~\ref{box:#1}}}}
\newcommand{\refboxt}[1]{\hyperlink{box:#1}{\textcolor{mycolorlink}{Box~\ref{box:#1}}}, page~\pageref{box:#1}}
\newcommand{\refboxp}[1]{\hyperlink{box:#1}{\textcolor{mycolorlink}{Box~\ref{box:#1}}} (page~\pageref{box:#1})}

\newcommand{\myciteauthor}[1]{\citeauthor{#1}}
\newcommand{\myciteyear}[1]{\hyperlink{autobib:#1}{\textcolor{mycolorcite}{\citeyear{#1}}}}
\newcommand{\mycitet}[1]{\myciteauthor{#1} (\myciteyear{#1})}
\newcommand{\mycitealt}[1]{\myciteauthor{#1} \myciteyear{#1}}

\newcounter{autofootnote}
\makeatletter
\def\@makefnmark{\hbox{\@textsuperscript{\hyperlink{autofnt:\theautofootnote}{\normalfont\@thefnmark}}}}

\renewcommand\footnoterule{%
	\kern-3\p@
	\hrule\@width.4\columnwidth
	\kern2.6\p@}
\makeatother

\let\oldfootnote=\footnote
\renewcommand{\footnote}[1]{\stepcounter{autofootnote}\oldfootnote{\samepage$\ $\linkup{autofnt:\theautofootnote}#1}} % redefine footnote to add a space, and avoid multiple page spawn.

\let\oldtoc=\tableofcontents
\renewcommand{\tableofcontents}{\setcounter{autotoc}{-1}\setlength\parskip{\myparskiptoc}\oldtoc\setlength\parskip{\myparskip}\setcounter{autotoc}{0}}
\let\oldlof=\listoffigures
\renewcommand{\listoffigures}{\setcounter{autolof}{0}\setlength\parskip{\myparskiptoc}\oldlof\setlength\parskip{\myparskip}\setcounter{autolof}{0}\setcounter{autotoc}{0}}
\let\oldlot=\listoftables
\renewcommand{\listoftables}{\setcounter{autolot}{0}\setlength\parskip{\myparskiptoc}\oldlot\setlength\parskip{\myparskip}\setcounter{autolot}{0}\setcounter{autotoc}{0}}

\makeatletter
\newcommand{\tocdotfill}[1]{\leaders\hbox{$\m@th\mkern #1 mu\hbox{.}\mkern #1 mu$}\hfill}

\renewcommand*{\l@chapter}[2]{%
	\ifnum \c@tocdepth >\m@ne
		\addpenalty{-\@highpenalty}%
		\vskip 1.0em\@plus\p@
		{\leftskip 0em\relax
		\rightskip \@tocrmarg% leave room for the page number
		\parfillskip -\rightskip% Ensure that the last line of the entry will be filled. Setting \parfillskip to a negative number prevents any overfull box messages.
		\parindent 0em\relax\@afterindenttrue
		\interlinepenalty\@M% Try and prevent breaks between lines in a multiple line entry.
		\leavevmode
		\@tempdima 2.5em\relax
		\advance\leftskip \@tempdima \null\nobreak\hskip -\leftskip
		{\bfseries %
		\begin{tikzpicture}[overlay, mycolordeco]%
			\fill (4pt,-4pt) -- ++(\linewidth-8,0pt) -- ++(0pt,16pt) -- ++(-\linewidth+8,0pt) -- cycle;
			\fill (\linewidth-4pt, 4pt) circle (8pt);
			\fill (4pt, 4pt) circle (8pt);
		\end{tikzpicture}%
		\hyperlink{autotoc:\theautotoc}{#1}\addtocounter{autotoc}{1}}\nobreak{\bfseries\tocdotfill{10000}}\hb@xt@\@pnumwidth{\hfil\bfseries \hyperlink{pag:#2}{#2}}\par}%
	\fi}

\renewcommand*{\l@section}[2]{%
	\ifnum \c@tocdepth >\z@
		\vskip \z@ \@plus.2\p@
		{\leftskip 1.0em\relax
		\rightskip \@tocrmarg
		\parfillskip -\rightskip
		\parindent 2.5em\relax\@afterindenttrue
		\interlinepenalty\@M
		\leavevmode
		\@tempdima 1.5em\relax
		\advance\leftskip \@tempdima \null\nobreak\hskip -\leftskip
		{\normalfont \hyperlink{autotoc:\theautotoc}{#1}\addtocounter{autotoc}{1}}\nobreak{\normalfont\tocdotfill{4.5}}\nobreak\hb@xt@\@pnumwidth{\hfil\normalfont \hyperlink{pag:#2}{#2}}\par}%
	\fi}

\renewcommand*{\l@subsection}[2]{%
	\ifnum \c@tocdepth >\@ne
		\vskip \z@ \@plus.2\p@
		{\leftskip 3.8em\relax
		\rightskip \@tocrmarg
		\parfillskip -\rightskip
		\parindent 4.0em\relax\@afterindenttrue
		\interlinepenalty\@M
		\leavevmode
		\@tempdima 2.0em\relax
		\advance\leftskip \@tempdima \null\nobreak\hskip -\leftskip
		{\normalfont \hyperlink{autotoc:\theautotoc}{#1}\addtocounter{autotoc}{1}}\nobreak{\normalfont\tocdotfill{4.5}}\nobreak\hb@xt@\@pnumwidth{\hfil\normalfont \hyperlink{pag:#2}{#2}}\par}%
	\fi}

\renewcommand*{\l@figure}[2]{%
	\vskip \z@ \@plus.2\p@
	{\leftskip 1.5em\relax
	\rightskip \@tocrmarg
	\parfillskip -\rightskip
	\parindent 1.5em\relax\@afterindenttrue
	\interlinepenalty\@M
	\leavevmode
	\@tempdima 3.3em\relax
	\advance\leftskip \@tempdima \null\nobreak\hskip -\leftskip
	{\normalfont \hyperlink{autolof:\theautolof}{#1}\addtocounter{autolof}{1}}\nobreak{\normalfont\tocdotfill{4.5}}\nobreak\hb@xt@\@pnumwidth{\hfil\normalfont \hyperlink{pag:#2}{#2}}\par}%
	}

\renewcommand*{\l@table}[2]{%
	\vskip \z@ \@plus.2\p@
	{\leftskip 1.5em\relax
	\rightskip \@tocrmarg
	\parfillskip -\rightskip
	\parindent 1.5em\relax\@afterindenttrue
	\interlinepenalty\@M
	\leavevmode
	\@tempdima 2.3em\relax
	\advance\leftskip \@tempdima \null\nobreak\hskip -\leftskip
	{\normalfont \hyperlink{autolot:\theautolot}{#1}\addtocounter{autolot}{1}}\nobreak{\normalfont\tocdotfill{4.5}}\nobreak\hb@xt@\@pnumwidth{\hfil\normalfont \hyperlink{pag:#2}{#2}}\par}%
	}

\makeatother

\usepackage{fancyhdr}
\newcommand{\pagenumber}{\thepage}

\fancypagestyle{normal}
	{
	\fancyhf{} % erase all fields in header and footer
	\fancyhead[LE]{\mychapname \thechapter}
	\fancyhead[CE]{\nouppercase{{\sc\leftmark}}}
	\fancyhead[CO]{\nouppercase{\rightmark}}
	\fancyhead[R]{\linkup{pag:\thepage}} % target for toc
	\fancyfoot[LE,RO]{\pagenumber}
	}

\fancypagestyle{nochapnumber}
	{
	\fancyhf{}
	\fancyhead[CE]{\nouppercase{{\sc\leftmark}}}
	\fancyhead[CO]{\nouppercase{\rightmark}}
	\fancyhead[R]{\linkup{pag:\thepage}}
	\fancyfoot[LE,RO]{\pagenumber}
	}

\fancypagestyle{frontmatter}
	{
	\fancyhf{}
	\fancyhead[C]{\nouppercase{{\sc\leftmark}}}
	\fancyhead[R]{\linkup{pag:\thepage}}
	\fancyfoot[LE,RO]{\pagenumber}
	}
		
\fancypagestyle{plain} % redefine the plain style, used in new chapter pages.
	{
	\fancyhf{}
	\fancyhead[R]{\linkup{pag:\thepage}}
	\fancyfoot[LE,RO]{\pagenumber}
	}

\renewcommand{\chaptermark}[1]{\markboth{{#1}}{}} % redefine the content of the chapter mark used by fancyhdr (must be after fancyhdr first call)
\newcommand{\linkads}[2][]{\ifthenelse{\equal{\printerversion}{1}}{\hphantom{~{\footnotesize\href{#2}{\ifthenelse{\equal{#1}{}}{[ADS}{[#1}~\includegraphics[scale=0.6]{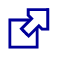}~]}}}}{~{\footnotesize\href{#2}{\ifthenelse{\equal{#1}{}}{[ADS}{[#1}~\includegraphics[scale=0.6]{figs/deco_link.eps}~]}}}}

\renewenvironment{thebibliography}[1]{
	\renewcommand{\bibname}{General bibliography} % set the title
	\markboth{\bibname}{} % set the headers
	\addcontentsline{toc}{chapter}{\bibname} % add to TOC (not done because of the section* change in natbib options)
	\cdp
	\pagestyle{frontmatter}
	\begin{oldthebibliography}{#1}
	\setlength{\columnsep}{40pt}
	\setlength{\parskip}{10pt}
	\setlength{\itemsep}{0pt}
	}{
	\end{oldthebibliography}}
\renewenvironment{thebibliography}[1]{
	\clearpage % star it on a new page (= no floats in the biblio)
	\renewcommand{\bibname}{References} % set the title
	\addcontentsline{toc}{section}{\bibname} % add to TOC
	\begin{oldthebibliography}{#1}
	\setlength{\parskip}{5pt}
	\setlength{\itemsep}{0pt}
	}{\end{oldthebibliography}}

\newcommand{\aj}{AJ}% Astronomical Journal
\newcommand{\araa}{ARA\&A}% Annual Review of Astron and Astrophys
\newcommand{\apj}{ApJ}% Astrophysical Journal
\newcommand{\apjl}{ApJ}% Astrophysical Journal, Letters
\newcommand{\apjs}{ApJS}% Astrophysical Journal, Supplement
\newcommand{\apss}{Ap\&SS}% Astrophysics and Space Science
\newcommand{\aap}{A\&A}% Astronomy and Astrophysics
\newcommand{\aapr}{A\&A~Rev.}% Astronomy and Astrophysics Reviews
\newcommand{\baas}{BAAS}% Bulletin of the AAS
\newcommand{\mnras}{MNRAS}% Monthly Notices of the RAS
\newcommand{\pasp}{PASP}% Publications of the ASP
\newcommand{\nat}{Nature}% Nature
\newcommand{\bain}{Bull.~Astron.~Inst.~Netherlands}% Bulletin Astronomical Institute of the Netherlands
\newcommand{\U}[1]{\ensuremath{\mathrm{~#1}}} % define unit
\newcommand{\erg}{\U{erg}}
\newcommand{\ergs}{\U{erg~s}}
\newcommand{\yr}{\U{yr}}
\newcommand{\Myr}{\U{Myr}}
\newcommand{\Gyr}{\U{Gyr}}
\newcommand{\pc}{\U{pc}}
\newcommand{\kpc}{\U{kpc}}
\newcommand{\Mpc}{\U{Mpc}}
\newcommand{\msun}{\U{M}_{\odot}}
\newcommand{\kms}{\U{km\ s^{-1}}}
\newcommand{\hi}{H{\sc i} }

\newcommand{\vect}[1]{\ensuremath{\boldsymbol{\mathrm{#1}}}}
\newcommand{\vectu}[1]{\ensuremath{\boldsymbol{\hat{\mathrm{#1}}}}}
\newcommand{\matr}[1]{\ensuremath{\mathbf{#1}}}

\newcommand{\chaphead}[1]
	{
	\begin{flushleft}
	\begin{picture}(0,0)
		\linethickness{1.5pt}
		\put(177,3){\line(-1,0){400}}
		\put(188,0){$\mathcal{O}$}
		\put(197,0){\emph{\textsf{verview}}}
		\put(285,3){\line(-1,0){40}}
		\qbezier(285,3)(300,3)(300,-12)
		\put(300,-12){\line(0,-1){12}}
	\end{picture}
	
	\begin{minipage}{300pt}
		\setlength{\arrayrulewidth}{1.5pt}
		\begin{tabular}{@{}p{280pt}@{}p{19.25pt}@{}|}
		#1 & \\
		\end{tabular}
	\end{minipage}

	\begin{picture}(0,0)
		\linethickness{1.5pt}
		\put(300,12){\line(0,1){12}}
		\qbezier(300,12)(300,-3)(285,-3)
		\put(285,-3){\line(-1,0){400}}
	\end{picture}
	\end{flushleft}
	\vspace{30pt}
	}

\begin{document}
\sloppy

\pagestyle{empty}
\begin{picture}(0,0)
\put(-10,-220){\includegraphics{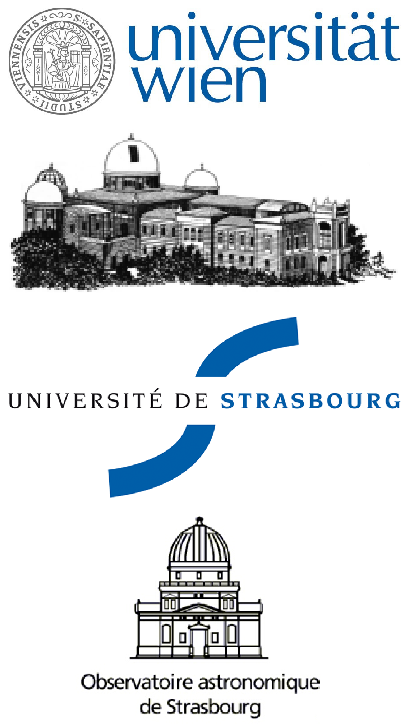}}
\end{picture}

\begin{flushright}
{\large
\vspace{-0.5cm}
Thesis presented to obtain the titles of\\
\vspace{0.3cm}
Doctor rerum naturalium\\
der Universit\"at Wien\\
\&\\
Doctor Philosophi\ae\\
de l'Universit\'e de Strasbourg\\

\vspace{0.3cm}

in Astrophysics\\}

\vspace{1.7cm}

{\Large by Florent \textsc{Renaud}}
\end{flushright}

\vspace{1cm}

\begin{center}
{\Huge Dynamics of the Tidal Fields and\\Formation of Star Clusters\\in Galaxy Mergers\\}
\end{center}

\vspace{1.5cm}

\begin{flushright}
{\small Defended on 16 July 2010 in Vienna, Austria}
\end{flushright}

\vspace{1.5cm}

\begin{center}
\begin{tabular}{@{}r@{\ \ }r@{ }ll}
%\arrayrulecolor{mycolorlogo}
\multicolumn{4}{@{}l}{{\large Examining board}}\\
\hline
President:&	Werner &\textsc{Zeilinger}&	Institut f\"ur Astronomie, Vienna \\
Referees:&	Thomas &\textsc{Lebzelter}&	Institut f\"ur Astronomie, Vienna \\
	&	Romain &\textsc{Teyssier}&	Institute for Theoretical Physics, Z\"urich\\
	&	Herv\'e &\textsc{Wozniak}&	Observatoire Astronomique, Strasbourg \\
Examiners:&	Pierre-A. &\textsc{Duc}&	SAP-CEA, Saclay \\
	&	Rainer &\textsc{Spurzem}&	Zentrum f\"ur Astronomie, Heidelberg \\
Supervisors:&	Christian &\textsc{Boily}&	Observatoire Astronomique, Strasbourg \\
	&	Christian &\textsc{Theis}&	Institut f\"ur Astronomie, Vienna \\
\hline
\end{tabular}
\end{center}

\cleardoublepage

\pagestyle{frontmatter}\setcounter{page}{1}\pagenumbering{roman}
\cdp
\chapter*{Abstracts}
\markboth{{\sc Abstracts}}{}

\vfill
%%%%%%%%%%%%%%%%%%%%%%%%%%%%%%%%%%%%%%%%%%%%%%%%%%
\section*{English abstract}

In interacting galaxies, strong tidal forces disturb the global morphology of the progenitors and give birth to the long stellar, gaseous and dusty tails often observed. In addition to this destructive effect, tidal forces can morph into a transient, protective setting called compressive mode. Such modes then shelter the matter in their midst by increasing its gravitational binding energy.

This thesis focuses on the study of this poorly known regime by quantifying its properties thanks to numerical and analytical tools applied to a spectacular merging system of two galaxies, commonly known as the Antennae galaxies. $N$-body simulations of this pair yield compressive modes in the regions where observations reveal a burst of star formation. Furthermore, characteristic time- and energy scales of these modes match well those of self-gravitating substructures such as star clusters and tidal dwarf galaxies. Comparisons with star formation rates derived from hydrodynamical runs confirm the correlation between the location of compressive modes and sites where star formation is likely to show enhanced activity. Altogether, these results suggest that the compressive modes of tidal fields plays an important role in the formation and evolution of young clusters, at least in a statistical sense, over a lapse of $\sim 10$ million years. Preliminary results from simulations of stellar associations highlight the importance of embedding the clusters in the evolving background galaxies to account precisely for their morphology and internal evolution. 

These conclusions have been extended to numerous configurations of interacting galaxies and remain robust to a variation of the main parameters that characterize a merger. We report however a clear anti-correlation between the importance of the compressive mode and the distance between the galaxies. Further studies including hydrodynamics are now underway and will help pin down the exact role of the compressive mode on the formation and later survival of star clusters. Early comparisons with such computations suggest that compressive modes act as catalysts or triggers of star formation.

\vfill
\newpage
\phantom{}\vfill
%%%%%%%%%%%%%%%%%%%%%%%%%%%%%%%%%%%%%%%%%%%%%%%%%%
\section*{R\'esum\'e en fran\c{c}ais}

Dans les galaxies en interaction, de colossales forces de mar\'ee perturbent la morphologie des prog\'eniteurs pour engendrer les longs bras d'\'etoiles, gaz et poussi\`eres que l'on observe parfois. En plus de leur effet destructeur, les forces de mar\'ee peuvent, dans certain cas, se placer dans une configuration protectrice appel\'ee mode compressif. De tels modes prot\`egent alors la mati\`ere en leur sein, en augmentant son \'energie de liaison.

Cette th\`ese se concentre sur l'\'etude de ce r\'egime peu connu en quantifiant ses propri\'et\'es gr\^ace \`a des outils num\'eriques et analytiques appliqu\'es \`a un spectaculaire syst\`eme de galaxies en fusion, commun\'ement appel\'e les Antennes. Des simulations $N$-corps de cette paire de galaxies montrent la pr\'esence de modes compressifs dans les r\'egions o\`u les observations r\'ev\`elent un sursaut de formation stellaire. De plus, les temps et \'energies caract\'eristiques de ces modes correspondent \`a ceux de la formation de sous-structures autogravitantes telles que des amas stellaires et des naines de mar\'ee. Des comparaisons avec les taux de formation stellaire d\'eriv\'es de simulations hydrodynamiques confirment la corr\'elation entre les positions des modes compressifs et les sites o\`u la formation des \'etoiles est certainement amplifi\'ee. Mis bout-\`a-bout, ces r\'esultats sugg\`erent que les modes compressifs des champs de mar\'ee jouent un role important dans la formation et l'\'evolution des jeunes amas, au moins d'un point de vue statistique, sur une \'echelle de temps de l'ordre de dix millions d'ann\'ees. Des r\'esultats pr\'eliminaires de simulations d'associations stellaires soulignent l'importance de plonger les amas dans leur environnement galactique en \'evolution, pour tenir compte pr\'ecis\'ement de leur morphologie et \'evolution interne.

Ces conclusions ont \'et\'e \'etendues \`a de nombreuses configurations d'interaction et restent robustes aux variations des principaux param\`etres caract\'erisant les paires de galaxies. Nous notons cependant une nette anti-corr\'elation entre l'importance du mode compressif et la distance entre ces galaxies. De nouvelles \'etudes incluant les aspects hydrodynamiques sont maintenant en cours et aideront \`a pr\'eciser le r\^ole exact du mode compressif dans la formation et la survie des amas d'\'etoiles. Les premi\`eres comparaisons avec de telles simulations sugg\`erent que les modes compressifs agissent en tant que catalyseurs ou amorces de la formation stellaire.

\vfill
\newpage
\phantom{}\vfill

%%%%%%%%%%%%%%%%%%%%%%%%%%%%%%%%%%%%%%%%%%%%%%%%%%
\section*{Deutsch Zusammenfassung}

Starke Gezeitenkr\"afte in wechselwirkenden Galaxien st\"oren die Morphologie dieser Systeme und f\"ordern die Entstehung ausgedehnter und h\"aufig beobachteter Filamente aus Sternen, Gas und Staub. Neben diesem zerst\"orerischen Effekt k\"onnen Gezeitenkr\"afte auch zu einem vor\"ubergehenden stabilisierenden Zustand, den sogenannten kompressiven Moden f\"uhren. Durch die Erh\"ohung der gravitativen Bindungsenergie der vorhandenen Materie wird diese dann von \"außeren gravitativen Einfl\"assen abgeschirmt.

Die vorliegende Arbeit besch\"aftigt sich mit diesen wenig untersuchten Zust\"anden durch Quantifizierung ihrer Eigenschaften mittels numerischer und analytischer Methoden, angewandt auf ein spektakul\"ares System verschmelzender Galaxien, bekannt als die Antennengalaxien. N-K\"orper Simulationen dieses Galaxienpaares ergeben kompressive Moden in denselben Regionen wo Beobachtungen eine erh\"ohte Sternentstehung  zeigen. Die charakteristischen Zeit- und Energieskalen dieser Moden \"ahneln au\ss{}erdem stark denen selbstgravitierender Substrukturen wie Sternhaufen oder sogenannten tidal dwarfs. Vergleiche mit Sternentstehungsraten aus hydrodynamischen Simulationen best\"atigen die Korrelation zwischen der Lage der kompressiven Moden und Bereichen mit erh\"ohter Sternentstehung. Zusammenfassend ist zu sagen, dass diese Resultate darauf hinweisen, dass kompressive Moden von Gezeitenfeldern statistisch betrachtet eine wichtige Rolle bei der Entstehung und Entwicklung junger Sternhaufen in einem Zeitraum von 10 Millionen Jahren spielen. Vorl\"aufige Resultate von Simulationen stellarer Assoziationen zeigen die Bedeutung der Einbettung dieser Haufen in die sich entwickelnden Muttergalaxien, um deren Morphologie und interne Entwicklung zu begr\"unden.

Diese Schlussfolgerungen wurden auf zahlreiche Konfigurationen wechselwirkender Galaxien erweitert und liefern bei Variation der charakteristischen Parameter verschmelzender Galaxien das gleiche Ergebnis. Es ist jedoch eine klare Anti-Korrelation zwischen der Wichtigkeit der kompressiven Moden und dem Abstand der Galaxien zueinander zu erkennen. Weitere hydrodynamische Studien sind nun in vollem Gange und werden dazu beitragen, den genauen Einfluss der kompressiven Moden auf die Entstehung und dem sp\"aterem \"Uberleben von Sternhaufen festzulegen. Aktuelle Vergleiche mit solchen Berechnungen weisen darauf hin, dass kompressive Moden als Katalysatoren oder Ausl\"oser von Sternentstehung angesehen werden k\"onnen.

\vfill
\cdp
\chapter*{Preamble}
\markboth{{\sc Preamble}}{}

%%%%%%%%%%%%%%%%%%%%%%%%%%%%%%%%%%%%%%%%%%%%%%%%%%
\section*{Prologue}

One of the particularities of this PhD is that it has been co-supervised. Indeed, I worked with two different advisors in two different institutes in two different countries. Advantages of such an organization are indubitably the richness of the ideas and advices I collected from both sides. Of course, sometimes I had to deal with points of view that may be incompatible, but generally, this helped me to learn how to make decisions and to stand up for them.

\begin{figure}[h]
\begin{center}
\includegraphics{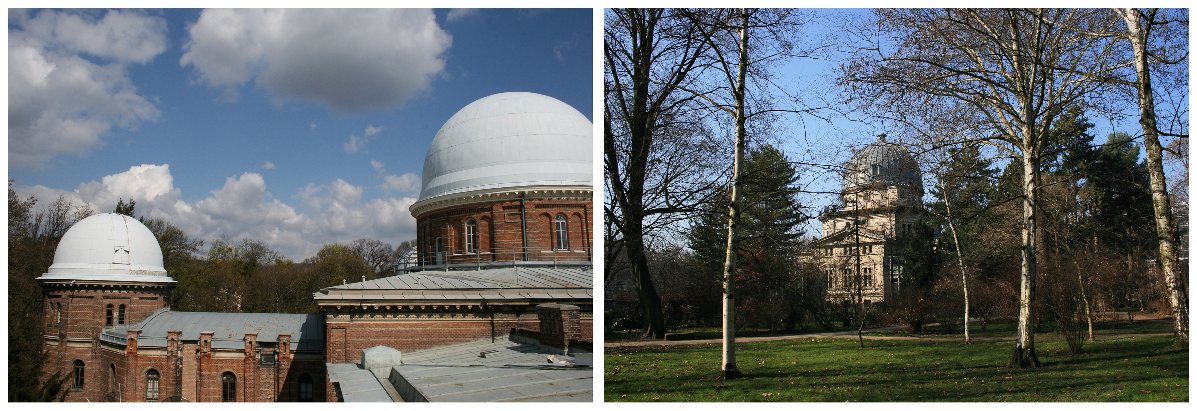}\\
\emph{Left:} Das Institut f\"ur Astronomie der Universit\"at Wien (\"Osterreich).\\\emph{Right:} L'Observatoire Astronomique de Strasbourg (France).
\end{center}
\end{figure}

This also requires much more financial means than a classical PhD and a good flexibility in the schedule. Fortunately for me, both Institutes and Universities supported this collaboration and provided the resources to make it a wonderful experience.

The work presented here begun within the framework of a previous PhD thesis ``Approche num\'erique de la dynamique et de l'\'evolution stellaires appliqu\'ees \`a la fusion galactique''\footnote{``Numerical approach of stellar dynamic and evolution applied to galactic mergers''. The manuscript (in french) can be downloaded: \href{http://tel.archives-ouvertes.fr/tel-00184822}{http://tel.archives-ouvertes.fr/tel-00184822\ifthenelse{\equal{\printerversion}{1}}{}{~\includegraphics[scale=0.6]{figs/deco_link.eps}}}}, led by Jean-Julien Fleck under the supervision of Christian Boily at the Observatoire de Strasbourg (2004-2007). Their pioneer methods have been optimized to extend their conclusions about the tidal field and the star clusters.

%%%%%%%%%%%%%%%%%%%%%%%%%%%%%%%%%%%%%%%%%%%%%%%%%%
\section*{How to read this document}

This manuscript aims to present the scientific motivations, methods, results, questions answered and questions raised during my PhD. It has been conceived to introduce this work but also to archive the results and the mathematical derivations. That is, the General Introduction gives a broad review about the major topics developed in the next Chapters, while the Appendices gather technical and analytical work, and give some hints for the derivation of several equations.

\mybox{
Supplementary ``boxes'' like this one give more details on technical and/or mathematical points. They can be skipped during the first reading.
}

In the electronic version of this manuscript, hyperlinks point to all referenced figures, tables, equations and sections. Within a citation, clicking on the year will direct you to the General Bibliography where you will find a link to the ADS page (which will open in you browser) of the publication. In addition, it is possible to click on the acronyms, and go where they have been defined for the first time. In this (illustrative) example: 
\begin{center}
\vspace{-12pt}
\begin{minipage}{10cm}
Once you have seen \reffigp{antennae_now}, you may be interested in looking at \reftab{antennae} and find out more about the \acr{notarg}{SFE}{star formation efficiency} in \refsec[formation]{sfe_massloss} or in \mycitet{Renaud2009}.
\end{minipage}
\end{center}
\vspace{-12pt}
you can click on ``Figure~\ref{fig:antennae_now}'', ``Table~\ref{tab:antennae}'', ``SFE'', ``Section~\ref{cha:formation}.\ref{sec:sfe_massloss}'' and ``2009''. Don't forget to use the {\tt Back} function of your favorite PDF reader to go back to the link!

%All over the present document, the following mathematical notations have been employed:
%\begin{center}
%\begin{minipage}{14cm}
%\begin{tabular*}{0.45\linewidth}{c@{\extracolsep{\fill}}l}
%$G$ 		& gravitational constant\\
%$M$		& mass\\
%$\rho$		& density\\
%$\phi$		& gravitational potential\\
%$\matr{T}$	& tidal tensor\\
%$E$		& energy\\
%$\Omega$	& potential energy\\
%$K$		& kinetic energy\\
%$R$		& radius\\
%\end{tabular*}
%\begin{tabular*}{0.45\linewidth}{|c@{\extracolsep{\fill}}l}
%$r$		& radial coordinate\\
%$\vect{r}$	& position vector \\
%$v$		& velocity\\
%$\gamma$	& acceleration\\
%$t$		& time\\
%$\delta^{ij}$	& =1 if $i=j$, 0 else\\
%$\vectu{e}$	& unit vector\\
%$\sigma$	& dispersion\\
%$\epsilon$	& star formation efficiency,\\
%		& smoothing length\\
%$e$		& eccentricity\\
%\end{tabular*}
%\end{minipage}
%\end{center}

\vspace{1cm}
\begin{flushright}
\emph{The author\\
somewhere in the Alps, between Vienna and Strasbourg\\
Wednesday, 17 February 2010 (1:15 am)}

\end{flushright}

\cdp
\chapter*{Acknowledgments}
\markboth{{\sc Acknowledgments}}{}

One of the unknown aspects of being a co-supervised PhD student, is that the ``Acknowledgments'' part of your manuscript could easily be longer than any other chapter.

In 2005, I was looking for a two-months internship in an institute, somewhere in France, to work in the vague field of astrophysics. Christian Boily welcomed me that summer and taught me patiently all the basic knowledge about galaxies, mergers and numerical simulations. On the last day of this short study, I met Christian Theis who was visiting us in Strasbourg, and who showed a great interest in my modest contribution. Two years later, as I was in California fighting with a computer that did not want to reproduce Stephan's Quintet, I asked Christian (Boily... oh boy! why do they share the same first-name?) about having the opportunity to continue in this fascinating domain he had introduced me to. This is when Christian suggested to work with Christian. At that time, I was not expecting that my first act as a young scientist would be to give them nicknames, to stop this constant confusion.

Christian Boily, aka CMB (what a name for an astrophysicist, don't you think?), you have taught me so such about so many topics (not always related to the thesis...) that nobody could count the number of whiteboard pens you wore out. You have been so kind and patient with me when I was asking thousands of (not always clever) questions, you helped me to solve so many problems in science and life that, more than being a mentor, you became a friend.

Christian Theis, first of all, you welcomed me in Vienna and always made me feel as part of a team, even when my german skills had reached their limits with ``Ein gro\ss{} Bier bitte''. You gave me so many good advices and very good input in my work and life, day after day. Your wisdom taught me a lot during these years. You gave me the freedom I needed to make my own mistakes and learn from them. I hope that we will be able to continue to work together, making breathtaking movies that will be showed in planetariums all around the world!

With your very different styles of supervision, Christian and Christian, you have formed the best combination of advisors I could have imagined when I applied to this PhD. Your approaches have been complementary and you have allowed me to pick up the best from both. You truly have transmitted your passion to me.

And because a PhD is much more than a student and his advisors, I would like to say a big thank you ...

... to the members of the jury of this thesis who have kindly accepted to read the present manuscript and judge my work. Your reports have all been very constructive.

... to the people that I have met here and there during visits and meetings, with whom I had great chats and, for some of them, began nice collaborations. Thanks in particular to 
%Clare~D., Curt~S., Damien~C., Fr\'ed\'eric~B., Genevi\`eve~P., \hbox{Jeong-Sun~H.}, Kevin~X., Mark~G., Phil~A., Simon~K. and Thorsten~N. for their active support.
P.~Appleton, F.~Bournaud, D.~Chapon, C.~Dobbs, M.~Gieles, J.-S.~Hwang, S.~Karl, T.~Naab, G.~Parmentier, C.~Struck and K.~Xu for their active support.

... to all the persons who provide us with high quality codes and tools. Special thanks to S.~Aarseth, W.~Dehnen, V.~Springel, R.~Spurzem, P. Teuben, R. Teyssier and the teams of the Astrophysics Data System (ADS).

%... to the teams of the NASA's Astrophysics Data System Bibliographic Services who help us everyday to save time and focus on science.

... to the Viennese people: Jeannette, Katharina, Laura and Milada, going through the Austrian paperwork would not have been possible without the helpful hands you gave me. And for the nice discussions and the fun we had in restaurants, our offices or around the (by now traditional) pizzas, many thanks to Adam, Ana, Angela, Armin, Bastian, Christoph, Deiter, Denise, Gerhard, Hanns, Harsha, Hossein, Ingo, Julia, Markus, Mikola, Pedro, Simone, St\'ephane, Sylvia, Thomas, Verena and Werner. D\u{e}kuji, gracias, thank you, \begin{picture}(26,0)\put(0,-4){\includegraphics[scale=1.2]{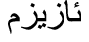}}\end{picture}, \begin{picture}(40,0)\put(0,-1.5){\includegraphics[scale=1.2]{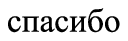}}\end{picture}, obrigado, grazie, merci, danke ihr allen! Special thanks to Hannes and Paul for the continuous\begin{picture}(0,0)\put(3,2){\line(1,0){40}}\end{picture} support fun they provided.

... \`a mes amis Strasbourgeois qui, \`a d\'efaut de m'avoir accueilli dans cette belle ville que je connaissais d\'ej\`a, ont \'egay\'e mes journ\'ees et m'ont enrichi de leur savoirs. Pour avoir r\'esolu un \`a un tous les petits probl\`emes, un grand merci \`a Estelle, Jean-Yves, Sandrine et Thomas. Un merci amical aux habitants du canap\'e vert, amateurs de caf\'e ou de picnic sur les bancs du parc; en particulier Ariane, Caroline (avec qui j'ai d\'ecouvert les myst\`eres de la poussi\`ere et des soustractions agricoles), Dominique et Rodrigo. Enfin, merci aux adeptes du BigBoss et des gardes-contre, pour leur bonne hummeur: Alexandre, Alexis, Anjali, Brent, Brice, Ciro, Cyril, Fabien, FX, George, Jean-Julien, Lionel, Maxime, Nicolas et Stéphane.

And last, but not least... Impossible pour moi d'oublier celui qui, adepte du cumul des mandats, a \'et\'e mon coll\`egue, voisin, projectionniste, logeur, chauffeur, psy, relecteur, voiturier, conseiller,  voyagiste, serveur, banquier, D.J., co-pilote, secr\'etaire, confident, cuisinier ... et qui restera mon meilleur ami. Morgan, tu as partag\'e mes fous-rires et mes d\'elires, et tu as su m'apporter un soutient ind\'efectible tout en restant un ami sinc\`ere. Au moment o\`u tu d\'ecouvres ces lignes, tu dois avoir la t\^ete pleine de r\'ef\'erences bibliographiques, titres de chapitres, figures et sch\'emas. Alors, pour la fin de ta th\`ese et ta carri\`ere en g\'en\'eral, je te souhaite tout le succ\`es que tu m\'erites, en attendant la publication du Fouesneau \& Renaud 201x dont nous avons si souvent parl\'e en lui attribuant un taux de citations ... astronomique~!
\tableofcontents\cdp\listoffigures\cdp\listoftables\cdp

%%%
\setcounter{page}{1}\pagenumbering{arabic}
\chanonumber{General Introduction}
\chaphead{This Chapter introduces briefly the historical ideas of galaxy mergers and star clusters, and the physical concepts we will refer to in the next parts.}

%%%%%%%%%%%%%%%%%%%%%%%%%%%%%%%%%%%%%%%%%%%%%%%%%%
\sect{Galaxy mergers}

%%%%%%%%%%%
\subsect{How to classify the unclassified?}
For a long time after their discovery, galaxies were thought to be clouds or nebulae, named island universes, suggesting that they were isolated and without any interaction (see e.g. \mycitealt{Hubble1929}). From Edwin Hubble (\myciteyear{Hubble1936}) to the recent Galaxy Zoo project (\mycitealt{Raddick2007}), an important body of work has aimed at classifying the galaxies according to their morphology, on the sky. Hubble's classification (or Hubble's pitch fork diagram, see \reffig{hubble_fork}) distinguishes between two major families: the ellipticals and the spirals.

\fig{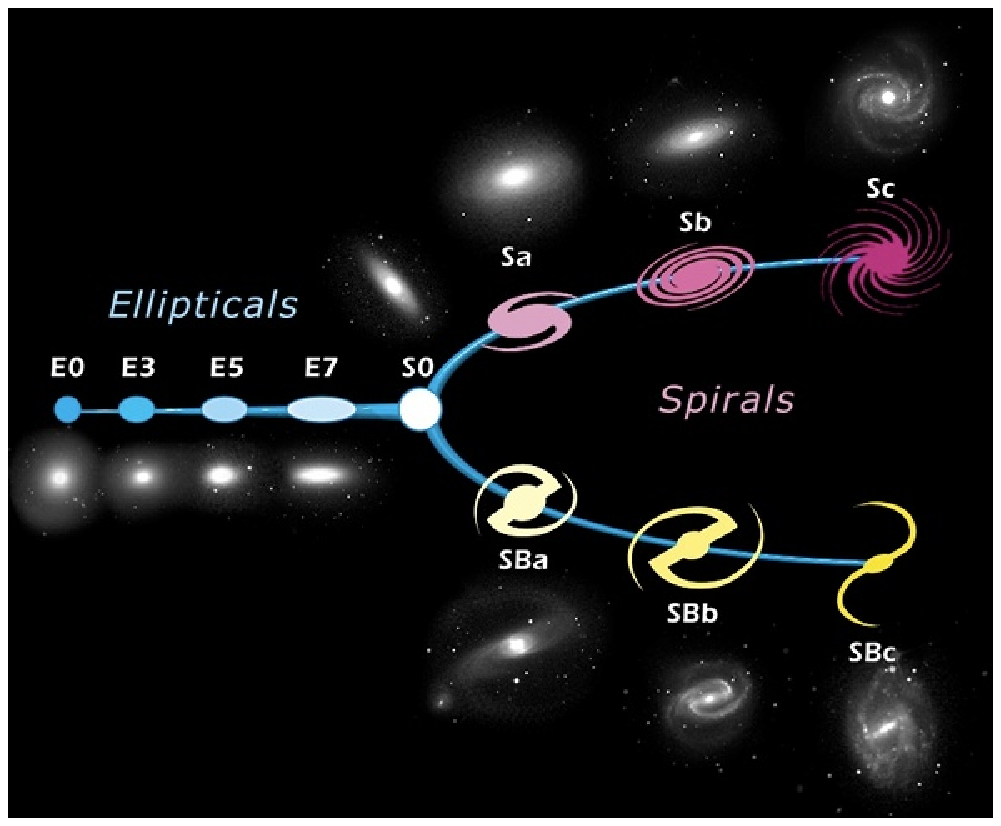}{Hubble's classification of galaxies}
{Hubble's classification of galaxies associates one or two letters for each family (E for the elliptical, S for the spirals, SB for the barred spirals) and, either a number to quantify either their axis ratio (ellipticals) or a letter for the opening angle of their spiral arms. (Picture adapted from Hubble Heritage Project website.)}

This nomenclature has been very efficient for most of the galaxies and is still used nowadays. However, as the observational techniques and accuracy increased with time, more and more galaxies did not fit into this classification. Many peculiar morphologies showing extended arms lead to uncertainties, but it appears that a lot of them could be multiple galaxies, so closely packed that they would affect each other's shape.

It rapidly became clear that the ``island universes'' were far from being isolated, as they are often associated in groups and clusters. Indeed, the environment plays a crucial role on the morphology of these galaxies. As two (or more) of them move close to each other, the gravitation brings them even closer and, if their relative velocity is low enough (i.e. less than the escape velocity), they finally merge into one single galaxy. Observations of various pairs of interacting galaxies suggested an evolutionary sequence (often called the Toomre sequence; see \reffig{toomre_sequence}), in which each step represents a well defined dynamical stage (\mycitealt{Toomre1977}; \mycitealt{Joseph1985}; \mycitealt{Hibbard1996}; \mycitealt{Laine2003}).

\fig{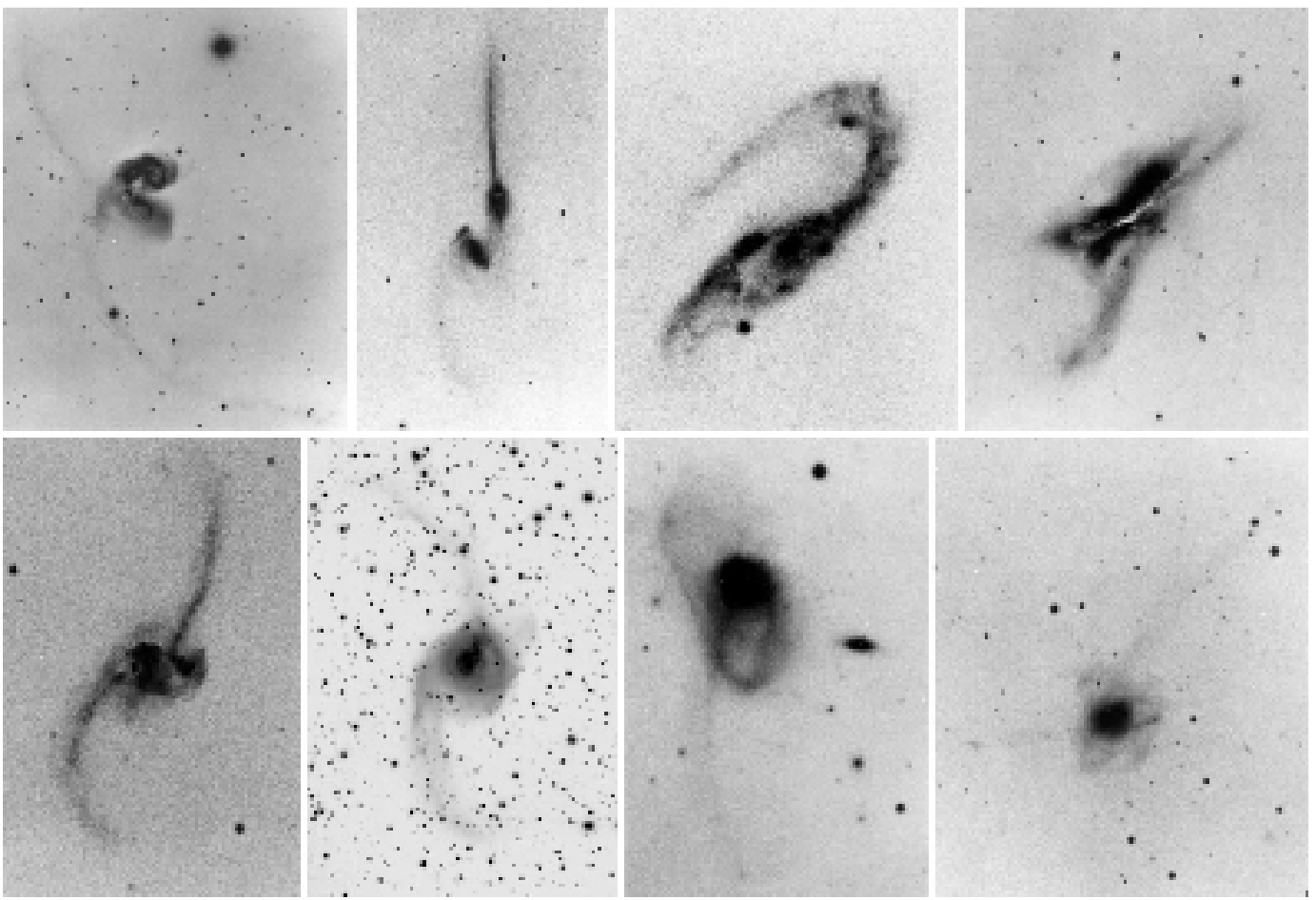}{Toomre's sequence}
{The Tommre sequence (\mycitealt{Toomre1977}) represents the supposed evolution of interacting galaxies. It starts with the early phases, when progenitors have just begun to interact, shows intermediate stages and finishes with the merger phase. From left to right, \emph{top}: NGC~4038/39 (the Antennae), NGC~4676 (the Mice), NGC~3509, NGC~520; \emph{bottom}: NGC~2623, NGC~3256, NGC~3921, NGC~7252. (Images from the Arp Atlas, montage by J. Hibbard.)}

Many phenomena are involved in such a process and thus mergers are wonderful laboratories for the astrophysicist. Among these phenomena, one counts tides, dynamical friction, shocks, star formation... An outstanding goal of this thesis is to weave them together to paint a more accurate picture of mergers. 

%%%%%%%%%%%%%%%%%%%%%%%
\subsect{A telescope meets a computer}

Obviously, the dynamics of such systems is of prime importance to understand the history of these galaxies. Our knowledge of the motion of the progenitor galaxies and their initial morphology has to be as accurate as possible. On the one hand, the only information we dispose of is the integrated light of the galaxies, as viewed form the Earth. Therefore, we miss the third dimension. On the other hand, the radial velocity can be measured, and gives us an idea about the motion along the line of sight, but not along the other directions. Furthermore, even if astronomy has been one of the first sciences in the history of mankind, accurate data has been collected over a very short period (few centuries) in regard to the timescales of galaxy mergers. And thus, we also miss the fourth dimension: time.

Therefore the challenge of understanding interacting galaxies consists in understanding phenomena that have occurred far away, over a very long time, just by getting snapshots of the objects, from our perspective, at the present time. The role of observers is to provide the theoreticians with as rich data as possible, through photometry and spectroscopy. Then, we build models that (partially) reproduces the physics and try to interpret the observations. Whereas the models will \emph{never} be perfect, they may allow to explore (up to a certain level) the unreachable quantities one needs to fully understand an object.

The first step would be an analytical job. By making strong assumptions, one can mathematically derive some values of interest for simple models. However, the complexity of the problem often forces the theoretician to make unrealistic hypotheses to simplify the mathematical developments. In the era of computers, we have a new way to solve the problems. The laws of physics that have been tested in laboratories are implemented as numerical treatments and applied to a set of initial conditions. The solution one obtains is no longer analytical, but numerical. At this point, the complexity becomes a second order problem while we have to face the precision versus speed issue. Indeed, once the physical laws have been implemented on the computer, the accuracy of the results is directly linked with the precision and the resolution of the computation. Usually, both can be increased at the cost of a longer calculation. \refcha{numerical} describes the numerical techniques used during this thesis and details the tests performed to validate the results.

With telescopes providing data of higher and higher resolution, the theoreticians must adapt the techniques they implement and use more and more state-of-the-art (super)computers to recover the physics of smaller and smaller systems.

%%%%%%%%%%%%%%%%%%%%
\subsect[lightbulbs]{When stars were light-bulbs}

In the case of galaxy mergers, simplifying assumptions are difficult to make: the distribution of matter offers no plane of symmetry, the masses are not negligible and the Keplerian laws of motion do not apply. All the usual simplifications are not possible because of the complexity level of the problem. Therefore, the use of simulations rapidly appeared vital to understand these systems.

\fig{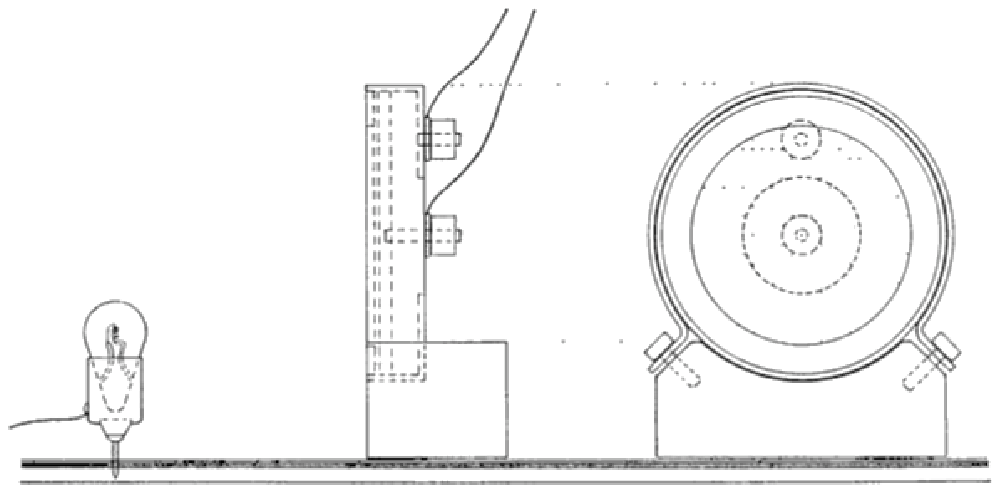}{Holmberg's first simulation}
{A light bulb and a photocell used by Holmberg as source and measurement tool of his proxy for gravitation. (Figure~1 of \mycitealt{Holmberg1941}.)}

For his first attempt, \mycitet{Holmberg1941} did not have computers. But he knew that both electromagnetic and gravitational interactions scale with the distance as $r^{-2}$. So he used 37 light-bulbs to mimic gravitational sources and some galvanometers (see \reffig{holmberg}) to measure the total amount of light at various points, i.e. the proxy for the gravitational force. Therefore, he was able to determine the direction and intensity of his pseudo gravitational force and then move the sources according to the equations of motion. 

Two decades later, the first digital computer simulations used 100 particles to reproduce stellar systems in equilibrium to validate the integrators (\mycitealt{vonHoerner1960}; \mycitealt{Pfleiderer1961}; \mycitealt{Aarseth1963}; \mycitealt{Eneev1973}) but the major breakthrough occurred in the seventies with the work of \mycitet{Toomre1972}. They set disks of particles to reproduce two disk galaxies. Each galaxy consists in a point mass surrounded by test (i.e. mass-less) particles. Therefore, only the two-body problem is to be solved, which can be done analytically through Kepler's laws of motion. This approach, called restricted simulation, allows to consider the effect of the potential (created by two bodies only) on a large number of test particles that do not influence the gravitational potential well. This gives an overview of the evolution of the system without costly computation.

\fig{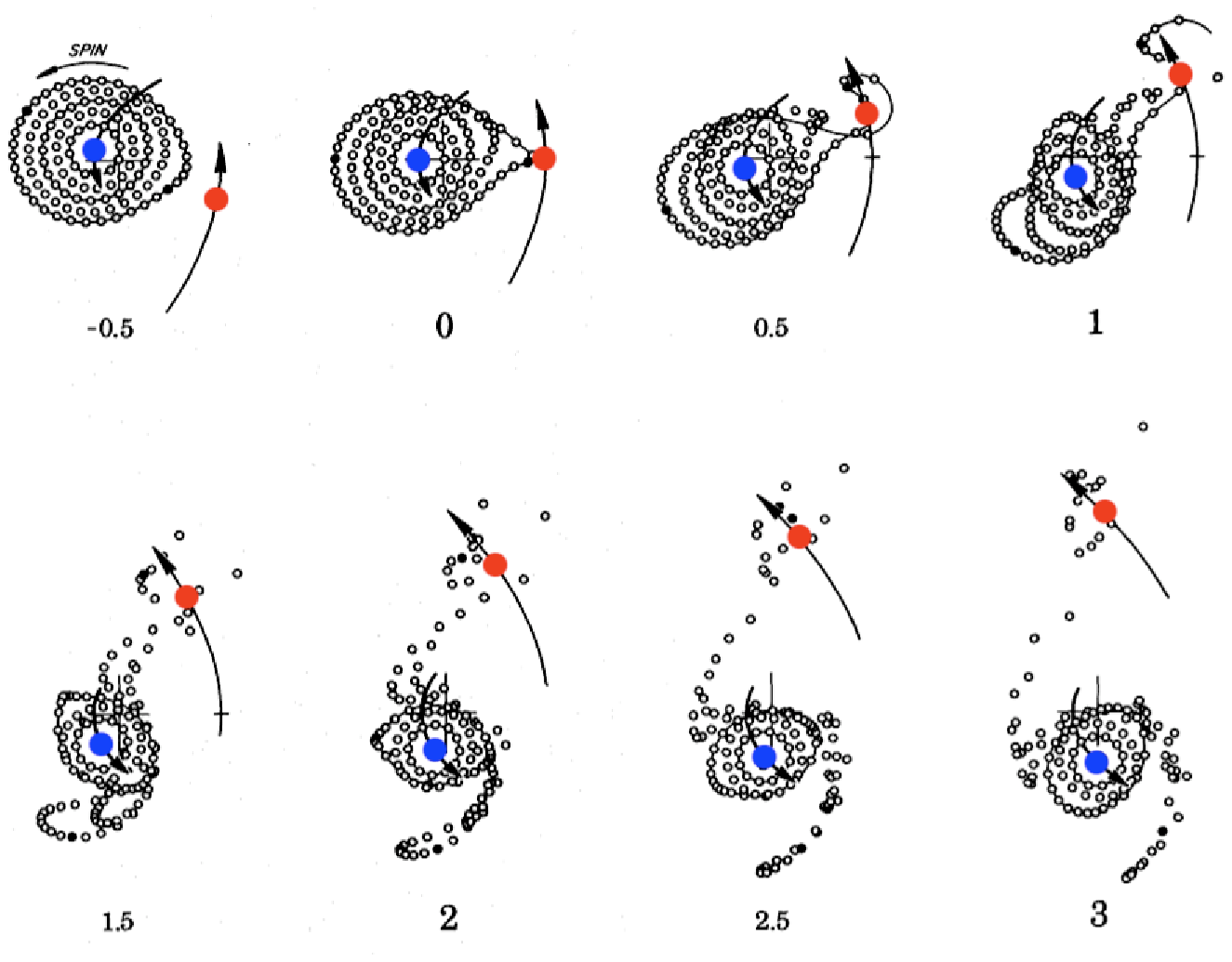}{Toomre \& Toomre simulations}
{Parabolic passage of one massive companion (red) near another galaxy (blue) surrounded by 120 test particles (empty circles). The mass of the companion is one quarter of that of the primary. The numbers indicate the time in numerical units, with $t = 0$ for the pericenter. (Adaptation of Figure~4 of \mycitealt{Toomre1972}.)}

Using several different configurations, \mycitet{Toomre1972} characterized the role of many parameters of the merger (like e.g. the mass ratio, the orientation of the spin of the progenitors...), and clearly showed that the perturbation of the gravitational potential of a disk by a point-mass intruder creates a bridge of test particles between the two galaxies and a tail in the opposite side of the disk, as shown in \reffig{toomre}.

These features are very common in merging galaxies, and not only there...

%%%%%%%%%%%%%%%%%%%%%%%%%%%%%%%%%%%%%%%%%%%
\sect{Tides}

%%%%%%%%%%%%%%%%%
\subsect{On Earth as in Heaven}
Twice a day, a roll of water comes to the coast and makes the level of the sea go up and down (see \reffig{fundy}). Everyone knows that this physical phenomenon, called tide, is mainly due to the Moon but only a few understands it. The oceanic tides should be considered as a combination of many effects, including the attraction of the Moon, but also, the one from the Sun, the oceanic currents, the geometry of the coast and the bottom of the sea. Among them, the gravitational attraction of the Moon and the Sun plays a similar role to what occurs in galaxy mergers.

\fig{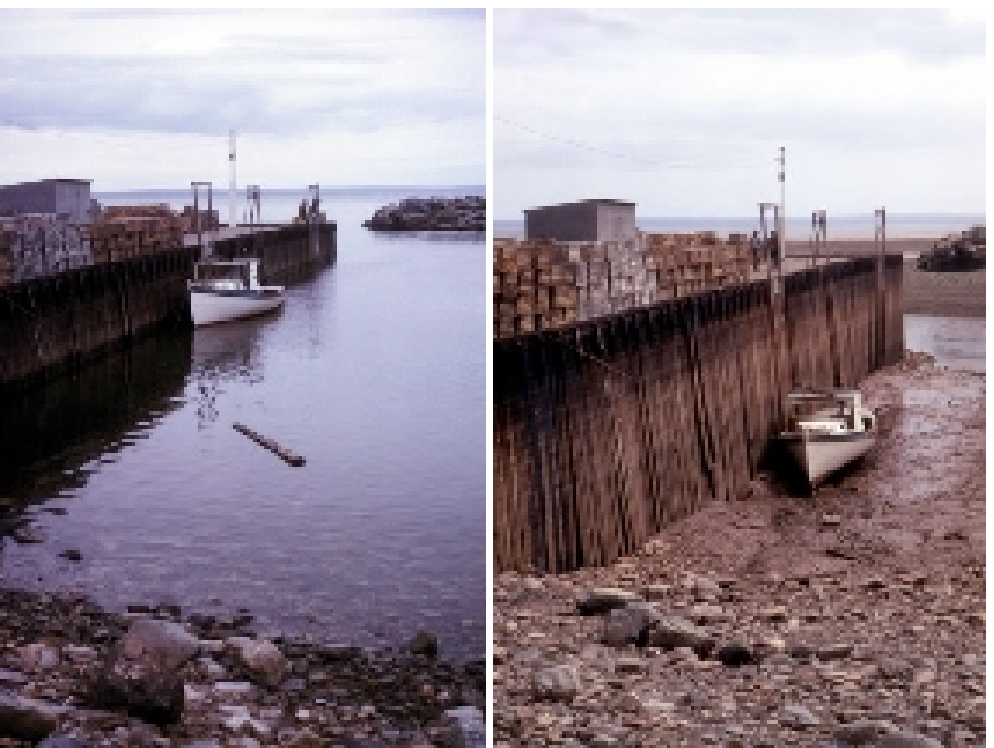}{Bay of Fundy}
{On the Earth, the gravitational potential of (mainly) the Moon deforms the shape of the oceans. According to the position of the Earth-Moon axis, we can experience high tide (\emph{left}) or low tide (\emph{right}), like in the bay of Fundy (between New Brunswick and Nova Scotia; photos by S. Wantman).}

So, what happens? Let's concentrate on the Earth and the Moon. The gravitation (as described by Newton) between the Moon and an element of the seas acts proportionally to their masses ($M$ and $m$) and the inverse square of their distance $r$:
\eqn{
F = G\frac{M m}{r^2}.
}
Consider five points (A to E) on the Earth (\reffig{earthmoon}). The force on a given mass element is stronger in A than at any other point while the one in C is the weakest. In the reference frame of the Earth, one has to subtract the force at the center of the Earth (in E). The differential force one obtains acts ``outward'' along the Earth-Moon axis (A and C), and "inward" along the perpendicular axis (B and D). That is why the seas are ``pulled away'' from the Earth in A and C, which makes for ``high tide'', while B and D are places of ``low tides''.

\fig{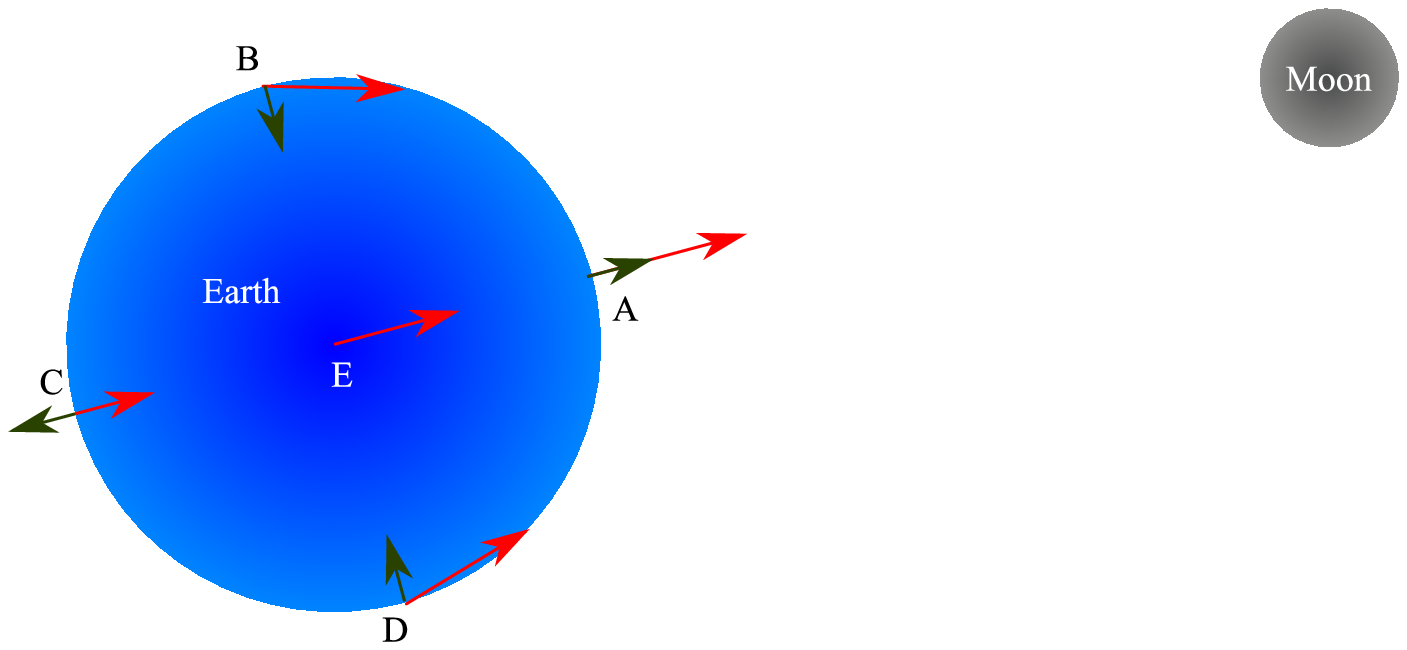}{Earth-Moon tides}
{The attraction of the Moon (red arrows) is stronger in A than in E and C. The differential force (green arrows) is therefore opposite in A and in C.}

This effect also applies to the atmosphere and the continents (\mycitealt{Lindzen1969}). In the last case, the inertia and the friction make it almost invisible. Because the disturbance is less important than the binding energy of the Earth, the planet remains stable and binds its oceans, continents and atmosphere. Obviously, the Earth has a similar effect on the Moon.

The same effect exists for much larger systems. In the case of galaxies, the material (stars, gas, dust ...) is bound by the gravitational potential of the galaxy itself. By modifying this potential, an external mass can disturb the equilibrium. The binding energy of galaxies is much smaller than that of planets, thus a perturbation like the close passage of an intruder galaxy can strip off their material over large distances. This explains the long plumes we observe in mergers.

%%%%%%%%%%%%%%%%%%%%%%%%
\subsect[prograderetrograde]{Prograde or retrograde, that is the question}
The mass ratio and pericenter distance of the two galaxies define the strength of the potential of the intruder with respect to the potential of the main galaxy. However, the tidal effect would be more efficient if it lasts for a long time. The duration of the perturbation corresponds to the time a mass element of the galaxy A sees the potential of the galaxy B, and thus is linked to the relative velocity of the two progenitors.

For a given value of the velocity (which depends on the orbits of the galaxies), two cases can happen. In the reference frame of the galaxy A, the internal angular momentum (i.e. the spin) of B can be aligned or anti-aligned with its orbital momentum. In the first case, called prograde encounter, the relative velocity of the material of B close to the galaxy A is low, and thus, the tidal effect of A on B acts longer. \emph{A contrario}, during a retrograde encounter, this velocity is higher and the effect shorter. 

\mybox[pro_retrograde]{
In the reference frame of the galaxy A, the orbital motion of the galaxy B, which has a rotational velocity $\Omega$, is characterized by a linear velocity
\eqn{
\vect{v}_\mathrm{orb} = r_\mathrm{AB} \ \vectu{e}_r \times \Omega \ \vectu{e}_z.
}
For simplicity, let's assume that the internal and orbital motions of B are coplanar, i.e. both rotational vectors are (anti-)aligned with $\vectu{e}_z$.
\begin{center}
\includegraphics{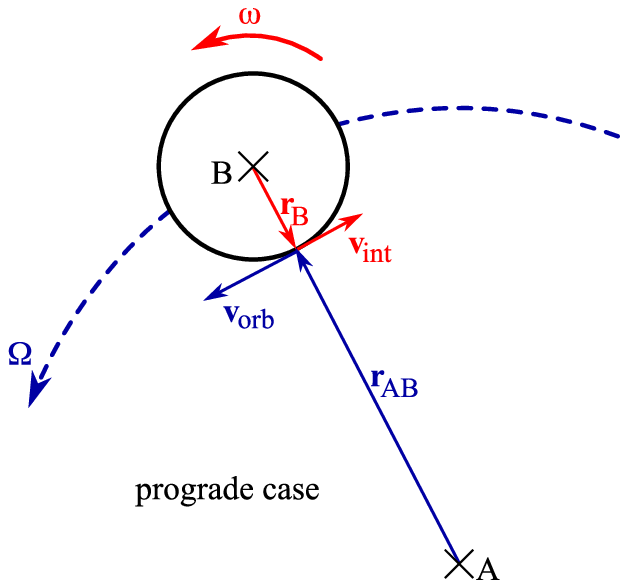}
\end{center}
Therefore, the vector of the internal rotation can either be $+\omega \vectu{e}_z$, or $-\omega \vectu{e}_z$. Thus, the velocity of an element of mass of the galaxy B, situated at a distance $\vect{r}_\mathrm{B}$ relative to the center of its host galaxy is
\eqn{
\vect{v}_\mathrm{int} = \left( - r_\mathrm{B} \ \vectu{e}_r \right) \times \left(\pm \omega \ \vectu{e}_z\right),
}
\follow}\mybox[nocnt]{
\noindent and the norm of total relative velocity reads
\eqn{
\left| \vect{v}_\mathrm{orb} + \vect{v}_\mathrm{int}\right| = r_\mathrm{AB} \ \Omega \mp r_\mathrm{B} \ \omega,
}
where the sign of the second term depends on the alignment of $\vect{\omega}$ with $\vectu{e}_z$, i.e. with $\vect{\Omega}$. For a prograde encounter, the spin ($\vect{\omega}$) and the orbital motion ($\vect{\Omega}$) are coupled (i.e. aligned). Therefore, the relative velocity is lower ($r_\mathrm{AB} \ \Omega - r_\mathrm{B} \ \omega$) than for a retrograde encounter ($r_\mathrm{AB} \ \Omega + r_\mathrm{B} \ \omega$).
}

As the result of the encounter, the tidal structures would be much more pronounced in a prograde case, as illustrated in \reffig{proretrograde}.

\fig{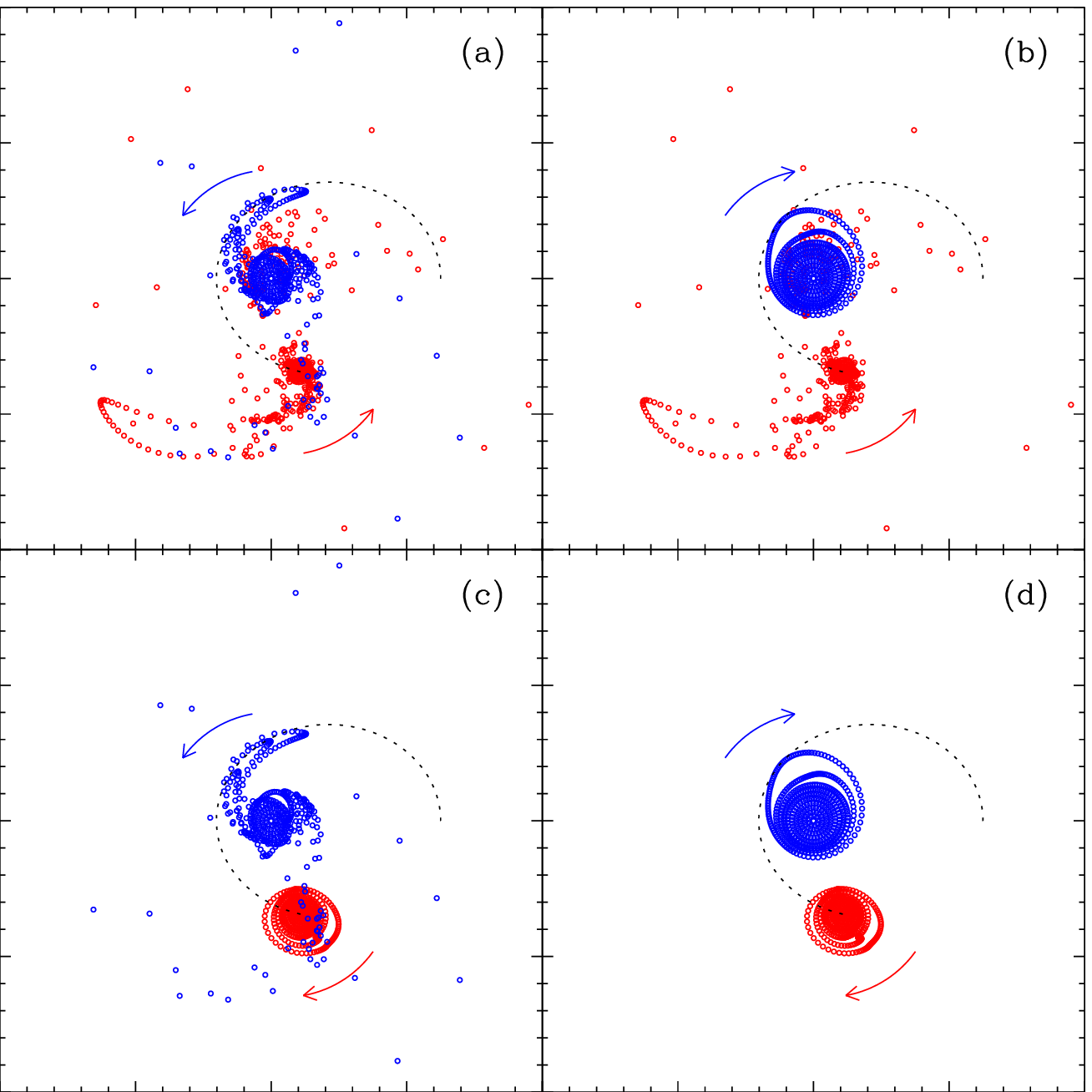}{Prograde vs. retrograde encounters}
{Toomre-like restricted simulations of prograde-prograde (a), retrograde-prograde (b), prograde-retrograde (c) and retrograde-retrograde (d) encounters between a major galaxy (blue) and an intruder (red). The mass ratio is 1:4; each galaxy is made of 500 particles; the intruder is set on an eccentric orbit (black dotted line). Arrows indicate the spin of the two disks. The orbital angular momentum is the same in all configurations. Large tidal features originating from a disk are only visible in the prograde case while a retrograde encounter leaves the galaxy fairly unaffected.}

Simple arguments allows to predict the global behavior of a pair of interacting galaxies. However, the precise, fine description of the tidal field requires some mathematical definitions.

%%%%%%%%%%%%%%%%%%%%%%%%
\subsect{Tidal tensor}
In the previous paragraphs, we have seen that tides arise from the gravitational attraction of a body on a spatially extended object. Hence, it is natural to introduce the spatial derivative of the force to characterize them. Let's consider, for example, a small galaxy D (``dwarf'') orbiting around a bigger, spherical one\footnote{Cosmological simulations show that this situation is extremely frequent and may be a fundamental step of the creation of the galaxies, within the hierarchical scenario. In the nearby Universe, the Magellanic clouds orbiting around the Milky Way are an example.} noted MW. According to Newton's second theorem, we can consider MW as a point mass, situated at its center of mass. The net acceleration $\vect{a}$ of a star $\alpha$ of the dwarf is the sum of the external effect of MW at its position $\vect{r}_\alpha$, and the one of all other stars in D:
\eqn{
\vect{a}_\alpha = \vect{a}_\mathrm{ext} (\vect{r}_\alpha) + \sum_{\beta \neq \alpha} \vect{a}_\beta(\vect{r}_\alpha).
}
Because the center of mass of D undergoes only the acceleration of MW, we can reproduce the differential scheme we used for the Earth (\reffig{earthmoon}) by switching to the reference frame of the dwarf galaxy:
\eqn{
\vect{a}_{\alpha,\mathrm{D}} = \left[ \vect{a}_\mathrm{ext}(\vect{r}_\alpha) - \vect{a}_\mathrm{ext}(\vect{r}_\mathrm{D}) \right] + \sum_{\beta \neq \alpha} \vect{a}_\beta(\vect{r}_\alpha).
}
We note that the distance $\vect{\delta} = \vect{r}_\alpha - \vect{r}_\mathrm{D}$ between $\alpha$ and the center of D is much smaller than the distance to MW. Therefore, one can linearize the previous expression and get
\eqn[deltatide]{
\vect{a}_{\alpha,\mathrm{D}} = \vect{\delta}\ \partial \vect{a}_\mathrm{ext} + \sum_{\beta \neq \alpha} \vect{a}_\beta(\vect{r}_\alpha).
}
Using the Einstein's summation convention, we have
\eqn{
a^i_{\alpha,\mathrm{D}} = \delta^j \ \partial^j \left(a^i_\mathrm{ext}\right) + \sum_{\beta \neq \alpha} a^i_\beta(r_\alpha),
}
where the effect of the external potential on the dwarf galaxy is described by the term
\eqn[introtide]{
T^{ji} \equiv \partial^j \left(a^i_\mathrm{ext}\right)
}
which is the $j,i$ term of the $3\times3$ matrix $\matr{T}$, called tidal tensor. Knowing its components, it is possible to compute the effect of the tidal field at any position within the dwarf. Because $a^i_\mathrm{ext}$ can be derived from the gravitational potential $-\partial^i \phi_\mathrm{ext}$, we note that the tensor is symmetric
\eqn[symmetrictensor]{
T^{ji} = - \partial^j \partial^i \phi_\mathrm{ext} = - \partial^i \partial^j \phi_\mathrm{ext} = T^{ij}.
}
Examples of the tidal tensors of classical density profiles are given in \refapp{triplets}.

%%%%%%%%%%%%%%%%%
\subsect[compressive_mode]{Introduction to compressive tides}
Aside from being symmetric, the tensor is real-valued and thus can be set in an orthogonal form. Three eigenvalues $\{\lambda_i\}$ measure the strength of the tidal field along the associated eigenvectors. The trace of the tensor reads
\eqn{
\mathrm{Tr}(\matr{T}) = \sum_i \lambda_i = -\partial^i \partial^i \phi_\mathrm{ext} = -\nabla^2 \phi_\mathrm{ext}.
}
Poisson's law gives a link with the local density $\rho$ (see \refsect[basic_concepts]{poisson}):
\eqn[poissontrace]{
\mathrm{Tr}(\matr{T}) = - 4\pi G \rho \le 0.
}
This means that the trace of the tidal tensor cannot be positive, and thus that it is impossible to compute simultaneously three positive eigenvalues. Remains the possibilities of two, one or no positive eigenvalues, as noted by \mycitet{Dekel2003}. In the Earth-Moon case (recall \reffig{earthmoon}), we saw that the differential forces act ``outward'' along the planet-satellite axis, while they are orientated ``inward'' in the other two directions, giving two negative $\lambda$'s. In analogy with solid bodies, the tides exert a stress along the Earth-Moon axis, and a compression along the remaining axes. As for those of the stress tensor, the eigenvalues of the tidal tensor denote a compression when negative and an extension when positive, along the associated eigenvector.

\fig{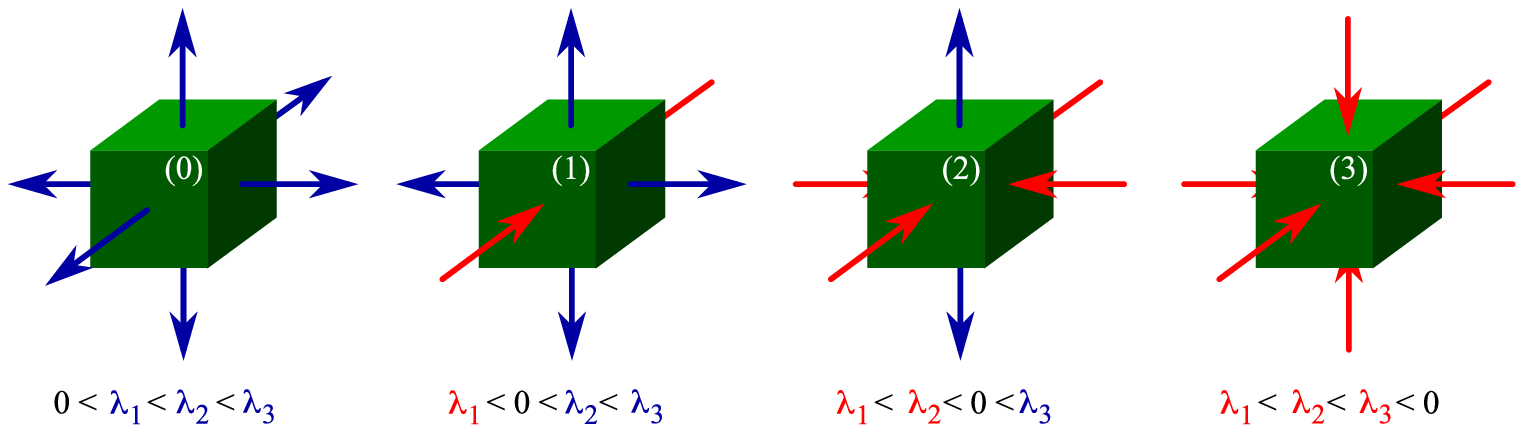}{Compressive vs. extensive modes}
{Different cases of tidal effects on an element (here, a green cube). In the configuration (0), all eigenvalues $\lambda_i$ are positive and the tides act as an extensive effect (blue arrows). In the case (1) (respectively (2)), the tides are compressive along one (respectively two) direction(s) (red arrows). In the last case, the tides are fully compressive: all the eigenvalues are negative.}

\reffig{cube_compressive} sums up all the configurations, labelled by the number of negative eigenvalues. Remember that the first case (0) is not possible in the Newtonian gravity, as it is linked to a negative density. Cases (1) and (2) are well known situations in, e.g. clusters of galaxies (\mycitealt{Valluri1993}) or spiral waves (\mycitealt{Gieles2007a}). The last case, where all three eigenvalues are negative has been overlooked in the literature. In this configuration, the tides are \emph{fully} compressive. In the rest of the present manuscript, we will refer to such a case as \emph{compressive tidal mode}. We will come back to this important point in the next Chapters.

Compressive or not, the mathematical representation of the tides allows us to describe the large scale effects on smaller scale objects, like dwarf galaxies or star clusters.

%%%%%%%%%%%%%%%%%%%%%%%%%%%%%%%%%%%%%%%%%%%%%%%%%%
\sect{Star clusters}

%%%%%%%%%%%%%%%%%
\subsect{Collection of stars}
Any astronomer will tell you that clusters of galaxies contain galaxies, which contain clusters of stars. Indeed, as the galaxies are often organized in clusters, the stars that form the galaxies are themselves grouped in clusters. Discovered first during the 17th century, many \emph{star} clusters have been catalogued by Charles Messier in 1774, without knowing that these objects were groups of stars. This update came later, in 1791 when William Herschel looked at M5 with his 40 feet-long telescope. He saw lots of stars in the cluster and even a core so dense that he could not resolve all its components.

With more and more powerful observational techniques, it has been possible to detect numerous clusters and to resolve them into stars in the Milky Way and nearby galaxies (e.g. Andromeda). Two types of clusters revealed themselves (see \reffig{m35}):

\fig{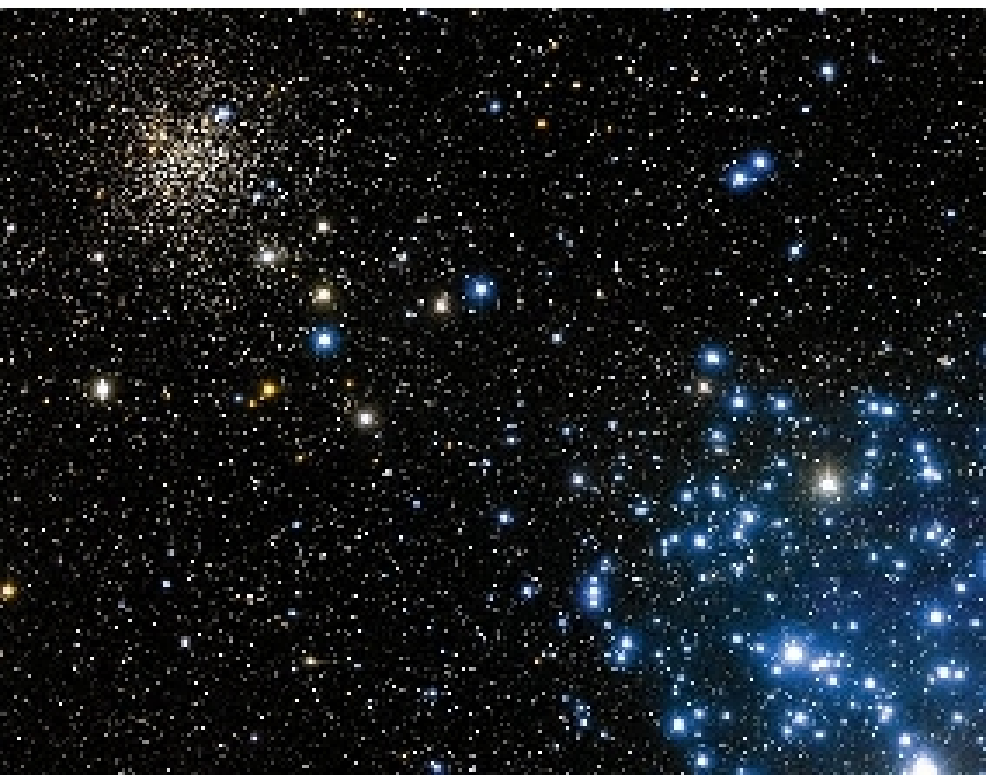}{Globular and open clusters}
{Two star clusters in the constellation of Gemini. NGC 2158 (\emph{top left}) is a globular cluster of $\sim 1.5 \Gyr$ while M35 (\emph{bottom right}) is an open cluster 10 times younger. The blue massive stars in NGC 2158 have already exploded as supernovae when many of them are visible in M35. (Image by J.-C. Cuillandre, CFHT.)}

\begin{itemize}
\item the globular clusters present a smooth spherical shape and a very high density in their center. They are often old ($\sim 1-10 \Gyr$) and very massive ($\sim 10^4-10^6 \msun$; see \mycitealt{Meylan1997} for a review);
\item the open clusters are much younger than the globulars, more metal-rich, and have low masses, as for the Pleiades or the Hyades clusters ($\lesssim 5000 \msun$).
\end{itemize}

Despite their differences, the two classes of star clusters share similar sizes, which leads to a large spread in their central density ($10^1 - 10^6 \msun\pc^{-3}$). This historical classification mainly relies on observations of clusters of the Milky Way, a galaxy that does not undergo a strong burst of star formation. However, in galaxies showing high \acr[s]{}{SFR}{star formation rates}, recent observations revealed the existence of young globular clusters, having the density and mass of globulars but a metallicity and mass function similar to those of open clusters (see e.g. \mycitealt{Holtzman1992}; \mycitealt{Whitmore1993}; \mycitealt{OConnell1994}; \mycitealt{Watson1996}; \mycitealt{MaizApellaniz2001}). The high mass-to-light ratio for these young stellar populations makes the clusters detectable over large distances, that is, in other galaxies than our Milky Way. They are suspected to be at the formation stage of the classical globular clusters, as those we observe in our Galaxy, provided they survive to the possible destructive effects. Indeed, we know that most of the stars form in clusters (\mycitealt{Lada2003}) but, in the Milky Way, only 1\% of them are still members of a cluster. This suggests a strong destruction factor, from internal and external causes. We will further explore this point in this manuscript.

%%%%%%%%%%%%%%%%%
\subsect{So many radii}

In the twentieth century, the growing body of observational data allowed a precise modeling of star clusters. The mass profile has been described by e.g. \mycitet{Plummer1911}, \mycitet{Jeans1915}, \mycitet{Michie1963}, \mycitet{King1962} or \mycitet{King1966}, followed by evolutionary studies (see e.g. \mycitealt{Spitzer1940}; \mycitealt{Henon1961}). These theories have led to a large number of physical quantities, among which the radii, which may be confusing, even today. For the sake of clarity, the definitions of some of these quantities are given below, considering the example of a star cluster orbiting in a galaxy.
\begin{itemize}
\item The tidal radius refers to the radius where the density of the cluster drops to zero. It is well defined for analytical profiles like the empirical \mycitet{King1962} or theoretical \mycitet{King1966} profiles, but very difficult to determine observationally. It is important to note that even the isolated objects have a tidal radius, which has \emph{stricto sensu} no relation with tides.
\item The Jacobi limit corresponds to the boundary of the region inside which the cluster is gravitationally dominant. For simplicity, it is often defined as the distance from the center of the cluster to the first Lagrangian point. In this case, it might be called Jacobi radius (see e.g. \mycitealt{Baumgardt2010}), but also Roche sphere or Hill sphere. Rigorously however, the Jacobi limit is not spherical. Outside this limit, the tidal forces due to the galaxy strip the material off, thus creating a cut-off limit. Therefore, the Jacobi limit is necessarily larger than or equal to the tidal radius.
\end{itemize}
Slightly different definitions also exist: the Roche limit usually applies to planets or satellites. It describes the limit where the tidal forces due to a larger object overcome the cohesion forces of the planet, thus destroying it. The Roche lobe refers to stars or binary stars where the tidal field is negligible but replaced by the thermodynamic pressure. When a star fills its Roche lobe, its material escapes from the gravitational bound.

In the following, we will focus on the properties of a star cluster and thus, using the historical definitions, we will refer to the radius where the density drops to zero as the tidal radius, and to the radius depending on the tidal field of the host galaxy as the Jacobi limit.

%%%%%%%%%%%%%%%%%
\subsect[clusters_formation]{Formation and environment}

If the general process of the formation of a cluster is well known, many details remain unclear and thus, out of reach of numerical simulations. We do know that a cluster forms from a giant molecular cloud ($\gtrsim 10^4 \msun$) that cools down (to a temperature of $\lesssim 10 \U{K}$), fragments and then collapses. Therefore, the temperature, chemical composition, density and pressure must control this process. In addition, shocks (\mycitealt{Barnes2004}), turbulence (\mycitealt{Elmegreen1997}; \mycitealt{MacLow2004}), ram pressure (\mycitealt{Kapferer2009}) and magnetic fields (\mycitealt{Verschueren1990}) are also regulators of the formation and must be correctly parametrized in the simulations. With so many knobs requiring a fine tuning, most of the studies (observational or theoretical, recent or older) focused on the formation of clusters from a global point of view, characterizing their location (\mycitealt{Mihos1993}), timescale (\mycitealt{Elmegreen2000}), rate (\mycitealt{Schmidt1959}; \mycitealt{Kennicutt1998}), or efficiency (\mycitealt{Rownd1999}; \mycitealt{diMatteo2007}), in isolated or interacting galaxies.

After these phases leading to a proto-cluster, the first massive stars form in a clump of gas which they rapidly ($\sim 1 \Myr$) blow away thanks to radiation and stellar winds. Note that all the gas from the molecular cloud is not converted into stars: the fraction of mass actually processed is called the \acr{}{SFE}{star formation efficiency} $\epsilon$:
\eqn[sfe]{
\epsilon = \frac{M_\star}{M_\mathrm{total}}
}
(generally of the order of a few percent), which leaves a fraction $(1-\epsilon)$ in the gaseous phase in the young cluster.

\fig{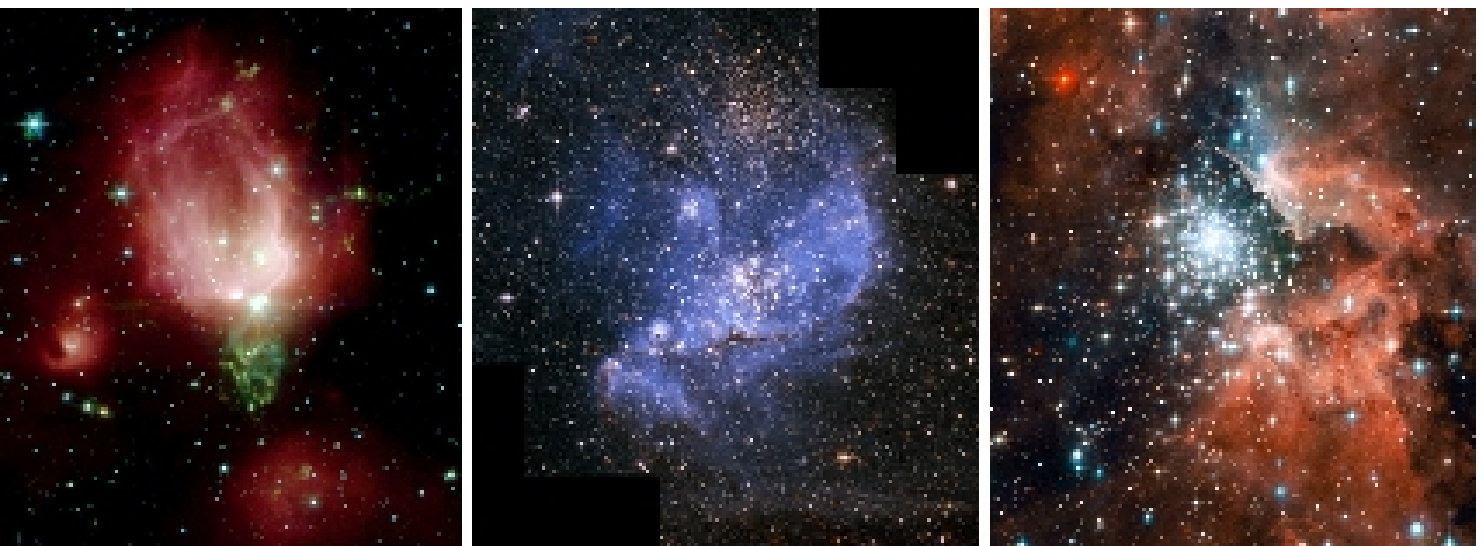}{Birth of a star cluster}
{First steps in the life of a star cluster. \emph{Left}: the first 130 stars of NGC~7129 formed in a gaseous and dusty nursery which they begin to spread away, in a bubble shape. \emph{Middle}: in the Small Magellanic Cloud, the young stars of NGC~346 have already expelled a fraction of the surrounding gas but are still embedded in it. \emph{Right}: in NGC~3603, the cluster has now swept out most of its gaseous content. (Images by T. Megeath, Spitzer; A. Nota, HST; J. Ma\`iz Apell\`aniz, HST.)}

At this point, stars themselves are not directly visible, but via the re-emission of their \acr{}{UV}{ultraviolet} light by the dust, in the \acr{}{IR}{infrared} wavelengths. The process continues with the destruction of the dust by intense \acr{}{UV}{ultraviolet} radiation (\mycitealt{Draine1979}), the formation of lower mass stars and the explosions of the first \acr[e]{}{SN}{supernovae} (\mycitealt{Castor1975}). As a result, all the gas left-over from the formation of the stars is widely expelled after $\sim 10^7 \yr$. If the cluster survives these steps and remains bound by its own gravitation despite this mass-loss (see \mycitealt{Goodwin1997}; \mycitealt{Geyer2001}; \mycitealt{Boily2003a}, \myciteyear{Boily2003b}; \mycitealt{Baumgardt2007}), it can begin its life in a gas-free region. \reffig{clusterbirth} sums up these major phases.

%%%%%%%%%%%%%%%%%
\subsect{Observations and completeness}

\fig{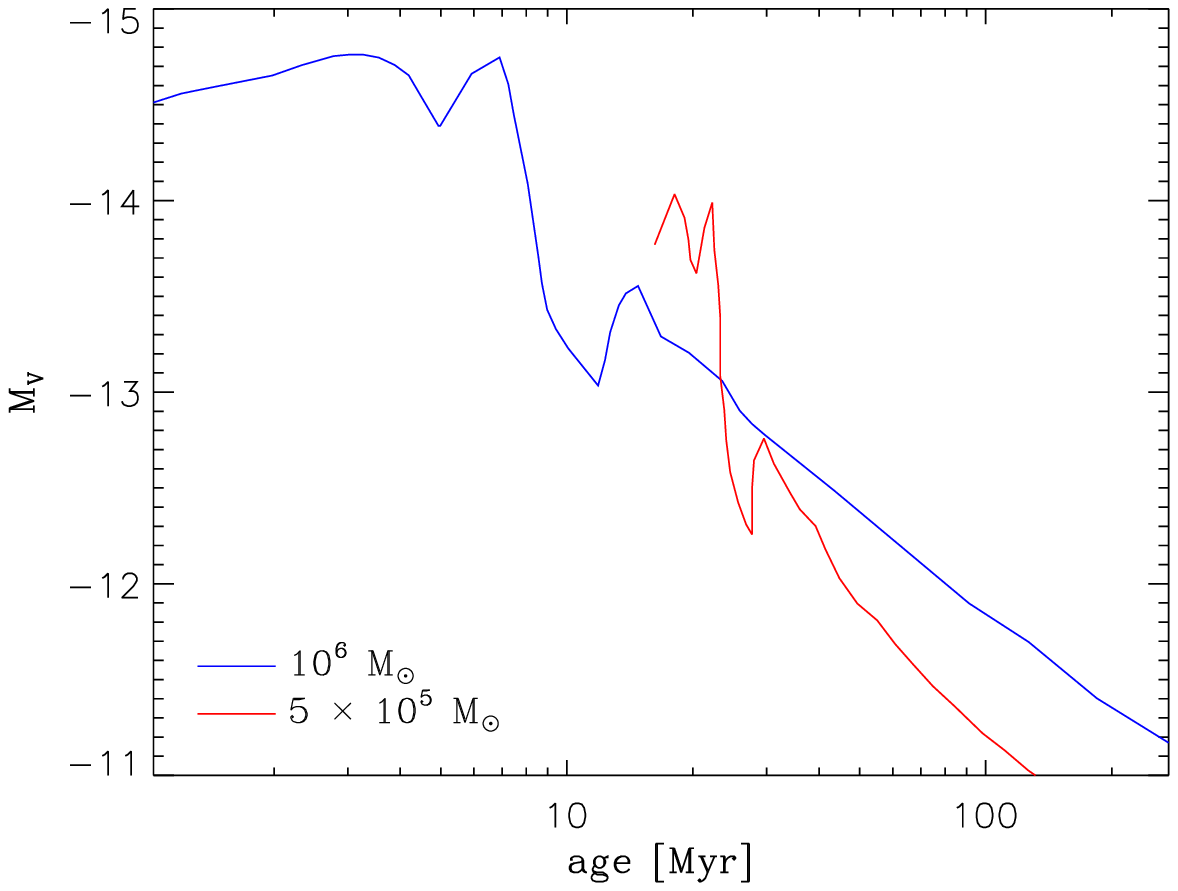}{Luminosity of a star cluster}
{Evolution of the visual magnitude of clusters of mass $10^6 \msun$ (blue) and $5\times 10^5 \msun$ (red). Here, the $5\times 10^5 \msun$ cluster forms $15 \Myr$ after the more massive one. (Data points from \mycitealt{Weidner2004}.)}

The link between observational and theoretical studies is however biased when the detection limits of the fainter clusters are reached. Once a star cluster is formed, the stars will evolve so that the total luminosity decreases, as shown by \mycitet{McCrady2003}, \mycitet{Weidner2004} and summarized in \reffig{weidner}.

This underlines two points that one must not forget when interpreting the observations: 
\begin{itemize}
\item The brightest cluster is not always the more massive one. Indeed, the $5\times 10^5 \msun$ cluster shown in \reffig{weidner} yields a higher luminosity than the other one (twice more massive) for $\sim 10 \Myr$. A constant mass-to-light ratio can only be assumed for clusters of similar ages.
\item Old clusters (for all masses), yield low luminosities and may not be detected by the instruments. When searching for statistical results on a population of clusters, one has to take the completeness limit of the sample into account and correct for the faint clusters.
\end{itemize}
In the end, the determination of the mass and age functions of clusters within a galaxy may be extremely biased by these two points because the mass estimate can yield significant errors, and the age distribution may miss the high ages (i.e. low $M_V$) end. We will see in \refcha{antennae} that extra caution must be taken when interpreting these observational estimates.

\mybox[cluster_disruption]{
The lifetime of a star cluster depends on the disrupting effects it may experience, both internal and external. In this way, the rate of death would depend on the initial mass of the cluster or in other words, its binding energy (\mycitealt{Spitzer1958}; \mycitealt{Baumgardt2003}). Observational data however, seem to be more controversial. Despite the difficulty to probe the low mass end of the mass function (because of the limit of detection), two \emph{empirical} models have been proposed: 
\begin{itemize}
\item a mass dependent disruption model (MDD, \mycitealt{Boutloukos2003}) gives the disruption timescale $t_\mathrm{dis} \propto M^\gamma$ with $\gamma \simeq 0.62$.
\item a mass independent disruption model (MID, \mycitealt{Fall2005}).
\end{itemize}
Both theoretical and numerical ($N$-body) studies favor the MDD model (see e.g. \mycitealt{Baumgardt2003}; \mycitealt{Gieles2008}). For instance, an external (extensive) tidal field truncates the cluster to the Jacobi radius, which can be linked to the half-mass radius (\mycitealt{Ostriker1972}) and finally, to the mass (\mycitealt{Larsen2004}). Therefore, the disruption of the cluster due to the tidal field is linked to the mass.

The MID model has been introduced to explain the evolution of the mass function of the Antennae galaxies (that we will see in more details in the \refsect[antennae]{antennae_clusters}). Then, it has been extended to other galaxies like the Small Magellanic Cloud (\mycitealt{Chandar2006}) or M33 (\mycitealt{Sarajedini2007}).
\follow}\mybox[nocnt]{
However, the completeness problem in these galaxies biases the age distribution of the clusters toward a $\propto \tau^{-1}$ law (\mycitealt{Gieles2007b}), which forbids to conclude on the role of the mass in the disruption.

Again, one must be careful when addressing statistical issues like the mass, luminosity or age functions, because the sample considered might not be complete. In any case, one should not forget that the Antennae widely differ from the other galaxies used in the studies of both models in the sense that, as a major merger, the environment of the clusters is very different from those existing in quiescent galaxies. This last point is explored further in the \refcha{antennae}.

Note that both models assume a power law for the cluster initial mass function (CIMF): $\Psi(M) = dN / dM \propto M^\beta$, with $\beta \simeq -2.0$. However, a different CIMF has recently been proposed by \mycitet{Larsen2009} and \mycitet{Gieles2009}, with the form of a \mycitet{Schechter1976} function:
\eqn{
\Psi(M) \propto M^\beta \exp{\left(-\frac{M}{M_\mathrm{c}}\right)}
}
where $\beta \simeq -2.0$ is the usual slope of a power law CIMF and $M_\mathrm{c}$ characterizes the exponential drop of the mass function ($M_\mathrm{c} \simeq 2 \times 10^5 \msun$ for quiescent spiral galaxies, see \mycitealt{Larsen2009}). \mycitet{Gieles2009} discussed in detail the parameters and the implications of such a function on the properties of the cluster populations.
}

Nowadays, the knowledge of star clusters, and especially their early life, in our Galaxy or others (\mycitealt{Harris1991}), appears to be the key to the understanding of as various phenomena as the formation of stars and the evolution of galaxies (see \mycitealt{Brodie2006} for a recent review). Several authors explored the dynamics or the stellar populations of clusters to derive cause-to-effect relations between the clusters themselves, internal (e.g. stellar evolution in \mycitealt{Hurley2000}, mass segregation in \mycitealt{White1977} or mass loss in \mycitealt{Hills1980}) and external (e.g. passage through the galactic disk or a spiral arm  as in \mycitealt{Gieles2007a} or the perturbation due to the tidal field described in \mycitealt{Lamers2005}) processes. Recent literature suggests to link the formation of the clusters to their environment, i.e. major events in the life of their host galaxy. Indeed, clusters and their stellar populations can trace back the history of their host (see the inversion methods of \mycitealt{Ocvirk2006}), what is, in particular for mergers, a fundamental aspect of our understanding. Both galactic and cluster scales are coupled as one strongly influences the other through stellar winds, bubbles, chimneys, chemical enrichment or tides. The present document focuses on the last point.

%%%%%%%%%%%%%%%%%%%%%%%%%%%%%%%%%%%%%%%%%%%%%%%%%%
\sect{This thesis}

We have shown above that large-scale interactions between galaxies lead to the development of tides that act at cluster-scale. Putting all these points together, one naturally ends with the title of the present thesis:
\begin{center}
\emph{Dynamics of the Tidal Fields \& Formation of Star Clusters in Galaxy Mergers}.
\end{center}

The work presented in this document tries to bring new elements to the giant jigsaw puzzle of interacting galaxies and stars clusters. Using numerical experiments, it aims to show that the large scale structures of galaxies give rise to phenomena that directly act on the smaller scales structures, like the star clusters. Among these effects, we concentrate on the tidal field in interacting galaxies and detail its bimodal nature by introducing and extensively developing the concept of compressive tides. This thesis suggests that the compressive tidal mode participates in the formation and early life of star clusters by preventing their dissolution and even helping their creation. This hypothesis is confirmed by an intensive comparison between our numerical results and the observations of a well-known pair of galaxies (the Antennae), and then extended to other systems. This document is organized as follow:

\begin{itemize}
\item The first Chapter presents the numerical methods used during this work, to model galaxy mergers and to compute the effects of the tides. It first briefly explains the concept of $N$-body simulations and details the algorithms used. The numerical technics and codes are then presented, as well as the test suite developed to validate the methods.

\item Their applications are shown in the second Chapter. There, we present the Antennae galaxies, from an observational point of view and then, the numerical model we created. We develop the results obtained about the tidal field of this model. 

\item The strong link found between the existence of a compressive tidal mode and the formation of star clusters and tidal dwarf galaxies is detailed in the third Chapter, thanks to a comparisons with the physical quantities that monitor star formation.

\item Chapter IV extends these conclusions to a broader range of parameters, showing how robust the results are. Many cases of interacting galaxies are covered, with no particular link to observed systems. The Chapter particularly presents the influence of the distance between the galaxies during the interaction.

\item In Chapter V, we take one step backward and explores the link between the mass profile of the dark matter halos and the tidal field. We show that the choice of the profile strongly influences the character of the tidal field which in the end may, or not, support the formation of substructures.

\item The numerous questions raised during this PhD and the perspectives they bring are presented in the Conclusions.

\item Finally, the Appendices bring more details to some points of the main text. The interested reader will find there an analytical complement (with many detailed derivations) that may help her/him to better understand this manuscript.
\end{itemize}

\vspace{1cm}
\begin{center}
\makebox[0.7cm][s]{G a s}\\
\makebox[1.25cm][s]{T i d e s}\\
\makebox[1.8cm][s]{S t a r s}\\
\makebox[2.35cm][s]{G r a v i t y}\\
\makebox[2.9cm][s]{C l u s t e r s}\\
\makebox[3.45cm][s]{G a l a x i e s}\\
\makebox[4cm][s]{C o m p u t e r s}\\
\vspace{0.7cm}
Now that all the characters have been introduced,\\
let's begin the/to play.
%\makebox[5cm][s]{Now that all the characters}\\
%\makebox[5cm][s]{}\\
%\makebox[5cm][s]{}\\
\end{center}

%%%%%%%%%%%%%%%%%%%%%%%%%%%%%%

%%%%
\def\mychapname{Chap. } % used in the header
\cha{numerical}{Numerical techniques}
\chaphead{This Chapter presents the numerical methods used in this work. It explains the choices made and the tests performed to validate the study.}

%%%%%%%%%%%%%%%%%%%%%%%%%%%%%%%%%%%%%%%%%%%%%%%%%%
\sect{$N$-body simulations}

%%%%%%%%%%%
\subsect[smoothing]{Smoothing the singularity}
In astrophysics, the $N$-body gravitational problem usually consists in solving the equations of motion of a system of $N$ particles using the Newtonian description of the gravitation. The acceleration $\vect{a}_i$ of the $i$-th particle is simply given by
\eqn[nbody]{
\vect{a}_i = \sum^N_{j \neq i} \frac{G\ m_j}{\left| \vect{r}_j - \vect{r}_i \right|^3}\left( \vect{r}_j - \vect{r}_i \right) -\vect{\nabla} \phi_\mathrm{ext}(\vect{r}_i),
}
where $\phi_\mathrm{ext}$ represents an optional external potential (which we will consider null hereafter). Once the initial conditions in position and velocity are set, the solution can be found by solving the $N$ systems of non-linear second order ordinary differential equations
\eqn{
\left\{
\begin{array}{lcl}
\displaystyle \vect{a}_i & = \displaystyle \frac{\partial \vect{v}_i}{\partial t}\\
\\
\displaystyle \vect{v}_i & = \displaystyle \frac{\partial \vect{r}_i}{\partial t},
\end{array} \right.
}
via e.g. an Euler or a leap-frog integration scheme. In the last case, the solutions in position $x_i$ and velocity $v_i$ are updated with
\eqn[leapfrog]{
\left\{
\begin{array}{lcl}
\displaystyle x_i^{t+1} & = x_i^t + v_i^{t+1/2} \ dt \\
\\
\displaystyle v_i^{t+1/2} & = v_i^{t-1/2} + a_i^t \ dt.
\end{array} \right.
}

However \refeqn{nbody} yields a singularity when the distance $\left| \vect{r}_j - \vect{r}_i \right|$ between two particles approaches zero. If such a situation occurs, an integration with a constant timestep $dt$ would make the $\left( a_i^t \ dt \right)$ term extremely large, leading to (unrealistic) high velocities and finally unbinding close particles. The obvious solution would be to use an adaptive timestep, that gets smaller when the distance between two particles becomes too short. All the other particles would then evolve with the timestep of the encounter (i.e. several order of magnitude smaller than usual). Because this situation may occur repeatedly in the system, this can considerably slow down the calculation.

A more efficient solution, invented by \mycitet{Aarseth1963}, consists in modifying the law of gravitation on the very small scales, by introducing a smoothing (or softening) length\footnote{This approach is similar to the \acr{notarg}{SPH}{smoothed particle hydrodynamics} method that smoothes the physical properties over a scalelength.} $\epsilon$, in \refeqn{nbody} (with $\phi_\mathrm{ext} = 0$)
\eqn{
\vect{a}_i = \sum^N_{j \neq j} \frac{G\ m_j}{\left( \left| \vect{r}_j - \vect{r}_i \right|^2 + \epsilon^2 \right)^{3/2}}\left( \vect{r}_j - \vect{r}_i \right).
}
With this method, the singularity does not exist as the denominator never vanishes. This avoids using very small timesteps, but at the price of introducing an error in the force. Note that $\epsilon$ can be used in other functional forms, called kernels, that change the expression of the approximated Newtonian force and can ensure, e.g., continuous derivatives (see \mycitealt{Dehnen2001} for a nice introduction, and the \refsect[triplets]{kernels} of the present document for some analytical expressions).

Small values of $\epsilon$ render a noisy system because of the finite number of bodies used. At the opposite, large values increase the bias in the expression of the force. The value of the smoothing parameter had traditionally been chosen in an \emph{ad hoc} way for a long time, until that \mycitet{Merritt1996} and \mycitet{Athanassoula1998} proposed an objective method minimizing both sources of error. The optimal smoothing length decreases with the number $N$ of particles, roughly as $N^{-0.44}$ (fine details depending of the actual distribution of particles within the system).

%%%%%%%%%%%
\subsect[nbody_units]{$N$-body versus physical units}
In \reffig{plummer}, we show a typical output from an $N$-body simulation. Can you tell if the system modeled is a globular cluster, an elliptical galaxy or even a cluster of galaxies? As one rapidly notices, the axes of the figure do not have units like kpc, therefore answering the question is not possible.

In the 1980's, with a growing number of simulations, the $N$-body unit system has been introduced (\mycitealt{Heggie1986}) in order to allow for an easy comparison of the models from various authors. It sets
\eqn{
G = M = -4E = 1,
}
where $G$ is the gravitational constant, $M$ is the total mass and $E$ the energy. The -4 factor for $E$ corresponds to a virial radius set to unity (see \refsect[virial]{virialradius} for details). One can deduce the units of distance $d$ and time $t$:
\eqn[dimensionalization]{
d & = -\frac{GM^2}{4E}\\
t & = \sqrt{\frac{d^3}{GM}} \nonumber \\
& = G \sqrt{\frac{M^5}{-64E^3}}.
}

\fig{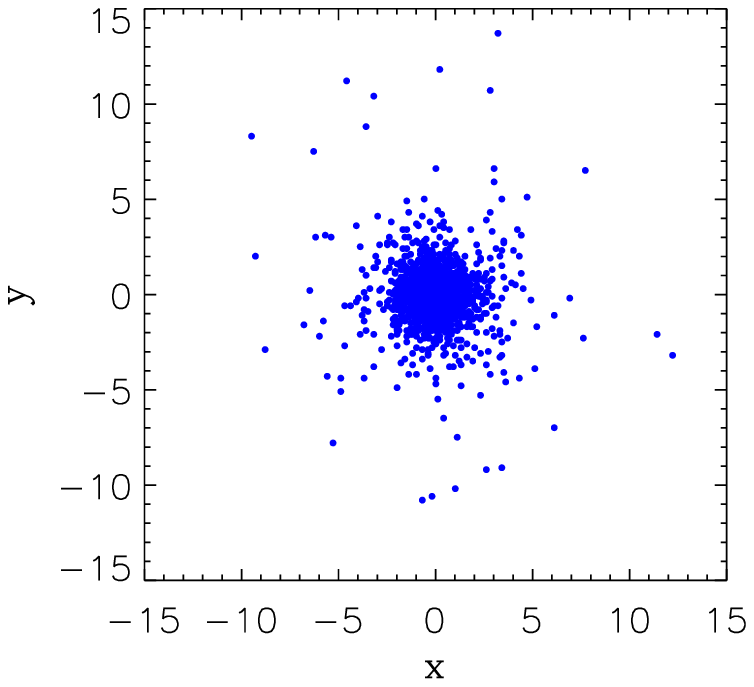}{$N$-body realization of a Plummer sphere}
{$N$-body realization of a Plummer sphere ($N = 5\,000$). The axes of this Figure do not have explicit units. Instead of physical units (like kpc), we have kept the $N$-body units system.}

We note that this system of units is not physically constrained, as soon as a dimensionless quantity like $GMt^2/d^3$ is left unchanged. Indeed, one can set the size of \reffig{plummer} to $20 \pc$ and its mass to $10^5 \msun$. In that case, one of the 5\,000 particles weights $20 \msun$ and one spatial unit represents $0.7 \pc$. This leads to a time unit of $\approx 3 \times 10^4 \yr$. However, by setting the values to bigger scales, e.g. a size of $30 \kpc$ and a mass of $10^{11} \msun$, one gets a timescale of $\approx 1.5 \Myr$. In the first case, \reffig{plummer} would represent a globular cluster, while in the second, it would more likely be an elliptical galaxy. That is why purely gravitational simulations\footnote{The introduction of hydrodynamical aspects like a cooling function adds a constraint on the timescale.} are freely scalable, and thus must be correctly documented to avoid misinterpretations. 

In the modeling of galaxies, the stellar encounters are not relevant regarding to the lifetime of the object, and thus, such systems can be considered collisionless.

\mybox[relax_times]{
Depending on the scales considered in a system, the role of each physical phenomenon may not be of the same importance. For instance, the dynamical timescale is linked to the typical size and velocity:
\eqn{
t_\mathrm{dyn} = \frac{R}{v},
}
and is approximately equal to the free-fall time:
\eqn{
t_\mathrm{dyn} \approx t_\mathrm{ff} = \sqrt{\frac{3\pi}{32G\rho}}.
}
This must be compared to the relaxation timescale and to the lifetime of the object
\eqn{
t_\mathrm{relax} \approx 0.1 \frac{N}{\ln(N)} t_\mathrm{dyn}
}
(see \mycitealt{Binney1987}). The following table gives order of magnitudes for the timescales of several systems:
\begin{center}
\begin{minipage}{8cm}
\begin{tabular*}{\linewidth}{l@{\extracolsep{\fill}}c@{\extracolsep{\fill}}c}
\hline\hline
system			& $t_\mathrm{dyn}$ [yr]	& $t_\mathrm{relax}$ [yr] \\
\hline
globular cluster	& $10^5$			& $10^9$ \\
galaxy			& $10^8$			& $10^{16}$ \\
galaxy cluster		& $10^9$			& $10^{11}$\\
\hline
\end{tabular*}
\end{minipage}
\end{center}
In the case of a globular cluster, the relaxation time is comparable to the lifetime of the cluster and thus, stellar encounters should not be neglected. The (violent) relaxation processes influence the internal properties of the cluster via, e.g. mass segregation (\mycitealt{White1977}).
\follow}\mybox[nocnt]{
For galaxies however, the relaxation time is much larger than the typical lifetime ($\sim 10^{10} \yr$), which indicates that they are collisionless systems in general. In the case of clusters of galaxies, both the relaxation time and the lifetime are comparable and thus, the conclusion on the collisional nature of such systems must take into account the details of their structure.}

The present Chapter describes the numerical methods and the associated tests, which are not linked to any physical object. Therefore, the numerical units will be used here, before switching to the physical system in the next Chapters.

%%%%%%%%%%%
\subsect[treecodes]{Barnes, Hut and Dehnen meet Newton under a (apple) tree}
For an entire system made of $N$ particles, the direct method requires to compute $(N-1)$ forces on each of the $N$ particles, i.e. $N\ (N-1) \sim N^2$ calculations. For physical systems like galaxies which require a large number of particles, the $N^2$ scaling becomes the major limitation to the high resolution studies. Nowadays, the largest simulations counts up to $10^{10}$ particles (see e.g. the Millennium-II simulation, \mycitealt{Boylan2009}) and obviously take advantage of other techniques.

Among many methods, we quickly describe here the tree code, which has been intensively used during this thesis. \mycitet{Barnes1986} introduced an inventive way of accelerating the computation considerably. Their idea is to group particles in branches (or cells) and to consider only the interaction between leaves and branches.

\fig{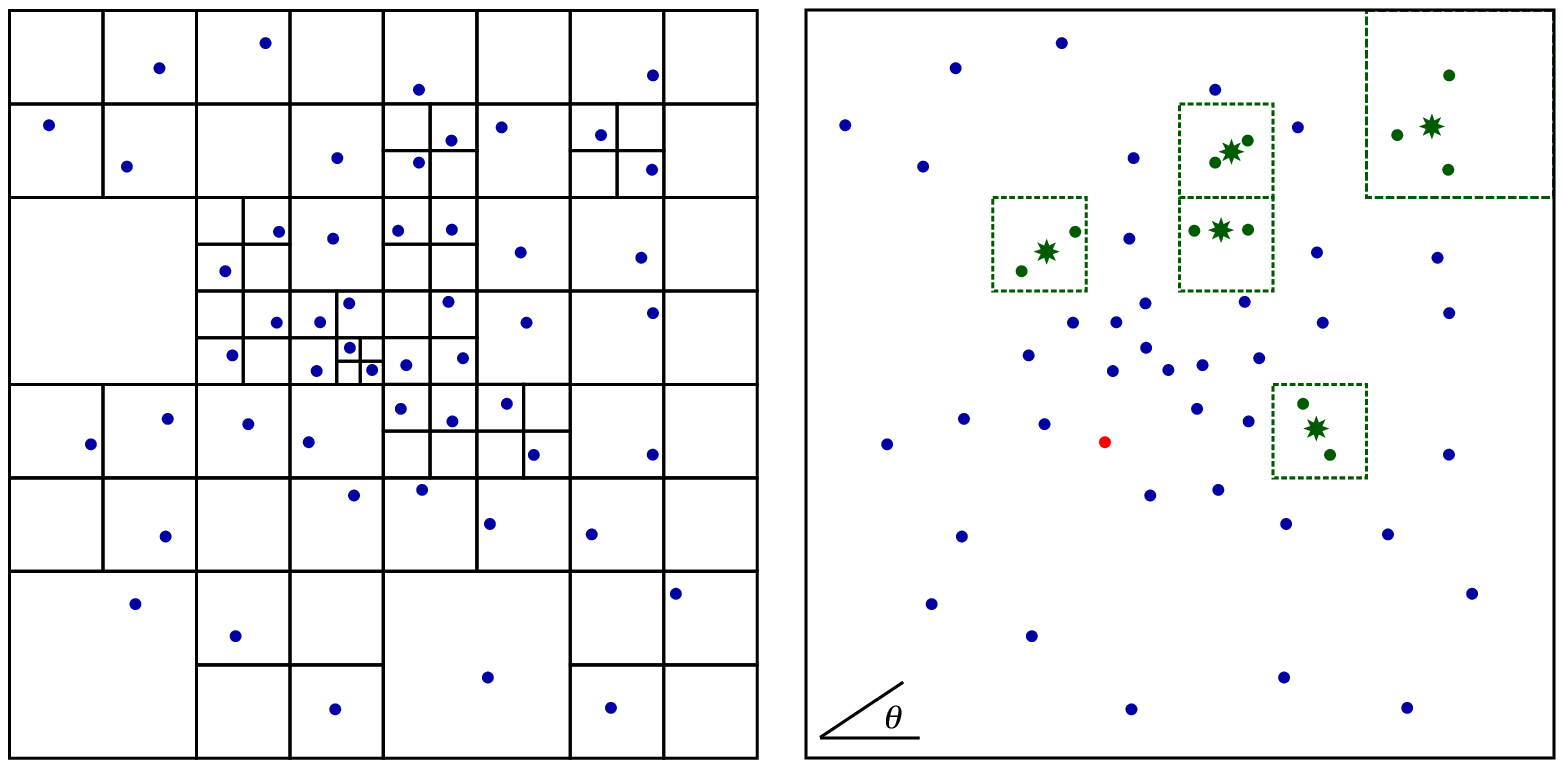}{Tree code}
{\emph{Left}: the entire space (here shown in 2D, for simplicity) is split into virtual cells and subcells. A (sub)cell is divided into daughter (sub)subcells if it contains more than one particle. These one-particle cells are called leaves. \emph{Right}: then, the acceleration on a given particle (red dot) is computed thanks to an angular criterion $\theta$ (see a schematic definition at the bottom-left corner of the right panel, and the text for details). If a group of cells (called branch, green squares on the figure) is viewed under a small angle from the red particle, then it is far enough and its effect can be approximated. All the leaves it contains are replaced by a virtual particle (green star) whose mass is simply the sum of the masses of all leaves, and situated in their center of mass. If a branch is too close to the red particles, then its leaves are left as it and considered separately, as in a direct code. In other words, if a group of particles is far and small enough from an external point, then the gravitational effect of these particles is taken into account as a whole, instead of separately, which speeds up the computation.}

First, the space is divided into virtual cells according to the following recursive scheme: if a cell contains more than one particle, it is divided into daughter subcells of exactly the half width (see \reffig{tree}, left). A cell with only one particle is called a leaf, while cells that have been split into subcells are branches. A pseudo (i.e. virtual) particle is then assigned to each branch, placed at its center of mass and contains the total mass of its cell. All the non-virtual particles in the system feel the effect of the pseudo-particles (instead of all the other real particles) if they are situated far enough (see \reffig{tree}, right).

A tolerance parameter $\theta\ (\sim 1)$ is used to define the distance criterion. A cell (branch) of a size $\ell$, at a distance $D$ of a given particle (leaf) is considered as a group if $\ell/D < \theta$. If not, its subcells are examined with the same criterion, and this recursively until (i) the criterion is fulfilled or, (ii) the subcell is a leaf (containing only one particle). (Note that the case $\theta = 0$ corresponds to a direct summation code, as the criterion on $\theta$ is never fulfilled.)

The total number of interactions is thus considerably reduced. This tree code brings the scaling down to $N \log{(N)}$, which accelerates the computation (and allows higher resolutions).

Few years later, \mycitet{Dehnen2000} improved the tree code by considering cell-cell instead of particle-cell interactions. When building the tree, each particle ``remembers'' in which cell it belongs. Then, all the particles in the same cell get the same effect from external cells. \reffig{direct_tree_codes} describes the three main algorithms: the direct summation, the tree code and Dehnen's code.

\fig{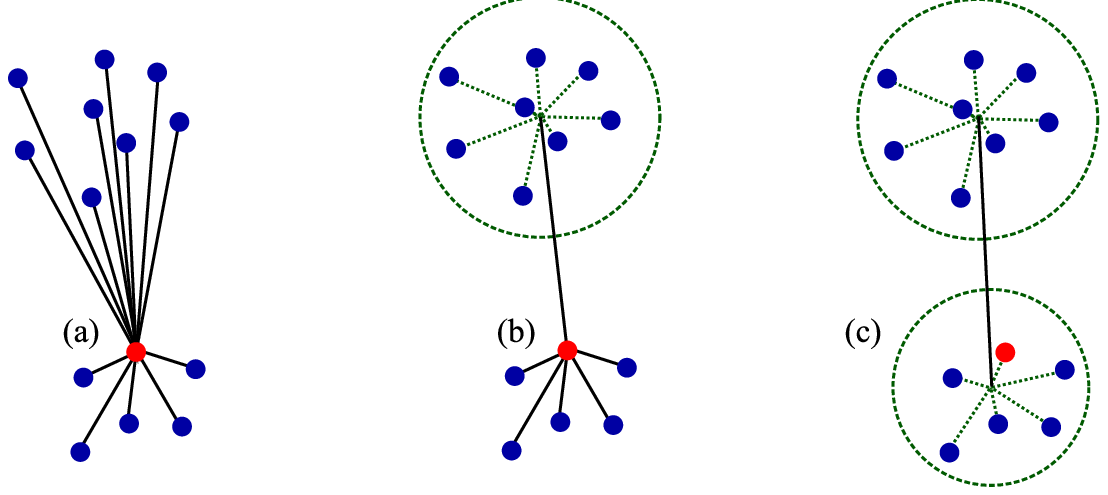}{Direct vs. tree codes}
{The direct summation (a) considers the interaction between all particles, whatever their respective distance is. The classical tree code (b) gathers particles at long distance into a single cell, while close particles remain treated as in the direct code. Dehnen's version of the tree code (c) takes advantage of the symmetry of the force and computes the interaction between the cells themselves.}

Although the actual gain of this method depends on the initial conditions of the system and on the rules used to build the tree, it has been noted that the scaling was reduced to $N$ (see \reffig{dehnen}). Recent tests revealed that Dehnen's code (called {\tt gyrfalcON}\footnote{GalaxY simulatoR using the Force ALgorithm with Complexity $\mathcal{O}(N)$.}) running on a single processor, competes with implementations of classical $N$-body codes distributed on up to heigh equivalent nodes (\mycitealt{Fortin2006}). Because technical details of these codes are not in the scope of this manuscript, we refer the reader to their introducing papers and the references therein (\mycitealt{Barnes1986}; \mycitealt{Dehnen2000}; \mycitealt{Dehnen2001}).

\fig{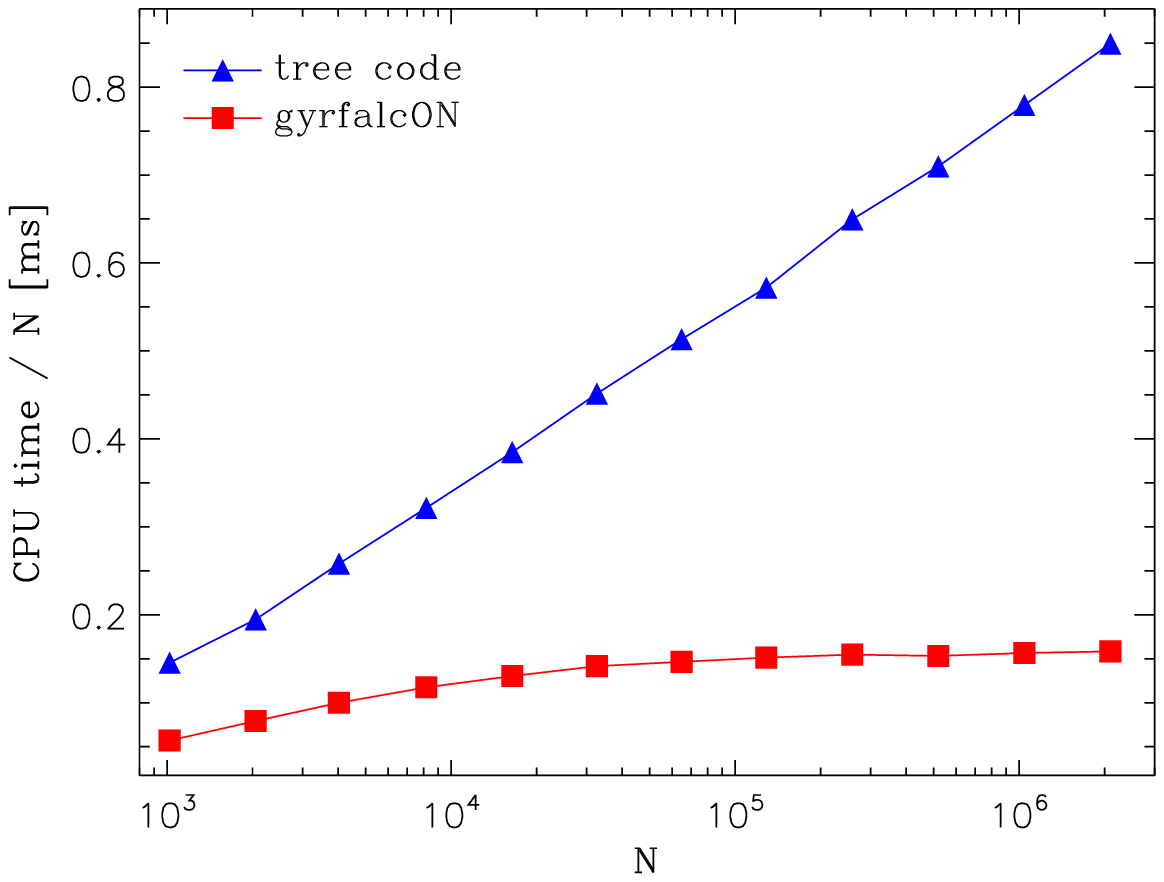}{Performances of the tree codes}
{Performances of the tree code {\tt gyrfalcON} (red) plotted as \acr{notarg}{CPU}{central process unit} time per body versus $N$, and compared to the classical tree code (blue). (Data points from \mycitealt{Dehnen2000}.)}

%%%%%%%%%%%
\subsect[nemo]{Finding Nemo}
All the codes presented above are publicly available through a toolbox (\mycitealt{Barnes1988b}; \mycitealt{Teuben1995}) called {\tt Nemo}\footnote{\href{http://bima.astro.umd.eu/nemo/}{http://bima.astro.umd.eu/nemo/\ifthenelse{\equal{\printerversion}{1}}{}{~\includegraphics[scale=0.6]{figs/deco_link.eps}}}} that contains hundreds of smallish and bigger codes used in classical $N$-body modeling. All these tools can be used in a shell, giving the parameters at the command-line or within an Unix shell script, as shown in \reffig{nemo}. It is therefore possible (and rather easy) to create two galaxy models, merge them into one snapshot, give them initial positions and velocities and make them collide, via e.g. {\tt gyrfalcON}. Visualization and diagnostic tools are also available.

\fig{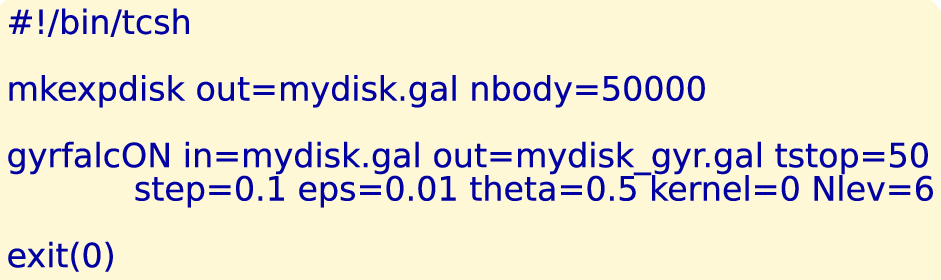}{Example of a Nemo script}
{Example of a Nemo script. This {\tt tcsh} script chains the commands to setup the problem and solve it. It first creates a model of an exponential disk with 50\,000 particles thanks to the command {\tt mkexpdisk} and saves it in the binary file {\tt mydisk.gal}. Then, it integrates the orbits of the particles with Dehnen's code {\tt gyrfalcON}, over $50/0.1 = 500$ timesteps (50 $N$-body time units), giving the values set by the user for the parameters (eps, theta, kernel, Nlev), and finally stores the result in the {\tt mydisk\_gyr.gal} file. Obviously, external programs can be inserted into the script to perform additional tasks.}

This study made intensive use of the {\tt Nemo} package, which has helped to concentrate on the parameter space and the physics. However, the numerical parameters used by the codes have been chosen after many tests and checks which ensured that approximations and errors were under control (see \refsec[numerical]{testsgyrfalcon}).

%%%%%%%%%%%
\subsect{Visualization}
As mentioned before, {\tt Nemo} does contain visualization tools. However, for the purpose of this thesis, it appeared that they would not be flexible and powerful enough to allow a deep investigation. Therefore, we developed a brand new code whose only task is to render the $N$-body systems modeled. Its name, {\tt Trackpart}, came from its very first aim: it has been designed to plot the trajectory of selected particles and to go back and forth in time, tracking them (see \reffig{trackpart}).

\fig{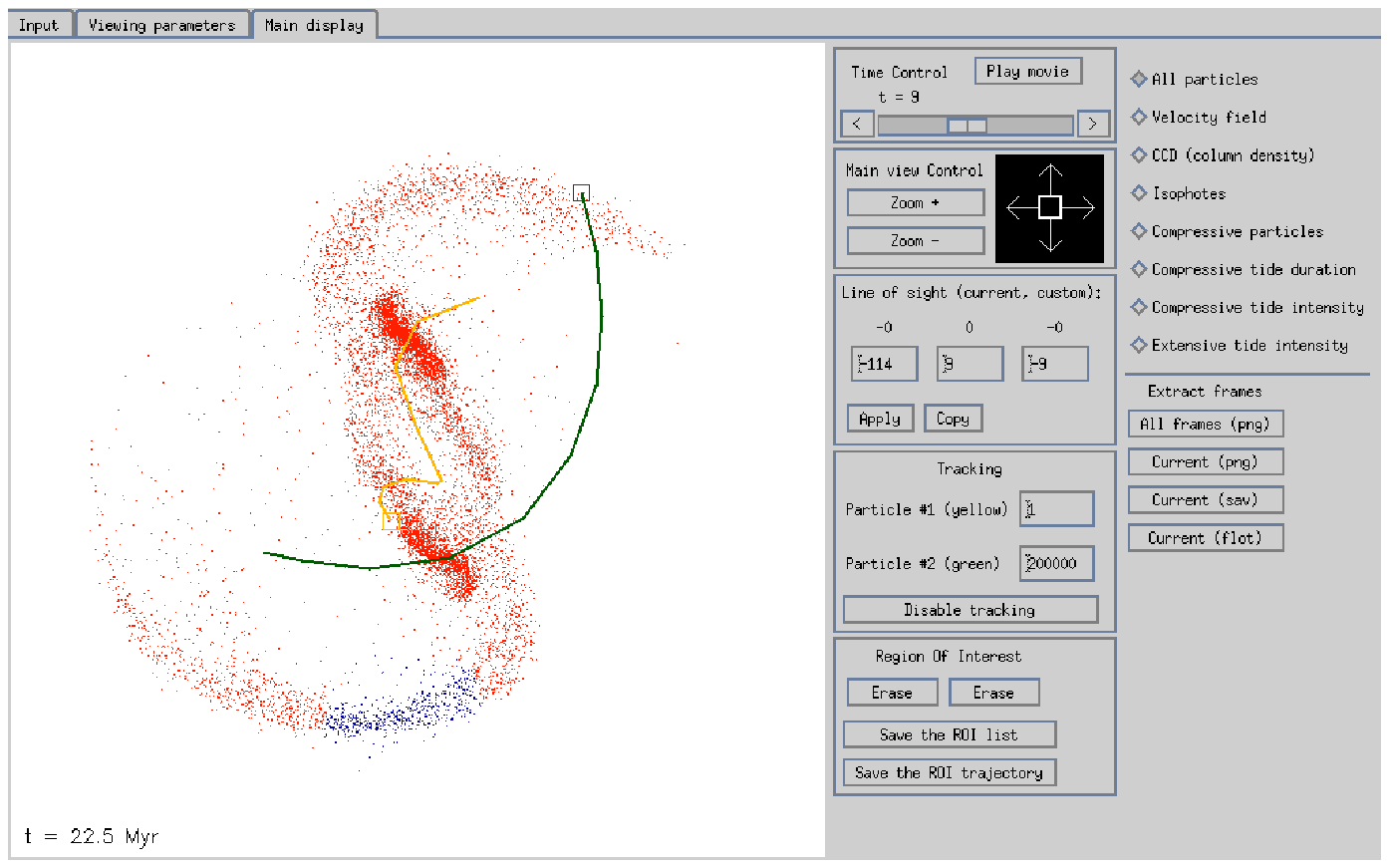}{Trackpart}
{The Trackpart interface allows to 3D-rotate an $N$-body model, to zoom-in or -out on details, or select a region of interest, tag the particles in it and follow them along time (as done for the blue part of the tidal tail in this screenshot) or plot the orbits of particles (orange and green here). Other visualization modes (velocity field, column density ...) are also available.}

{\tt Trackpart} has always been developed in the scope that external modules or function could easily been added. That is why we chose IDL as programming language, a well-known tool in the astrophysics community. Starting from scratch, we created routines dedicated to the 3D viewing (xyz-rotation with the mouse, zoom, translation) and rendering devices that efficiently compute \emph{on-the-go} all the quantities needed (column density, line-of-sight velocity, isocontours...), depending on the 3D orientation. A graphical user interface has been built to read the input files, create images and movies quickly and easily. Most of the images of $N$-body simulations in this manuscript have been prepared using this tool.

%%%%%%%%%%%
\subsect[testsgyrfalcon]{Testing the gravity}
In $N$-body simulations, the most accurate result will come from a direct summation code with a very small timestep. As we have seen above, this would be an \emph{extremely} long computation and is thus compensated thanks to approximations. Obviously, both these slight changes in the physical behavior of the system and the numerical treatment introduce errors in the final result. Hence, the parameters tuning the approximations must be chosen carefully, to get the best combination between accuracy and speed (keeping in mind that the solution will always be an arbitrary choice and thus depends on what is important for the user).

First, we compare the results given by {\tt gyrfalcON} with those of a classical tree code (called {\tt hackcode1}). For this experiment, we place ourselves in a critical situation: the face-on collision of two disks. Let's consider two parallel identical exponential disks, each made of $N = 50\,000$ particles, separated by a distance of five times their characteristic radius. Without initial velocity, their mutual gravitational attraction makes them collapse on each other, as plotted on \reffig{faceon}. We expect the integrator to make the largest error on the computation of the accelerations when the disks go from quasi-isolated stage to violent collision, in other words, for a sharp evolution of the density. By quantifying the order of magnitude of the errors made in this case, one can evaluate the quality of the computation.

\fig{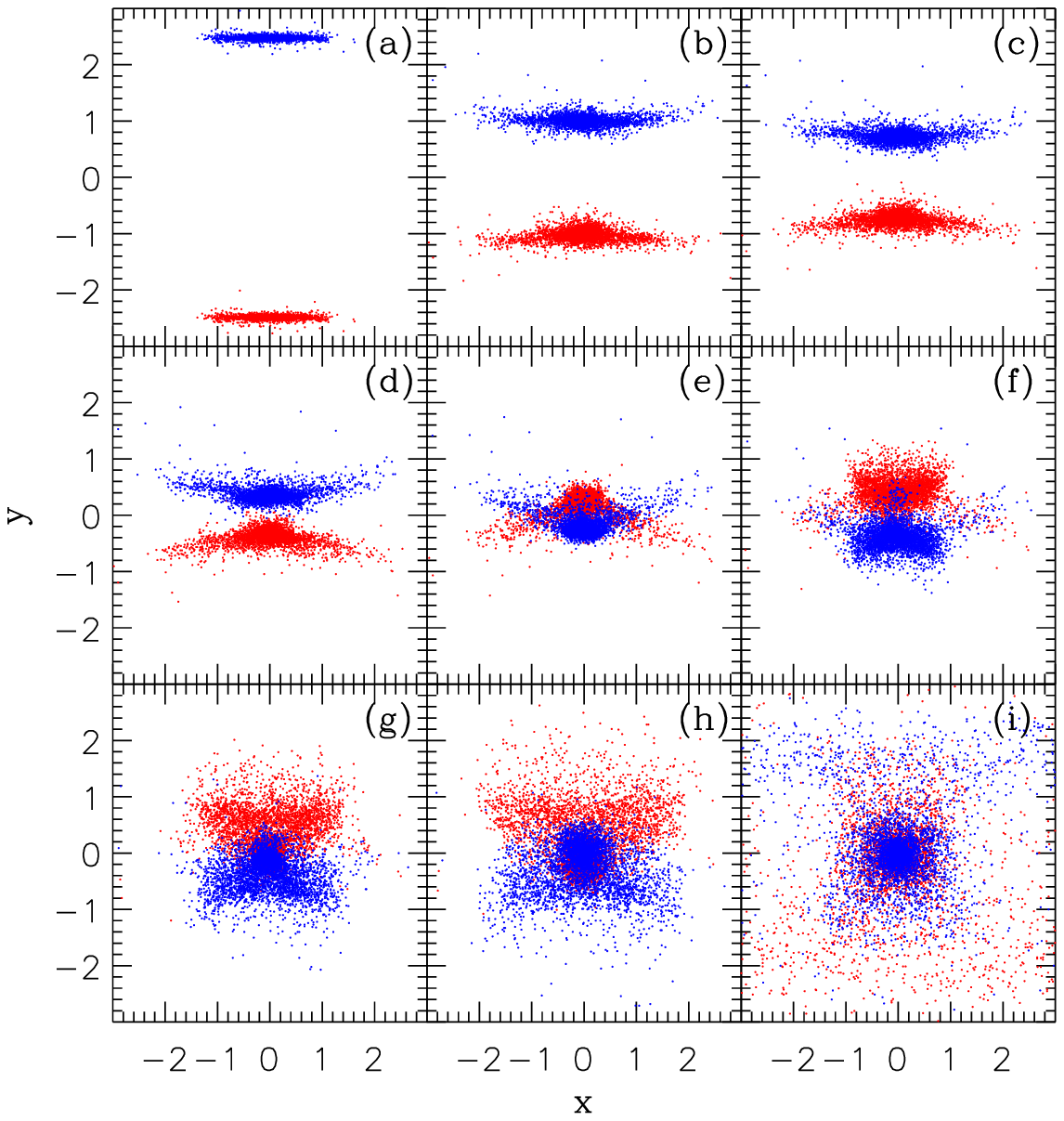}{Face-on collision of two disks}
{Major steps in the evolution of a face-on collision between two exponential disks. Initially (panel a), the disks are separated by 5 times their characteristic radius and without initial velocity. As the gravity gathers them (panels b, c and d), the central region of a disk goes deeper into the potential well of the other, while the external parts are less rapidly affected. After the first collision (panel e), the disks are dynamically heated and become thicker. The remnants slow down considerably (panels f and g) and fall back on each other (panel h) before being indiscernible. The panels corresponds to the computational times 1.0, 8.0, 8.5, 9.0, 9.5, 10.0, 10.5, 11.0 and 18.0.}

We choose to monitor the thickness of one of the disks. For each snapshot, we extract the $z$-coordinate of every particle and compute the standard deviation $\sigma_z$. Results are shown on \reffig{faceon_sigmaz}. The long distance interaction first warps the disk ($t \lesssim 9$, panels a b c and d on \reffig{faceon}). At $t \approx 10$ and later, the thickening is accelerated by the collision which unbinds a fraction of the particles, resulting in a increase for $\sigma_z$.

\fig{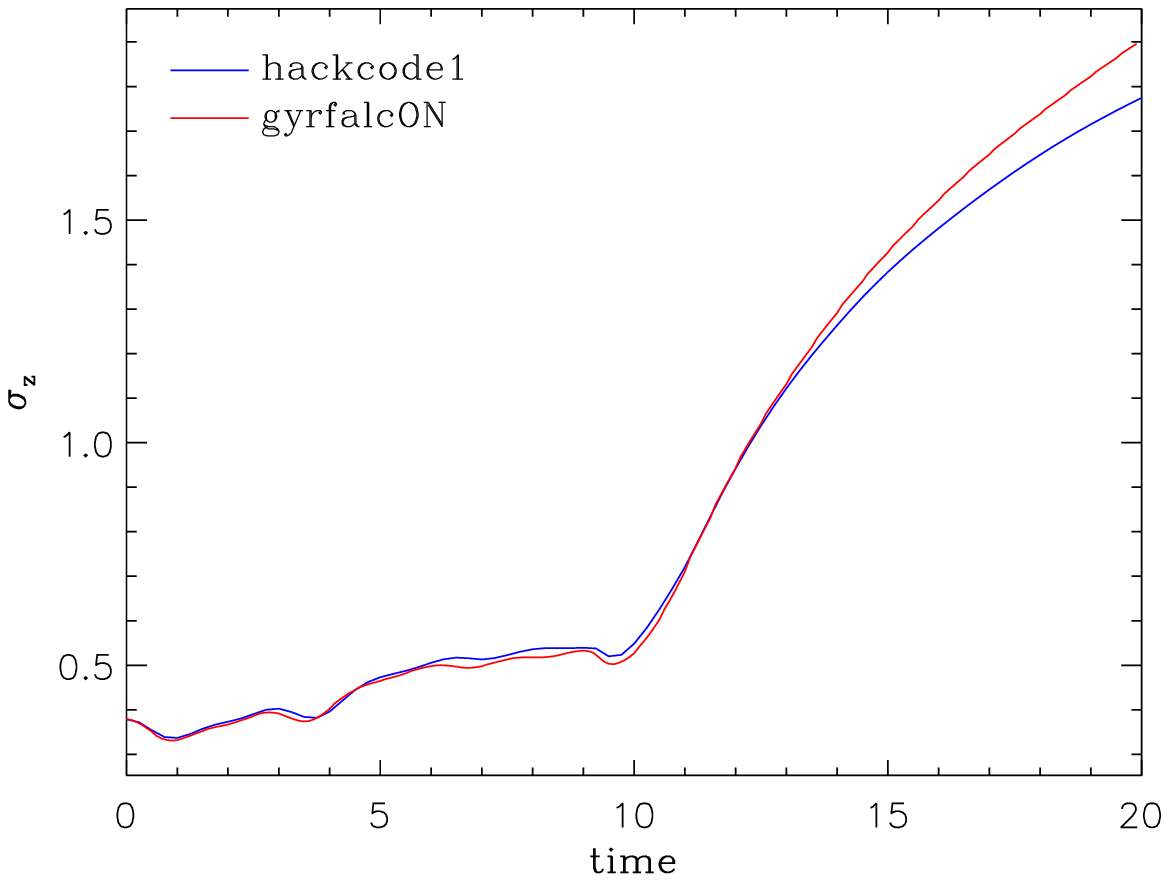}{Thickness of a disk during a face-on collision}
{Evolution of the thickness of one of the disks during the face-on collision shown in \reffig{faceon}. The thickness is measured as the standard deviation $\sigma_z$ of the $z$-coordinates of the particles. Both a classical tree code ({\tt hackcode1}) and Dehnen's code ({\tt gyrfalcON}) show similar results, within an error range of $\sim 5\%$. The increasing of $\sigma_z$ is well explained by the collision (see text for details).}

The two tree codes give comparable results (with relative differences of less than $\sim 5\%$). However, this does not mean that their results are physically correct. To check this, we monitor the conservation of energy and angular momentum while tuning the parameters of {\tt gyrfalcON} (see \reffig{faceon_energy}), like the smoothing length $\epsilon$, the softening kernel or the tolerance parameter $\theta$.

\fig{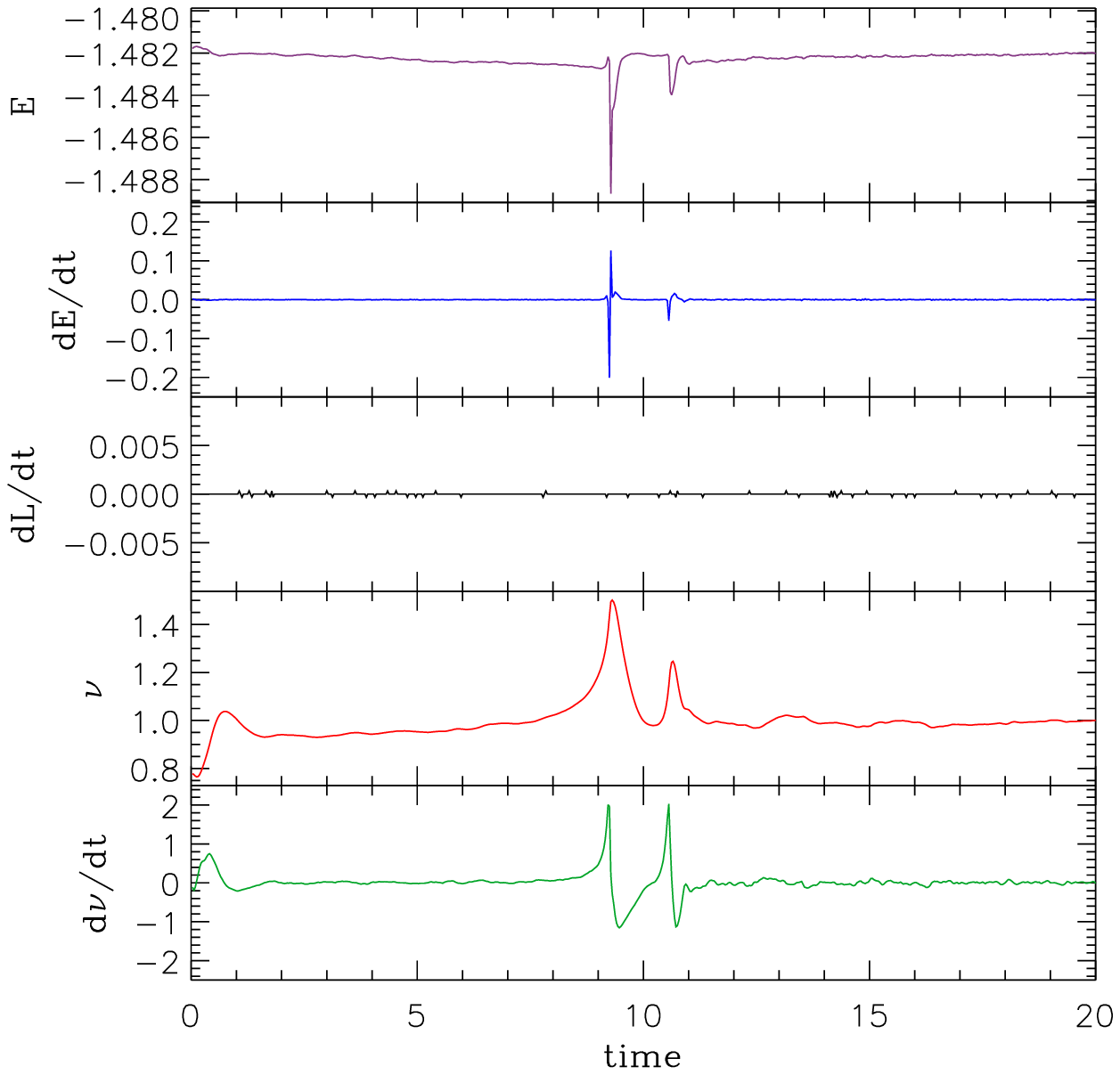}{Conservation of $E$, $L$ and $\nu$ during a face-on collision}
{Variations of the total energy $E = K + \Omega$ (with $K$ and $\Omega$, the kinetic and gravitational energies), angular momentum $L$ and virial ratio $\nu = -2K / \Omega$ during a face-on collision, as plotted on \reffig{faceon}. We note that the system is not virialized initially, but only after $t\approx 2$. The energy and the angular momentum are well conserved since the beginning of the simulation, showing that the code is well parametrized. However, the two collisions are clearly visible in the energy plot ($t \approx 9.2$ and $t \approx 10.5$, panels e and g on \reffig{faceon}). Here, the code faces the problem of the finite timestep, and is forced to sacrifice the accuracy to keep the computation fast. Anywhere else, the energy is conserved, with an relative error smaller than $10^{-3}$. We also note that the simulation shows the same level of energy and virial ratio before and after the collision. The virial ratio shows some fluctuations, but always less than $10\%$, which reflects the physical evolution of the system.}

Except for a brief moment during the collision phases, the energy is well-conserved\footnote{Note that the angular momentum is \emph{de facto} conserved by the intrinsic symmetry of the tree built by {\tt gyrfalcON}.}. When tuning the parameters, we noted that the accuracy remains at a similar level for $\theta \in [0.4, 0.6]$, but the computation is sped up for large values of $\theta$. This confirms that the default set of parameters ($\theta = 0.4$, kernel $P1$, see \refeqnt{kernel_p1}) allows a correct usage of the code, at this resolution (i.e. $N\sim 10^4-10^5$). For higher number of particles however, $\epsilon$ must be changed, according to the relation of \myciteauthor{Merritt1996} (\myciteyear{Merritt1996}, see also \refsec[numerical]{smoothing} of the present document).

The errors due to the collision are mainly due to the finite timestep used in the leap-frog scheme: when the acceleration becomes large, the potential energy is not properly converted into kinetic energy and the total energy $E$ is not \emph{exactly} conserved, as visible in the upper panel of \reffig{faceon_energy}. In other words, the code is too slow in updating the physical quantities, with respect to the underlying physics. This effect can be reduced by using individual adaptive timesteps. In this case, when a particle satisfies a certain criterion (e.g. it lies in a region of high density), its proper timestep is reduced so that time sampling becomes a less critical aspect of the integration. {\tt gyrfalcON} offers to play with adaptive timestep. After some rapid tests, it appeared that this was not critical and that the default timestep scheme (with small errors but rapid) was sufficient for our usage.

The broad conclusions of these tests are that (i) the smoothing length $\epsilon$ has to be chosen according to the number of particles, (ii) the tolerance parameter $\theta$ can be set $\approx 0.4$, and (iii) the kernel and the adaptive timestep are second-order parameters regarding to our study. With its very high efficiency, the code {\tt gyrfalcON} is perfectly suited for the simulations of interacting galaxies of this thesis.

%%%%%%%%%%%%%%%%%%%%%%%%%%%%%%%%%%%%%%%%%%%%%%%%%%
\sect[method_tt]{Computing the tidal tensor}

Once the $N$-body simulation is done, we have a proper description of the mass distribution, i.e. the gravitational potential. The next step of our study consists in evaluating the tidal tensor at various points of the simulation, to see how the tidal field evolves in space and time.

%%%%%%%%%%%
\subsect{Method}

To compute the tidal tensor in our simulations, we start with the positions and masses of the particles at a given time, as input data. Let's consider a particle P from our simulation. According to \refeqnp{introtide}, we simply need to compute the terms $\partial^j (a^i_\mathrm{ext})$. To do so, a pseudo-particle A (i.e. mass-less) is placed on each side of a cube of size $2\delta$ centered on P, as shown in \reffig{cube}. Then, the net acceleration from \emph{all} the $N$ particles in the simulation is computed at the positions of the pseudo-particles.

\fig{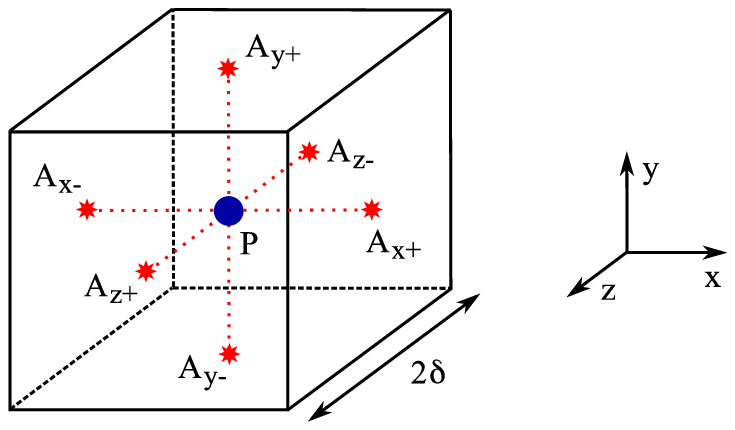}{Computation method of the tidal tensor}
{Cube used in the computation of the tidal tensor. The central point P is one of the particles of the $N$-body simulation, while the six points A are pseudo-particles, added to compute the tensor.}

We note that, in this case, we have not retrieved $a^i_\mathrm{ext}$ yet, but the total acceleration. Thus, we need to subtract the ``internal'' effects (i.e. the acceleration due to P at the positions of the A particles), that we do explicitly knowing the kernel and the distance $\delta$ between P and A. Once done, we pair the pseudo-particles of two opposite sides of the cube and perform the first-order differencing. For example, for the x-axis
\eqn[finitedifference]{
T^{xi} = \frac{a^i_\mathrm{ext}(A_{x+}) - a^i_\mathrm{ext}(A_{x-})}{2\delta}.
}

After doing this along the three directions, we get the nine components of the tidal tensor, in the base $\{\vectu{e}_x,\vectu{e}_y,\vectu{e}_z\}$ which is arbitrary. Therefore, the next step is to diagonalize the tensor and get the three eigenvalues $\{\lambda_i\}$ and the associated eigenvectors $\{\nu_i\}$.

\mybox[tidal_code]{
Technically, the code first reads a snapshot (called source) from the simulation to get the positions and masses of all P particles. Then, it builds a new snapshot, called sink, containing only the pseudo-particles (A). Both source and sink are given to {\tt getgravity}, a {\tt Nemo} tool that uses the same libraries as {\tt gyrfalcON}, with the same parameters as {\tt gyrfalcON} which has integrated the orbits in the simulation. Then, {\tt getgravity} retrieves the three components of the acceleration for all sink particles. Knowing $\delta$, $\epsilon$, the kernel used in the functional form of the gravity and the mass of the particle P, the effect of the central particle P on A is explicitly computed and subtracted from the total acceleration. The first-order difference scheme is then applied to retrieve only the upper part of the tensor (which is symmetrical). These six components are passed to the direct method of \mycitet{Kopp2008} that solves the characteristic equation of the tensor and derives efficiently and quickly the eigenelements. Note that the errors introduced by this method are generally close to machine precision and thus, well below these due to the computation of the accelerations. The experiment is then repeated for all snapshots. The complete algorithm is schematically summarized on \reffig{algo}.

A slightly modified version of this algorithm has also been implemented: instead of centering the cubes on particles of the simulation, they are placed on a regular 3D grid, independent of the mass distribution of the $N$-body model. In this case, there is no need to subtract the effect of a central particle in the cube.
}

\fig{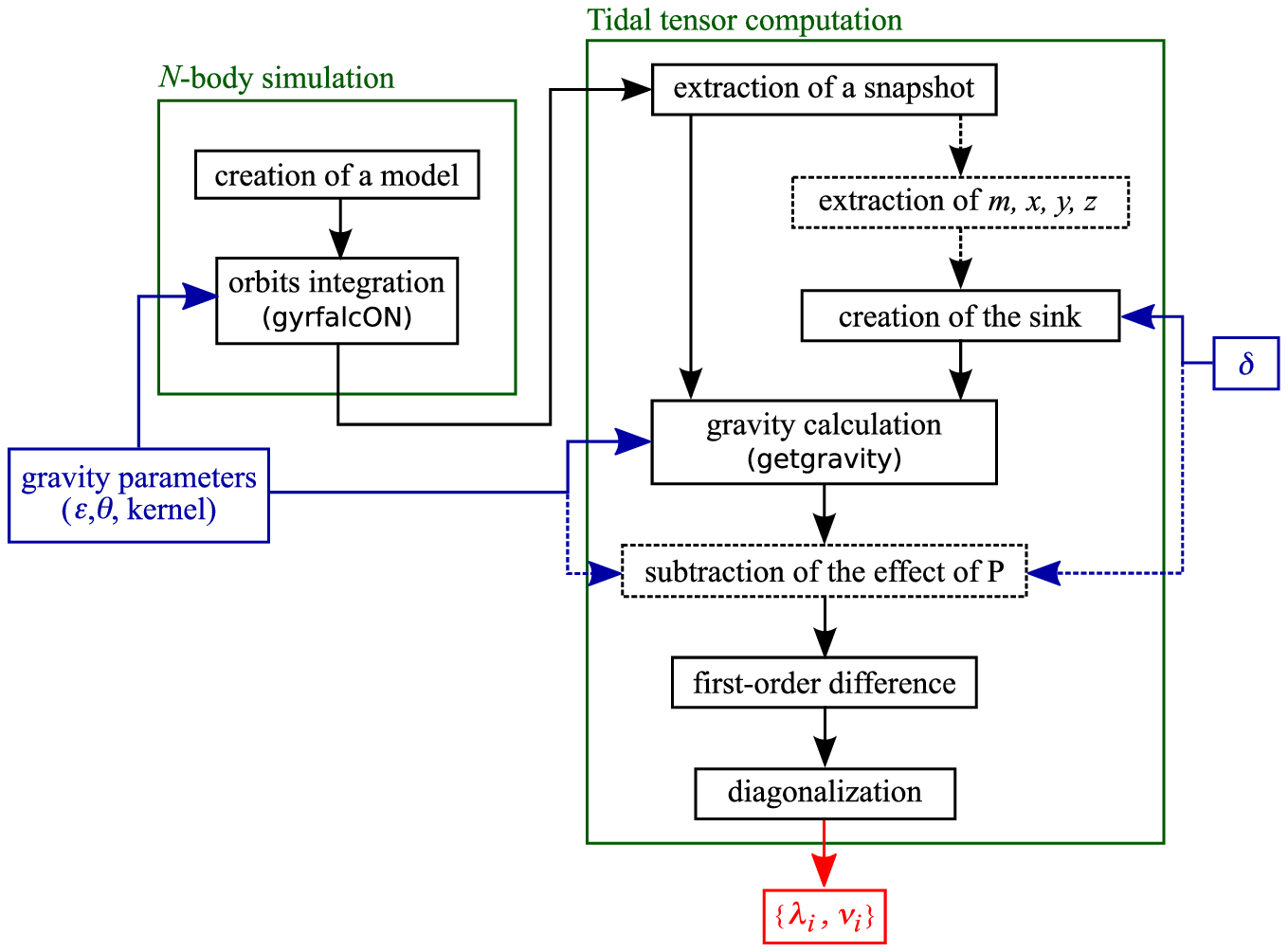}{Algorithm for the computation of the tidal tensor}
{Schematic algorithm of the computation of the eigenelements $\{\lambda_i, \nu_i\}$ from the creation of the $N$-body model. Steps represented with dashed lines are skipped in the grid-based version of the code.}

In the end, while the gravitational potential is resolved over the length $\epsilon$, the resolution of the tidal field is $\delta$.

%%%%%%%%%%
\subsect{Tests}

The only parameter of this method, $\delta$, must be chosen carefully to minimize errors. This problem is similar to the choice of the optimal smoothing length $\epsilon$. When setting the value of $\delta$, one has to cope with two sources of errors:
\begin{itemize}
\item Large values: the cube tends to include all the particles that build the external potential. The approximation of the distant source (see \refeqnt{deltatide}) is not valid.
\item Small values: the $N$-body rendition of the mass distribution fails to resolve the fine structure of the potential. In addition, round-off errors in the calculation of the finite-difference are introduced when the term $\left[a^i_\mathrm{ext}(A_{x+}) - a^i_\mathrm{ext}(A_{x-})\right]$ in \refeqn{finitedifference} becomes close to the numerical precision.
\end{itemize}

Therefore, we immediately have a lower limit, set by the accuracy criterion on the potential, with $\delta > \epsilon$. Practically, the round-off errors only kick in when $\delta$ is much lower than $\epsilon$, meaning that this will not influence our choice. But the problem of finding a suitable upper limit remains. To address this question, we ran tests on $N$-body renditions of well-known potentials and compared the results obtained with different values of $\delta$ to the analytical solutions derived in \refsecp[triplets]{ttexamples}.

The simplest case is the well-known Plummer model whose potential is
\eqn{
\phi(r) = -\frac{GM}{\sqrt{r_0^2+r^2}}.
}
Here, $M$ is the total mass and $r_0$ the core radius. However, this profile yields an important issue for its $N$-body rendition: it does not have a truncation radius. Indeed, its finite mass spreads over the entire space, which is unrealistic. That is why the models we use (since \reffig{plummer}) are artificially truncated. They are set up by the tool {\tt mkplummer} (within the {\tt Nemo} package) which considers that $99.9\%$ of the total mass is enclosed within a radius of $22.8042468$. Knowing from \refeqnp{plummerintegratedmass} that $99.9\%$ of the mass corresponds to a radius of $\approx 38.7 r_0$, we deduce that $r_0 \approx 0.589$ and can now normalize the results to compare them with the theory.

Let's begin with the $N$-body rendition of the potential. First, {\tt gyrfalcON} computes the gravitational potential at the positions of the $N=500\,000$ particles of our Plummer model, which we compare to the theoretical value. Results are shown in \reffig{potential_plummer}.

\fig{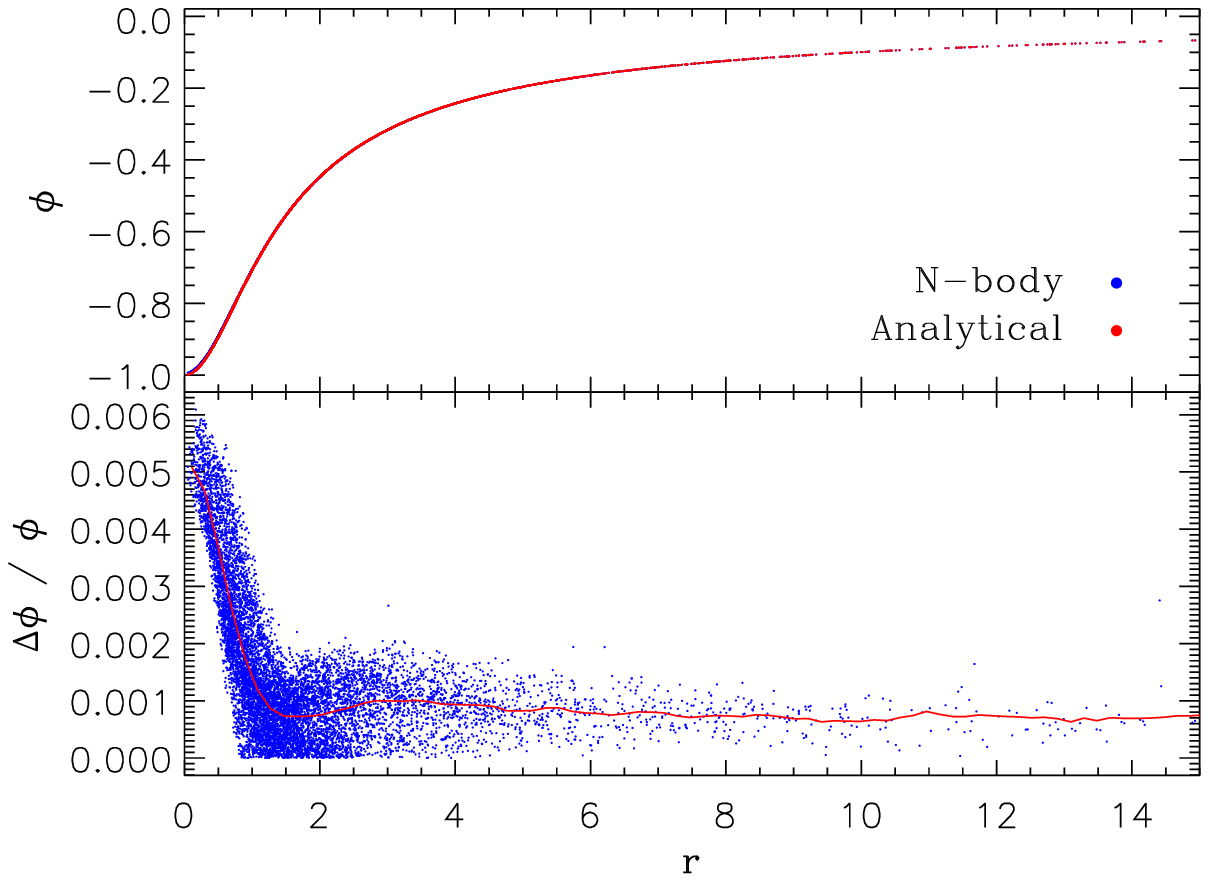}{Numerical potential of a Plummer sphere}
{\emph{Top:} Measured (blue) and theoretical (red) potential of an $N$-body realization of a Plummer sphere ($N=500\,000$). \emph{Bottom:} Relative error between the potential measured and the theoretical value. The red curve corresponds to the mean error per radial bin. For clarity, only 10\,000 points are plotted in this figure.}

Note that the standard deviation between the numerical and the theoretical values is $\sigma_\phi \approx 10^{-3}$. The relative error is everywhere lower that $0.6\%$ and thus, the potential is very well rendered. Next is the acceleration:
\eqn{
|\vect{a}(r)| = GM\frac{r}{\left(r_0^2+r^2\right)^{3/2}}.
}
\reffig{force_plummer} plots the norm of the acceleration at the positions of the particles and, once again, the comparison with theoretical value. In this case, one has $\sigma_a \approx 3 \times 10^{-3} \approx 3 \sigma_\phi$ and the relative errors reach $5\%$ locally. With the $N=500\,000$ particles of our test case, we have set $\epsilon = 0.01$. Note that the passage from the potential to the acceleration increases the level of error by one order of magnitude.

\fig{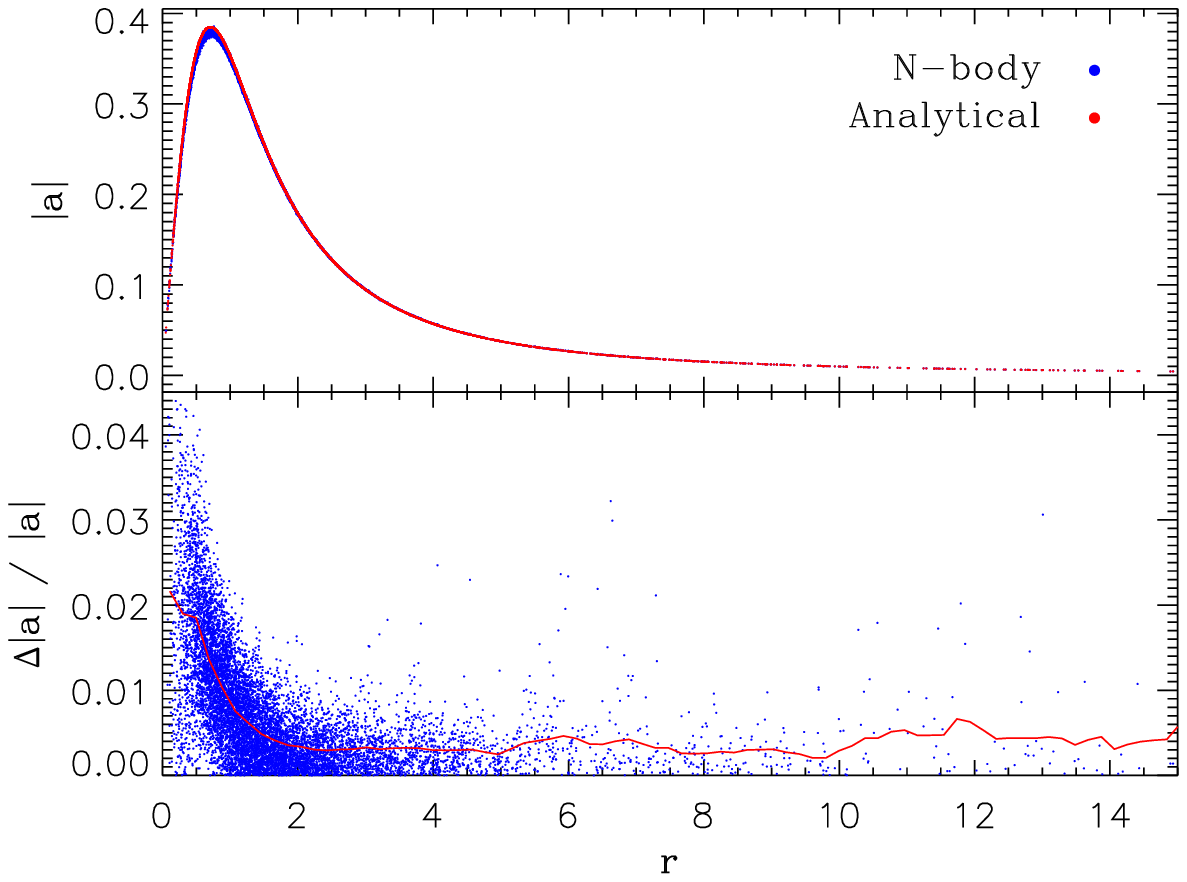}{Numerical acceleration of a Plummer sphere}
{Same as \reffig{potential_plummer}, but for the norm of the acceleration.}

Although the same experiment is possible for the tidal tensor, the analytical derivation of its eigenvalues is much more involved\footnote{The determinant of the tidal tensor, which is the product of its eigenvalues could also be used but again, its analytical derivation may be long, even for simple cases.}. Instead, we concentrate on the $T^{xx}$ term of the tensor, for $y=z=0$, i.e. through the center of the model:
\eqn{
T^{xx}(r) = -GM \frac{r_0^2 -2r^2}{\left(r_0^2+r^2\right)^{5/2}}
}
(see \refsect[triplets]{plummer} for a detailed derivation). Obviously, one will not find particles with $y=z=0$ in the simulation, so we need to use the grid-based version of the code (recall \refbox{tidal_code}) to compute the tensor along the x-axis. At a given resolution (i.e. for a certain number of particles), we vary the size of the cube $\delta$, using the same parameters to compute the gravity as in {\tt gyrfalcON}. \reffig{tt_plummer} and \reffig{tt_plummer_zoom} show the results.

\fig{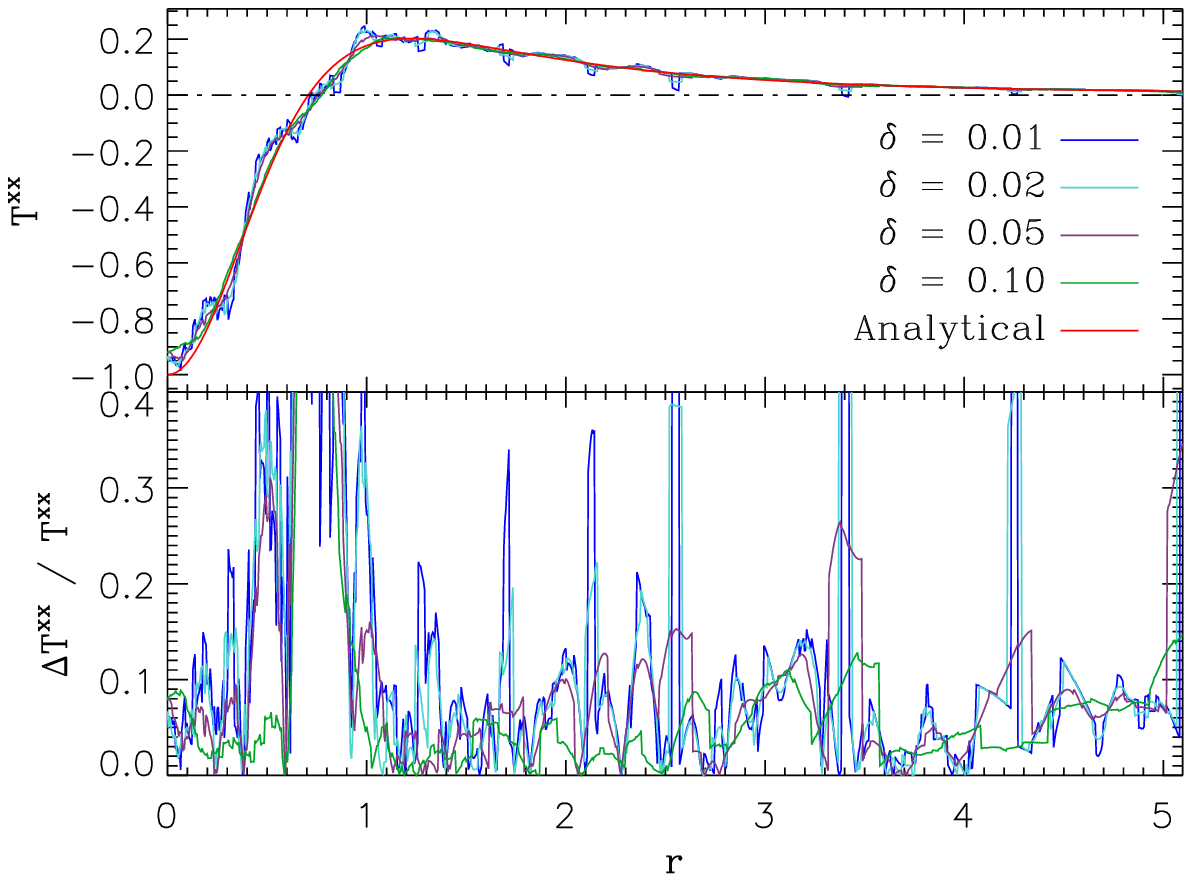}{Numerical tidal tensor of a Plummer sphere}
{\emph{Top:} $T^{xx}$ term of the tidal tensor of a Plummer sphere. The tensor is computed at 1\,000 positions along the radial axis. \emph{Bottom:} Relative error between numerical and theoretical values. Note that the largest values correspond to extremely small absolute values of the tensor, and thus are not directly relevant as a measurement of quality.}

\fig{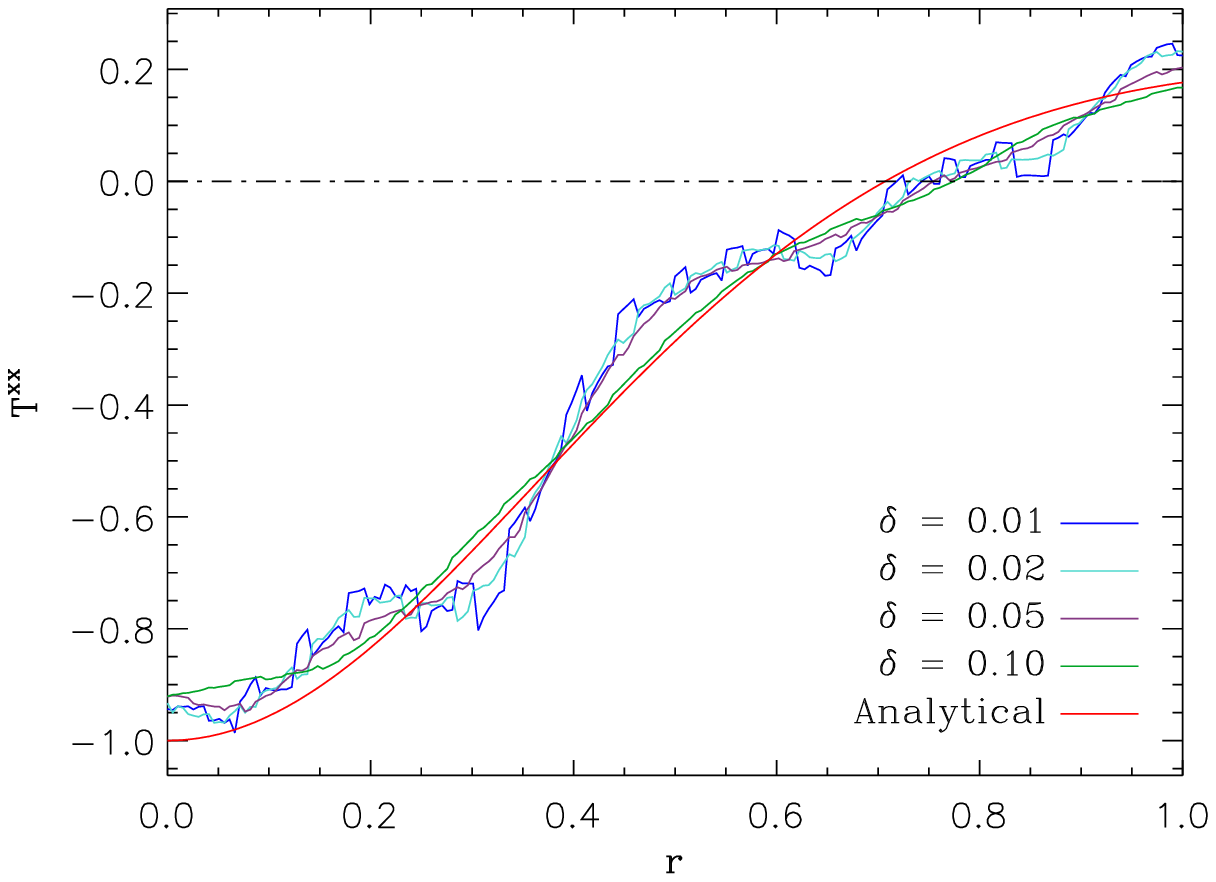}{Numerical tidal tensor of a Plummer sphere (zoomed-in)}
{Zoom-in on the inner region of \reffig{tt_plummer}. Large values of $\delta$ are generally smoother (less noisy) than small values but further from the analytical solution (red curve).}

\tab{5cm}{tt_std}
{$\sigma_T$ for several $\delta$'s}
{c@{\extracolsep{\fill}}c}
{
$\delta$ & $\sigma_T ~[\times 10^{-2}]$ \\
\hline
0.01 & $3.3$ \\
0.02 & $2.8$ \\
0.05 & $2.2$ \\
0.10 & $1.8$ \\
}{}

The standard deviations are given in \reftab{tt_std}. As for the passage from the potential to the acceleration, the computation of the tidal tensor from the acceleration increases the level of error by one order of magnitude. This means that our method used to compute the tensor has a quality comparable to those of the tree code, used to get the acceleration.

\reffig{tt_plummer} exhibits small, yet significant errors which are mainly visible for $r \approx$ 2.1, 2.5, 3.4 and 4.3. The origin of these errors is a problem which is discussed in details in \refapp{glitches}.

As predicted, large cubes smooth the potential and thus the terms of the tidal tensor, while smaller cubes do not correctly reproduce the fine structures. However, with a too large cube, the finite-difference approximation is not valid anymore and an error is made in our calculation of the tensor. As visible in \reffig{tt_plummer_zoom}, the curves corresponding to large $\delta$ are smooth and regular but quite far from the analytical value. \emph{A contrario}, the small values yield noisy curves, but on average closer to the theoretical solution\footnote{Note that, while the noise comes from the finite differences scheme, the systematic errors (i.e. the noiseless signal) are mainly due to the approximation made on the value of the force, when using a kernel.}.

Once again, the choice is arbitrary. An intermediate value like $\delta = 0.05$ seems to take into account both sources of error, reasonably. As mentioned before, the size of the cube strongly depends on the local description of the potential, that is the softening length $\epsilon$. In the previous example, we had $\epsilon = 0.01$ and $\delta = 0.05$. This suggests $\delta = 5 \epsilon$.

In principle, the mean density in a cube must be independent of the resolution, i.e. $N\delta^3$ must be constant. With this assumption and the relation for $\epsilon$ from \mycitet{Merritt1996}, it is possible to test the validity of our empirical relation $\delta = 5\epsilon$, as shown in \reffig{cube_resolution}.

\fig{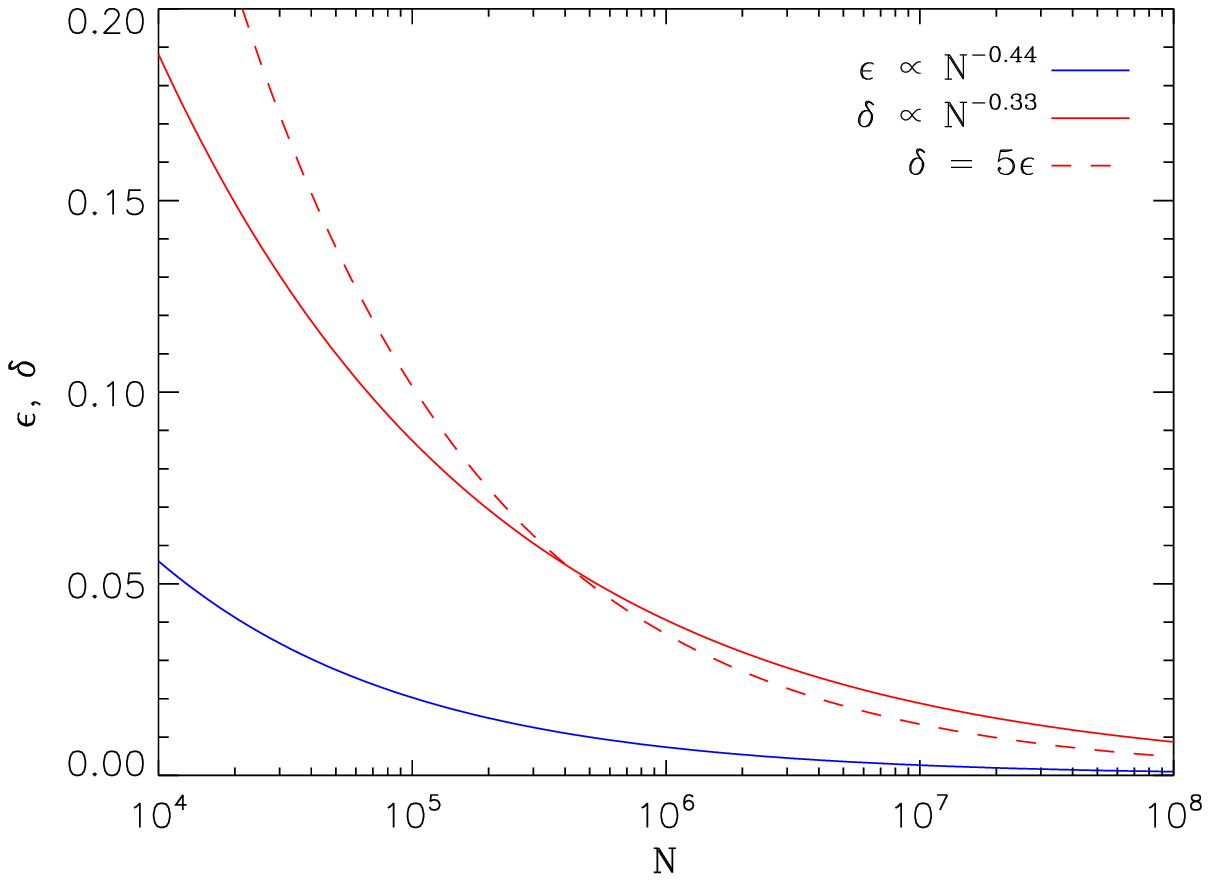}{Relation between $\epsilon$, $\delta$ and $N$}
{The relation giving the optimal value for the softening length $\epsilon$ (blue curve) from \mycitet{Merritt1996} is compared to those assuming a constant density in the cubes (red), and to the empirical scaling law (red dashed). The proportionality factors have been chosen to get the same value for $\delta$ at a resolution of $N=500\,000$ particles.}

We note that the empirical relation matches those from the density of the cubes over a couple of decades, but the deviation becomes too large for small numbers of particles. The differences introduced when using one or the other relation are comparable to the changes for $\delta$, already discussed in \reffig{tt_plummer} and \reffig{tt_plummer_zoom}.

One must not forget that all scaling laws above depend on the actual distribution of the mass. Therefore, one should not rely too strongly on them and introduce some flexibility. Obviously, the precise determination of the optimal values of $\epsilon$ and $\delta$ is out-of-reach for non analytical systems. That is why, knowing that the values chosen would rarely be optimal, we decide to simply use the quantities $(N\epsilon^3)$ and $(N\delta^3)$ as constants, in the range of our typical resolutions ($\sim 10^{5-6}$ particles).

%%%%%%%%%%%
\subsect[alternative_method]{Alternative method}

In addition to the finite differences method presented above, another simple approach is possible. In an $N$-body system, the gravitational potential at the point of coordinates $(x,y,z)$ is given by the sum of individual contributions of all the particles:
\eqn{
\phi(x,y,z) = \sum_{k=1}^{N} \varphi_k(x,y,z),
}
where $\varphi_k$ is the potential of the $k$-th particle, evaluated at $(x,y,z)$. Thus, the components of the tidal tensor at this position read
\eqn[alternative_method]{
T^{ij}(x,y,z) & = -\partial^i \partial^j \phi(x,y,z) \nonumber \\
& = - \partial^i \partial^j \sum_{k=1}^{N} \varphi_k(x,y,z) \nonumber \\
& = \sum_{k=1}^{N} -\partial^i \partial^j \varphi_k(x,y,z) \nonumber \\
& = \sum_{k=1}^{N} \tau^{ij}_k(x,y,z),
}
where $\tau_k$ is the tidal tensor of the $k$-th particle. Here, $T^{ij}$ can be interpreted as the sum of the individual contributions of the particles to the tidal tensor.

A simple approach uses a point-mass potential for the particles, which leads to a tidal tensor of the form
\eqn{
\tau^{ij}_k(x,y,z) = \frac{GM}{r^5}\left( 3x_ix_j - \delta^{ij} r^2\right).
}
However, as we have seen above, a kernel is often used to smooth the potential of the particles (see \refsect[triplets]{kernels}). Once the tidal tensor associated with the right kernel is known, one has to sum the contribution of all the particles, at the position $(x,y,z)$. Finally, the total tidal tensor $\matr{T}$ can be set in diagonal form, as before.

The obvious advantage of this method is to eliminate the errors due to the finite differences scheme, as presented in the last Section. However, the previous implementation makes intensive use of Dehnen's tree code to compute the accelerations on the cubes. Therefore, it can be applied efficiently to high resolution simulations, for a small computational cost. It is also possible to implement a tree method for our alternative approach. However, the limited time of this thesis did not allow such an exploration and we have only implemented a (time-wise costly) direct method.

The results for our Plummer test case are shown in \reffig{alternative_method_tt}. We also illustrate here the effect of the kernel. The finite difference scheme (which uses a kernel $P1$, see \refeqnt{kernel_p1}) is a noisy version ($\sigma = 0.048$) of the associated alternative method ($\sigma = 0.040$), where maximum errors are found where the gradient of the acceleration is important (see \reffig{force_plummer}). However, the computational cost of the direct, alternative method does not worth the gain in precision for the solution.

\fig{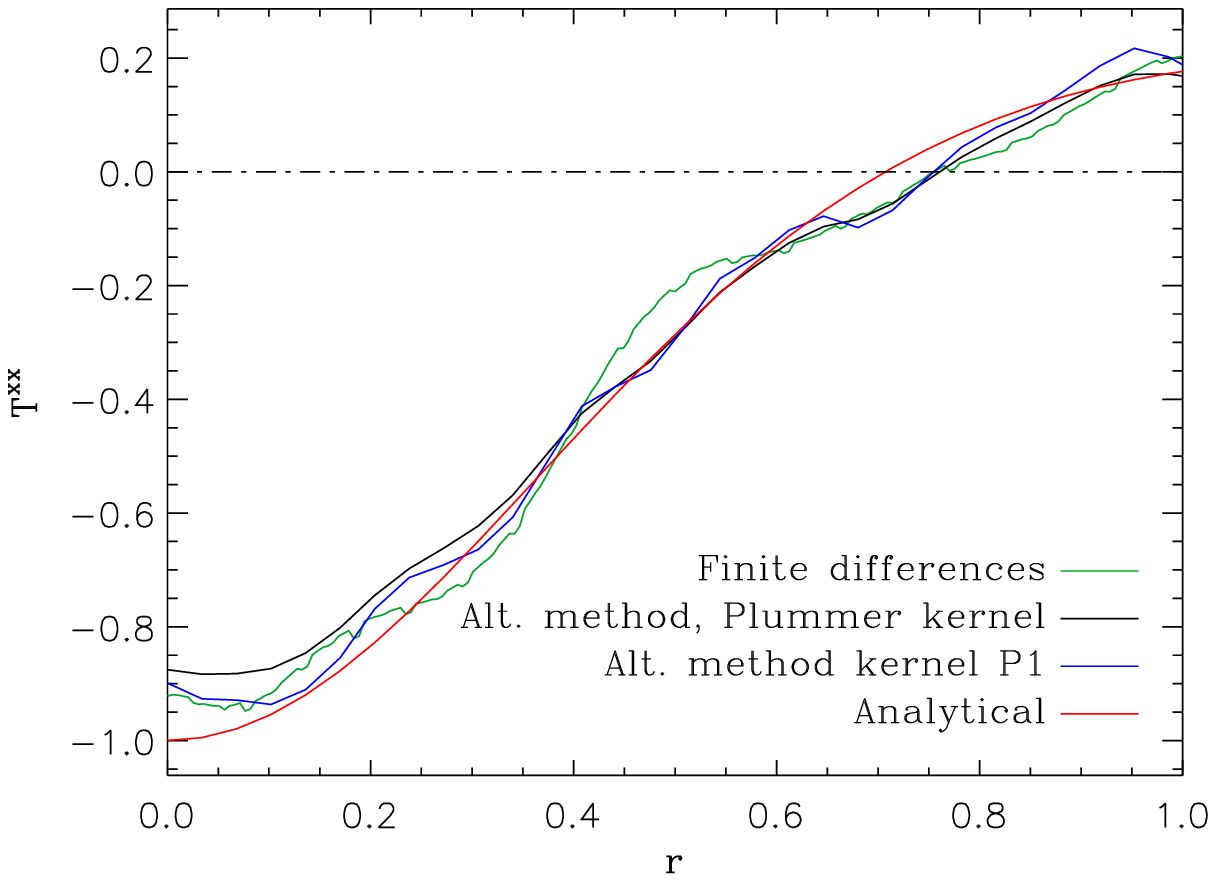}{Alternative method to compute the tidal tensor}
{Comparison between the finite differences methods and the alternative method, for again a Plummer sphere with $N=500\,000$. The black curve represents the tidal tensor computed with a Plummer kernel for the particles, while the blue curve denotes a $P1$ kernel (see text). The green curve is the result found with finite differences (kernel $P1$ and $\delta = 0.05$).}

The results presented in the following Chapters have all be obtained with the first-order differencing scheme.

%%%%%%%%%%%%%%%%%%%%%%%%%

\cha{antennae}{The Antennae galaxies}
\chaphead{This Chapter aims to introduce the galaxy merger which has been used as an observational reference throughout this work. Then, it details the results presented in \mycitet{Renaud2008}, and in the Sections~4 and 6 of \mycitet{Renaud2009}.}

%%%%%%%%%%%%%%%%%%%%%%%%%%%%%%%%%%%%%%%%%%%%%%%%%%
\sect{A traditional merger}

%%%%%%%%%%%
\subsect{A long time ago, in two galaxies (not so?) far far away...}
The main difference between astrophysics and the other fields of science is that any experiment is impossible to realize. Therefore, our theories rely on the simulations that we constantly compare with the observations of as many objects as possible. To study a phenomenon (galaxy mergers in our case), one has to collect a lot of high precision data to validate the theories. That is why the most observed objects are often the most reproduced numerically.

For mergers, the natural candidate is the Antennae galaxies (NGC~4038/39, \reffig{antennae}), named after the peculiar shape of their long tails, already observed in photographic plates by \mycitet{Duncan1923}\footnote{Back then, the two galaxies were still considered as a single \emph{nebula}.} who noted ``two faint extensions, like antennae, [...] both concave toward the west'' (see also \mycitealt{Schweizer1978}). Indeed, the Antennae are the closest major merger observed, and thus one of the most studied\footnote{During the 30 months of this PhD, the ADS system has referenced over 170 publications mentioning the Antennae, i.e. one every 5 days!}.

\fig{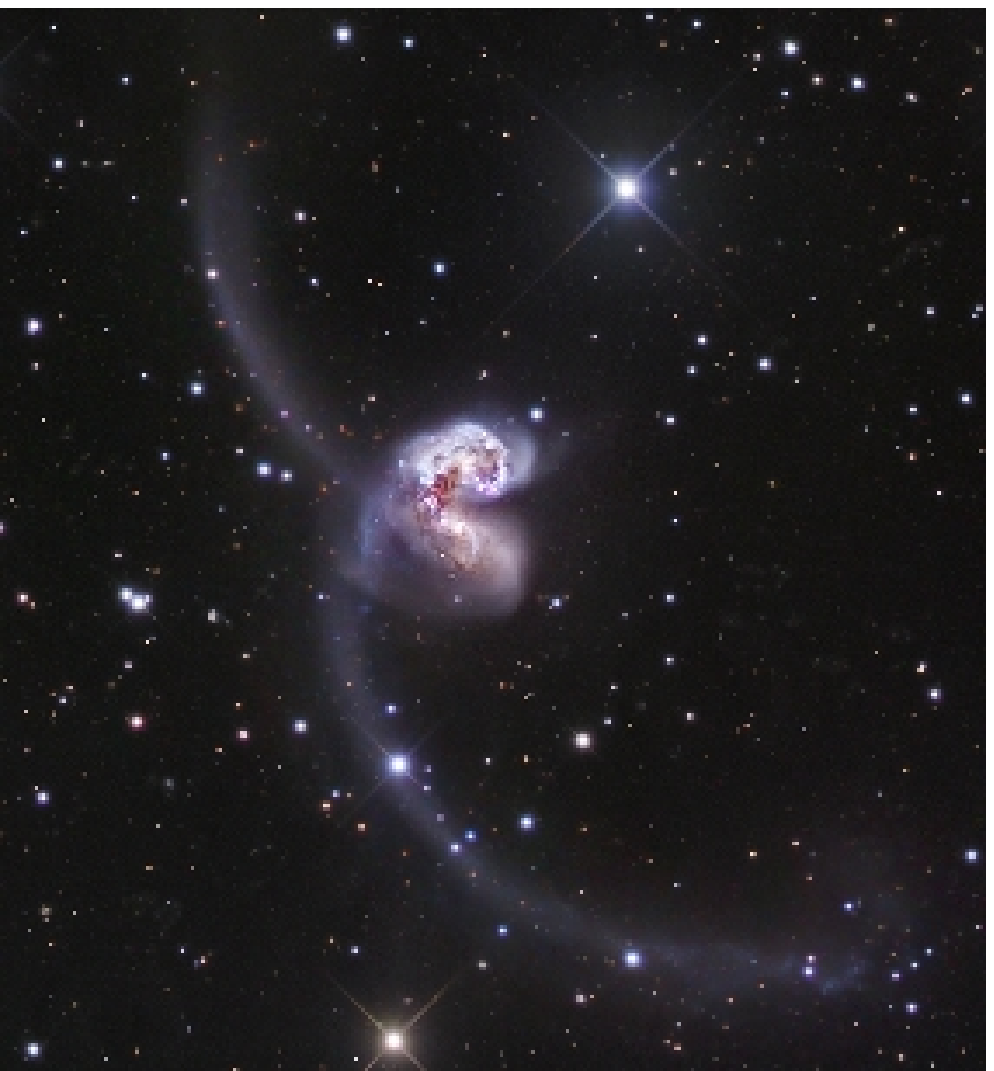}{Tidal tails of the Antennae}
{The Antennae show two extended tidal tails as well as a complex web of structures in the central region. North is up, east is left. (Image by R. Gendler.)}

The distance to this system has been subject to discussions over the last decades. Such an estimate is of paramount importance as it is used to derive almost all the other physical quantities (like the age, the size and even the mass). Many studies established distances ranging from $6 \Mpc$ to $30 \Mpc$ (see e.g. \mycitealt{Rubin1970}; \mycitealt{Whitmore1995}; \mycitealt{Fabbiano2001}; \mycitealt{Hibbard2001}; \mycitealt{Zezas2002}; \mycitealt{Saviane2004}). This controversy is mainly due to the large errors introduced by every method.

Up to recently, most of the authors agreed on a value of $19.2 \Mpc$, derived from the recession velocity (\mycitealt{Whitmore1999}). \mycitet{Saviane2008} used photometry of the tip of the red giant branch to determine a distance of $13.3 \pm 1.0 \Mpc$, much shorter than the commonly used value. Few months before (on December 18, 2007), \mycitet{Drake2007} had detected a new type-Ia \acr{}{SN}{supernova} (called SN~2007sr) in the southern tail of the Antennae. \mycitet{Schweizer2008} used it as a standard candle to find a distance of $22.3 \pm 2.8 \Mpc$, in agreement with models of large scale flows of galaxies (\mycitealt{Tonry2000}). Clearly, the debate is still open.

In the following, we adopt the common value of $19.2 \Mpc$, which gives a total extension of the Antennae (from the tip of one arm to the one of the other) of $\sim 110 \kpc$.

\mybox{
If one assumes a different distance for the Antennae, the intrinsic scales will change. Let $D = d' / d$ be the ratio between a distance $d'$ (e.g. $13.3 \Mpc$) and the value adopted $d$ (here $19.2 \Mpc$). Starting from the observational data, the apparent diameter and the observed magnitude of an object must be the same for all distances. This implies that the other physical quantities (e.g. size, mass, timescale) are different. The invariance of the apparent diameter gives the change in length:
\eqn{
R' = R \ D.
}
Usually, the absorption term is implicitly included in the apparent magnitude. Here, we explicitly write the absorption $A_\star$ as an additional factor to the magnitude \emph{without absorption}, $m_\star$, to get the observed apparent magnitude:
\eqn{
m_\mathrm{obs} = m_\star + A_\star,
}
i.e. the quantity measured by the telescope. With the definition of the distance modulus, one can link $m_\mathrm{obs}$ to the absolute magnitude $M_\star$
\eqn{
m_\mathrm{obs} = M_\star + A_\star + 5\log{\left(\frac{d}{10 \pc}\right)}.
}
$m_\mathrm{obs}$ is also invariant, therefore
\eqn{
m_\mathrm{obs} & = m'_\mathrm{obs} \nonumber \\
M_\star - M'_\star & = \left( A'_\star - A_\star \right) + 5\log{\left(D\right)}.
}
The term $A_\star$ includes all the effects of absorption, i.e. from the atmosphere (for ground based telescopes), the Milky-Way, the intergalactic medium and the inner absorption of the object itself. When varying the distance to the Antennae, only the intergalactic contribution will change. For instance, when going from $13 \Mpc$ to $19 \Mpc$, the column density of the intergalactic medium increases by $6 \Mpc$. However, the densities measured there are so low that one can reasonably simplify the problem by assuming that the difference is negligible (\mycitealt{Schmidt1975} derived the order of magnitude of $10^{-3} \U{mag\ Mpc^{-1}}$), in other words $A'_\star - A_\star \approx 0$.

The absolute magnitude is also linked to the luminosity $L_\star$ through
\eqn{
M_\star = -2.5 \log{\left(L_\star\right)} + k,
}
where $k$ is a constant.
\follow}\mybox[nocnt]{
The mass-to-light ratio $\Upsilon \equiv M / L_\star$ does not vary with the distance $d$, thus we have
\eqn{
M_\star - M'_\star & = -2.5 \log{\left(\frac{M}{\Upsilon}\right)} + k + 2.5 \log{\left(\frac{M'}{\Upsilon}\right)} - k \nonumber \\
5\log{\left(D\right)} & = 2.5 \log{\left(\frac{M'}{M}\right)} \nonumber \\
M' & = M \ D^2,
}
which is the equivalent relation for the mass. Then, for the density ($\rho \propto M / R^3$) and the timescale ($t \propto \rho^{-1/2}$), we get
\eqn{
\rho' & = \rho \ D^{-1} \\
t' & = t \ D^{1/2}.
}

In other words, if $d' < d$, the density of a cluster is higher if the Antennae are situated at $d'$ than if they are at $d$.

It is usual (see \refbox{schmidt}, below) to consider that the stars are formed when the density of the interstellar medium overcomes a threshold $\rho_\mathrm{c}$, which must be independent of the distance between the galaxies and the observer. Therefore, $\rho_\mathrm{c}$ is reached more easily in dense galaxies, or in our case, when the Antennae are close. That is, if the Antennae are situated at a distance $d' = 13.3 \Mpc$ instead of $d = 19.2 \Mpc$, the formation of stars must be much more efficient.

Indeed, consider that we want to form a cluster of mass $M$ (the same for all distances), from a molecular cloud of size $\ell_1$ (respectively $\ell'_1$) and density $\rho$ (respectively $\rho'$). We need to condense the cloud from its original size $\ell_1$ to a smaller object of size $\ell_2$ and density
\eqn{
\rho_\mathrm{c} & = c \ \rho \nonumber \\
\frac{M}{\ell_2^3} & = c \frac{M}{\ell_1^3} \nonumber \\
\ell_2 & = c^{-1/3} \ell_1
}
where $c$ is a constant. The laws of physics describing the formation of stars are independent of the distance and therefore
\eqn{
\rho'_\mathrm{c} & = \rho_\mathrm{c} \nonumber \\
c' \ \rho' & = c \ \rho \nonumber \\
c' & = c \ D.
}
\follow}\mybox[nocnt]{
The mass being also independent of $d$, we have
\eqn{
\rho' {\ell'_1}^3 & = \rho \ell_1^3 \nonumber \\
\ell'_1 & = \ell_1 \ D^{1/3},
}
which finally leads to 
\eqn{
\ell'_2 = \ell_2.
}
Once we have these relations, let's have a look at the influence of $d$ on the strength of the tidal field denoted by $\lambda$, which has the dimension of $G\rho$, that is
\eqn{
\lambda' = \lambda \ D^{-1}.
}
The tidal energy per mass element for the cloud of size $\ell_1$ is
\eqn{
E_\mathrm{tides, 1} = -\frac{1}{2} \lambda \alpha \ \ell_1^2
}
(see \refsect[virial]{virial_tides}), which means that
\eqn{
E'_\mathrm{tides, 1} = E_\mathrm{tides, 1} \ D^{-1/3}.
}
Once the stars have formed, the region to consider has the size $\ell_2$ and the tidal energy becomes
\eqn{
E'_\mathrm{tides, 2} = E_\mathrm{tides, 2} \ D^{-1}.
}
When compared to the gravitational energy $\Omega$ per unit mass, we have
\eqn{
\frac{E'_\mathrm{tides, 1}}{\Omega'_1} = \frac{E_\mathrm{tides, 1}}{\Omega_1}
}
for the cloud, and
\eqn{
\frac{E'_\mathrm{tides, 2}}{\Omega'_2} = \frac{E_\mathrm{tides, 2}}{\Omega_2} \ D^{-1}
}
for the cluster. 

In the end, the role of the tidal field remains the same or is slightly underestimated if the distance to the Antennae drops from $19.2 \Mpc$ to $13.3 \Mpc$. Note, however, that all order of magnitude computations are not affected by these uncertainties.
}

%%%%%%%%%%%
\subsect{Meet a strange animal}

During a symposium in (formerly West-) Germany, \mycitet{Schweizer1978} referred to the Antennae as an ``animal from the zoo of Alar Toomre''. He categorized the galaxies in the subspecie of the ``animals which always huddle in pairs, each individual having one long tail''. Even if the long tidal tails are the first visual signature of the Antennae, most of the high and very high resolution observations have been performed in the central part of these galaxies. They revealed an incredibly rich and complex web of substructures, a sign of a strong activity in the last $\sim 10^7 \yr$ (\reffig{antennae_center}).

\fig{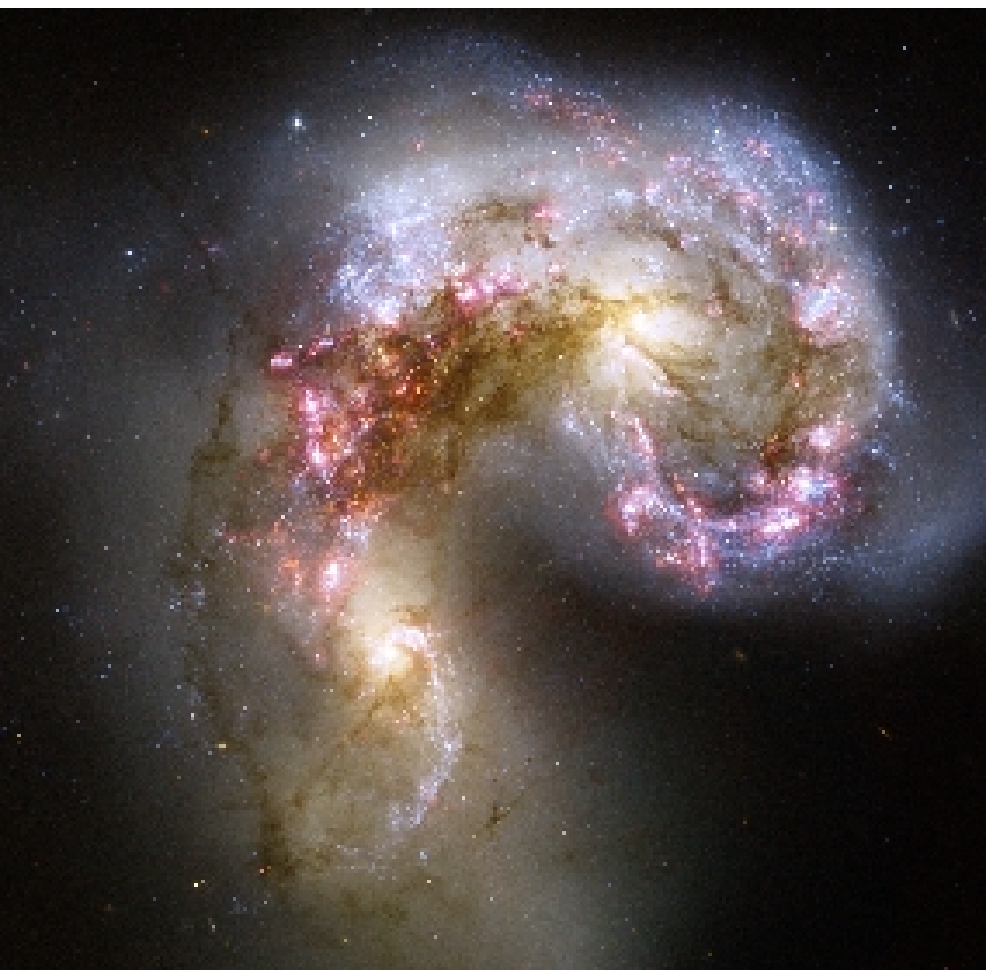}{Nuclei of the Antennae}
{The central part of the Antennae gathers the two nuclei of the progenitor galaxies as well as numerous star clusters created during their encounter. The area between the two nuclei filled with dust (brown) and molecular clouds (red/pink) is called the overlap region. Note that the tidal tails (not shown here) crosses, so that the southern tail is associated to the northern nucleus (NGC~4038) while the northern smaller arm is linked to the southern core (NGC~4039). (Image by B. Whitmore, HST.)}

Velocity maps and early simulations of the Antennae suggest that the two nuclei have already undergone a passage next to each other and that we are observing the second one\footnote{Note however that this contradicts the classification of the Antennae as the first step of the Toomre sequence (see \reffigt{toomre_sequence}) while the next step (represented by the Mice) clearly shows an undergoing first passage. See also \mycitet{Karl2010}.}. These two passages would be responsible for the fireworks we see today. One might notice the major features of this region: the arc-shape structure in the northern part surrounding the nucleus of NGC~4038, and the overlap region between the two cores. In both areas, a lot of star clusters are visible. It is likely that they originate from the tidal interaction, because they show ages much younger than those of the galaxies. This recent and violent episode of star formation, called starburst, would be typical for interacting galaxies. However, the exact recipe of this formation is still subject to debate, because of the complexity of such a system.

\fig{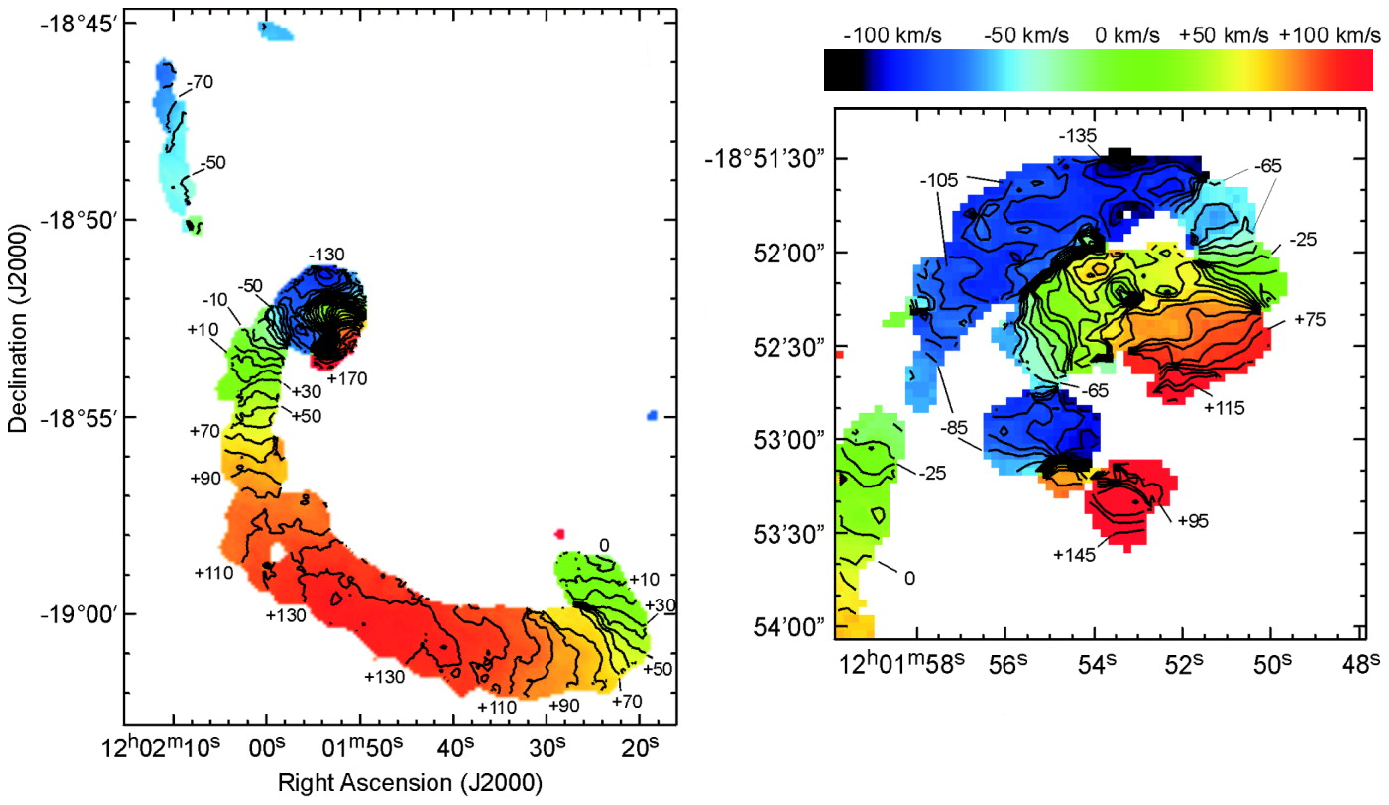}{HI maps of the Antennae}
{Intensity-weighted \hi velocity map of the Antennae (\emph{left}) and their central region (\emph{right}). The color scale is the same for both images, with green indicating velocities close to systemic. On the right-hand side panel, one can distinguish the rotation of the nuclei, especially for NGC~4039, in the bottom part. (Adaptation of Figures 2 and 9 of \mycitealt{Hibbard2001}.)}

In order to better understand the Antennae, one has to gather morphological and kinematical data, from the numerous different observational sources. The work of \hbox{\mycitet{Hibbard2001}} brings many clues, through \hi maps. E.g., in \reffig{hibbard}, one can easily distinguish the two progenitors thanks to steep velocity gradients. Note also the peculiar distribution of \hi in the northern tail, much shorter than the southern. This suggests that the major merger is not a symmetric object, and that NGC~4038 (northern nucleus, southern tail) dominates the present distribution of matter.

The merger has been observed in all the other wavelengths: e.g. radio (\mycitealt{Hummel1986}), far-\acr{}{IR}{infrared} (\mycitealt{Bushouse1998}), mid-\acr{}{IR}{infrared} (\mycitealt{Wang2004}), near-\acr{}{IR}{infrared} (\mycitealt{Brandl2005}), \acr{}{UV}{ultraviolet} (\mycitealt{Hibbard2005}) and X-ray (\mycitealt{Baldi2006}). With such a mine of informations available, drawing the portrait of the Antennae is out of the scope of the present document. Important facts about the star clusters and the \acr[s]{}{TDG}{tidal dwarf galaxies} are quickly presented in the following. However, the reader is strongly encouraged to consult these papers and the references therein for more details.

%%%%%%%%%%%
\subsect[antennae_clusters]{Observations of star clusters...}
Over the last two decades, both observational and theoretical studies have shown that intense episodes of star formation take place in interacting galaxies. As the closest sample, the star clusters of the Antennae have been explored by several authors who have detected some 1000 of them in the central region of the merger (\mycitealt{Whitmore1995}; \mycitealt{Meurer1995}; \mycitealt{Mengel2005}; \mycitealt{Whitmore2007}; \mycitealt{Bastian2009}; \mycitealt{Whitmore2010}).

Using deep images in the $V$ band, \mycitet{Zhang1999} derived the mass function of these clusters and found an important discrepancy with respect to those of other galaxies (\reffig{zhang99}). Indeed, the usual distribution of magnitudes of old globular clusters follows a Gaussian law, which gives a log-normal shape for the mass function when assuming a constant mass-to-light ratio (\mycitealt{Harris1991}). In the Antennae however, the mass function seems to be best described by a power law $M^\beta$ of index $\beta \approx -2$ for the range $10^4 \msun < M < 10^6 \msun$ (solid curves in \reffig{zhang99}).

\fig{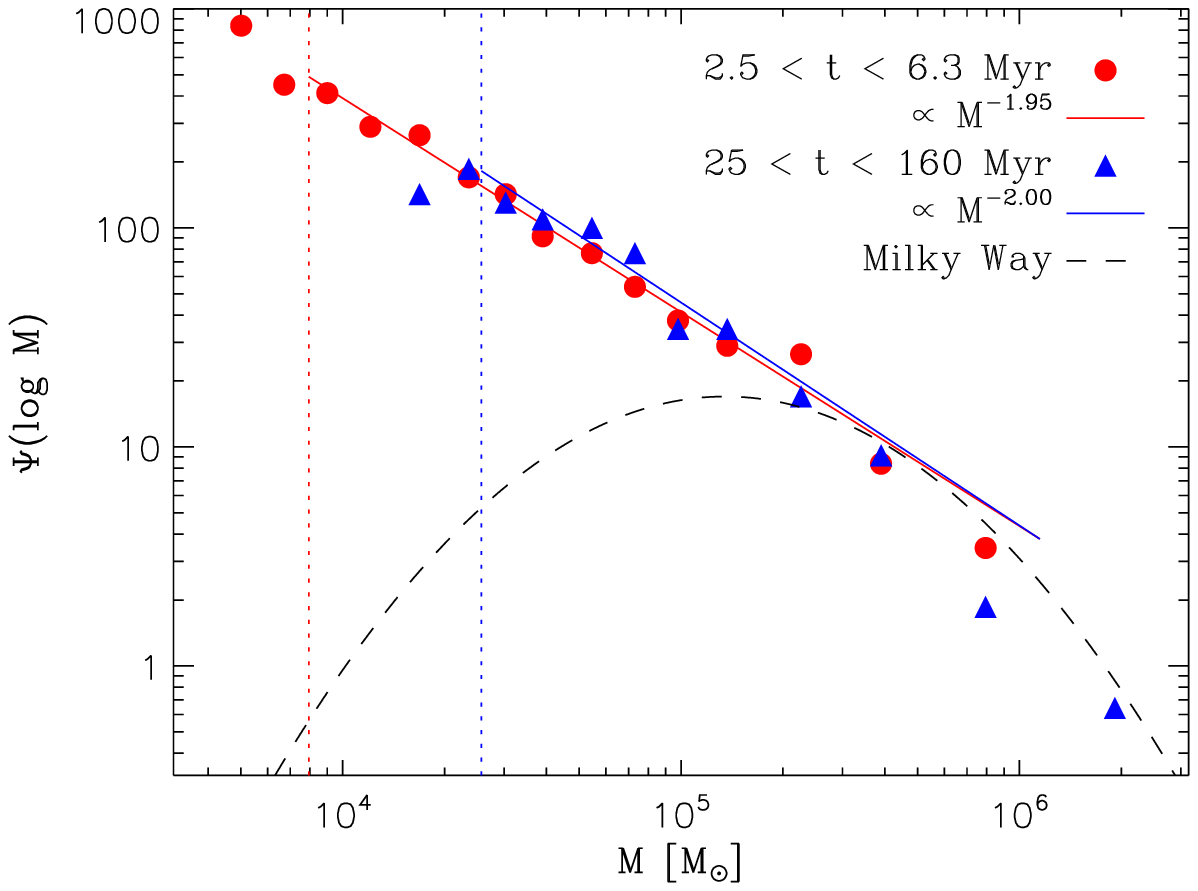}{Mass function of the clusters}
{Mass function of the Antennae star clusters, for two age bins. The $50\%$ completeness limit is indicated by vertical dashed lines. The two solid lines represent power-laws $\Psi(M) \propto M^\beta$. The black dashed curve shows the log-normal mass function, for the clusters of the Milky Way, derived from a Gaussian distribution of magnitudes in the $V$ band ($<M_V> = -7.3$ ; $\sigma(M_V) = 1.2$, see Table~2 of \mycitealt{Harris1991}) with a mass-to-light ratio $M/L_V = 2$. (Data points from \mycitealt{Zhang1999}.)}

Theoretical studies suggested that mass-loss through two-body relaxation, tidal shocks and stellar evolution might be responsible for a dynamical transition of the mass function, from an universal power-law to a log-normal distribution, as observed in e.g. the Milky Way (\mycitealt{Fall1977}; \mycitealt{Vesperini1998}; \mycitealt{Fall2001}). This can be explained by the dissolution of the low mass associations ($M < 10^4 \msun$) within the first $10^{7-9} \yr$ of their life, and a mild mass-loss of the more massive clusters, shifting to intermediate masses ($10^{4-5} \msun$). Ultimately (over $\sim 10^{10} \yr$), this leads to the mass function observed in the Milky Way for older objects. The effects responsible for the dissolution and mass-loss of the clusters can be either internal (violent relaxation) or external (via the tidal field), as presented in \refsecp{clusters_formation}.

\mycitet{Fall2005} studied the age of the clusters by comparing their magnitudes in different bands ($U, B, V, I$) with those from stellar population models (\mycitealt{Bruzual2003}). After applying a correction for completeness\footnote{Observational data are often biased by incompleteness, due to technical (e.g. limit in magnitude of the detector) or physical (like dust obscuration) reasons. Sometimes however, it is possible to correct for completeness (see e.g. \mycitealt{Whitmore1999}; \mycitealt{Anders2007} for applications to the Antennae galaxies).} and an age binning, they derived the feature-less age distribution $dN/d\tau \propto \tau^{-1}$ shown in green in \reffig{fall05}, with a median age of $\sim 10 \Myr$.

\fig{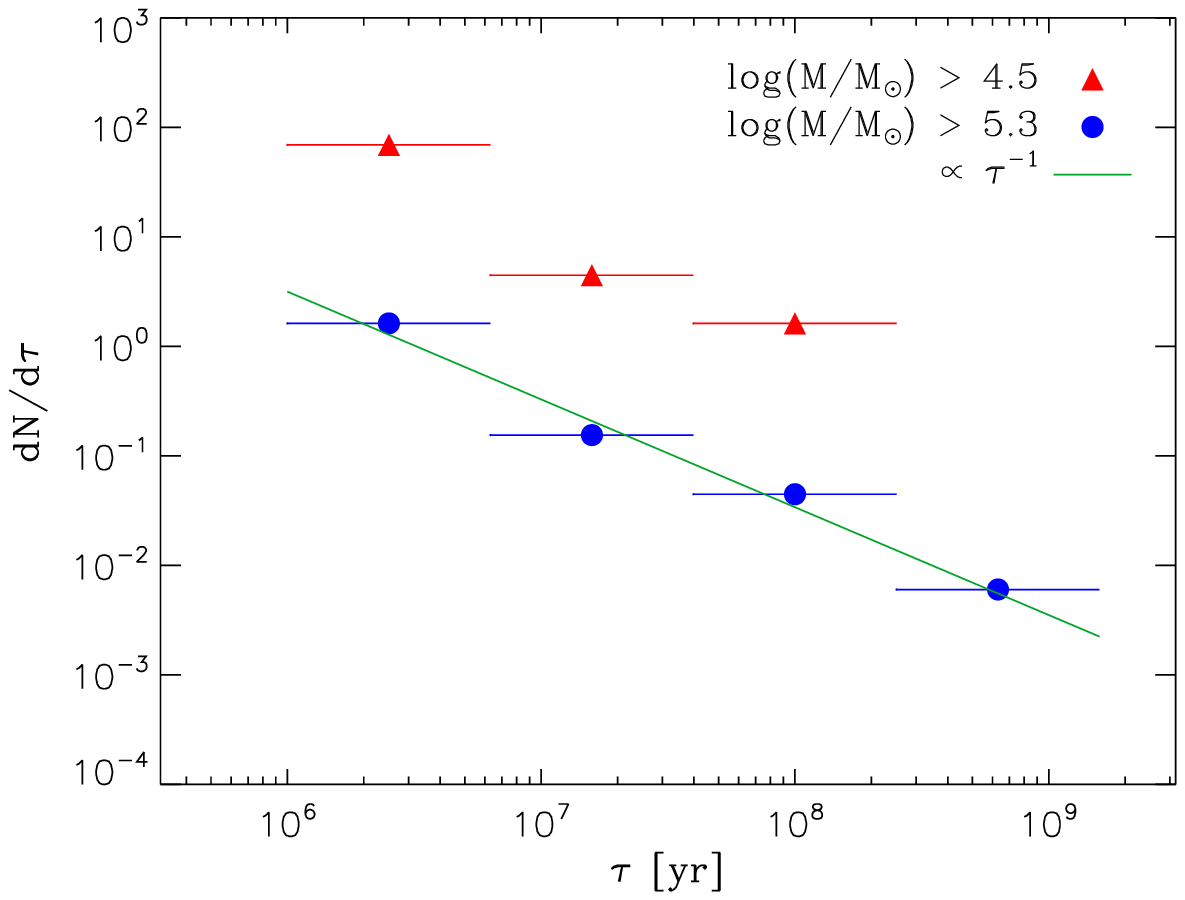}{Age distribution of the clusters}
{Age distribution of the clusters in the Antennae, based on different mass criteria. The age binning is showed by horizontal bars. The green line represents a distribution $dN/d\tau \propto \tau^{-1}$. (Data points from \mycitealt{Fall2005}.)}

This can be interpreted by considering the birth and death rates of the clusters. The feature-less distribution is incompatible with discrete episodes of star formation, well separated in time, without any destruction. However, it is possible to assume a constant star formation rate\footnote{This assumption is clearly an oversimplification, especially in the case of mergers, as noted by \mycitet{Whitmore2010}. The evolution of the (non-constant) \acr{notarg}{SFR}{star formation rate} is discussed in \refcha{formation}.} combined with a certain disruption rate. In order to get the $\tau^{-1}$ distribution, $\sim 90\%$ of the clusters must dissolve at each time decade. In other words, the younger bin is continuously filled by the constant formation rate. As time flows, $10\%$ of its content is shifted to older age bins while the other $90\%$ are destroyed. As a consequence, less and less clusters reach the oldest ages, which gives out the measured distribution. This rapid dissolution, often called ``infant mortality'', has also been observed for clusters in the Milky Way (\mycitealt{Lada2003}) and the Small Magellanic Cloud (\mycitealt{Chandar2006}), over the first $\sim 100 \Myr$. As mentioned before, this dissolution results from both internal and external effects, among which the tidal field of the host galaxy. That is why the age distribution of the star clusters is an important question that we will address in the following, from the standpoint of galaxy mergers.

%%%%%%%%%%%
\subsect{... and tidal dwarf galaxies}

While star clusters are mainly located in the central region of the Antennae, it is worth noticing that other structures exist in the tidal tails. \mycitet{Schweizer1978} and \mycitet{Mirabel1992} were the first to detect a pair of massive \hi clouds undergoing star formation, near the tip of the southern tail (see \reffig{antennae_uv}). This has been interpreted as \acr[s]{}{TDG}{tidal dwarf galaxies} exhibiting properties similar to those of dwarf irregulars or blue compact dwarf galaxies (see e.g. \mycitealt{Duc1999}). 

\fig{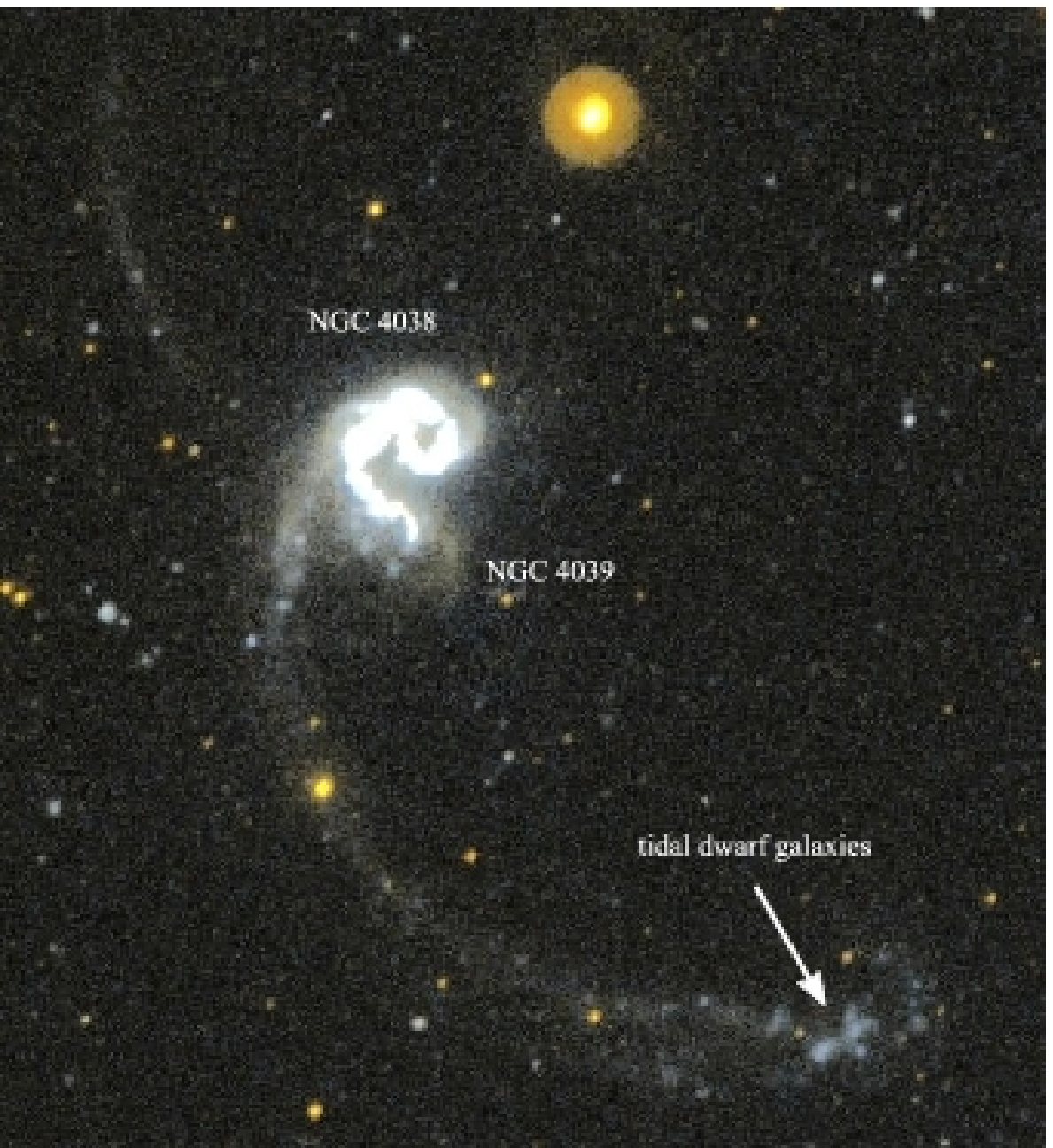}{UV image of the Antennae}
{Composite image of far-\acr{notarg}{UV}{ultraviolet} (blue) and near-\acr{notarg}{UV}{ultraviolet} (red) observations of the Antennae by $GALEX$ (\mycitealt{Martin2005}), clearly showing the \acr[s]{notarg}{TDG}{tidal dwarf galaxies} candidates near the tip of the southern tail. (Adaptation of Figure~1.a of \mycitealt{Hibbard2005}.)}

These observations left open the question pertaining to the origin of these blue regions. To address it, \mycitet{Hibbard2005} used \acr{}{UV}{ultraviolet} data and optical images to study the ages of the stellar populations in these dwarf galaxies (\reffig{hibbard05}). They noted that a continuous formation of stars is very unlikely in these areas. The different densities of gas in the two tails have led to different star formation epochs which explains the spanning of relative age between the two tidal structures.

\fig{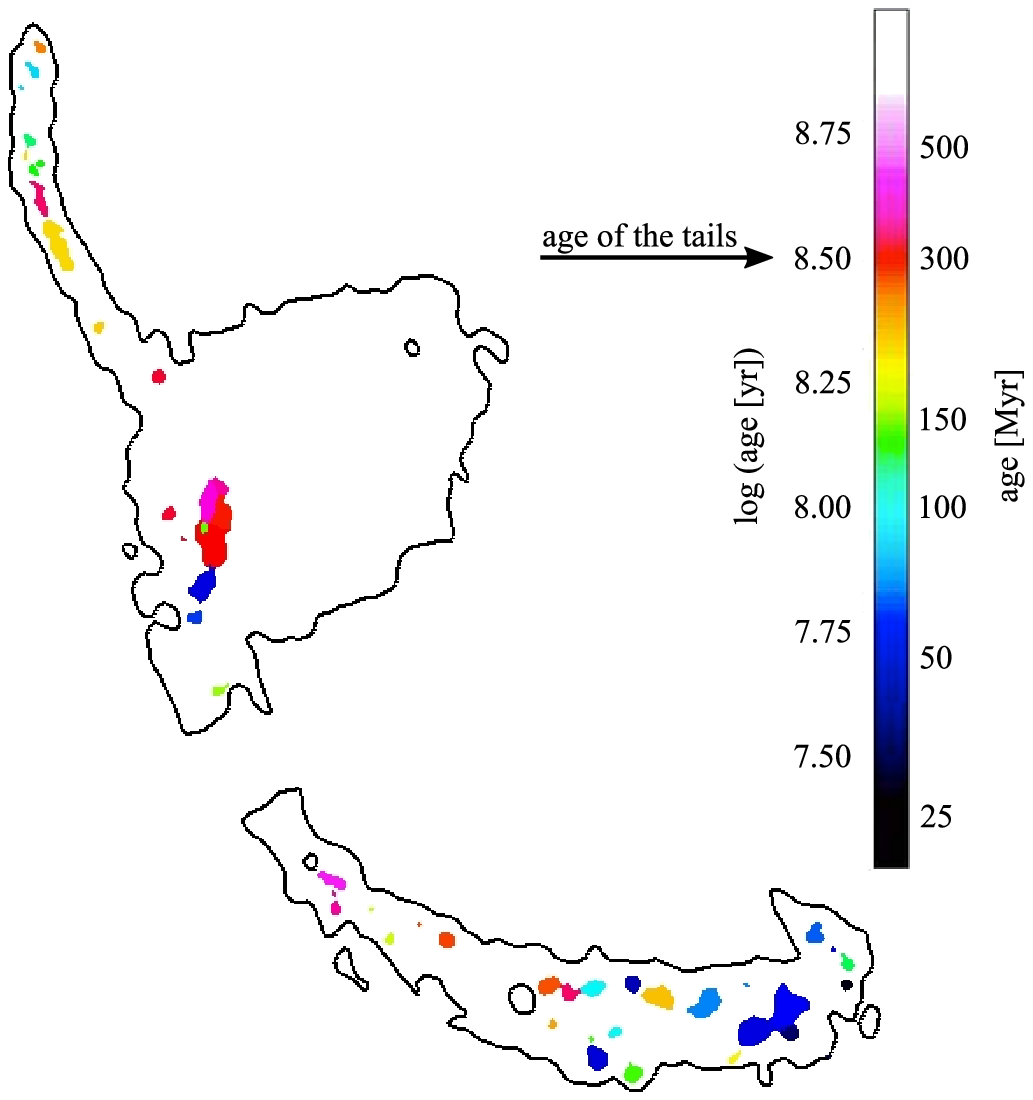}{Age of tidal structures}
{Intensity-weighted ages of the tidal regions derived from \acr{notarg}{SSP}{simple stellar population} thanks to (far-\acr{notarg}{UV}{ultraviolet} -- near-\acr{notarg}{UV}{ultraviolet}) and (far-\acr{notarg}{UV}{ultraviolet} -- V) data. A gradient is visible along the southern tail, from old populations close to the nucleus to younger near the tip. The pattern is more complex within the northern tail. The contour line represents a surface density of $27.5 \U{mag\ arcsec^{-2}}$ in the near-\acr{notarg}{UV}{ultraviolet} band. The arrow on the color bar shows the dynamical age of the southern tail from the simulation of \myciteauthor{Barnes1988a} (\myciteyear{Barnes1988a}, see the next Section). (Adaptation of Figure~2.c of \mycitealt{Hibbard2005}.)}

In the southern tail, \acr{}{SSP}{simple stellar population} models reveal that most stars have formed before the tail itself (\mycitealt{Hibbard2005}). However, structures visible in \acr{}{UV}{ultraviolet} yield younger ages than the dynamical age of the arm (estimated via numerical simulations), suggesting a recent episode of star formation. Therefore, the stars observed in the tails are likely to be of two origins: (1) the material of the progenitor disks which has been expelled during the tidal interaction and (2) an \emph{in situ} formation after the creation of the tails. As the bluest region, the \acr[s]{}{TDG}{tidal dwarf galaxies} are the youngest objects in the arm, and consist of two populations of young ($4-30 \Myr$) and intermediate ($80-100 \Myr$) ages.

Both central and tail regions of the Antennae are pristine laboratories for the study of the formation of star clusters and \acr[s]{}{TDG}{tidal dwarf galaxies}. Clearly, the tidal field plays a major role in the evolution of these populations. Therefore, a 4D (space and time) description of the galaxies will bring important clues to interpret the observations. This is only feasible via numerical simulations.

%%%%%%%%%%%
\subsect{Previous simulations of the Antennae}
Since the first observations, the morphology of the Antennae raised many questions about their origin. In their precursor work, \mycitet{Toomre1972} made the first attempt to reproduce the galaxies with simulations (\reffig{toomre_barnes_antennae}, top panel). 

\fig{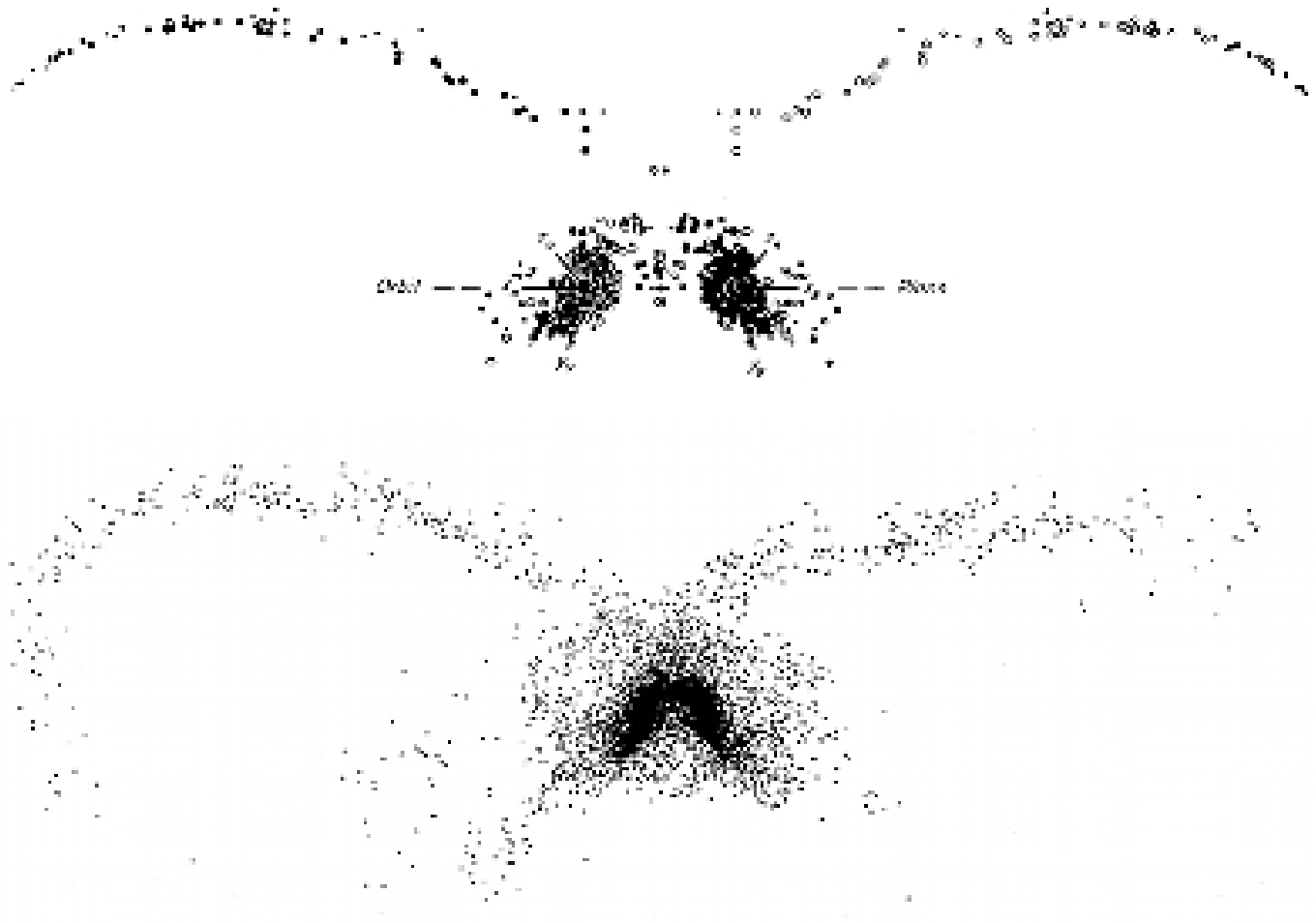}{First simulations of the Antennae}
{\emph{Top:} First simulation of the Antennae galaxies by \mycitet{Toomre1972}. Filled or open symbols distinguish the origin of the particles. (Figure~23 of \mycitealt{Toomre1972}.) \emph{Bottom:} First self-consistent simulation of the Antennae (\mycitealt{Barnes1988a}). Contrary to the restricted simulation of \mycitet{Toomre1972}, this model accounts for the orbital decay and better reproduces the inner region of the merger. (Figure~2 of \mycitealt{Barnes1988a}.)}

As mentioned in the \refsecp{lightbulbs}, these simulations are not self-consistent as they only compute the influence of two point-masses on pseudo-particles. However, this approach manages to reproduce the long tails and proves that they originate from a tidal interaction. The main problem of this model resides in its symmetry, visible in the extension of the tails. The authors also noted a too large separation between the nuclei and suggested to take orbital decay into account to reach a better configuration. Indeed, their point masses stay in their initial Keplerian orbit, as no phenomenon makes them lose energy. Such restricted simulations cannot model the transfer of energy through dynamical friction (see \refsect[basic_concepts]{dynamical_friction}), as the pseudo-particles do not create any gravitational drag.

The first self-consistent simulation of the Antennae came 16 years later, from \myciteauthor{Barnes1988a} (\myciteyear{Barnes1988a}, see \reffig{toomre_barnes_antennae}, bottom panel). This model has been used to show the role of the \acr{}{DM}{dark matter} halos, which act as ``a sink for binding energy and angular momentum as the visible galaxies spiral together'', so that the nuclei are closer together.

\mycitet{Mihos1993} performed the first simulation of the Antennae including hydrodynamics. With a gaseous component in addition to the stars and \acr{}{DM}{dark matter}, the simulation shows a dissipative part, used to form new stars (see \reffig{mihos}).

\fig{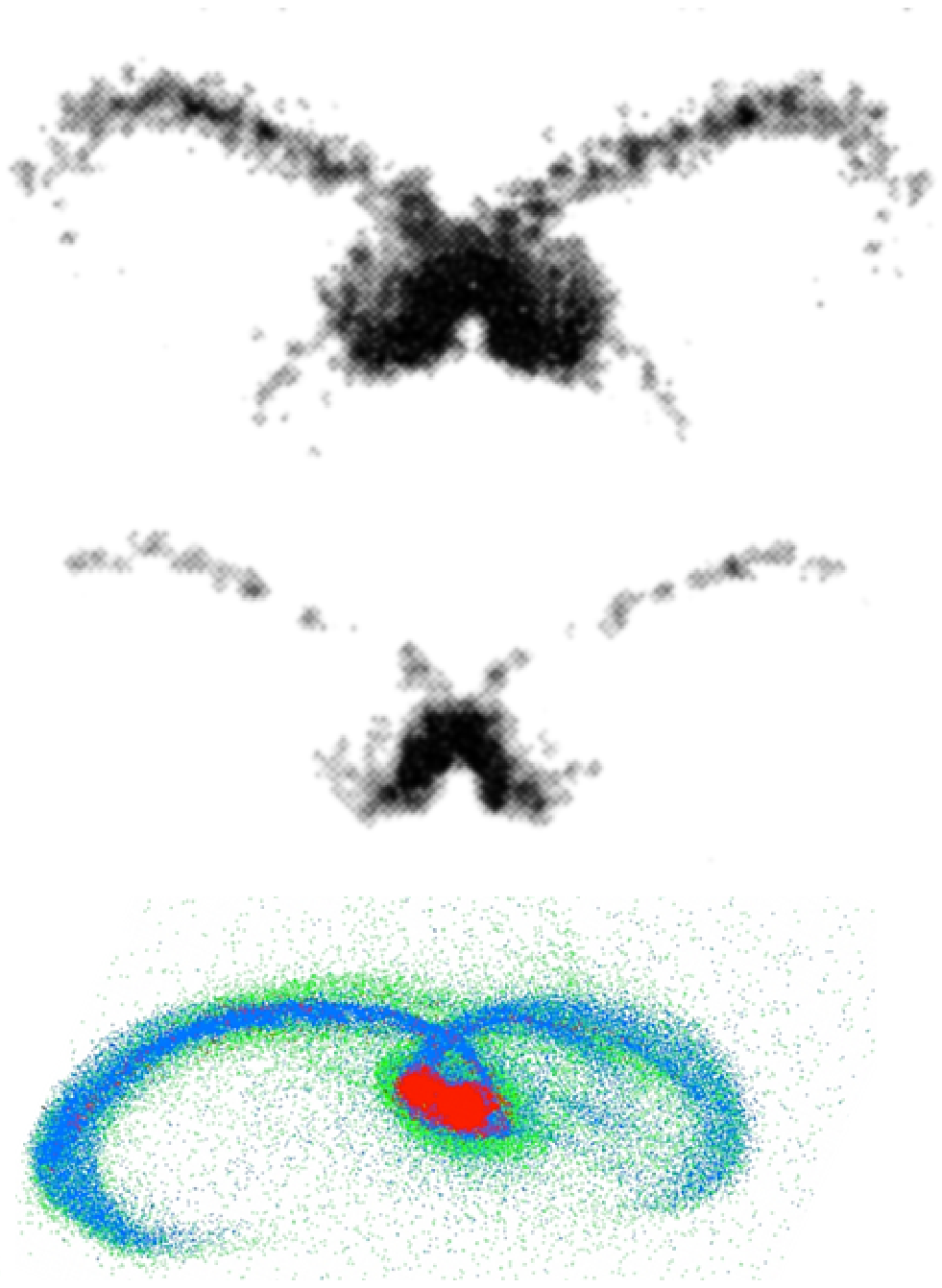}{SPH simulations of the Antennae}
{\emph{Top:} First hydrodynamic simulation of the Antennae by \mycitet{Mihos1993}. \emph{Middle:} map of the star formation rate from the same simulation. (Adaptation of Figure~8 of \mycitealt{Mihos1993}.) \emph{Bottom:} Recent \acr{notarg}{SPH}{smoothed particle hydrodynamics} simulation by \mycitet{Karl2008}. Old stars are green, new stars are red and the gas is shown in blue. (Adaptation of Figure~2 of \mycitealt{Karl2008}.)}

\mybox[schmidt]{
In hydrodynamic simulations, a ``recipe'' is needed to convert an element of gas into an element of stellar material. This prescription is often based on the empirical relation of \mycitet{Schmidt1959}, who established a link between the \acr{notarg}{SFR}{star formation rate} per surface unit $\dot{\Sigma}_\star$ and the observed gas surface density $\Sigma_\mathrm{g}$:
\eqn{
\dot{\Sigma}_\star \propto \Sigma_\mathrm{g}^n.
}
\follow}\mybox[nocnt]{
This relation has been tuned by \mycitet{Kennicutt1998}, who determined $n = 1.4 \pm 0.15$. In the simulations, the Schmidt-Kennicutt law is converted to describe the \acr{notarg}{SFR}{star formation rate} as a function of the gas volume density:
\eqn{
\dot{\rho}_\star \propto M \rho_\mathrm{g}^n,
}
where $M$ is the mass of the initial cloud (\mycitealt{Mihos1993}). The normalization factors are usually chosen to set the \acr{notarg}{SFR}{star formation rate} of an isolated galaxy to $\sim 1 \msun \yr^{-1}$. Other approaches (\mycitealt{Katz1992}; \mycitealt{Springel2000}) use the timescale for star formation, which is linked to the time needed by the gas cloud to collapse:
\eqn{
\dot{\rho}_\star \propto \frac{\rho_\mathrm{g}}{\left(G\rho_\mathrm{g}\right)^{-1/2}},
}
which gives a Schmidt law with $n=3/2$.
}

Such a method provides interesting clues about the location and age of newly formed structures. However, the physical details of the collapse of the molecular cloud and the other competing processes are not described, as noted by \mycitet{Barnes2004}.

Many other models have been created, most of them based on the orbital parameters of \mycitet{Toomre1972}. Today, some of the \acr{}{CPU}{central process unit} power is dedicated to the research of a better set of parameters, before performing the simulation itself (see e.g. \mycitealt{Karl2008}). In the next Section, we present our model of the Antennae and explain why such parameter studies are more than welcome!

%%%%%%%%%%%%%%%%%%%%%%%%%%%%%%%%%%%%%%%%%%%%%%%%%%
\sect{Numerical model}

In order to explore the tidal field of the Antennae, one needs a 3D representation of its potential. As discussed before, the gaseous component of galaxies does not strongly influence the gravitational potential, because of its relatively small mass. Therefore, stars and \acr{}{DM}{dark matter} seem to be the only components required in the modeling of the gravitational field of the merger. As our simulations are collisionless, we use {\tt Nemo} and {\tt gyrfalcON} for the setup and the integration of the orbits (see \refsect[numerical]{nemo} and \refsect[numerical]{testsgyrfalcon}). This Section details and justifies the choices made for the main parameters.

%%%%%%%%%%%
\subsect{Individual galaxies}
During a merger, the tidal stripping mostly affects the dynamically cold components of the galaxies, i.e. their disk. Indeed, the observations of long tidal tails in the Antennae suggest that the progenitors are spiral galaxies, instead of ellipticals. Within the {\tt Nemo} package, the tool {\tt magalie} (\mycitealt{Boily2001}) creates a baryonic component made of an exponential disk and a \mycitet{Hernquist1990} bulge embedded in an isothermal \acr{}{DM}{dark matter} halo, in a similar way as \mycitet{Hernquist1993} who presented a method to build multi-component models of galaxies.

In the observations of the tidal tails, we noted an asymmetry between the two galaxies: the northern tail (associated with NGC~4039, the southern nucleus) shows a shorter extension than its counterpart. This suggests that the initial disk of NGC~4039 was more compact than the other one. However, the luminosity and size of the two nuclei are similar, which, in a first approximation, implies that they have similar masses. Hence, the radius of the disk of NGC~4039 has to be smaller than the one of NGC~4038.

In our model, the halo component is made of \acr{}{DM}{dark matter} \emph{only}. However, it is clear that typical spiral galaxies also have a baryonic extension of a similar shape, called spheroid. Therefore, our bulge component from {\tt magalie} is much more extended than a classical bulge, and can be mentally divided into an actual bulge and a spheroid of lower density. Thanks to this trick, all the baryonic matter is included in the disk and bulge components.

The integration of the model of a single progenitor (i.e. in isolation) with {\tt gyrfalcON} reveals that it is not initially virialized. Indeed, small scale relaxation effects occur for about a crossing time, while the overall morphology is not affected by the redistribution of the kinetic energy. To avoid any artifact, the isolated model is first left evolving to a stable virialized system, and then extracted to be used as initial configuration.

%%%%%%%%%%%
\subsect{Orbital parameters}
Once the parameters of the progenitors have been set, one can focus on the geometry of the merger. The aim here is to determine how the two galaxies interact with each other, in term of orbits, relative mass and size, inclination and so on.

As presented above, the two progenitors have similar masses and thus, the Antennae are considered as a major merger (i.e. a pair of galaxies with a mass ratio greater than 3:1). For simplicity, we assume that their rotation curves are also similar, which implies that no scaling (in mass, velocity or size) has to be done to the progenitors. We also keep the same parameters for the disks, bulges and halos (except the radius of the disk, as mentioned before).

To determine the orbit of the galaxies, one has to set the initial conditions as differential position and velocity of the progenitors. Obviously, there are no strong constraints on the space of the orbital parameters because the only criteria are the observed velocities and the morphology at present time, i.e. two quantities that depend on the line of sight and the initial inclination of the disks with respect to the orbital plane. That is, the only way to set the initial value of an orbital parameter is to assume that all the others are known, to integrate the model until it best matches the observations, and to iterate with another parameter. However, if one plays with, say, the orbital eccentricity, it is very likely that the inclination that would give the best match will change too. In other words, all the variables are inter-dependent and cannot be tuned sequentially, one after the other.

Nevertheless, one has a few clues from observations and a basic knowledge of tidal interactions. For example, the large extension of both arms favors a prograde-prograde encounter (see \refsect{prograderetrograde}). This constrains the inclination of the disks relatively to the orbital plane. Furthermore, the two nuclei are very close to each other, suggesting that they recently underwent or will soon undergo a pericenter passage. Because the tidal tails had time to expand to their current positions, this passage is not the first one. After the first passage, the nuclei came back close to each other (as observed), meaning that the orbit is bound. However, one should not conclude that the initial eccentricity has to be smaller than unity. Indeed, in order to create the big tails, each progenitor underwent a close passage through the material of the other. Then, a fraction of the initial orbital energy has been transfered to the individual particles via dynamical friction (see an example in \reffigt{dynamical_friction_orbits}). Therefore, even a Keplerian orbit with an eccentricity slightly greater than unity could lead to the bound orbit required by the Antennae data.

%%%%%%%%%%%
\subsect{Simulation of the merger}
It would be possible to explore the parameter space through a grid of models, but the number of combinations and the computational power needed to perform a single simulation forbid such a method\footnote{Several authors proposed to explore the parameter space with restricted simulations, that are obviously much faster to run than self-consistent ones (see e.g. \mycitealt{Theis2001}; \mycitealt{Barnes2009}). However, these simulations are still facing problems in the recovering of the orbital decay, an important property of mergers.}. One could also imagine to consider the observations as the initial state and go backward in time, up to a moment where the galaxies are isolated. Although possible in principle, such an approach would require a very accurate knowledge of quantities like the velocities perpendicular to the line of sight, or the mass distribution, which are not measurable via observations.

\tab{10cm}{antennae}{Parameters of the Antennae model}
{l@{\extracolsep{\fill}}c@{\extracolsep{\fill}}c}
{
Parameter & NGC~4038 & NGC~4039\\
\hline
\multicolumn{3}{l}{Numbers of particles}\\
$N_{\mathrm{total}}$ & $7 \times 10^5$ & $7 \times 10^5$ \\
$N_{\mathrm{disk}}$ & $2 \times 10^5$ & $2 \times 10^5$ \\
$N_{\mathrm{bulge}}$ & $1 \times 10^5$ & $1 \times 10^5$ \\
$N_{\mathrm{halo}}$ & $4 \times 10^5$ & $4 \times 10^5$ \\
\hline
\multicolumn{3}{l}{Scalelengths}\\
$h_{\mathrm{disk, vertical}}$ & $0.2$ & $0.2$ \\
$r_{\mathrm{disk}}$ & $1.0$ & $1.0$ \\
$r_{\mathrm{bulge}}$ & $1.0$ & $1.0$ \\
$r_{\mathrm{halo}}$ & $7.0$ & $7.0$ \\
\hline
\multicolumn{3}{l}{Cut-off radii}\\
$c_{\mathrm{disk, vertical}}$ & $2.0$ & $2.0$ \\
$c_{\mathrm{disk}}$ & $5.0$ & $3.0$ \\
$c_{\mathrm{bulge}}$ & $1.0$ & $1.0$ \\
$c_{\mathrm{halo}}$ & $7.0$ & $7.0$ \\
\hline
\multicolumn{3}{l}{Masses}\\
$m_{\mathrm{disk}}$ & $1.0$ & $1.0$ \\
$m_{\mathrm{bulge}}$ & $1.0$ & $1.0$ \\
$m_{\mathrm{halo}}$ & $5.0$ & $5.0$ \\
\hline
\multicolumn{3}{l}{Toomre parameters}\\
$Q$ & $1.5$ & $1.5$ \\
\hline
\multicolumn{3}{l}{Disk inclinations}\\
$\theta_x$ & $60^{\circ}$ & $-60^{\circ}$ \\
$\theta_y$ & $0^{\circ}$ & $30^{\circ}$ \\
\hline
\multicolumn{3}{l}{Initial coordinates}\\
$(x,y)$ & $(6.0, 6.0)$ & $(-6.0, -6.0)$ \\
$(v_x,v_y)$ & $(-0.5, -0.25)$ & $(0.5, 0.25)$\\
}
{\note The parameters are given in numerical units ($G=1$) with the conversion factors $4.4 \kpc$, $1.9 \times 10^2 \kms$ and $3.6 \times 10^{10} \msun$ for the length, velocity and mass respectively. The Keplerian equivalent orbit has an eccentricity of 0.96, a distance to pericenter of 1.31 ($= 5.78 \kpc$) and a semi-major axis of 17.74 ($= 78.06 \kpc$).}

Hence, the approach adopted here is an iterative method where each parameter is tuned manually until the simulations give an acceptable match to observations (morphology and kinematics). The quality of the match is determined by eye. Note that the goal of this study is not to produce a better model for the Antennae than those from the literature, but simply to find a set of parameters which will, in the end give a good description of the tidal field. The final set of parameters, presented in \reftab{antennae}, has been obtained after running $\sim 100$ low resolution attempts. \reffig{antennae_op} and \reffig{antennae_los} display the morphology of the system, as the logarithm of the surface density, a good way to mimic the CCD images from observations (see \mycitealt{Dubinski2008} for an introduction to visualization of $N$-body simulations). \emph{In fine}, our model is similar to that of \mycitet{Barnes1988a}, except that he used lighter King models for the bulges. The choice made here allows a finer description of the central region of the merger, closer to observations.

\fig{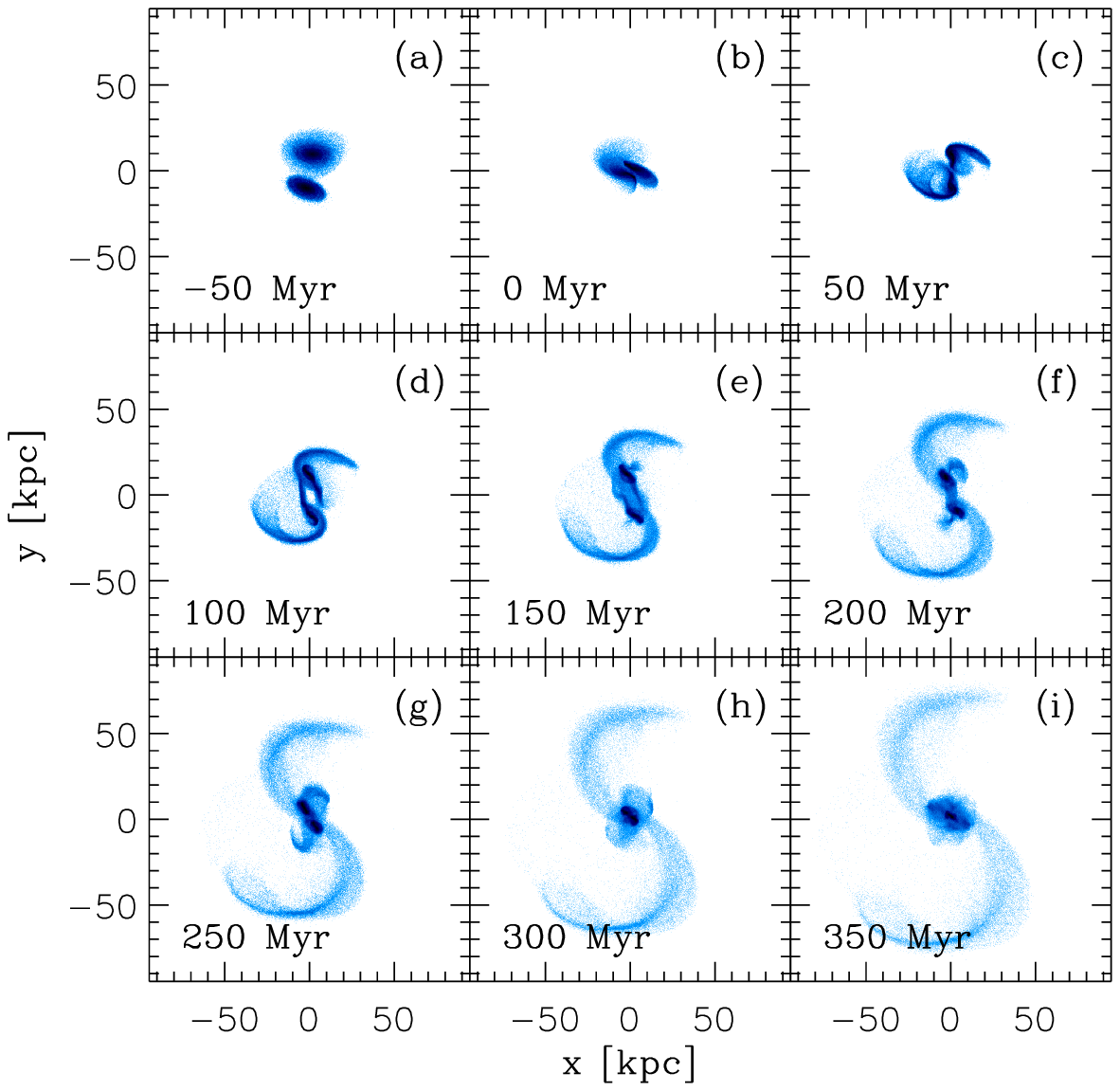}{Simulation of the Antennae, in the orbital plane}
{Morphology of the Antennae model, shown in the orbital plane, every $50 \Myr$. The shades of blue represent the logarithmic surface density. $t = 0 \Myr$ corresponds to the first pericenter passage.}

\fig{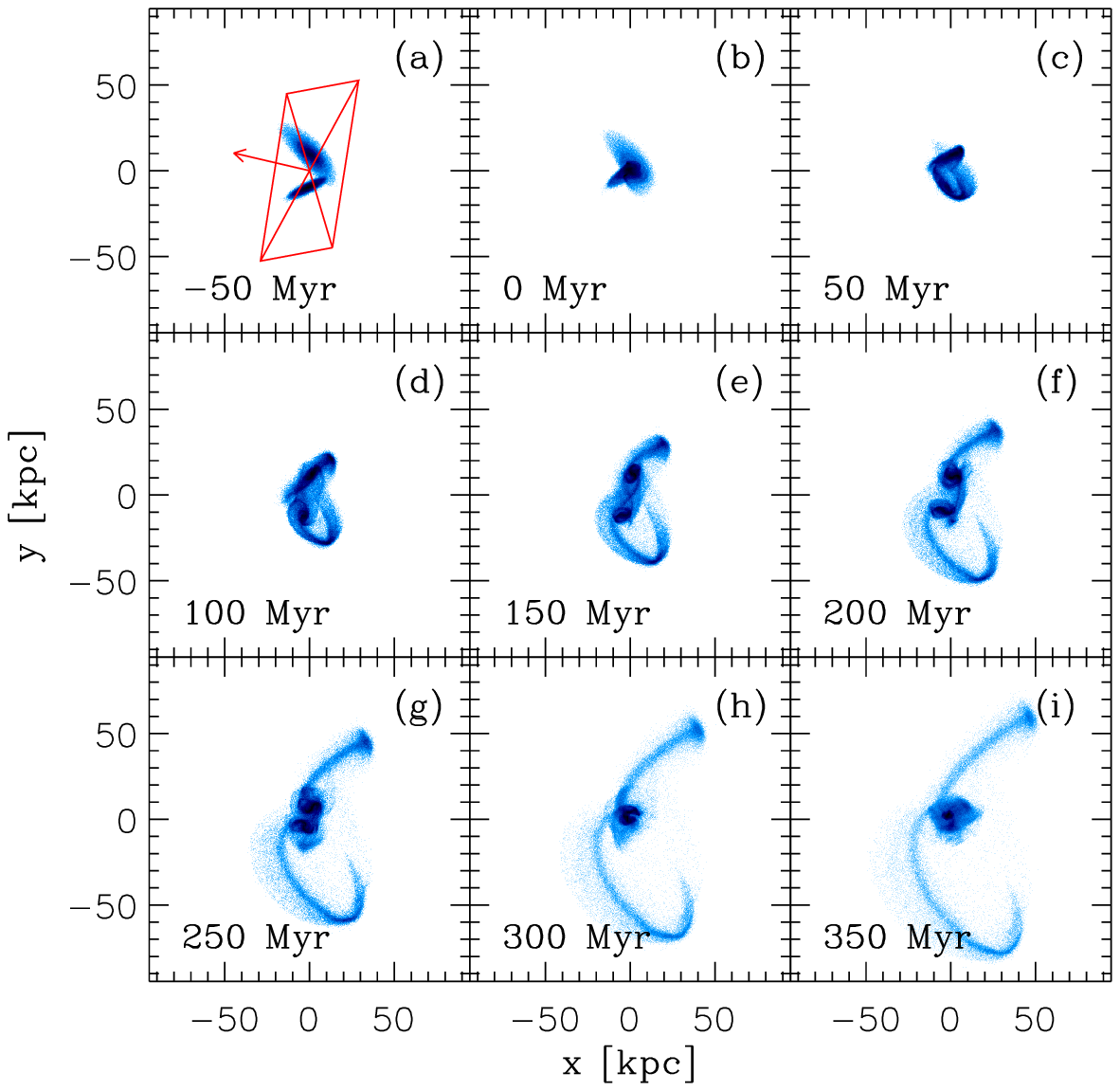}{Simulation of the Antennae, in the plane of the sky}
{Same as \reffig{antennae_op} but in the plane of the sky. The orbital plane is shown in red in the panel a. The arrowhead marks the position of the point $(0,0,+50\kpc)$ in the orbital plane (i.e. foreground), which is now behind the plane of the sky (the arrow is pointing away from the reader). The best match with the observations is obtained for $t=300 \Myr$ (panel h, see also \reffig{antennae_now}).}

The entire parameter space exploration and the simulations themselves are done using numerical units. By comparing the final result with observations, it is possible to determine the conversion factors. Observational data constrain well the size of an object and its velocity along the line of sight. The mass and time however, are not directly measured. That is why the non-dimensionalization used here is based on the spatial extension of the tidal tails and the peak in the velocity field measured by \mycitet{Hibbard2001}. With these factors and the relations given in \refsecp[numerical]{nbody_units}, one can retrieve the equivalent factors for the mass and time, and deduce the physical resolution of the simulation. In this case, each particle represents $\sim 2\times 10^5 \msun$ and a snapshot corresponds to $2.5 \Myr$. (Hereafter, the origin of time $t=0$ corresponds to the first pericenter passage.) The smoothing length used covers a radius of $40 \pc$. These numbers match well the physical size and mass of a typical star cluster, which means that a particle in this simulation represents a single cluster (whose internal dynamics are out of reach).

The best match with observations is reached $300 \Myr$ after the first pericenter (\reffig[h]{antennae_los} and \reffig{antennae_now}). Despite the too long northern tail, the model displays a central region showing most of the structures of visible in e.g. HST images. The velocity field (along the line of sight) is shown in \reffig{antennae_velocity}. The general structures are well reproduced, notably the gradient along the southern tail, up to the tip which shows a ``folding'' pattern. We also remark that each nucleus yields a $v_z$ component of opposite sign to those of the related tail, as previously noted by the observations (see \reffig{hibbard}). On a smaller scale, the simulation fails to show the internal rotation of the nuclei. A multicomponent distribution is visible in the central region, showing the complexity of the system that is currently undergoing an efficient mixing of the matter.

\fig{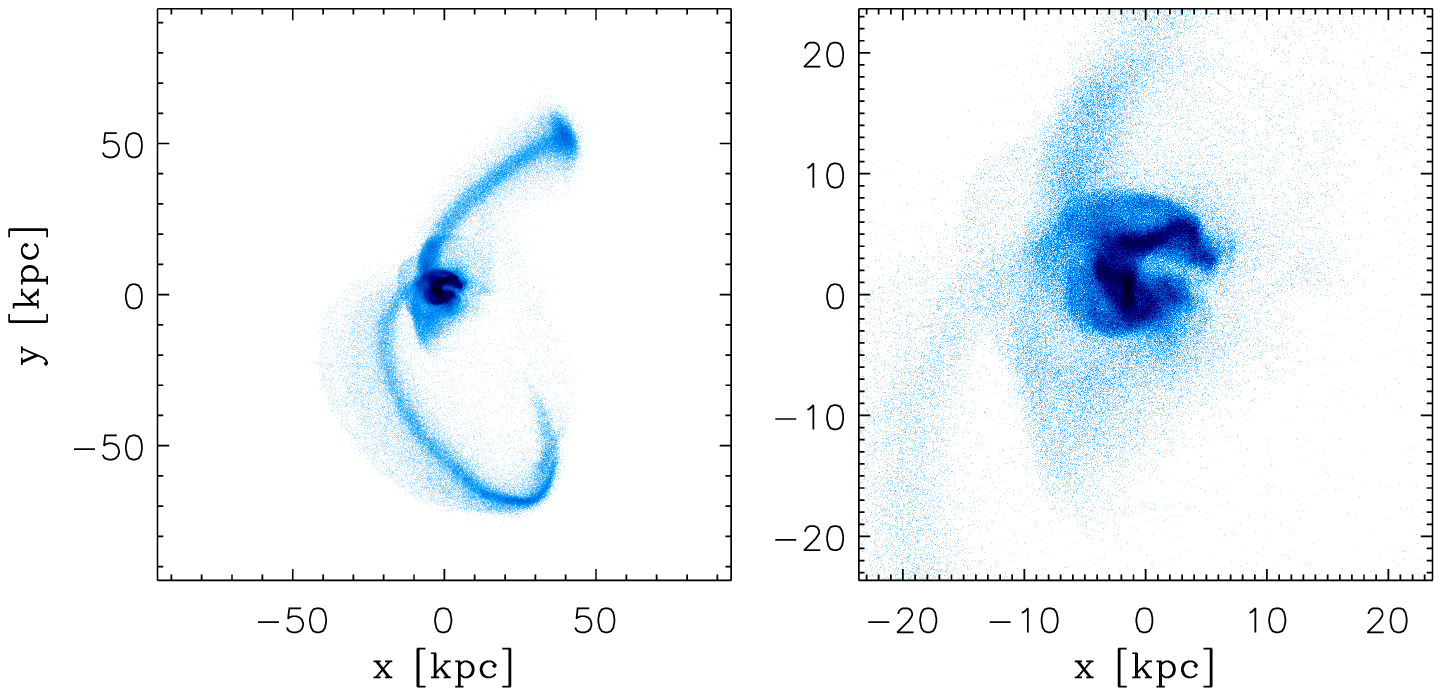}{Model of the Antennae, at present time}
{Morphology of the Antennae model for $t = 300 \Myr$, at large scale (\emph{left}), and in the central region (\emph{right}). The southern tail matches well the observations, although the northern one is too extended. In the central region, the origin of the tails is clearly visible, as well as the main features: the two nuclei, the overlap region and the arc shape in the northern part.}

\fig{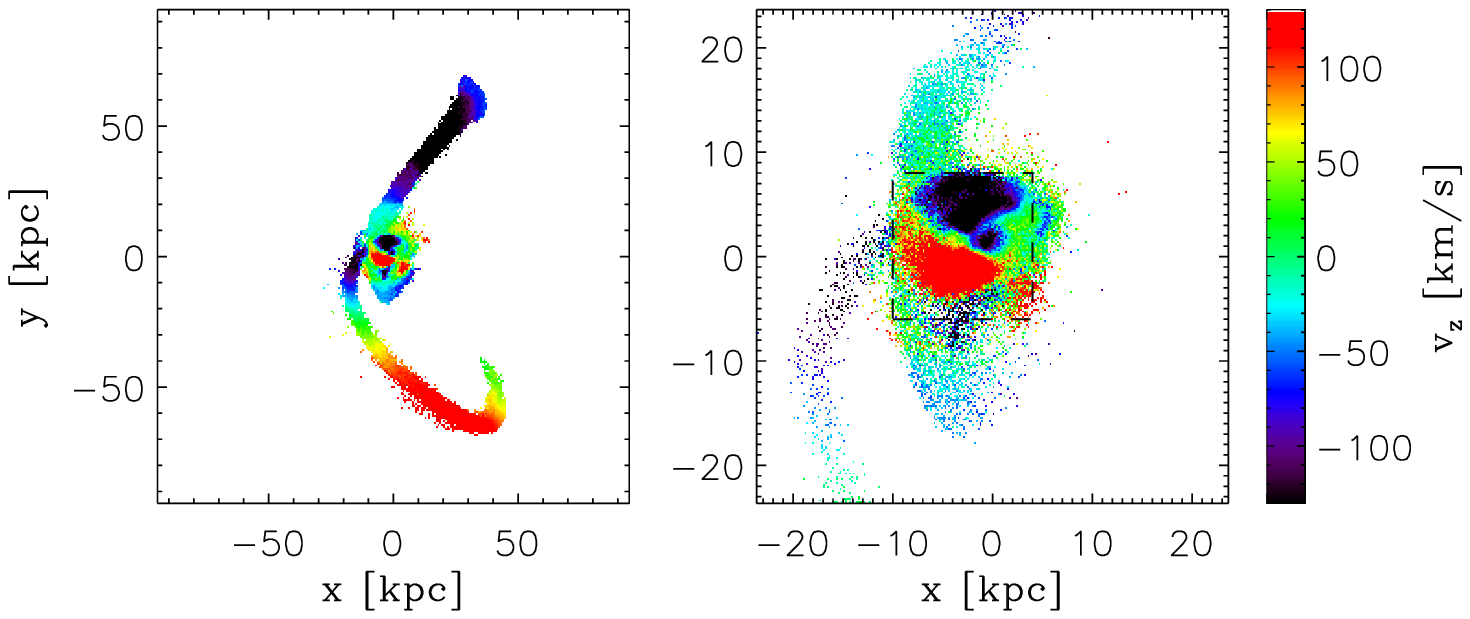}{Velocity field}
{The velocity field along the line of sight of the Antennae model (\emph{left}) and a zoom-in on the central region (\emph{right}), at present time. The dashed square represents approximately the size of the right panel of \reffig{hibbard}. The kinematics of the tails are well reproduced by the simulation but the internal rotation of the nuclei is not clearly visible. Note that the peak velocity matches \emph{de facto} the observations because it has been chosen for the non-dimensionalization of the simulation. Therefore the quality of the model is estimated from the gradients along the tails, and in the central region.}

The good match with observations for both the morphology and the kinematics gives us confidence in this model for the next steps of our study. Before that, we need to characterize the simulation with more details, in particular its evolution in time.

\reffig{antennae_op_traj} shows the orbits of the progenitors in the \emph{orbital plane}, corresponding to the trajectories of the center of mass of the most bound particles in each galaxy. Indeed, by considering the center of mass of all the particles, one would take into account the tidal tails, which would lead to an offset in position while the nuclei merge (see an example on \reffigt{dynamical_friction_orbits_zoom}). This effect disappears when selecting the particles that stay in the nuclei throughout the simulation\footnote{The center of density or potential give good results too, even when considering all the particles. However these quantities are determined at high computational cost.}. As a complement, \reffig{antennae_dist} displays the evolution of the three-dimensional distance between the two centers of mass. The two pericenter passages are clearly visible as local minima for the distance. This plot tells us that the current configuration ($t = 300 \Myr$) is reached just after the second passage. One can also remark that the dynamical friction forbids a third separation of the two nuclei, which rapidly merge. The evolution of the distance between galaxies in mergers is of paramount importance when considering the \acr{}{SFR}{star formation rate} or the effect of the tides, as we will see below.

\fig{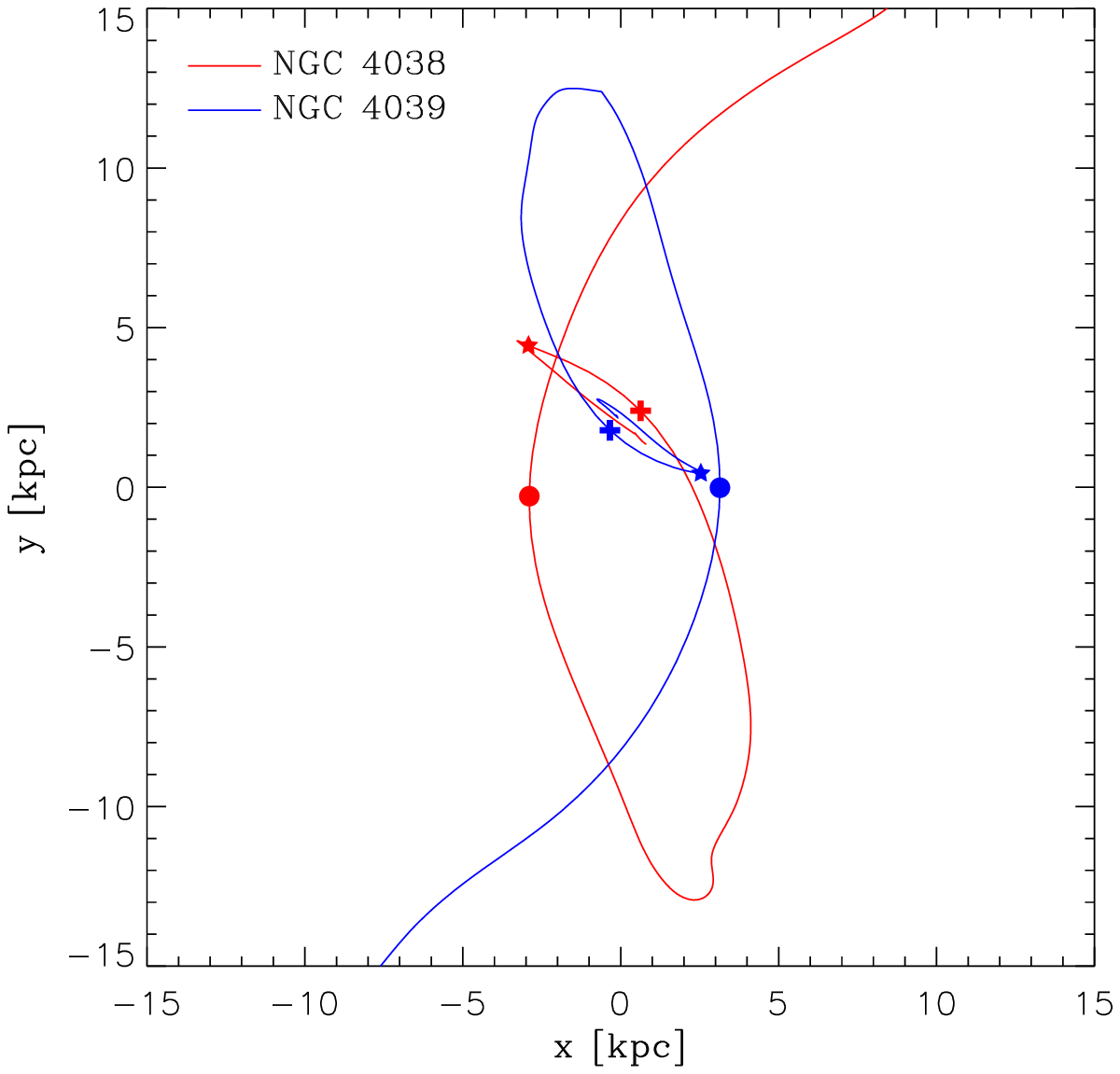}{Trajectories of the progenitors}
{Trajectories of the centers of mass of the most bound particles of the two progenitors, again in the orbital plane. The pericenter passages are marked with $\bullet$ ($t=0 \Myr$, \reffig[b]{antennae_op}) and $+$ ($t=285 \Myr$). The configuration closest to observations is shown with $\star$ ($t=300 \Myr$, \reffig[h]{antennae_op}).}

\fig{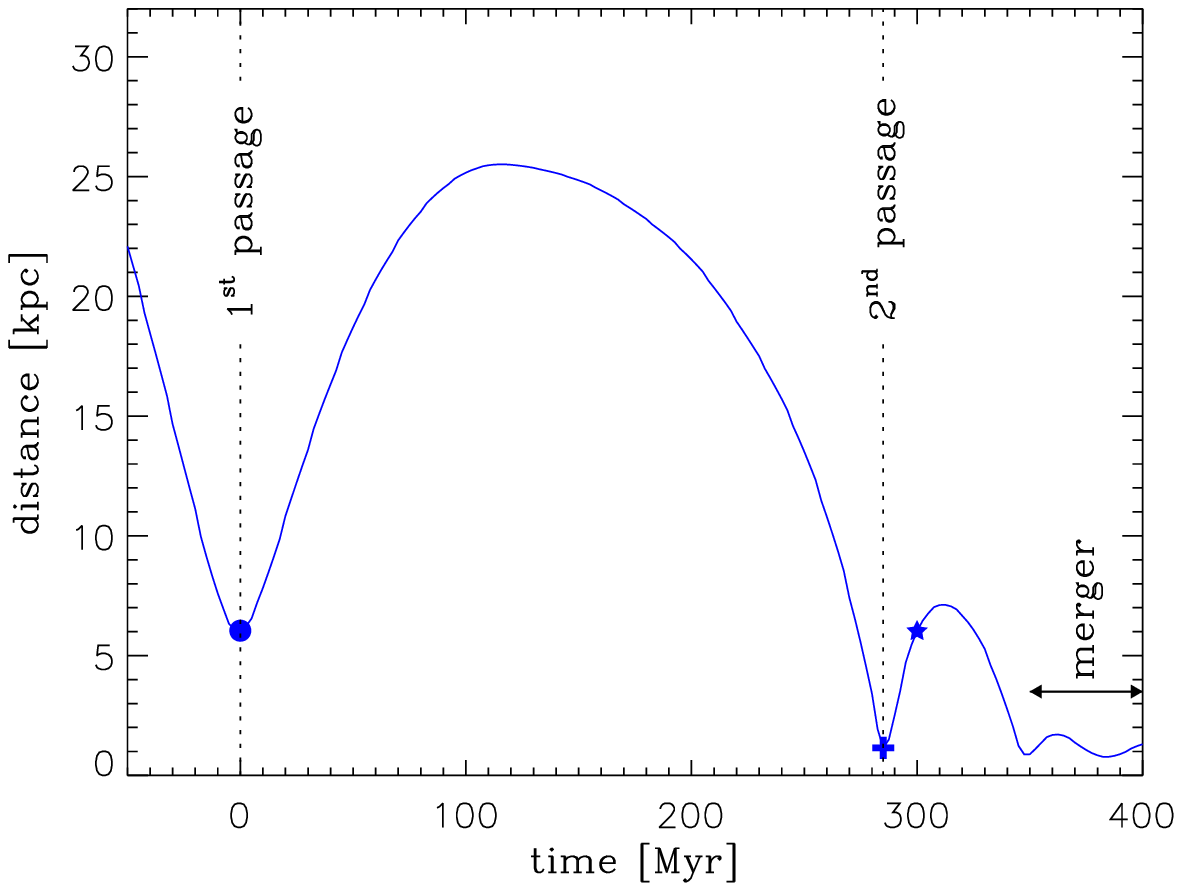}{Evolution of the distance between the nuclei}
{Evolution of the distance between the two progenitor galaxies. Again, the pericenter passages and the current configuration are marked with $\bullet$, $+$ and $\star$, respectively. The merger phase is clearly visible after $t \simeq 350 \Myr$, when the distance shows a damped oscillatory evolution.}

%%%%%%%%%%%%%%%%%%%%%%%%%%%%%%%%%%%%%%%%%%%%%%%%%
\sect{Tidal field of the progenitors}

The main topic of this thesis is to quantify the effect of the tides on the baryonic matter. For this reason, we concentrate our exploration on the stellar particles, i.e. the disks. The results presented in the following make intensive use of the method to compute the tidal tensor at the position of the particles of the disks (recall \refsect[numerical]{method_tt}). With the resolution adopted here, the cubes used for the finite difference scheme have a size $\delta \approx 220 \pc$, i.e. well beyond the half-mass radius of typical clusters.

Before studying the tidal field of the merger, it is important to get familiar with those of the progenitors. Indeed, if every object creates its own tidal field, it is the gathering of two of them that will build an interesting pattern, responsible for e.g. the formation of the tidal tails. That is why, in our approach, it is important to distinguish the intrinsic tidal field from the one due to external objects (a second galaxy in this case). It is also a good opportunity to become familiar with the tools and diagnostics we will use to study the statistics of the tidal field of much more complex systems. In this Section, we consider the model of NGC~4038 in isolation. Note that, despite a small difference in the radial parameter of the disk, NGC~4039 yields comparable results.

%%%%%%%%%%%
\subsect{Compressive nucleus}

\reffig{isolated_potential} (top panel) shows the radial profile of the potential $\phi$ of one of the progenitors of the Antennae model, taken in isolation. The associated mathematical functional form shows a change of curvature from convex to concave at a radius $r_\mathrm{c} \approx 1.3 \kpc$ (vertical red line). As the second derivative of the potential, the tidal tensor $\matr{T}$ is the negative of its Hessian matrix, so a change of curvature in $\phi$ implies a change of sign in $\matr{T}$. Therefore, in its diagonal form, the dominant term of $\matr{T}$ goes from negative to positive with increasing radius. The other eigenvalues being negative everywhere, the tides are fully compressive (see \refsect{compressive_mode}) in the inner $\approx 1.3 \kpc$ region. On the bottom panel of \reffig{isolated_potential}, one reads that the corresponding mass represents $\sim 2.5\%$ of the total mass of the disk (i.e. $\sim 9.0 \times 10^8 \msun$).

\fig{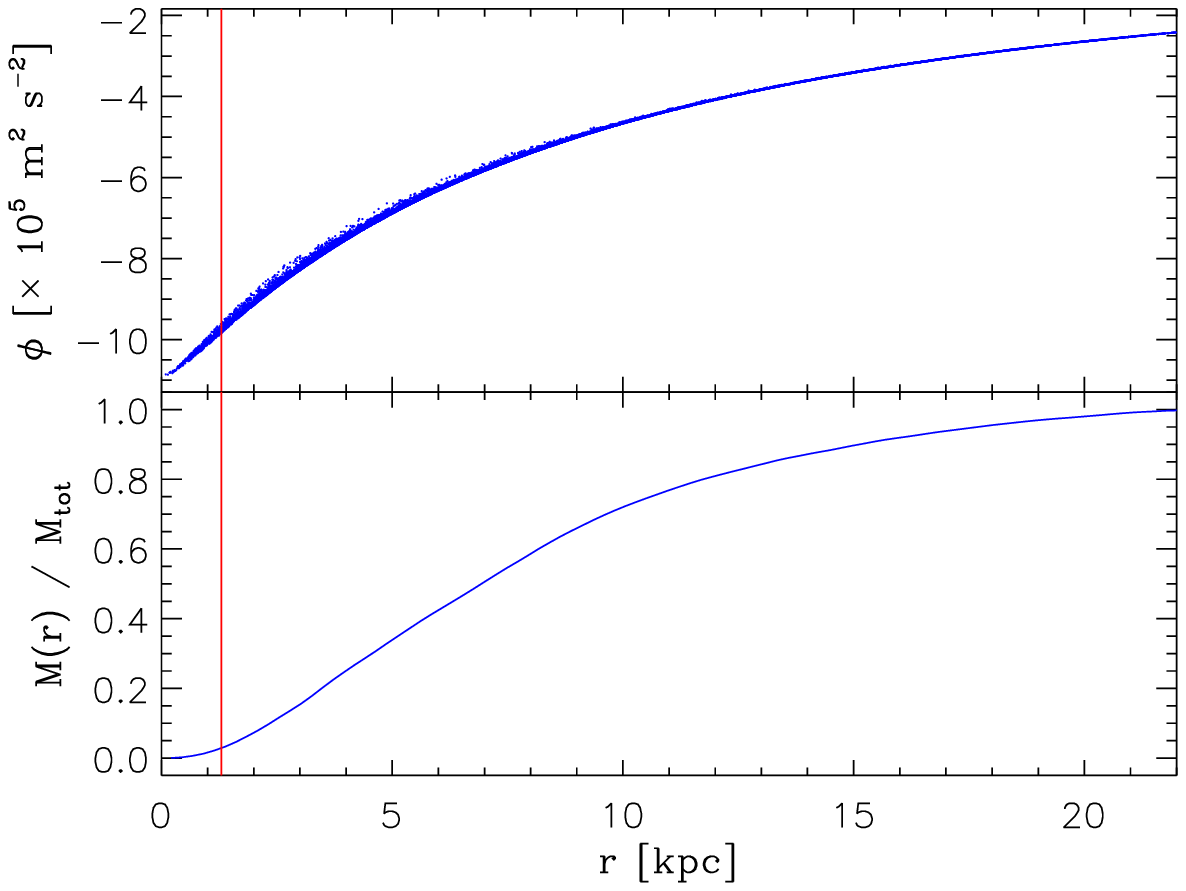}{Gravitational potential and integrated mass of a progenitor}
{\emph{Top:} Gravitational potential of the disk particles of one progenitor galaxy. A change in the curvature at $r_c \approx 1.3 \kpc$ is marked by a red line: using mathematics terminology, the potential $\phi(r)$ is a convex function for $r \le r_\mathrm{c}$, while it is concave elsewhere. In other words, the second derivative is positive in the inner region, which corresponds to a compressive tidal field. \emph{Bottom:} Normalized integrated mass of the disk. The radius $r_\mathrm{c} \approx 1.3 \kpc$ encloses $\sim 2.5\%$ of the total mass.}

%%%%%%%%%%%%
\subsect{Three classes of orbits}

These figures depend on the shape of potential of the progenitor, and thus should not vary with time. However, it is important to understand that only the \emph{statistics} of the compressive modes are constant but the evolution of the tidal field of individual particles is not. In a disk, the mean motion of the particles is the rotation. But radial and vertical components also exist, voiding the orbit of a given particle to be a perfect circle. In this Section, we detail the impact of the orbital parameters of the particles into the statistics of the tidal field, in term of duration of the compressive mode.

In projection on the plane of the disk, the possible cases for orbits are illustrated in \reffig{orbits_compressive_region}. The particle on orbit A has an energy and angular momentum high enough to stay permanently outside of the compressive region. On the orbit B, the particle shows a similar level of energy, but a lower angular momentum. That is why it has a small pericenter distance and thus, enters and leaves the compressive area. Finally, the orbit C has a low energy level and remains in the compressive region.

\fig{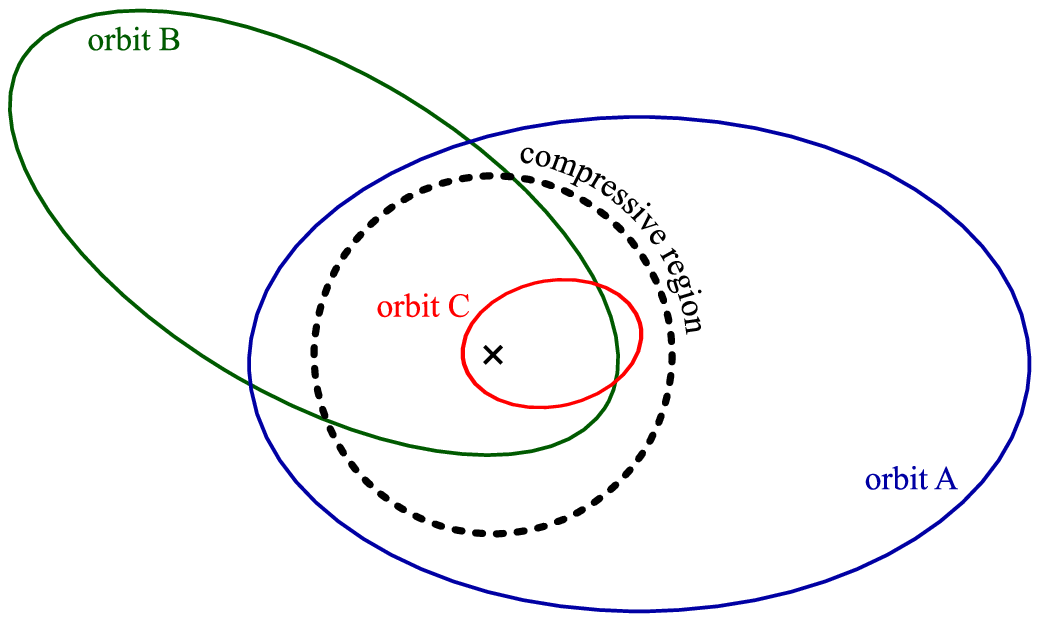}{Examples of orbits of particles}
{The three possible cases of orbits of particles in a disk, with respect to the compressive radius (dashed black circle). (The eccentricities have been exaggerated for ease of illustration.) Orbit A (blue) is entirely outside the radius of the compressive region while orbit B (green) enters and leaves it. Orbit C is enclosed by the compressive radius. The $\times$ sign marks the center of the disk. Note that the vertical components of the orbits are not taken into account explicitly, for simplicity.}

Therefore at a given time $t$, the fraction of the particles in compressive mode (i.e. closer than $\sim 1.3 \kpc$ to the center of the galaxy) are made of large binding energy (class C) and lower binding energy but low angular momentum (class B) particles. At $t+dt$, all the particles of the class C are still in this compressive region but some of the class B have left or entered it. Statistically, the total mass fraction in compressive mode remains $\sim 2.5\%$ but the particles which build it have partially changed. \reffig{isolated_r_l_e} plots the total energy $E$ and the angular momentum $L$ of the particles of the disk, as in \mycitet{Hemsendorf2002}. One reads the minimum energy $E_\mathrm{c} \approx -6.9 \times 10^{39} \erg$ and the maximum angular momentum $L_\mathrm{c} \approx 5.0 \times 10^{53} \ergs$ at the boundary of the compressive region ($r_\mathrm{c} \approx 1.3 \kpc$). With these values, it is possible to plot the diagnostic \reffig{isolated_l_e}, and to distinguish the three classes of orbits.

\fig{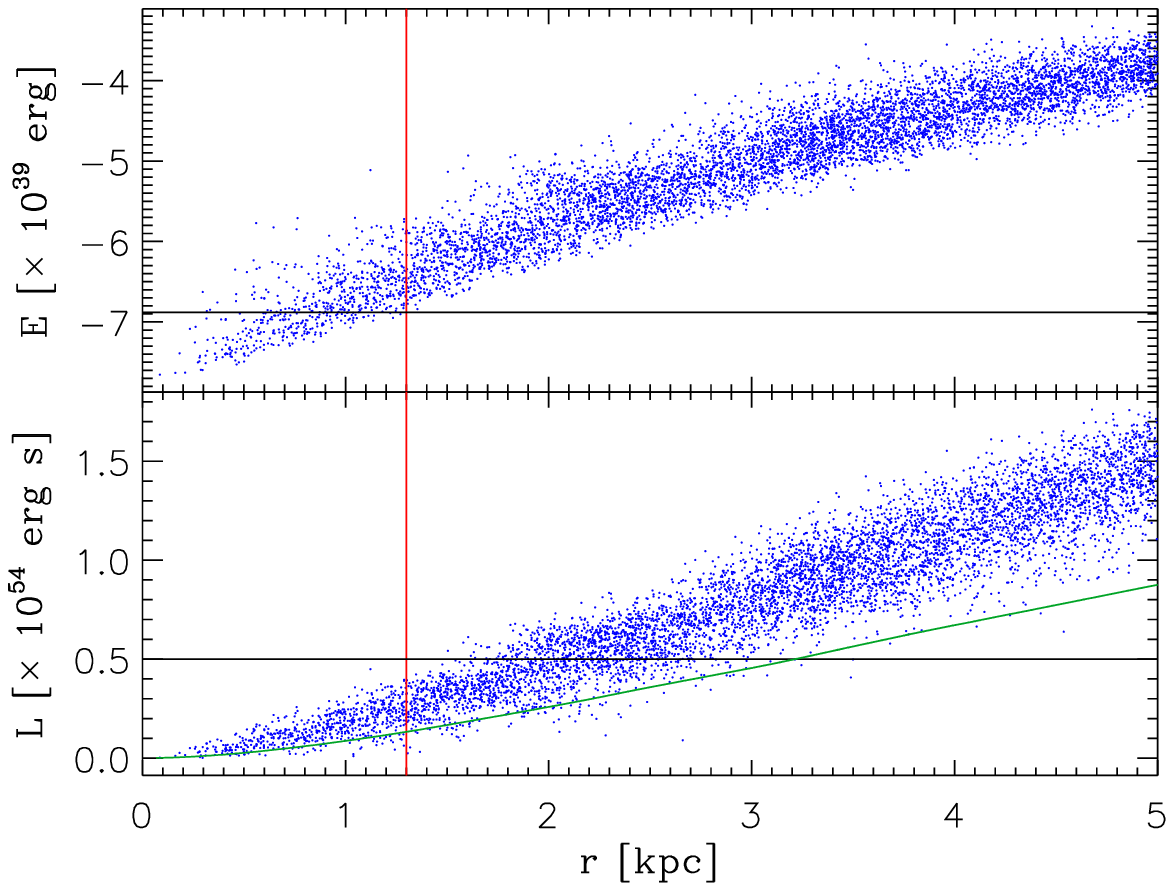}{Energy and angular momentum of the particles}
{Total energy (\emph{top}) and angular momentum (\emph{bottom}) of the disk particles. The vertical red line marks the boundary $r_\mathrm{c}$ of the compressive region. Horizontal lines show the level of energy $E_\mathrm{c}$ and angular momentum $L_\mathrm{c}$ at this boundary, used in the following diagnostic. The green curve represents the theoretical angular momentum of circular orbits.}

\fig{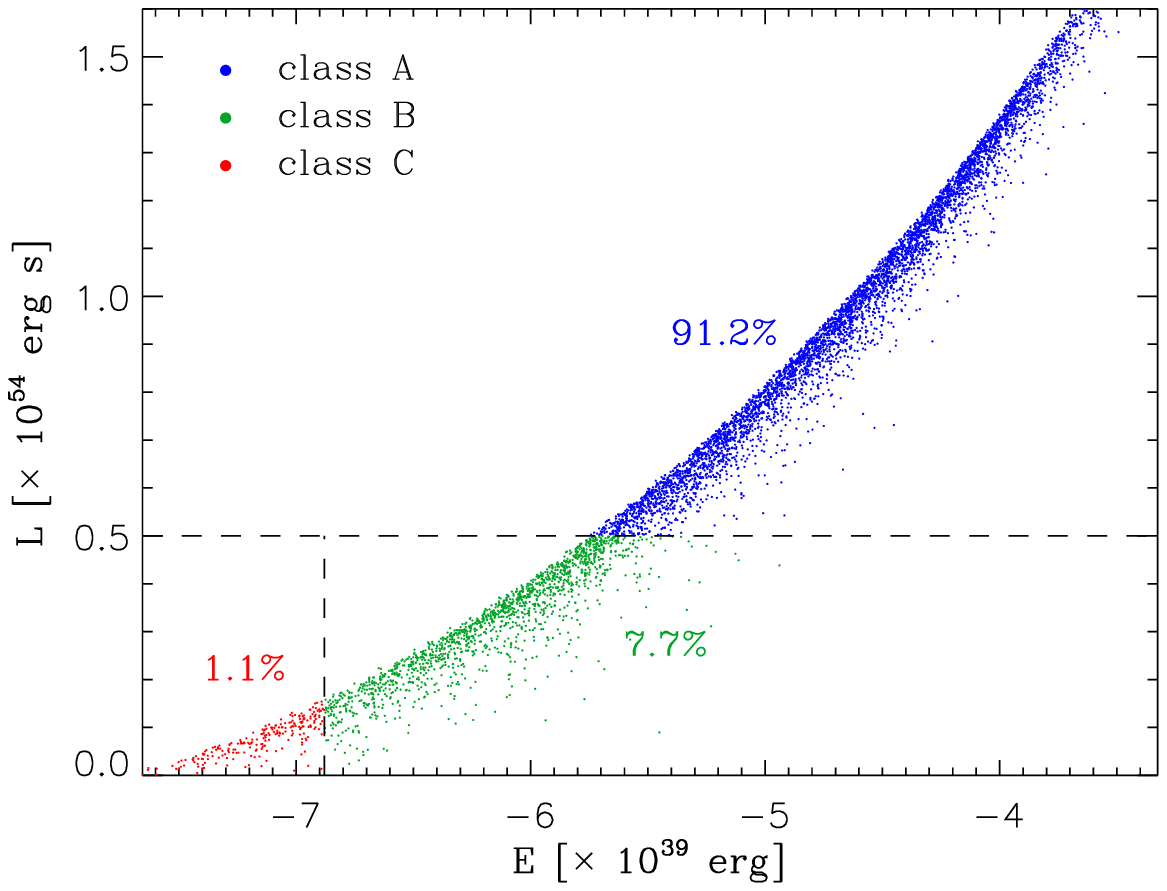}{Classes of orbits}
{The three classes of orbits A (blue), B (green) and C (red) are distinguishable in the energy-angular momentum plane. (Note that only the particles within the central $5 \kpc$ are displayed.) The value of energy $E_\mathrm{c}$ and angular momentum $L_\mathrm{c}$ at the boundary of the compressive region are marked with dashed lines. The class A particles have $E > E_\mathrm{c}$ and $L > L_\mathrm{c}$ and thus never enter into the compressive region. The B class gathers particles with a high energy level ($E > E_\mathrm{c}$) but a low angular momentum ($L < L_\mathrm{c}$) which allows them to enter the central region for a finite amount of time. The C particles have a low energy ($E < E_\mathrm{c}$) and low angular momentum ($L < L_\mathrm{c}$) and hence, remain in the compressive volume. For each class, the percentage indicates the fraction of the total disk mass.}

\emph{In fine}, when considering the timescale of the compressive mode, one has to take into account the contribution of the three classes: class A yields a null period in compressive mode, class B shows a characteristic time $\tau$, depending on the details of the orbits of its particles, and class C has an arbitrarily long time in compressive mode. To estimate $\tau$, we compute the free-fall time of a particle through the compressive region and get $\tau \approx 37 \Myr$. We note that this value is well approximated by the period of a quarter of circular orbit at the radius $r_\mathrm{c}$:
\eqn{
\frac{T_\mathrm{circ}}{4} & = \frac{2\pi r_\mathrm{c}^{3/2}}{4 \sqrt{GM(r<r_\mathrm{c})}} \nonumber \\
& \approx 37 \Myr.
}

\mybox{
It is important to make the difference between the time needed by a particle left at $r = r_\mathrm{c}$ without an initial velocity to reach $r = 0$, and the time the sphere of radius $r_\mathrm{c}$ will take to collapse if its kinetic energy is instantaneously set to zero. In the first case the mass of the inner region (source of gravity) depends on the position of the particle, and thus evolves with time, while in the later case, this mass is kept constant. For a homogeneous sphere of density $\rho_0$, one can show that the first timescale is
\eqn{
\sqrt{\frac{3\pi}{16 G \rho_0}},
}
and the second (more often used in the literature) is 
\follow}\mybox[nocnt]{
\eqn{
\sqrt{\frac{3\pi}{32 G \rho_0}}.
}
In our case, we are interested in characterizing the motion of a particle inside the compressive region, thus the first timescale.
}

The real time spent in compressive mode must be shorter than the time needed by a particle with no initial velocity to get from $r_\mathrm{c}$ to the center and then to $r_\mathrm{c}$ again (enter, cross and exit the compressive zone), i.e. approximately twice our estimate of the free-fall time: $\approx 75 \Myr$. 

This can be interpreted by considering the circular orbit of a particle situated at $r_\mathrm{c}$, plus an epicyclic variation which makes it oscillate around its mean radius. That is, we characterize the fraction of the orbital motion during which the particle goes inside the compressive region, over the total orbital period.

In this simple approach, we have neglected the two-body relaxation, the orbital decay and possible resonance phenomena. Including these second order effects would modify our conclusions for some particles but not for the overall statistics. This assumption has been confirmed by the relative stability of the previous results, when applying the same methods to different snapshots (and thus different epochs). \reffig{isolated_compressive_fraction_long} shows the evolution of the mass fraction in compressive mode over $10 \Gyr$, i.e. much longer than the dynamical times of merging systems. A systematic increase ($0.05 \% \Gyr^{-1}$) is visible however, and can be attributed to secular evolution of the galaxy (possibly a slow transfer of angular momentum through a weak bar mode). In the following, this weak effect has been neglected.

\fig{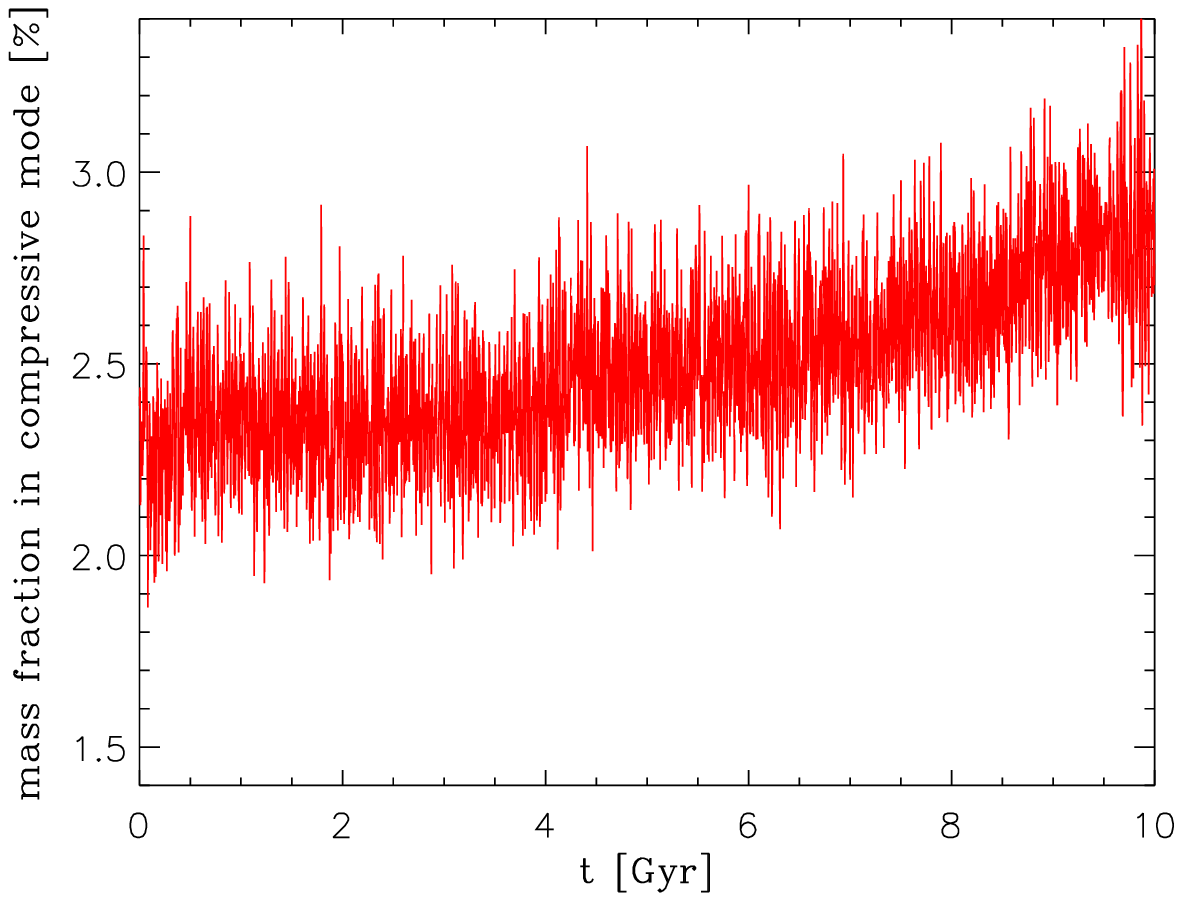}{Evolution of the mass fraction in compressive mode}
{Long term evolution of the mass fraction in compressive mode. (Here again the time resolution is $2.5 \Myr$.) Over $10 \Gyr$, the fraction changes by a few $0.1\%$ only, from its initial value.}

%%%%%%%%%%%%
\subsect[lus_tt]{Statistical timescale of compressive modes}

One seeks a statistical description of the period spent in compressive mode to quantify the idea introduced above. In \mycitet{Renaud2008}, we defined two quantities to distinguish between the two characteristics of the time evolution of the tidal field: the \acr{}{lus}{longest uninterrupted sequence} (\emph{\acr{notarg}{lus}{longest uninterrupted sequence}}) and the \acr{}{tt}{total time} (\emph{\acr{notarg}{tt}{total time}}). For a given particle, \emph{\acr{notarg}{lus}{longest uninterrupted sequence}} is the duration of the longest continuous episode in compressive mode, while \emph{\acr{notarg}{tt}{total time}} is the sum of all compressive episodes, i.e. the total time during which the particle stands in a compressive tidal field. For example, consider a particle in extensive mode for $80 \Myr$, going to compressive mode for $40 \Myr$, extensive again for $30 \Myr$ and finally compressive during $50 \Myr$. Its \emph{\acr{notarg}{lus}{longest uninterrupted sequence}} is $50 \Myr$ while its \emph{\acr{notarg}{tt}{total time}} equals $40+50 = 90 \Myr$. In any case, one has \emph{\acr{notarg}{lus}{longest uninterrupted sequence}} $\le$ \emph{\acr{notarg}{tt}{total time}}, the equality being reached when the particle experiences only one or zero compressive event. Therefore, the three classes of orbits yield different statistical timescales: class A shows \emph{\acr{notarg}{lus}{longest uninterrupted sequence}} = \emph{\acr{notarg}{tt}{total time}} = 0, class B has $0 <$ \emph{\acr{notarg}{lus}{longest uninterrupted sequence}} $\le$ \emph{\acr{notarg}{tt}{total time}} and class C yields the largest values of \emph{\acr{notarg}{lus}{longest uninterrupted sequence}} (which also equals \emph{\acr{notarg}{tt}{total time}}).

\reffig{isolated_lus} plots the distributions of \emph{\acr{notarg}{lus}{longest uninterrupted sequence}} and \emph{\acr{notarg}{tt}{total time}}, in term of mass fraction\footnote{The reader might take the time to get familiar with this kind of figure, as it will be often used in the next Sections and Chapters.}: for a given time interval $\Delta t$, the curve gives the mass fraction of the disk that yields a \emph{\acr{notarg}{lus}{longest uninterrupted sequence}} (or \emph{\acr{notarg}{tt}{total time}}) longer than or equal to $\Delta t$. The plot is normalized by considering only the non-constantly compressive particles (i.e. classes A and B). Therefore, the curves on these plots are always decreasing with a maximum value of $100\%$ for $\Delta t = 0$ (since any compressive mode lasts more than $0 \Myr$).

\fig{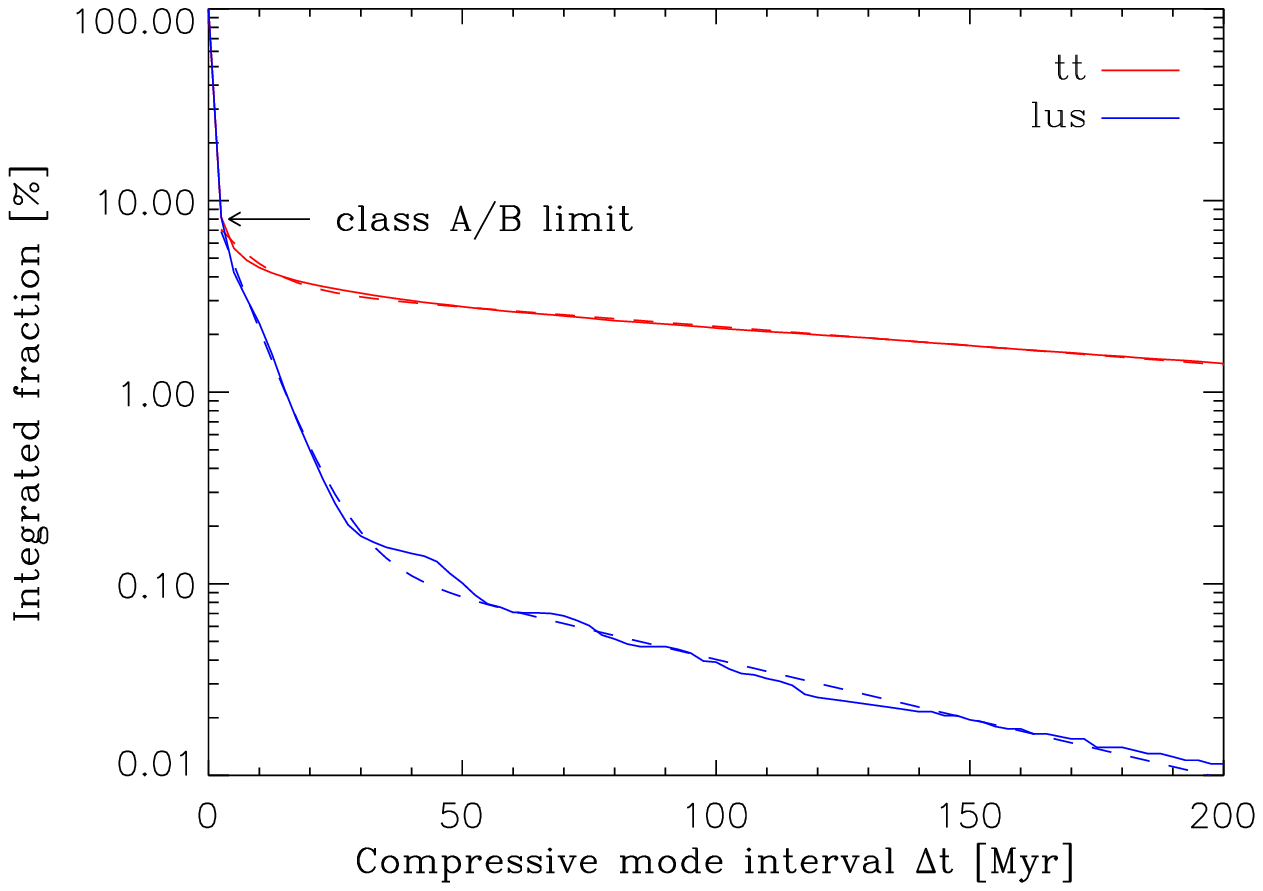}{Distribution of \emph{lus} and \emph{tt} for an isolated galaxy}
{\emph{\acr{notarg}{lus}{longest uninterrupted sequence}} and \emph{\acr{notarg}{tt}{total time}} distributions of compressive mode, integrated for $1.1 \Gyr$. The arrow indicates the time interval where the class B particles become important. The dashed lines corresponds to the double exponential fit to the numerical results (see \refeqn{doubleexp} and \reftab{lus_tt_fit_isolated}).}

In \reffig{isolated_lus}, the \emph{\acr{notarg}{lus}{longest uninterrupted sequence}} and \emph{\acr{notarg}{tt}{total time}} distributions are equal for the period of $\Delta t < 2.5 \Myr$ (i.e. one snapshot, the time resolution) and $\sim 92 \%$ of the mass. This matches well our estimates for the class A. The remaining $\sim 8\%$ are the class B particles, with a \emph{\acr{notarg}{lus}{longest uninterrupted sequence}} longer than one snapshot. The long periods reflect the fluctuations previously mentioned (e.g. two-body relaxation) and thus corresponds only to a small number of particles.

Both distributions can be fitted over the interval $]0, 200] \Myr$ by double exponential laws of the form
\eqn[doubleexp]{
f_1 \exp{\left(-\frac{t}{\tau_1}\right)} + f_2 \exp{\left(-\frac{t}{\tau_2}\right)},
}
where $f_1$ and $f_2$ are constants, and $\tau_1$ and $\tau_2$ are characteristic timescales. \reftab{lus_tt_fit_isolated} lists the best fit parameters and the residual error $\sigma$, as shown with dashed line in \reffig{isolated_lus}. These double exponential laws suggest two regimes defined by the time intervals $\Delta T_1$ and $\Delta T_2$ during which one exponential dominates the other\footnote{In \mycitet{Renaud2008} and \mycitet{Renaud2009}, the distributions have been split into short and long term regimes, and each has been fitted separately. In the present document, the distributions are fitted with \refeqn{doubleexp} over the entire range $]0,200] \Myr$. This avoids the approximation in the definition of the transition zone between $\Delta T_1$ and $\Delta T_2$, and generally gives better fits. Therefore, the parameters presented here differ from those in our previous publications. The overall distributions are similarly reproduced however, and the conclusions in all documents are not affected by this change.}. For the \emph{\acr{notarg}{lus}{longest uninterrupted sequence}} distribution, the transition occurs at $\Delta t \approx 27 \Myr$, where the distribution shows a knee in the log-normal plane.

\tab{12cm}{lus_tt_fit_isolated}{Parameters of the \emph{lus} and \emph{tt} distributions}
{l@{\extracolsep{\fill}}c@{\extracolsep{\fill}}c@{\extracolsep{\fill}}c@{\extracolsep{\fill}}c@{\extracolsep{\fill}}c}
{
Distribution & $f_1$ & $\tau_1$ [Myr] & $f_2$ & $\tau_2$ [Myr] & $\sigma$ \\
\hline
\emph{lus} & 10.1 & 6.2 & 0.2 & 69.9 & 0.16 \\
\emph{tt} & 5.0 & 7.6 & 3.5 & 216.1 & 0.15 \\
}{}

Note that these results depend on the duration of integration. Indeed, when integrating over a long time, one allows the particles to stand more often in a compressive mode. For quiescent systems, like our isolated progenitor, the \emph{\acr{notarg}{lus}{longest uninterrupted sequence}} is set by quantities that are roughly conserved over time ($E$ and $L$). With no perturbation, the particles remain on similar orbits and thus have the same \emph{\acr{notarg}{lus}{longest uninterrupted sequence}} for each revolution. The \emph{\acr{notarg}{tt}{total time}} however necessarily increases with time and leads to a flatter distribution. \reffig{isolated_lus_integration} illustrates this point by showing the \emph{\acr{notarg}{lus}{longest uninterrupted sequence}} and \emph{\acr{notarg}{tt}{total time}} distributions for different integration periods. The \emph{\acr{notarg}{lus}{longest uninterrupted sequence}} varies for no more than $1\%$ while differences for \emph{\acr{notarg}{tt}{total time}} reach one order of magnitude for $\Delta t = 200 \Myr$. In any case however, the distributions are still well fitted with double exponential laws and the change of regime between $\Delta T_1$ and $\Delta T_2$ remains at approximately the same $\Delta t$.

\fig{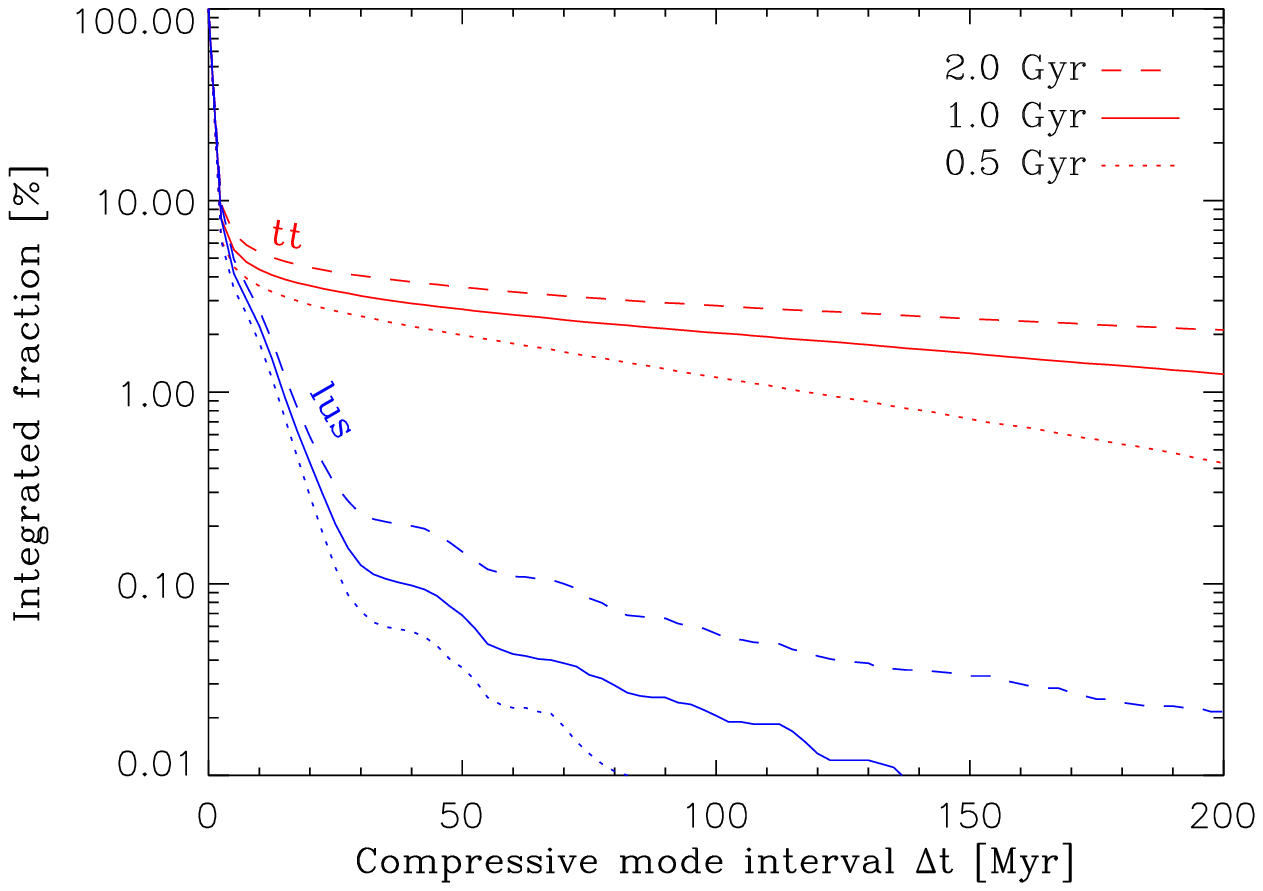}{Dependance of \emph{lus} and \emph{tt} on the integration time}
{\emph{\acr{notarg}{lus}{longest uninterrupted sequence}} and \emph{\acr{notarg}{tt}{total time}} distributions of compressive mode, for several integration times ($0.5 \Gyr$, $1.0 \Gyr$ and $2.0 \Gyr$). As expected, the long integration periods yield a higher integrated fraction, as one gives more opportunities to the particles to go into the compressive region.}

%%%%%%%%%%%%%%%%%%%%%%%%%%%%%%%%%%%%%%%%%%%%%%%%%%
\sect{Tidal field of the merger}

In the previous Section, we have characterized the properties of the tidal field of the progenitor galaxies of our Antennae model. It is now possible to apply the same method to the merger itself and quantify its effect on the tidal field, with respect to isolated galaxies. In a first approach, we focus on a binary test for the nature of the tides: compressive or extensive. To do so, the sign of the maximum eigenvalue of the tidal tensor for all the particles of both disks is examined. 

%%%%%%%%%%%%
\subsect{Mass fraction in compressive mode}
For an isolated progenitor, the mass fraction in compressive mode depends on the number of particles lying in the central region, i.e. $\sim 2.5\%$. We have seen that this number is statistically relevant and remains stable for an arbitrarily long time. During the course of the merger however, the fast evolving shape of the global potential is expected to influence this value by creating local, off-nuclear cores and thus additional compressive regions.

\reffig{antennae_histo} shows the evolution of the mass fraction in compressive mode, from the isolation stage to the merger phase. The relation between major interaction events and the increase of this fraction is clearly shown in this plot as peaks (see also \reftab{events}).

\fig{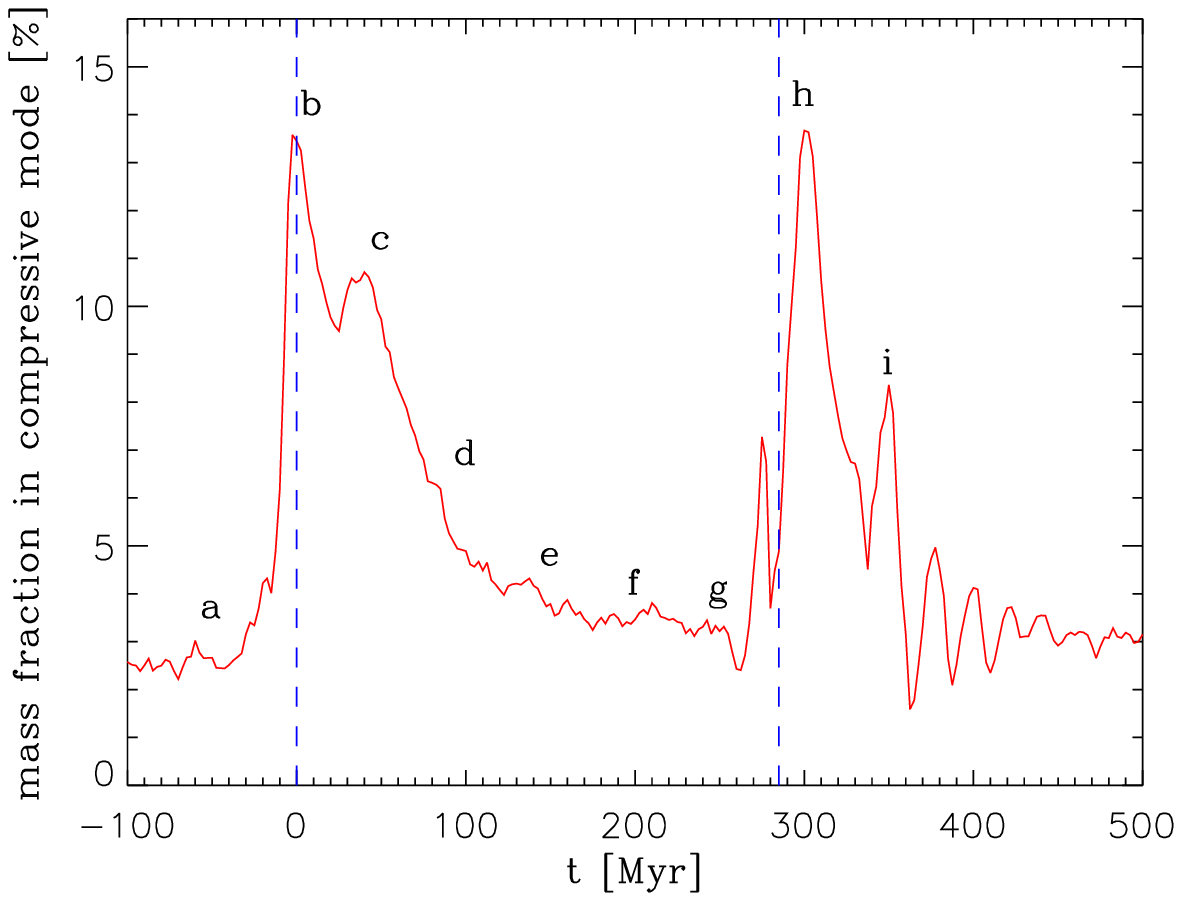}{Mass fraction in compressive mode}
{Evolution of the mass fraction in compressive mode through the course of the merger of the Antennae galaxies. The two blue dashed lines indicate the first and second pericenter passages.}

\tab{8cm}{events}{Interaction events}
{c@{\extracolsep{\fill}}c@{\extracolsep{\fill}}l}
{
Id.$^\star$ & t [Myr] & Event \\
\hline
a & -50 & isolation\\
b & 0 & first passage\\
c & 50 & creation of the bridges\\
d & 100 & maximum separation\\
e & 150 & expansion of the tails\\
f & 200 & expansion of the tails\\
g & 250 & expansion of the tails\\
h & 300 & second passage, now\\
i & 350 & merger phase\\
}{$^\star$ The Id letters refer to the panel identifiers in \reffig{antennae_op}, \reffig{antennae_los} and \reffig{antennae_compressive}, as well as time markers in \reffig{antennae_histo}.}

Both passages lead to a rapid\footnote{The rise begins a few Myr before the pericenter passage itself. A compressive event appears just before the distance reaches its minimum, because of the extended size of the disks. The shift in time between the beginning of the compressive peak and the maximum corresponds to the time needed for a galaxy to fly from the edge to the center of the other.} increase of the mass fraction in compressive mode, by a factor $\approx 5$ compared to the level in isolation. The decreasing rate then depends on the configuration of the merger. When the galaxies fly away from each other (steps d, e, f and g), the fraction is slowly and smoothly decreasing, while it changes much more quickly during the merger phase (step i). This last phase is well visible as a fairly regular damped oscillation. We note that the fraction is comparable for both isolation regimes: before interaction ($t \lesssim -20 \Myr$) and after the merger ($t \gtrsim 450 \Myr$).

As for an isolated progenitor, the final fraction ($\sim 3\%$) remains stable for a very long time. It maintains however a mild oscillation (mostly covered by the noise) with the same period than the oscillatory merger phase ($\sim 20 \Myr$), well visible in the frequency domain: \reffig{antennae_histo_psd} shows the \acr{}{PSD}{power spectral density} of the mass fraction in compressive mode.

\fig{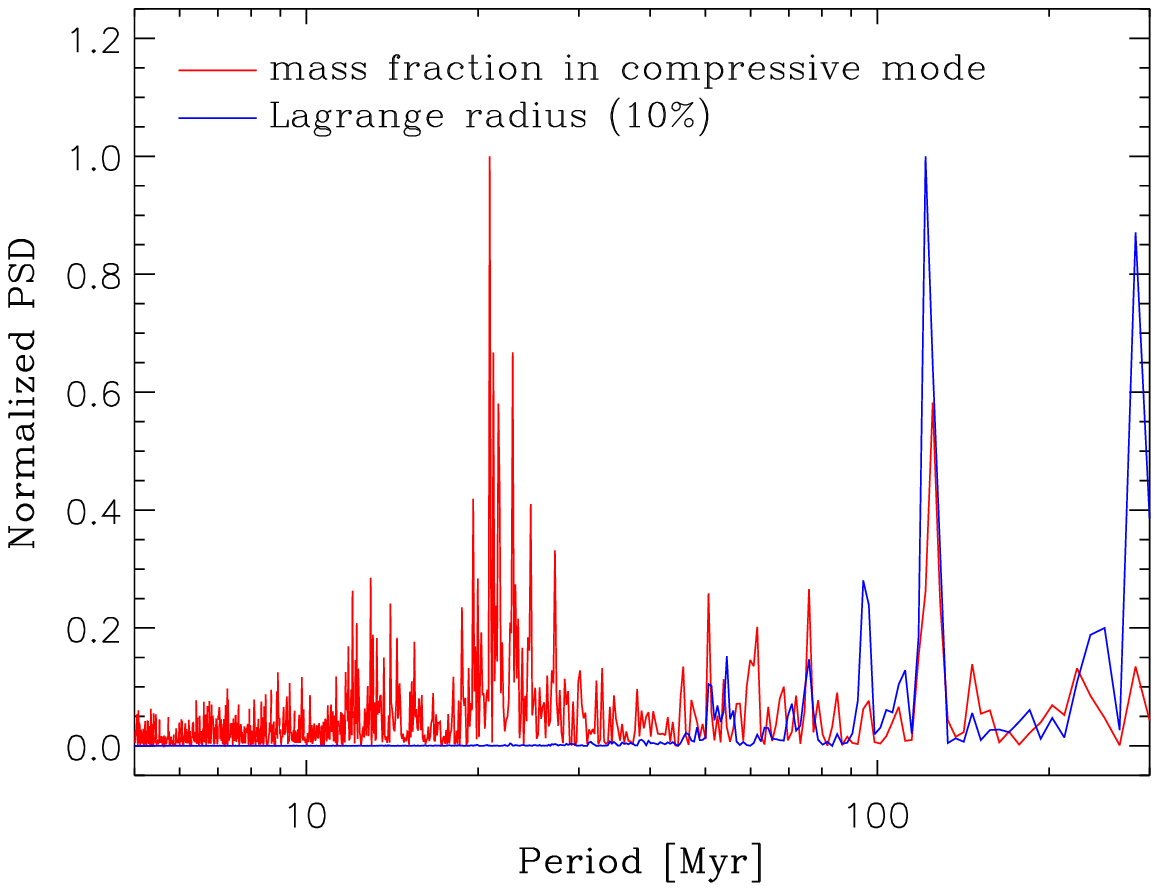}{PSD of the mass fraction in compressive mode}
{\acr{notarg}{PSD}{Power spectral density} of the mass fraction in compressive mode (red) after the merger phase ($t \in [475, 4725] \Myr$). The main peak at $20 \Myr$ (surrounded by a forest of smaller yet significant ones) corresponds to the oscillation of the two nuclei, already visible in previous figures. Secondary peaks ($52, 75, 125, 145 \Myr$) are also visible in the \acr{notarg}{PSD}{power spectral density} of the $10\%$ Lagrange radius (blue), which indicates a ``breathing'' of the merger remnant.}

%%%%%%%%%%%%
\subsect[antennae_spatial_distrib]{Spatial distribution of the compressive tidal mode}
In isolation, the tidal field shows a compressive mode in the central region of the galaxy. In the course of a merger, the tidal forces responsible for the creation of the large scale structures like bridges and tails are likely to disturb the character of the tidal field on similar scales and build new compressive regions.

\fig{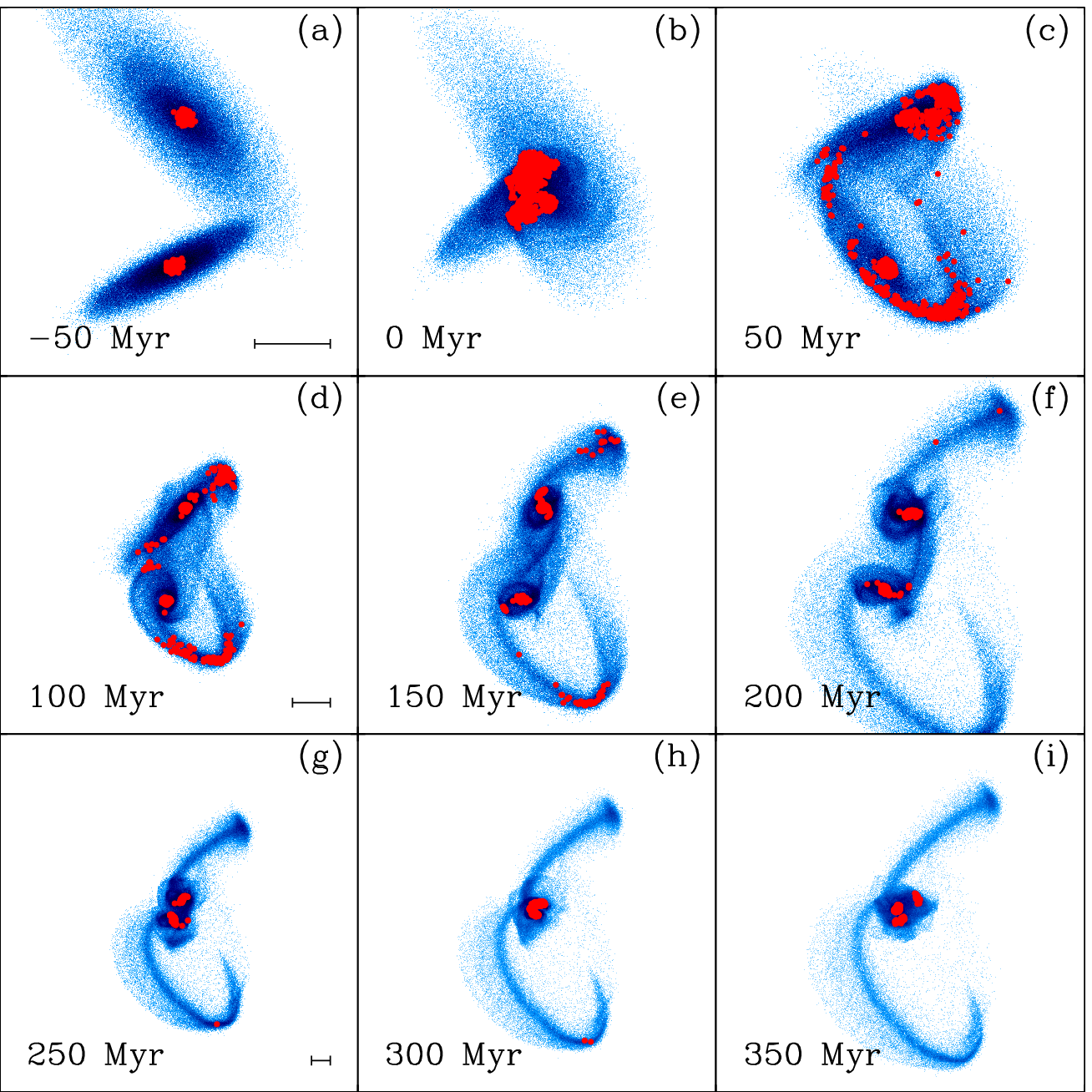}{Spatial distribution of the compressive mode}
{Spatial distribution of the particles in compressive tidal mode in the Antennae model, in the plane of the sky. The particles in compressive mode are marked with red dots. On each row, the black line represents $10 \kpc$. Again for the sake of clarity, only one out of 40 red dots is displayed. (Figure~3 of \mycitealt{Renaud2009}.)}

\reffig{antennae_compressive} displays the morphology of the merger and marks particles standing in compressive mode with red dots. The panel (a) clearly shows the intrinsic compressive region discussed above ($r_\mathrm{c} \approx 1.3 \kpc$), as the progenitors can still be considered isolated. During the first pericenter passage (b), the compressive regions expand to the local cores in the potential, where the two disks overlap. The tails that grow while the progenitors fly away from each other (panels c, d, e and f) contain several compressive regions, well separated from the nuclei of their host galaxy. The largest of those is situated near the tip of both tails, i.e. close to the position of the observed \acr[s]{}{TDG}{tidal dwarf galaxies} (recall \reffig{antennae_uv}). Note that the tidal bridges also display compressive areas, but significantly less than the tails. A transient spiral pattern already existing in the morphology of the merger is partially reproduced in the map of the compressive mode for $t\approx 200 \Myr$ (see the inverted S-shape near the southern nucleus in panel f).

\fig{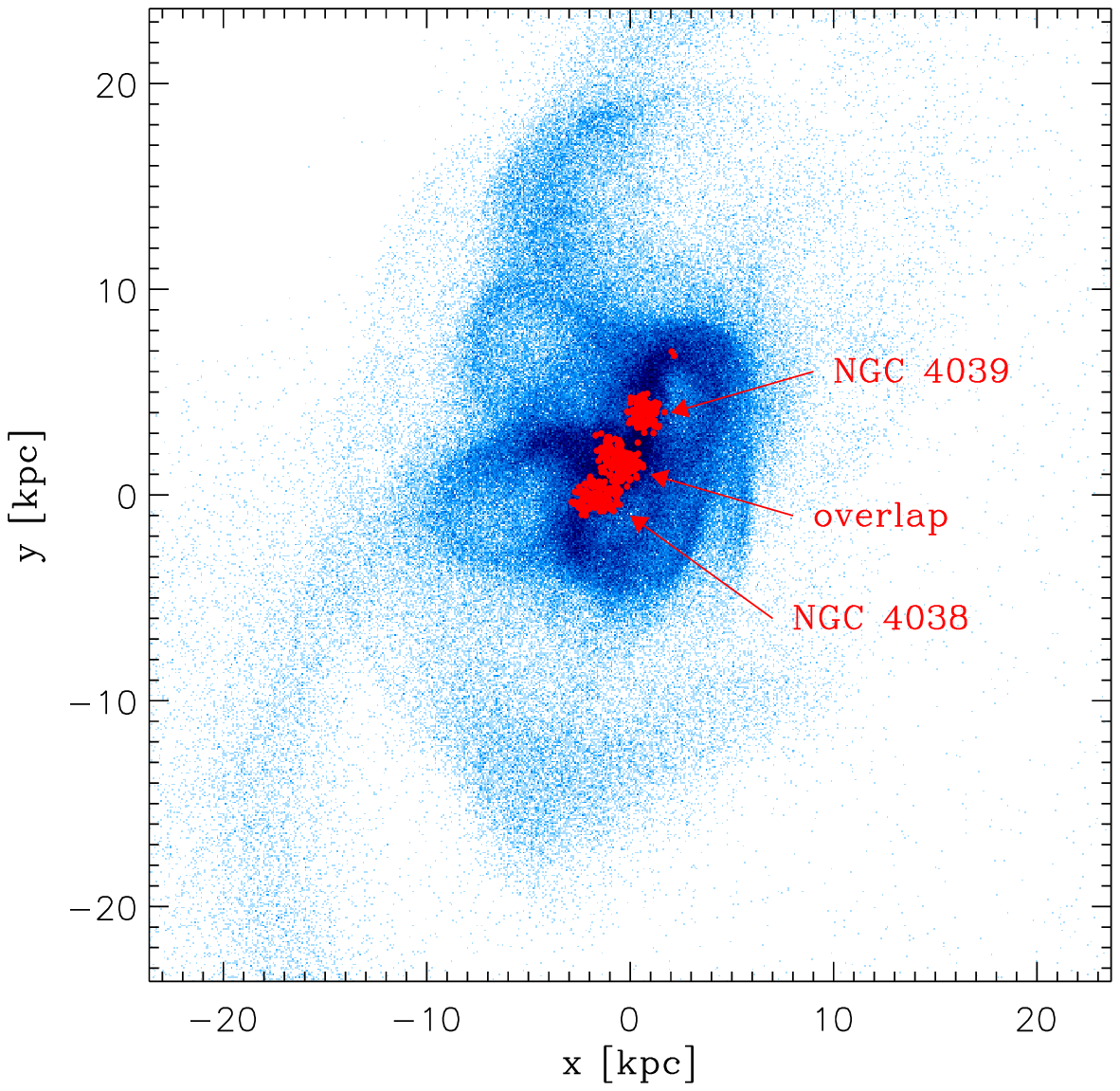}{Central compressive regions, before the second passage}
{Map of the central compressive regions for $t = 270 \Myr$, i.e. $10\Myr$ before the second pericenter passage. The spiral arms formed during the first passage are clearly visible around the two nuclei, as pseudo loops in this projection.}

During the second passage (panels g and h), the overlap of the two disks rises a large compressive region (see \reffig{antennae_compressive_218}). Because of the loss of their orbital angular momentum, the two galaxies spiral in on each other, leading to an oscillatory phase visible in \reffig{antennae_dist}. The same happens for the compressive regions. At the end of the merger phase, we note the presence of two zones, visible on \reffig[i]{antennae_compressive}, expanding radially in a similar fashion than a wave. Finally, all the off-center compressive areas vanish within a few Myr, but the newly formed central nucleus remains tidally compressive.

It appears that, during the course of the merger, compressive regions match the locations of over-densities (overlap, nuclei) but also intermediate and low densities (bridges, tails). \reffig{antennae_compressive_density} displays the ratio between the column density of compressive particles and the one of all particles, for NGC~4038 at $t = 150 \Myr$. The gradients visible in this plot denotes that the compressive mode is independent of the density in the tidal structures but shows a correlation with the densest regions in the nuclei (a similar pattern also exist for the northern galaxy).

\fig{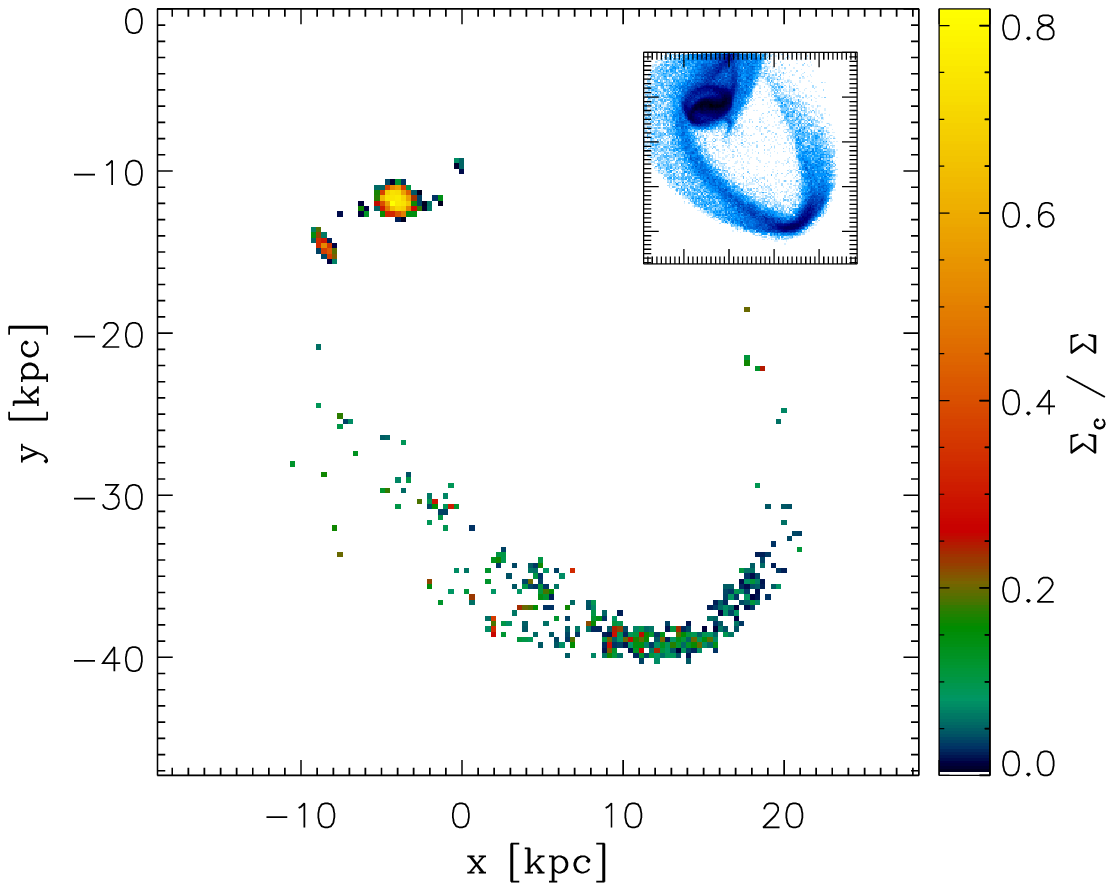}{Density of compressive particles}
{Ratio of the column density of the compressive particles $\Sigma_\mathrm{c}$ to the total column density $\Sigma$ of the southern galaxy in the plane of the sky at $t=150 \Myr$ (see the equivalent morphology in the thumbnail). The column densities are computed on a grid made of $300 \pc$-wide squared cells. In the nucleus, most of the particles are compressive but the fraction is much smaller in the tidal tail. Note that a ratio of unity is reach for the \emph{volume} density in the nucleus, but not for the column density, because of foreground and background particles that are not in compressive mode.}

\reffig{antennae_compress_map} does not only shows the compressive regions but the entire tidal field, computed in a regular grid, for the central part of the Antennae. It also overlays the positions the young cluster candidates observed by \mycitet{Mengel2005}. Most of these clusters lie in the red, compressive areas of the map, i.e. in the overlap region and the western-loop. Even when taking the projection effect into account, the strong correlation visible between the \emph{young} clusters and the compressive regions points out the role of the tidal field in the early phases of the life of a star cluster. Still the question of the timescale of these modes persists.

\fig{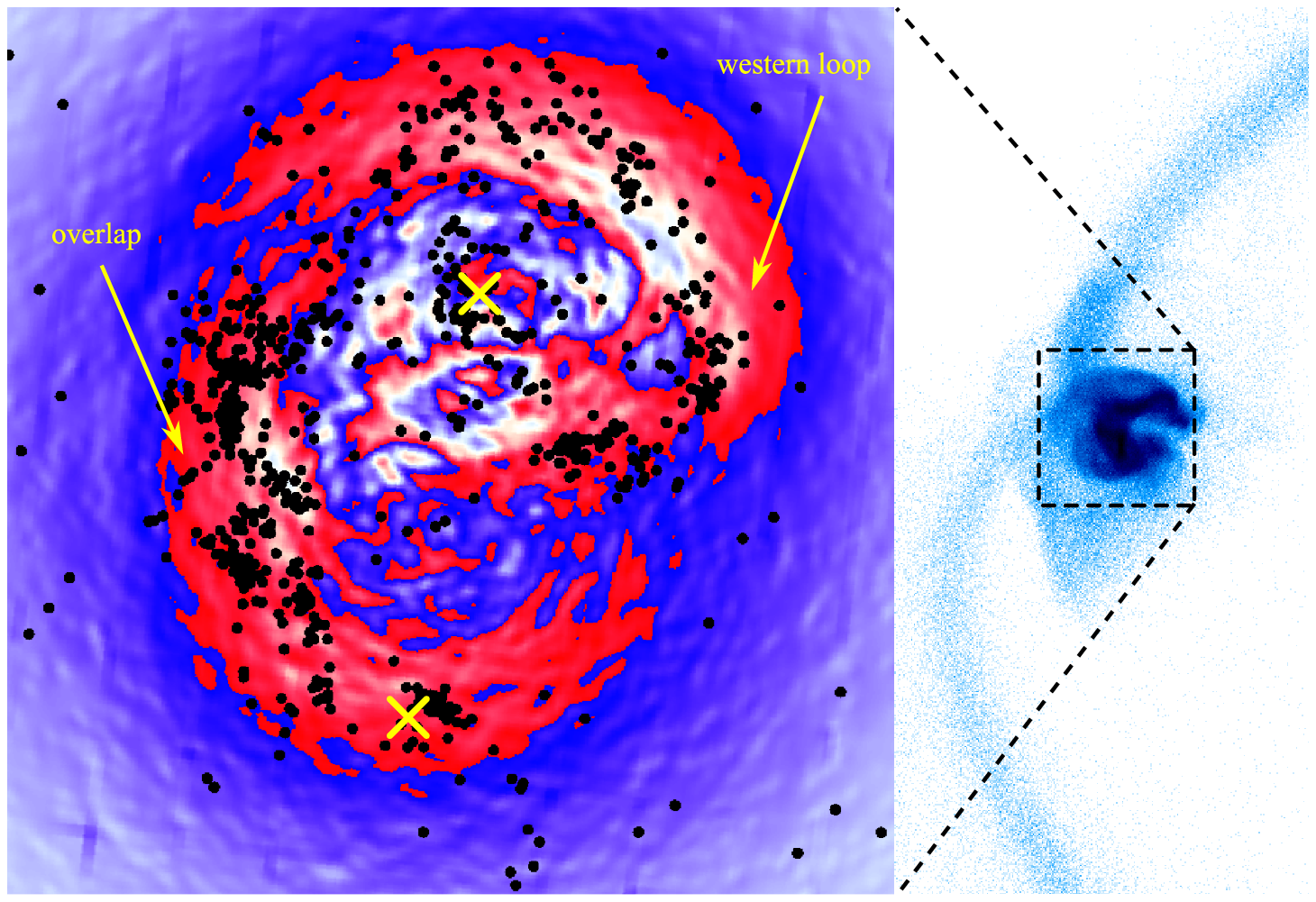}{Tidal field of the central part of the Antennae}
{Tidal field of the central part of the Antennae galaxies for $t=300 \Myr$. The blue to red colors label extensive to progressively more compressive mode. White corresponds to a neutral tidal field. The black dots indicate the positions of the young cluster candidates identified by \myciteauthor{Mengel2005} (\myciteyear{Mengel2005}, see their Figure~1). The map covers a $\sim 12 \kpc$ wide square, parallel to the plane of the sky, $\approx 2.2 \kpc$ backward from the plane containing the global center of mass. The nuclei of the two galaxies are marked with yellow crosses. Arrows roughly indicate the two major off-nuclear regions: the overlap and the western-loop. (Adaptation of Figure~2 of \mycitealt{Renaud2008}.)}

%%%%%%%%%%%
\subsect{Age distribution of the compressive tides}

One may argue that the good match between the positions of the observed young clusters and the existence of compressive mode might be a coincidence. A comparison between the duration of both phenomena (compressive tidal field and star formation) would be another point to confirm or not the supposed link between the two. To do so, we compare the time during which this mode is operative, to the typical timescales for the formation of the star clusters, i.e. a few Myr. We do this statistically, thanks to the \emph{\acr{notarg}{lus}{longest uninterrupted sequence}} and \emph{\acr{notarg}{tt}{total time}} previously introduced.

\reffig{antennae_period} compares the \emph{\acr{notarg}{lus}{longest uninterrupted sequence}} distribution of the merger with the one of an isolated progenitor. The interaction clearly induces more short-term compressive events ($\Delta t \lesssim 50 \Myr$) than for the progenitors taken in isolation. Here again, the \emph{\acr{notarg}{lus}{longest uninterrupted sequence}} distribution can be fitted with a double exponential law (recall \refeqn{doubleexp}) with the parameters given in \reftab{antennae_lus}. The timescale for the first period ($\tau_1$) remains very similar to the isolated case (only the mass fraction changes), however $\tau_2$ is significantly higher for the merger. Note that the position of the knee marking the transition between $\Delta T_1$ and $\Delta T_2$ is shifted from $\approx 27 \Myr$ to $\approx 50 \Myr$. All together, these points underline that the interaction drives many short-term compressive events but also a few much longer episodes than in a quiescent disk. 

\fig{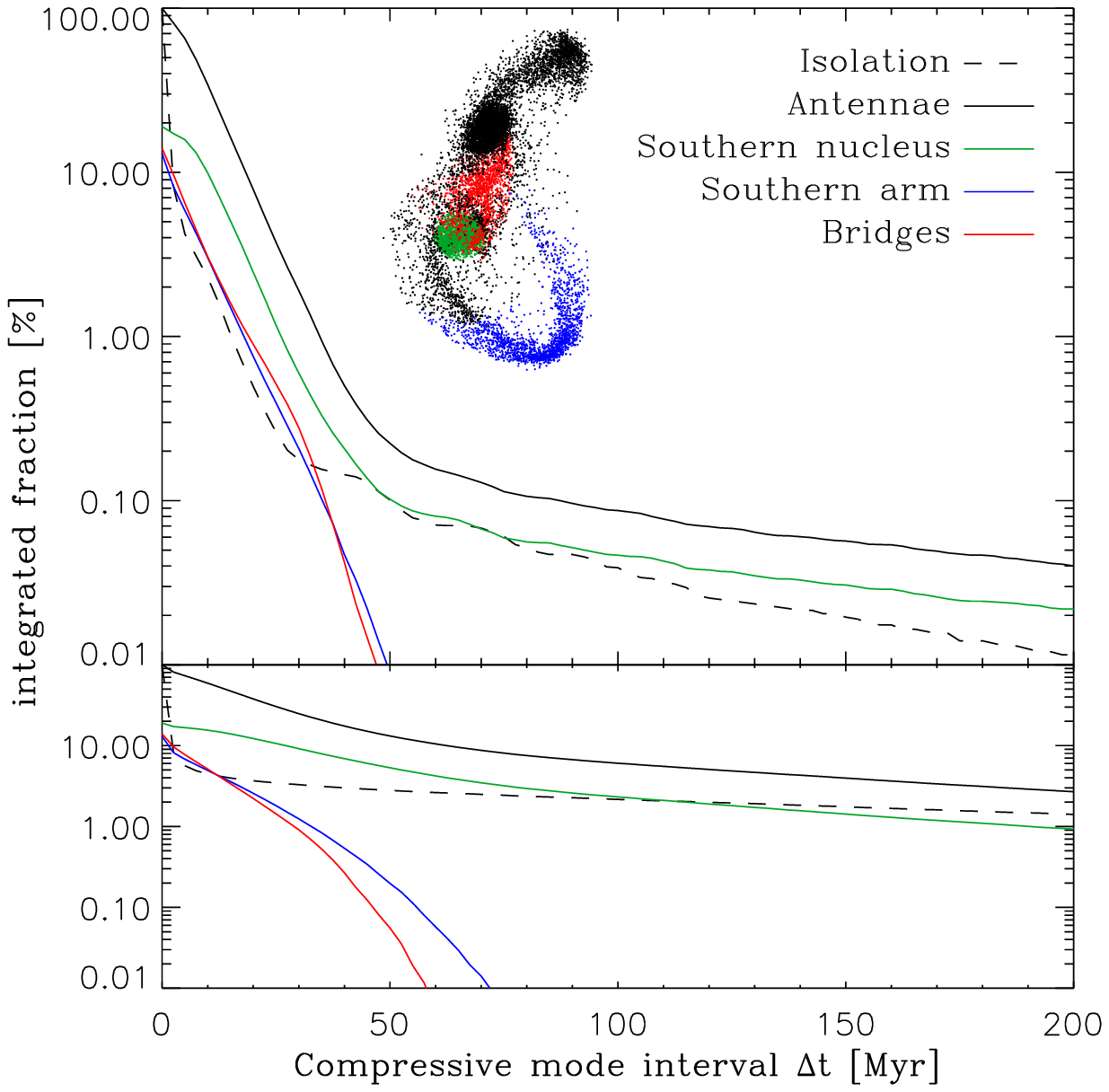}{\emph{lus} and \emph{tt} distributions in the Antennae}
{\emph{\acr{notarg}{lus}{longest uninterrupted sequence}} (\emph{top}) and \emph{\acr{notarg}{tt}{total time}} (\emph{bottom}) distributions in the Antennae (solid black line), compared to a progenitor in isolation (dashed black, as in \reffig{isolated_lus}). Several regions of the Antennae have also been extracted, as shows in the thumbnail (for $t = 150 \Myr$). For these subsets, the integrated fraction is normalized to the entire merger. Indeed, the value read for $\Delta t = 0$ is the mass fraction corresponding to the selected region: 19\% for the southern nucleus (NGC~4038), 13\% for the southern arm and 14\% for the bridges. Note that the part of the arm closest to the nucleus has not been selected because it rapidly falls back into the central region and thus, biases the statistics. In all cases, the integration has been performed for $t \in [-100, 1000] \Myr$.}

\tab{10cm}{antennae_lus}{Parameters of the \emph{lus} distributions}
{l@{\extracolsep{\fill}}c@{\extracolsep{\fill}}c@{\extracolsep{\fill}}c@{\extracolsep{\fill}}c}
{
Region & $f_1$ & $\tau_1$ [Myr] & $f_2$ & $\tau_2$ [Myr] \\
\hline
Isolation & 10.1 & 6.2 & 0.2 & 69.9 \\
Antennae & 137.2 & 6.8 & 0.2 & 119.3 \\
Southern nucleus & 32.1 & 7.3 & 0.1 & 120.3 \\
Southern arm & 12.3 & 7.4 & - & - \\
Bridges & 21.3 & 5.6 & - & - \\
}{}

The underlying reason for these differences resides in the interaction between the two galaxies, and thus depends on the tidal structures. To investigate this idea, different regions of the merger have been explored separately. Clearly, the double exponential distribution is due to the nuclei of the galaxies which yield long compressive episodes. The short-term features on the other hand come mainly from the transient tidal structures like the tidal tails and bridges, which do not show any compressive episodes over $\Delta T_2$. The shorter timescales $\tau_1$ however, are of the same order of magnitude everywhere.

It is worth noticing that the periods reported here ($\tau_1 \approx 10^7 \yr$, knee at $\Delta t \approx 50 \Myr$) match well the timescales of star cluster formation. These numbers even suggest that the compressive mode could be active as long as the early phase of the life of a young cluster, before the tidal field switches back to the more usual extensive mode. The consequences of this change and their links with the very survival of these stellar associations will be discussed further.

%%%%%%%%%%%%
\subsect{Realizations of the statistics}

We have introduced the concepts of \emph{\acr{notarg}{lus}{longest uninterrupted sequence}} and \emph{\acr{notarg}{tt}{total time}} distribution to get a statistical description of the duration of the compressive tidal mode, applied to each mass element of our simulations. However, any statistics is a collection made of particular cases and it is important to keep in mind that the actual behavior of the tidal field for a given particle may not be well represented by the statistics presented above. In the following, we extract some typical cases for the evolution of the tidal field, in order to complete the picture we have built with statistic tools.

\reffig{antennae_part_orbits} displays a sample of seven orbits, with dots marking the compressive events along the orbits. Particles labeled A and B stay in the vicinity of the nucleus of NGC~4038. Particle A even remains in the intrinsic compressive region of the galaxy and thus, shows an almost uninterrupted compressive episode. By contrast, B only yields a relatively short compressive event, during the first pericenter passage. C goes to the bridges before falling back into the nucleus. Similarly, D falls back during the merger phase, an episode clearly denoted by a compressive sequence after the second passage. On a larger scale, E is ejected to the tidal arm by the first interaction but is captured back by the nucleus before the second. F and G remain in the tidal tails, with G being part of the \acr[s]{}{TDG}{tidal dwarf galaxies} region spotted in \refsec[antennae]{antennae_spatial_distrib}.

\fig{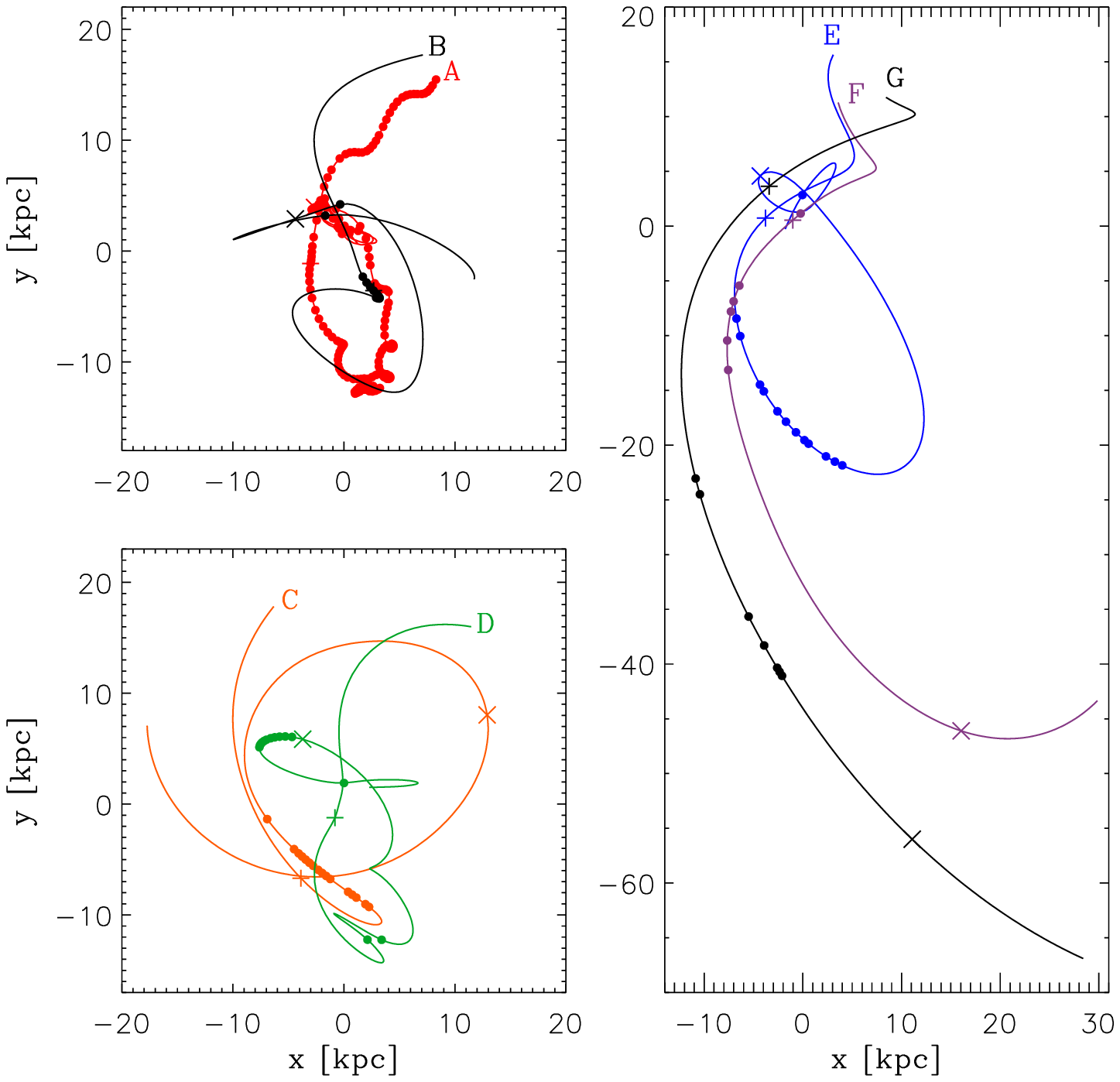}{Sample of seven orbits}
{Sample of orbits (in the orbital plane) of seven particles in the central region (\emph{left}) and at large scale (\emph{right}). A compressive event for a given particle is shown by a dot on its orbit. The initial position of each particle is labeled by a letter. The pericenter passages of the progenitors are marked with crosses ($+$ and $\times$, respectively) on every orbit.}

Such a plot shows the tidal history of individual particles or, in other words, some realizations of the statistics presented before. The major families of orbits are well summarized in this Figure which allows us to get a broader picture of how the tidal field evolves. In addition, this kind of approach is the starting point of the simulation of star clusters, using the tidal field derived here as boundary conditions, as presented in the next Chapter.

%%%%%%%%%%%%%%%%%%%%%%%%%%%%%%%%%%%%%%%%%%%%%%%%%%

\cha{formation}{Tides and star formation}
\chaphead{In this Chapter, the results on the tidal field of the Antennae galaxies are applied to the context of the formation of star clusters. We highlight the links between the scalelengths of star formation and those of the galactic tides, in which the clusters are embedded. Some of this work has been presented in \mycitet{Renaud2008} and \mycitet{Renaud2009}.}

%%%%%%%%%%%%%%%%%%%%%%%%%%%%%%%%%%%%%%%%%%%%%%%%%%
\sect{Timing and energy of a compressive mode}

In the previous Chapter, we have seen that the compressive mode of the tidal field shows properties similar to those of the formation of star clusters and \acr[s]{}{TDG}{tidal dwarf galaxies} in the Antennae galaxies. First, the location of the compressive regions in our simulation corresponds surprisingly well with the sites of numerous candidate young clusters. Second, the duration of these modes is of the same order as the timescale of star formation. These correlations are a strong hint that the tidal field, and especially its compressive modes, play a significant role in the formation of star clusters. It is therefore of prime importance to compare quantitatively the tides with other mechanisms involved in the formation of clusters.

%%%%%%%%%%%
\subsect{The compressive mode as a destructive effect}

As already suggested, the tidal field plays a role in the formation of the star clusters. It may also participate in their destruction either through:
\begin{itemize}
\item the expansion of the external layers triggered by gas mass-loss and enhanced by extensive tidal mode,
\item tidal shocks, as introduced by \myciteauthor{Spitzer1958} (\myciteyear{Spitzer1958}, see also \mycitealt{Spitzer1987}, \mycitealt{Kundic1995} and \mycitealt{Gnedin1999}).
\end{itemize}

\mybox{
In the context of star clusters, a tidal shock was first described by \mycitet{Spitzer1958} as the dynamical heating of a cluster going through a galactic disk. Let's consider the density profile of a radially infinite disk, i.e. only depending on the height. A cluster centered on $Z_\mathrm{c}$ undergoes the acceleration $g(Z_\mathrm{c})$ from the disk. A star at the height $z$ with respect to the center of the cluster (\emph{ergo} at $Z = z + Z_\mathrm{c}$ with respect to the disk) has the acceleration $g(Z)$. In the reference frame of the cluster, its acceleration is therefore:
\eqn{
\frac{d^2 z}{d t^2} = \frac{d v_\mathrm{z}}{d t} & = g(Z) - g(Z_\mathrm{c}) \nonumber \\
& \approx z \left.\frac{dg}{dZ}\right|_{Z_\mathrm{c}}. 
}
The velocity of the cluster ($\sim 100 \kms$) and the thinness of the disk ($\sim 100 \pc$) yield a crossing time of $\sim 1 \Myr$, i.e. much shorter than the crossing time of the stars within the cluster. That is, it is possible to neglect the proper motion of the stars during the tidal shock, which is the framework of the impulsive approximation. Therefore, $z$ remains constant and $dZ$ can be replaced $Vdt$ (where $V$ is the constant orbital velocity) which gives
\eqn{
\Delta v_\mathrm{z} & = \frac{z}{V}\Delta g \nonumber \\
& = -2 g_\mathrm{m} \frac{z}{V},
}
where $g_\mathrm{m}$ is the value of the acceleration taken at large heights (i.e. where the density is low). The velocity after the shocks becomes
\eqn{
\vect{v}' = \vect{v} + \vect{\Delta v_\mathrm{z}}, 
}
and the new kinetic energy per unit mass is
\eqn{
E' & = \frac{1}{2} \left( \vect{v}^2 + 2 \vect{v}\cdot\vect{\Delta v_\mathrm{z}} + \vect{\Delta v_\mathrm{z}}^2  \right) \nonumber \\
\Delta E & = \vect{v}\cdot\vect{\Delta v_\mathrm{z}} + \frac{1}{2} \vect{\Delta v_\mathrm{z}}^2.
}
\follow}\mybox[nocnt]{
The first term can be rewritten as $\vect{v_\mathrm{z}}\vect{\Delta v_\mathrm{z}}$ and yields a null average because of the symmetry in the distribution of $v_\mathrm{z}$. As a result, the average change in energy becomes
\eqn{
\left<\Delta E\right> = 2 g^2_\mathrm{m} \frac{z^2}{V^2} > 0,
}
which translates into a gain in energy for the stars, thus a dynamical heating of the cluster. This approach also applies when a cluster passes through a spiral arm, as noted by \mycitet{Gieles2007a}.
}

Note that when a cluster is tidally shocked (as in the formalism of \mycitealt{Spitzer1987}), the tidal forces only exist along one dimension and are actually compressive. However, the gain in energy of the cluster members does not depends on the sign of the tidal field and thus, the compressive modes have a destructive effect in this \emph{impulsive} fashion. In the \emph{adiabatic} regime however, the compressive modes act as a protective cocoon, by contrast with the extensive modes which enhance the dissolution (see \refsec{sfe_massloss} and \refsec{nbody6} below).

%%%%%%%%%%%
\subsect{Age of the compressive events}

The statistical study of the \emph{\acr{notarg}{lus}{longest uninterrupted sequence}}- and \emph{\acr{notarg}{tt}{total time}} distributions for the isolated progenitors and the Antennae is key to understand and describe the evolution of the tidal field. However, as noted in the previous Chapter, the actual history of an individual particle is a much more complex realization of these regimes: a particle can enter the compressive regime at the time of, say, a pericenter passage and remain in it for a certain period of time. After that, it may enter an extensive region for e.g. several $\times 10^7 \yr$, before immersing itself in a compressive mode again, and so on (recall the examples given in \reffigt{antennae_part_orbits}). As a result, the interpretation of \emph{\acr{notarg}{lus}{longest uninterrupted sequence}}- and \emph{\acr{notarg}{tt}{total time}} profiles independently is not sufficient to determine the net effect of the tidal field on a given particle along its orbit. As shown below, a combination of the two distributions provides a better statistics of the total age of the compressive mode that mimics the repeated switch from extensive to compressive modes.

The age distributions of both \emph{\acr{notarg}{lus}{longest uninterrupted sequence}} and \emph{\acr{notarg}{tt}{total time}} have been well-fitted with double exponential laws, with short ($\tau_1$) and long ($\tau_2$) timescales, corresponding to an early regime $\Delta T_1$ and a later phase $\Delta T_2$. A plausible interpretation is that the compressive history is first continuous during the short interval $\Delta T_1$. Then, when the tidal field evolves more significantly during the longer time lapse $\Delta T_2$, the tides go through a succession of \emph{discrete} compressive episodes, described by the \emph{\acr{notarg}{tt}{total time}} statistics. Hence, it would seem likely that the compressive event is continuous over the short timescale of \emph{\acr{notarg}{lus}{longest uninterrupted sequence}} (i.e. $\tau_\mathrm{lus,1}$), and that it breaks up into several episodes over the longer timescale of \emph{\acr{notarg}{tt}{total time}} (i.e. $\tau_\mathrm{tt,2}$). Consequently, the proxy for the actual time spent in compressive mode is a combination\footnote{This differs from the results presented in \mycitet{Renaud2009}, where we have explicitly split the \emph{\acr{notarg}{lus}{longest uninterrupted sequence}} and \emph{\acr{notarg}{tt}{total time}} distributions into a short ($]0,50] \Myr$) and a long ($[50,100] \Myr$) timescale. This has given two single exponential fits per distribution. The approach followed here does not make this separation as our distributions are directly fitted by double exponential laws over the entire time interval. The separation between the interval is no longer arbitrary and thus, the quality of the fits is better with this new method.} of the \emph{\acr{notarg}{lus}{longest uninterrupted sequence}} distribution over the interval $\Delta T_1$, and of the \emph{\acr{notarg}{tt}{total time}} distribution over the interval $\Delta T_2$:
\eqn{
f_1 \exp{\left(-\frac{t}{\tau_\mathrm{lus,1}}\right)} + f_2 \exp{\left(-\frac{t}{\tau_\mathrm{tt,2}}\right)}.
}

\fig{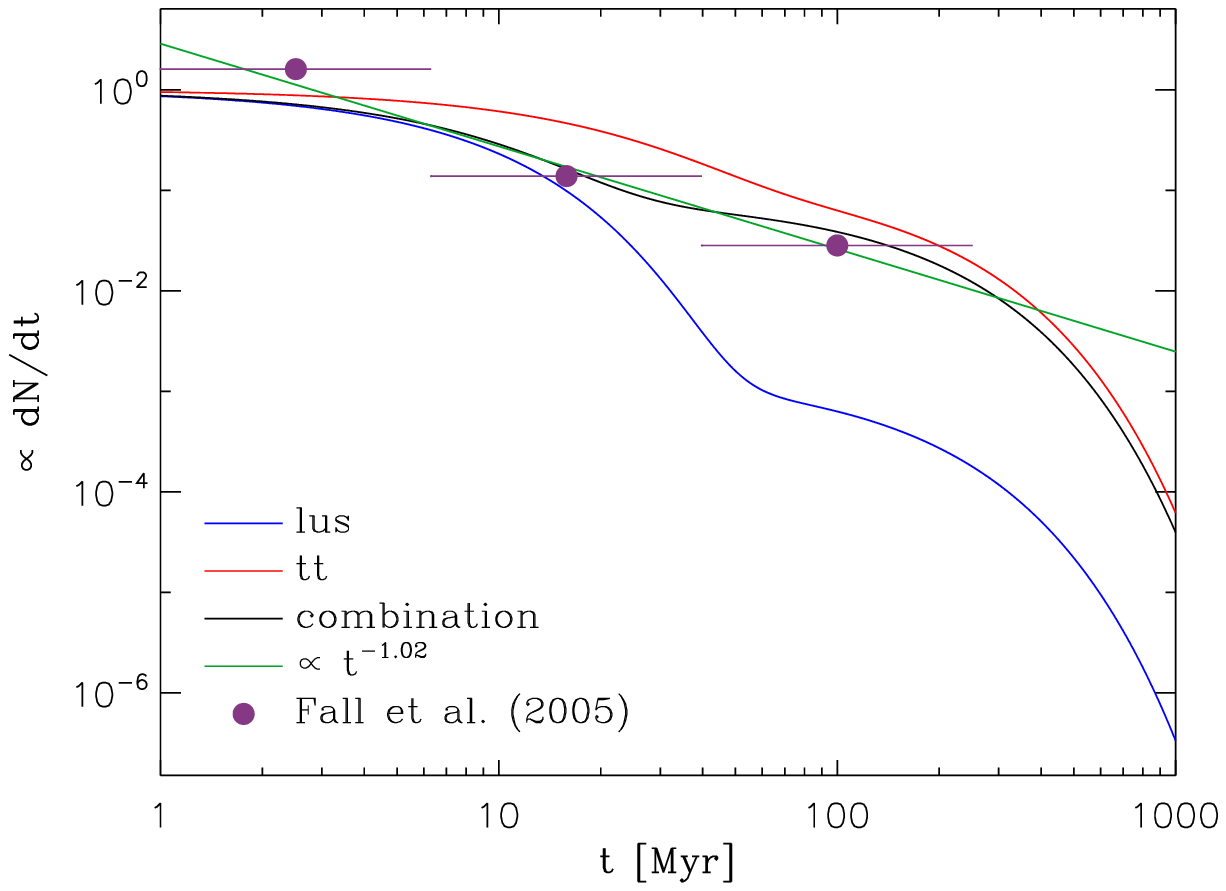}{Age distribution of the compressive modes}
{Age distribution of the compressive modes, as a combination of the \emph{\acr{notarg}{lus}{longest uninterrupted sequence}} and \emph{\acr{notarg}{tt}{total time}} statistics for the short and long timescales, respectively. This new distribution is well fitted with a power law of index $\sim -1$ over the first $300 \Myr$ (green). The purple dots are the data points from \mycitet{Fall2005}, as in \reffigp{fall05}.}

\reffig{antennae_age} displays the \emph{\acr{notarg}{lus}{longest uninterrupted sequence}}- and \emph{\acr{notarg}{tt}{total time}} distributions plus the new combination in the log-log plane. The latter can be fitted with a power law $dN/dt \propto t^{-1.02}$, over the range $[5,100] \Myr$ (with a residual $\sim 10^{-2}$). We note that the power law does not match our combination for ages $t \gtrsim 300 \Myr$, which corresponds to the time lapse between the two pericenter passages: longer time intervals include the pre- and post-interaction history, when the tidal field is derived from an equilibrium configuration.

When considering the possible destructive effect of the compressive mode, our new combination of the \emph{\acr{notarg}{lus}{longest uninterrupted sequence}}- and \emph{\acr{notarg}{tt}{total time}} distributions can reflect both the formation and destruction rates of the clusters. That is, it is possible in a first approximation to identify the duration $t$ of the compressive events with the age $\tau$ of the clusters. In that case, the distribution derived in \reffig{antennae_age} gives the number of clusters per age bin, and thus can directly be compared to the observational data. Surprisingly, the slope of our power law is very similar to the one obtained by \mycitet{Fall2005} from observations of the Antennae, who found $\tau^{-1}$ (recall \refsect[antennae]{antennae_clusters}). As a result, the duration of the compressive modes matches the age function of the star clusters, up to $\sim 300 \Myr$. For older ages, our interpretation misses the secular evolution of the clusters due to internal phenomena, that rule their dissolution rate, in particular when the galaxies are quasi-isolated (i.e. for $t \gtrsim 300 \Myr$).

Even if the link between our proxy and the empirical $\tau^{-1}$ distribution is not a full chain of reasoning, the striking similarity of the two is still noteworthy. For the age interval considered here, taking into account the gravitational effect of the compressive tidal mode can account for the formation and dissolution rates of the clusters observed in the Antennae galaxies. Therefore, it is very likely that the evolution of the tides (going through a series of on/off compressive modes and shocks) influences the shape of the mass function and the dissolution rate of the clusters. As a consequence, the tides play a role in the empirical concept of ``infant mortality''.

%%%%%%%%%%%
\subsect{Strength of the tides}

Up to now, we have mainly concentrated on the binary character of the tidal field (extensive or compressive), i.e. the very existence of the compressive modes. However, to compare its energy input with other mechanisms, one needs to quantify the strength of the tides for the various regions of the galaxies, at different steps of the merger. This will help to determine in which situation (where in the galaxy and for what duration), the tidal field is the most important, on the cluster scale.

\reffig{lambda_distrib} shows the distribution function of the the maximum eigenvalue of the tidal tensor from our Antennae simulation, at three different epochs. At any stage in the evolution of the merger, this distribution is by far not symmetric with respect to zero (which bounds fully compressive and extensive modes). Indeed, most of the particles have extensive ($\lambda_\mathrm{max} > 0$) yet weak tides, while the smaller compressive mass fraction shows larger (negative) values. When the two progenitors are still well-separated ($t = -100 \Myr$), the compressive part of the distribution (left-hand wing) contains a relatively small fraction of the mass. This fraction increases however during the interaction and reaches highly negative values. The merger clearly acts on the distribution by ``expanding'' its wings to very negative values and, in a similar but milder way, to large positive values (extensive modes). Knowing that a given particle goes through a series of compressive/extensive switches (recall \reffigt{antennae_part_orbits}), one may expect fast transitions from the left-hand side to the right-hand side of this distribution, and \emph{vice versa}. During the major steps of the interaction, the dynamical time of the merger is shorter, which corresponds to a rapid motion of the clusters along their orbit (for those in the central region). That is, the switch from compressive to extensive (or the opposite) is more rapid and frequent than in the quiescent periods. The expansion and flattening of the distribution of the maximum eigenvalue however, suggest that the amplitude of theses modes can be higher at the times of the pericenter passages. As a result, the tidal modes would be shorter but more intense than during a quiet phase.

\fig{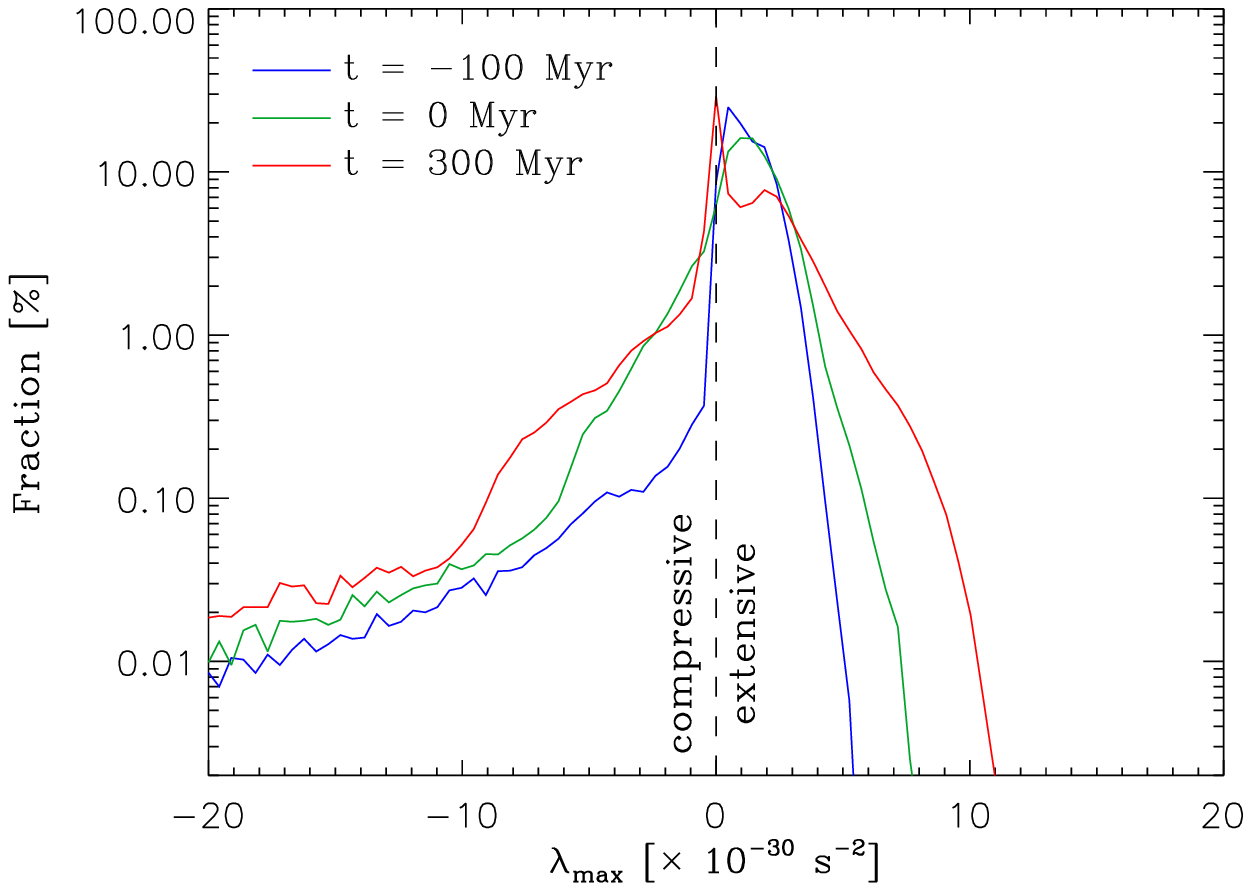}{Distribution of the max. eigenvalue of the tidal tensor}
{Distribution function of the maximum eigenvalue of the tidal tensor of the Antennae model before interaction (blue), at the first passage (green) and during the merger phase (red). Negative values denote a compressive mode.}

\reffig{lambda_space_distrib} completes the picture by showing the distribution of the maximum eigenvalue along the y-axis space coordinate. The Figure helps to understand that the extremely compressive regions are found mainly in the nuclei, while the outer volumes yield compressive tides of amplitudes comparable to those of extensive modes. Note that only the maximum eigenvalue is shown here, which means that the minimum value associated with a compressive mode ($\lambda_\mathrm{min} < \lambda_\mathrm{max} < 0$) has an even higher absolute value.

\fig{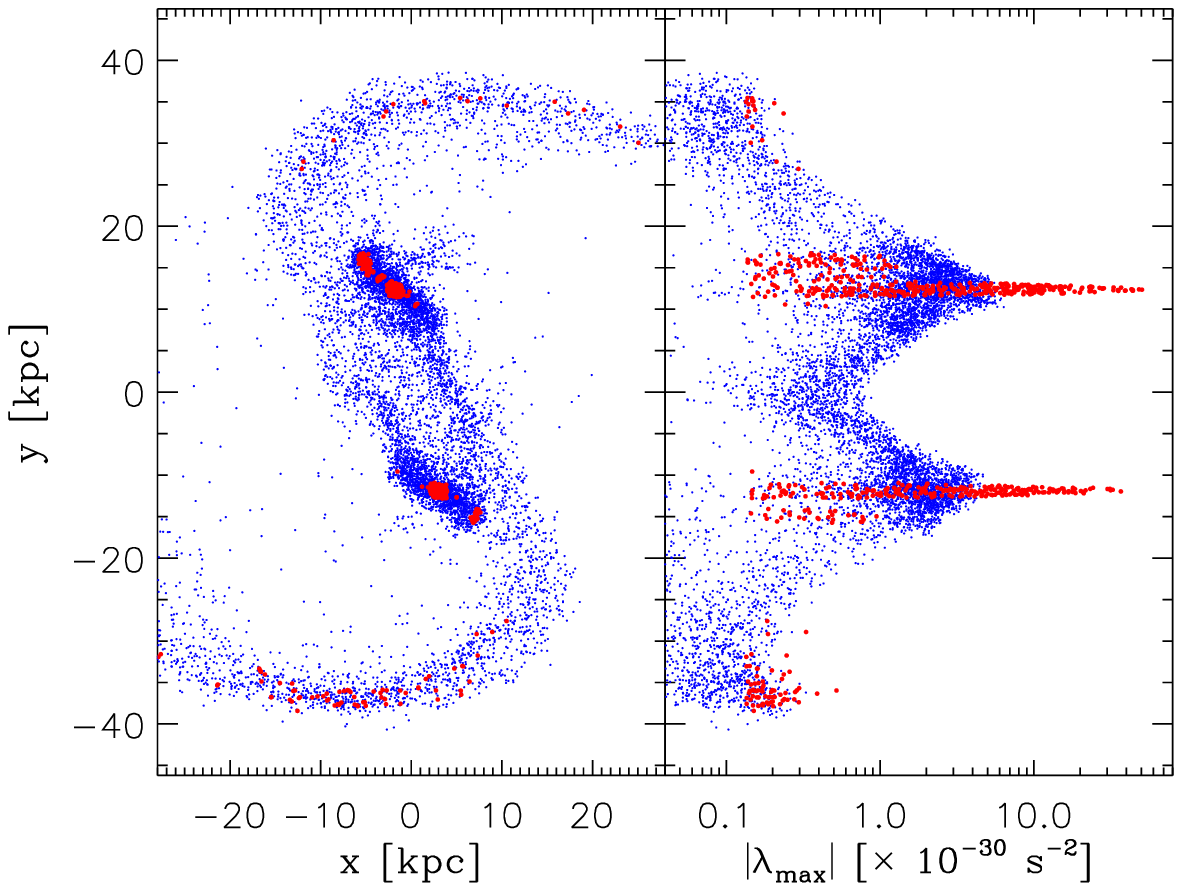}{Spatial distribution of the max. eigenvalue}
{\emph{Left:} spatial distribution of the particles in compressive (red) and extensive (blue) mode, in the orbital plane at $t = 150 \Myr$. \emph{Right:} absolute value of the maximum eigenvalue as a function of $y$.}

%%%%%%%%%%%
\subsect{Energy}

Knowing the strength of the field, we are now able to quantify the energy input by the tides. One can compare the role of the tidal field with that of other phenomena, via an estimate of the energy involved. Let's consider a cluster of mass $M$ embedded in an isotropic tidal field, characterized by the unique eigenvalue $\lambda$. The associated energy is given by:
\eqn[energy_tides]{
E_\mathrm{tides} & = -\frac{1}{2}\lambda \int_0^M r^2 dm \nonumber \\
& = -\frac{1}{2}\lambda \alpha \ M \ R_\mathrm{t}^2,
}
where $R_\mathrm{t}$ is the radius where the density of the cluster drops to zero, and $\alpha$ is a dimensionless quantity enclosing the details of the mass distribution (see \refsect[virial]{virial_tides}). The resolution of our simulations ($M = 10^5 \msun$ and $R_\mathrm{t} = 10^2 \pc$) corresponds well to the typical mass and size of a star cluster (\mycitealt{Meylan1997}). One can assume a certain mass distribution and set $\alpha$ (generally of the order of unity, see \refsect[virial]{alpha}) in \refeqn{energy_tides}. As seen in \reffig{lambda_distrib}, the typical values for $\lambda$ are $\sim 10^{-30} \U{s^{-2}}$, which translates to a tidal energy of $\sim 10^{48-49} \erg$. This does not compare well with the kinetic energy released by a type-II \acr{}{SN}{supernova} blast ($10^{51} \erg$) which would, therefore expand similarly with or without a tidal field. The winds of O-type stars may, however, be affected. Several authors evaluated the energy related to this process at the level of $10^{46-48} \erg$ (\mycitealt{Cappa1998}; \mycitealt{Martin2008}), i.e. comparable or perhaps smaller than the energy of the tides. That is, a gas flow driven by massive stars would be slowed down significantly by the external effect of the tides and probably even kept bound within the cluster. In this case, the mass-loss due to stellar winds would be considerably reduced, which may prevent the early dissolution of the young cluster.

In the \acr[s]{}{TDG}{tidal dwarf galaxy} region (recall \reffigt[e]{antennae_compressive}), the compressive tidal mode acts on much shorter times but over an extended region. By applying the same estimates from \refeqn{energy_tides} to such a volume ($M = 10^9 \msun$ and $R_\mathrm{t} = 10 \kpc$), one gets $10^{56-57} \erg$, i.e. much more than the energy release of a \acr{}{SN}{supernova} event. This value can be compared with a (rough) estimate of the number of type-II \acr[e]{}{SN}{supernovae} that could form in such a region. Let's assume an initial gas mass of $M_\mathrm{gas} = 10^9 \msun$, converted into stars with a \acr{}{SFE}{star formation efficiency} of $\epsilon \sim 10 \%$. Then, consider a \mycitet{Salpeter1955} \acr{}{IMF}{initial mass function} of the form $\xi(m) \propto m^{-2.35}$, and get the mass fraction associated with the massive stars ($m \in [10,120] \msun$) which will die as type-II \acr[e]{}{SN}{supernovae}:
\eqn{
f_\mathrm{m, SNe-II} = \frac{\displaystyle \int_{10 \msun}^{120 \msun} m \ \xi(m) \ dm}{\displaystyle \int_{0.1 \msun}^{120 \msun} m \ \xi(m) \ dm} \approx 13 \%.
}
Note that the slope of the Salpeter \acr{}{IMF}{initial mass function} for the high mass end is flatter than the one proposed by more recent works. For instance, \mycitet{Kroupa2001} suggested $\xi(m) \propto m^{-2.7}$ for $M > 3 \msun$, which would lead to less \acr[e]{}{SN}{supernovae} and thus a smaller energy release into the \acr{notarg}{ISM}{interstellar medium} (see also \mycitealt{Scalo1998}). That is, the Salpeter \acr{}{IMF}{initial mass function} gives an upper limit on the number of \acr{}{SN}{supernova} events, thus setting the maximum energy level to compete with the tidal field.

With a mean mass of $\overline{m} = 22 \msun$ for the \acr[e]{}{SN}{supernovae}, one has $M_\mathrm{gas} \ \epsilon \  f_\mathrm{m,SNe-II} \ / \ \overline{m} \approx 10^6$ type-II \acr[e]{}{SN}{supernovae} formed in the region considered. The associated instantaneous energy release is therefore $10^{57} \erg$ and compares well with the energy of the tidal field\footnote{This estimate misses the details of the \acr{notarg}{SFE}{star formation efficiency} and the \acr{notarg}{IMF}{initial mass function}, as well as the actual effect of the feedback (not simultaneous in time and space for all \acr{notarg}{SN}{supernova} blasts) on the \acr{notarg}{ISM}{interstellar medium} (turbulence, external pressure ...).}. In other words, the compressive tidal field can compete with the usual stellar feedback (winds for the smallest star cluster forming regions, and supernovae explosions for the bigger \acr{}{TDG}{tidal dwarf galaxy} forming volumes) and alter the dynamical and chemical evolution of these regions, by changing their binding energy.

%%%%%%%%%%%%%%%%%%%%%%%%%%%%%%%%%%%%%%%%%%%%%%%%%%
\sect{Star formation}

In this Section, we take one step further and include the tidal field in the chain of processes that form stars. To understand its role on the microphysics, one must recall the general concept of star formation. Stellar associations are formed by the collapse of a molecular cloud, when the contracting effects (gravitation, external pressure, shocks) overwhelm the internal pressure.  In a first approximation, this can be encapsulated in the Jeans formulation summarized in \refbox{jeans}.

\mybox[jeans]{
The Jeans criterion describes the stability of an homogeneous sphere which is infinitesimally compressed from a density $\rho_0$ to a new density $\rho_1$. By solving the fluid equations and taking the gravitational potential $\phi$ into account, one gets the wave equation
\eqn{
\frac{\partial^2 \rho_1}{\partial t^2} - v_\mathrm{s}^2 \ \nabla^2 \rho_1 - 4\pi G \rho_0 \rho_1 = 0,
}
where $v_\mathrm{s}$ is the speed of sound in the medium (see \mycitealt{Binney1987} for the detailed derivation). From this follows the dispersion relation between the frequency $\omega$ and the wavenumber $k$ of the solution $\rho_1(x, t) \propto \exp{(\imath[kx - \omega t])}$:
\eqn{
\omega^2 = v_\mathrm{s}^2 k^2 - 4\pi G \rho_0.
}
That is, the growing solution for the perturbation exists if $\omega^2 < 0$, corresponding to an instability. Therefore, the compressed sphere is unstable if
\eqn{
k^2 < \frac{4\pi G \rho_0}{v_\mathrm{s}^2},
}
or, in term of wavelength
\eqn{
\lambda^2 > \frac{\pi v_\mathrm{s}^2}{G \rho_0} = \lambda^2_\mathrm{J}.
}
$\lambda_\mathrm{J}$ is called the Jeans length.
\follow}\mybox[nocnt]{
 This corresponds to the Jeans mass
\eqn{
M_\mathrm{J} = \frac{4\pi}{3} \rho_0 \left(\frac{\lambda_\mathrm{J}}{2}\right)^3,
}
which defines the upper limit for objects stable to gravitational collapse.
In other words, a gas cloud more massive than $M_\mathrm{J}$ has a proper gravitation that makes it collapse. In the first approximation of a homogeneous and non-turbulent medium, the Jeans criterion can be used to determined the conditions of the collapse of molecular clouds, i.e. the early phases of star formation.
}

At smaller scale, the molecular cloud fragments into small clumps (see \reffig{klessen98}) that continue to accrete the gas reservoir and cool through radiative transfer (\mycitealt{Klessen1998}, \mycitealt{Klessen2001}). Turbulence and magnetic field play important roles in this process (see \mycitealt{Shu1987} and \mycitealt{MacLow2004} for reviews), but are out of the scope of this document.

\fig{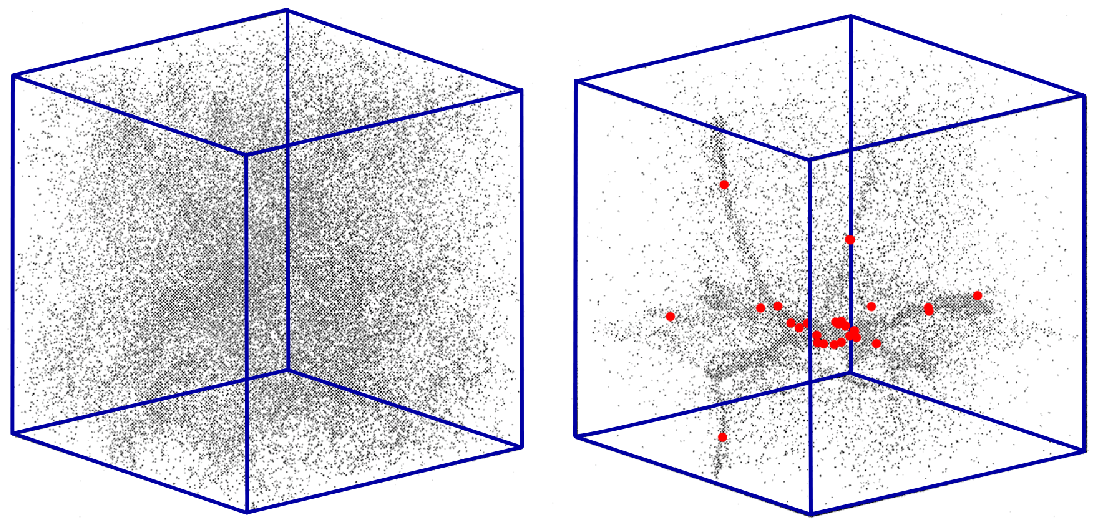}{Fragmentation of a molecular cloud}
{Fragmentation of a molecular cloud (\emph{left}) into proto-stellar cores (red, \emph{right}) that accrete the surrounding material. The situation on the right-hand side snapshot is reached after two free-fall times. (Adaptation of Figure~1 of \mycitealt{Klessen1998}.)}

In practice at cluster scale, the numerical simulations mimic the Schmidt-Kennicutt empirical law (recall \refboxt{schmidt}) by setting a star formation rate depending on the gas (volume) density
\eqn{
\dot{\rho}_\star \propto \rho_\mathrm{g}^{1.5} \quad \textrm{if } \rho_\mathrm{g} > \rho_\mathrm{c} \qquad \textrm{and} \qquad \dot{\rho}_\star = 0 \quad \textrm{else},
}
$\rho_\mathrm{c}$ being a critical gas density (see also \mycitealt{Springel2000}, \mycitealt{Springel2003} and \mycitealt{Bate2003}). Additional processes complete the model by regulating the star formation. In particular, stellar (\acr{}{SN}{supernova}) feedback is transfered to the \acr{}{ISM}{interstellar medium} either via small kicks in the kinetic energy of the surrounding particles in \acr{}{SPH}{smoothed particle hydrodynamics} (\mycitealt{Navarro1993}), and/or through a heating of the medium (\mycitealt{Springel2000}; \mycitealt{Kim2009}). The cooling process can also be included, often thanks to the functions of \mycitet{Sutherland1993} for collisional ionization equilibrium.

%%%%%%%%%%%
\subsect{Galactic to cluster scales coupling}

During its collapse and fragmentation, a molecular cloud does not keep a regular shape but rather a filamentary structure including dense seeds (recall \reffig{klessen98}). At a slightly more advanced stage, the young cluster will probably have a spheroidal profile, before getting a spherical shape, typical of globulars. This means that a young cluster is far from being spherically symmetric. Therefore, its host galaxy creates a torque that can modify the internal dynamics of the cluster. The goal of this Section is to describe this torque which couples the galactic scale to cluster-size elements.

This problem strongly relies on angular dependencies for both the cluster and the galaxy. To ease the mathematical derivations in the general case, it is preferable to use spherical coordinates, as opposed to the more usual cartesian base. Hence, our idea is to decompose the quantities required to compute the torque in the base of spherical harmonics\footnote{Interestingly, the historical origin of spherical harmonics comes from Pierre-Simon de Laplace who, in \emph{M\'ecanique C\'eleste} (1782), computed the gravitational potential of a set of point-masses, using the work of Adrien-Marie Legendre. The name ``spherical harmonics'' first appeared in \emph{Treatise on Natural Philosophy} (1867) by William Thomson (aka Lord Kelvin) and Peter Guthrie Tait.}. In addition, \myciteauthor{Anderson1992} (\myciteyear{Anderson1992}, see also \mycitealt{Kawai1998}) provides a way to evaluate a potential at the position $\vect{r}$ from the value $\phi(a\vect{s})$ it takes on the surface of a sphere, called Anderson's sphere, of radius $a$:
\eqn[anderson_inside]{
\phi(\vect{r}) = \frac{1}{4\pi} \int_S \sum_{n=0}^{\infty} (2n+1) \left(\frac{r}{a}\right)^n P_n\left(\frac{\vect{s}\cdot\vect{r}}{r}\right) \phi(a\vect{s})\ ds
}
for $r < a$, and
\eqn[anderson_outside]{
\phi(\vect{r}) = \frac{1}{4\pi} \int_S \sum_{n=0}^{\infty} (2n+1) \left(\frac{a}{r}\right)^{n+1} P_n\left(\frac{\vect{s}\cdot\vect{r}}{r}\right) \phi(a\vect{s})\ ds
}
for $r > a$, where $P_n$ denotes the $n$-th Legendre polynomial. This allows us to compute the potential of either the galaxy or the cluster at any point, from its values on the Anderson's sphere, given by the spherical harmonics decomposition. (A step-by-step derivation is available in \refapp{harmonics}, with conventions and notations.)

For example, let's compute the torque created by a point-mass galaxy on a cluster. One needs the force of the galaxy on the stars of the cluster, i.e. minus the gradient of the potential of the galaxy, at the positions of the stars. This can be done through the evaluation of the potential of the galaxy on the Anderson's sphere which includes the entire cluster (see \reffig{anderson}).

\fig{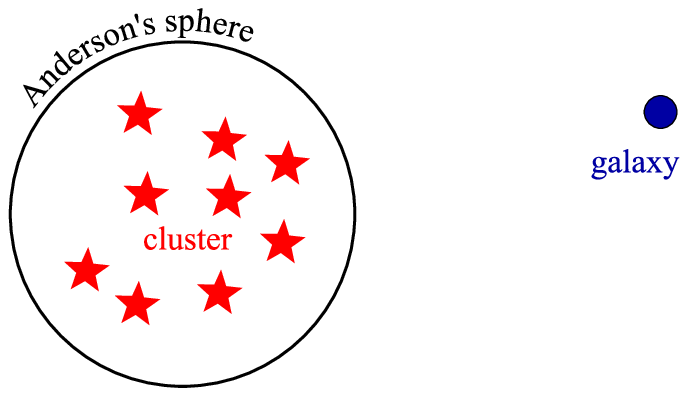}{Anderson's sphere}
{A point mass galaxy creates a torque on a star cluster. Setting the Anderson's sphere around the cluster eases the computation of either the potential of the galaxy at the positions of the stars, or the potential of the cluster at the position of the galaxy.}

Therefore, the torque created by the galaxy on the cluster is
\eqn{
\vect{\Gamma_\mathrm{G/c}} = - \int_V \rho_\mathrm{c}(\vect{r}) \left[\vect{r} \times \vect{\nabla}\phi_\mathrm{G}(\vect{r})\right] dV,
}
with $\rho_\mathrm{c}$ the density of the cluster, and $\phi_\mathrm{G}$ the potential of the galaxy given by \refeqn{anderson_inside}, the stars being \emph{inside} the sphere.

Note that, because of Newton's law of reciprocal actions, it is also possible to compute the same torque from the potential of the cluster and the density of the galaxy (as in \refapp{harmonics}):
\eqn{
\vect{\Gamma_\mathrm{G/c}} = -\vect{\Gamma_\mathrm{c/G}} = \int_V \rho_\mathrm{G}(\vect{r}) \left[\vect{r} \times \vect{\nabla}\phi_\mathrm{c}(\vect{r}) \right] dV.
}
This time however, one needs to evaluate the potential at the position of the galaxy, i.e. \emph{outside} the Anderson's sphere, and thus it must be computed thanks to \refeqn{anderson_outside}. In the case of a point mass galaxy, the integral reduces to the value of the integrand at the position of the galaxy. For more complex distributions, one can decompose the density $\rho_\mathrm{G}$ in spherical harmonics on a shell centered on the cluster, but bigger than the Anderson's sphere. \emph{In fine}, the problem is reduced to the computation of the torque created by one shell on another, concentrical one. The analytical solution in the general case is given in \refsecp[harmonics]{spher_harm_general_case}.

To summarize, there are two ways of solving this problem: from the decomposition in spherical harmonics of
\begin{itemize}
\item the density of the galaxy and the potential of the cluster,
\item the density of the cluster and the potential of the galaxy.
\end{itemize}

Configurations with a high degree of symmetry have been used as test cases of the formulas and their numerical implementation. The next step is to use the mass distributions obtained from our simulations of merging galaxies, and from simulations of star clusters.

In the logical continuation of our numerical exploration, we choose to evaluate the potential of the galaxy on a sphere of similar size as the cubes used in the computation of the tidal tensor (recall \refcha{numerical}).

\mybox{
Technically, the potential of the entire galaxy is computed on a spherical grid thanks to {\tt getgravity}. These values are passed to the tool {\tt S2Kit} (\mycitealt{Healy2003}) which computes the coefficients of the spherical harmonics decomposition. Then, our code applies \refeqnp{density_discrete} by summing on all the particles of the cluster to get the total torque created by the galaxy. 

Note that when the spherical grid is centered on a particle of the galaxy model, this central particle yields only a monopole term for the potential, and thus no torque on the cluster. Therefore, there is no need to subtract the value of its potential, by contrast with our method for the tidal tensor (recall \refboxt{tidal_code}).

In practice, the sum on the spherical harmonics can be truncated to the sixth order, the coefficients of higher orders being negligible.
}

To generate an anisotropic distribution of the mass of a proto-cluster (as in \reffig{klessen98}), we set up configurations based on the cold-collapse of spherically symmetric systems. A small number of particles ($\sim 10^3$) is deliberately chosen to obtain a Poisson noise that favors the formation of substructures like clumps. Note that the turbulence modes present in fragmenting molecular clouds (see e.g. \mycitealt{MacLow2004} and references therein) yield a more complex noise spectrum.

First, a homogeneous sphere made of $\sim 2000$ particles is created and left with no internal velocity. Its cold-collapse rapidly gives out clumps of matter, as shown in the left-hand side of \reffig{cold_collapse}. Second, a Plummer sphere is truncated to 90\% of its total mass ($\sim 2000$ particles too) and also undergoes a cold-collapse (right-hand side of \reffig{cold_collapse}). Note that the density profile of the Plummer sphere implies a more centrally dominated collapse, the outskirts being more slowly affected than the core. Because of orbit crossing, the structure obtained is more elongated and resembles a bar.

\fig{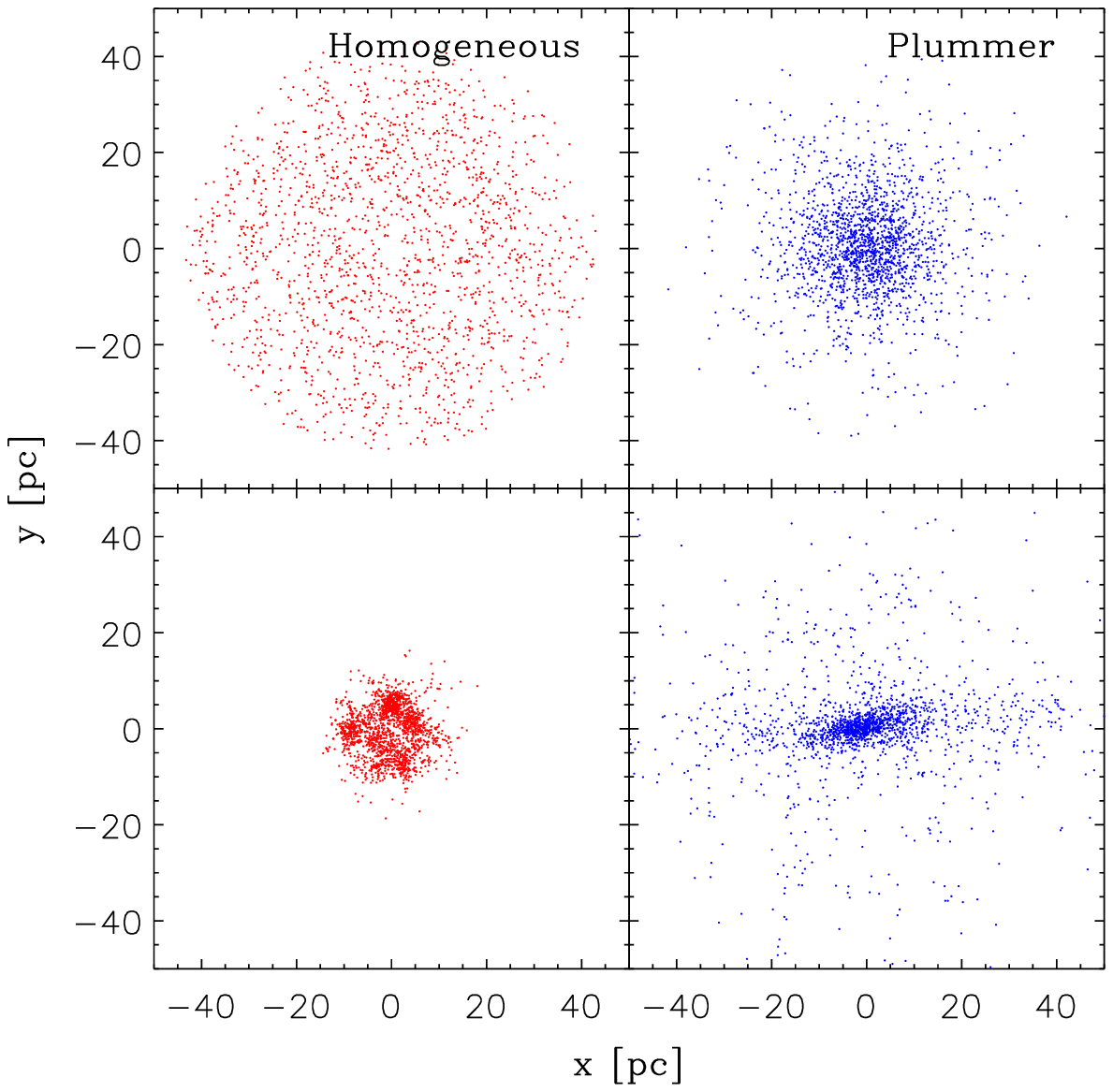}{Cold collapse}
{A homogeneous distribution (\emph{top-left}) forms clumps of matter (\emph{bottom-left}) during its cold-collapse. The same transformation applied to a truncated Plummer sphere (\emph{top-right}) yields a bar (\emph{bottom-right}).}

These two transient structures have degrees of asymmetry close to those seen in the simulations of the collapse of a molecular cloud and thus, are expected to yield comparable torque when embedded in a host galaxy. To compare their response to the scale coupling described above, \reffig{torque_cold_collapse} plots the norm of the torque exerted by a point-mass galaxy situated at $10 \kpc$ and $10^6$ times more massive than the cluster. The corresponding statistics are given in \reftab{stat_torque}. Clearly, the bar mode (ex-Plummer) yields a stronger torque than the clumps (ex-homogeneous) and has a higher dispersion. This is explained by a weak torque when the bar is aligned with or perpendicular to the galaxy, and a much higher value for an oblique inclination, while the clumps always give out a comparable torque. 

\fig{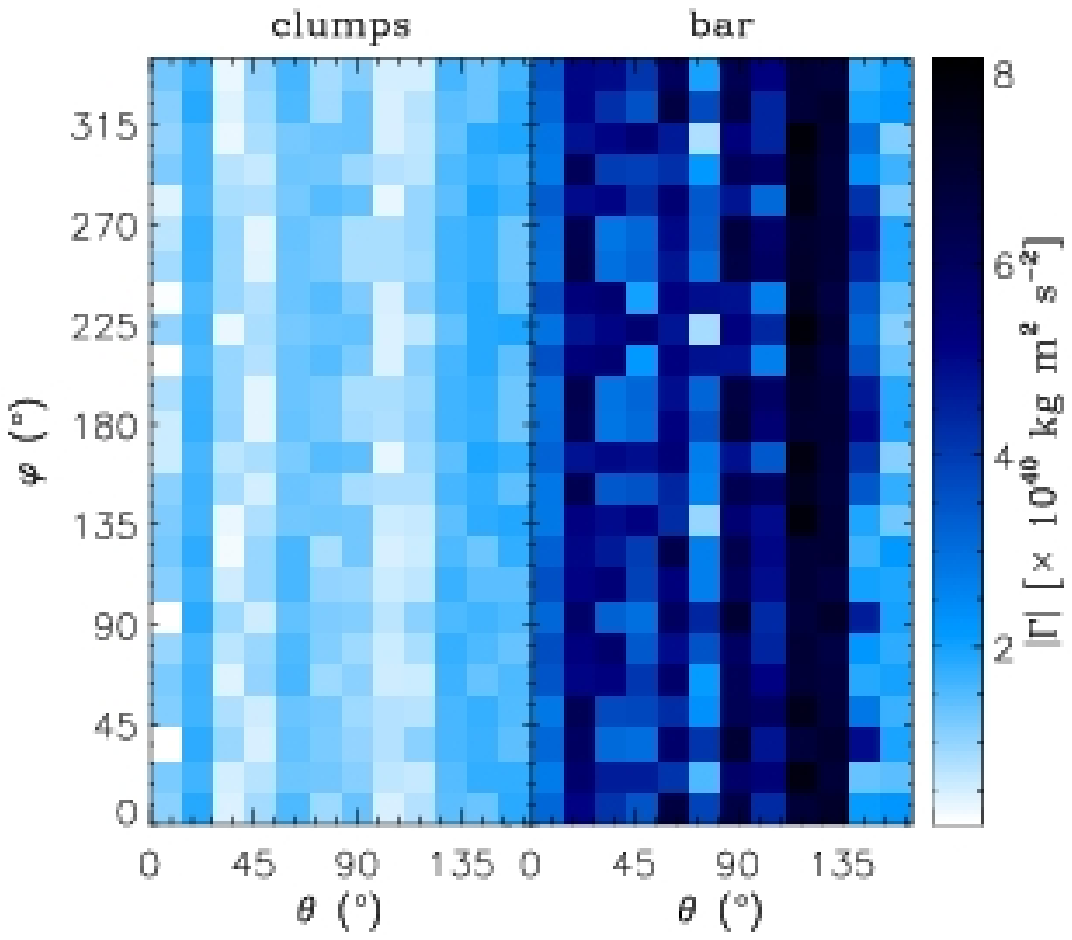}{Torque on clumps and bar}
{Map of the torque created by a point-mass galaxy on the mass distributions shown in \reffig{cold_collapse} as a function of their orientation ($\theta, \varphi$) with respect to the galaxy.}

\tab{12cm}{stat_torque}{Statistics of the torques}
{l@{\extracolsep{\fill}}c@{\extracolsep{\fill}}c}
{
\multirow{2}*{Model} & mean & standard deviation\\
&  [$\times 10^{40}\U{kg\ m^2\ s^{-2}}$] &  [$\times 10^{40}\U{kg\ m^2\ s^{-2}}$]\\
\hline
Clumps (ex-homogeneous) & $1.1$ & $0.4$\\
Bar (ex-Plummer) & $4.5$ & $1.8$\\
}{}

\mybox{
The effect of a torque can be compared with the auto-gravitation of the proto-cluster, in terms of energy. An \emph{order of magnitude estimate} of the equivalent kinetic energy of the torque is given from the time derivative of the angular momentum:
\eqn{
\Gamma & = \frac{dL}{dt} \nonumber \\
& \approx m \frac{\langle r v \rangle}{\Delta t},
}
where $\langle x \rangle$ denotes the mean value of $x$, so that
\eqn{
E_\mathrm{torque} & \approx \frac{1}{2} m \langle v \rangle^2 \nonumber \\
& \approx \frac{\Gamma^2 (\Delta t)^2}{2 \langle r \rangle m}.
}
If one assumes that the galaxy creates a torque $\Gamma = 10^{40} \U{kg\ m^2\ s^{-2}}$ during $\Delta t = 1 \Myr$ on a proto-cluster of mass $m = 10^5 \msun$ and half-mass radius $\langle r \rangle = 5 \pc$, the equivalent energy is $E_\mathrm{torque} \sim 10^{44} \erg$. For comparison, the potential energy of the same proto-cluster would be
\eqn{
\Omega \approx \frac{Gm^2}{2 \langle r \rangle} \sim 10^{50} \erg.
}
This means that a typical torque due to a galaxy on a proto-cluster does not compete with the gravitational energy and cannot modify its binding energy in a significant way. The orbital structure however, could be modified.
}

To conclude, a point-mass galaxy can induce a stronger torque on an elongated proto-cluster than on a clumpy one. Such a torque may create or enhance the internal rotation of the cluster and thus reorganize its population. Note however that the torques computed are instantaneous and that the evolution of the proto-cluster, as well as the one of its environment, would change the orientation and/or the distribution of the matter, and thus the net effect of the torque. 

To go further, the point-mass galaxy is replaced by our model of the Antennae, presented in \refcha{antennae}. Its potential is decomposed on an Anderson's sphere of radius $a = 220 \pc$ (i.e. the size of the cubes used to compute the tidal tensor) placed at the beginning of the compressive episode along the orbit of the particle D (next to the green $\times$ sign on \reffigt{antennae_part_orbits}), i.e. $305 \Myr$ after the first pericenter passage of the progenitors and in the western loop. This corresponds to a star forming region observed in the Antennae galaxies. For the cluster, the distributions obtained by cold-collapse are scaled to get a total mass of $2 \times 10^5 \msun$ and a mean radius of $\sim 5 \pc$.

The main source of gravitation is now much closer and more massive than in our point-mass example: $\sim 1\kpc$ instead of $10 \kpc$ and twice the mass (corresponding to two nuclei). As a consequence, the torque computed is stronger, in average, by a factor $\sim 200$. To have a fair comparison of the values of the torques, the axes of reference for $\theta$ and $\varphi$ have been chosen so that the dipole of the potential of the Antennae (i.e. the direction of minimum potential) has the same angular position than the point-mass galaxy in the previous example.

\fig{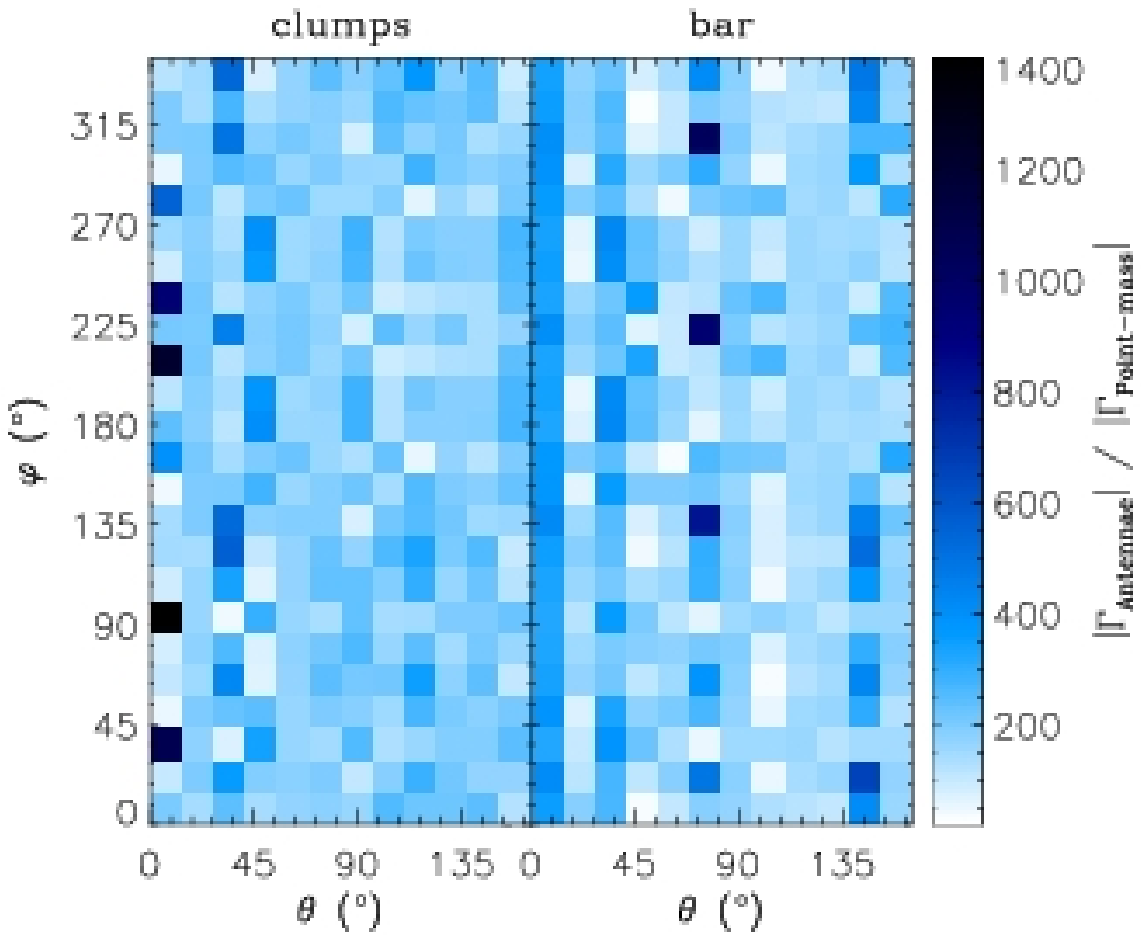}{Comparison of the torque from point-mass and the Antennae}
{Map of the ratio of the torque created by the Antennae and a point-mass galaxy on the mass distributions shown in \reffig{cold_collapse}. See text for details.}

Because the torque partially depends on the mass distribution of the cluster (e.g. possible alignment of the bar with the galaxy), we find similarities in the angular map of the torque created by the Antennae and in \reffig{torque_cold_collapse}. The ratio of the two is plotted in \reffig{torque_ratio}. As estimated before, the mean values of the ratio are 207 and 203 for the clumps and the bar respectively, and the standard deviations reach 140 and 128. This shows that the torque on the bar depends slightly less on the mass distribution of the galaxy than the one on the clumps. This can also be interpreted as a more robust response of an elongated structure in a varying potential (orientation and strength), e.g. along the orbit of a cluster-size element.

It is possible to estimate the equivalent energy of these torques. The proto-cluster being much closer to the galaxy than in the point-mass experiment, the time during which the torque may be effective is much shorter. Using an estimate from the circular velocity, one finally gets an energy $\sim 5\times 10^2$ times higher than the one obtained for the point mass. However, these estimates have a strong dependance with, e.g. the distance and mass of the galaxy. Therefore, such orders of magnitude remain imprecise. For a finer description of the coupling between the galactic and cluster scales, numerical simulations are required. In this matter, a star cluster must be simulated using the external effect of the galaxy, as we do in the \refsec{nbody6}.

%%%%%%%%%%%
\subsect[sfe_massloss]{Star formation efficiency}

The existence of a tidal field introduces a new term in the expression of the virial theorem. (\refapp{virial} details its derivation.) Therefore, the criterion for the survival of a young cluster after its gas expulsion is changed. To illustrate this, we repeat the analytical work of \mycitet{Hills1980} and \mycitet{Boily2003a} made for clusters in isolation, in the case of an external tidal field (see \refsect[virial]{sfe_hills}).

In isolation, a young cluster dissolves if the velocity dispersion of its stars is not balanced by the gravitational well. If it is the case, the cluster is not able to keep its material bound. When the very first stars form in the cluster, their velocities are set according to the gravitational potential, built by both the stars and the gas. When the gas spills over to the galactic \acr{}{ISM}{interstellar medium}, the stars remain with the same velocity dispersion, while the potential well has become shallower. In other words, the escape velocity is reduced to a level possibly below that of individual stars. This depends on the fraction of the mass lost, i.e. the mass fraction of the gas, with respect to the total mass of the cluster. This quantity comes from the \acr{}{SFE}{star formation efficiency} $\epsilon$:
\eqn{
M_\mathrm{gas} & = M - M_\star \nonumber \\
& = M (1-\epsilon).
}
Hills found that the cluster survives if $\epsilon > 50\%$, i.e. if at least half of the initial mass is converted into stars (see also \refsect[virial]{sfe_hills_isolation}).

When the same proto-cluster is embedded in a tidal field, the dynamics is changed. Again in \refsecp[virial]{sfe_hills}, we have extended the previous derivation by taking into account the additional effect of an isotropic tidal field (i.e. represented by an unique eigenvalue $\lambda$), starting from the work of \mycitet{Fleck2007}, and extending his results to other transformations.

In this case, the response of the cluster to the tidal field is a strong function of its density profile and thus, the calculation is more involved. It is very likely that this distribution changes after the gas expulsion. Unfortunately, a general solution does not exist. For simplicity, let's consider a proto-cluster which will undergo a homologous transformation, i.e. keeping the same shape but not the same amplitude\footnote{Keep in mind that is assumption is very rough: the external layers of the cluster are much more affected than its core which inevitably leads to a reorganization of the matter, and thus a change of the density profile.} of density $\rho_0$. We note $A$ the ratio of tidal to gravitational energies:
\eqn{
A = \frac{\lambda \alpha \ R^2_\mathrm{t}\ r_\mathrm{v}}{G M}
}
(see \refapp{virial} for the definitions and notations). \refeqnp{bound_impuls} tells us that, after an instantaneous expulsion of the gas, the cluster remains stable if
\eqn{
1 - 2\epsilon + \epsilon \left(\frac{r'_\mathrm{v}}{r_\mathrm{v}}\right)^{-1} - 4 A \left[ 1 + \frac{\alpha'}{\alpha} \left( \frac{{R'}_\mathrm{t}}{R_\mathrm{t}} \right)^2 \right] \le 0.
}
By homology, we have $\alpha = \alpha'$ and $r'_\mathrm{v} / r_\mathrm{v} = {R'}_\mathrm{t} / R_\mathrm{t}$, which leads to
\eqn{
%\epsilon \geq \frac{{R'}_\mathrm{t}}{R_\mathrm{t}} \frac{4A\left(\frac{{R'}^2_\mathrm{t}}{R^2_\mathrm{t}}\right)-1}{1-2\frac{{R'}_\mathrm{t}}{R_\mathrm{t}}}
\frac{\epsilon}{2} + \frac{{R'}_\mathrm{t}}{R_\mathrm{t}} \left(\frac{1}{2} - \epsilon - 2A \right) + 2A \left( \frac{{R'}_\mathrm{t}}{R_\mathrm{t}} \right)^3  \le 0.
}
This relation constrains the radius ${R'}_\mathrm{t}$ of the cluster after the gas expulsion, once one knows (i) the \acr{}{SFE}{star formation efficiency} $\epsilon$ and (ii) the strength of the tidal field, embedded in the term $A \propto \lambda$. \reffig{virial_sfe} gives some examples of expansion of the cluster, for several tidal fields. In extensive mode, a solution (i.e. a positive radius ${R'}_\mathrm{t}$) exists only for a certain range of \acr[s]{}{SFE}{star formation efficiencies}. Note that, without a tidal field ($A = 0$), one retrieves the solution of \mycitet{Hills1980} and the $\epsilon > 50\%$ limit. However, the most important conclusion is that the presence of a compressive field allows the cluster to reach a new equilibrium stage after the gas loss, for \emph{all} values of $\epsilon$.

In other words, when the gas is expelled, the kinetic energy of the stars is too high with respect to the rest of the energy to keep the cluster bound: it is supervirialized. Therefore, it must expand. Our analytical solution gives the range of possible expansions (y-axis on \reffig{virial_sfe}) that keep the cluster bound. In an extensive mode, a too severe mass-loss (low $\epsilon$) leads to the dissolution, while a higher \acr{}{SFE}{star formation efficiency} allows some ``bound'' expansions. However, if the tidal field is strong (large $A$), important expansions correspond to the dissolution of the cluster due to the extensive mode. By contrast, in compressive mode the tidal field cannot dissolve the cluster. Therefore, the expansion has only a lower limit, set by the kinetic energy of the stars.

\fig{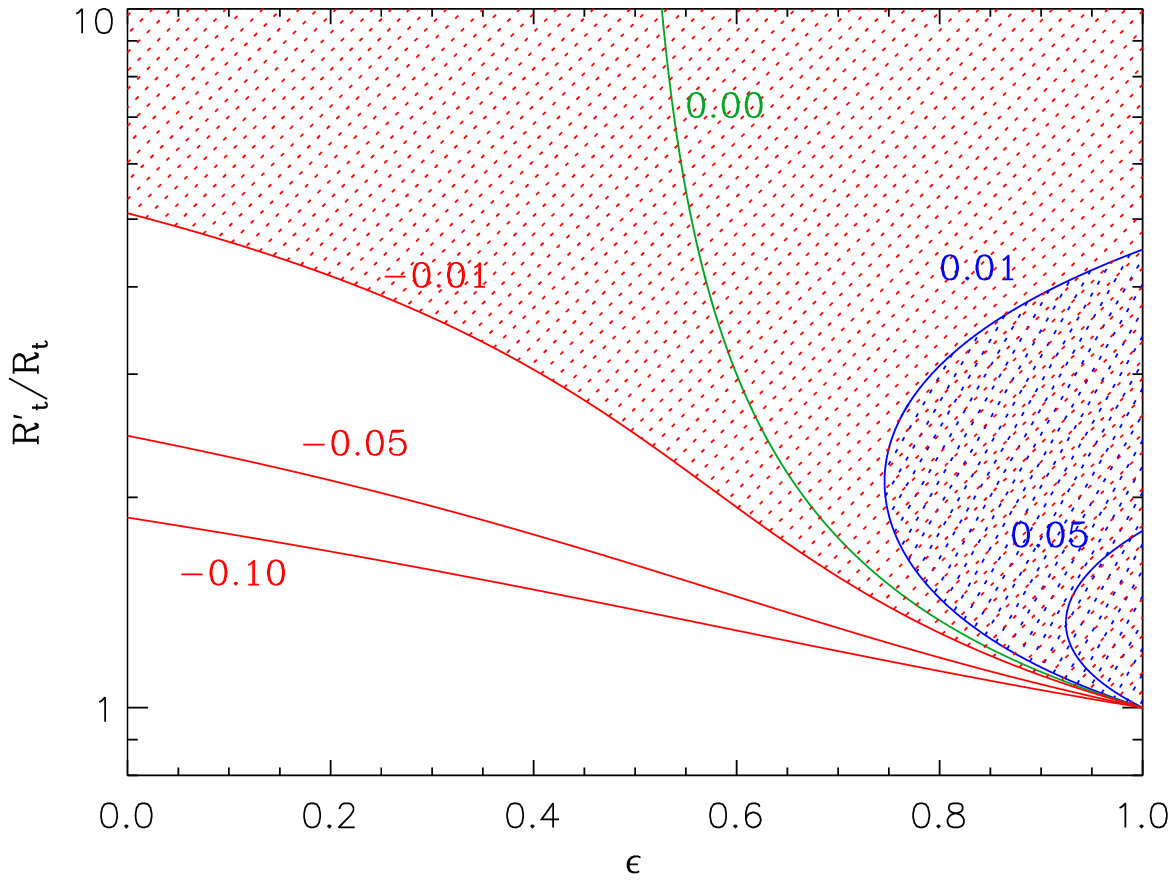}{Expansion of a young cluster}
{Expansion of a cluster after the expulsion of the gas, in an extensive tidal field ($\lambda > 0$, blue curves), without tides ($\lambda=0$, green) and in compressive mode ($\lambda < 0$, red), as a function of the star formation efficiency $\epsilon$. The numerical value on each curve is $A$, i.e. the ratio between the tidal and gravitational energies. The areas filled with red and blue dots correspond to bound configurations for clusters, for $A=-0.01$ and $A=0.01$, respectively.}

As seen in \reffigp{antennae_part_orbits}, a young cluster will likely experience a switch from compressive to extensive mode. If it forms in a compressive, supervirialized cocoon and then goes to extensive mode, it will likely dissolve. (This is visible on \reffig{virial_sfe} where the compressive mode offers more possible configurations than its extensive equivalent.) Such a process plays an important role in the ``infant mortality'' process, and thus affects the death rate of the clusters and their age distribution. Note that it is not straightforward to conclude on the mass function, because the link between the tidal energy and $M$ is only known once the mass profile of the cluster is set.

%%%%%%%%%%%
\subsect{Multiple main sequences}

In addition to possible mass loss, the repeated on/off evolution of the compressive tidal mode can influence the chemistry of the cluster. In this Section, we introduce an idea that may (partially) explain the existence of observed features in the color-magnitude diagram of several massive clusters. The reader should keep in mind that the following arguments are purely qualitative and speculative. The exploration of this point requires a detailed numerical treatment that deserves a study of its own. 

In our common understanding of star clusters, we assume an initially homogeneous chemical composition and a coeval formation of the stars. This idea is often referred to as the \acr{}{SSP}{simple stellar population}. However, recent observations of resolved massive star clusters have revealed multiple populations (see e.g. \mycitealt{Bedin2004}; \mycitealt{Piotto2009}; \mycitealt{Milone2009b}) in several cases, denoting a more complex star formation history and/or chemistry at the time of formation. These appear in color-magnitude diagrams as multimodality in sequences or branches, as visible in \reffig{milone}.

\fig{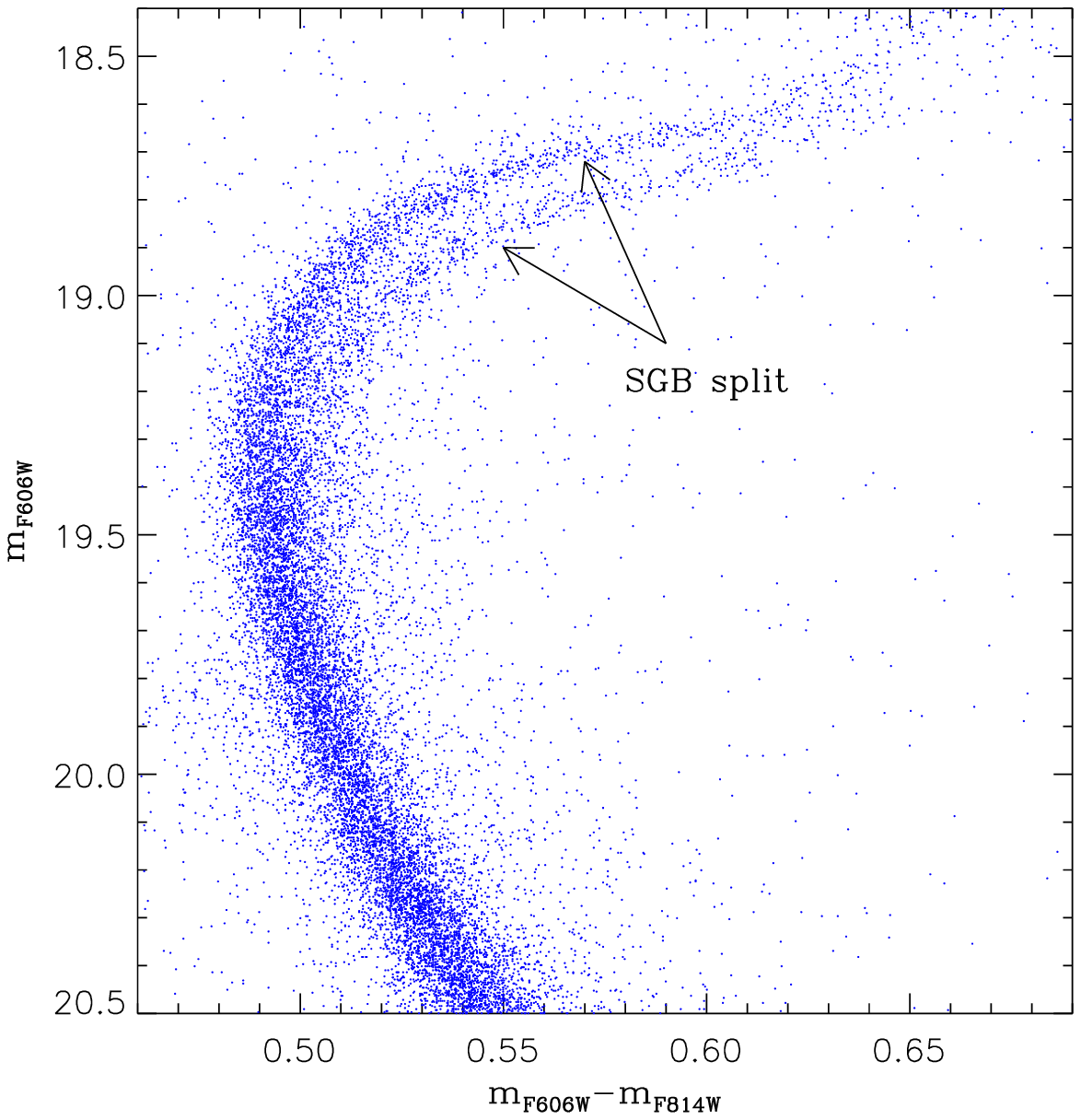}{Color-magnitude diagram of NGC~1851}
{Color-magnitude diagram of NGC~1851 made with HST ACS by \mycitet{Milone2008}. Two \acr[s]{notarg}{SGB}{sub-giant branches} are clearly visible in the upper part of the diagram. (Data points from \mycitealt{Milone2008}.)}

The first discovery of such a multiple \acr[s]{}{MS}{main sequence} has been made in $\omega$ Cen (the most massive cluster in the Milky Way), which displays a split into a dominant red and a secondary blue \acr[s]{}{MS}{main sequence}, at all radii. Spectroscopy shows that the bluer sequence has twice the light elements abundances (He, CNO) of the red one (\mycitealt{Piotto2005}). These peculiarities confirmed that $\omega$ Cen is an anomalous cluster, possibly the remnant of a dwarf galaxy (see \mycitealt{Lee1999}). However, the presence of multiple populations of stars has been detected in other, perhaps less peculiar, clusters in the Galaxy like NGC~2808 (\mycitealt{Piotto2007}), NGC~1851 (\mycitealt{Milone2008}) or others (not yet published, see \mycitealt{Piotto2009} for a list of preliminary results). Recently, clusters showing similar features in their color-magnitude diagrams have been found in the Small and Large Magellanic Clouds (\mycitealt{Glatt2008}), where it has even been mentioned that \emph{most} of the clusters contain multiple populations (detected via splits in the turn-off and \acr{}{SGB}{sub-giant branch} of their color-magnitude diagrams, see \mycitealt{Milone2009a} and \reffig{milone}).

The explanation of such observations is still debated. Because each cluster yields different properties of its branches, it is likely that the origin of the multiple populations is far from being unique. A first possibility would be the formation of new stars during the encounter between the cluster and a giant molecular cloud. \mycitet{Bekki2009} have shown that this scenario would explain the presence of two turn-offs, one due to the initial cluster and the other from the stars formed during the interaction with the cloud, at a different epoch.

Alternatively, \mycitet{PflammAltenburg2009} suggested that a massive cluster could accrete some of the gas met along its orbit through the galaxy. Again, this new gas would be processed into a new generation of stars, explaining both the age and abundances differences.

Another possibility involves the merging of two distinct star clusters. One can imagine that, in binary systems of clusters, a loss of angular momentum could lead to a merging of the two populations of stars into a single structure. However, the age difference between the two remains difficult to explain because both clusters are supposed to share the same origin, and thus to have similar properties, like the age of their stellar content (see e.g. \mycitealt{Dieball2002}).

These ideas are based on the formation of stars from an external source of gas. By contrast, one can imagine the creation of a secondary population from an \emph{internal} source of gas, i.e. the \emph{intra}cluster medium itself. This implies to be able to retain the ejecta of the stellar feedback (winds and \acr{}{SN}{supernova} blasts) within the cluster and introduces a ``self-pollution''. The enrichment of the \acr{}{ISM}{interstellar medium} within the cluster by the first stars would be the source of a new generation, with higher CNO (carbon, nitrogen, oxygen) abundances, as observed. The main issue of such a scenario is that the ejecta have to remain in the cluster instead of being blown away. Several theories suggest to consider the matter from the winds of asymptotic giant branch stars (\mycitealt{DErcole2008}) or massive rotating stars (\mycitealt{Decressin2007}; \mycitealt{Decressin2008}) whose winds are slow. After the violent gas removal due to type-II \acr[e]{}{SN}{supernovae}, these ejecta could gather in the cluster center and be transformed into stars.

In this manner, the presence of a compressive tidal mode could be the missing link that helps to bring and/or retain the matter within the cluster. Indeed, if the cluster moves through an external source of gas (a molecular cloud or a second cluster), the compressive field could increase the binding energy of the main cluster so that this external amount of matter is captured and retained. Similarly, the ejecta from the cluster members would be more easily kept bound.

On the one hand, the accretion from an external source would not be specific to certain chemical elements and thus, even heavy ones (e.g. iron) should be detected in the second generation of stars. With the retention of ejecta on the other hand, the new abundances should be more similar to those of the first, original generation. 

The fine details of these processes (winds, accretion, enrichment, impact of the tides) are not well understood yet and it is too early to pin down the exact effect of the tidal field on such a complex star formation history.

%%%%%%%%%%%
\subsect{Schmidt-Kennicutt and the compressive mode}

\fig{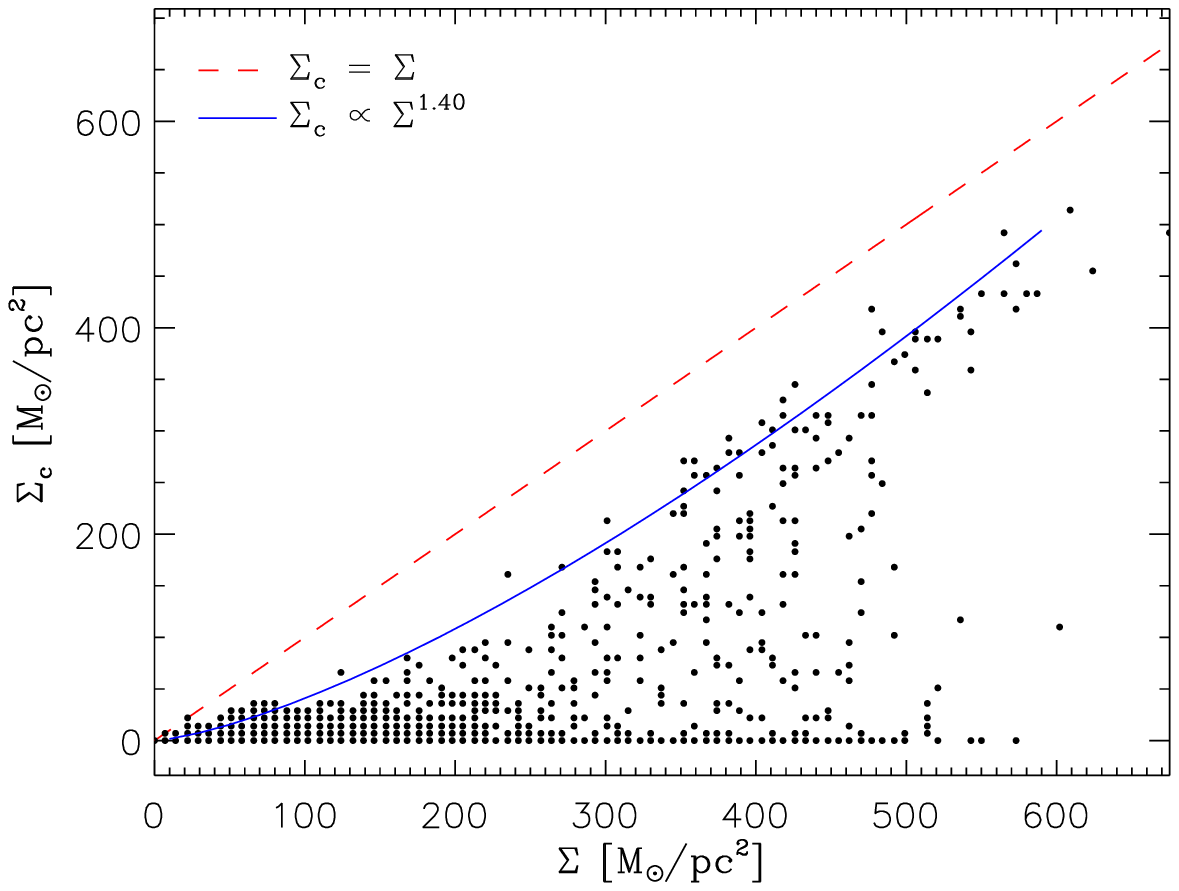}{$\Sigma_\mathrm{c}$ compared to $\Sigma$}
{Surface density of the particles in compressive mode $\Sigma_\mathrm{c}$ compared to those of all the particles $\Sigma$. The red dashed line marks the equality of the two quantities. The blue line shows the best fit of the upper envelope of the scattered distribution, as a power-law of index $\approx 1.40$. The apparent binning is due to an integer number of particles per pixel.}

As mentioned in \refboxp{schmidt}, the \acr{}{SFR}{star formation rate} can be empirically linked to the power $\approx 1.4$ of the surface density of the gas, via the Schmidt-Kennicutt relation. In addition, our results suggest to connect the formation of stars to the concept of compressive tidal mode. Therefore, it seems natural to seek a relation between the surface density of the gas and the compressive mode. To explore this further, we plot in \reffig{ks_surface} the two surface densities $\Sigma$ (all particles) and $\Sigma_\mathrm{c}$ (compressive particles only) for all the pixels of the map presented in \reffigp{antennae_compressive_density}.

Because one cannot find more particles in compressive mode than the total number of particles, the line $\Sigma_\mathrm{c} = \Sigma$ is the logical upper limit. We define the envelope of the points displayed here as the maximum value of $\Sigma_\mathrm{c}$ taken in bins of $20 \U{\msun/pc}^2$. It can be well adjusted with a power-law $\Sigma_\mathrm{c} \propto \Sigma^{1.40}$ (with the standard deviation $\sigma = 26.2 \U{\msun/pc}^2$), i.e. the same expression as the empirical Schmidt-Kennicutt relation. This means that the surface density of compressive modes is limited by the same relation that links the \acr{}{SFR}{star formation rate} with the surface density of the gas.

Note however that our relation $\Sigma_\mathrm{c} \propto \Sigma^{1.40}$ concerns the envelope and thus is a \emph{limit}: many points stand below this line on \reffig{ks_surface}. Such a dispersion does not exist in the Schmidt-Kennicutt relation, where all the data points are close to the empirical law. All together, these results suggest that (i) again, the compressive mode plays a role in the formation of star as it shows a similar relation with the density as the \acr{}{SFR}{star formation rate} does and (ii) the large dispersion in our relation appeals for a finer prescription for star formation, including sub-scale physics and hydrodynamics. Such a more detailed comparison between the evolution of the compressive modes and the \acr{}{SFR}{star formation rate} throughout the interaction history is feasible when considering hydrodynamical runs.

%%%%%%%%%%%
\subsect[hydro]{Hydrodynamical studies}

In terms of mass fraction as well as intensity, the compressive tidal field is clearly associated to the major events in the interaction history of the merger as revealed by \reffigp{antennae_histo}. One can also identify the peaks around the pericenter passages with those noted by several authors who studied the \acr[s]{}{SFR}{star formation rates} from simulations of merging or interacting galaxies. For instance, \mycitet{diMatteo2007} presented a statistical study of 240 \acr{}{SPH}{smoothed particle hydrodynamics} simulations of interacting galaxies and explored the evolution of their \acr[s]{}{SFR}{star formation rates} (see also \mycitealt{diMatteo2008}). They noted that the pericenter passages could be identified with sharp increases in the \acr{}{SFR}{star formation rate}, by a factor depending on the fine details of the morphological types of the progenitors and the orbital parameters\footnote{Note however that they suggested that not all interactions are to be systematically linked with a starburst activity.}.

In our own description, the impact of tidal stripping during close passages is diminished because a pressurized component is missing. In addition, we do not take into account the formation of stars at the first passage, that inevitably depletes the gas reservoir, leaving a lower gas mass available to form stars during the second passage. However, even without the response of the gas to the interaction, the evolution of the mass fraction in compressive mode (i.e. the relative importance of this mode) matches well the curve of the \acr{}{SFR}{star formation rate} versus time.

We have recently extended our palette of tools to explore the history of star formation, with two techniques that include hydrodynamics: \acr{}{AMR}{adaptive mesh refinement} and \acr{}{SPH}{smoothed particle hydrodynamics}. We have applied these tools to models of the Antennae, and we begin to compare their results with the properties of the compressive tidal mode.

%%%%
\subsubsection{With \acr{notarg}{AMR}{adaptive mesh refinement}}
An hydrodynamical study of our model of the Antennae is currently conducted in collaboration with researchers at the CEA\footnote{Commissariat \`a l'\'Energie Atomique, Saclay.} (PI: D.~Chapon). With the initial conditions presented in \reftabp{antennae}, a gaseous disk of 10\% of the baryonic mass has been added to both galaxies. Gravitational and hydrodynamical equations have been solved thanks to the \acr{}{AMR}{adaptive mesh refinement} code Ramses (\mycitealt{Teyssier2002}) with 14 levels of refinement (\reffig{ramses_amr}), which gives a maximal resolution of $12 \pc$. \reffig{ramses} shows the morphology of the system at $t \sim 100 \Myr$, i.e. between the two pericenter passages. Because of the additional gaseous mass and its response to the interaction, a small difference exists in the timing between the two passages: the \acr{}{AMR}{adaptive mesh refinement} run gives out a separation of $260 \Myr$ instead of $300 \Myr$ for the purely gravitational model.

\mybox{
The \acr{notarg}{AMR}{adaptive mesh refinement} technique allows to place the computational power where needed. The code starts with a regular box (in our case, a cube of $200 \kpc$) and checks for a refinement criterion (here, a certain density level). If the criterion is fulfilled, the cube is refined into cells, which are also examined for the same criterion and so on (in a similar way as a tree-code, recall \refsect[numerical]{treecodes}) until either the criterion is not fulfilled, or the maximum level of refinement is reached. In this way, it is possible to have an extremely high resolution in regions of interest, like cluster forming zones, and to keep a low resolution in e.g. empty spaces to keep the computation fast. \reffig{ramses_amr} illustrates this point.

Once the mesh has been built, the equations of (magneto-)hydrodynamics are solved in each cell. In addition, the stellar particles are evolved thanks to a traditional Poisson solver for the total gravity (gas + stars + \acr{notarg}{DM}{dark matter}).
}

\fig{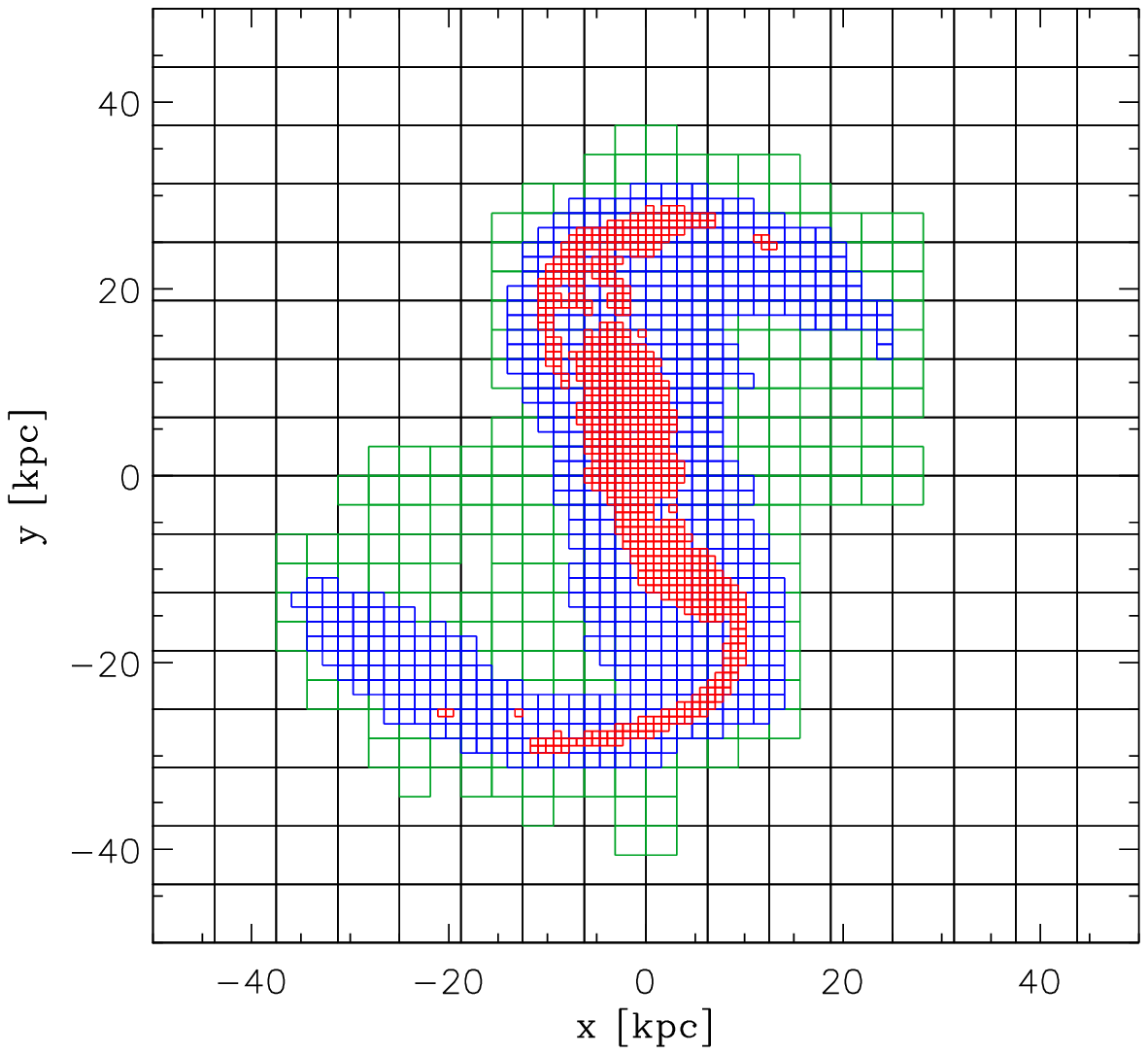}{AMR grid of the simulation of the Antennae}
{Grid used in the simulation of the Antennae galaxies, viewed in the orbital plane at $t\approx 100 \Myr$ (i.e. as in \reffigt[d]{antennae_op}). Black, green, blue and red correspond to the refinement levels 5, 6, 7 and 8. In total, the mesh counts $\sim 10^6$ cells (up to level 14). For graphical purposes, only the densest regions (in projection) are plotted, which explains that some cells seem to be not correctly split into four sub-elements in this Figure.}

\fig{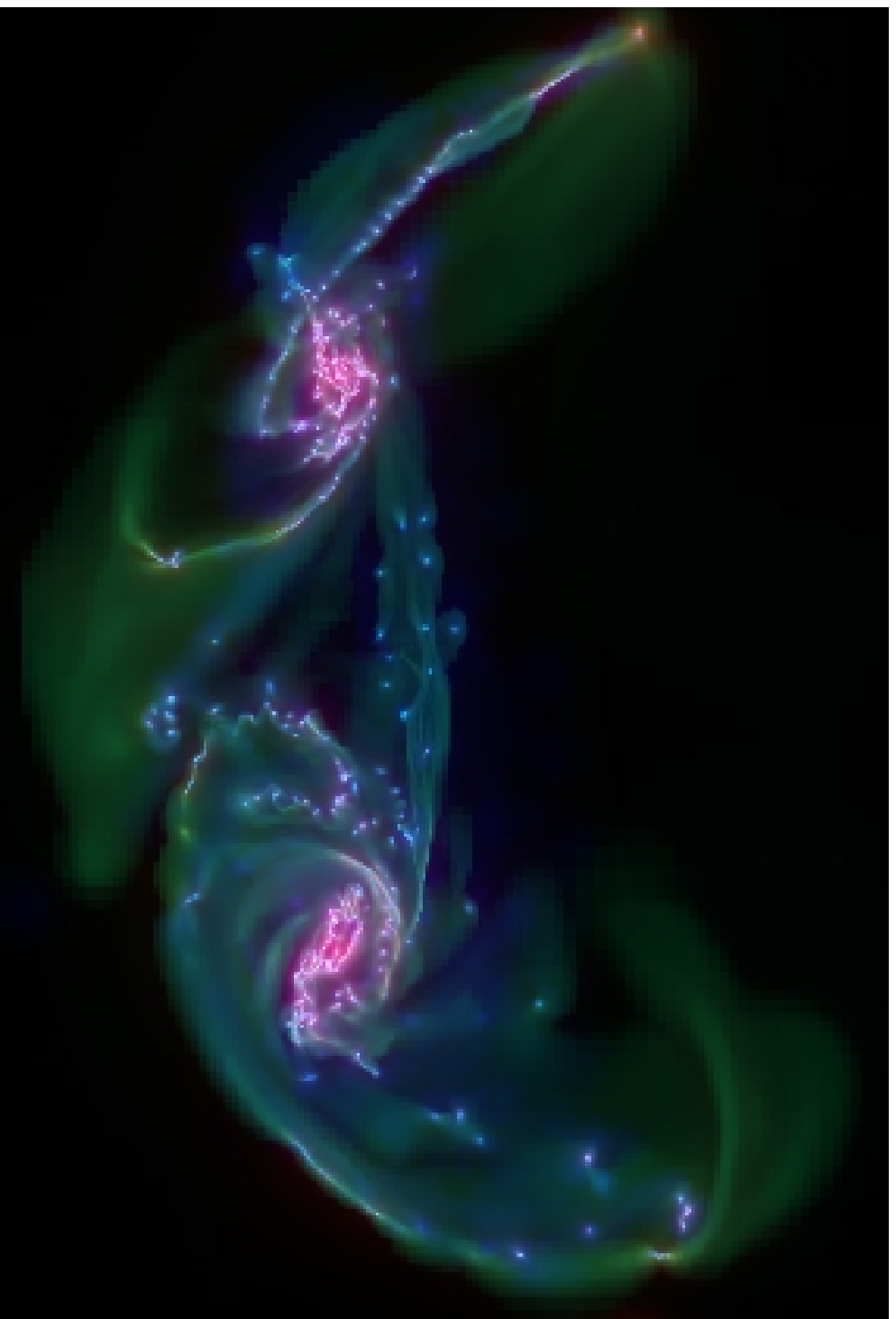}{AMR simulation of the Antennae}
{Hydrodynamical simulation of the Antennae with the Ramses code. This snapshot corresponds to $t \approx 100 \Myr$ (i.e. as in \reffigt[d]{antennae_compressive}). The density of gas is shown in green, red represents the old stars while blue denotes the newly formed stars. Thanks to the \acr{notarg}{AMR}{adaptive mesh refinement}, the largest pixels cover $\sim 1.5 \kpc$ while the smallest ones (e.g. the blue seeds of star formation in the bridges) represent $12 \pc$ only. (Image from D.~Chapon, private communication.)}

In the preliminary runs, no feedback nor magnetic field have been introduced. The star formation recipe is based on a Jeans mass taken equal to $2\times 10^5 \msun$ (matching the typical value of the massive clusters) via a polytropic equation of state\footnote{When collapsing, a molecular cloud reaches a density so high that its heat cannot be radiated away (\mycitealt{MacLow2004}). That is, it becomes opaque to radiative cooling and should no longer be considered isothermal ($\gamma =1$). That is why the equation of state adopts the polytropic form $P\propto \rho^\gamma$, with $\gamma > 1$.} $P\propto \rho^\gamma$. The formation rate of clusters is fixed at a high value, close to the one set by the numerical resolution limit, which gives a \acr{}{SFR}{star formation rate} that might be over-estimated. A deeper exploration of the parameters of the equation of state, the formation efficiency and the cooling function will be done in the next months.

\fig{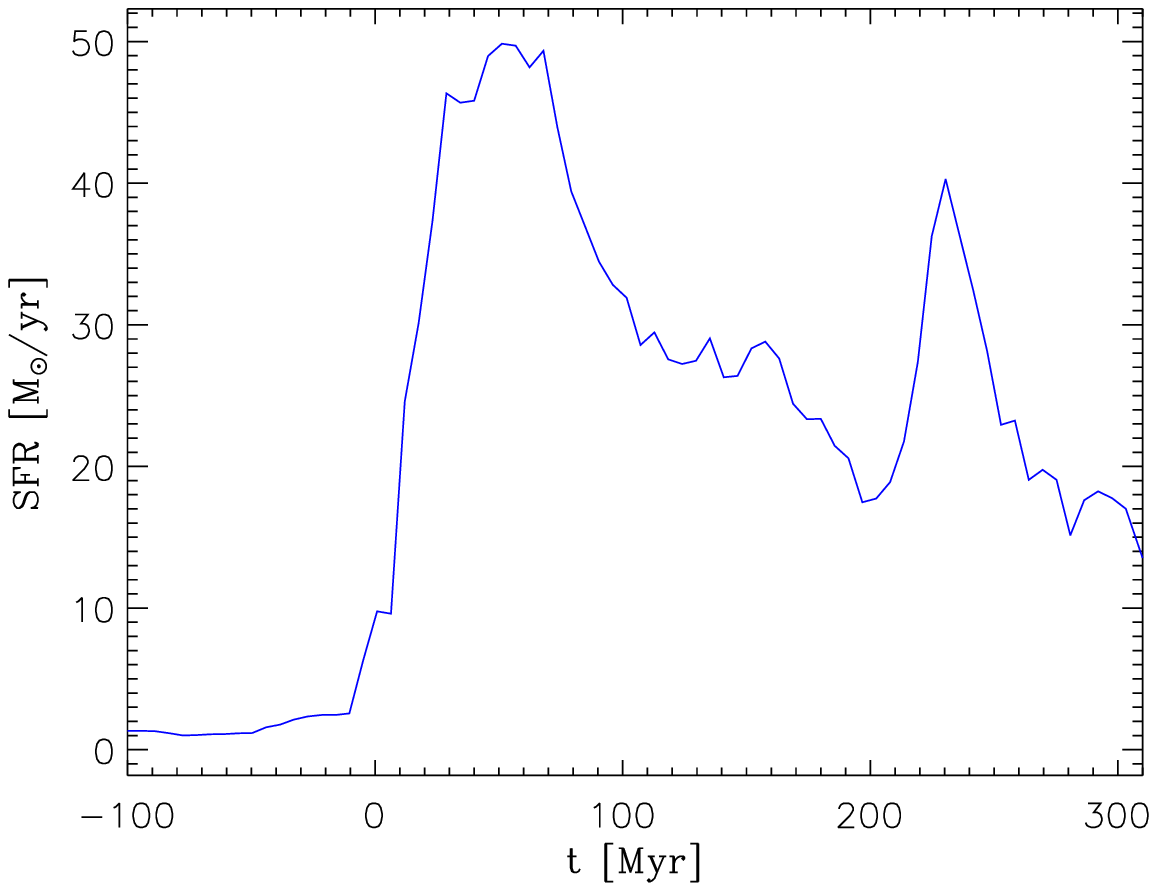}{SFR in an AMR simulation of the Antennae}
{\acr{notarg}{SFR}{Star formation rate} from an \acr{notarg}{AMR}{adaptive mesh refinement} simulation of the Antennae. See the text for details.}

\reffig{sfr_ramses} shows the evolution of the \acr{}{SFR}{star formation rate} during the simulation. Despite the orbital difference between the \acr{}{AMR}{adaptive mesh refinement} and the dry runs, the \acr{}{SFR}{star formation rate} matches surprisingly well the trend seen in the mass fraction in compressive mode (thumbnail in \reffig{sfr_sph}, below). Indeed, the first pericenter is clearly marked by a sharp peak, followed by a slow decrease up to the second passage. We note a difference however: the second major peak in the \acr{}{SFR}{star formation rate} is lower (by 20\%) than the first one. In comparison, the two peaks are of similar amplitude for the mass fraction in compressive mode (dry run). This difference can be explained when considering the dynamics of the gaseous component. During the first interaction, the gas reservoir has been depleted by (i) tidal stripping and (ii) its conversion into stars. In other words, a fraction of the gas has been spread out (recall the \hi tails detected by \mycitealt{Hibbard2001}), and the rest has been processed to form the stars, with a certain efficiency $\epsilon$. These two reasons combined explain that a smaller amount of gas is available at the second passage to form new stars. That is why the second peak in the \acr{}{SFR}{star formation rate} is lower than the first one. For the case of the dry run on the other hand, the only difference between the two events is the mass-loss due to the creation of the tails ($\sim 30\%$ of the total mass of the stellar disk). This should lead to a smaller second peak: however, because of the loss of angular momentum of the progenitors, the pericenter distance is much smaller (recall \reffigt{antennae_dist}), which enhances the fraction of compressive modes\footnote{This point will be developed in details in the next Chapter.}. In turn, the combination of these two effects (mass-loss and close passage) balance each other so that the second peak turns out to be of similar amplitude to the first one.

Even if our hydrodynamical study with Ramses is still at a preliminary phase and some tuning of e.g. the star formation recipe is required, we note a strong correlation with the dry run presented in the previous Chapter, and in particular between the evolution of the mass fraction in compressive mode and the one of the \acr{}{SFR}{star formation rate}. In addition to the good matches between the location of the compressive tidal field, their duration, their energy and the properties of the young clusters observed in the Antennae, this new relation with the \acr{}{SFR}{star formation rate} strongly suggests that the compressive mode is linked to the formation of substructures. An important work with our Ramses simulations will be done during the following months to fully explore and quantify the very first points we have highlighted here. One of the key question to be addressed pertains to the existence of a time delay between the creation of a compressive mode and the actual formation of proto-stellar seeds from the fragmentation of the \acr{}{ISM}{interstellar medium}.

Note that the \acr{}{AMR}{adaptive mesh refinement} technique has already been used to study star formation in interacting galaxies by \mycitet{Kim2009}. They set up a planar encounter of two disk galaxies and derived the evolution in time of the \acr{}{SFR}{star formation rate}, with a resolution of $2\times 10^3 \msun$. There again, peaks exist at the times of the close passage, denoting a shock-triggered formation, as mentioned by \mycitet{Barnes2004} with \acr{}{SPH}{smoothed particle hydrodynamics} runs. As a complement to this work, our new Antennae model with Ramses will provide a good sample to explore the details of the connections with the tidal field.

%%%%
\subsubsection{With \acr{notarg}{SPH}{smoothed particle hydrodynamics}}

In parallel with our \acr{}{AMR}{adaptive mesh refinement} study, a brand new \acr{}{SPH}{smoothed particle hydrodynamics} model of the Antennae galaxies has been presented in \mycitet{Karl2010}, using the code {\tt Gadget2} (\mycitealt{Springel2005a}). This model is part of a parameter study aimed at finding better initial conditions for the modeling of the Antennae galaxies. Note that the initial conditions differ from the purely gravitational run presented in \refcha{antennae}. In particular, the \acr{}{SPH}{smoothed particle hydrodynamics} model yields a longer orbital period. The resolutions however, are comparable. As visible in \reffig{sfr_sph}, the \acr{}{SFR}{star formation rate} is, again, enhanced at the times of the two pericenter passages. We note that the first peak is much less pronounced than the second one, and that the associated ``burst'' of star formation is diluted in time over $\sim 500 \Myr$.

\fig{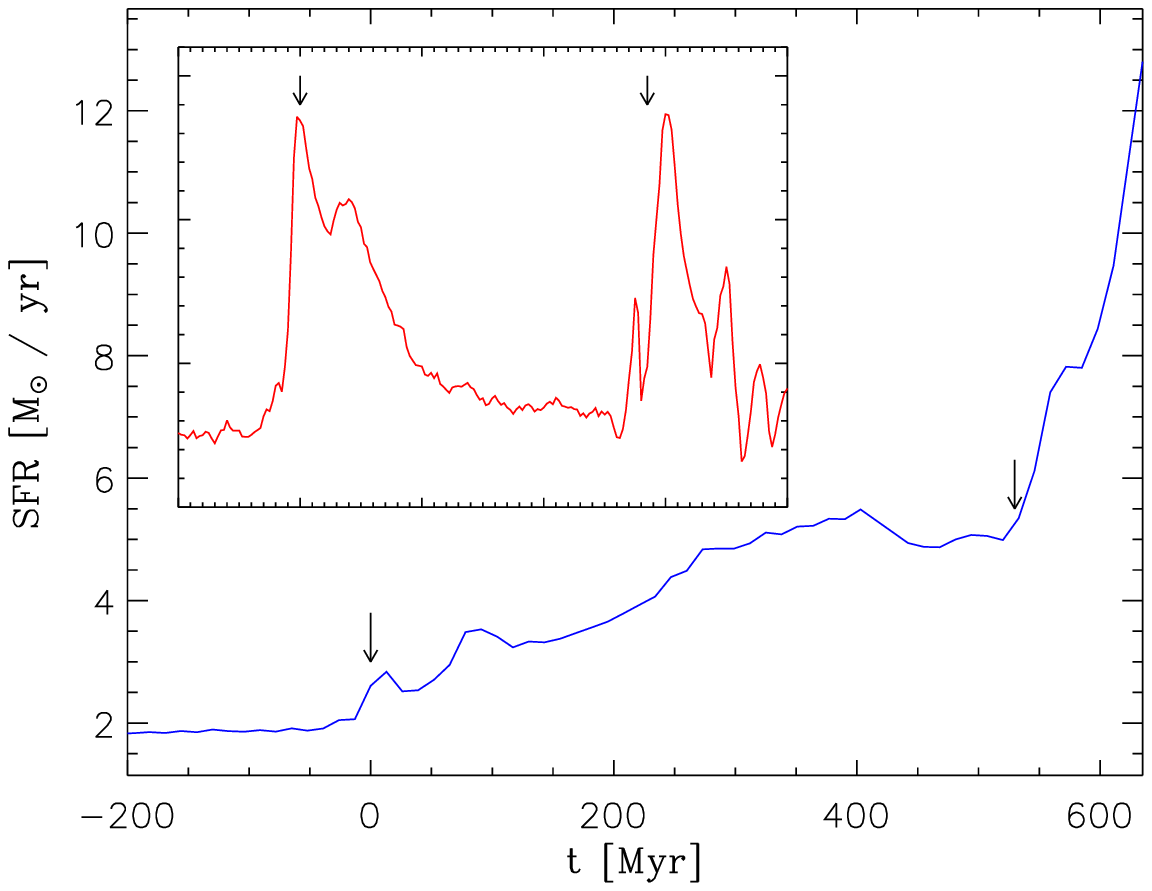}{SFR in a SPH simulation of the Antennae}
{\acr{notarg}{SFR}{Star formation rate} from the \acr{notarg}{SPH}{smoothed particle hydrodynamics} simulation of the Antennae presented in \mycitet{Karl2010}. Arrows indicate the two pericenter passages. The evolution of the \acr{notarg}{SFR}{Star formation rate} for $t > 600 \Myr$ is still being computed. The thumbnail shows the shape of the evolution of the mass fraction in compressive mode from the dry simulation (as presented in \reffigt{antennae_histo}), as a reminder.}

The reason for this is still under investigation but seems to be linked with the low spin of the disks which makes them more compact and thus, more resistant to the effects of the interaction. The gas is then slowly stripped and can fall back onto the disks during a much longer period, which translates into a diluted ``burst'' of star formation. In this case, the mild decrease of the \acr{}{SFR}{star formation rate} at $\sim 400 \Myr$ would correspond to the depletion of the intergalactic gas reservoir. For the second passage, the progenitors have already been affected which could mean a much more efficient and rapid formation of stars, as noted on \reffig{sfr_sph}.

Furthermore, the \acr{}{SPH}{smoothed particle hydrodynamics} model uses massive Hernquist \acr{}{DM}{dark matter} halos while the dry run and the \acr{}{AMR}{adaptive mesh refinement} model have lighter, isothermal halos. This additional difference may also play an important role in the formation of stars. The influence of the shape of the \acr{}{DM}{dark matter} profile is studied in \refcha{halos} in more detail.

All the concepts and results presented so far show a correlation between compressive tidal field and star formation, in our models of the Antennae galaxies. It is important however to confirm that our conclusions do not exclusively apply to this merger but can also be extended to a broad range of configurations. Therefore, an exploration of the influence of morphological and orbital parameters on the characteristic of the compressive tidal modes is presented in the next Chapter, to allow a comparison with the trends in the \acr{}{SFR}{star formation rate} noted by e.g. \mycitet{diMatteo2008}.

%%%%%%%%%%%
\sect[nbody6]{Stellar dynamics}

Before investitaging other galaxies, let's focus on the early life of a cluster, i.e. after the formation phase. In addition to triggering or strengthening star formation, the host galaxy can modify the morphology and internal evolution of a cluster through the tidal field. Thanks to the tidal tensor, it is possible to take into account the potential of the galaxy on the spatially extended cluster in a relatively simple, yet realistic way (i.e. for non constant and non isotropic tides). In particular, the tidal field seen by a cluster along a non regular orbit (like those found in mergers) changes much more rapidly than the ones typically described in the literature (see e.g. \mycitealt{Baumgardt2003}). Therefore, our method allows to explore the role of the large scales (galaxies) on the smaller scales (star clusters) for all kind of orbits.

However, the study of the evolution of a cluster requires a more precise description than the collisionless $N$-body approach we have followed so far. As presented in \refboxp{relax_times}, globular clusters are collisional systems. Therefore, a numerical treatment of close encounters between stars, and binaries must be used. Furthermore, one should account for stellar initial mass function, stellar evolution, feedback... The code used in our galactic exploration (i.e. {\tt gyrfalcON}) does not consider these aspects and thus, a different tool is needed. {\tt Nbody6} (see \mycitealt{Aarseth1999} and references therein) and its version for parallel computers {\tt Nbody6++} (\mycitealt{Spurzem2003}) fit our requirements as they allow to simulate star clusters and the evolution of their stellar contents.

Based on an idea of \mycitet{Fleck2007}, we have adapted {\tt Nbody6++} so that it can take an external tidal field into account via the associated tidal tensor. In this new code, called {\tt Titens}, the initial tensor is strictly equal to zero and then, it is rapidly grown to reach the values required\footnote{An immediate growth of the tensor would lead to a dramatic change of the energy of the system, which would make {\tt Titens} detect an anomaly and stop the computation.}. This ensures continuity and coherence of the initial setting of the cluster, which is built without external field. This growth is done over a couple of timesteps, i.e. $\sim 10^{4-5} \yr$.

\mybox{
{\tt Titens} reads the nine values of the tidal tensor from the galactic simulation, plus the associated time, in an external file. The timestep of {\tt Titens} ($\sim 10^4 \yr$) is much shorter than the one of {\tt gyrfalcON} ($\sim 10^6 \yr$) and thus, the tensor is badly sampled for a direct use in a stellar dynamics run. Therefore, its values are interpolated when needed by {\tt Titens}, using a linear scheme for the first snapshot and a quadratic interpolation for all the others.

Note that both the time and the tensor must be scaled properly, because {\tt gyrfalcON} and {\tt Titens} do not share the same system of units.
}

It is now possible to model a cluster with a given mass profile (e.g. Plummer, King, ...), an \acr{}{IMF}{initial mass function} and stellar evolution (\mycitealt{Hurley2000}). For the preliminary runs presented here, we focus on the effect of the tidal field and thus, temporarily switch-off stellar evolution. Nonce, all the stars in our clusters have the same mass. Therefore, the effects of a mass function such as a non-trivial luminosity function and mass segregation do not occur. That is, the role of the tidal field \emph{only} is made easier to detect and to analyze. Obviously, the full possibilities of {\tt Nbody6++} will be exploited right after these first studies.

%%%%
\subsect{In an isotropic tidal field}

For the first examples, five isotropic and constant tidal fields have been artificially set, with values typical of what we found in the galactic runs: 
\begin{itemize}
\item an isolated case: $\lambda = 0$, used as a reference,
\item a weak field (extensive or compressive): $\lambda = \pm 2\times 10^{-30}\U{s^{-2}}$,
\item and a strong field : $\lambda = \pm 6\times 10 ^{-30} \U{s^{-2}}$.
\end{itemize}
In addition, we have created a ``transition'' tidal field which is compressive for the first $25 \Myr$ and then becomes extensive forever, with the values of the weak regime in both cases.

Note that an isotropic extensive tidal tensor corresponds to a negative density (its trace being positive, see \refeqnt{poissontrace}). Although not physical, such cases allow a simple analysis and a direct comparisons with 1D simulations (e.g. Fokker-Planck, Monte-Carlo). The possible mass-loss due to the field will be, in this case, over-estimated.

\fig{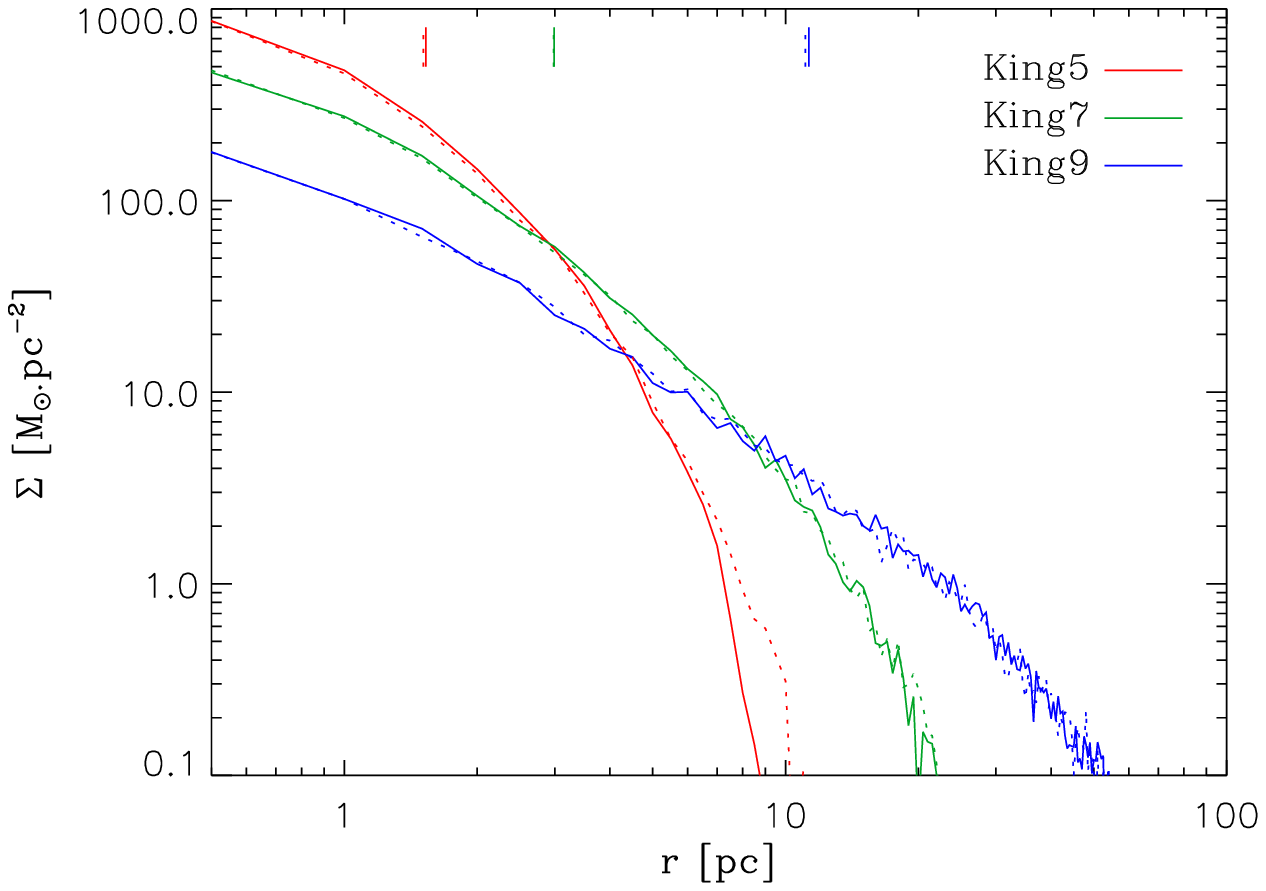}{Density profiles of the King models}
{Surface density profiles of the three King models in isolation, at initial time (solid curves) and after $100 \Myr$ (dotted curves). Only a small evolution due to relaxation is visible. Vertical ticks in the upper part mark the half-mass radii of the models.}

The effect of the tidal field depends on the density of the cluster. Therefore it is expected to be stronger in the outer parts of a mildly concentrated object than close to the center of a dense profile. To confirm this idea, King models with various parameters ($\Psi_0/\sigma^2$) have been set-up: for a given total mass, a small parameter corresponds to a peaked profile and thus, a small half-mass radius, while a higher number is associated to a more extended cluster (see \mycitealt{Binney1987}). \reffig{king_density_profile} sums up this by showing the surface density profiles of the three models we have used: King5, King7 and King9. Every cluster is made of $10^4$ stars of exactly $1 \msun$ each. The profiles do not change significantly over $100 \Myr$. The stability of the models in isolation is key in this study: it ensures that intrinsic phenomena (e.g. relaxation) do not bias the effect of the external field.

The evolution of some Lagrange radii is shown on \reffig{rlagr_isotropic_k7} for the King7 model. The inner regions are not strongly affected by the tidal fields used, and the associated Lagrange radii (e.g. 50\%) remain fairly independent of the strength of the tides. In the outer parts (70\%, 90\%) however, the extensive mode spreads efficiently the matter so that the ``size'' of the cluster increases. The compressive fields have a weaker, opposite effect on the radii but are still clearly distinguishable from the isolated case.

When embedded in the ``transition'' field (black curve), the cluster first gets a high potential energy: it becomes supervirialized. When the tides switch to extensive mode ($t > 25 \Myr$), the potential gets flatter and the velocity dispersion of the stars is too high to maintain equilibrium. Therefore the outer regions are violently expelled from the cluster: this translates into the growth of the 70\% and 90\% Lagrange radii, much quicker than for the constant extensive field of the same strength (solid blue curve).

\fig{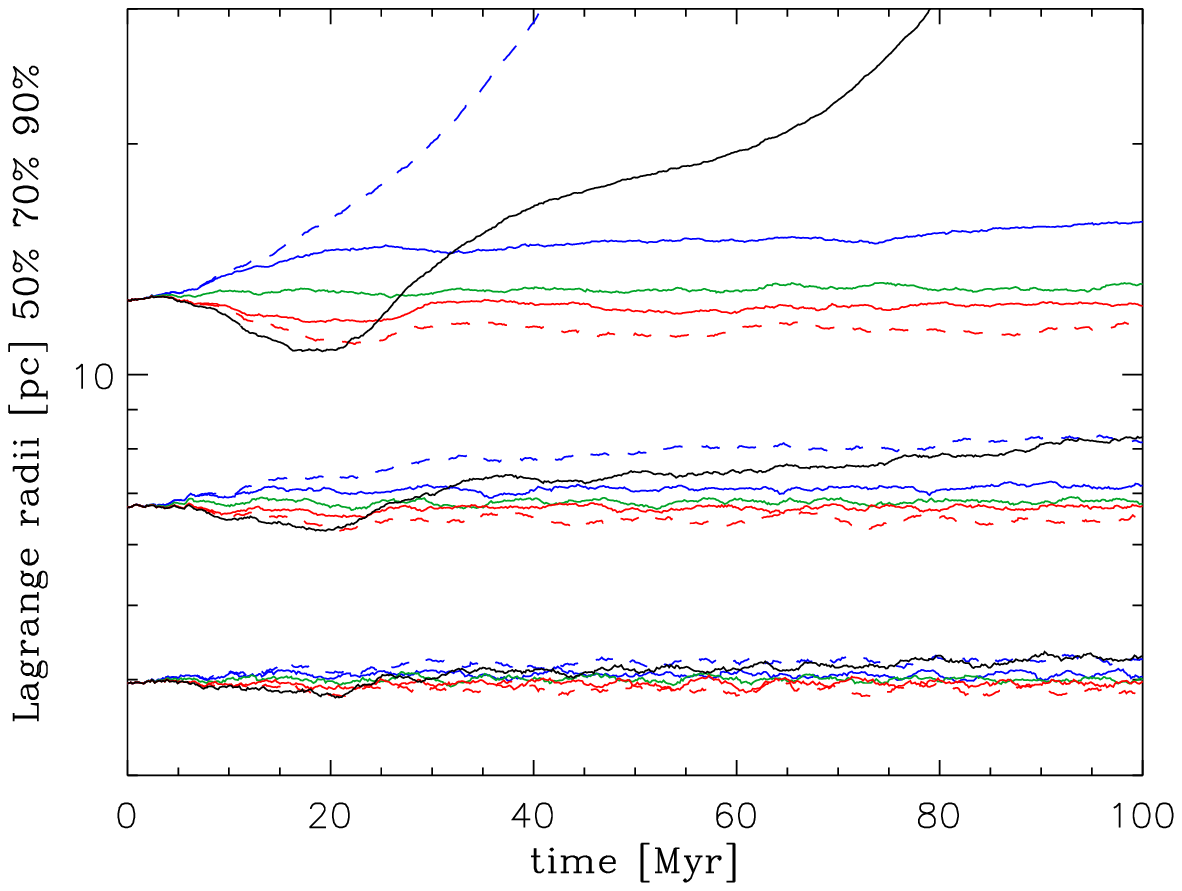}{Lagrange radii of a King7 model in a tidal field}
{Evolution of the 50\%, 70\% and 90\% Lagrange radii of the King7 model in isolation (green), in a weak field (solid curves) and in a strong field (dashed curves). Blue corresponds to the extensive fields while red denotes the compressive modes. The solid black lines are associated to the ``transition'' field (i.e. weakly compressive for $25 \Myr$ and weakly extensive afterwards.). See text for details.}

On the one hand, the same effect is even more dramatic for the King9 model, as shown in \reffig{rlagr_isotropic_k9}. In this case, the inner parts of the cluster are also affected and only the core ($\sim 10\%$ radii) remains fairly stable. On the other hand, the King5 model is too dense to be really modified by the tides. Even the outer regions keep the same size than in isolation: no effect of the field is visible on the 90\% Lagrange radius.

\fig{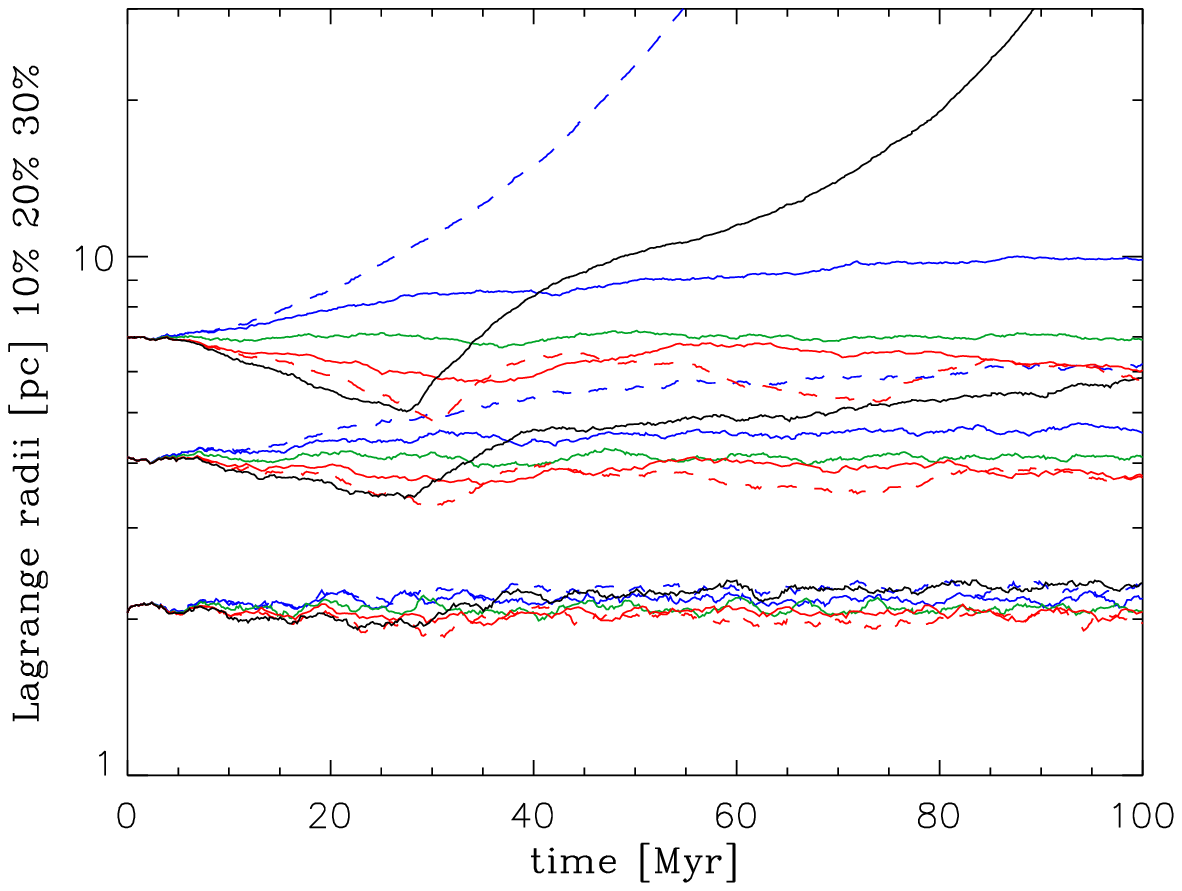}{Lagrange radii of a King9 model in a tidal field}
{Same as \reffig{rlagr_isotropic_k7}, but for the 10\%, 20\% and 30\% Lagrange radii of the King9 model.}

%%%%
\subsect{In the Antennae}

\fig{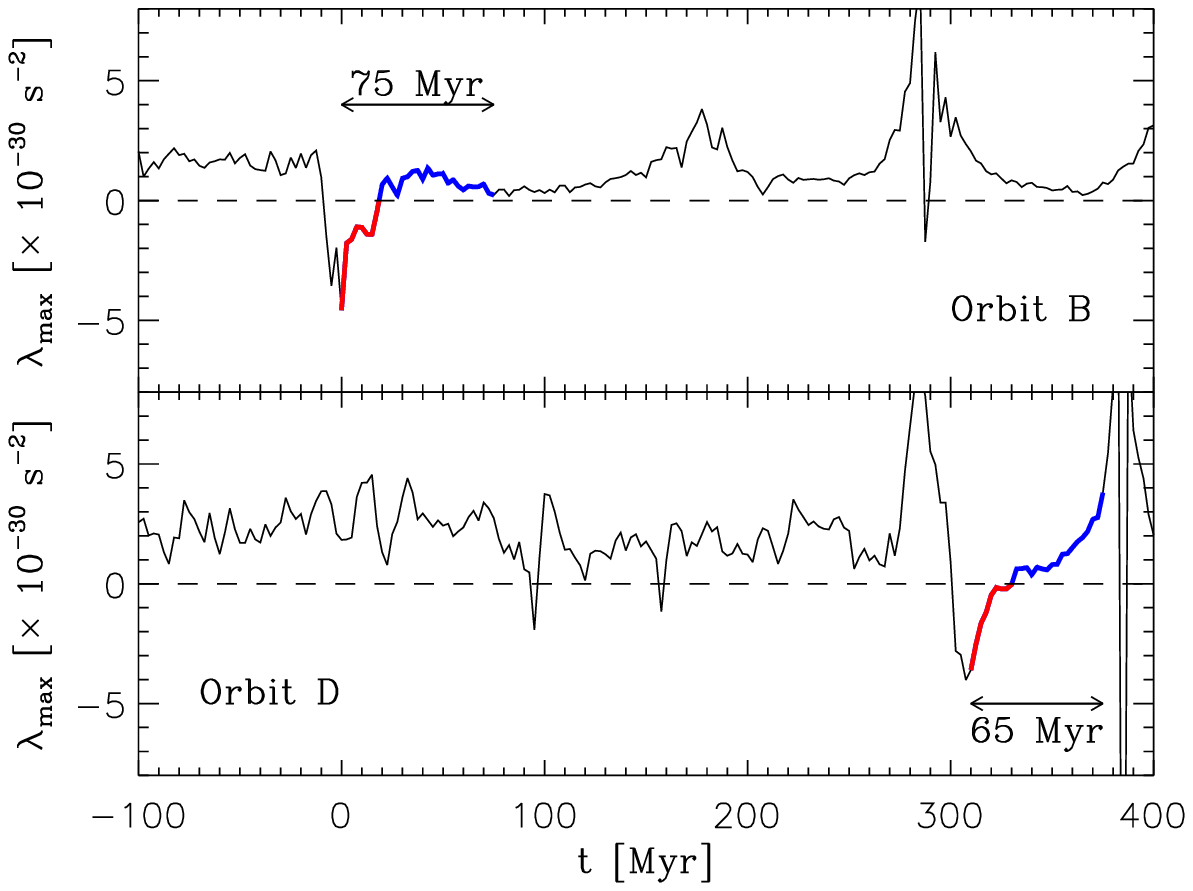}{Evolution of the max. eigenval. of the tidal tensor of the part. B and D}
{Evolution of the maximum eigenvalue of the tidal tensor of the particles B and D (see \reffigt{antennae_part_orbits}). The x-axis indicates the time of the galactic simulation. The colored parts correspond to the fields extracted and given to {\tt Titens}. Both start with a compressive mode (red) lasting $\approx 20 \Myr$, before going to an extensive regime (blue). In this last step, the field is stronger for the particle D than for B.}

These examples help to understand the behavior of a cluster when embedded in a simple tidal field. However, the previous Sections and Chapters revealed that merging galaxies yield tides that are far from being constant or isotropic. Therefore, we repeat the same experiment, with the same clusters, but this time in tidal fields obtained from the simulation of the Antennae (see \refcha{antennae}). The orbit of the particle B has been chosen (recall \reffigt{antennae_part_orbits}) because it has a long compressive episode induced by the first pericenter passage, and a moderated (yet significant) strength. The particle D gives out a comparable tidal field at the second passage, but with a stronger extensive regime. \reffig{orbits_partBD} displays the evolution of the maximum eigenvalue of the tidal tensors of these particles, and the time lapses considered for the {\tt Titens} runs. From now on, the time reference (t = 0) will be taken at the start of these lapses.

\fig{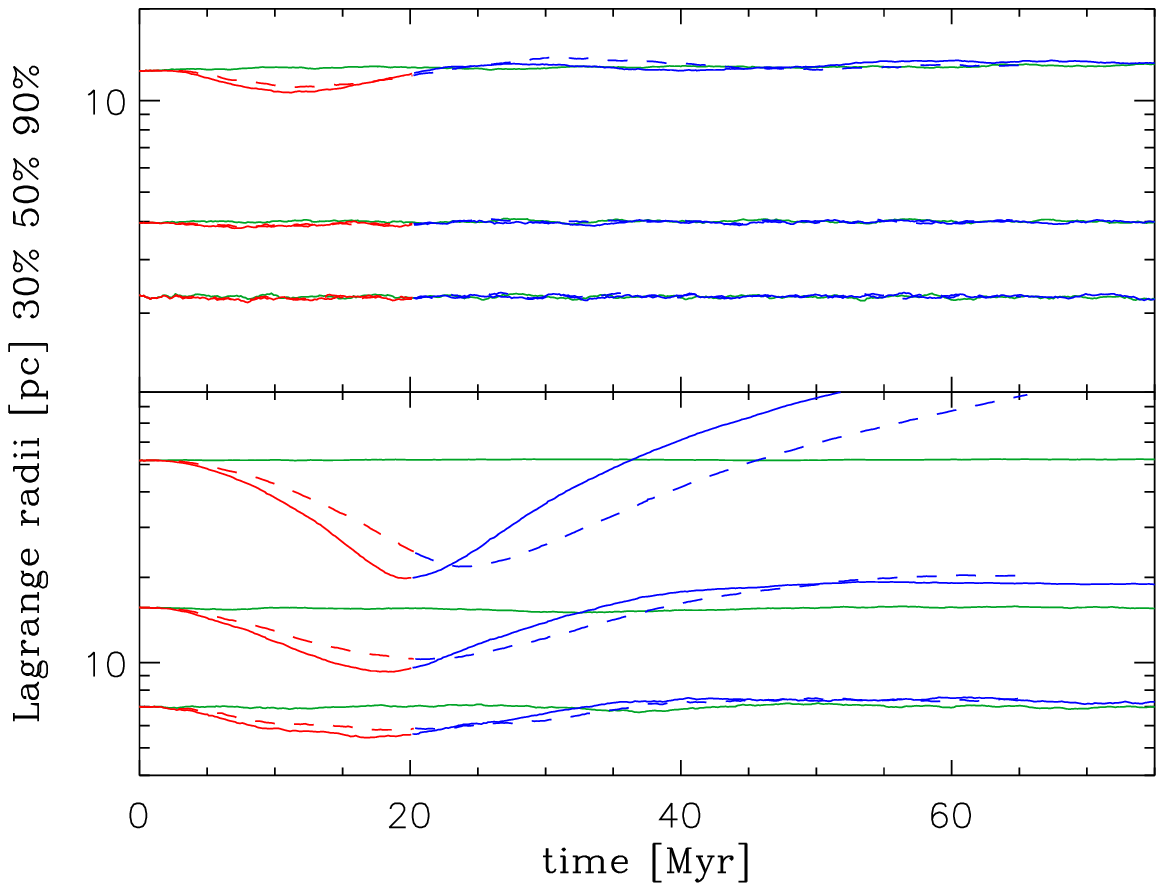}{Evolution of Lagrange radii of King models, along the orbits B and D}
{Evolution of the Lagrange radii of the King7 (top) and King9 (bottom) in isolation (green), along the orbits of the particles B (solid red/blue lines) and D (dashed lines). The parts of the curves in red denote a fully compressive mode (first $20 \Myr$).}

\reffig{rlagr_partBD} shows the evolution of the Lagrange radii in the inner parts (30\%, 50\%) and in the outer regions (90\%) along the two orbits, for the King7 and King9 models. As before, the King9 cluster is more deeply affected by the tidal field. The effect of the stronger extensive mode of the orbit D (with respect to those of the orbit B) is marked by a more pronounced expansion of the external layers of the clusters. However the general behaviors remain fairly comparable along both orbits.

\fig{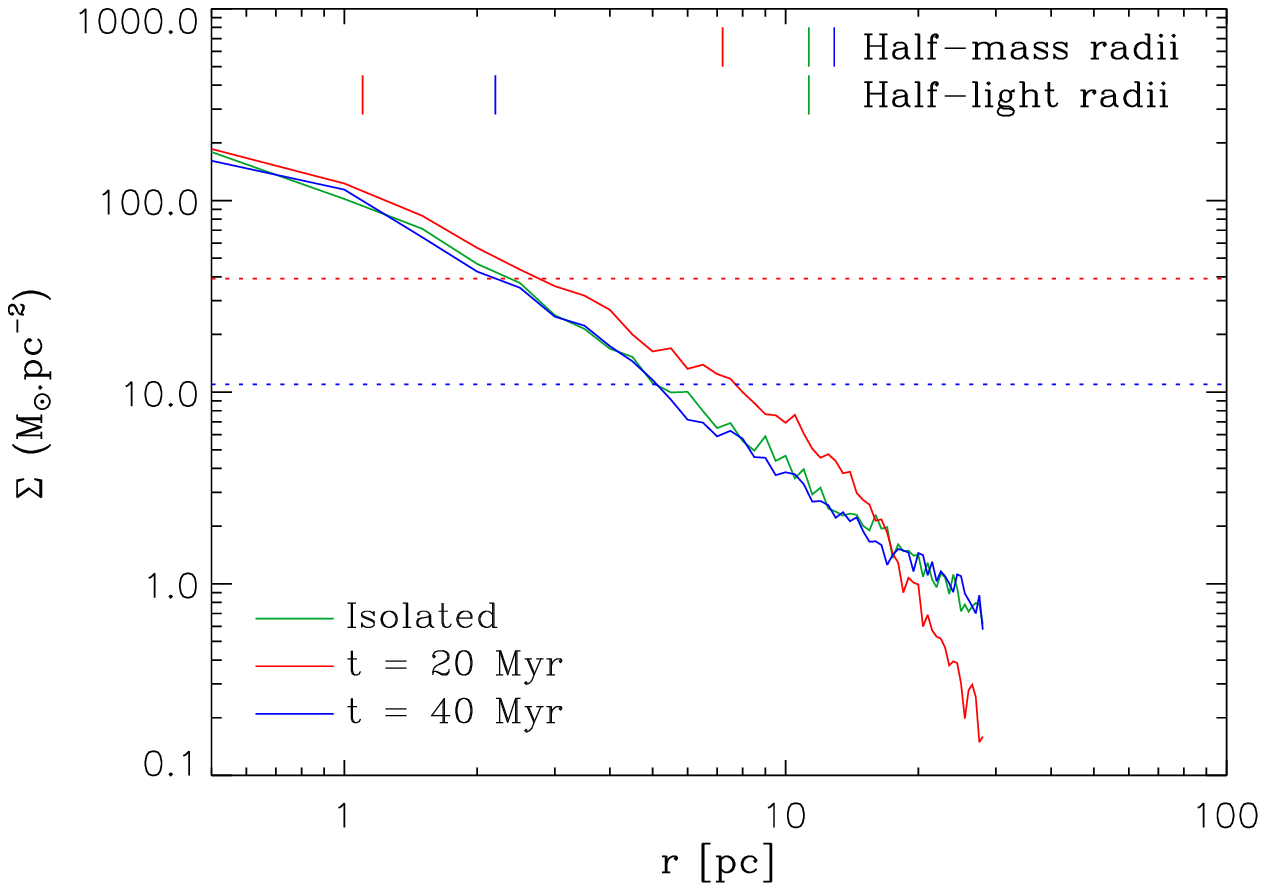}{Surface density profiles of the King9 model along the orbit B}
{Surface density profiles of the King9 model along the orbit B (at $t=20 \Myr$, red and $t=40\Myr$, blue) compared to the isolated case (green). The vertical ticks mark the half-mass and half-light radii. The horizontal dotted lines denote the level of the local surface density of the surrounding galaxy. Obviously, such a level does not exist in the isolated case.}

The violent expansion of the outer parts of the King9 model implies a change of its morphology and thus, a modification of its visual appearance. To illustrate this, we now take an observational perspective. The surface density profiles of the clusters are shown in \reffig{dens_profile_king9_orbitB}. Because all the stars in our runs have the same mass, this directly corresponds to a surface brightness profile. However, at these positions inside the Antennae galaxies, the surface brightness surrounding the cluster is far from being null and should be taken into account. To do so, the average density at the position of the cluster (i.e. along the orbits of B and D) is computed, and an equivalent brightness for the background is determined. (The equivalent surface density is shown as dotted lines in \reffig{dens_profile_king9_orbitB}.) We set the detection level at twice the value of the brightness of the background. Thus, all the outer, diffuse regions of the cluster are not detected.

\mybox{
The brightness of the background depends on the column density along the line of sight, and thus, will change with the orientation of the galactic model. To avoid this, we compute the \emph{local} volume density at the position of the cluster in the Antennae, thanks to the trace of the tidal tensor (recall \refeqnt{poissontrace}). Then, the volume density is integrated over the depth of the cube used to compute the tensor, and the surface density obtained is converted into a brightness, using the same scale as for the cluster itself.}

In the upper part of \reffig{dens_profile_king9_orbitB}, vertical ticks mark the half-mass and half-light radii: the half-mass radius takes into account all the stars of the initial cluster, while the half-light radius only considers the stars which have been detected.

\fig{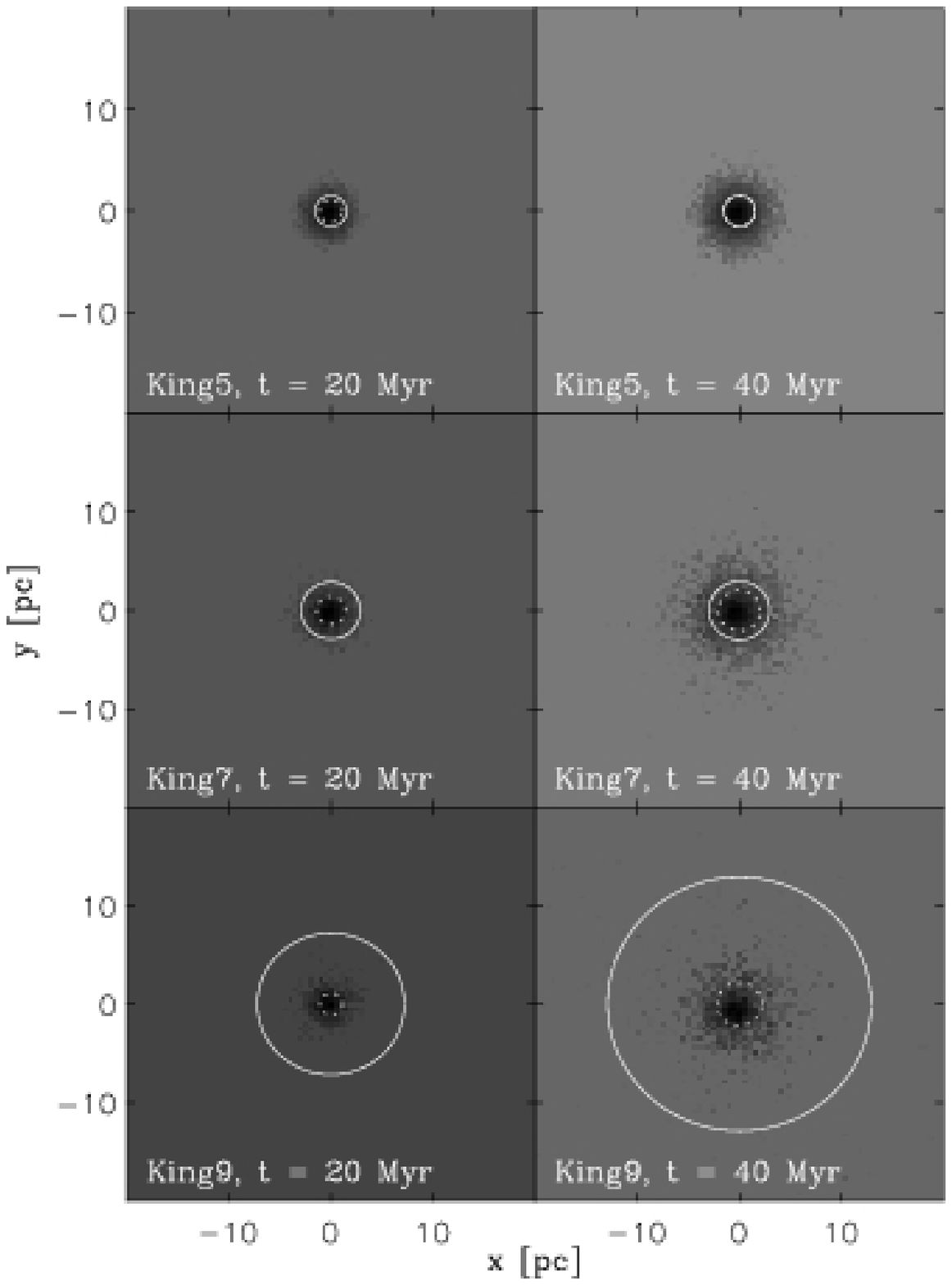}{Surface brightness of the three King models along the orbit B}
{Surface brightness of the three King models along the orbit B at the end of the compressive event ($t = 20 \Myr$, \emph{left}) and in the extensive mode ($t = 40 \Myr$, \emph{right}). The solid line circles mark the half-mass radii (considering all the stars) and the dotted circles show the half-light radii (from stars which brightness exceeds twice the level of the background). Each pixel covers $0.5 \times 0.5$ \U{pc^2}. The colors range from white (surface density = 0) to black (central surface density of the cluster).}

\reffig{ccd_orbitB} displays the surface brightness of the three King models embedded in the the tidal fields of the orbit B at $t = 20 \Myr$ (i.e. at the end of the compressive regime) and $t = 40 \Myr$ (in extensive mode). Again, the peaked King5 is almost not affected by the external field. Contrarily, the King7 and King9 models are strongly disturbed and their two radii increase. If one considers that a star which is not detected is in the field and not anymore in the cluster, the mass-losses at $t=40\Myr$ are $\approx 8\%$, $29\%$ and $76\%$ of the initial mass ($10^4 \msun$), for the King5, King7 and King9 models respectively.

Along the orbit of the particle D, the models give out a comparable response to the tidal field (\reffig{ccd_orbitD}). As already visible in \reffig{rlagr_partBD}, the half-mass radii remain similar to those found in along the orbit B. However, the half-light radius of the King9 cluster has been multiplied by a factor $\sim 2$. This is not to be linked with the dynamical evolution of the cluster, but with a much fainter background, i.e. a lower density. This last point demonstrates that extra caution should be taken when analyzing both the observations and the simulations of the morphology of the star clusters.

\fig{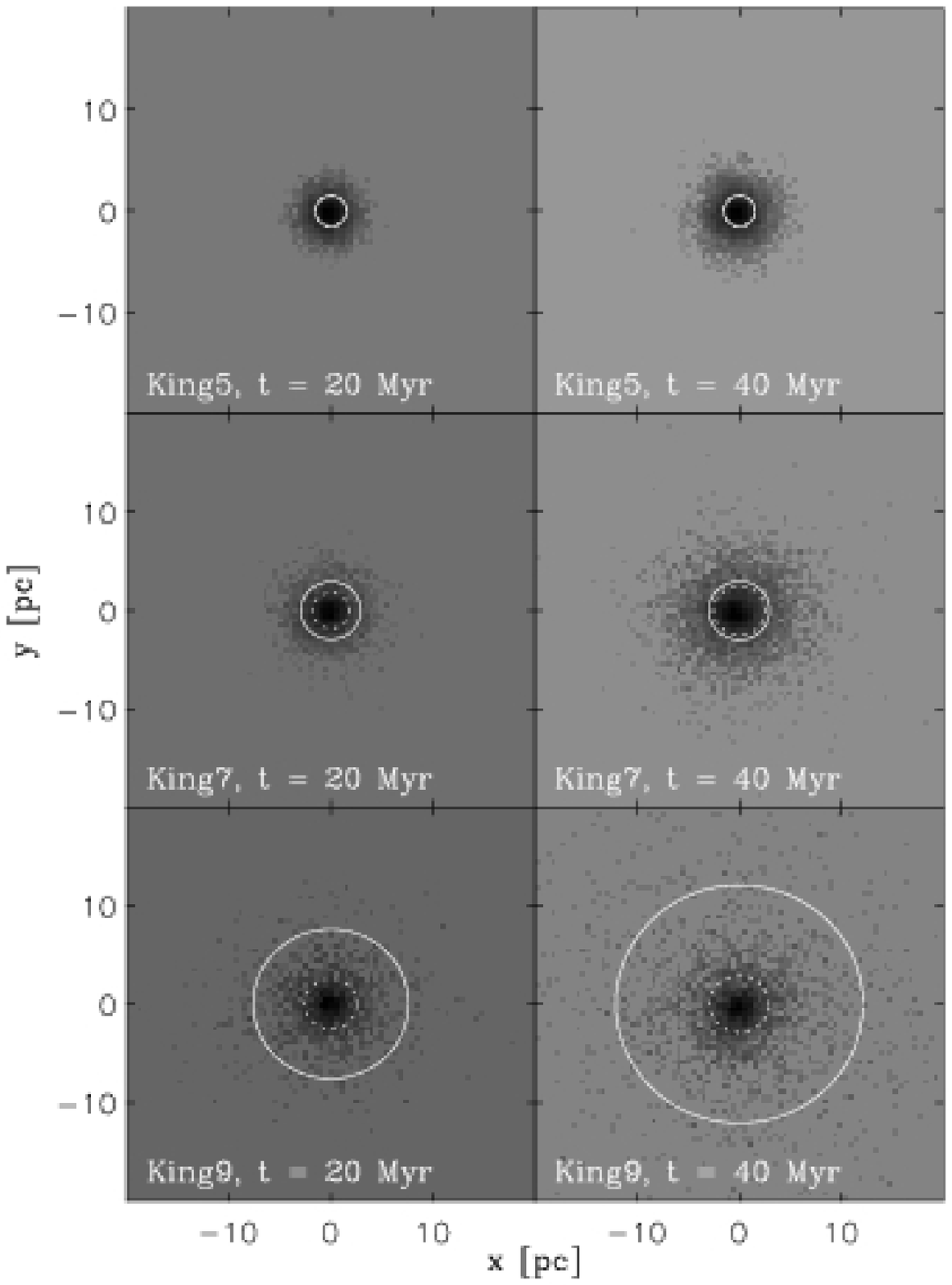}{Surface brightness of the three King models along the orbit D}
{Same as \reffig{ccd_orbitB}, but along the orbit D.}

%%%%
\subsect{Conclusions}

As expected since the previous Sections, the fast transition from a compressive tidal mode to an extensive one leads to severe disruption and mass-loss of clusters yielding mildly peaked density profiles. With these first results, one can get a broader picture of the combined effects at stake at cluster scale:
\begin{itemize}
\item As we have seen, a fast evolving tidal field impacts on the external layers of a cluster. This changes significantly the morphology, mass, luminosity profile and internal dynamics.
\item An initial mass function for the stars within the cluster also speeds up the evolution, via mass segregation (\mycitealt{White1977}; \mycitealt{Fleck2006}). The more massive stars migrate to the center, and then weaken it through violent feedback (winds and \acr{}{SN}{supernova} blasts). (This point will be explored in the next months, with {\tt Nbody6++} and {\tt Titens}.)
\end{itemize}
These two effects combined act in the central regions and the external layers of a cluster by modifying its evolution. Therefore, a relation with the mass function of the clusters in the Antennae, and in merging galaxies in general, could be established. The question of ``infant mortality'' should be better understood too.

Note also the importance of the binary/multiple fraction in the stellar population that is of prime importance when deriving velocity dispersions (see the recent works of \mycitealt{Kouwenhoven2008} and \mycitealt{Gieles2010}), and thus the dynamical stage of a cluster (in particular, supervirialized or not). This aspect will also be considered in the next steps of our study.

%%%%%%%%%%%%%%%%%%%%%%%%%%%%%%%%%%%%%%%%%%%%%%%%%%

\cha{survey}{Many interacting galaxies}
\chaphead{In this Chapter, we extend our conclusions about the Antennae galaxies to a broad range of configurations. Over the entire Chapter, the Antennae model presented before is considered as the reference for our parameter survey. We detail here the results initially presented in the Section 5 of \mycitet{Renaud2009}.}

%%%%%%%%%%%%%%%%%%%%%%%%%%%%%%%%%%%%%%%%%%%%%%%%%%
\sect{Need for a parameter survey}

During the first part of this thesis, we have focused on the Antennae galaxies. All the methods and diagnostics have been implemented with the view to compare the results obtained with direct observations of a \emph{real} object. However, it seems natural to wonder if this pair of galaxies is representative of a larger set of configurations.

Our results about the Antennae point out a strong correlation between the compressive tidal mode and the early life of star clusters. The next step of our study is thus to determine whether the same conclusions are applicable to a broad sample of interacting galaxies, and more precisely, which parameters are relevant for the growth of the compressive tides, and which are not. In the following Sections, we explore the role of several of these parameters by covering a significant volume of the entire parameter space, centered on those of the Antennae model (see \reffig{survey}). The main properties of the compressive tidal modes of these models are summarized in \reftab{parameter_survey}.

Our goal here is to compare the different models with the Antennae, that we consider and will refer to as the reference of this survey.

\fig{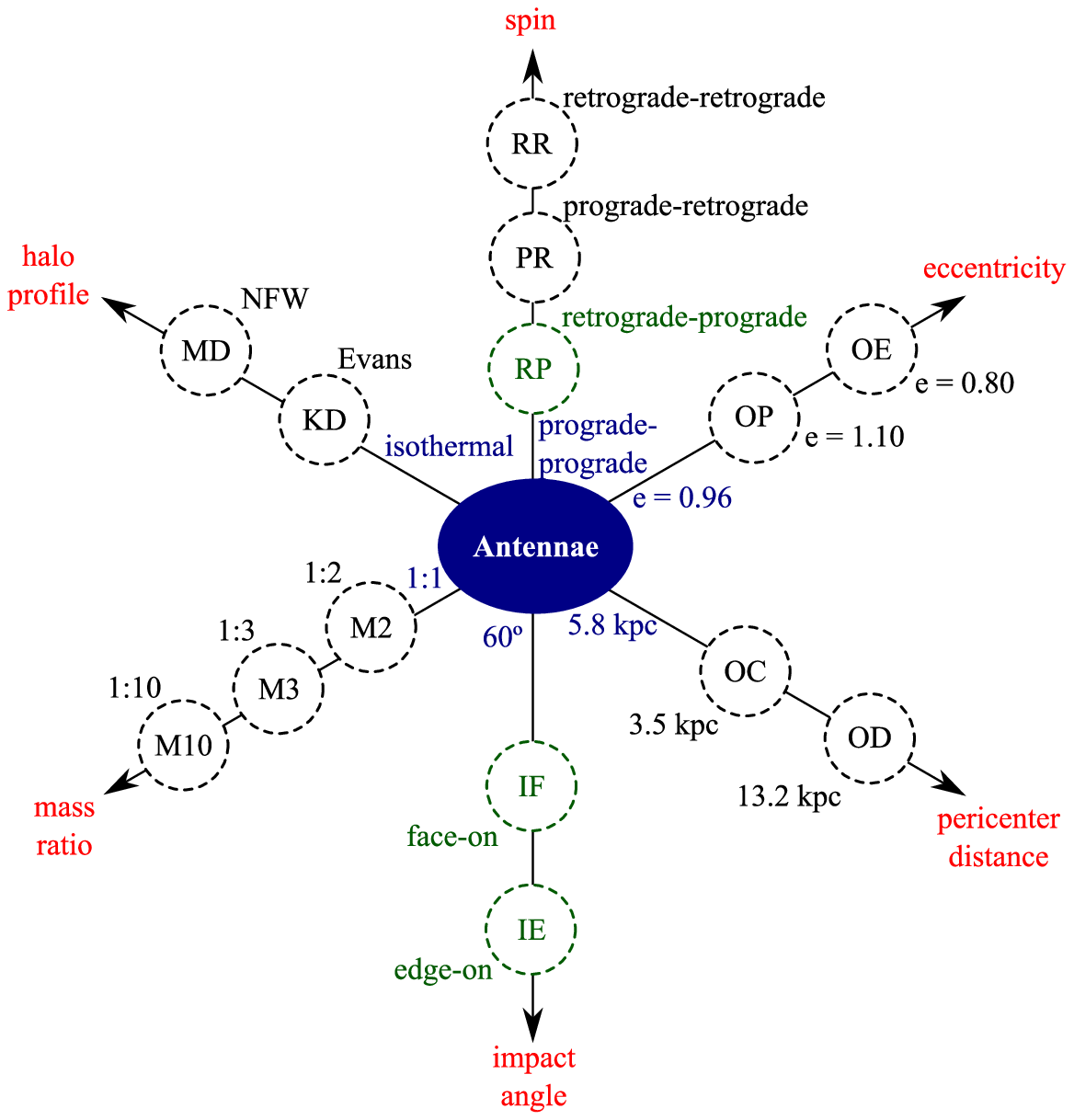}{Map of the parameter survey}
{Schematic map of the original parameter survey presented in \mycitet{Renaud2009}. The models in green have been added in the present document. Along each axis, a circles represents a model with a given value for the parameter considered.}

\tab{14cm}{parameter_survey}{Parameter survey}
{l@{\extracolsep{\fill}}c@{\extracolsep{\fill}}c@{\extracolsep{\fill}}c@{\extracolsep{\fill}}c}
{
Model & Key parameter & Peak$^\star$ 1 [\%] & Peak$^\dag$ 2 [\%] & ${\tau_1}^\ddag$ [Myr] \\
\hline
\multicolumn{2}{l}{Antennae (see \refcha{antennae})} & 11.5 & 11.5 & 6.7 \\
\hline
\multicolumn{5}{c}{Spin (see \refsec[survey]{survey_spin})}\\
\acr{notarg}{RP}{RP} & retrograde-prograde & 7.5 & 14.0 & 7.3 \\
\acr{notarg}{PR}{PR} & prograde-retrograde & 7.5 & 14.0 & 7.3 \\
\acr{notarg}{RR}{RR} & retrograde-retrograde & 7.0 & 18.5 & 6.7 \\
\hline
\multicolumn{5}{c}{Orbits (see \refsec[survey]{survey_orbits})}\\
\acr{notarg}{OP}{OP} & high eccentricity & 10.0 & 9.5 & 7.4 \\
\acr{notarg}{OE}{OE} & low eccentricity & 14.5 & 8.5 & 6.5 \\
\acr{notarg}{OD}{OD} & distant encounter & 3.5 & 9.5 & 7.3 \\
\acr{notarg}{OC}{OC} & close encounter & 15.5 & - & 5.5 \\
\hline
\multicolumn{5}{c}{Impact angle (see \refsec[survey]{survey_impact})}\\
\acr{notarg}{IF}{IF} & face-on collision & 39.5 & 12.5 & - \\
\acr{notarg}{IE}{IE} & edge-on collision & 21.0 & 8.5 & 7.0 \\
\hline
\multicolumn{5}{c}{Mass ratio (see \refsec[survey]{survey_mratio})}\\
\acr{notarg}{M2}{M2} & 2:1 mass ratio & 8.5 & 5.5 & 6.6 \\
\acr{notarg}{M3}{M3} & 3:1 mass ratio & 5.5 & 6.5 & 6.3 \\
\acr{notarg}{M10}{M10} & 10:1 mass ratio & 1.5 & 3.5 & 5.5 \\
\hline
\multicolumn{5}{c}{Progenitors model (see \refsec[survey]{survey_models})}\\
\acr{notarg}{KD}{KD} & \mycitet{Kuijken1995} & 4.5 & 1.5 & 3.3 \\
\acr{notarg}{MD}{MD} & \mycitet{McMillan2007} & 10.0 & 9.0 & 2.6 \\
}
{\note Based on Table~3 of \mycitet{Renaud2009}. \\ $^\star$ Mass fraction in compressive mode added by the first pericenter passage (compared to the isolation stage).\\ $^\dag$ Mass fraction in compressive mode added by the second major peak (merger phase).\\ $^\ddag$ Exponential timescale of the \emph{\acr{notarg}{lus}{longest uninterrupted sequence}} distribution for the short time interval $\Delta T_1$. Note that this is not the same definition than in \myciteauthor{Renaud2009} (\myciteyear{Renaud2009}; recall \refeqnt{doubleexp} of the present document.}

%%%%%%%%%%%%%%%%%%%%%%%%%%%%%%%%%%%%%%%%%%%%%%%%%%
\sect[survey_spin]{Spin}

As presented in \refsecp{prograderetrograde}, the spin-orbit coupling of interacting galaxies is of paramount importance for their dynamical evolution, in term of timescale and creation of tails. For a galaxy B orbiting a galaxy A, the transition from prograde to retrograde goes through all the inclination angles $i$ of the disk of B with respect to the orbital plane around A: for $i=0^\circ$, one has a prograde encounter, while $i=180^\circ$ yields a retrograde system. In the following, we extend this definition to intermediate angles ($i \in [0^\circ,90^\circ]$ and $i \in [90^\circ,180^\circ]$) so that a prograde encounter corresponds to an important perturbation of the disk by the tidal interaction, while a retrograde configuration only drives mild, transient perturbations.

\mycitet{Toomre1972} presented a study on the influence of the spin-orbit coupling and concluded that it is a key factor in the creation of tidal tails. In this Section, we propose a similar work, but for self-consistent disks embedded in \acr{}{DM}{dark matter} halos. Three configurations are considered: in the \acr{}{PP}{PP} (prograde-prograde) model, both galaxies have prograde encounters. This is the Antennae model presented in \refcha{antennae}. By flipping one of the disks, one gets the \acr{}{PR}{PR} (prograde-retrograde) or the \acr{}{RP}{RP} (retrograde-prograde) model. Note that the choice of flipping one disk or the other is not important, because of the high level of symmetry in the simulation. Finally, when the spins of both disks have been inverted, one obtains the \acr{}{RR}{RR} (retrograde-retrograde) model. The morphologies of the four models are shown in \reffig{ppprrprr_morphology}.

\fig{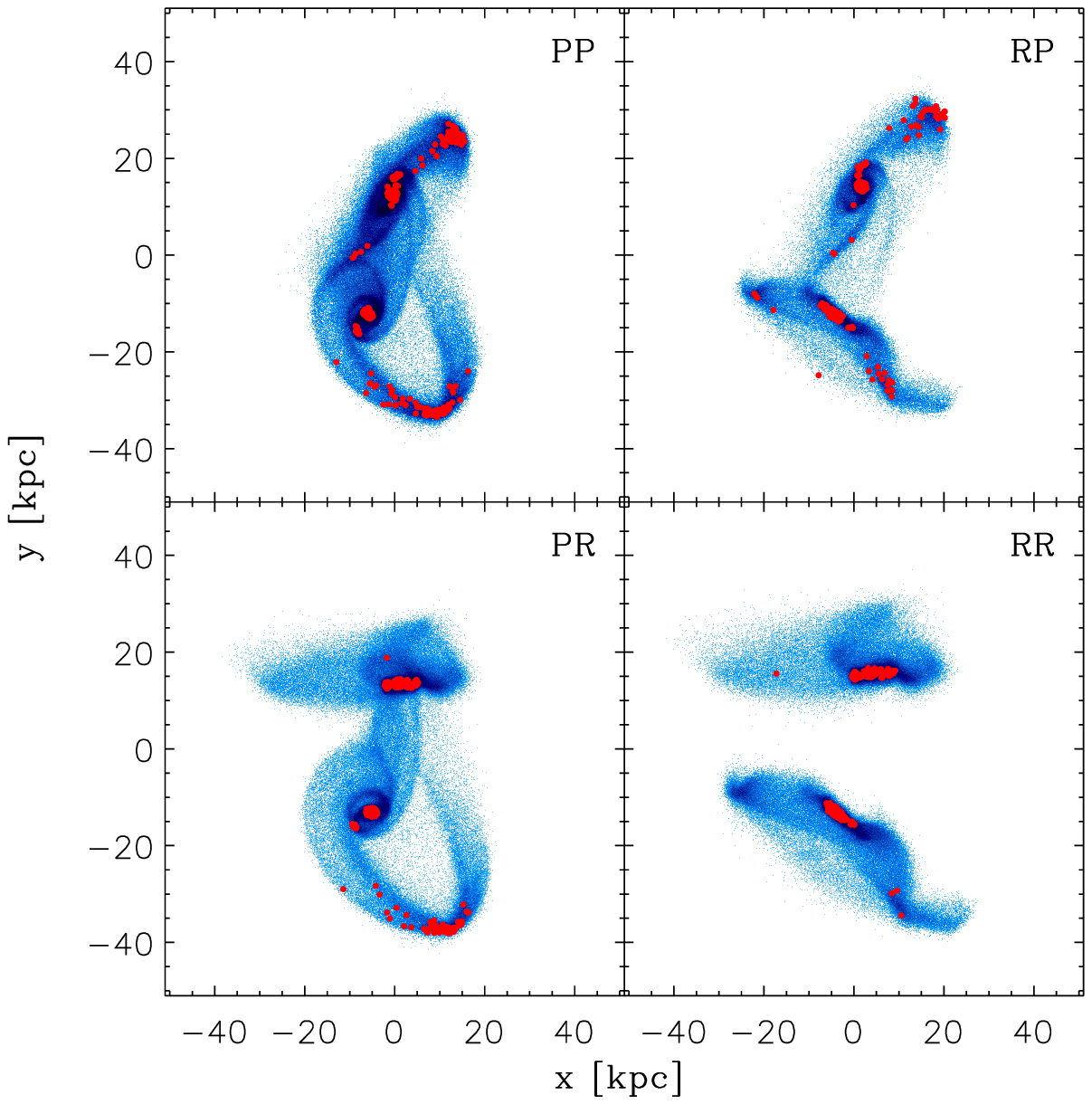}{Morphology of the PP, RP, PR and RR models}
{Morphology of the \acr{notarg}{PP}{PP}, \acr{notarg}{RP}{RP}, \acr{notarg}{PR}{PR} and \acr{notarg}{RR}{RR} models at approximately the same dynamical step (maximum separation of the progenitors). The link between the prograde configuration and the creation of tidal tails is clearly visible. Here again, the red dots represent particles in compressive mode.}

%%%%%%%%%%%
\subsect{Mass-loss}
The long tails existing in the Antennae galaxies are a strong hint that the encounter which created them was prograde for both disks. In this \acr{}{PP}{PP} model, each disk ejects $\sim 30\%$ of its mass in the form of tails, as already noted by \mycitet{Barnes1988a}. Note that this value depends on the geometry of the merger and varies with, e.g. the pericenter distance. However, if one only changes the disk inclination (by steps of $180^\circ$), the mass-loss remains at the level of $\sim 30\%$ (prograde) or $\sim 0$ (retrograde).

A significant fraction of these ejecta will progressively fall back onto the nuclei, which helps the dissolution of the tails, long after the merger stage ($\sim 1 \Gyr$). This means that the tidal tails are not completely unbound to the central part of the progenitor galaxies. However, during the long range interaction period (i.e. between the first passage and the merger), the dynamical mass of the galaxies are strongly affected by the creation of the arms, which directly influences their trajectories.

\fig{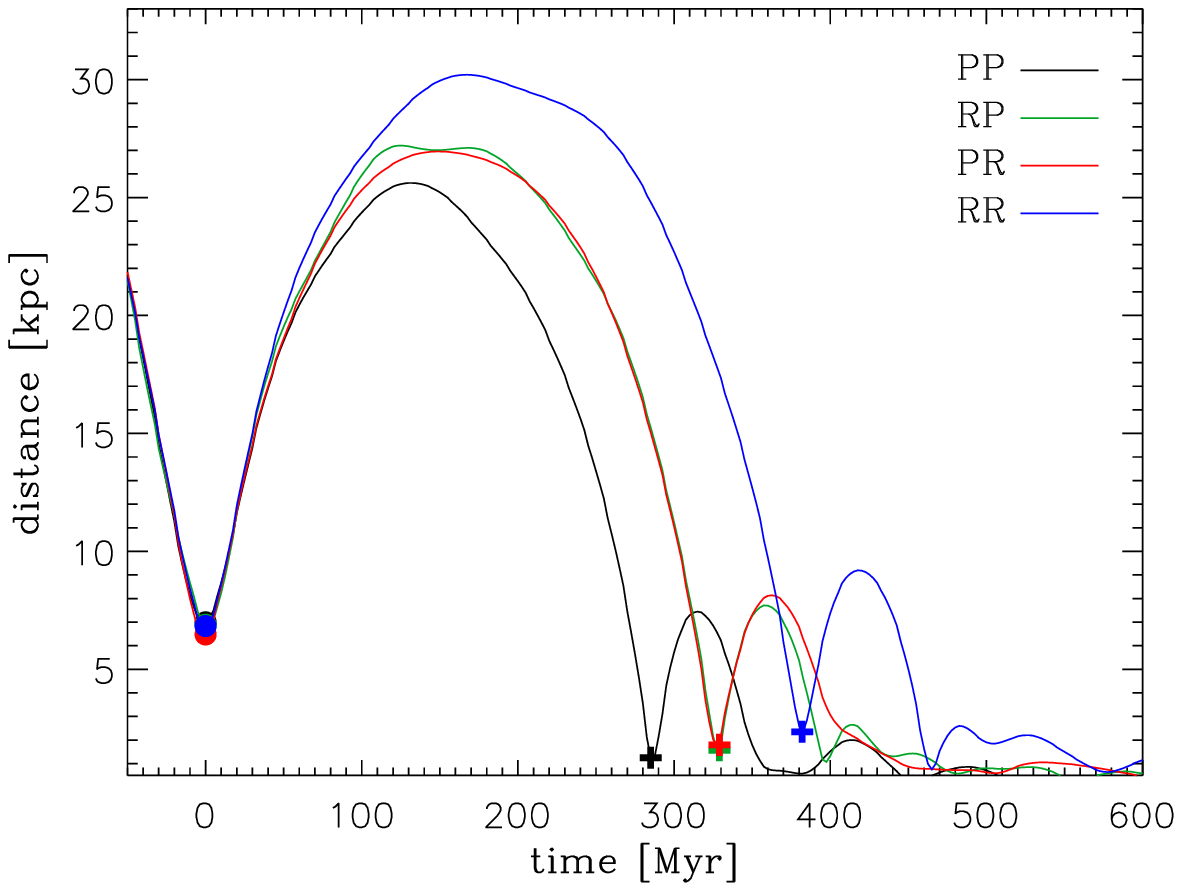}{Evolution of the distance in the PP, RP, PR and RR models}
{Evolution of the distance between the two galaxies, for the \acr{notarg}{PP}{PP} (black, as in \reffigt{antennae_dist}), the \acr{notarg}{RP}{RP} (green), the \acr{notarg}{PR}{PR} (red) and the \acr{notarg}{RR}{RR} (blue) models. \acr{notarg}{RP}{RP} and \acr{notarg}{PR}{PR} are very similar, due to the high level of symmetry of the progenitors. The differences in dynamical masses clearly affect the orbital period and separation. The two pericenter passages are marked with $\bullet$ and $+$ respectively.}

\reffig{ppprrprr_distance} shows the evolution of the distance between the two galaxies, for each model. First of all, no major difference is visible before the first pericenter passage ($t < 0$). Afterwards, the \acr{}{PR}{PR} and \acr{}{RP}{RP} models yield very comparable behaviors. For this reason, we will only consider the \acr{}{PR}{PR} model in the following. Because of different amounts of mass-loss, the time of the second pericenter is shifted by $\sim 50 \Myr$ per tail created, i.e. it occurs at $\sim 280 \Myr$ for \acr{}{PP}{PP} (two tails), $\sim 330 \Myr$ for \acr{}{PR}{PR} (one tail) and $\sim 380 \Myr$ for \acr{}{RR}{RR} (no tail). Keeping these dynamical data in mind, it is possible to make a link with the tidal field one expects to find in such systems.

%%%%%%%%%%%
\subsect{Compressive mode}

In \refcha{antennae}, we have seen that the major events in the evolution of the compressive tidal mode match the pericenter passages. Therefore, with the dynamics presented in the previous Section, one expects the mass fraction in compressive mode in the \acr{}{PR}{PR} and \acr{}{RR}{RR} models to be peaked at the time of their respective passages. Indeed, as shown in \reffig{spin_histo}, the previously mentioned shift of $50 \Myr$ is visible between the three models.

\fig{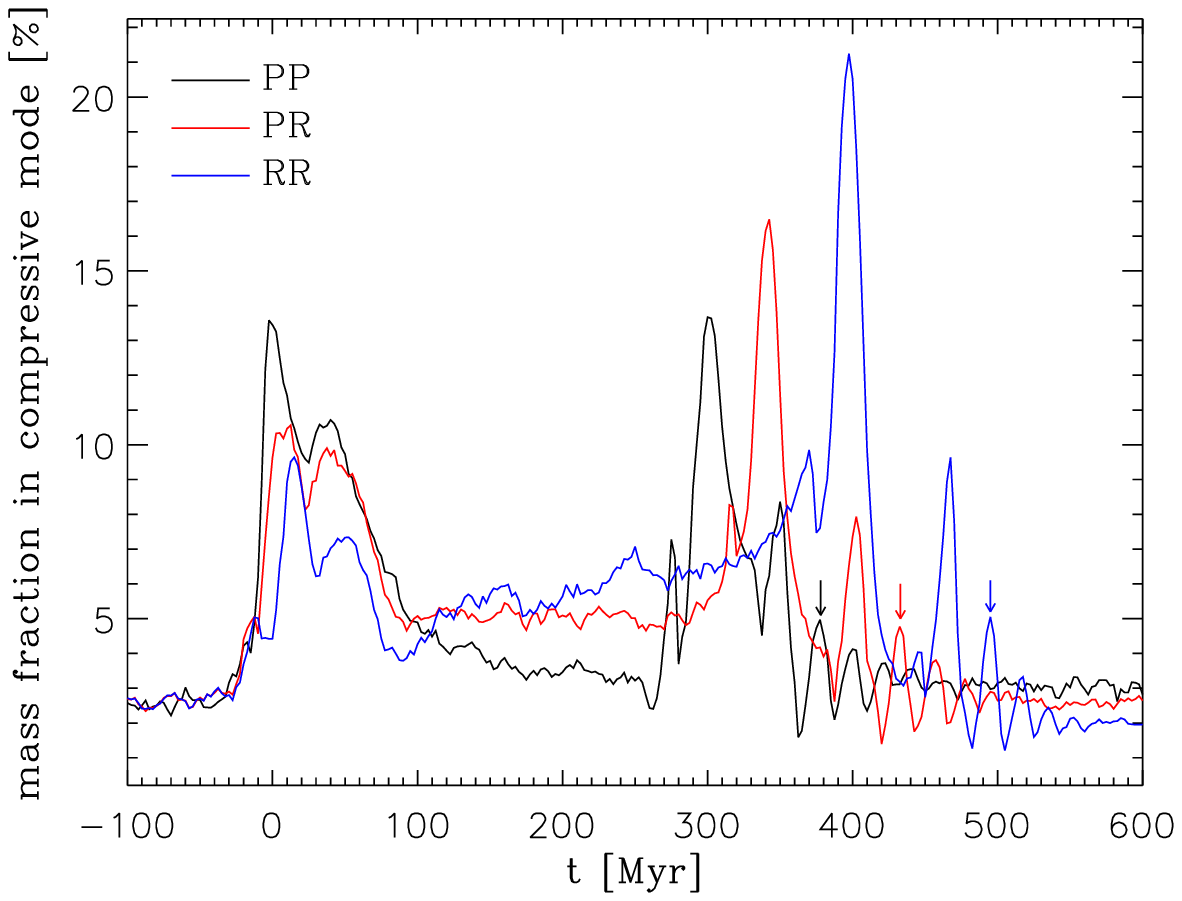}{Mass fraction in compressive mode for the PP, PR and RR models}
{Evolution of the mass fraction in compressive mode for the \acr{notarg}{PP}{PP}, \acr{notarg}{PR}{PR} and \acr{}{RR}{RR} models. As before, $t=0$ corresponds to the first pericenter passage. A clear shift in time is visible between the models and can be simply explained by dynamical considerations. (Adaptation of Figure~10 of \mycitealt{Renaud2009}.)}

The main differences in this Figure correspond to the second pericenter passage, when, in addition to being shifted in time, the peaks have different amplitudes. Again this can be explained by a dynamical analysis. First, consider that $30\%$ of the stellar mass of \acr{}{PP}{PP} have been ejected to the tails during the first passage and thus, will not be affected when the nuclei interact again. Let $M_0$ be the total initial stellar mass, and $M_1$, the mass that remains in the nuclei. In the \acr{}{PP}{PP} case, one has $M_1 = (1 - 0.3) M_0 = 0.7 M_0$. By reading the amplitude of the second peak, we see that $\sim 15\%$ of $M_0$, or $(15\% / 0.7) M_1 \approx 21\%$ of $M_1$ is in compressive mode. This fraction of the ``bound'' mass should be the same for the three models. For \acr{}{PR}{PR}, only one tail is created, so $M_1 = (1-0.3/2) M_0 = 0.85 M_0$. One expects $21\%$ of $M_1$ $= 21\% \times 0.85 M_0 \approx 17\%$ of $M_0$ to be in compressive mode at the second passage. This corresponds well to the value in \reffig{spin_histo}. The same way, no tail is created in \acr{}{RR}{RR}, so $M_1 \approx M_0$ and the expected fraction is $21\%$ of $M_0$, which also matches the value measured.

In addition, differences between the three models also exist between the two passages, corresponding to a re-arrangement of the matter within each galaxy according to proper internal dynamics. Elsewhere all models yield comparable behaviors, showing a quiet stage (intrinsic $3\%$), an oscillatory phase for the merger and finally a stable fraction with variations similar to the initial noise. All together, these points show that the major events in the compressive history of the mergers (peaks) are dominated by dynamical properties (mass and orbital period).

Remains the analysis of the timescale of the compressive events. Do the spins of the galaxies influence the time a cluster-size element spends in compressive mode? We have seen that the \emph{\acr{notarg}{lus}{longest uninterrupted sequence}} and \emph{\acr{notarg}{tt}{total time}} distributions depend on the time of integration, i.e. the time given to the particles to enter or exit the compressive regions (recall \refsect[antennae]{lus_tt}). In the case of the Antennae model, this time has been arbitrarily defined and an \emph{order of magnitude} for the timescale ($\sim 10^7 \yr$) has been obtained. For the comparison with the \acr{}{PR}{PR} and \acr{}{RR}{RR} configurations however, one requires a greater precision to highlight the influence of the spin on the timescale (which is expected to change slightly but remain at the same order of magnitude). To do so, the time of integration should be strictly the same for all three models. But as noted above, the dynamics of the systems strongly differs and thus, the times granted to the particles of the three models to enter and exit compressive zones are different.

\fig{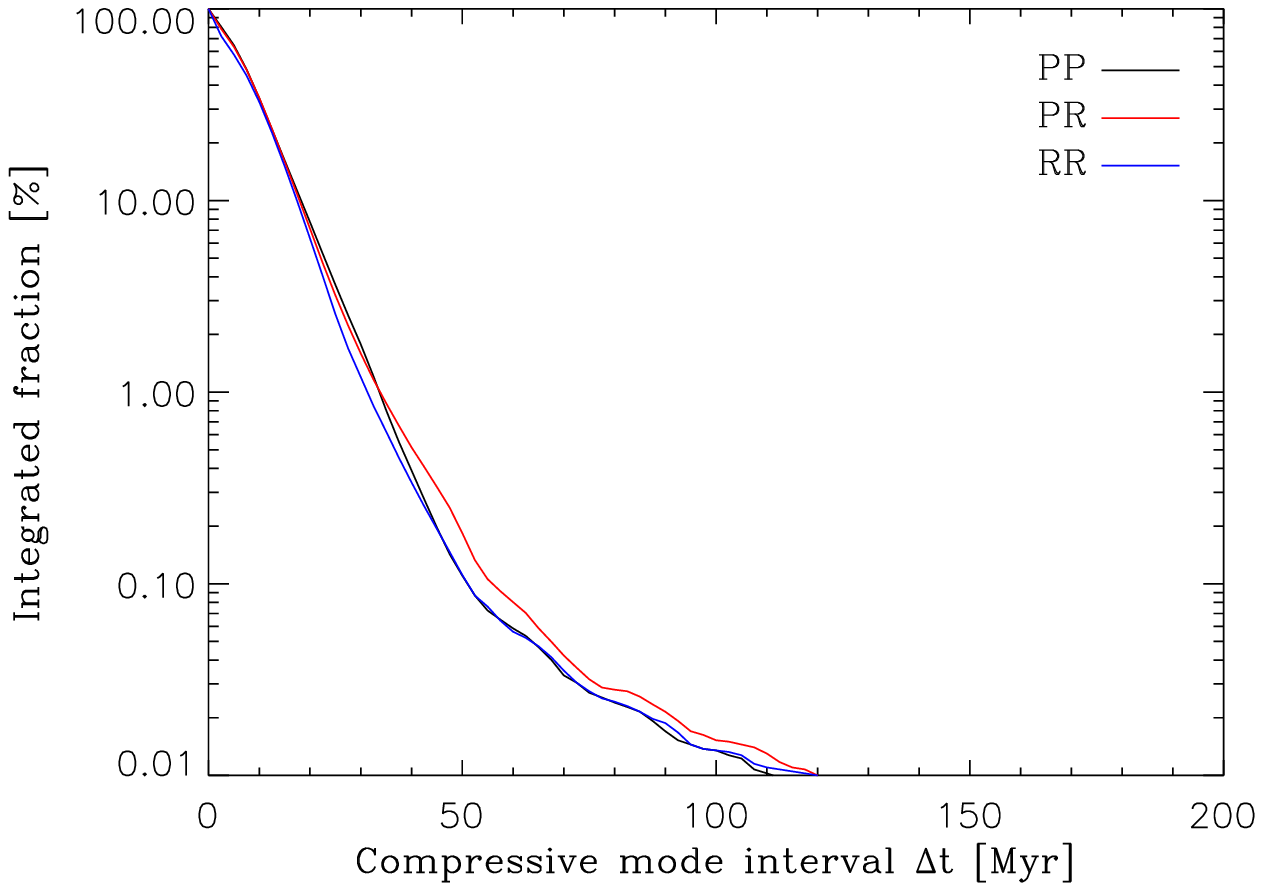}{\emph{lus} distributions for the PP, PR and RR models}
{\emph{\acr{notarg}{lus}{longest uninterrupted sequence}} distributions of the \acr{notarg}{PP}{PP}, \acr{notarg}{PR}{PR} and \acr{notarg}{RR}{RR} models, integrated from $t = -100 \Myr$ (isolated stage) to their respective merger phase. No clear trend is visible, which demonstrates that the orientation of the spin of the progenitors does not influence the time spent in compressive mode.}

To balance this, we stop the integration at the time of the third peak in the merger phase (see the arrows on \reffig{spin_histo}). This gives a good equivalent stage in the dynamical evolution of the merger. Note that it would have been preferable to stop later (e.g. at $1 \Gyr$, as in \refcha{antennae}) but it is not possible to precisely extract a dynamical stage from either the morphology (because the progenitors have merged) or the mass fraction in compressive mode. In this context, a comparison of the \emph{\acr{notarg}{tt}{total time}} distributions is not relevant (because of its sensitivity to the time of integration), and one should focus on \emph{\acr{notarg}{lus}{longest uninterrupted sequence}} only. \reffig{spin_period} shows that the differences between the three models represent less than $1\%$, and that all the properties of the distributions (double-exponential law, knee at $\sim 50 \Myr$) are conserved when changing the spin. We note that the characteristic times $\tau_1$ and $\tau_2$ remain the same for all three models (see \reftab{parameter_survey}). From \reffigp{antennae_period}, we know that the tidal tails yield similar timescales for the \emph{\acr{notarg}{lus}{longest uninterrupted sequence}} distribution than the nuclei do (over $\Delta T_1$). Therefore, the presence of zero, one, or two tail(s) does not have a major influence on the \emph{\acr{notarg}{lus}{longest uninterrupted sequence}} distribution, as visible when changing the spin.

%%%%%%%%%%%%%%%%%%%%%%%%%%%%%%%%%%%%%%%%%%%%%%%%%%
\sect[survey_orbits]{Orbits}

In observed pairs of interacting galaxies, both the orbital period and the pericenter distance take heterogeneous values, because of (i) a higher density of galaxies at high redshift and (ii) few constrains on the geometry of the encounters. The exploration of the role of different orbital configurations of interacting pairs is thus important in our study of the tidal field. For the next part of our survey, we focus on the orbit of the progenitors, keeping all the others parameters (e.g. mass, spin, initial shape) as in the Antennae model. Therefore, the only changes made here involve the eccentricity $e$ and the pericenter distance $d$.

Because of dynamical friction, the centers of mass of the galaxies do not have a Keplerian motion and thus, it is difficult to describe their orbits accurately. For the sake of simplicity, the orbits investigated here are defined thanks to a Keplerian equivalent: from the isolated stage, we measure the eccentricity and the pericenter distance that the galaxies would have if replaced by point-masses, i.e. without orbital decay. Note that these quantities do not correctly fit the real values after the first passage (see \refsect[basic_concepts]{dynamical_friction}) and are only used as an informational data to ease the comparisons between our models. \reftab{orbits} summarizes the properties of the five models presented here. (In total, 16 orbital configurations have been considered in this survey, but only five are presented here, for the sake of clarity.)

\tab{10cm}{orbits}{Orbital survey}
{l@{\extracolsep{\fill}}c@{\extracolsep{\fill}}c@{\extracolsep{\fill}}c}
{
Model & $e$ & $d$ [kpc] & remarks \\
\hline
Antennae & 0.96 & 5.8 & reference model \\
\acr{notarg}{OP}{OP}$^\star$ & 1.10 & 5.8 & almost parabolic \\
\acr{notarg}{OE}{OE} & 0.80 & 5.8 & low eccentricity \\
\acr{notarg}{OD}{OD} & 0.96 & 13.2 & distant encounter \\
\acr{notarg}{OC}{OC}$^\dag$ & 1.50 & 3.5 & close encounter \\
}
{\note Table~4 of \mycitet{Renaud2009}.\\ $^\star$ Orbital decay makes this orbit bound.\\ $^\dag$ High eccentricity is required to avoid a central collision that would hide the effect of the initially non-zero pericenter distance.}

%%%%%%%%%%%
\subsect{Eccentricity and distance}

\fig{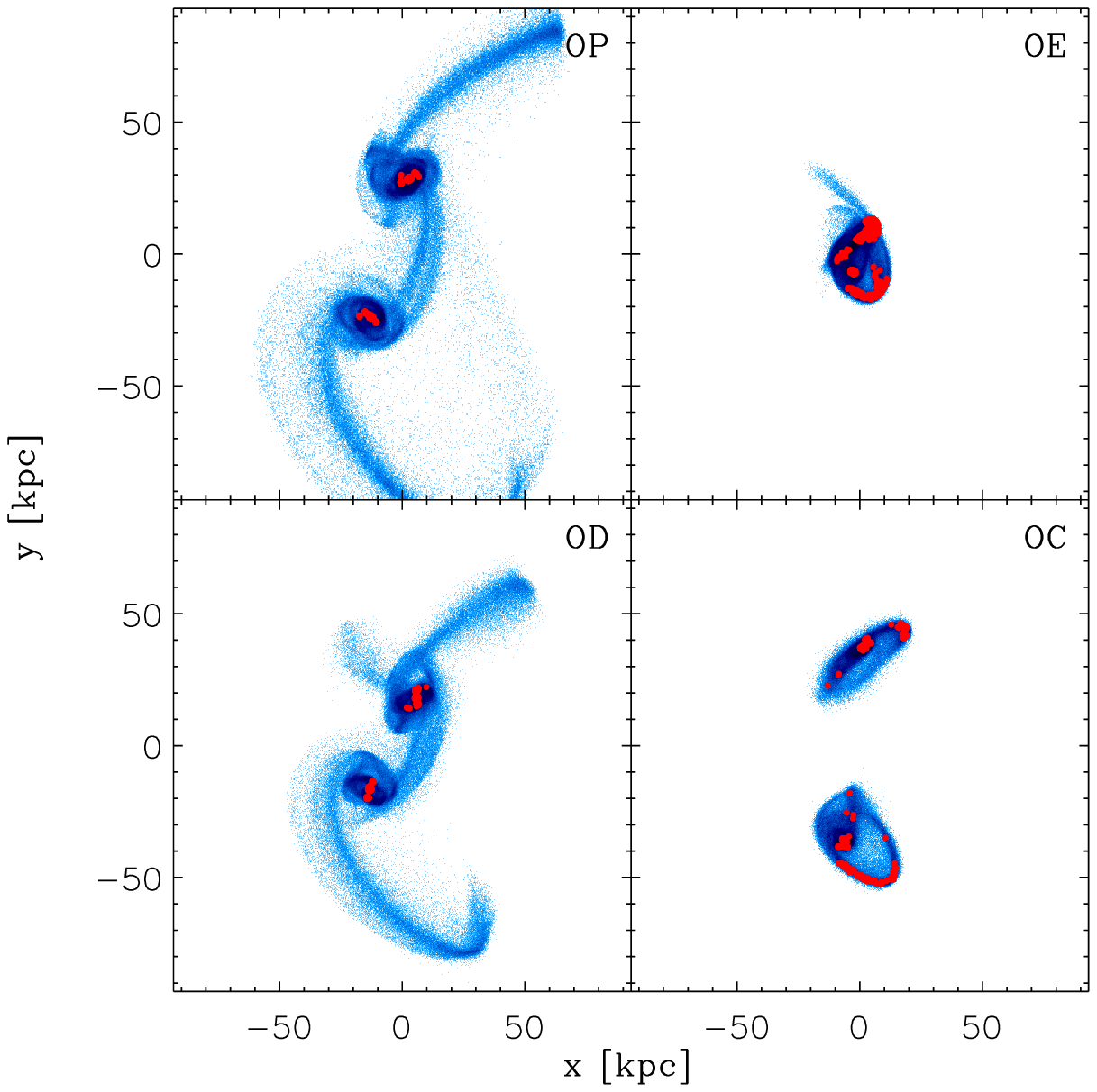}{Morphology of the OP, OE, OD and OC models}{Morphology of galaxies from our orbital survey, at approximately the same dynamical step (maximum separation of the progenitors). Note that the galaxies are not bound in the \acr{notarg}{OC}{OC} model. By contrast, \acr{notarg}{OE}{OE} yields a very short separation (see also \reffig{survey_orbits_distance}).}

\reffig{survey_orbits_morphology} shows the morphology of the systems and the locations of their compressive modes when the galaxies have reached their maximum separation. In parallel, \reffig{survey_orbits_distance} displays the evolution of the distance between the two galaxies, and the evolution of the mass fraction in compressive mode is shown in \reffig{survey_orbits_histo}. The \acr{}{OP}{OP} model develops very long tails as well as an important bridge. The compressive regions that exist in these tidal structures short after the first pericenter passage have already disappeared in the snapshot shown in \reffig{survey_orbits_morphology} ($t \simeq 450 \Myr$). The \acr{}{OE}{OE} run shows a very different configuration. Because of the low eccentricity of the orbit, a large fraction of the orbital angular momentum of the progenitors is transfered to individual particles during the first passage (which dynamically heats the galaxies). As a consequence, the merger occurs just after the first encounter, forbidding a large separation of the galaxies. Thus, the transient tidal features (mainly in the tails and bridges) still exist when the progenitors start to merge ($t\sim 150 \Myr$).

\fig{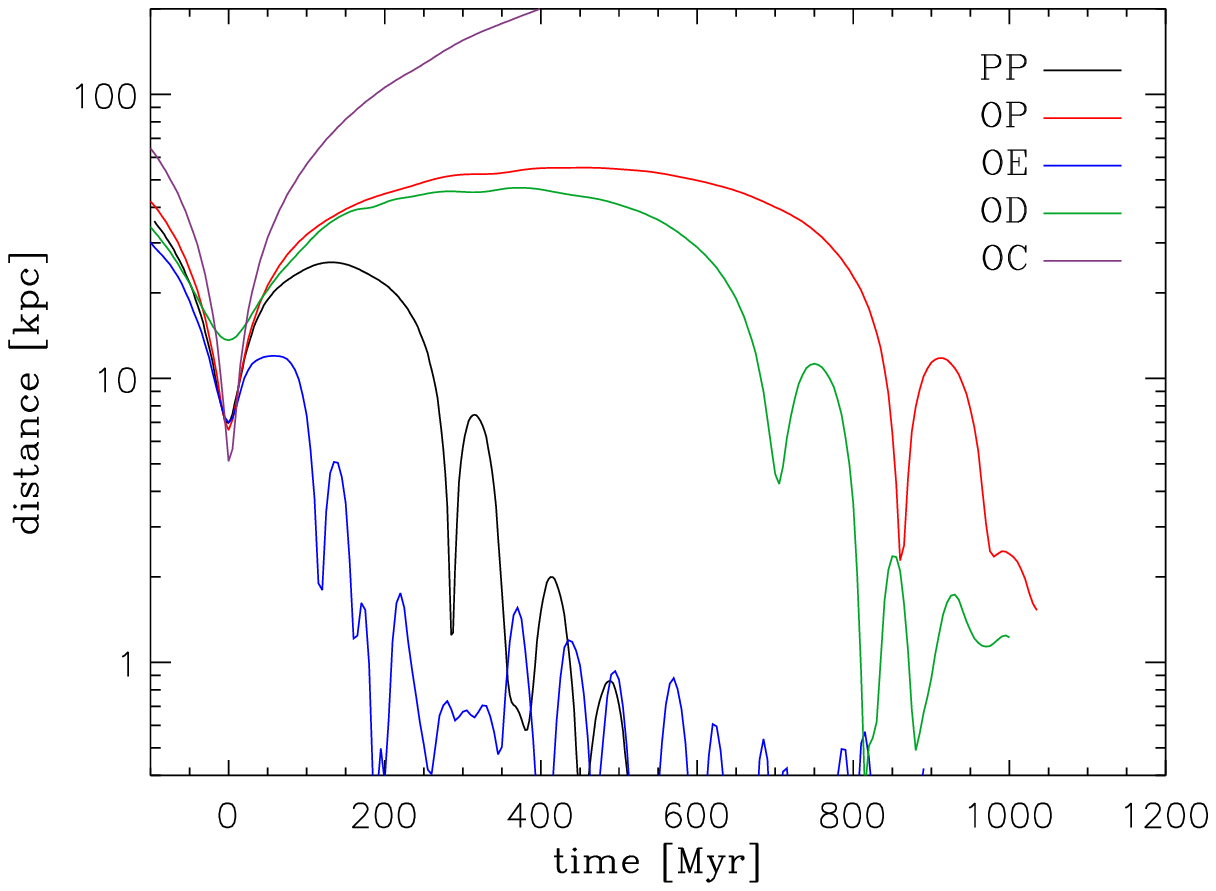}{Evolution of the distance in the OP, OE, OD and OC models}
{Evolution of the distance between the two galaxies for the models of our orbital survey. Note that the vertical-axis is in logarithmic scale. The merger phase (for all models but \acr{notarg}{OC}{OC}) is visible as damped oscillations.}

As expected, the peaks of compressive modes match the pericenter passages, and the merger phase is also clearly marked. As for the Antennae, the mass fraction in compressive mode returns to a quiescent level ($\sim 3\%$) between the interaction events and remains stable for an arbitrarily long time after the merger. The Antennae and the \acr{}{OP}{OP} models share a similar amount of mass in compressive mode at the first passage ($t\approx 0$) while \acr{}{OE}{OE} yields a significantly higher value. By contrast with the spin survey (see the previous Section), the change already occurs at the \emph{first} pericenter, i.e. at a stage when the disks are merely unchanged from their initial state. This demonstrates that the variation is not due to tidal structures, because they have not been created yet. In fact, in addition to the eccentricity, the relative velocity of the progenitors has a much lower value for \acr{}{OE}{OE} than in the Antennae and \acr{}{OP}{OP} cases, suggesting that large compressive regions have time to develop all over the disks during the first passage. To conclude, the eccentricity is of paramount importance to set the characteristics of the orbits (period, friction) but does not have a strong direct impact on the tidal field itself.

\fig{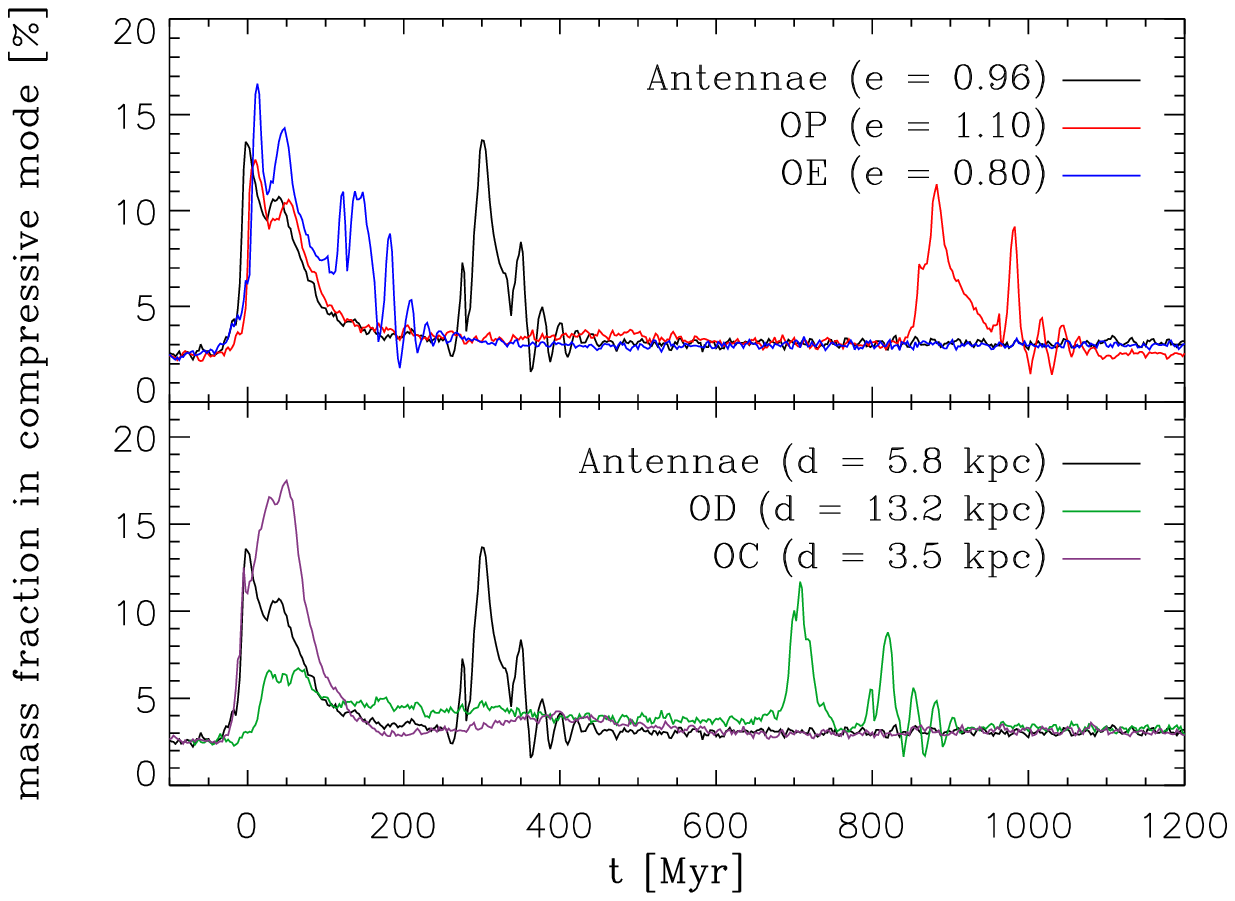}{Mass fraction in compressive mode for our orbital survey}
{\emph{Top:} Effect of the eccentricity on the mass fraction in compressive mode. \emph{Bottom:} Same but for the pericenter distance. (Adaptation of Figure~12 of \mycitealt{Renaud2009}.)}

\reffig{survey_orbits_histo} (bottom panel) clearly exhibits that the distant encounter (\acr{}{OD}{OD}) provokes a smaller amount of matter to enter in compressive mode than the close encounter (\acr{}{OC}{OC}). Furthermore, the second peak for \acr{}{OD}{OD} ($t\simeq 700 \Myr$) corresponds to a second passage, much closer ($\sim 5 \kpc$) than the first one ($\sim 13 \kpc$ at $t = 0$), and induces a higher mass fraction in compressive mode. This suggests that a close passage yields larger compressive regions than a more distant one.

%%%%%%%%%%%
\subsect[flybys_mergers]{Flybys and mergers}

To investigate the last point further, we have conducted a secondary orbital survey, varying the pericenter distance only. \reffig{survey_orbits_flybysmergers} presents several measurements of the amplitude of the first peak as a function of the distance, covering very close passages and more remote encounters. In the later case, the progenitors undergo a mild interaction and easily escape from their mutual gravitational attraction. Note that such a flyby configuration can be obtained with a high velocity, and yet close encounter. However, this survey does not include high-velocity encounters, which limits its scope to slow ($\sim 200 \kms$) field-like galaxies (as opposed to clusters of galaxies having typical velocity dispersions of $\sim 1000 \kms$).

\fig{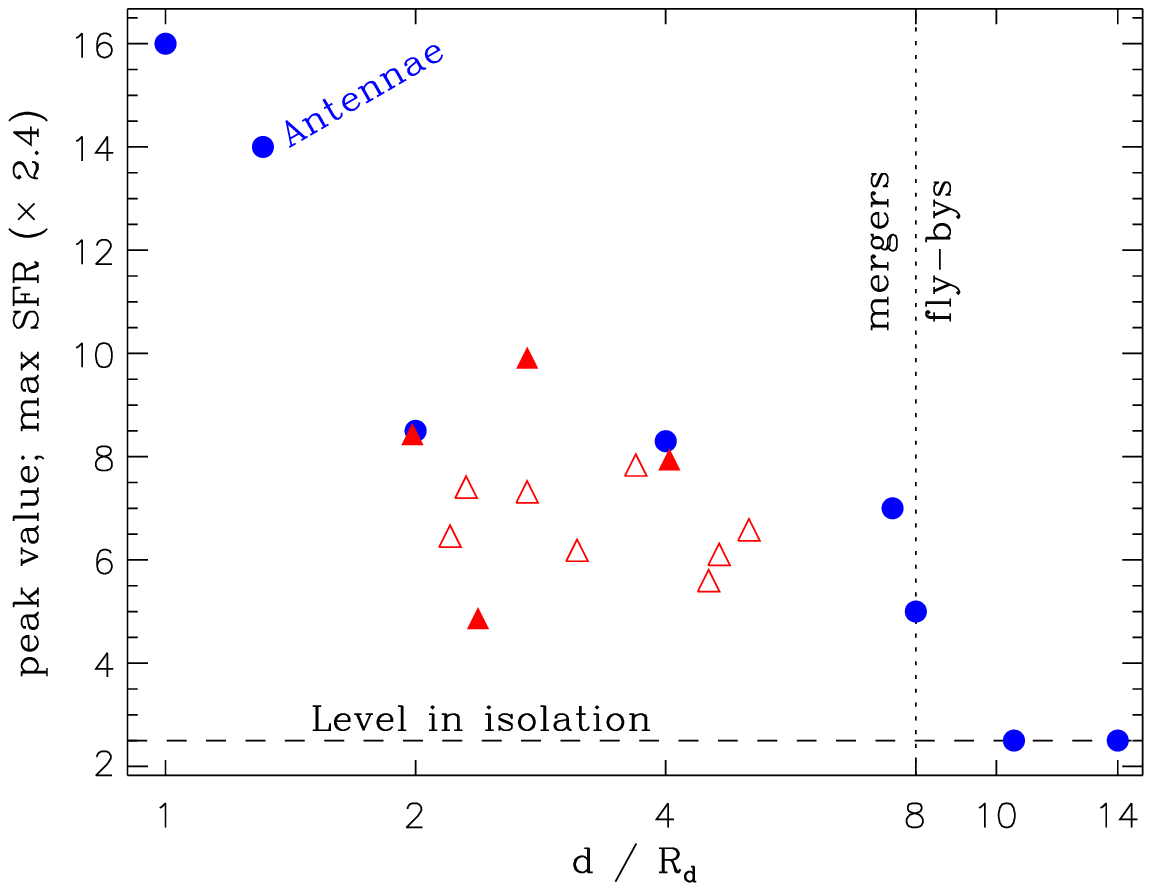}{Compressive modes vs. pericenter distance}
{Mass fraction in compressive mode at the first pericenter passage versus the impact parameter $d$, normalized to the disk scalelength $R_\mathrm{d}$ (blue dots). For these points, the vertical line marks the limit between flyby and merger galaxies. The red symbols are measurements of maximum \acr[s]{notarg}{SFR}{star formation rates} by \myciteauthor{diMatteo2007} (\myciteyear{diMatteo2007}, their Figure~21). Filled triangles represent the mergers while the open ones denote the flybys. Both can exist for all $d/R_\mathrm{d}$ because of different relative velocities. (Adaptation of Figure~13 of \mycitealt{Renaud2009}, with data points from \mycitealt{diMatteo2007}.)}

In the case of non-merging galaxies (right-hand side of \reffig{survey_orbits_flybysmergers}), the mass fraction in compressive mode remains at the same level than in isolation. That is, the interaction is too weak to induce compressive modes (other than intrinsic). When the distance decreases however, the fraction increases slowly before peaking sharply at $d/R_\mathrm{d} \sim 1$, i.e. for distances comparable to the scalelength of the progenitors' disks.

When linking the compressive modes to star formation, as in \refcha{formation}, it is interesting to compare this trend with the results from simulations containing gas. \mycitet{diMatteo2007} performed a large survey of interacting galaxies with the aim to study the \acr{}{SFR}{star formation rate}. The red triangles on \reffig{survey_orbits_flybysmergers} represent the maximum value of the \acr{}{SFR}{star formation rate} they measured, for direct (prograde-prograde) encounters of spiral galaxies (that they called ``L-L dir'') only. The distance is normalized to their disk scalelength ($R_\mathrm{d}$ is taken equal to the parameter $a_\star$ of their Miyamoto-Nagai models).

For flybys (open triangles), the \acr{}{SFR}{star formation rate} tends to slightly decrease when the distance increases. Although in the case of mergers, these authors noted that the \acr{}{SFR}{star formation rate} increases with the distance. Over the range $d/R_\mathrm{d} \sim 2-4$ however, the \acr{}{SFR}{star formation rate} seems to be fairly independent of the distance, at least for spiral-spiral mergers (filled triangles). It is important to note that the quantity considered here is the \emph{maximum} \acr{}{SFR}{star formation rate}, i.e. non necessarily the value at the first passage. Therefore, the value of the \acr{}{SFR}{star formation rate} from \mycitet{diMatteo2007} can correspond to either a burst in star formation, or a slow increase due to, say, gas infall. Unfortunately, this does not allow to link star forming episodes with tidal events, as we do. 

With surveys like the one of di Matteo, making comparisons between the time evolution of the \acr{}{SFR}{star formation rate} and the one of the compressive mode would certainly bring new lights on the exact role of the tidal field, with respect to gas dynamics like stripping or infall. The first steps of such a study have begun with our model of the Antennae and \acr{}{SPH}{smoothed particle hydrodynamics} and \acr{}{AMR}{adaptive mesh refinement} runs (see \refsect[formation]{hydro}).

%%%%%%%%%%%%%%%%%%%%%%%%%%%%%%%%%%%%%%%%%%%%%%%%%%
\sect[survey_impact]{Impact angle}

From the previous study of the influence of the spin and the pericenter distance, one knows that the importance of the compressive mode increases with (i) a high mass involved in the collision and (ii) a short range interaction. To go further on these points, two new configurations are presented in this Section\footnote{These new models are not part of the survey published in \mycitet{Renaud2009}.} (see \reffig{face_edge}). The first one (\acr{}{IF}{IF}) is the face-on collision of two spiral galaxies, with disks perpendicular to the orbital plane, as in \refsecp[numerical]{testsgyrfalcon}. (Note that the progenitors are not raw exponential disks anymore but the usual models used in the Antennae reference.) In the second one (\acr{}{IE}{IE}), the disks are aligned with the orbital plane, so that they first collide through their edges. These two models should not be seen as statistically relevant for real configurations, but rather as limits for the relative inclination of the disks.

\fig{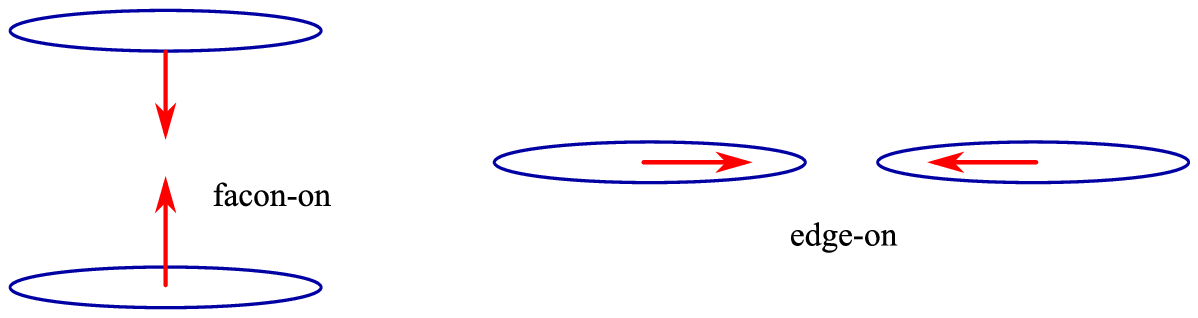}{Configuration of the IF and IE models}
{Schematic view of the configuration of the face-on (\acr{notarg}{IF}{IF}) and edge-on (\acr{notarg}{IE}{IE}) collisions. Red arrows indicates velocity vectors.}

In both cases, the pericenter distance is null and the disks overlap almost entirely at this time. That is, $\sim 100\%$ of the stellar mass sits in the configuration of a central collision, which is expected to rise large compressive regions (recall \reffig{survey_orbits_flybysmergers}). \reffig{survey_impact_histo} shows the evolution of the mass fraction in compressive mode of the two models, compared to the Antennae. Interestingly, a small, minor peak is visible at $t\approx 0$, i.e. when the two nuclei overlap. Hence, the time of the first pericenter passage does not \emph{exactly} correspond to the major peak in the mass fraction. However, the rapid increase occurs $\sim 10 \Myr$ after, when the nuclei are already separated by $\sim 4 \kpc$. This matches the creation of very large compressive regions in the tidal tails created shortly (but not immediately) after the passage. As before, these compressive structures disappear within a few $10^7 \yr$, and new ones are created during the second passage and the merger phase.

\fig{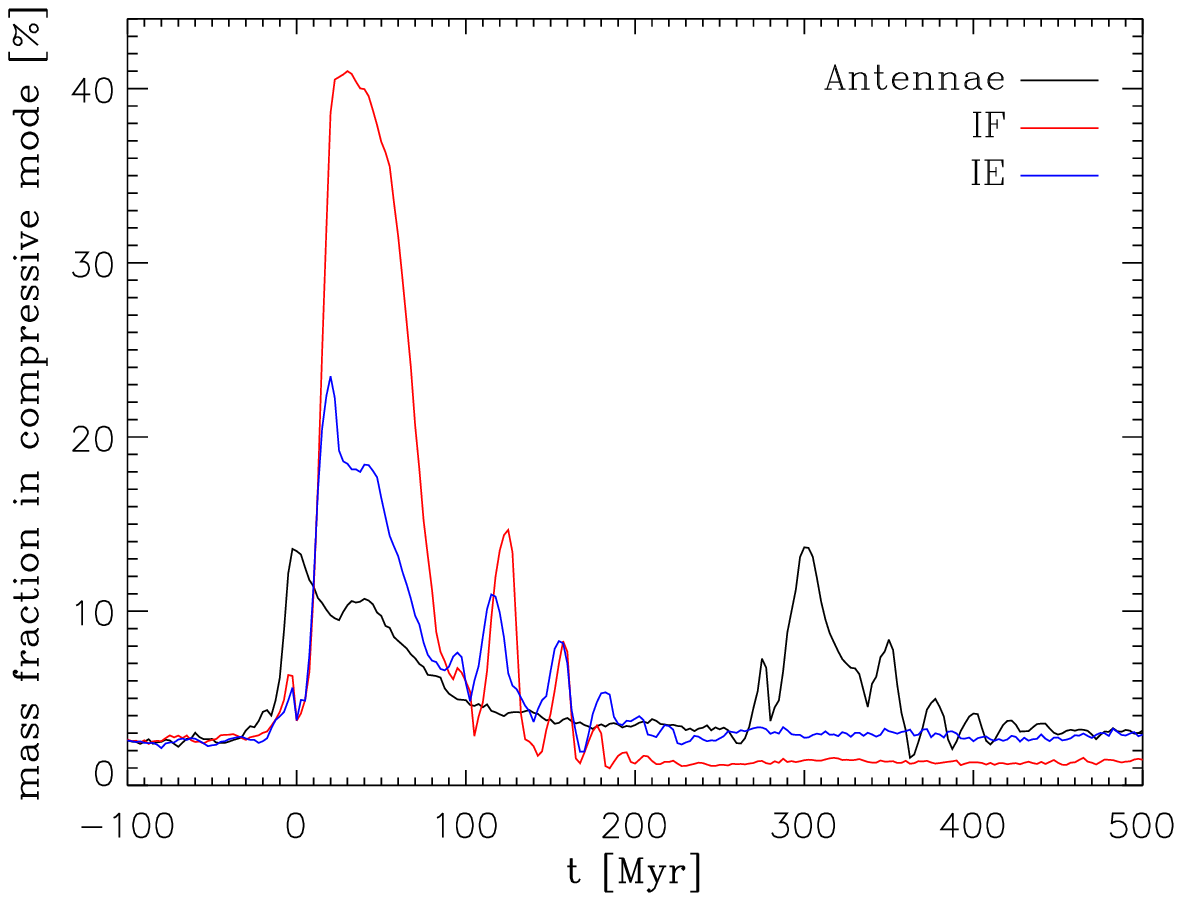}{Mass fraction in compressive mode in the IF and IE models}
{Evolution of the mass fraction in compressive mode for the face-on collision (\acr{notarg}{IF}{IF}) and the edge-on one (\acr{}{IE}{IE}).}

The amplitude of the peaks for both \acr{}{IF}{IF} and \acr{}{IE}{IE} models confirms that the compressive tidal mode scales with the amount of matter that goes ``deep'' into the potential wells. The \emph{\acr{notarg}{lus}{longest uninterrupted sequence}} timescales are comparable to those of the Antennae. However, the \acr{}{IF}{IF} model is badly fitted with the usual double exponential profile, because of a very large number of transient features (representing $\sim 40\%$ of the stellar mass and lasting for $\sim 5 \Myr$) in the tidal tails which expend at high velocity. Therefore, the \emph{\acr{notarg}{lus}{longest uninterrupted sequence}} statistics are biased by this short-period contribution.

%%%%%%%%%%%%%%%%%%%%%%%%%%%%%%%%%%%%%%%%%%%%%%%%%%
\sect[survey_mratio]{Mass ratio}

The \acr{}{CDM}{cold dark matter} theory suggests that the galaxy-size structures in the Universe form through the accretion of smaller aggregates, i.e. via repeated merger events (see e.g. \mycitealt{Peebles1982}; \mycitealt{Springel2005b}; \mycitealt{Stewart2008} and references therein). The fine details of the evolution of such objects rely on the mass ratio between the components of the merger. At a smaller scale and when including baryonic matter, this parameter also seems to be important in the study of interacting pairs of galaxies (\mycitealt{Naab2003}; \mycitealt{Bournaud2005}; \mycitealt{Naab2006}; \mycitealt{Johansson2009}). These works tell apart the major mergers (mass ratio greater than 3:1) and the minor mergers. However, it is important to precise which mass one refers to. \mycitet{Stewart2009} noted that observational works often use the baryonic mass (which can be measured via light) while the theoreticians consider the baryons and the \acr{}{DM}{dark matter}. In this study, we adopt the theoretical point of view and refer to the total mass (stars and \acr{}{DM}{dark matter}) when using the mass ratio.

For this section of our survey, we have considered a major merger (2:1), a minor merger (10:1) and an intermediate one (3:1) in addition to our Antennae reference (1:1), and kept all the other parameters (spin, orbit, shape) as in the reference. However, changing the mass of the galaxies modifies the gravitational energy, while the kinetic energy of their particles is left unchanged. This inevitably leads to instability because a galaxy of lower mass but same velocity dispersion will dissolve. To avoid this, the size and the velocity have been scaled in addition to the mass, so that the density and the virial ratio are kept constant. This choice maintains the intrinsic mass fraction in compressive mode at $\sim 3\%$ of the disk mass.

\mybox[rescale_galaxy]{
When changing the mass of an $N$-body system, one has to ensure that the internal dynamics are not affected. To keep the two-body heating low, the mass of individual particles should remain the same. Therefore, creating a low mass galaxy from an heavier one implies to reduce the number of particles. 
\follow}\mybox[nocnt]{
Furthermore, when modifying the mass from $M_1$ to $M_2$, one can preserve the density $\rho$ and the virial ratio $\nu$:
\eqn{
\rho_1 & = \rho_2 \nonumber \\
\frac{M_1}{\ell_1^3} & = \frac{M_2}{\ell_2^3},
}
and 
\eqn{
\nu_1 & = \nu_2 \nonumber \\
\frac{M_1}{\ell_1 \sigma_1^2} & = \frac{M_2}{\ell_2 \sigma_2^2},
}
where $\ell$ and $\sigma$ represent the length and velocity scales. This leads to
\eqn[resizing]{
\ell_2 = \ell_1 \left(\frac{M_2}{M_1}\right)^{1/3},
}
and
\eqn{
\sigma_2 = \sigma_1 \left(\frac{M_2}{M_1}\right)^{1/3}.
}
}

The morphology of the galaxies is shown in \reffig{survey_mratio_morphology}. On the one hand, for the major and intermediate mergers (\acr{}{M2}{M2} and \acr{}{M3}{M3}), the morphology and the position of the compressive modes in the southern galaxy (which has not been changed) remain very similar to those of the Antennae. In the minor merger case (\acr{}{M10}{M10}) however, the small intruder clearly has a too weak influence on the main disk which only exhibits a faint tidal tail. On the other hand, the modification of the aspect of the counterpart strongly depends on its relative mass: the low mass progenitors undergo mass loss via tidal stripping. In the same time, they induce radial perturbations in the main disk, which lead to the creation of transient features like spiral arms and even a bar. We note that such structures generally yield cores in the gravitational potential and thus compressive regions, as visible in the lower-right panel of \reffig{survey_mratio_morphology}. As a consequence, the mass fraction in compressive mode is expected to remain fairly low (close to isolation level) in the satellite galaxy but higher in the major progenitor, than in the nuclei of 1:1 encounters. 

\fig{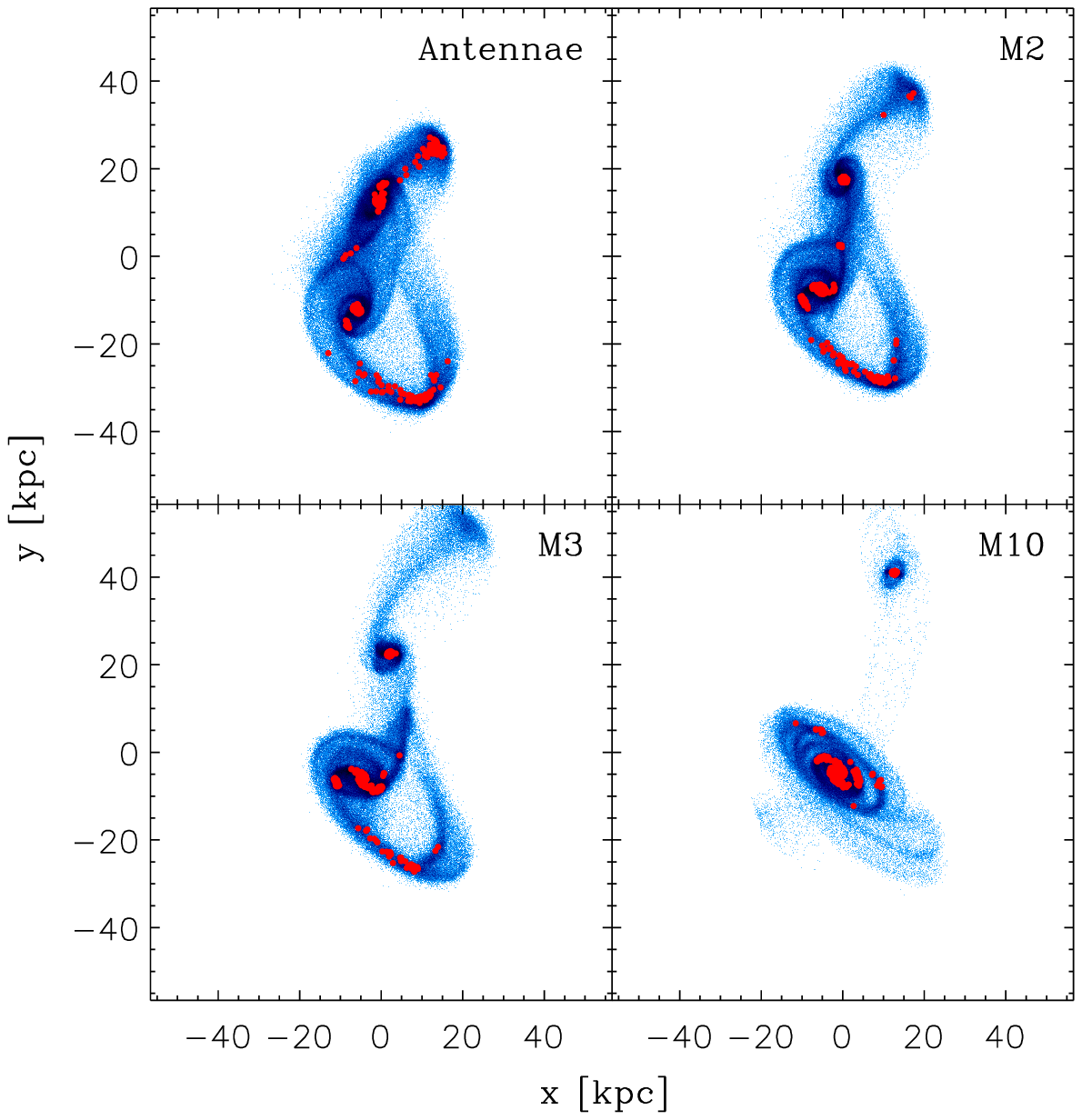}{Morphology of the Antennae, M2, M3 and M10 models}
{Morphology of the Antennae, \acr{notarg}{M2}{M2}, \acr{notarg}{M3}{M3} and \acr{notarg}{M10}{M10} models at approximately the same dynamical step (maximum separation of the progenitors). The minor merger leaves the main disk (south) merely unaffected. However, transient spiral structures are created during the interaction.}

This idea is confirmed by \reffig{survey_mratio_histo}. As always, the first passage is clearly marked by a peak. For the minor mergers (\acr{}{M3}{M3} and \acr{}{M10}{M10}), the decrease of the mass fraction between the two passages is replaced by a slow increase, due to the creation of the transient compressive structures (spiral and bar) mentioned above. In the \acr{}{M3}{M3} model, they still exist when the second passage occurs ($t\sim 425 \Myr$), while they have time to slowly vanish in \acr{}{M10}{M10} until the merger ($t \sim 900 \Myr$). In this case, the maximum amount of matter in compressive mode is reached \emph{between} two interaction events. However, it is still possible to tell it apart, as it does not correspond to a peaked, rapid change. In terms of timescale, this can be compared to the transient compressive modes found in the tidal tails and the bridges of the Antennae (recall \reffigt{antennae_period}).

\fig{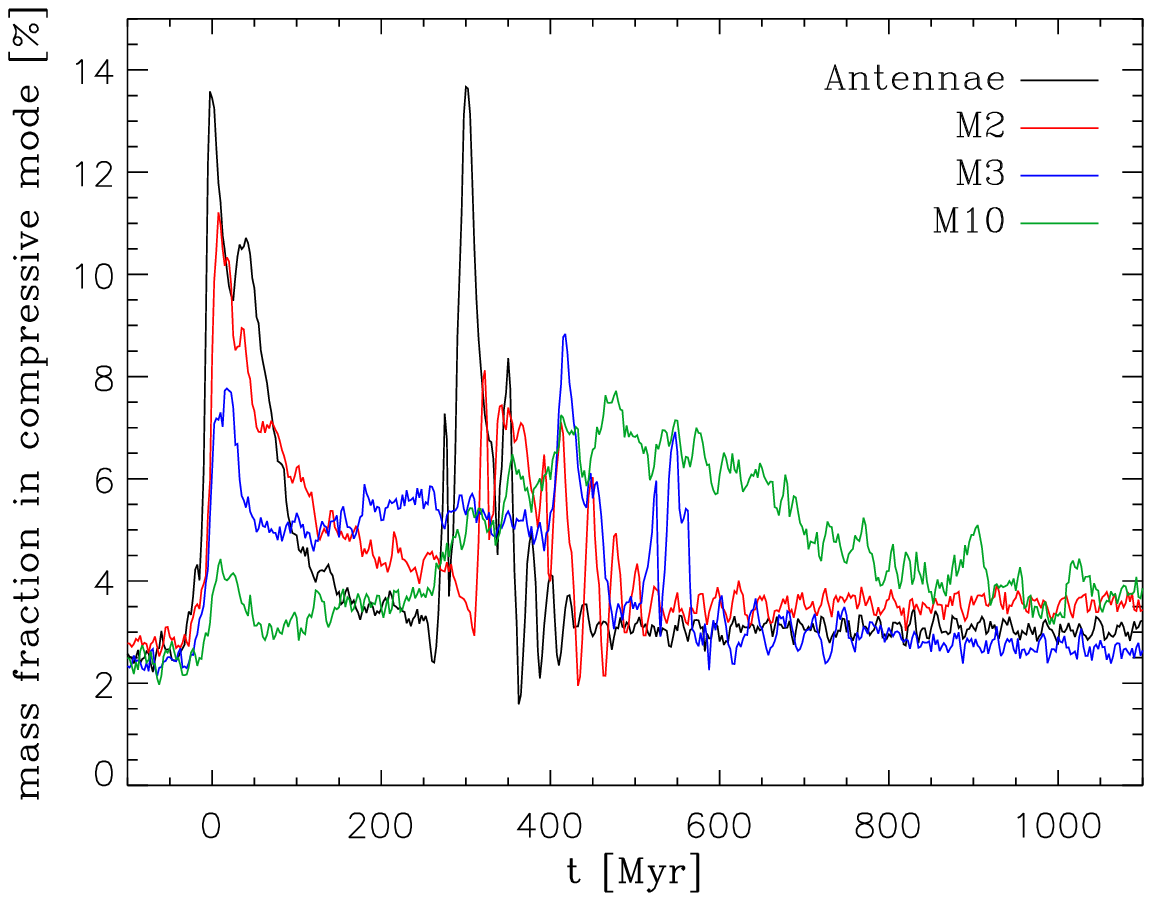}{Mass fraction in compressive mode in the M2, M3 and M10 models}
{Evolution of the mass fraction in compressive mode for the major (Antennae, \acr{notarg}{M2}{M2}), intermediate (\acr{notarg}{M3}{M3}) and minor (\acr{}{M10}{M10}) mergers of our survey. Despite very different dynamics, the relation between the interaction events and the peaks is still clear. (Adaptation of Figure~14 of \mycitealt{Renaud2009}.)}

\emph{In fine}, minor mergers induce many short-lived compressive regions, leading to a sharper \emph{\acr{notarg}{lus}{longest uninterrupted sequence}} distribution. As a result, the exponential timescale $\tau_1$ is slightly shorter for the minor mergers than for the major ones. By contrast with the orbital survey, these conclusions derive from internal dynamics and are not biased by the orbital period.

%%%%%%%%%%%%%%%%%%%%%%%%%%%%%%%%%%%%%%%%%%%%%%%%%%
\sect[survey_models]{Further compound models}

Up to now, we have always used an exponential disk, an \mycitet{Hernquist1990} bulge and an isothermal halo for our progenitors. However, the halo is suspected to influence the characteristics of the tidal field, because it dominates the gravitational potential of the galaxy. To verify this hypothesis, the last step of this parameter survey sets up different compound galaxy models. However, it is not possible to simply modify the parameters of the halo, leaving the other two components unchanged. In composite models, the global potential is built by the sum of the three parts. Therefore, changing one requires the adjustment of the others, in order to conserve the stability and coherence of the galaxy. That is why the disk and the bulge have to be modified, too. To do so, equilibrium distribution functions from the literature\footnote{The details on the construction of such distribution functions are out of the scope of the present document. The interested reader will find their derivations in the references presented below.} have been used to create two new composite models, simply keeping a circular velocity of $\sim 220 \kms$ at solar radius.

The first one (noted \acr{}{KD}{KD}) is a \myciteauthor{Kuijken1995} (\myciteyear{Kuijken1995}, model MW-A) model of a Milky Way-like galaxy. It gathers a 3D version of a \mycitet{Shu1969} disk, a \mycitet{King1962} bulge and a lowered \myciteauthor{Evans1993} (\myciteyear{Evans1993}; see also \mycitealt{Kuijken1994}) \acr{}{DM}{dark matter} halo. The $N$-body realization\footnote{Note that the conversion factors from $N$-body to physical units are slightly different than previously: $3.6 \kpc$, $5.5\times 10^{10} \msun$ and $14.2 \Myr$.} of this model have been done with the {\tt mkkd95} tool of the {\tt Nemo} package. \acr{}{KD}{KD} has slightly more massive halo ($\times 1.3$) and disk ($\times 1.5$) than the reference model. The scaling is so that the truncation radius of the disk is the same than in the reference.

\fig{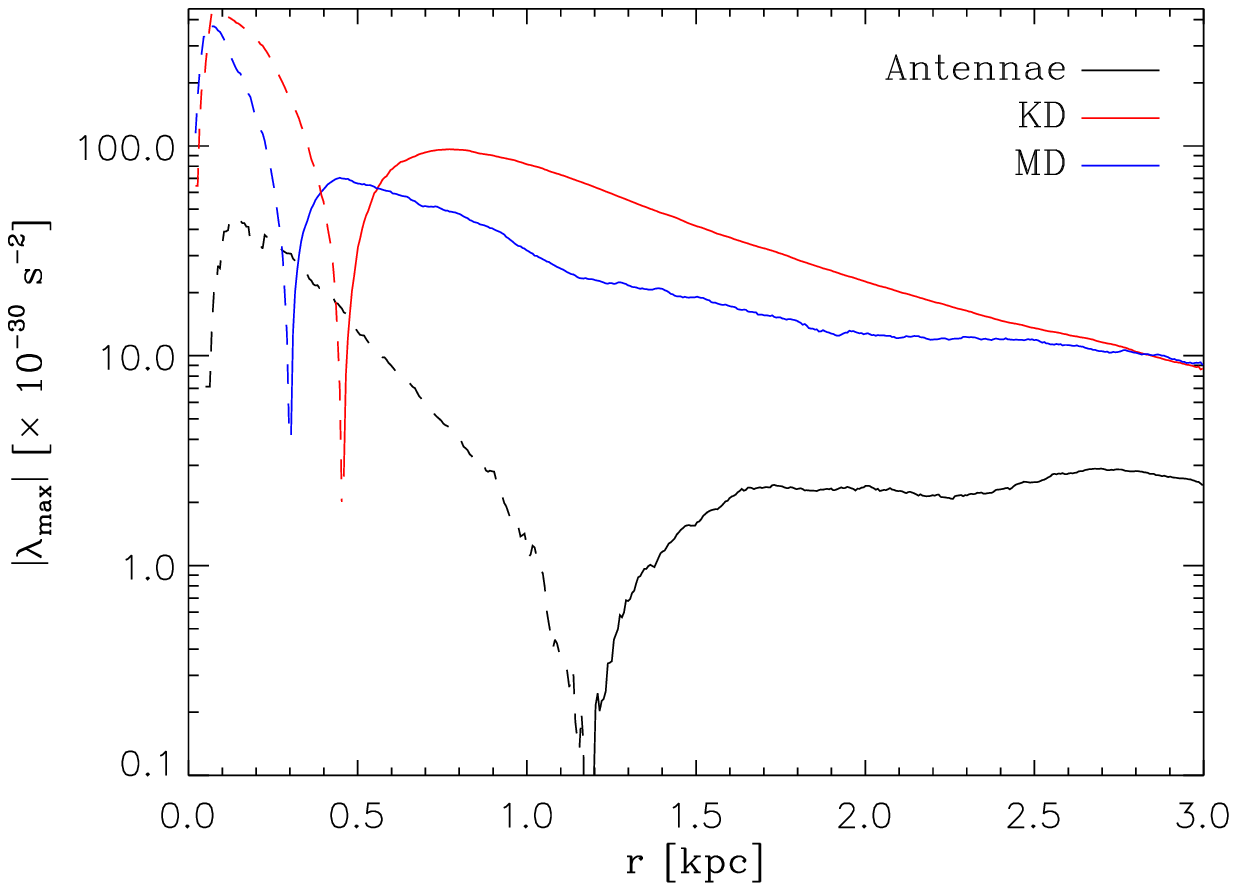}{Tidal profiles of our compound models}
{Absolute value of the maximum eigenvalue of the tidal tensor of the three compound models (Antennae, \acr{notarg}{KD}{KD} and \acr{notarg}{MD}{MD}, in isolation), as a function of radius. Dashed curves represent the negative (i.e. compressive) part of the profiles.}

These changes in the shape of the galaxy imply modifications of the potential and thus of the tidal profile too. \reffig{survey_models_tidal_profiles} plots the maximum eigenvalue of the tidal tensor as a function of the radius, for the galaxies taken in isolation. The intrinsic central compressive region exists for all models and is clearly marked on this plot. We note that its diameter and tidal strength vary significantly from one model to the other. That is, one may expect different results from the (by-now) traditional study of the tidal field of the merger. These differences are not visible in the location of the compressive regions, as shown in \reffig{survey_models_morphology}: the major areas of compressive tides (nuclei, bridges, tails) are retrieved in a comparable way than in the reference model.

\fig{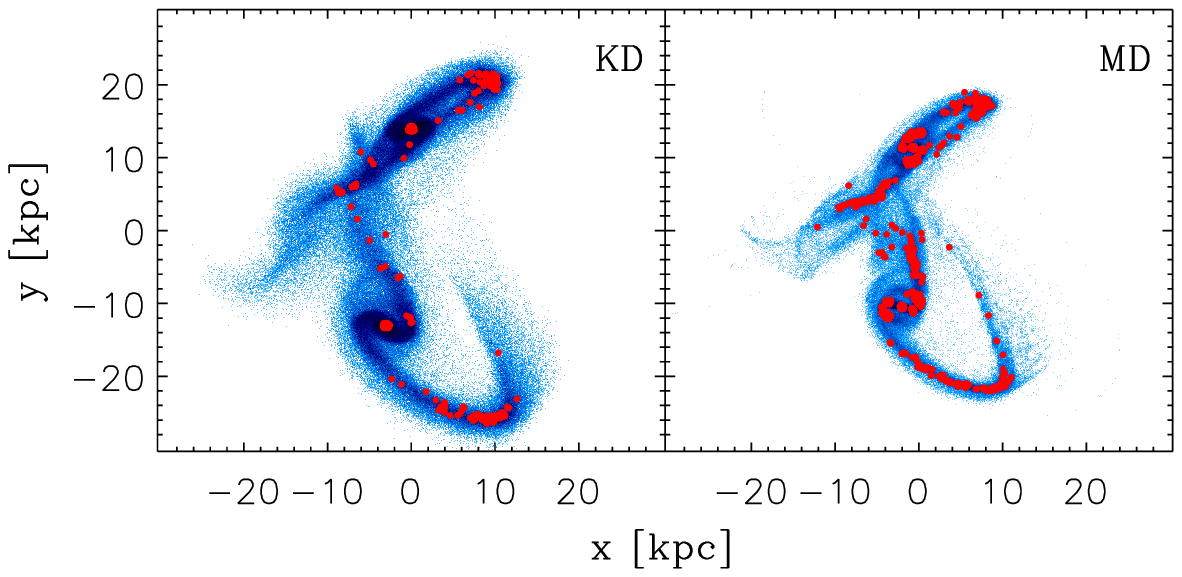}{Morphology of the KD and MD models}
{Morphology of the \acr{notarg}{KD}{KD} and \acr{notarg}{MD}{MD} models at approximately the same dynamical step (maximum separation of the progenitors).}

\fig{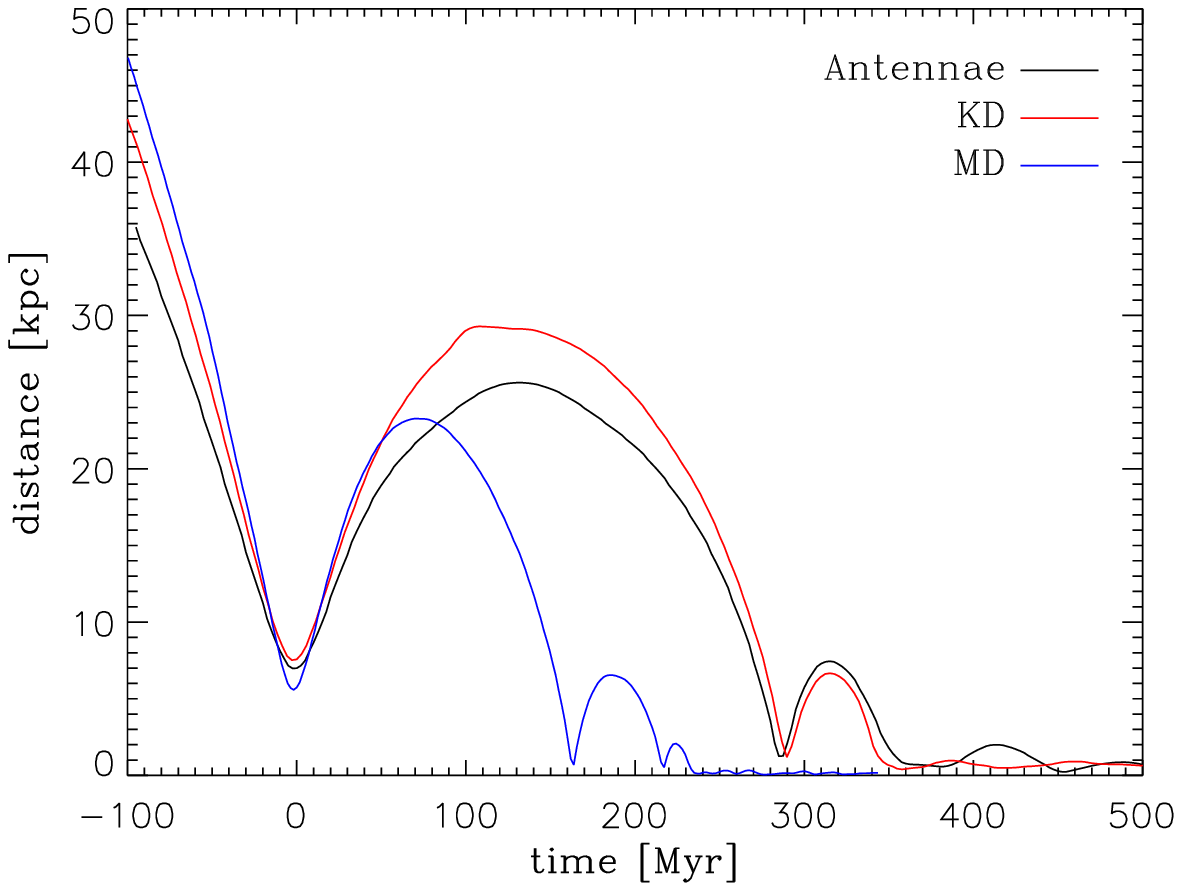}{Evolution of the distance in the KD and MD models}
{Evolution of the distance between the two galaxies for the \acr{notarg}{KD}{KD} and \acr{notarg}{MD}{MD} models. Except a faster evolution of the \acr{notarg}{MD}{MD} model, the global behaviors are comparable to those of the Antennae reference.}

\reffig{survey_models_distance} displays the evolution of the distance between the two progenitors, marking the usual steps: first passage, separation, second passage and merger. The three configurations lead to a very similar evolution, especially between the reference and \acr{}{KD}{KD}. In parallel, \reffig{survey_models_histo} shows the mass fraction in compressive mode. As expected, the intrinsic fraction (i.e. from isolation stage) is different for \acr{}{KD}{KD} from the Antennae case. This is explained by (i) a smaller central compressive region and (ii) a different density profile (or in other words, a different mass enclosed inside the compressive limit). Only $\sim 0.6\%$ of the disk's mass is intrinsically in compressive mode for \acr{}{KD}{KD}, instead of $\sim 3\%$ for the Antennae model. However, we note a very good correlation between the two models in the evolution of this fraction through the merger. In \acr{}{KD}{KD}, the peaks are better separated in time which suggests a shorter timescale for the compressive mode, and a more efficient orbital mixing of the particles. This is confirmed when investigating the \emph{\acr{notarg}{lus}{longest uninterrupted sequence}} distribution: a timescale of $\sim 3.3 \Myr$ is measured, instead of $\sim 7 \Myr$ (see \reftab{parameter_survey}). Here again, the double exponential distribution is very well retrieved.

\fig{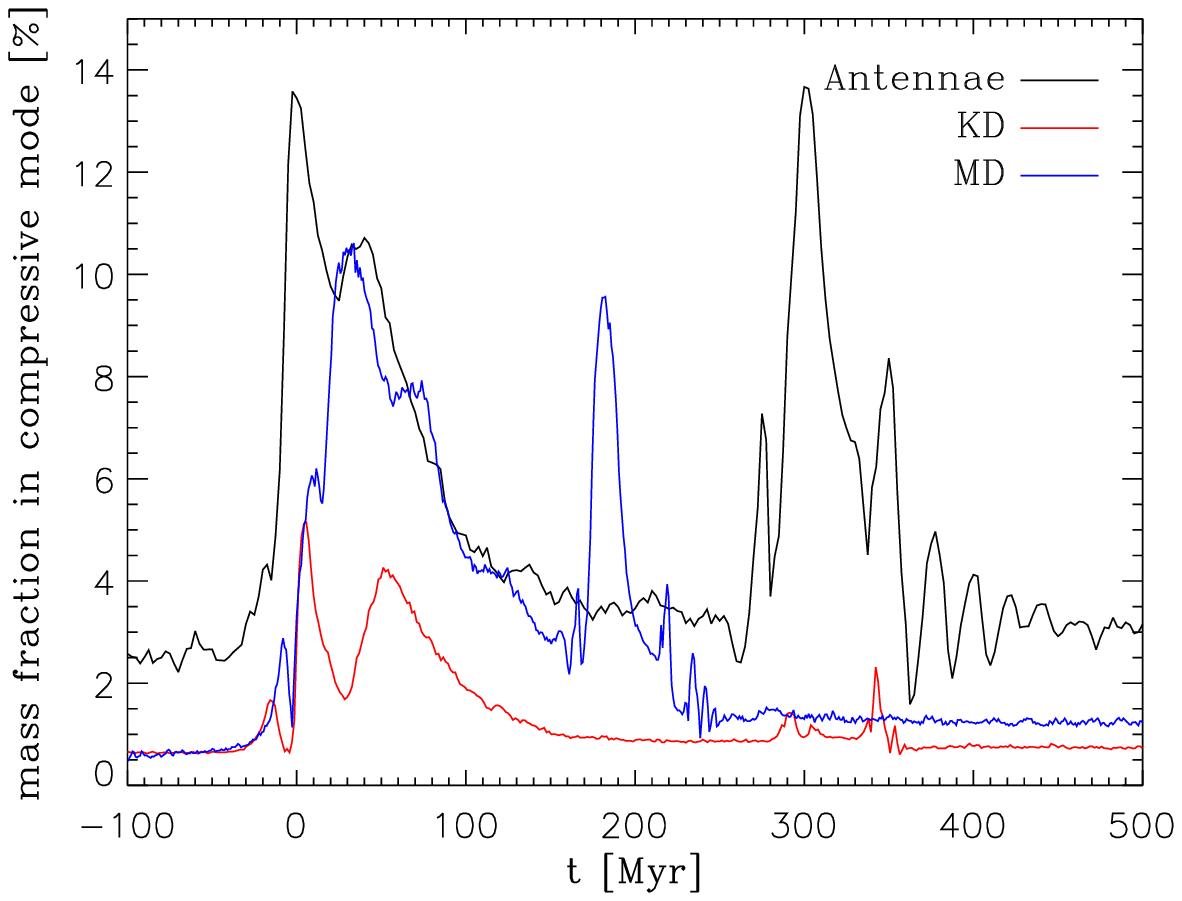}{Mass fraction in compressive mode for the KD and MD models}
{Evolution of the mass fraction in compressive mode for the three compound models. Differences in the intrinsic fraction ($t < 0$) and in the time lapse between the peaks exist, but the overall behavior is well conserved.}

The last model of our survey is taken from \myciteauthor{McMillan2007} (\myciteyear{McMillan2007}, hereafter \acr{}{MD}{MD}). This time, the halo is a massive \acr{}{NFW}{NFW} profile (\mycitealt{Navarro1997}). It is combined with an \mycitet{Hernquist1990} bulge and an exponential disk. The $N$-body model\footnote{As before, the conversion factor from $N$-body to physical units are different than in the Antennae case: $2.8 \kpc$, $6.0\times 10^{10} \msun$ and $8.8 \Myr$.} is created thanks to the {\tt mkgalaxy} tool within {\tt Nemo}. The major difference between this model and the reference is the extension of its halo: it shows a characteristic radius of $12 \kpc$ and is 24 times more massive than the one of the Antennae model.

As for \acr{}{KD}{KD}, the intrinsic mass fraction in compressive mode (see \reffig{survey_models_histo}) is quite low ($\sim 0.7\%$). In addition to the central compressive region, number of small, transient compressive zones exist all over the disk, because of a peculiar local distribution of the mass. Note that these features are smoothed out when averaging the tidal profile per radial bins, as in \reffig{survey_models_tidal_profiles}. The massive halo and the important dynamical friction reduce the orbital period by $\sim 130 \Myr$ (\reffig{survey_models_distance}), which forbids the mass fraction in compressive mode to return to its isolation level between the two passages, like in the \acr{}{M3}{M3} model presented in the previous Section.

We note that the \acr{}{NFW}{NFW} halo of a progenitor extends well beyond the limits of the disk, so that the entire stellar structure (initially the disk, and then the nucleus and the tails) sits in the inner part of the \acr{}{NFW}{NFW} profile (see \refsect[triplets]{nfw}). There, the density profile of the halo \emph{only} ($\rho \propto r^{-1}$) yields less extensive tides than in the outer regions ($\rho \propto r^{-3}$), meaning that the additional effect of the disk and the bulge would more easily create compressive zones than in the external part of the halo.

As in the \acr{}{KD}{KD} model, the timescale of compressive modes is shorter than the one of the Antennae ($\sim 2.6 \Myr$) but again, the double exponential law is well reproduced.

These two models confirm that the importance of compressive modes in term of mass fraction depends on the distribution of the matter in the progenitors. However, its evolution seems fairly independent of the internal structure of the galaxies, as the behavior of the tidal field follows the chain of events in the interaction. To go further in the investigation of the role of the halo \emph{only}, \refcha{halos} presents an analytical derivation of the tidal fields of typical halos, derived from cosmological simulations.

%%%%%%%%%%%%%%%%%%%%%%%%%%%%%%%%%%%%%%%%%%%%%%%%%%
\sect{Conclusions of the survey}

The survey presented in this Chapter allows an exploration of the role of the major parameters of a galaxy merger. The overall conclusion has three folds:
\begin{itemize}
\item the formation of compressive regions is systematically triggered by major interaction events (passages, merger phase). Flybys do not yield a compressive tidal field.
\item the mass fraction in compressive mode is strongly anti-correlated with the pericenter distance: a close encounter induces a large amount of matter to sit in such modes while a distant one has a milder influence.
\item the duration of these modes depends on the dynamics of the merger and of the individual galaxies, but always yields the same order of magnitude ($10^7 \yr$).
\end{itemize}
Finally, the conclusions drawn in \refcha{antennae} about the special case of the Antennae galaxies can be extended to a broad range of parameters. The orders of magnitudes derived apply to many (if not all) configurations and, more detailed descriptions follow the ``rules of thumb'' mentioned above.

%%%%%%%%%%%%%%%%%%%%%%%%%

\cha{halos}{Tidal field and dark matter halos}
\chaphead{In this Chapter, we widen our field of view to larger scales. Using the results of cosmological studies, we investigate the influence of the density profile of galactic halos on the character of the tidal field. This work is summarized in \mycitet{Renaud2010a}.}

%%%%%%%%%%%%%%%%%%%%%%%%%%%%%%%%%%%%%%%%%%%%%%%%%%
\sect{Dark matter halos}

%%%%%%%%%%%
\subsect{Missing mass}
The early measurements of velocity dispersions of stars (\mycitealt{Oort1932}) and galaxies in clusters (\mycitealt{Zwicky1933}) revealed that the corresponding dynamical mass is $\sim 10-100$ greater than what is derived from the light: the ``luminous'' mass can not explain the dynamics of astrophysical objects (in the framework of the Newtonian physics). Therefore, it has been naturally suggested that this ``missing mass'' is not luminous, or in other words that \acr{}{DM}{dark matter} exists in the Universe. Many decades after, the nature of this matter remains unknown and puzzling for physicists (see \mycitealt{Trimble1987}; \mycitealt{Carr1994} for reviews). It is supposed that an important part of its mass would be in a form of non-baryonic matter\footnote{These candidate particles are called WIMPs (weakly interacting massive particles) because they should have a very small cross-section to avoid interaction with the ``classical'' baryons that would make them easier to detect.}, with the rest being non-luminous baryons (black holes, brown dwarf stars, planets). Hereafter, we take the shortcut of calling ``baryonic'', all the non dark matter.

Detailed studies on individual galaxies revealed that their luminous mass (stars and gas) is embedded in a \acr{}{DM}{dark matter} halo (see e.g. \mycitealt{Freeman1970}). However, the distribution of this material is not well-constrained by observations, \emph{de facto}. That is the reason why one has to assume a shape for the density profile of the \acr{}{DM}{dark matter} halo, when modeling galaxies (isolated or interacting). In the previous Chapters, we have considered isothermal (\refcha{antennae}), Evans and \acr{notarg}{NFW}{NFW} (\refsect[survey]{survey_models}) profiles, via distribution functions found in the literature. Other profiles can be studied analytically and/or numerically in principle, but the distribution functions of composite models (i.e. containing a stellar disk and bulge in addition to the \acr{}{DM}{dark matter} halo) are generally not available. Furthermore, the equilibrium conditions do not couple a distribution of the baryonic matter for every halo and thus, a disk embedded in a given halo cannot (in principle) be exported to another one. That is, measurements of the effect of the halo on a property of a galactic system (like the tidal field, as discussed in \refsect[survey]{survey_models}) is often hidden by (if not buried in) the effect of different disks and bulges.

In this Chapter, we circumvent this issue by investigating the properties of the \acr{}{DM}{dark matter} halo \emph{only}, with an analytical method. The goal of this approach is to highlight the differences existing between the profiles, in particular for the character of the tidal field (extensive or compressive).

%%%%%%%%%%%
\subsect{Cosmological simulations, cores and cusps}
The bottom-up scenario suggests that structures (and among them, the galaxies) form through the repeated accretion of clumps of matter. Following this idea, cosmological simulations have been able to reproduce the organization of galaxies in groups and clusters, as observed (see e.g. \mycitealt{Tegmark2004}; \mycitealt{Springel2005b}). With such numerical studies, it is possible to retrieve the density profiles of the halos, and possibly derive an analytical definition. That is how \myciteauthor{Navarro1997} (\myciteyear{Navarro1997}, hereafter \acr{}{NFW}{NFW}) fitted their simulated \acr{}{DM}{dark matter} halos (once they have reached equilibrium) with the formula
\eqn{
\rho_\mathrm{N}(r) = \frac{\rho_{\mathrm{N},0}}{ \frac{r}{r_{\mathrm{N},0}} \left( 1 + \frac{r}{r_{\mathrm{N},0}}\right)^2},
}
where $\rho_{\mathrm{N},0}$ is a characteristic density and $r_{\mathrm{N},0}$ a scale radius. This profile provides a good approximation on a wide range of radii and masses (as noted by e.g. \mycitealt{Navarro1997}; \mycitealt{Navarro2004}), a strong hint that it is universal. Comparable results have been obtained by \myciteauthor{Moore1999} (\myciteyear{Moore1999}, hereafter \acr{}{M99}{M99}) who found
\eqn{
\rho_\mathrm{M}(r) = \frac{\rho_{\mathrm{M},0}}{\left(\frac{r}{r_{\mathrm{M},0}}\right)^{1.5}\left[1+\left(\frac{r}{r_{\mathrm{M},0}}\right)^{1.5}\right]}.
}
Let's first focus on the logarithmic slope of these profiles, defined as 
\eqn{
-\beta(r) \equiv \frac{d\ln(\rho)}{d\ln(r)}.
}
\reffig{navarro2004} plots this slope as a function of $r$. Both profiles yield $\beta = 3$ in their outer regions (i.e. for $r \gg r_0$), and a finite value  (1.0 and 1.5, respectively) for $r \ll r_0$. However, more recent high resolution runs noted a small, yet systematic deviation of simulations from the analytical description (\mycitealt{Navarro2004}; \mycitealt{Boily2006}; \mycitealt{Hansen2006}). Indeed, there is no sign of convergence toward a finite value of $\beta$ in the inner region, as already suggested by the data of \myciteauthor{Navarro2004} (\myciteyear{Navarro2004}, dashes on \reffig{navarro2004}). In other words, several models define a density profile with a central cusp denoted by $\beta(r=0) \ge 1$, while simulations tend to predict $\beta(r=0) = 0$, i.e. cored halos (see \mycitealt{deBlok2010}).

\mybox[corecusp]{
What distinguishes a core from a cusp? In spherical symmetry, a profile yields a central core if the slope of its potential at $r=0$ is null. In mathematical terms, a cored distribution verifies
\eqn{
\lim_{r\to 0}\left(\frac{d\phi(r)}{dr}\right) & = 0 \nonumber \\
\lim_{r\to 0} \left(r^{-2}\int_0^r x^2 \rho(x) \ dx \right) & = 0 \nonumber \\
\int_0^r x^2 \rho(x) \ dx & \stackrel{r\to 0}{\sim} \ r^k \qquad \textrm{with } k > 2.
%\frac{d \ln\left( \int_0^r x^2 \rho(x) \ dx \right)}{d \ln(r)} & > 2 \nonumber \\
}
Thanks to the Maclaurin expansion (i.e. Taylor series at $r=0$) of the integrand, one has
\eqn{
\rho(r) & \stackrel{r\to 0}{\sim} \ r^{k-3} \nonumber\\
%\frac{d \ln\left(\rho(r)\right)}{d \ln(r)} & > -1 \nonumber\\
\beta(r=0) & < 1.
}
Therefore, all density profiles with a logarithmic slope $\beta$ strictly smaller than unity correspond to cored potential. All the other cases yield a cusp.
}

\fig{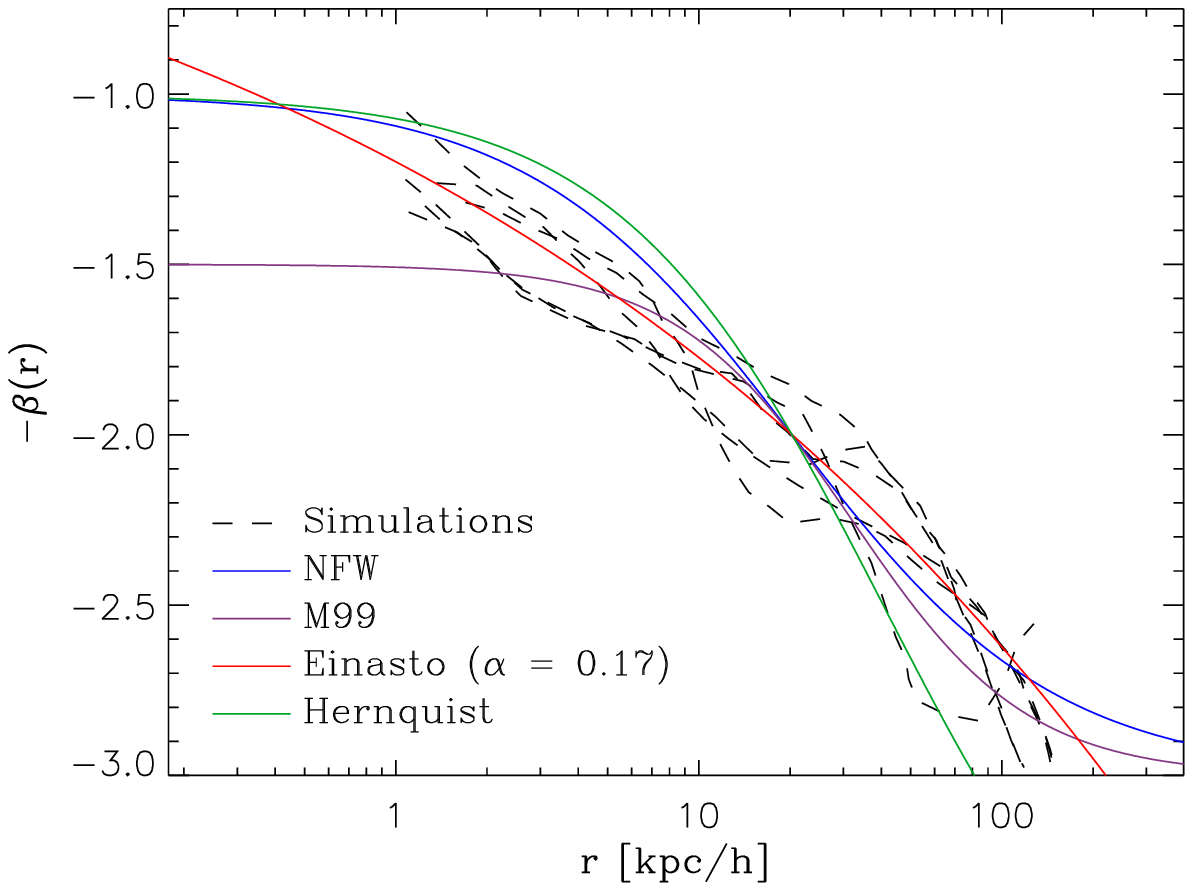}{Logarithmic slope of density profiles}
{Logarithmic slope of the density profiles of galactic halos ($r_0 = 20\ h^{-1}\kpc$) from cosmological simulations (\mycitealt{Navarro2004}, dashes), compared to the \acr{notarg}{NFW}{NFW} (blue), \acr{notard}{M99}{M99} (purple) and Einasto (red) analytical curves. The slope of the Hernquist (green) profile is also plotted, for reference only. (Simulation data points from \mycitealt{Navarro2004}.)}

A more sophisticated profile\footnote{The reference is difficult to find but we refer the reader to \mycitet{Cardone2005} who proposed a detailed investigation of the properties of the Einasto density profile.}, called \mycitet{Einasto1965}, better matches the simulated halos (\mycitealt{Navarro2004}):
\eqn{
\rho_\mathrm{E}(r) = \rho_{\mathrm{E},0} \exp{\left\{-\frac{2}{\alpha}\left[\left(\frac{r}{r_{\mathrm{E},0}}\right)^{\alpha}-1\right]\right\}}.
}

This profiles yields a logarithmic slope which is a power-law of radius\footnote{Such a function of the \emph{volume} density is equivalent to the \mycitet{Sersic1968} law on \emph{surface} density (see \mycitealt{Merritt2005} for a discussion on this similarity).}: $\beta(r) \propto r^\alpha$, so that the central part exhibits a core. When applied to the simulations, the best fit is found for the index $\alpha = 0.17$ (\mycitealt{Navarro2010}), a value that we adopt hereafter. Note that \mycitet{Stadel2009} found $\alpha = 0.15$. The conclusions presented below are not affected by such a change and thus, we leave the discussion on the value of $\alpha$ to the cosmologists.

Unfortunately, the difference between cored and cuspy profiles raises at small radii only and it is still difficult (with current resolutions) to distinguish them. In \refsec[halos]{halos_tidal}, we show that their tidal fields however, are very different.

%%%%%%%%%%%
\subsect{$\gamma$-family}
Before these simulations, \mycitet{Dehnen1993} and \mycitet{Tremaine1994} analytically defined a family of density profiles as
\eqn{
\rho_\gamma(r) = \frac{(3-\gamma) \ M}{4\pi} \ \frac{r_{\gamma,0}}{r^\gamma \ (r+ r_{\gamma,0})^{4-\gamma}},
}
where $r_{\gamma,0}$ is a scale radius and $\gamma \in [0,3[$ a parameter. Note that for $\gamma=1$ and $\gamma = 2$, one retrieves the \mycitet{Hernquist1990} and \mycitet{Jaffe1983} profiles, respectively. This definition has been first used to describe elliptical galaxies and bulges, but has been extended to \acr{}{DM}{dark matter} halos later on (see e.g. \mycitealt{Merritt1996}; \mycitealt{Jeon2009}). Its main advantage for our topic is that it yields a wide range of logarithmic slopes: $\beta(r=0) = \gamma$, from cored configurations ($\gamma < 1$) to strong cusps.

%%%%%%%%%%%
\subsect[comparisons]{Making comparisons}
In order to compare the properties of several \acr{}{DM}{dark matter} halos profiles, one has to ensure that they are properly scaled. Indeed, all the analytical forms use scale-free parameters ($\rho_0$, $r_0$, $M$) that do not always represent the same physical quantity. It is therefore of prime importance to establish a relation between them, before making any comparisons. Being halos of galaxies, all the profiles should have a comparable impact on the dynamics of the embedded disk. For this reason, we define a reference radius $r_\mathrm{s}$ containing $1/5$ of the total mass (see e.g. the model of the Milky Way in \mycitealt{Binney2008}), so that the circular rotation speed of the baryonic material at this point remains fairly unchanged. To do so, the integrated mass is computed for each profile, which gives the total mass when evaluated at an infinite radius. The mathematical derivations are detailed in \refsecp[triplets]{ttexamples}. Then the equation of the mass profile is solved to find the relation between $r_0$ and $r_\mathrm{s}$.

In the case of Einasto, one finds (numerically) that $r_{\mathrm{E},0} \approx 0.36 \ r_\mathrm{s}$. Then, $\rho_{\mathrm{E},0}$ can be determined:
\eqn{
\rho_{\mathrm{E},0} & \approx M \frac{2^{(3/\alpha)} \ \alpha^{(1-3/\alpha)}}{4\pi \ (0.36 \ r_\mathrm{s})^3 \ \Gamma\left(\frac{3}{\alpha}\right)}\exp{\left(-\frac{2}{\alpha}\right)} \\
& \approx 0.14\  \frac{M}{r_\mathrm{s}^3}.
}

Repeating this exercise for the $\gamma$-family, one gets
\eqn{
r_{\gamma,0} = r_\mathrm{s}\left[5^{1/(3-\gamma)} -1 \right].
}
Thanks to \refeqnp{dehnen_mass} and \refeqnp{einasto_mass}, \reffig{integrated_mass} shows the mass profiles for Einasto (red) and Hernquist ($\gamma = 1$, green).

\fig{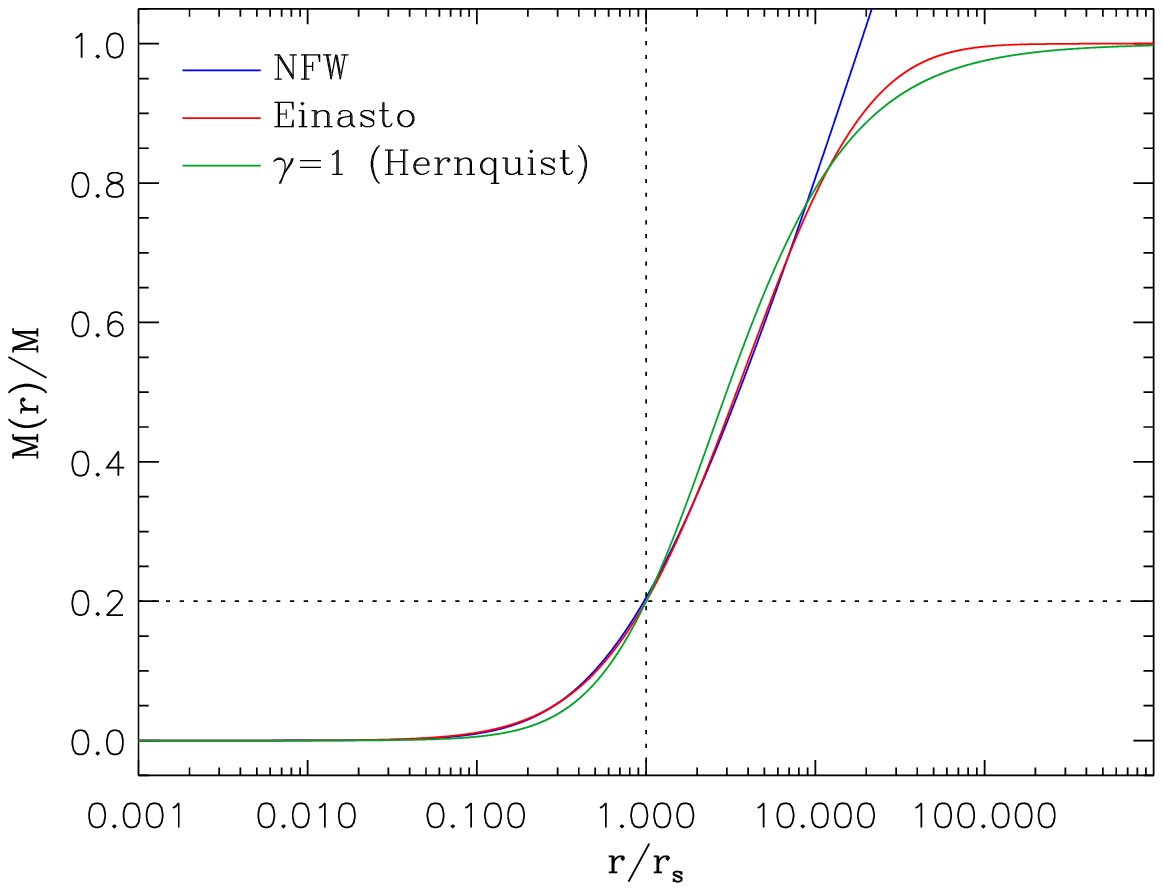}{Mass profile of the halos}
{Integrated mass of the \acr{notarg}{NFW}{NFW}, Einasto and Hernquist profiles as a function of the radius, normalized to $r_\mathrm{s}$ so that $M(r_\mathrm{s}) = M / 5$.}

The case of \acr{}{NFW}{NFW} is more problematic, as its mass (see \refeqnt{nfw_mass}) integrates to infinity, which means that a total mass cannot be defined. To overcome this, one must recall that the \acr{}{NFW}{NFW} and Einasto profiles are comparable over a certain range of radii. Therefore, it is possible to set the scale of \acr{}{NFW}{NFW} via the one of Einasto. \reffig{integrated_mass} plots the best fit of the Einasto mass profile with the analytical form of a \acr{}{NFW}{NFW}, which gives
\eqn{
r_{\mathrm{N},0} & \approx 0.34 \ r_\mathrm{s} \nonumber \\
& \approx 0.9 \ r_{\mathrm{E},0},
}
and
\eqn{
\rho_{\mathrm{N},0} \approx 4.5 \ \rho_{\mathrm{E},0}.
}
The divergence for $r \gg 1$ leading to an infinite mass is clearly visible but one also notices that the two profiles are remarkably comparable for smaller radii.

\fig{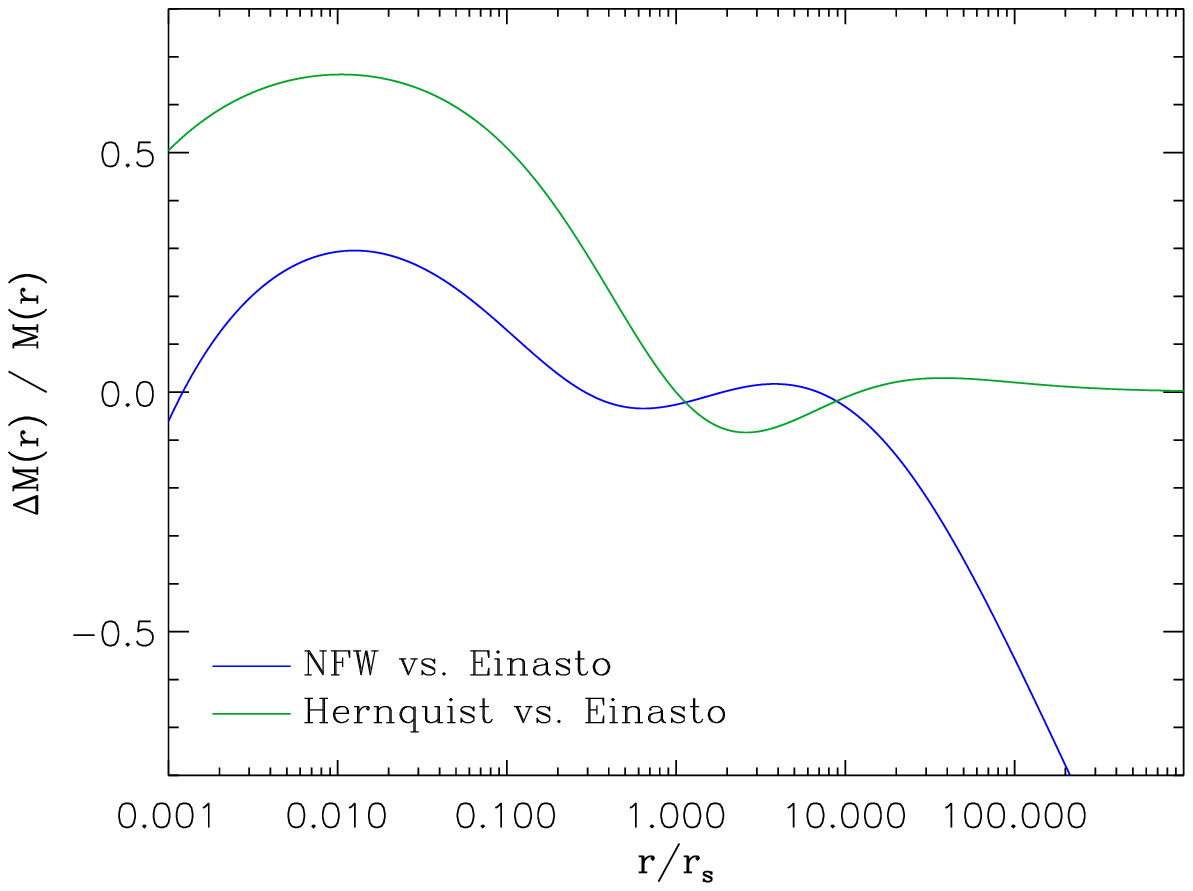}{Relative difference between the mass profiles}
{Relative difference between the integrated masses of the \acr{notarg}{NFW}{NFW} and the Einasto profiles (blue), and between Hernquist and Einasto (green).}

\reffig{integrated_mass_error} displays the relative difference between the three profiles: for Einasto and \acr{}{NFW}{NFW}, it reaches the level of $30\%$ for $r \approx 40 \ r_\mathrm{s}$ only, i.e. for $\approx 0.61 \ M$. Therefore, it is still difficult to distinguish between both models from simulations, which are limited in resolution. However, these small differences strongly impact on the associated tidal field.

%%%%%%%%%%%%%%%%%%%%%%%%%%%%%%%%%%%%%%%%%%%%%%%%%%
\sect[halos_tidal]{Tidal field}

%%%%%%%%%%%
\subsect{In isolation}
With scaling relations between the models, it is possible to go further in their comparison and exhibit their tidal field. The analytical derivations are given in \refsecp[triplets]{ttexamples}. For simplicity, we will use $M = 1$ and $r_\mathrm{s} = 1$ in the following.

First, it is noticeable that compressive tides only exist in the core of a potential, therefore, in the region where the logarithmic slope of the density profile is smaller than unity (recall \refbox{corecusp}). With \reffig{navarro2004}, it immediately appears that \acr{}{NFW}{NFW}, \acr{}{M99}{M99}, Hernquist and the isothermal profiles do not have compressive tidal modes, while Einasto does in its inner part.

As a confirmation, \reffig{analytical_tensor_halos_profiles} displays the value $T^{xx}$ of the tidal tensor of these profiles, as a function of radius. As expected, only the Einasto profile yields compressive tides at small radii. Note that even if this central part is very small\footnote{In the typical case of the Milky Way, one has $r_\mathrm{s} \sim 10 \kpc$. Therefore the compressive region of the equivalent Einasto profile would cover $\sim 100 \pc$ only.}, the overall shape of the tidal profile strongly differs from the one of e.g. \acr{}{NFW}{NFW}, though their density profiles are comparable.

\fig{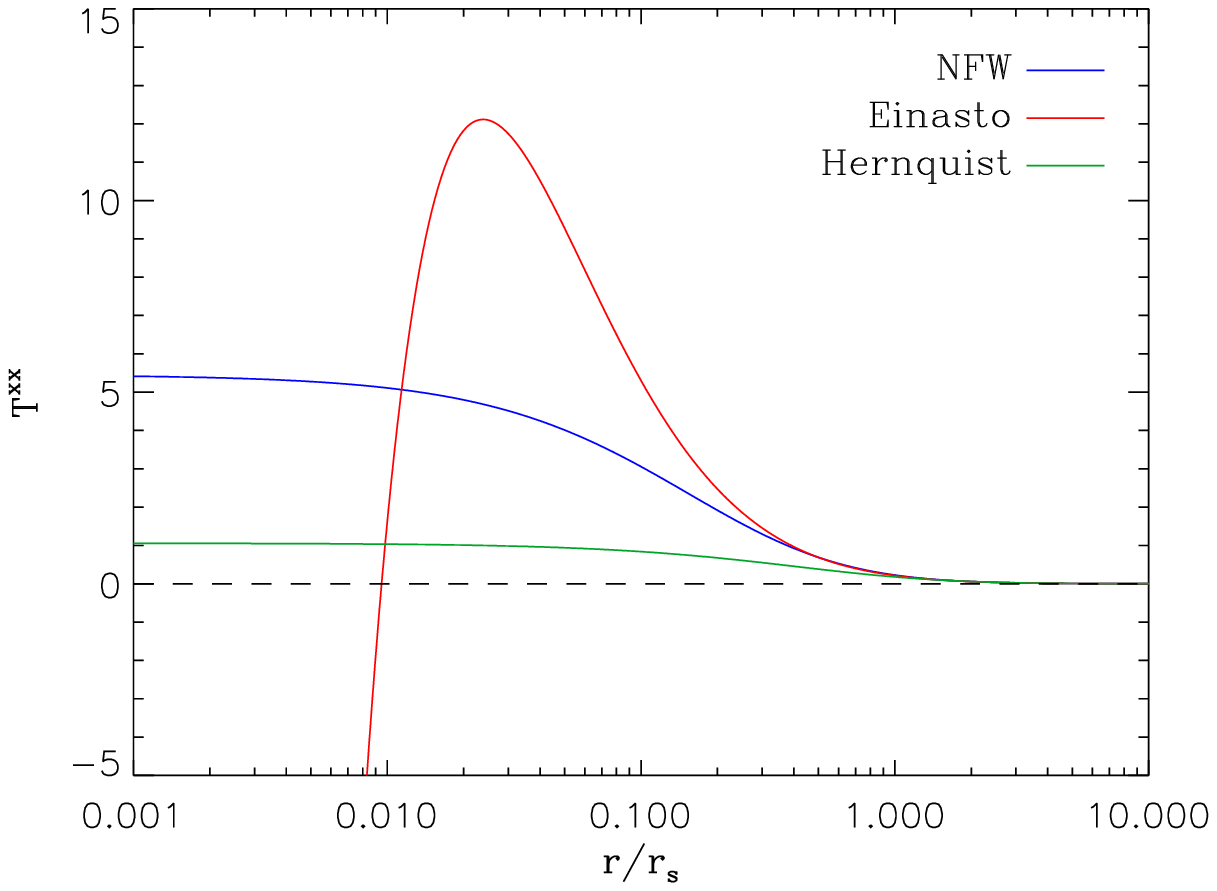}{Tidal profiles of the halos}
{Tidal profile of the \acr{}{NFW}{NFW}, Einasto and Hernquist profiles. The horizontal dashed line marks the transition between extensive ($>0$) and compressive ($<0$) tides.}

\fig{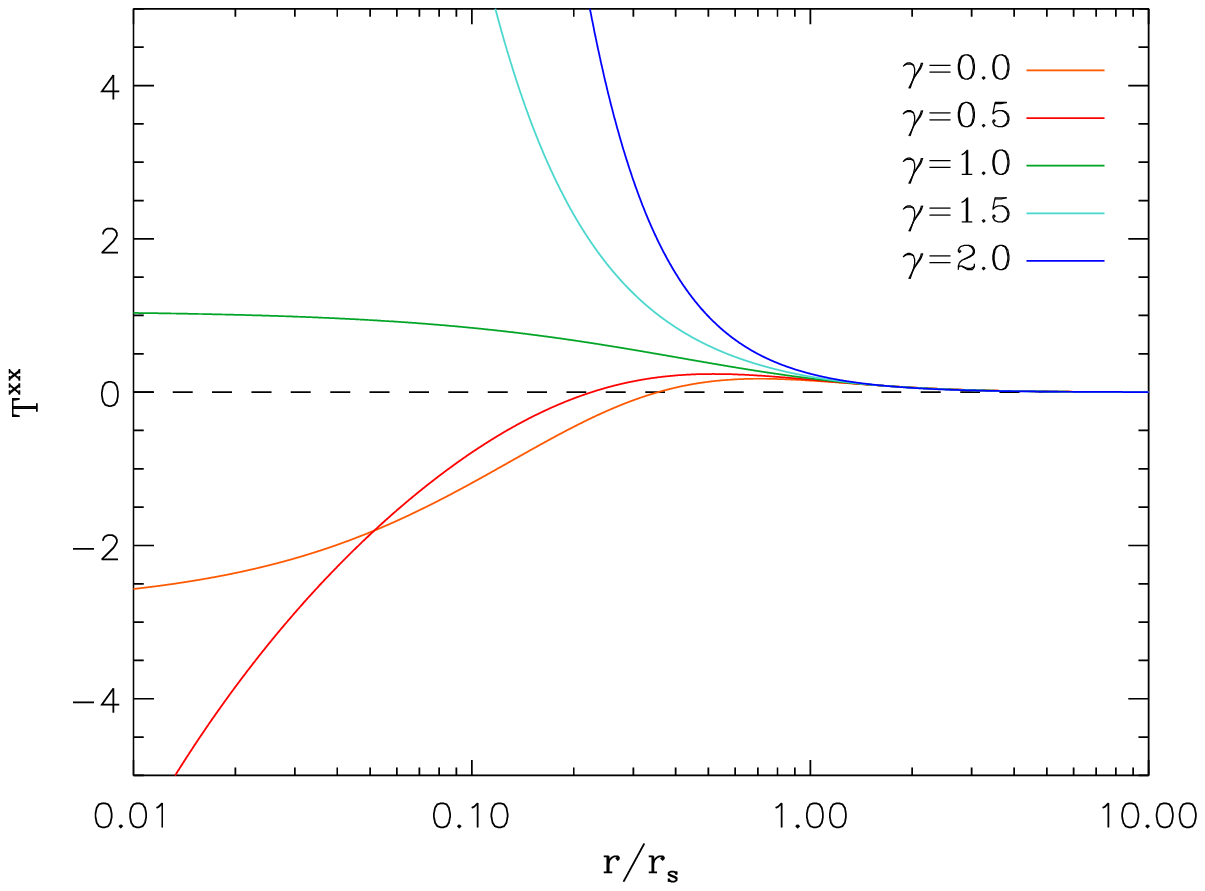}{Tidal profiles of the $\gamma$-family}
{Same as \reffig{analytical_tensor_halos_profiles}, but for some models of the $\gamma$-family.}

The same way, \reffig{analytical_tensor_gamma_profiles} shows the tidal profile of several members of the $\gamma$-family. Compressive tides only exist if $\gamma = \beta(r=0) < 1$. In these cases (corresponding to cored potentials), the compressive central region covers a radius of $r_{\gamma,0}\ (1-\gamma)/2$. In all the other cases, the tides are extensive everywhere. Again, a broad range of configurations for the tidal field is accessible from a restricted amount of density profiles.

%%%%%%%%%%%
\subsect{In symmetrical pairs}
The analysis of spherically symmetric mass distributions reveals that the group of profiles commonly used for modeling the \acr{}{DM}{dark matter} halos spawns very different tidal fields. As presented in the previous Chapters, the importance of the tides is highlighted when galaxies interact. Therefore, this Section repeats the analysis made in the previous one, but when two profiles are involved. Unfortunately, the symmetry is lost and thus, a purely analytical derivation of the tidal field becomes (very) challenging\footnote{The derivation of the tidal tensor can still be done using the superposition principle for the potential, and the distributivity of its second derivative over the addition. However, the global tensor yields an azimuthal dependence that would be averaged if considering the radial profile only.}. The first possible solution is to perform $N$-body simulations. However, such a method introduces noise in the rendition of the potential, which propagates in the computation of the tidal tensor. Furthermore, the Einasto and \acr{}{NFW}{NFW} profiles would be hardly distinguishable at typical resolution, and thus their differences would be buried in the $N$-body noise. Finally, it is not always easy to construct distribution functions for all theoretical models, as mentioned before.

The alternative adopted here consists in a semi-analytical method: the tidal tensor of each individual halo is calculated analytically (i.e. by hand) and then implemented as a function in a numerical code. Using the parameters of the global system (masses, sizes and positions of the halos), the code sums the individual tidal tensors at every point in space to get a global tensor and finally set it in diagonal form. Therefore the tidal field is computed for a \emph{static} configuration (see \refsec[halos]{static} for a discussion on this point and the limits of this approach). We focus on configurations close to those of the first pericenter passage of interacting galaxies: the overlap of two identical potentials, one shifted by a distance $d$ with respect to the other.

\fig{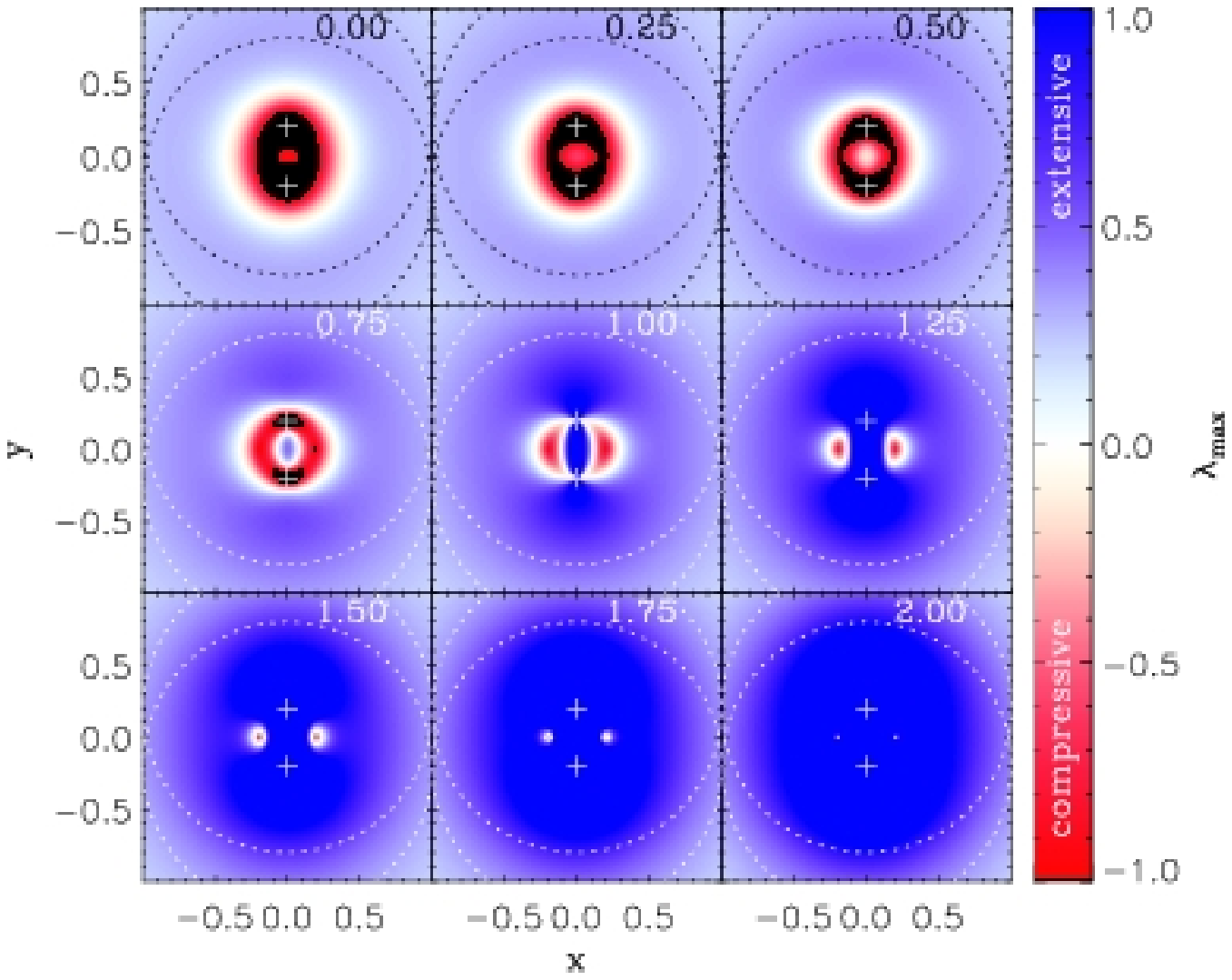}{Map of $\lambda_\mathrm{max}$ for the $\gamma$-family}
{Map of the maximum eigenvalue of the tidal tensor for two identical $\gamma$ profiles separated by a distance $d=0.4\ r_\mathrm{s}$ and centered on the plus signs. Dotted circles mark the radii $r_\mathrm{s}$. The number in the top-right corner of each panel represents the value of $\gamma$. Black is used to enhance the contrast and denotes extremely compressive areas ($\lambda_\mathrm{max} < -1$). (Figure~1 of \mycitealt{Renaud2010a}.)}

\reffig{dehnen_map} plots the maximum eigenvalue $\lambda_\mathrm{max}$ of the total tensor of the $\gamma$ profiles, with red denoting fully compressive tides and blue, extensive regions. In the immediate vicinity of the centers (white plus signs), the tidal field is as in isolation, i.e. a compressive mode only exists for $\gamma < 1$. At larger distances however, non-intrinsic red zones appear, due to the overlap of the two halos. Interestingly, these zones are visible up to $\gamma \simeq 2$, while isolated models are totally extensive for $\gamma \geq 1$. This indicates that the overlap of the potentials changes the very \emph{character} of the field. Note that these regions are not situated between the wells but on the perpendicular axis. Therefore, these compressive modes may impact on the (possibly) embedded baryonic matter in different ways, depending on its alignment with respect to the orbit of the merger.

\fig{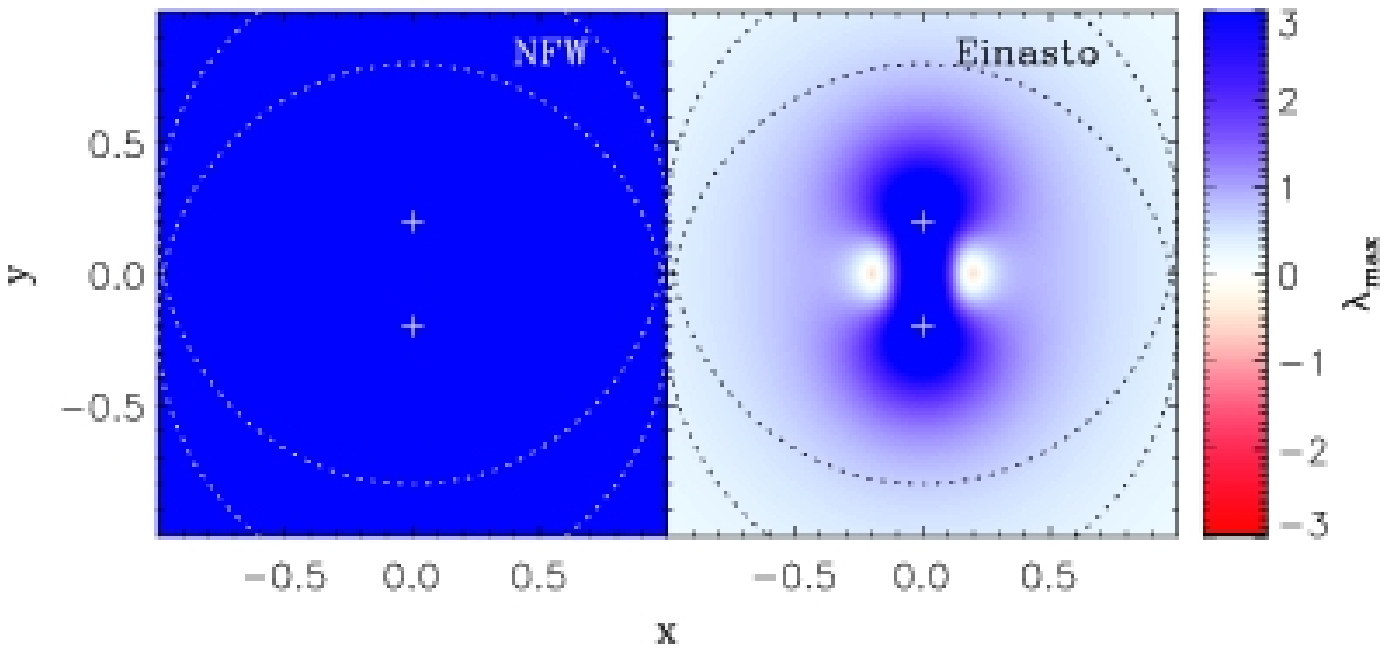}{Map of $\lambda_\mathrm{max}$ for NFW and Einasto}
{Same as \reffig{dehnen_map} but for two \acr{notarg}{NFW}{NFW} (\emph{left}) and two Einasto (\emph{right}) profiles. Note that the amplitude of the color bar has changed, to enhance the contrast. Clearly, the character of the tidal field (compressive or extensive) strongly depends on the details of the profile chosen. (Adaptation of Figure~2 of \mycitealt{Renaud2010a}.)}

In \reffig{nfw_einasto_map} the experiment is repeated but with two \acr{}{NFW}{NFW} and two Einasto profiles. As before, the intrinsic compressive region of the Einasto model is found at the center of the wells. Note that its diameter ($\sim 10^{-2}\ r_\mathrm{s}$, recall \reffig{analytical_tensor_halos_profiles}) corresponds to $\approx 0.5 \U{mm}$ on the Figure. Therefore, it is hidden behind the plus sign\footnote{In the electronic version of the manuscript, the reader can zoom-in on the sign and discover a small, yet visible red and black pattern. With the highest zoom factor, it is even possible to detect an asymmetry: the extremely compressive part (black) is preferentially found opposite to the second well, while the red zone faces it.}. In addition, the patterns found in \reffig{dehnen_map} for e.g. $\gamma =1.25$ are approximatively reproduced, in sharp contrast with the \acr{}{NFW}{NFW} case that does not yield any compressive region.

To conclude, the differences in the intrinsic tidal fields of Einasto and \acr{}{NFW}{NFW} already noted in \reffig{analytical_tensor_halos_profiles} are strengthened when another, identical profile is involved: secondary compressive regions exist between the Einasto wells while the entire space remains in extensive mode with \acr{}{NFW}{NFW} halos.

%%%%%%%%%%%
\subsect{Other configurations}
The results obtained in the previous Section can be extended to other distances $d$. First, we have checked that \acr{}{NFW}{NFW} profiles never give out compressive tidal modes, for any separation of the wells. Secondly, the upper row of \reffig{einasto_distance_mratio_map} plots the map of $\lambda_\mathrm{max}$ for two Einasto halos closer (left panel) and more distant (right) than in the previous case. The size and amplitude of the non-central compressive regions show a clear dependence with $d$, as expected according to the results of \refsecp[survey]{flybys_mergers}. 

To pursue the investigation of other configurations, the mass ratios $\mu = M_1 / M_2$ can also be changed, by setting a different weight for the two tidal tensors: $\matr{T} = \matr{T}_1 + \mu \ \matr{T}_2$.
(Note that this method allows us to scale the \acr{}{NFW}{NFW} halo although it has an infinite mass.) With the view of modeling galactic halos, the radial scale ($r_0$) is adjusted so that the average density is kept constant (recall \refeqnt{resizing}). As expected, the \acr{}{NFW}{NFW} halos remain entirely in extensive mode. The bottom-row of \reffig{einasto_distance_mratio_map} shows the results for Einasto profiles with the mass ratios 3:1 (left panel) and 10:1 (right panel). The symmetry of the compressive regions is obviously broken but the overall topology is well conserved. These regions stand closer to the lightest halo (the upper one on the Figure) than to the major one because the extensive modes are weaker there, and can easily be compensated or even overcome.

\fig{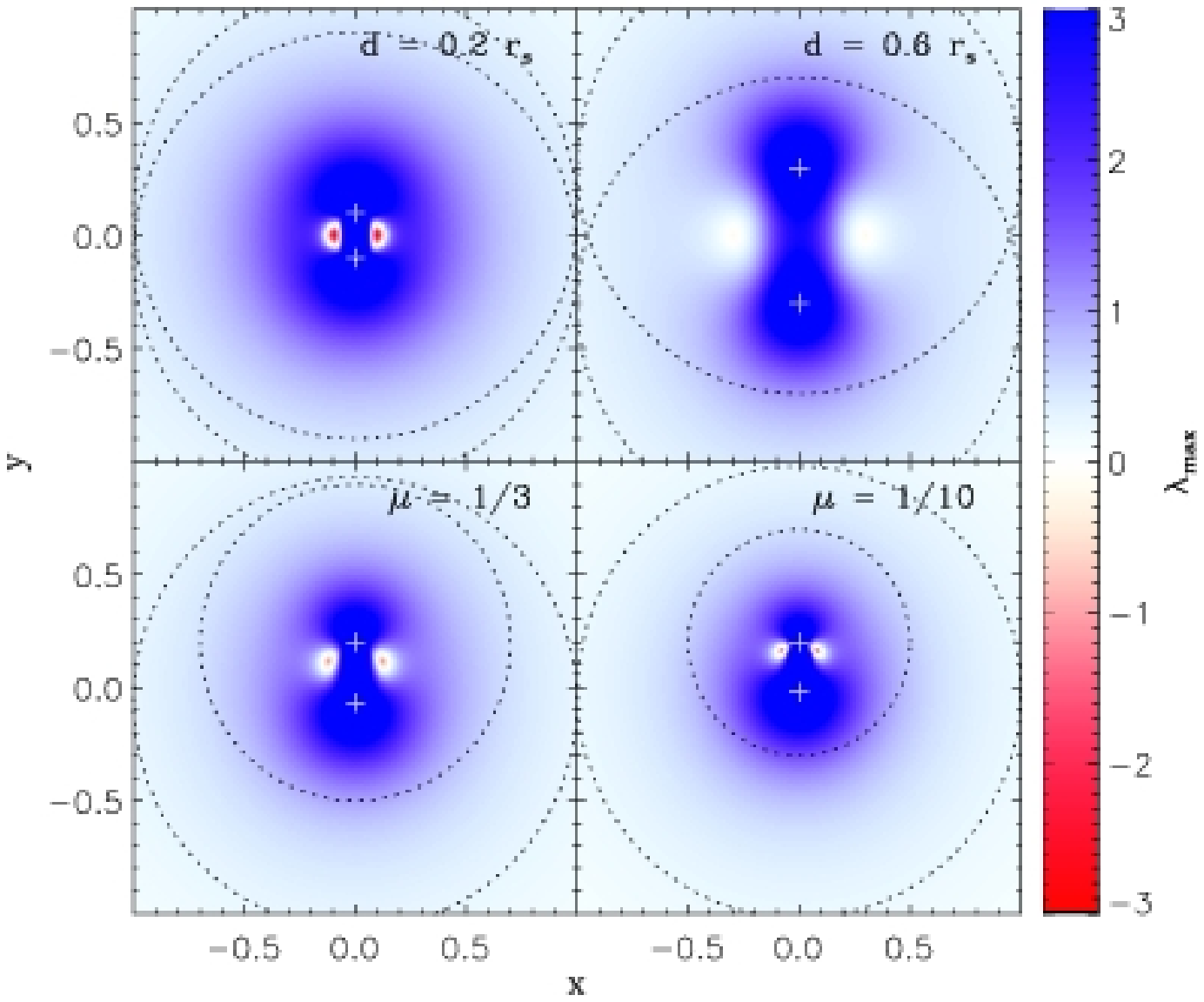}{Map of $\lambda_\mathrm{max}$ for several distances and mass ratios}
{Same as previous but for Einasto profiles separated by $d=0.2\ r_\mathrm{s}$ (\emph{top-left}) and $d=0.6\ r_\mathrm{s}$ (\emph{top-right}). In the bottom row, the distance remains $d = 0.4 (r_{\mathrm{s},1}+r_{\mathrm{s},2})/2$ but with a mass ratio of 3:1 (\emph{bottom-left}) and 10:1 (\emph{bottom-right}). In these cases, the lighter halo is the upper one. The global center of mass is always set at the origin.}

The amplitude of both the compressive and the extensive modes is comparable in all three cases presented (1:1, 3:1 and 10:1). This confirms that the strength of the tidal field mainly depends on the distance between the structures and is fairly invariant with the mass ratio.

The situation depicted above corresponds to well-defined events in cosmological simulations when for instance, a sub-halo falls onto a main object. In that case, the compressive modes that develop around its trajectory may influence the dynamics of the material already present in these regions.

%%%%%%%%%%%%%%%%%%%%%%%%%%%%%%%%%%%%%%%%%%%%%%%%%%
\sect{Validity and limits of the method}

%%%%%%%%%%%
\subsect{Role of the baryons}
In this Chapter, we have focused on configurations applicable to interacting spiral galaxies for which the separation is a fraction of the diameter of the disks. At these scales, a sole description of \acr{}{DM}{dark matter} is not enough to address the question pertaining to the dynamics of the system. Therefore, baryonic physics should be included, the same way it is done in the previous Chapters, by taking the bulges and the disks into account. However, the purpose of this study is to show that the halos alone already influence the character of the tidal field over $\sim 10 \kpc$ scales.

The results presented demonstrate that, on the one hand the choice of the shape of the halo has a limited impact on the overall density profile but, on the other hand it strongly affects the properties of the tidal field. When putting these facts in the context of star formation (recall \refcha{formation}), the \acr{}{NFW}{NFW} profile does not favor the formation of clusters or \acr[s]{}{TDG}{tidal dwarf galaxies}, though the comparable Einasto profile does. Note that we do not claim that a given profile would \emph{prevent} star formation, but simply that its tidal field creates a less favorable environment with respect to another \acr{}{DM}{dark matter} distribution. The additional baryonic physics would set the fine details of these processes, on the base of the conclusions drawn here.

%%%%%%%%%%%
\subsect[static]{Static versus live}
This semi-analytic approach uses the superposition of two spherically symmetric halos separated by a distance of the same order of magnitude than their characteristic scale. Because this situation is comparable to what happens in interacting galaxies, one may ask whether the deformation of one halo due to the tidal field of the other can really be neglected. Indeed, the phenomenon responsible for the creation of the tidal tails also influences the morphology of the \acr{}{DM}{dark matter}\footnote{in a milder way though, because the velocity dispersion of the halo is more important than the one of the disk.}.

One of the assumptions of this study is that the static configurations presented resemble the one of a \emph{first} pericenter passage, thus before the halos are strongly affected by the interaction. To verify this last point, \reffig{hernquist_superposition} compares the maps of $\lambda_\mathrm{max}$ of a ``real'' live merger to the one from a static equivalent. In a first experiment, two Hernquist potentials are rendered with $5 \times 10^5$ particles each, and placed on an elliptical orbit. The equations of motions are solved by {\tt gyrfalcON} (see \refcha{numerical}) up to the first pericenter passage, when the tidal field is computed (panel a). For the second experiment, the two halos are replaced by their initial mass profiles ($N$-body again) at their positions at the first passage, enforcing \emph{de facto} the spherical symmetry (panel b). The two maps are hardly distinguishable, which confirms that the halos are not morphologically evolved \emph{before the first passage}. On panel c, the result of our semi-analytical approach for the same configuration is shown, which helps to appreciate the level of numerical noise in the previous two cases.

\fig{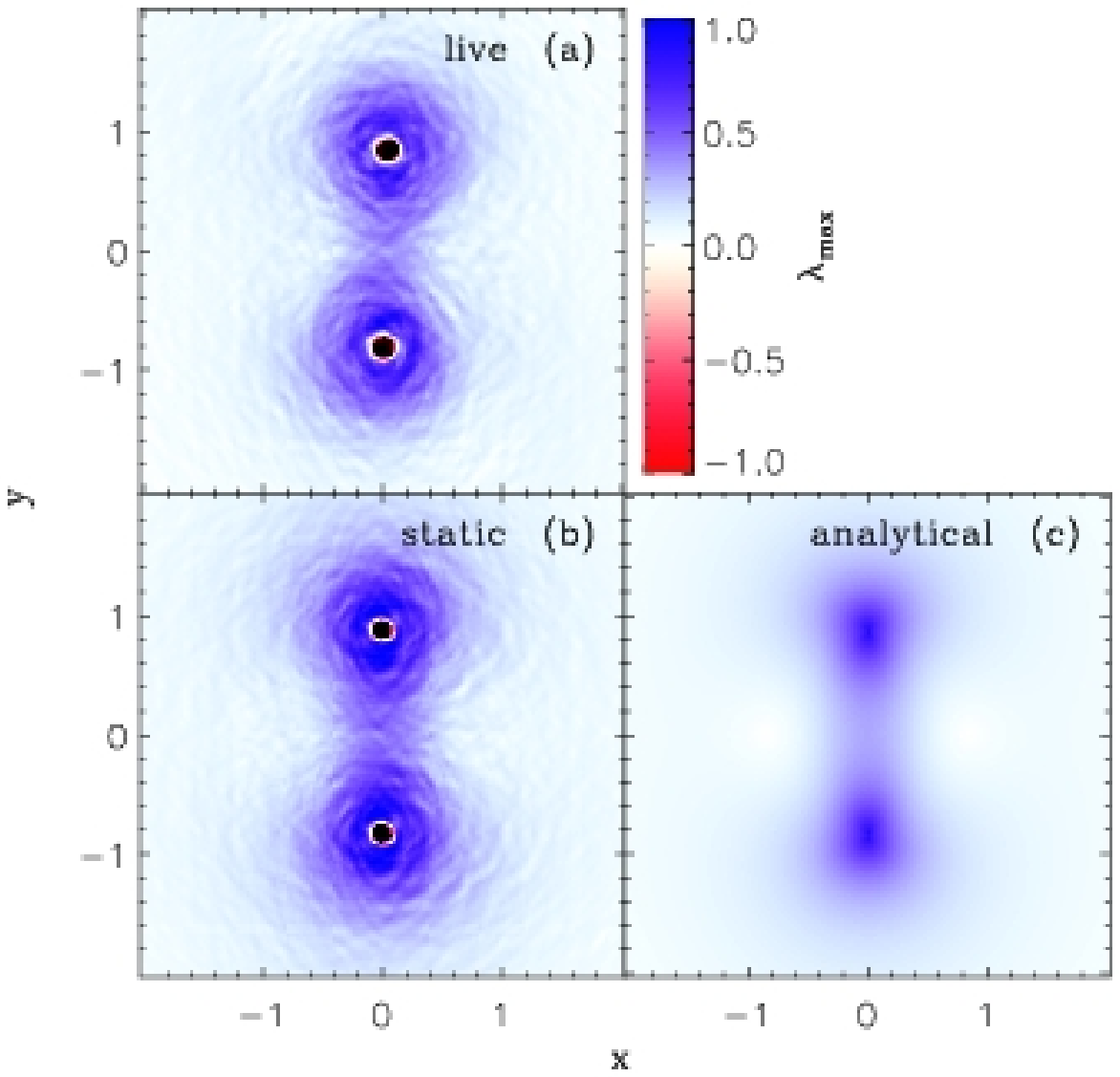}{Live, static and analytical}
{Map of $\lambda_\mathrm{max}$ for two live (a), static (b) and analytical (c) Hernquist potentials. The three experiments yields comparable results (except the numerical noise), which demonstrates that the halos are not morphologically affected by the interaction prior to their first passage. Note that the difference visible in the central part of the wells comes from a mathematical discontinuity at $r=0$ (see \refeqnt{tt_hernquist}) that cannot be reproduced numerically.}

%%%%%%%%%%%
\subsect{Triaxial halos}
The \acr{}{DM}{DM} halos extracted from cosmological simulations are not perfectly spherical objects but triaxial, ranging from oblate to prolate (see e.g. \mycitealt{Davis1985}; \mycitealt{Hayashi2007} and references therein). This asymmetry becomes important when considering a disk in rotation inside the halo: its plane should lie close to the one defined by the axes of the halo. Not taking this into account would false the interpretation of the velocity curves of such disks and, in the end may even bias the conclusions on the inner slope of the \acr{}{DM}{dark matter} profiles (see \mycitealt{Hayashi2006} on the cusp-core problem).

For a preliminary study like the one presented in this Chapter, we have restricted the investigation to spherically symmetric distributions, i.e. described by a monopole term. It is possible however to extrapolate to quadrupoles, or ``bared'' systems  by combining two identical potentials A and B, separated by a distance $\epsilon \ll r_\mathrm{s}$, thus forming an elongated structure. This new object has been used as one of the halos involved in our previous configurations, the other being a third, spherical potential C. To keep the mass ratio between the halos equal to unity, the A and B profiles have been downscaled by a factor two in mass (and $2^{1/3}$ in size, recall \refboxt{rescale_galaxy}). The distance between the barycenters of AB and C is set to $d = 0.4 \ r_\mathrm{s} \gg \epsilon$, as before. 

\fig{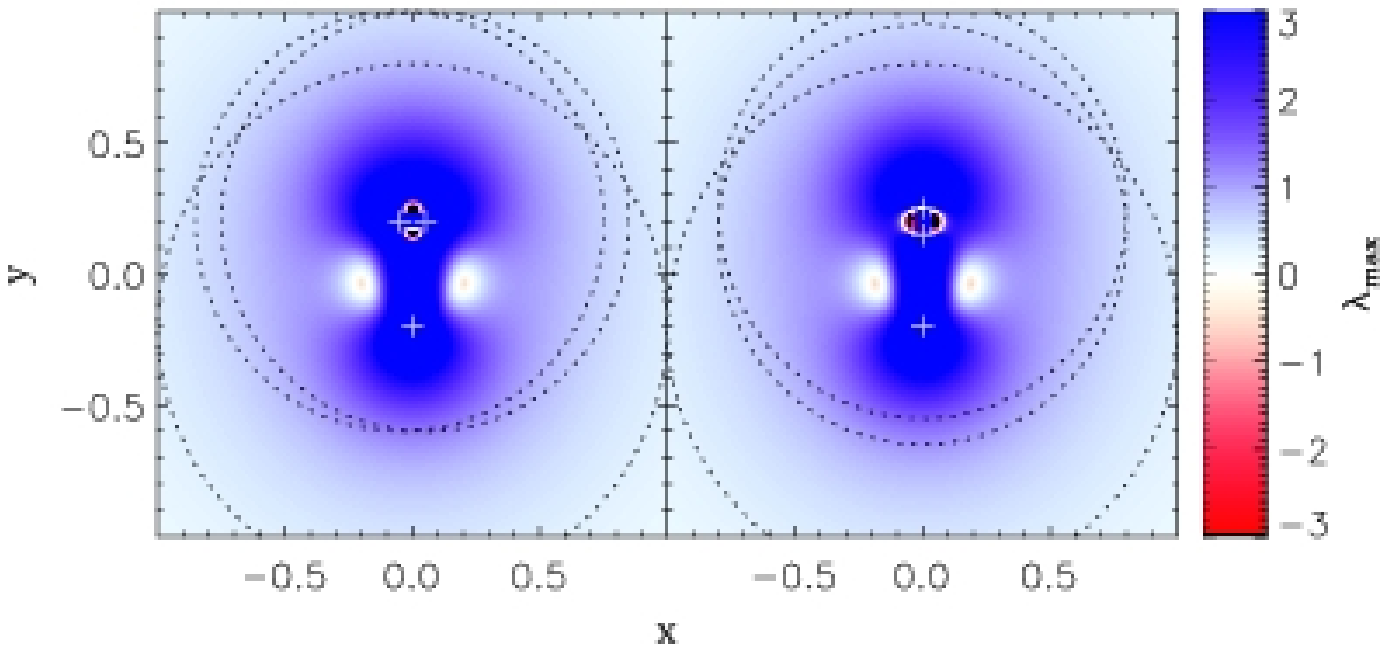}{Triaxial halos}
{Map of $\lambda_\mathrm{max}$ when an elongated structure is anti-aligned (\emph{left}) and aligned (\emph{right}) with another potential. The patterns of compressive zones remain comparable to those given out by spherical models.}

\reffig{einasto_triaxial_map} displays the usual map when the bar is anti-aligned and aligned with C. For both orientations, the resulting map is equivalent to the one of two pairs of potentials with respective separations $\epsilon$ and $d$. The morphology of the compressive regions slightly changes with the orientation of the bar. However, the large scale structures remain fairly unaffected by the triaxiality. Separations up to $\epsilon \sim d$ have also been explored to confirm the robustness of these points. In conclusion, the presence of a triaxial halo modifies the size and shape of the tidal regions (extensive and compressive) but does not influence the very character of the field.

%%%%%%%%%%%%%%%%%%%%%%%%%%%%%%%%%%%%%%%%%%%%%%%%%%
\sect[mond]{Modified dynamics}

At the beginning of this Chapter, we have introduced to concept of \acr{}{DM}{dark matter} to explain velocities higher than expected from the luminous mass in the context of Newtonian dynamics. However, instead of changing the mass, it is also possible to change the laws of physics. This is the line pursued by the \acr{}{MOND}{modified Newtonian dynamics} theory (\mycitealt{Milgrom1983}; \mycitealt{Bekenstein1984}). It proposes that the intensity of the gravitational force $\vect{F}$ decreases as the inverse of the distance ($r^{-1}$) in the weak field regime, while it stays Newtonian ($r^{-2}$) elsewhere:
\eqn{
\vect{F} = m \ \mu\left(\frac{a}{a_0}\right) \vect{a} \qquad\textrm{with}\quad \mu(x) = \left\{
\begin{array}{l}
1\quad\textrm{if } x \gg 1 \\
x\quad\textrm{if } x \ll 1,\\
\end{array}\right.
}
where $a$ is the acceleration and $a_0 \sim 1.2 \times 10^{-10} \U{m\ s}^{-2}$ the critical acceleration (see \mycitealt{Sanders2002} for a detailed review). In this case, the addition of dark matter to the ``classical'' mass is not required\footnote{at galactic scale. Several studies show that \acr{}{DM}{dark matter} is still needed in the \acr{}{MOND}{MOND}ian prescription at the scale of galaxy clusters (see \mycitealt{Sanders2002}).} and a typical spiral galaxy only consists of bare disk and bulge.

Recently, \mycitet{Tiret2007} presented a model of the Antennae galaxies, replacing the Newtonian gravitation with \acr{}{MOND}{MOND}. The morphology obtained is in good agreement with observational data, as shown in \reffig{tiret}.
\fig{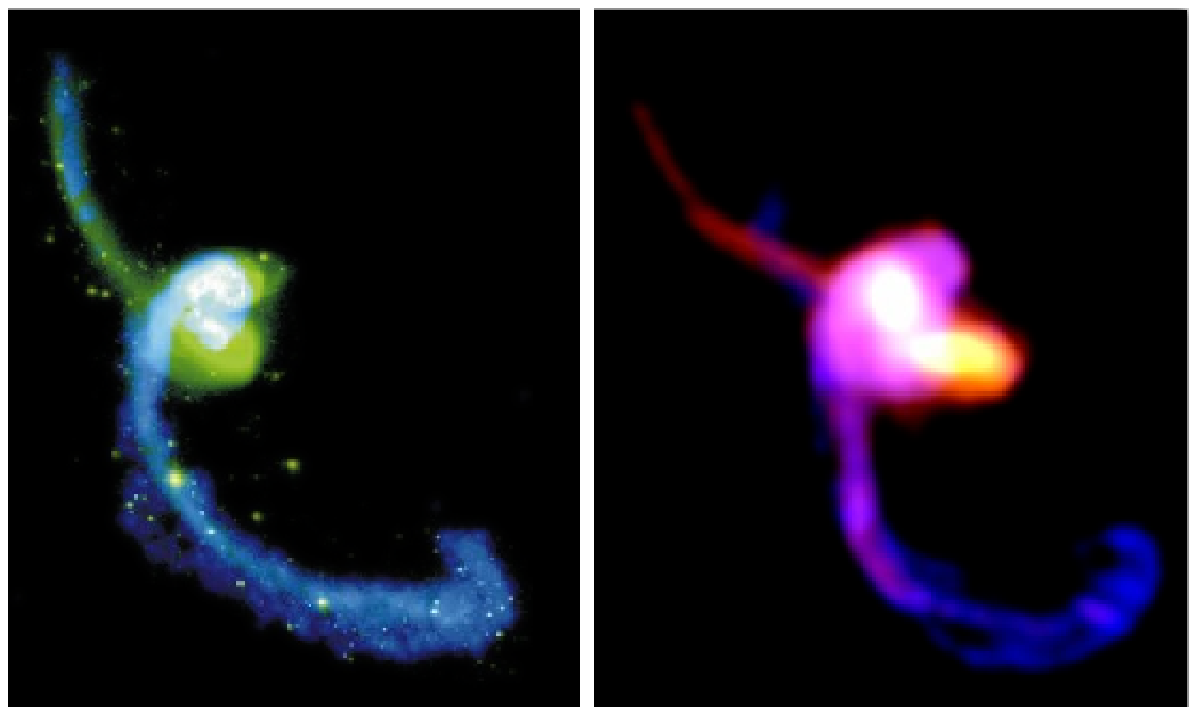}{Simulation of the Antennae with MOND}
{\emph{Left:} observations of the Antennae from \mycitet{Hibbard2001}. Green represents stars and the gas is blue. \emph{Right:} simulation with \acr{notarg}{MOND}{MOND} by \mycitet{Tiret2007}. Stars are shown in yellow/red shades and the gas is still blue. (Adaptation of Figure~2 of \mycitealt{Tiret2007}.)}

The tidal field of galaxies within the framework of \acr{}{MOND}{MOND} is expected to be different than the one found with Newtonian dynamics. However, nothing excludes \emph{a priori} the existence of compressive regions of similar shape or even intensity. To investigate this interesting point, a project in collaboration with H. Zhao (University of St Andrews) has been started, with the aims of (i) computing the tidal tensor thanks to \acr{}{MOND}{MOND} in the configuration of the Antennae obtained with Newtonian physics (e.g. those presented in \refcha{antennae}) and (ii) repeating the experiment on a fully \acr{}{MOND}{MOND}ian model. Lot of tests need to be performed first and unfortunately, it is yet too soon to present the results here.

%%%%%%%%%%%%%%%%%%%%%%%%%

%%%%
\chanonumber{Conclusions \& perspectives}

\vspace{1.0cm}
\begin{flushright}
\emph{Every new beginning comes from some other beginning's end.}\\
--- Seneca
\end{flushright}
\vspace{0.3cm}

%%%%%%%%%%%%%%%%%%%%%%%%%%%%%%%%%%%%%%%%%%%%%%%%%%
\sect{Conclusions}

During this study, we have focused on purely gravitational simulations of interacting galaxies and their tidal field. A collection of numerical tools has been developed to monitor the physical quantities related to this field. Thanks to it, we have shown that a special mode of the tides, called compressive mode, exhibits striking similarities with the properties of young substructures like star clusters and tidal dwarf galaxies.
\begin{itemize}
\item The evolution of the mass in compressive mode is strongly correlated to the star formation rate. Peaks exist in both quantities when the interaction is at its climax.
\item The compressive regions are situated at the positions of observed clusters and tidal dwarf galaxies.
\item Their typical durations ($\sim 10^7 \yr$) are comparable to the age of these structures.
\item The corresponding energy (from $\sim 10^{46}$ to $ 10^{57}\erg$) is sufficient to have a significant influence on the dynamical and chemical evolution of these objects.
\item The importance of the compressive mode decreases for mild, distant interactions, though it is of prime importance for mergers.
\end{itemize}

It has also been shown that, if the details of the statistics of the compressive tides vary with the configuration of the galaxies and their encounter, these conclusions are robust over a wide range of parameters.

It is now clear that the compressive tidal mode plays a non-negligible role in the formation and early evolution of star clusters. This demonstrates that a link exists between the large scales ($\sim 10^{1-2} \kpc$) where galactic dynamics set the morphology and kinematics of stellar and gaseous structures, and the smaller ones ($\sim 10^{1-2} \pc$) for which the hydrodynamics takes over and determines the formation and chemical evolution of stellar associations.

The details of such a link and the importance of the tidal field with respect to the other phenomena involved still need to be quantified thanks to high-resolution hydrodynamical simulations of giant molecular clouds but also young stellar associations. This next step requires a precise tuning of the parameters ruling the formation of stars and thus, will be done with extra-caution during the next months thanks to smoothed particle hydrodynamics and adaptive mesh refinement runs. Furthermore, a study of the dynamical evolution of a cluster embedded in a tidal field is underway. We are now introducing initial mass function and stellar evolution to model a broad range of internal phenomena in addition to the external effect of the tides. These simulations will help us to pin down the survival or destruction conditions of a star cluster, during the first $100 \Myr$ of its life.

Hopefully, all these projects will bring answers to many of the questions raised by the present work.

%%%%%%%%%%%%%%%%%%%%%%%%%%%%%%%%%%%%%%%%%%%%%%%%%%
\sect{Now and next}

In the continuation and aside from this work, I have conducted several projects related to the topic of this PhD. For some of them, the first results have just become available and are the starting points of very exciting explorations.

%%%%%%%%%%%
\subsect{Star formation in M51}
The Antennae galaxies have been chosen as observational reference because they are the closest major merger known. However, other mergers deserve as much attention as them, like for instance the grand design galaxy M51 and its companion NGC~5195. In this object, the tidal field has already played a crucial role in the formation of large scale structures like tails and spirals. It is then reasonable to question whether the smaller scales are also affected by the tides and in which fashion. 

\fig{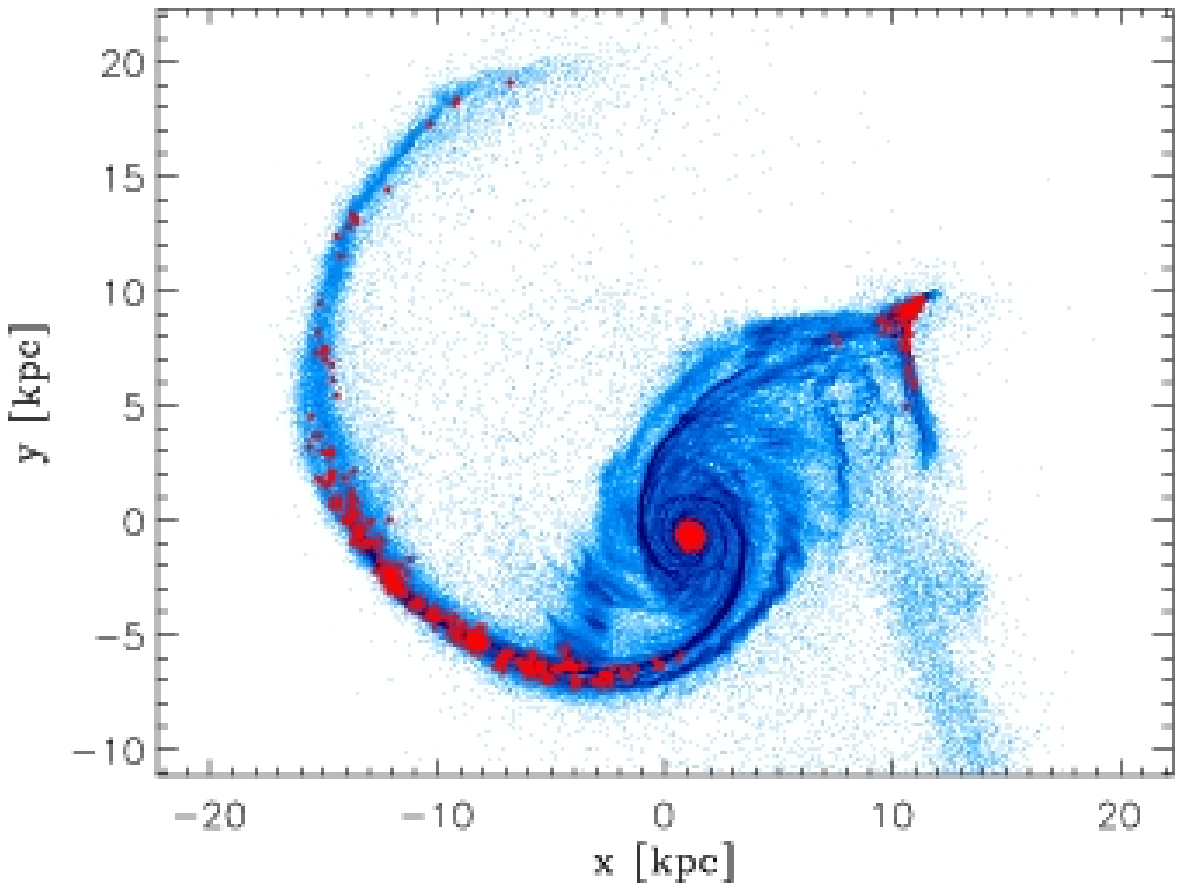}{Simulation of M51}
{Morphology of M51, $180 \Myr$ after the first passage of the intruder. Blue represents the surface density of the stars \emph{and} the gas. As usual, red shows the particles in compressive mode. The intruder NGC~5195 is a point whose mass is one third of the the one of M51.}

To address these points, I use the smoothed particles hydrodynamics simulation presented in \mycitet{Dobbs2010} to compute the tidal field and look for compressive regions. \reffig{m51} shows the very first results. Nodes of compressive modes exist along the main tidal tail, which suggests to push further the analysis of their properties, with respect to e.g. the Jeans mass or length. Interestingly, we also note that the spiral arms (which have been created during the encounter) are free of compressive zones. 

Thanks to the presence of gas in the model, we will monitor the regions of star formation and check for links with those of compressive tides. We will also study the evolution of the star formation rate through the major steps in the history of the interaction.

%%%%%%%%%%%
\subsect{$N$-body simulation of the Stephan's Quintet}
What applies to two galaxies should work for five. To confirm this basic statement, I focused on Stephan's Quintet, a compact group of five galaxies that shows a very complex web of tidal structures and hydrodynamical features, e.g. the largest shock known in the Universe. Up to now, many scenarios based on observations in several wavebands have been proposed to explain each structure. In this project, I suggest a new approach that consists in splitting the interaction history into well-defined and separated events and to link them sequentially to get the global ``storyboard'' of the group.

\fig{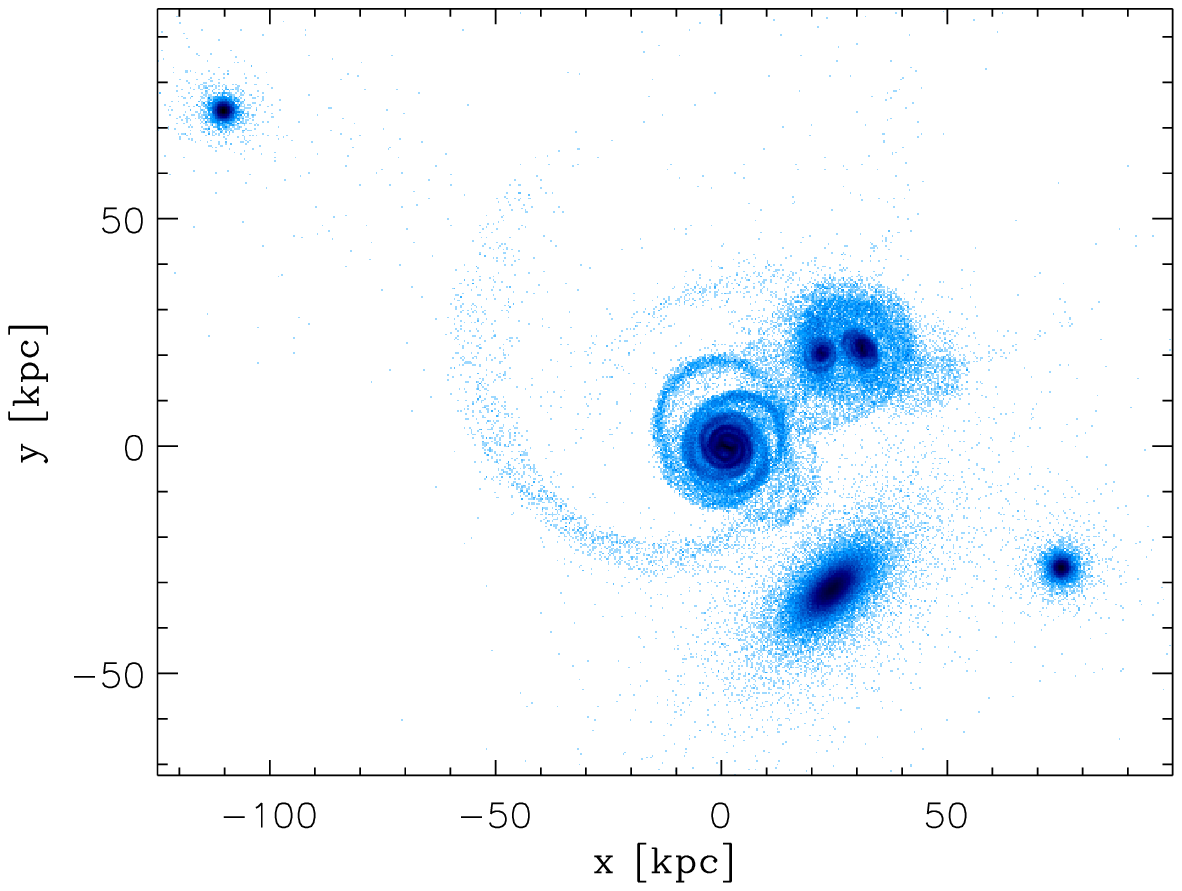}{Numerical simulation of the Stephan's Quintet}
{$N$-body simulation of the five members of Stephan's Quintet, plus the foreground galaxy in the lower part of the Figure. Tidal tails and inner structures match well the observations, despite some issues on the exact positions of the progenitors. (Adaptation of Figure~10 of \mycitealt{Renaud2010b}.)}

To verify this hypothesis and constrain the possible scenarios, I have performed more than 3\,000 simulations of the quintet to find a set of parameters that reproduces the morphology and the kinematics of the entire group (see \reffig{sq}). This manual exploration of the parameter space has shown that several scenarios found in the literature are very unlikely while others allow to fairly reproduce the observations. These results will be published soon and hopefully will give new lights on the observations to better constrain their interpretation.

The next step of this work will take these initial conditions and include the gaseous component needed to re-create the shock. This will be part of a larger project (gathering American and French observers and theoreticians) that has already started with the PhDs of Jeong-Sun Hwang in Iowa and Pierre Guillard in Paris (see \mycitealt{Hwang2010}; \mycitealt{Guillard2010}).

Obviously, the tidal field of the quintet will also be computed!

%%%%%%%%%%%%%%%%%%%%%%%%%

%%%%
\appendix\def\mychapname{Appendix } % used in the header
\cha{basic_concepts}{Basic concepts}
\chaphead{This Appendix briefly introduces some of the physical concepts or relations used in the main text.}

%%%%%%%%%%%%%%%%%%%%%%%%%%%%%%%%%%%%%%%%%%%%%%%%%%
\sect{Density \& potential}

When considering analytical definitions of mass profiles, it is extremely useful to get the gravitational potential from the density and \emph{vice versa}. In this Section, we detail the derivations of the relations between the two quantities.

%%%%%%%%%%%%%%%%%%%%%%%%%
\subsect[poisson]{From the density to the potential: Poisson's law}
The gravitational force on a particle of mass $m$ at the position $\vect{r} = x \vectu{e}_x + y \vectu{e}_y + z \vectu{e}_z$ due to an element of mass $\delta m$ situated at $\vect{r}' = x' \vectu{e}_x + y' \vectu{e}_y + z' \vectu{e}_z$ is
\eqn{
\delta  \vect{F}(\vect{r}) = G \ m \ \frac{\vect{r}'-\vect{r}}{|\vect{r}'-\vect{r}|^3} \ \delta m,
}
which can also be written thanks to the density $\delta m = \rho(\vect{r}') \ d^3r'$. By summing on all mass elements, one gets the total force
\eqn[forcepoisson]{
\vect{F}(\vect{r}) = G \ m \int \frac{\vect{r}'-\vect{r}}{|\vect{r}'-\vect{r}|^3} \ \rho(\vect{r}') \ d^3r'.
}
First, we notice that
\eqn{
\vect{\nabla} \left( \frac{1}{|\vect{r}'-\vect{r}|} \right) = \frac{(x'-x) \vectu{e}_x + (y'-y) \vectu{e}_y + (z'-z) \vectu{e}_z}{|\vect{r}'-\vect{r}|^3} = \frac{\vect{r}'-\vect{r}}{|\vect{r}'-\vect{r}|^3}.
}
Therefore, 
\eqn{
\vect{F}(\vect{r}) & = G \ m \int \vect{\nabla} \left( \frac{1}{|\vect{r}'-\vect{r}|} \right) \ \rho(\vect{r}') \ d^3r' \nonumber \\
& = m \ \vect{\nabla} \left( G \int \frac{\rho(\vect{r}')}{|\vect{r}'-\vect{r}|} \ d^3r' \right).
}
We now define the gravitational potential
\eqn[potential]{
\phi(\vect{r}) \equiv - G \int \frac{\rho(\vect{r}')}{|\vect{r}'-\vect{r}|} \ d^3r',
}
and write
\eqn{
\vect{F}(\vect{r}) = - m \ \vect{\nabla} \phi(\vect{r})
}
and 
\eqn{
\vect{\nabla} \cdot \vect{F}(\vect{r}) = - m \ \left( \vect{\nabla} \cdot \vect{\nabla} \right) \phi(\vect{r}) = - m \ \nabla^2 \phi(\vect{r}).
}

Back to \refeqn{forcepoisson}, we can take the divergence of the force and substitute it with the Laplacian of the potential
\eqn[poissondivergence]{
\vect{\nabla} \cdot \vect{F}(\vect{r}) & = G \ m \int \vect{\nabla} \cdot \left( \frac{\vect{r}'-\vect{r}}{|\vect{r}'-\vect{r}|^3} \right) \ \rho(\vect{r}') \ d^3r' \nonumber \\
- \nabla^2 \phi(\vect{r}) & = G \int \vect{\nabla} \cdot \left( \frac{\vect{r}'-\vect{r}}{|\vect{r}'-\vect{r}|^3} \right) \ \rho(\vect{r}') \ d^3r'.
}

\mybox{
We note that 
\eqn{
\vect{\nabla} \cdot \left( \frac{\vect{r}'-\vect{r}}{|\vect{r}'-\vect{r}|^3} \right) & = \frac{\partial}{\partial x} \left( \frac{x'-x}{\left[(x'-x)^2+(y'-y)^2+(z'-z)^2\right]^{3/2}} \right) \nonumber \\
& \quad  + \frac{\partial}{\partial y} \left( \frac{y'-y}{\left[(x'-x)^2+(y'-y)^2+(z'-z)^2\right]^{3/2}} \right) \nonumber \\
& \quad  + \frac{\partial}{\partial z} \left( \frac{z'-z}{\left[(x'-x)^2+(y'-y)^2+(z'-z)^2\right]^{3/2}} \right) \nonumber \\
& = -\frac{3}{|\vect{r}'-\vect{r}|^3}+3\frac{(x'-x)+(y'-y)+(z'-z)}{|\vect{r}'-\vect{r}|^5} \nonumber \\
& = -\frac{3}{|\vect{r}'-\vect{r}|^3}+3\frac{(\vect{r}'-\vect{r}) \cdot (\vect{r}'-\vect{r})}{|\vect{r}'-\vect{r}|^5} \nonumber \\
& = 0 \qquad \forall \ \vect{r}' \ne \vect{r}.
}
}

The integrand in \refeqn{poissondivergence} is null for $\vect{r}' \ne \vect{r}$. Therefore, we can take the density out of the integral to get
\eqn{
- \nabla^2 \phi(\vect{r}) = G \ \rho(\vect{r})\ \int \vect{\nabla} \cdot \left( \frac{\vect{r}'-\vect{r}}{|\vect{r}'-\vect{r}|^3} \right)  \ d^3r'.
}
Then, we change the reference frame of the divergence from $\vect{r}$ to $\vect{r}'$:
\eqn{
\nabla^2 \phi(\vect{r}) = G \ \rho(\vect{r})\ \int \vect{\nabla}_{\vect{r}'} \cdot \left( \frac{\vect{r}'-\vect{r}}{|\vect{r}'-\vect{r}|^3} \right)  \ d^3r'.
}
Thanks to the divergence theorem, we also have
\eqn{
\nabla^2 \phi(\vect{r}) = G \ \rho(\vect{r})\ \int \frac{\vect{r}'-\vect{r}}{|\vect{r}'-\vect{r}|^3} \ d^2\vect{s}.
}
Here, we can write
\eqn{
d^2\vect{s} = |\vect{r}'-\vect{r}| (\vect{r}'-\vect{r}) d\Theta,
}
where $d\Theta$ is the infinitesimal solid angle. Then, we get
\eqn{
\nabla^2 \phi(\vect{r}) = G \ \rho(\vect{r})\ \int d\Theta,
}
and finally the Poisson's law
\eqn[poisson]{
\nabla^2 \phi(\vect{r}) = 4\pi G \ \rho(\vect{r}),
}
which becomes
\eqn{
\frac{1}{r^2}\left[\frac{\partial}{\partial r} \left( r^2 \frac{\partial \phi(r)}{\partial r}\right)  \right]  =  4\pi G \ \rho(r)
}
for spherically symmetric systems.

%%%%%%%%%%%%%%%%%%%%%%%%%%
\subsect{From the potential to the density}
To invert Poisson's law and derive the density from the gravitational potential, we consider spherically symmetric systems only. At a given radius $r$, we can consider two regions: a sphere (inside) and a infinite shell (outside). By superposition, we can write
\eqn{
\phi(r) =  \phi_\mathrm{in}(r) + \phi_\mathrm{out}(r) \qquad \forall\ r \in [0,\infty[.
}
Newton's first theorem stipulates that the force due to a shell of matter is null, everywhere inside the shell. Therefore, the potential there must be constant and we can estimate it at any point. E.g., $\phi_\mathrm{out}(r) = \phi_\mathrm{out}(0)$.

By replacing $d^3\vect{r}'$ with $4\pi x^2 dx$ in the definition of the gravitational potential (\refeqn{potential}), and applying the formula for $r=0$, one gets
\eqn{
\phi_\mathrm{out}(r) = -4\pi G \int_0^{\infty}{x\rho_\mathrm{out}(x)\ dx}.
}
However, by construction, $\rho_\mathrm{out}(x<r) = 0$. Thus, 
\eqn{
\phi_\mathrm{out}(r) =  -4\pi G \int_r^{\infty}{x\rho_\mathrm{out}(x)\ dx}.
}
Furthermore, Newton's second theorem tells us that the force from the inside region is equivalent to the one of all its matter concentrated in a point-mass at the center. We obtain this enclosed mass by summing the masses of infinitesimal shells
\eqn{
\phi_\mathrm{in}(r) = -\frac{G M(<r)}{r} = -\frac{4 \pi G}{r} \int_0^r{x^2\rho_\mathrm{in}(x)\ dx}.
}
And finally, the global potential is
\eqn[rhotophi]{
\phi(r) =  -4\pi G \left( \frac{1}{r} \int_0^r{x^2\rho(x)\ dx} + \int_r^{\infty}{x\rho(x)\ dx} \right).
}
Note that by applying the Laplace operator to this expression, and considering the limit $\phi(\infty) \to 0$, one quickly finds Poisson's law.

%%%%%%%%%%%%%%%%%%%%%%%%%%%%%%%%%%%%%%%%%%%%%%%%%%
\sect[dynamical_friction]{Dynamical friction}

In restricted simulations, a galaxy is represented by a point-mass surrounded by pseudo-particles. Therefore, the orbit of the galaxy follows Kepler's laws of motion. However, an interesting fact about galaxy mergers is that they actually merge! Both galaxies loose energy when they collide and spiral on each other until they finally merge. This loss of orbital energy is called dynamical friction.

\fig{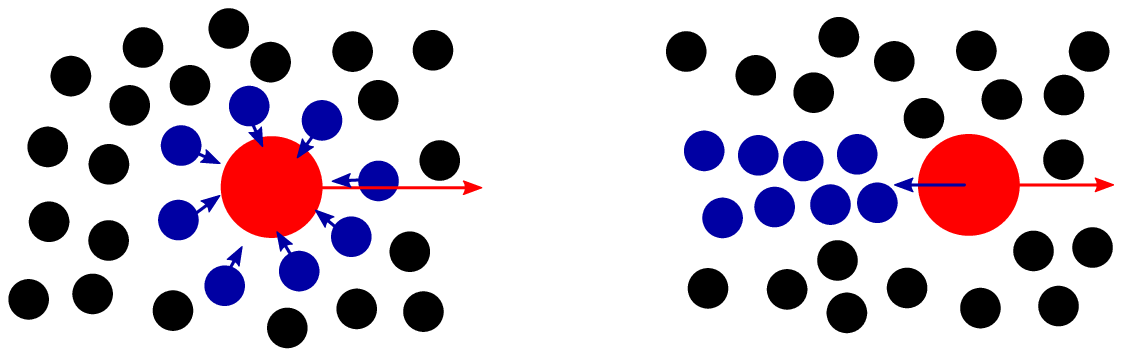}{Dynamical friction}
{An object (red) moves in a field of stars uniformly distributed (black). The closest stars (blue) are attracted toward the position of the object, but as it moves away (\emph{right}), these stars group into a retarding wake and attract the object back, which slows it down.}

Consider an object flying through a population of stars uniformly distributed (\reffig{dynamical_friction}). The closest stars are attracted by this object and tend to group around it. However, because of the proper motion of the flying object, this gathering of mass actually occurs behind it. This leads to a non-uniform distribution of the mass that drags the object backward, as a retarding wake. Finally, the motion of the object is slowed down: it looses kinetic energy (which is transfered to the nearby stars).
 
For galaxy mergers, that implies that the motion of an intruder through a primary galaxy will be strongly affected by dynamical friction. The initial kinetic energy of the intruder will be lost and its orbit would bend. An intruder initially set on a Keplerian parabolic orbit (eccentricity $e=1$) will effectively have a bounded orbit ($e < 1$) after it has experienced dynamical friction, through the material of another galaxy. \reffig{dynamical_friction_orbits} and \reffig{dynamical_friction_orbits_zoom} compare the orbits of two \mycitet{Plummer1911} spheres (blue) with the equivalent Keplerian orbit (red). The effect of dynamical friction in visible in the orbital decay of the Plummer spheres, showing that they have lost a fraction of their initial orbital kinetic energy (see also \reffig{dynamical_friction_distance}).

\fig{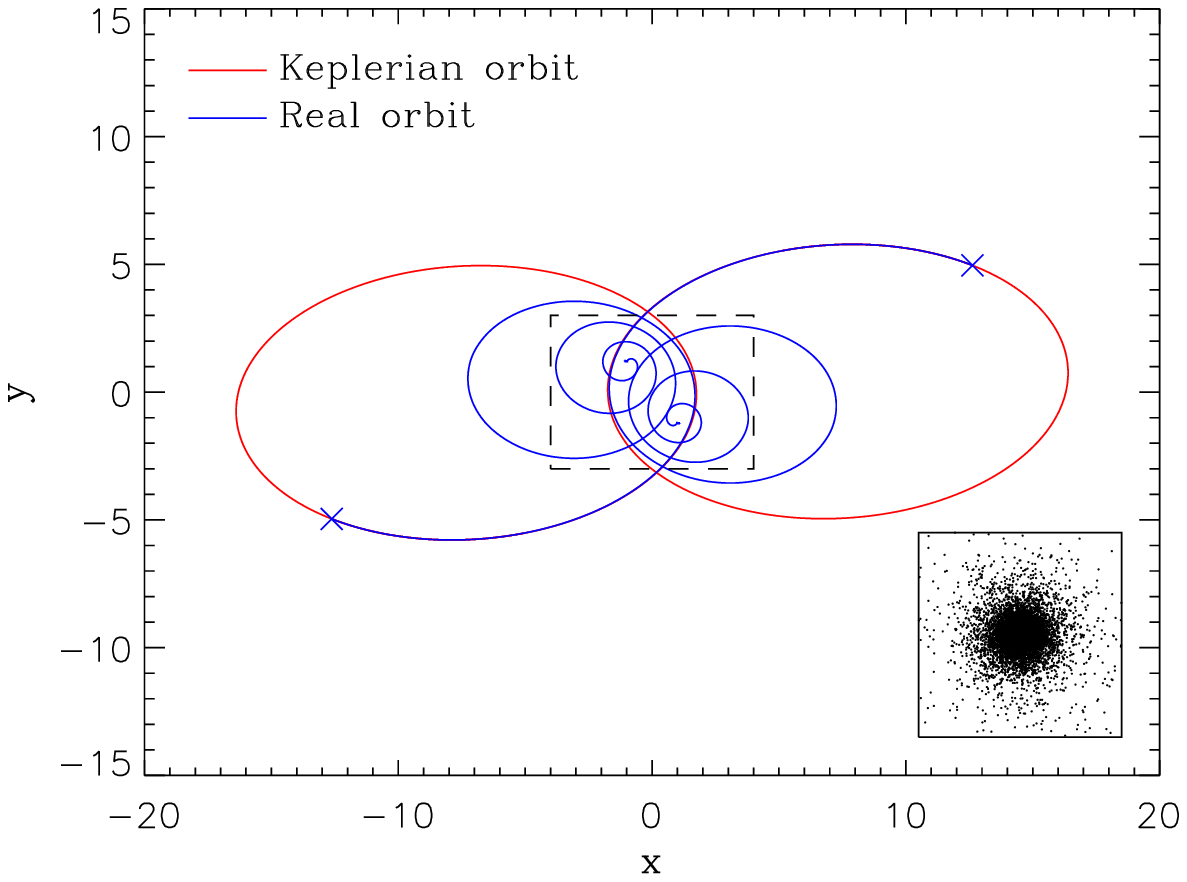}{Orbits of objects with and without dynamical friction}
{Orbit of the centers of mass of two Plummer spheres ($N=10\,000$, see their actual size in the thumbnail) initially set on a Keplerian orbit (of eccentricity $e=0.5$, red) at the position marked with $\times$. See \reffig{dynamical_friction_orbits_zoom} for a zoom-in on the central region (dashed rectangle).}

\fig{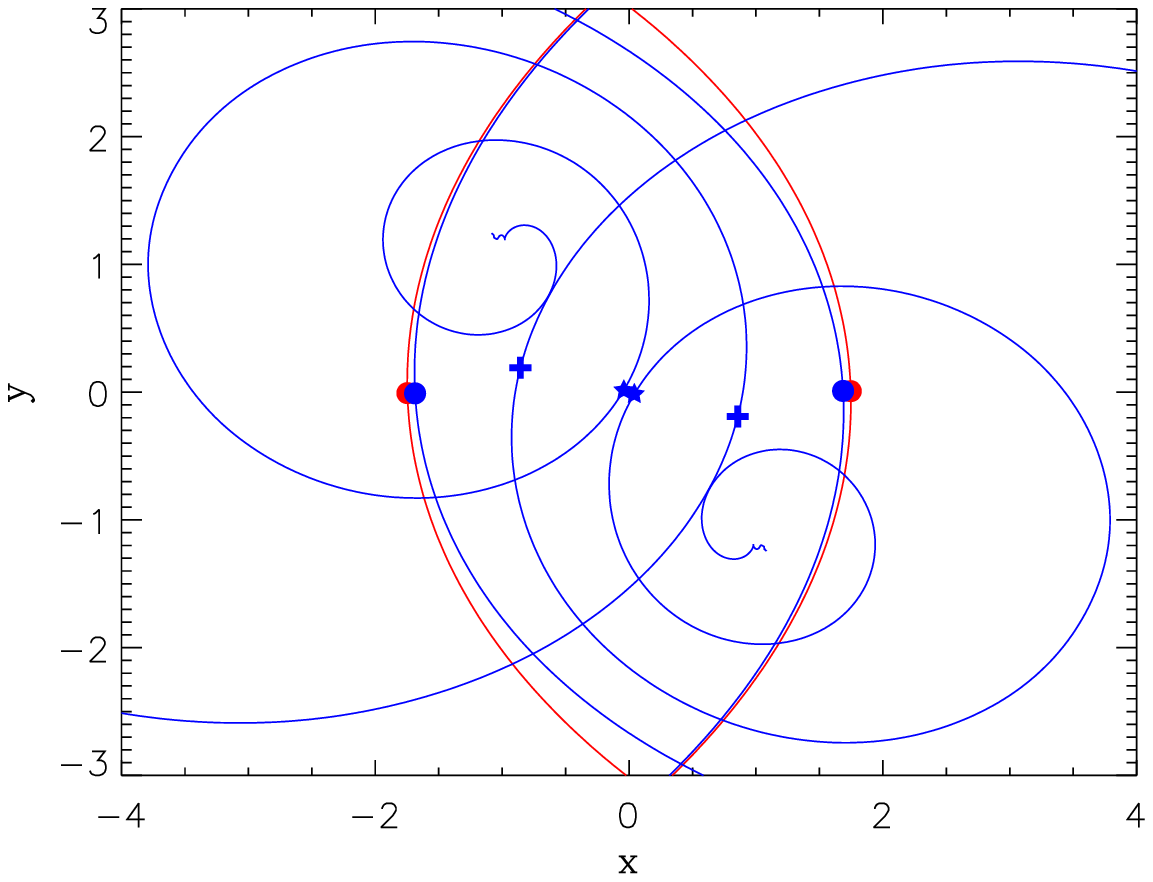}{Zoom-in on the orbits}
{Zoom-in on the central region of \reffig{dynamical_friction_orbits}. The pericenter passages are marked with $\bullet$, $+$ and $\star$, respectively. Note that the two objects do not spiral around the same point at the end of the simulation. This is because they undergo a very close passage which scatters an important fraction of their particles, leading to an offset in the computation of their centers of mass.}

\fig{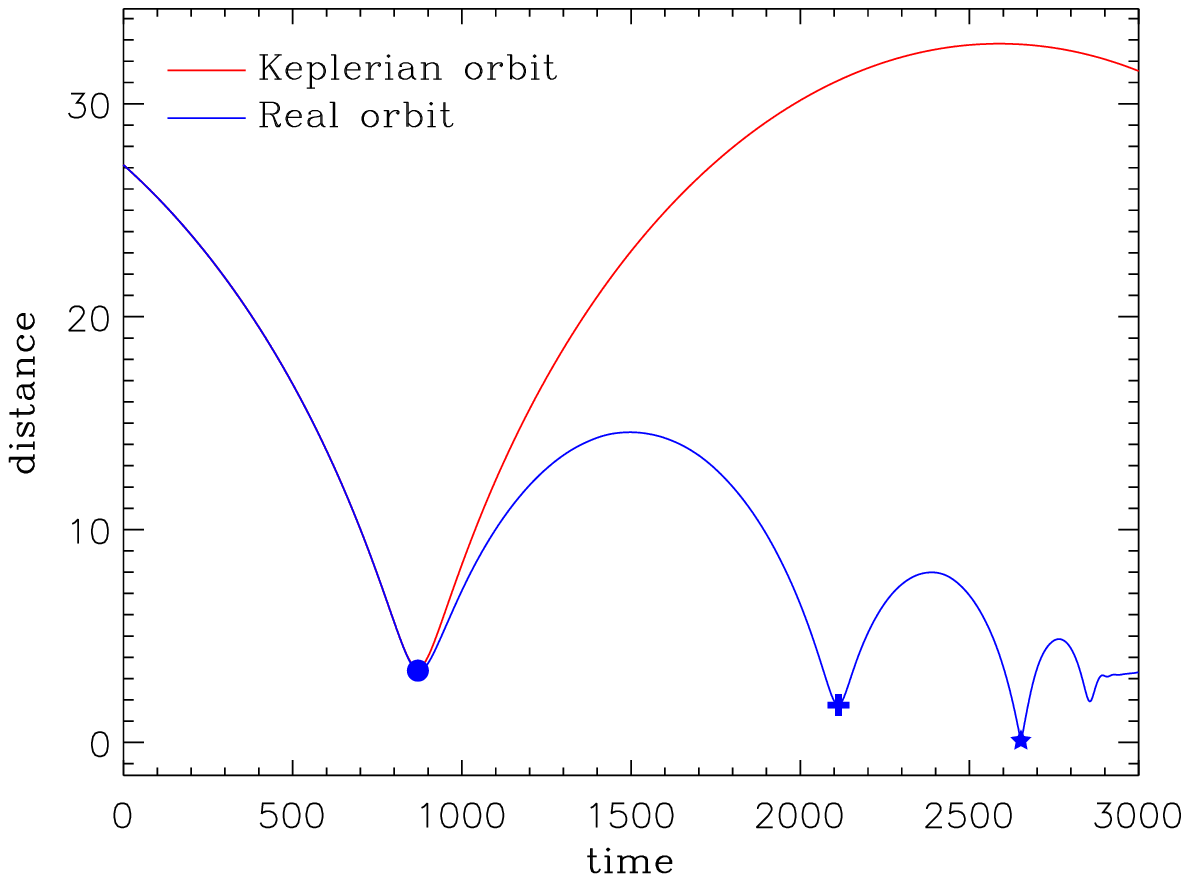}{Distance between two objects with and without dynamical friction}
{Evolution of the distance between the centers of mass of the Plummer spheres (see \reffig{dynamical_friction_orbits}). As in \reffig{dynamical_friction_orbits_zoom}, the symbols ($\bullet$, $+$, $\star$) correspond to the pericenter passages. The two configurations give out almost identical results up to the first pericenter passage. Afterward, the dynamical friction redistributes the orbital energy and the amplitude of the orbits decreases. The non-regular evolution of the distance after $t=2700$ corresponds to the scatter of particles (mentioned earlier) after the very close passage at $t \approx 2650$.}

This phenomenon has been studied for many years. \mycitet{Chandrasekhar1943} gave an mathematical expression for the velocity change the moving body undergoes:
\eqn{
\vect{F}_\mathrm{dyn} = -4\pi \ln{(\Lambda)}\frac{G^2M^2\rho}{v^3}\left(\mathrm{erf}(X) - \frac{2X}{\sqrt{\pi}}e^{-X^2} \right) \vect{v},
}
where $v$ is the velocity of the body, $M$ the mass considered (body + medium) and $\rho$ the density of the surrounding medium. $X = v / (\sigma \sqrt{2})$ denotes the ratio of the velocity to the one of a Maxwellian distribution with dispersion $\sigma$. The term $\ln{(\Lambda)}$, called Coulomb logarithm (by analogy with the theory of plasmas), is a constant depending on the maximum impact parameter between the bodies and their typical relative velocity. Its usual value ranges from unity to $\sim 10$.

Even if this mathematical description suffers from strong simplifications\footnote{For example, it neglects the self-gravity of the medium, in particular in the retarding wake that forms behind the moving body. In addition, the value $\ln{(\Lambda)}$ is rather difficult to compute and is often arbitrary chosen.} it is possible to implement it in restricted simulations and thus correct the motion of the galaxies. Although, a proper description of the structures forming during a collision requires self-gravity of this regions. Therefore, instead of the fast restricted runs, consistent $N$-body simulations are generally performed on powerful computers. In this case, each particle has a non-null mass and fully participates in the global potential, and the Keplerian solution is not valid anymore\footnote{An analytical solution exists for the two-body problem only, in the general case.}.

\cha{triplets}{Density-potential-tidal tensor triplets}
\chaphead{This Appendix details the analytical derivations of the main dynamical quantities of several density profiles. The notations $G$, $\phi$, $\rho$, $\matr{T}$, $r$, $M$ and $M(r)$ refer to the gravitational constant, the potential, the density, the tidal tensor, the radial coordinate, the total mass, and the mass enclosed in a sphere of radius $r$, respectively.}

%%%%%%%%%%%%%%%%%%%%%%%%%%%%%%%%%%%%%%%%%%%%%%%%%%
\sect[ttexamples]{Classical cases}

For simplicity, all the core or characteristic radii are designed by $r_0$, while $\rho_0$ represents the characteristic density. Note however that the physical definition of these quantities may vary from one model to the other, as discussed in \refsecp[halos]{comparisons}.

%%%%%%%%%%%%%
\subsect[plummer]{Plummer}
We consider the \mycitet{Plummer1911} potential given by 
\eqn[plummerpot]{
\phi(r) = -\frac{GM}{\sqrt{r_0^2+r^2}}.
}
Using Poisson's law (\refeqnt{poisson}), we get the density
\eqn[plummerdensity]{
\rho(r) = \frac{3Mr_0^2}{4\pi \left(r_0^2+r^2\right)^{5/2}}.
}
To get the integrated mass, one can solve
\eqn{
M(r) = 4\pi \int_0^r x^2 \rho(x)\ dx = \frac{3M}{r_0} \int_0^r \frac{\frac{x^2}{r_0^2}}{\left(1+\frac{x^2}{r_0^2}\right)^{5/2}}\ dx
}
(which is possible through the change of variable $\tan{\theta} = x/r_0$), or simply note that the system must obey Newton's second theorem
\eqn{
-\frac{GM(r)}{r^2} = -\nabla \phi(r),
}
and get (much quicker!)
\eqn[plummerintegratedmass]{
M(r) = \frac{M r^3}{ \left( r_0^2 + r^2\right)^{3/2}}.
}
We deduce that the radius $r_0$ encloses $\approx 35\%$ of the total mass.

Starting with the potential (\refeqn{plummerpot}) and using Einstein's summation convention, we can derive the tidal tensor (see \refeqnt{symmetrictensor}):
\eqn{
T^{ij} = GM \partial^i \left( - \frac{x_j}{\left(r_0^2+r^2\right)^{3/2}}  \right)
}
and then, 
\eqn[plummertt]{
T^{ij} = -GM \frac{ \delta^{ij}\left(r_0^2+r^2\right) - 3 x_i x_j }{\left(r_0^2+r^2\right)^{5/2}}
}
(with $\delta^{ij} = 1$ if $i=j$ and 0 otherwise), which gives, for $r = x$ (i.e. $y=z=0$):
\eqn{
T^{xx}(r) = -GM \frac{r_0^2 -2r^2}{\left(r_0^2+r^2\right)^{5/2}}.
}

%%%%%%%%%%%%%
\subsect{Logarithmic potential}
This profile is defined by
\eqn{
\phi(r) = \frac{1}{2}v_0^2 \ln{\left(1+\frac{r^2}{r_0^2}\right)},
}
where $v_0$ denotes the circular velocity for the outer regions of the profile. The density is given by Poisson's law
\eqn{
\rho(r) = v_0^2 \frac{r^2 + 3r_0^2}{\left(r^2+r_0^2\right)^2}
}
and the integrated mass comes from Newton's second theorem
\eqn{
M(r) = v_0^2 \frac{r^3}{G \left(r^2+r_0^2\right)}.
}
We note that this mass integrates to infinity. Thus, it is impossible to use the term $M$ in our calculations. For the tidal tensor, we have
\eqn{
T^{ij} = -v_0^2 \partial^i \left(\frac{x_j}{r^2+r_0^2}\right)
}
and then
\eqn{
T^{ij} = -v_0^2 \frac{\delta^{ij} \left(r^2+r_0^2\right) -2 x_ix_j}{\left(r^2+r_0^2\right)^2},
}
which gives, for $r = x$
\eqn{
T^{xx}(r) = v_0^2 \frac{r^2 - r_0^2}{\left(r^2+r_0^2\right)^2}.
}

%%%%%%%%%%%%%
\subsect{Hernquist}
The same can be done for the \mycitet{Hernquist1990} profile:
\eqn{
\phi(r) = -\frac{GM}{r_0+r}.
}
One can easily find the density
\eqn{
\rho(r) = \frac{Mr_0}{2\pi r \left(r_0+r\right)^3}.
}
Once again, the integrated mass is found thanks to Newton's second theorem
\eqn{
M(r) = \frac{Mr^2}{\left( r_0 + r\right)^2}.
}
We compute the tidal tensor
\eqn{
T^{ij} = -GM \partial^i \left(  \frac{x_j}{\left( r_0 + r \right)^2 r}  \right),
}
and get 
\eqn{
T^{ij} = -\frac{GM}{\left( r_0 + r \right)^3} \left[ \left(\delta^{ij}-\frac{x_ix_j}{r^2}\right) \frac{r_0+ r}{r} - \frac{2x_ix_j}{r^2} \right],
}
which gives, for $r = x$
\eqn[tt_hernquist]{
T^{xx}(r) = \frac{2GM}{\left(r_0+r\right)^3}
}
(see also \refsect[halos]{halos_tidal}, in the framework of the $\gamma$-family with $\gamma = 1$).

%%%%%%%%%%%%%
\subsect{$\gamma$-family}
\mycitet{Dehnen1993} and \mycitet{Tremaine1994} defined the $\gamma$-family of density-potential pairs as follows:
\eqn{
\rho(r) = \frac{(3-\gamma) M}{4\pi} \frac{r_0}{r^\gamma(r_0+r)^{4-\gamma}}
}
and
\eqn{
\phi(r) = \left\{ \begin{array}{ll} 
\displaystyle{ -\frac{GM}{r_0(2-\gamma)} \left[1-\left(\frac{r}{r_0+r}\right)^{(2-\gamma)}\right] } & \textrm{if }\gamma \neq 2 \\
\\
\displaystyle{ \frac{GM}{r_0} \ln{\left(\frac{r}{r_0+r}\right)} } & \textrm{if }\gamma = 2,
\end{array} \right.
}
which is equivalent to the \mycitet{Hernquist1990} profile for $\gamma = 1$, and to the \mycitet{Jaffe1983} profile for $\gamma = 2$. The integrated mass is therefore
\eqn[dehnen_mass]{
M(r) = \frac{M \ r^{3-\gamma}}{(r_0+r)^{3-\gamma}}.
}
For $\gamma \ne 2$, the associated tidal tensor is
\eqn{
T^{ij}_{\gamma \neq 2} & = -GM\partial^i \left[ x_j r^{-\gamma} \left(r_0+r\right)^{\gamma-3} \right] \nonumber \\
& = -\frac{GM}{r^{\gamma} \left(r_0+r\right)^{3-\gamma}} \left[ \delta^{ij} - \gamma \frac{x_ix_j}{r^2} + (\gamma - 3) \frac{x_ix_j}{\left(r_0+r\right) r} \right].
}
We note that the expression of $T^{ij}_{\gamma \neq 2}$ is equal to $T^{ij}_{\gamma = 2}$. So, we have the general expression
\eqn{
T^{ij} = -\frac{GM}{r^{\gamma} \left(r_0+r\right)^{3-\gamma}} \left[ \delta^{ij} - x_ix_j \frac{3r+r_0\gamma}{\left(r_0+r\right) r^2} \right]
}
for all $\gamma$'s, which gives, along the x-axis
\eqn{
T^{xx}(r) = GM\frac{2r + r_0(\gamma - 1)}{r^{\gamma} \left(r_0+r\right)^{4-\gamma}}
}
(see also \refsect[halos]{halos_tidal}).

%%%%%%%%%%%%%
\subsect[nfw]{NFW}
The density of the \acr{}{NFW}{NFW} (\mycitealt{Navarro1997}) profile is given by
\eqn{
\rho(r) = \frac{\rho_0 r_0}{ r \left( 1 + \frac{r}{r_0}\right)^2}.
}
Using \refeqnp{rhotophi}, we get
\eqn{
\phi(r) = - 4 \pi G \rho_0 r_0  \left( \frac{1}{r} \int_0^r{\frac{x\ dx}{\left(1 + \frac{x}{r_0}\right)^2}} + \int_r^{\infty}{\frac{dx}{\left(1 + \frac{x}{r_0}\right)^2}} \right).
}
Recall that the first integral (that we can solve with, e.g., a partial integration) is the integrated mass. Finaly, we obtain
\eqn{
\phi(r) = - 4 \pi G \rho_0 r_0^2   \frac{r_0}{r} \ln{\left(1+\frac{r}{r_0} \right)}
}
and
\eqn[nfw_mass]{
M(r) = 4\pi \rho_0 r_0^3 \left[ \ln{\left(1+\frac{r}{r_0}\right)} - \frac{r}{r+r_0} \right].
}
As for the logarithmic potential, the mass of an \acr{}{NFW}{NFW} profile integrates to infinity. The tidal tensor is given by
\eqn{
T^{ij} = 4 \pi G \rho_0 r_0^3 \partial^i \left[\frac{x_j}{r^2\left(r_0 + r\right)} - \frac{x_j \ln{\left(1+\frac{r}{r_0}\right)}}{r^3}\right],
}
and finally
\eqn{
T^{ij} & = 4 \pi G \rho_0 r_0^3 \left[ \left(3 x_i x_j\ r^{-5} - \delta^{ij}\ r^{-3}\right)\ \ln{\left(1+\frac{r}{r_0}\right)} \right. \nonumber \\ 
& \quad  \left. + \frac{\delta^{ij}}{r^2\left(r+r_0\right)} - \frac{x_i x_j}{r^3\left(r+r_0\right)^2} - \frac{3 x_ix_j}{r^4\left(r+r_0\right)} \right],
}
which gives, for $r = x$
\eqn{
T^{xx}(r) = 4 \pi G \rho_0 \left(\frac{r_0}{r}\right)^3 \left[ 2 \ln{\left(1+\frac{r}{r_0}\right)} - \frac{3r^2 + 2 r r_0}{\left(r+r_0\right)^2} \right]
}
(see also \refsect[halos]{halos_tidal}).

%%%%%%%%%%%%%
\subsect{Einasto}
Less classical than the previous ones, the Einasto profile (\mycitealt{Einasto1965}; \mycitealt{Cardone2005}) has recently been presented as a better match to simulated dark matter halos (\mycitealt{Navarro2010}). The density-potential pair is defined by
\eqn{
\rho(r) = \rho_0 \exp{\left\{-\frac{2}{\alpha}\left[\left(\frac{r}{r_0}\right)^{\alpha}-1\right]\right\}}
}
\eqn{
\phi(r) = -\frac{GM}{r_0 \Gamma\left(\frac{3}{\alpha}\right)} \left\{r_0 \frac{\Gamma\left(\frac{3}{\alpha}\right) - \Gamma\left(\frac{3}{\alpha};\frac{2r^{\alpha}}{\alpha r_0^{\alpha}}\right)}{r} + \left(\frac{2}{\alpha}\right)^{1/\alpha} \left[ \Gamma\left(\frac{2}{\alpha}\right) + \Gamma\left(\frac{2}{\alpha};\frac{2r^{\alpha}}{\alpha r_0^{\alpha}}\right) \right] \right\}
}
where $\Gamma(x)$ and $\Gamma(x,y)$ are the gamma function and incomplete gamma function, defined by
\eqn{
\Gamma(x) = \Gamma(x,0) \qquad \textrm{and} \qquad \Gamma(x, y) = \int_y^{\infty}t^{x-1}e^{-t}\ dt .
}
The integrated mass is again derived from the potential thanks to Newton's second theorem
\eqn[einasto_mass]{
M(r) = M \frac{\Gamma\left(\frac{3}{\alpha}\right) - \Gamma\left(\frac{3}{\alpha};\frac{2r^{\alpha}}{\alpha r_0^{\alpha}}\right)}{\Gamma\left(\frac{3}{\alpha}\right)},
}
with
\eqn{
M = \frac{4\pi \ \rho_0 \ r_0^3}{\alpha} \left(\frac{2}{\alpha}\right)^{-3/\alpha} \Gamma\left(\frac{3}{\alpha}\right) \exp{\left(\frac{2}{\alpha}\right)}.
}

\mybox{
To compute the tidal tensor, we note that
\eqn{
\partial^j \Gamma\left(\frac{\theta}{\alpha};\frac{2r^{\alpha}}{\alpha r_0^{\alpha}}\right) = -x_j 2^{\theta/\alpha} \alpha^{1-\theta/\alpha} r_0^{-\theta} r^{\theta-2} \exp{\left(-\frac{2r^{\alpha}}{\alpha r_0^{\alpha}}\right)}.
}
}
Hence,
\eqn{
T^{ij} & = \frac{GM}{r_0 \Gamma\left(\frac{3}{\alpha}\right)} \partial^i \left\{ -x_j r^{-3} r_0 \Gamma\left(\frac{3}{\alpha}\right) + x_j 2^{3/\alpha} \alpha^{1-3/\alpha} r_0^{-2} \exp{\left(-\frac{2r^{\alpha}}{\alpha r_0^{\alpha}}\right)} \right. \nonumber\\
& \quad  \left. + x_j r^{-3} r_0 \Gamma\left(\frac{3}{\alpha};\frac{2r^{\alpha}}{\alpha r_0^{\alpha}}\right) -x_j 2^{3/\alpha} \alpha^{1-3/\alpha} r_0^{-2} \exp{\left(-\frac{2r^{\alpha}}{\alpha r_0^{\alpha}}\right)} \right\} \nonumber \\
& = \frac{GM}{\Gamma\left(\frac{3}{\alpha}\right)} \partial^i \left\{ x_j r^{-3} \left[ \Gamma\left(\frac{3}{\alpha};\frac{2r^{\alpha}}{\alpha r_0^{\alpha}}\right) - \Gamma\left(\frac{3}{\alpha}\right)  \right] \right\}.
}
Then, we have
\eqn{
T^{ij} & = \frac{GM}{\Gamma\left(\frac{3}{\alpha}\right)}  \left\{ \frac{\delta^{ij}-3r^{-2}x_ix_j}{r^3} \left[ \Gamma\left(\frac{3}{\alpha};\frac{2r^{\alpha}}{\alpha r_0^{\alpha}}\right) - \Gamma\left(\frac{3}{\alpha}\right)\right] \right. \nonumber\\
& \quad  \left. - x_ix_j r^{-2} 2^{3/\alpha} \alpha^{1-3/\alpha} r_0^{-3} \exp{\left(-\frac{2r^{\alpha}}{\alpha r_0^{\alpha}}\right)} \right\}.
}
Again, for $r = x$ we have
\eqn{
T^{xx}(r) = \frac{GM}{\Gamma\left(\frac{3}{\alpha}\right)} \left\{ \frac{-2}{r^3} \left[ \Gamma\left(\frac{3}{\alpha};\frac{2r^{\alpha}}{\alpha r_0^{\alpha}}\right) - \Gamma\left(\frac{3}{\alpha}\right)\right] - 2^{3/\alpha} \alpha^{1-3/\alpha} r_0^{-3} \exp{\left(-\frac{2r^{\alpha}}{\alpha r_0^{\alpha}}\right)} \right\}
}
(see also \refsect[halos]{halos_tidal}).

%%%%%%%%%%%%%%%%%%%%%%%%%%%%%%%%%%%%%%%%%%%%%%%%%%
\sect[kernels]{Working with kernels}

When using $N$-body simulations, a physical object is represented by a discrete distribution of particles. To limit the effects of this numerical technique, the properties of each particle (in our cases, the gravitational potential) are smoothed over a scalelength $\epsilon$. The functional form that indicates the way of smoothing is the kernel.

It is important to be familiar with the kernels when using them for the calculation of the gravitational force, or the tidal tensor: the kernel is explicitely used when substracting the effect of the particle in the center of a finite differences cube (see \refsect[numerical]{method_tt}) and for the computation of the tensor itself in the alternative method (see \refsect[numerical]{alternative_method}).

Dehnen's tree code ({\tt gyrfalcON}, \mycitealt{Dehnen2000}) proposes four different kernels, derived from density profiles. Kernel $P0$ is the classical Plummer form, detailled in the previous Section (with $r_0 = \epsilon$). The next one, $P1$ is a modified version of Plummer. Its density reads
\eqn{
\rho_{P1}(r) = \frac{15M \ \epsilon^4}{8 \pi\left(\epsilon^2 + r^2\right)^{7/2}}.
}
Using again \refeqn{rhotophi}, one gets
\eqn[kernel_integrals]{
\phi_{P1}(r) = -\frac{15 \ GM}{2\epsilon} \left[ \frac{\epsilon}{r} \int_0^{r/\epsilon}\frac{x^2 \ dx}{\left(1 + x^2\right)^{7/2}} + \int_{r/\epsilon}^{\infty}\frac{x \ dx}{\left(1 + x^2\right)^{7/2}} \right].
}
The first integral can be solved by setting $x = \tan{\theta}$ and using trigonometry relations:
\eqn{
\int_0^{r/\epsilon}\frac{x^2 \ dx}{\left(1 + x^2\right)^{7/2}} = \int_0^{\arctan(r/\epsilon)} \left(\cos^3\theta - \cos^5\theta \right) \ d\theta.
}

\mybox{
It is interesting to know that $(\forall \ n > 0)$
\eqn{
\int \cos^n\theta \ d\theta & = \sin\theta \ \cos^{n-1}\theta + (n-1) \int \sin^2\theta \cos^{n-2}\theta \ d\theta \nonumber \\
& = \sin\theta \ \cos^{n-1}\theta + (n-1) \left[ \int \cos^{n-2}\theta \ d\theta - \int \cos^n\theta \ d\theta \right] \nonumber \\
& = \frac{1}{n} \left[\sin\theta \ \cos^{n-1}\theta + (n-1) \int \cos^{n-2}\theta \ d\theta \right].
}
}

We finally find
\eqn{
\int_0^{r/\epsilon}\frac{x^2 \ dx}{\left(1 + x^2\right)^{7/2}} = \frac{r^3 \ \left(5\epsilon^2+2r^2\right)}{15\left(\epsilon^2 + r^2\right)^{5/2}}.
}
Simpler, the second integral in \refeqn{kernel_integrals} is
\eqn{
\int_{r/\epsilon}^{\infty}\frac{x \ dx}{\left(1 + x^2\right)^{7/2}} = \frac{\epsilon^5}{5\left(\epsilon^2 + r^2\right)^{5/2}},
}
which gives
\eqn[kernel_p1]{
\phi_{P1}(r) = -\frac{GM \left(3\epsilon^2+2r^2\right)}{2\left(\epsilon^2 + r^2\right)^{3/2}}.
}
From this simple form, it is easy to find the tidal tensor
\eqn{
T^{ij}_{P1} = \frac{GM}{2\left(\epsilon^2 + r^2\right)^{7/2}} \left[ x_ix_j \left(6r^2 + 21\epsilon^2\right) - \delta^{ij} \left(2r^4+5\epsilon^4+7r^2\epsilon^2\right) \right],
}
which can be implemented as $\tau^{ij}$ in the alternative method (see \refeqnt{alternative_method}).

The potentials of the other kernels (not used in this work) can be derived similarly:
\eqn{
\rho_{P2}(r) & = \frac{15M}{16\pi} \left[ \frac{7\epsilon^6}{\left(\epsilon^2 + r^2\right)^{9/2}}-\frac{2\epsilon^4}{\left(\epsilon^2 + r^2\right)^{7/2}} \right] \\
\phi_{P2}(r) & = -\frac{GM \left(9\epsilon^4+10\epsilon^2r^2+4r^4\right)}{4\left(\epsilon^2 + r^2\right)^{5/2}},
}
and
\eqn{
\rho_{P3}(r) & = \frac{105M}{64\pi} \left[ \frac{9\epsilon^8}{\left(\epsilon^2 + r^2\right)^{11/2}}-\frac{4\epsilon^6}{\left(\epsilon^2 + r^2\right)^{9/2}} \right] \\
\phi_{P3}(r) & = -\frac{GM \left(45\epsilon^6+70\epsilon^4r^2+56\epsilon^2r^4+16r^6\right)}{16\left(\epsilon^2 + r^2\right)^{7/2}}.
}
Note that both $P2$ and $P3$ yield negative densities in their outer parts, corresponding to super-newtonian forces. \reffig{kernels} displays the density profiles and the associated potentials.

\fig{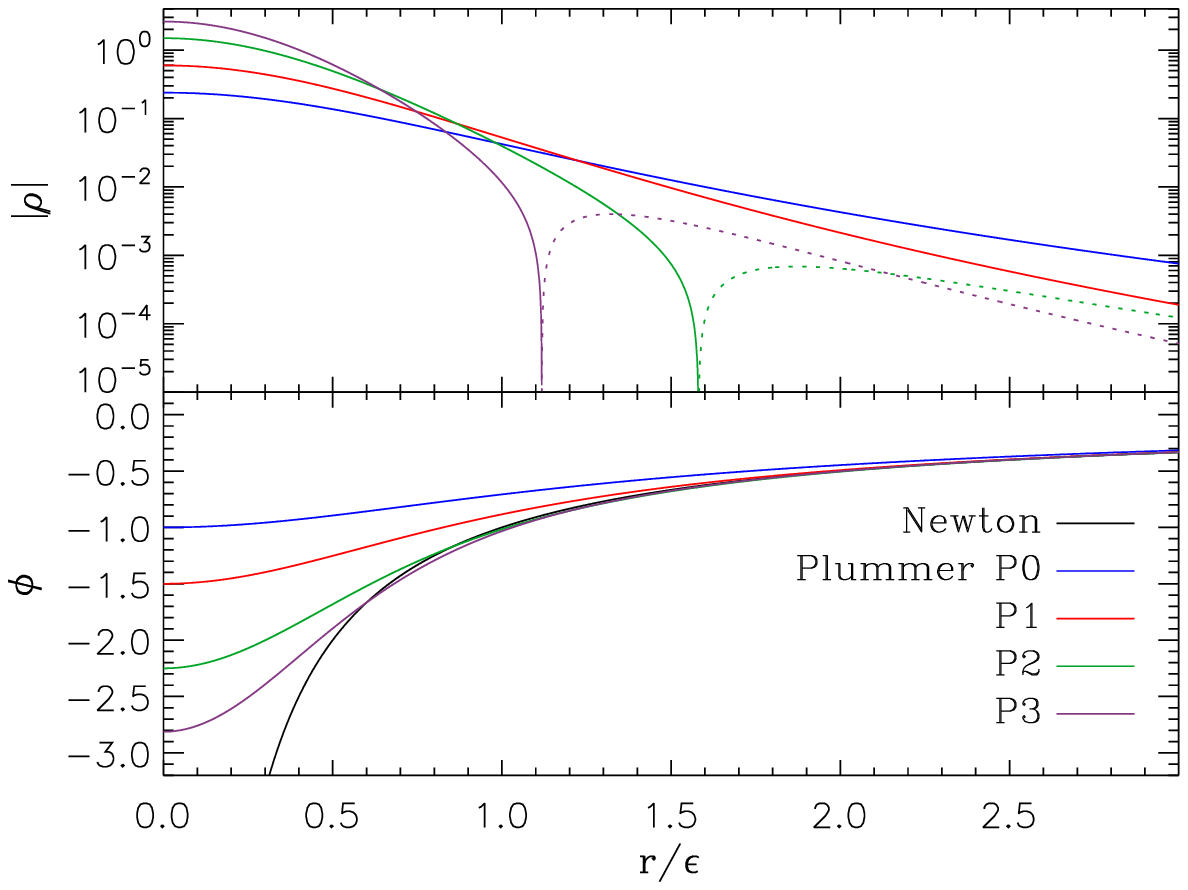}{Softening kernels}
{\emph{Top:} Density profiles (absolute value) of the four kernels implemented in {\tt gyrfalcON}. The negative part (outer regions) of $P2$ and $P3$ is shown with dashed lines. \emph{Bottom:} Associated potentials compared with the Newtonian solution (i.e. no smoothing). The same total mass has been used in all cases.}

%%%%%%%%%%%%%%%%%%%%%%%%%%

\cha{glitches}{Glitches in the gravity}
\chaphead{In this Appendix, we investigate the sources of errors appearing in the numerical computation of the tidal tensor.}

%%%%%%%%%%%%%%%%%%%%%%%%%%%%%%%%%%%%%%%%%%%%%%%%%%
\sect{The problem}

In \refcha{numerical}, we have presented a finite differences method to compute the tidal tensor via the accelerations measured on a cube. Using a $N$-body realization of a Plummer sphere, we have shown that the profile of the dominant term of the tensor is well reproduced, despite a certain level of noise. In the \reffigp{tt_plummer} however, some glitches are visible for well-defined radii. They are more visible in the zoomed-in version presented in \reffig{glitches}.

\fig{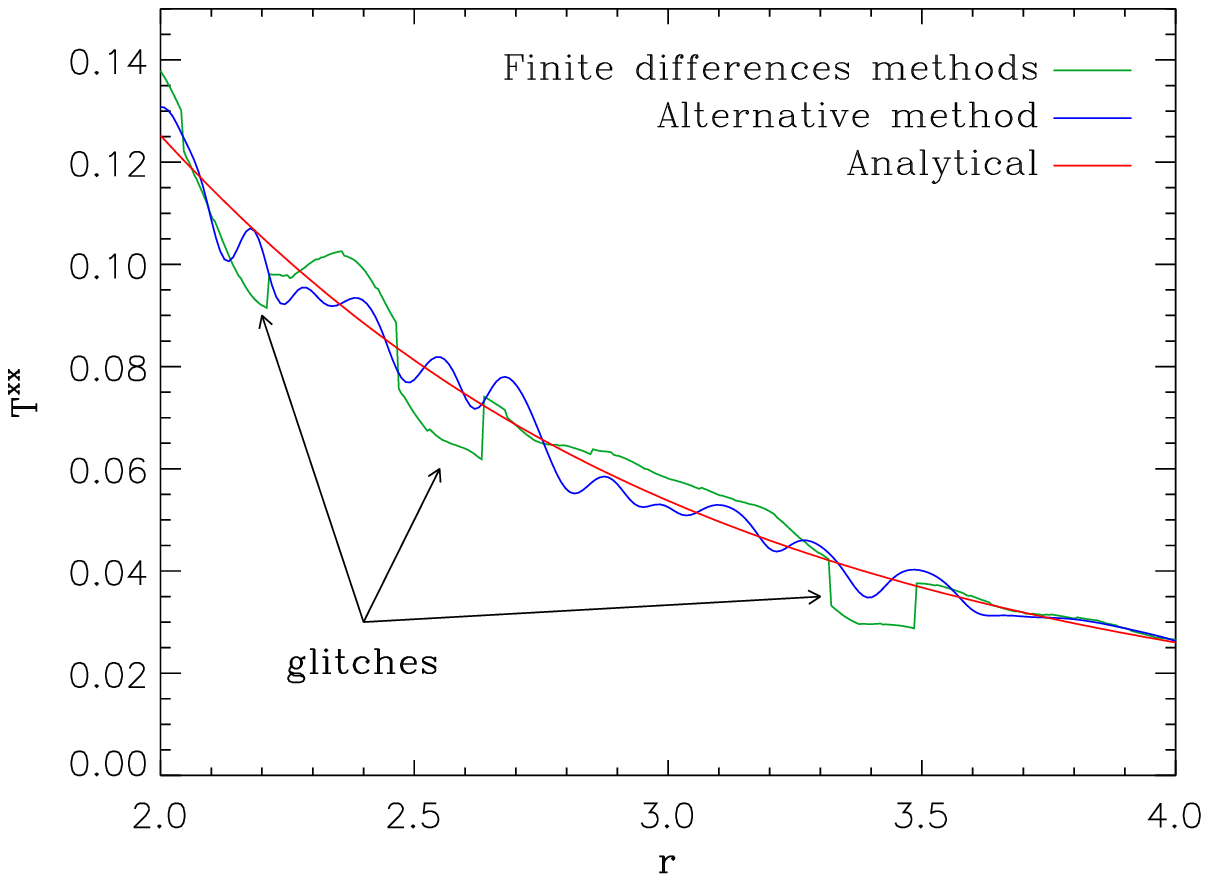}{Glitches in the tidal tensor}
{$T^{xx}$ term of the tidal tensor computed with the finite differences method and alternative method, compared with the analytical solution.}

The errors mainly consist in a systematic, brutal shift from the expected correct solution to a lower value of the tidal term, over small radial ranges. Note that they do not exist in the alternative method (see \refsect[numerical]{alternative_method}) which simply sums the contributions of all individual particles to the tidal tensor.

The origin of this problem has been investigated by several means, presented below.

%%%%%%%%%%%%%%%%%%%%%%%%%%%%%%%%%%%%%%%%%%%%%%%%%%
\sect{Possible sources}

The fact that other methods give solutions without these shifts indicates that the their origin is independent of the $N$-body aspect of the system itself. Therefore, the rendition of the potential is exonerated from the list of possible causes. Remains the finite differences scheme and the computation of the acceleration.

The first order difference method implemented to compute the derivative of the acceleration introduces a numerical noise. However one would expect such a noise to be present everywhere in the computation and to yield a continuous evolution along, say, the radial axis. The phenomenon observed here is clearly different as it appears locally, and through discrete changes. That is, the glitches can not originate from the differencing scheme and therefore come from a problem due to the computation of the acceleration. Furthermore, the effect tends to disappear when the tolerance angle $\theta$ goes to zero, as suggested on \reffig{glitches_theta}, and to completely disappear for a direct summation method (called alternative method, recall \reffig{glitches}).

\fig{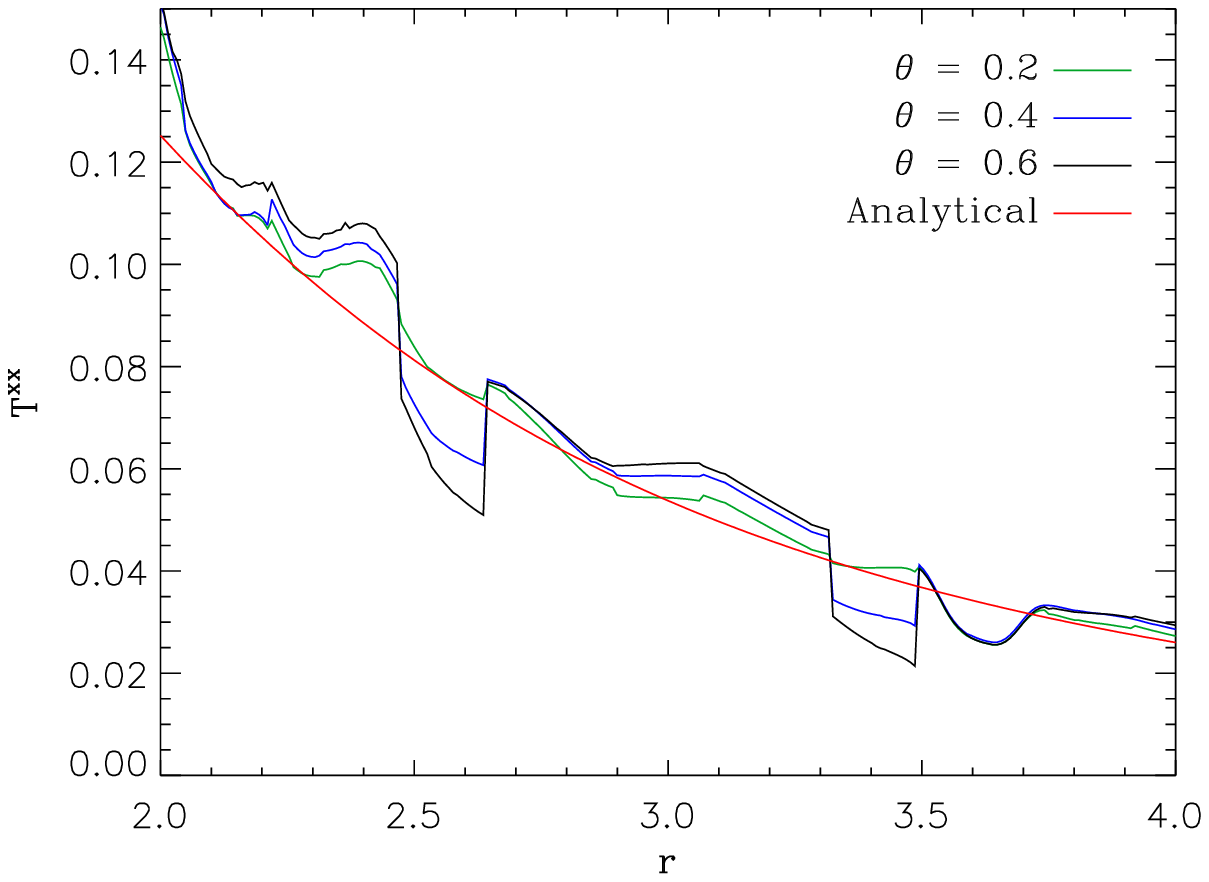}{Influence of $\theta$}
{Same as previous, but for different values of the tolerance parameter $\theta$ used in the building of the tree.}

Note however that a small tolerance parameter implies that more and more particles are considered for the direct summation computation, which weakens the power of the treecode. In the following (as well as for the rest of the study), we chose to sacrifice a small amount of precision in order to keep high computational performances, and set $\theta = 0.4$.

Hence, it is possible to narrow the field of possibilities to the tree aspect of the code and the way the accelerations are computed between the (sub)cells. To explore this in more details, one has to compare the value of the acceleration given by {\tt gyrfalcON} with those from other codes. In this matter, \reffig{acc_gyr_vs_hack} displays the relative error in the evaluation of the acceleration, between Dehnen's code ({\tt gyrfalcON}) and a classical treecode, {\tt hackcode}.

\fig{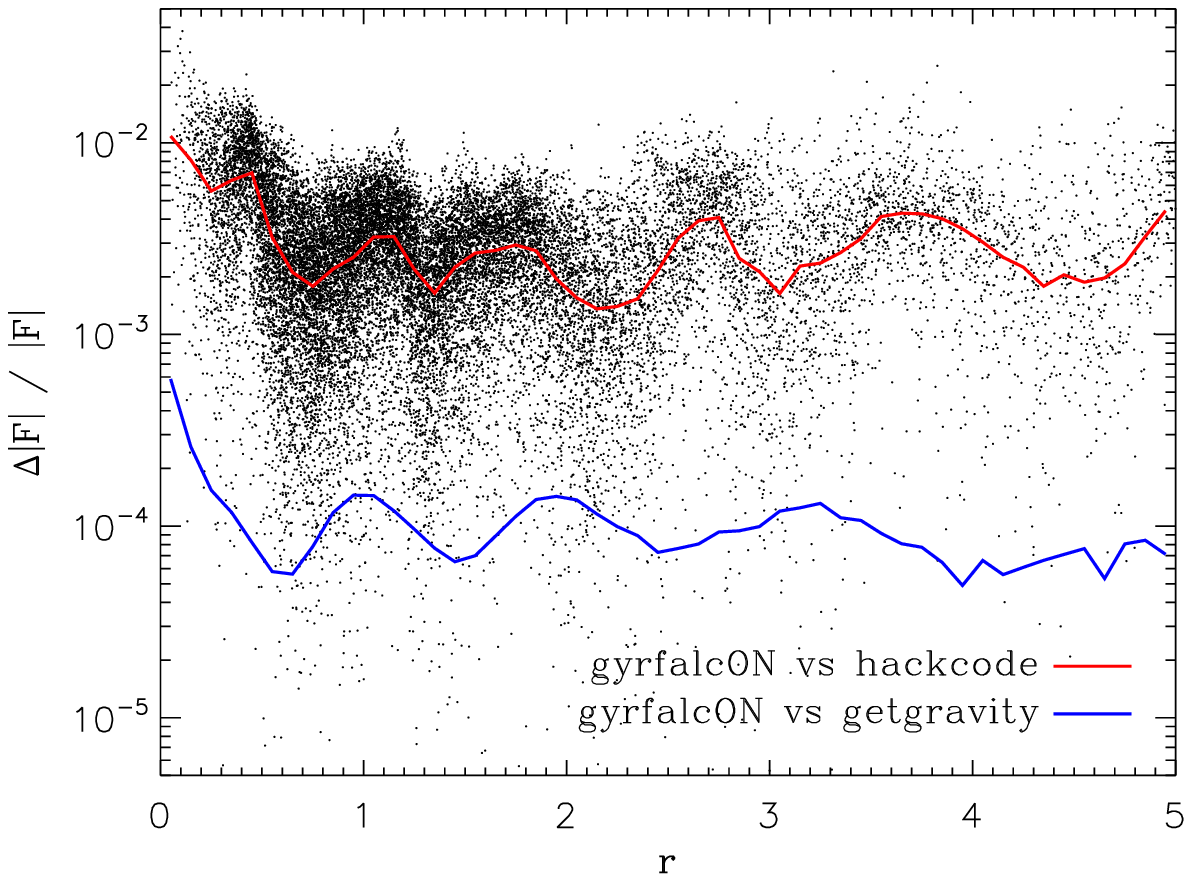}{Differences in the value of the acceleration}
{Relative error in the evaluation of the acceleration between {\tt gyrfalcON} and {\tt hackcode} (classical treecode). The solid lines indicates the median value of the relative error between the different tools.}

Errors are of the order of magnitude of 0.1\% for the modulus of the acceleration \linebreak($|\vect{a}| = \sqrt{a_x^2+a_y^2+a_z^2}$). The same study on individual components ($a_x$, $a_y$ or $a_z$) reveals that, for a very small number of particles ($\sim 10$ over $10^5$), these errors can reach $10\,000 \%$, but are compensated by a high accuracy for the other components.

In the same time, \reffig{acc_gyr_vs_hack} reveals a small, yet significant difference between the accelerations computed by {\tt gyrfalcON} or {\tt getgravity} (the tool used to compute the acceleration on the fictive cube around each stellar particle). As parts of the {\tt falcON} suite, both tools use the same libraries to create the tree and compute the accelerations, which makes the measured difference very strange. An explanation for this has not been found during this work and the problem had have to be overcome.

%%%%%%%%%%%%%%%%%%%%%%%%%%%%%%%%%%%%%%%%%%%%%%%%%%
\sect{Partial solution}

Re-writing Dehnen's code was clearly not the goal of this study. Instead, it would have been possible to use the alternative method to compute the tidal tensor with the highest possible accuracy. Indeed, the only error introduced by this method is the softening kernel of the particles, a default which can \emph{not} be avoided in $N$-body simulations. However, the direct summation has a very high cost of \acr{}{CPU}{central process unit} time and does not allow the statistical study (i.e. with a large number of particles and simulations) that we have conducted.

The glitches in the {\tt getgravity} method do not concern a large fraction of the particles. In addition, it is likely (yet, impossible to check) that the problem does not exist for more complex configurations (where the tree is forced to be non-symmetric), like the ones used in this work. Finally, the glitches have a limited amplitude that may not compromise the determination of the sign of the eigenvalues, in other words, the compressive character of the tidal field. To be on the safe side however, we decided to set a ``gray-zone'' of width 0.14: if the \emph{maximum} eigenvalue of the tidal tensor sits inside this range ($-0.07 \le \lambda_\mathrm{max} \le 0.07$), the study of the tidal field is inconclusive. That is, the compressive tidal mode gathers the particles yielding a tidal tensor with a maximum eigenvalue less than -0.07 (in numerical units).

\cha{harmonics}{Torques on spherical harmonics}
\chaphead{This Appendix gives the detailed mathematical derivation of the torque created by a galaxy on one of its clusters. Each Section presents a special configuration of both the galaxy and the cluster, starting from a simple example and going to more general (and involved) cases.}

%%%%%%%%%%%%%%%%%%%%%%%%%%%%%%%%%%%%%%%%%%%%%%%%%%
\sect[textbook_torque]{A dipole cluster around a point mass galaxy}

The goal of this derivation is to get a mathematical representation of the effect of a host galaxy on a non-spherical star cluster, say, a spheroid. A non-isotropic tidal field would then induce a torque on the cluster and possibly trigger (or enhance, or slow down) its rotation.

To illustrate our approach, let's consider a textbook example where the cluster is made of two point-mass $m_1 = m_2 = m_\mathrm{c}/2$ separated by a distance $2a$ along the vertical axis, and where the galaxy is a point mass $m_\mathrm{G}$ situated at distances $r_1$ and $r_2$ from the cluster points (see \reffig{dipole_potential}).

\fig{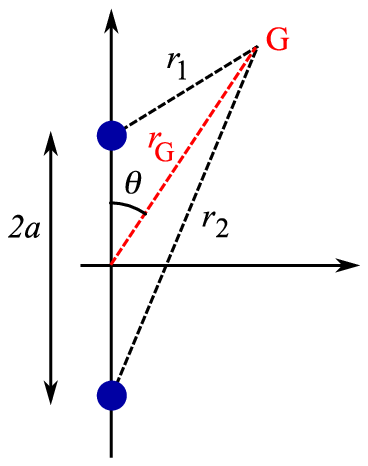}{Configuration of a dipole cluster}
{A cluster is represented by two points (blue) situated at distances $r_1$ and $r_2$ of a point mass galaxy G. Such a configuration has an axial symmetry, which makes it independent of $\varphi$, in spherical coordinates.}

The potential of the cluster evaluated at the position of the galaxy is
\eqn{
\phi_\mathrm{c}(\vect{r_\mathrm{G}}) = -\frac{G m_\mathrm{c}}{2 r_1} - \frac{G m_\mathrm{c}}{2 r_2}.
}
The distances are evaluated with trigonometric relations in spherical coordinates:
\eqn{
r_1^2 & = r_\mathrm{G}^2 + a^2 - 2ar_\mathrm{G}\cos{\theta}\\
r_2^2 & = r_\mathrm{G}^2 + a^2 + 2ar_\mathrm{G}\cos{\theta},
}
and can be expanded in Legendre polynomials $P_n$ (see \refeqn{legendre_poly} below) to give
\eqn{
\frac{1}{r_1} & = \frac{1}{r_\mathrm{G}}\sum_{n=0}^{\infty} \left(\frac{a}{r_\mathrm{G}}\right)^n P_n(\cos{\theta})\\
\frac{1}{r_2} & = \frac{1}{r_\mathrm{G}}\sum_{n=0}^{\infty} \left(-\frac{a}{r_\mathrm{G}}\right)^n P_n(\cos{\theta})
}
where $r_\mathrm{G}>a$, so that
\eqn{
\phi_\mathrm{c}(\vect{r_\mathrm{G}}) & = -\frac{G\ m_\mathrm{c}}{r_\mathrm{G}} \sum_{n=0}^{\infty} \left(\frac{a}{r_\mathrm{G}}\right)^{2n} P_{2n}(\cos{\theta}).
}
By truncating the sum for $n > 1$, one gets the monopole and quadrupole terms (the dipole being zero):
\eqn[textbook_potential]{
\phi_\mathrm{c}(\vect{r_\mathrm{G}}) & \approx -\frac{G m_\mathrm{c}}{r_\mathrm{G}} \left[ 1 + \frac{a^2}{2r_\mathrm{G}^2} (3\cos^2{\theta} - 1) \right].
}
The torque induced by the galaxy on the cluster about the origin (i.e. the center of mass of the cluster) is therefore
\eqn[textbook_torque]{
\vect{\Gamma_\mathrm{G/c}} = -\vect{\Gamma_\mathrm{c/G}} & \approx - m_\mathrm{G} \ \vect{r_\mathrm{G}} \times \left[-\vect{\nabla}\phi_\mathrm{c}(\vect{r_\mathrm{G}})\right] \nonumber \\
& \approx 3\frac{G m_\mathrm{c} m_\mathrm{G} \ a^2}{r_\mathrm{G}^3}\cos{\theta}\sin{\theta}\ \vectu{\varphi}.
}
Note that for $a = 0$, one retreives the case of a monopole cluster which yields no torque.

Such a result can be obtained because of the simplicity of the setting. This corresponds well to the case of globular clusters which exhibit a regular shape. For younger objects like proto-clusters which have more filamentary structure, a more detailed derivation is required and the use of spherical harmonics is preferable.

%%%%%%%%%%%%%%%%%%%%%%%%%%%%%%%%%%%%%%%%%%%%%%%%%%
\sect{A cylindrical cluster}

We now write the expression of the torque for a mass distribution of the cluster symmetric with respect to the $\theta = 0$ axis (i.e. depending on the radial and latitudinal but not longitudinal coordinates).

\mybox[spherical_harmonics]{
We adopt here the Condon-Shortley convention, as in \mycitet{Binney1987}. A spherical harmonic is defined by:
\eqn{
\Psi_\ell^m(\theta,\varphi) = \sqrt{\frac{2\ell + 1}{4\pi}\ \frac{(\ell-|m|)!}{(\ell+|m|)!}}P_\ell^{|m|}(\cos{\theta})\ e^{\imath m \varphi}\times \left\{ \begin{array}{ll}(-1)^m & \textrm{ for } m \geq 0 \\ 1 & \textrm{ for } m < 0 \end{array}\right.,
}
where $P_\ell^{|m|}$ is the associated Legendre function of the first kind defined by:
\eqn{
P_\ell^m(x) = \left(1-x^2\right)^{m/2} \frac{d^m P_\ell^0(x)}{dx^m},
}
with $P_\ell^0$ the Legendre polynomial, itself defined by:
\eqn[legendre_poly]{
P_\ell^0(x) = P_\ell(x) = \frac{1}{2^\ell \ell!} \frac{d^\ell \left(x^2-1\right)^\ell}{dx^\ell}.
}
For negative degrees, one has the relation
\eqn{
\Psi_\ell^{-m}(\theta,\varphi) = (-1)^m \ \Psi_\ell^{m*}(\theta,\varphi),
}
where the asterisk denotes the complex conjugation of the harmonic $\Psi$.
}

Consider a spheroidal cluster of potential $\phi_\mathrm{c}(\vect{r})$ embedded in a galaxy represented by a density $\rho_\mathrm{G}(\vect{r})$. The net torque of the galaxy on the cluster is the opposite of the one from the cluster on the galaxy, given by the time derivative of the angular momentum $\vect{L_\mathrm{G}}$:
\eqn{
\vect{\Gamma_\mathrm{G/c}} = -\vect{\Gamma_\mathrm{c/G}} & = -\frac{d\vect{L_\mathrm{G}}}{dt} \nonumber\\
& = \int_V \rho_\mathrm{G}(\vect{r}) \left[\vect{r} \times \vect{\nabla}\phi_\mathrm{c}(\vect{r}) \right] dV.
}
In our approach, both the density of the galaxy and the potential of the cluster are decomposed in spherical harmonics on concentric spherical shells, to ease the evaluation of their gradients. \myciteauthor{Anderson1992} (\myciteyear{Anderson1992}, see also \mycitealt{Kawai1998}) proposed a fast multipole method (or FMM) to express the potential of the cluster at the position $\vect{r}$, from its potential $\phi_\mathrm{c}(a\vect{s})$ on the surface of a sphere of radius $a < r$:
\eqn[Anderson]{
\phi_\mathrm{c}(\vect{r}) = \frac{1}{4\pi}\sum_{n=0}^{\infty} (2n+1) \left(\frac{a}{r}\right)^{n+1} \int_S P_n\left(\frac{\vect{s}\cdot\vect{r}}{r}\right) \phi_\mathrm{c}(a\vect{s})\ ds,
}
where $P_n$ is the $n$-th Legendre polynomial, and $ds = \sin{\theta_\mathrm{s}}\ d\theta_\mathrm{s}\ d\varphi_\mathrm{s}$. We decompose $\phi_\mathrm{c}(a\vect{s})$ in spherical harmonics:
\eqn[potential_cluster_spherharm]{
\phi_\mathrm{c}(a\vect{s}) = \frac{G m_\mathrm{c}}{a}\sum_{k=0}^{\infty} C_k^0\ \Psi_k^{0}(\theta_\mathrm{s},\varphi_\mathrm{s}),
}
(The upper 0 index of $\Psi$ translates the independence of the potential on $\varphi_\mathrm{s}$.) The argument of the Legendre polynomial in \refeqn{Anderson} is $(\vect{s}\cdot\vect{r})/r = \cos{\alpha}$, with $\alpha$ the angle between direction vectors $\vect{s}$ and $\vect{r}$ which define the $(\theta_\mathrm{s},\varphi_\mathrm{s})$ and $(\theta,\varphi)$ angles, respectively . This can be re-writen as
\eqn{
P_n(\cos{\alpha}) = \frac{4\pi}{2n+1} \sum_{j=-n}^{n} \Psi_n^{j}(\theta,\varphi)\ \Psi_n^{j*}(\theta_\mathrm{s},\varphi_\mathrm{s}).
}
Therefore, one has
\eqn{
\int_S P_n\left(\frac{\vect{s}\cdot\vect{r}}{r}\right) \phi_\mathrm{c}(a\vect{s}) ds = & \frac{G m_\mathrm{c}}{a} \frac{4\pi}{2n+1} \nonumber \\
&  \sum_{k=0}^{\infty}\sum_{j=-n}^{n} C_k^0\  \Psi_n^{j}(\theta,\varphi)  \int_S \Psi_k^{0}(\theta_\mathrm{s},\varphi_\mathrm{s})\ \Psi_n^{j*}(\theta_\mathrm{s},\varphi_\mathrm{s})\ ds.
}
The orthogonality between the spherical harmonics implies
\eqn{
\int_S \Psi_k^{0}(\theta_\mathrm{s},\varphi_\mathrm{s})\ \Psi_n^{j*}(\theta_\mathrm{s},\varphi_\mathrm{s})\ ds = \delta_k^n \delta_j^0,
}
and thus
\eqn[phi_cluster_cylindrical]{
\phi_\mathrm{c}(\vect{r}) = & \frac{G m_\mathrm{c}}{a} \sum_{n=0}^{\infty} \left(\frac{a}{r}\right)^{n+1} C_n^0\ \Psi_n^{0}(\theta,\varphi).
}
Computing the gradient, one has
\eqn{
\vect{r} \times \vect{\nabla}\phi_\mathrm{c}(\vect{r}) & = -\frac{1}{\sin{\theta}}\frac{\partial \phi_\mathrm{c}(\vect{r})}{\partial \varphi}\ \vectu{\theta} + \frac{\partial \phi_\mathrm{c}(\vect{r})}{\partial \theta}\ \vectu{\varphi} \\
& = \frac{G m_\mathrm{c}}{a} \sum_{n=0}^{\infty} \left(\frac{a}{r}\right)^{n+1} C_n^0\ \frac{\partial \Psi_n^{0}(\theta,\varphi)}{\partial \theta}\ \vectu{\varphi},
}
and
\eqn{
\frac{\partial \Psi_n^{0}(\theta,\varphi)}{\partial \theta} & = \sqrt{\frac{2n+1}{4\pi}}\frac{d P_n(\cos{\theta})}{d \theta} \nonumber \\
& = -\sqrt{\frac{2n+1}{4\pi}}\sin{\theta}\ \frac{d P_n(\cos{\theta})}{d \cos{\theta}} \nonumber \\
& = - \sqrt{\frac{2n+1}{4\pi}}P_n^1(\cos{\theta}).
}
The index $=1$ of the associated Legendre function implies that the term $n=0$ vanishes. Physically, this means that the monopole term of the potential of the cluster does not participate in the torque induced by the galaxy. Finally, one gets
\eqn[torque]{
\vect{\Gamma_\mathrm{G/c}} = - \frac{G m_\mathrm{c}}{a}  \sum_{n=1}^{\infty} \sqrt{\frac{2n+1}{4\pi}} C_n^0 \int_V \rho_\mathrm{G}(\vect{r})\left(\frac{a}{r}\right)^{n+1} P_n^1(\cos{\theta})\ \vectu{\varphi} \ dV.
}

%%%%%%%%%%%%%%%%%%%%%%%%%
\subsect{First case: point mass galaxy}

The simplest case is to consider the host galaxy as a point-mass at the position $\vect{r}_\mathrm{G}$. Thus, the density reads $\rho_\mathrm{G}(\vect{r}) = m_\mathrm{G}\ \delta(\vect{r}-\vect{r}_\mathrm{G})$. \refeqn{torque} becomes
\eqn[pointmassgalaxy]{
\vect{\Gamma_\mathrm{G/c}} = - \frac{G m_\mathrm{c} m_\mathrm{G}}{a}  \sum_{n=1}^{\infty} \sqrt{\frac{2n+1}{4\pi}} C_n^0 \left(\frac{a}{r_\mathrm{G}}\right)^{n+1} P_n^1(\cos{\theta})\ \vectu{\varphi}.
}

Back to our textbook example (\refsec{textbook_torque}), we can identify the potential of the cluster at $\vect{r_\mathrm{G}}$ to its decomposition in spherical harmonics (\refeqn{phi_cluster_cylindrical}) up to the quadrupole, with the help of \refbox{spherical_harmonics}:
\eqn{
\phi_\mathrm{c}(\vect{r_\mathrm{G}}) \approx \frac{G m_\mathrm{c}}{a} \left[ C_0^0 \frac{a}{r_\mathrm{G}} \sqrt{\frac{1}{4\pi}} + C_1^0  \frac{a^2}{r_\mathrm{G}^2}  \sqrt{\frac{3}{4\pi}}\cos{\theta} + C_2^0  \frac{a^3}{r_\mathrm{G}^3}  \sqrt{\frac{5}{16\pi}}(3\cos^2{\theta}-1)\right],
}
and get, by comparison with \refeqn{textbook_potential},
\eqn{
C_0^0 = -\sqrt{4\pi} \quad ; \quad C_1^0 = 0 \quad ; \quad C_2^0 = -\sqrt{\frac{4\pi}{5}}.
}
By using these values in \refeqn{pointmassgalaxy}, one retrieves the result of \refeqn{textbook_torque}.

%%%%%%%%%%%%%%%%%%%%%%%%%
\subsect{Second case: a two point galaxy}

If now the galaxy is made of two points symetrically distributed with respect to the cluster, i.e. at the positions $(r_\mathrm{G},\theta_\mathrm{G},\varphi_\mathrm{G})$ and $(r_\mathrm{G},\pi-\theta_\mathrm{G},\pi+\varphi_\mathrm{G})$ the density becomes
\eqn{
\rho_\mathrm{G}(\vect{r}) = \frac{m_\mathrm{G}}{2} \frac{\delta(r-r_\mathrm{G})}{r_\mathrm{G}^2} \left[ \frac{\delta(\theta-\theta_\mathrm{G})}{\sin{\theta_\mathrm{G}}} \delta(\varphi - \varphi_\mathrm{G}) + \frac{\delta(\theta - \pi + \theta_\mathrm{G})}{\sin{(\pi - \theta_\mathrm{G})}} \delta(\varphi - \pi - \varphi_\mathrm{G}) \right].
}
Such a configuration has new symetrical properties, which allow us to verify our calculations.

\mybox{
One has to be careful when using Dirac functions in spherical coordinates. In this system, a volume element is not rectangular because of the Jacobian matrix used in for the transformation of coordinates from the cartesian base. This implies that the point $\vect{r_\mathrm{G}} = (r_\mathrm{G},\theta_\mathrm{G},\varphi_\mathrm{G})$ is \emph{not} represented by
\eqn{
\delta(\vect{r}-\vect{r_\mathrm{G}}) \ne \delta(r-r_\mathrm{G})\delta(\theta-\theta_\mathrm{G})\delta(\varphi-\varphi_\mathrm{G}),
}
but by
\eqn{
\delta(\vect{r}-\vect{r_\mathrm{G}}) = \frac{\delta(r-r_\mathrm{G})}{r_\mathrm{G}^2}\frac{\delta(\theta-\theta_\mathrm{G})}{\sin{\theta_\mathrm{G}}}\delta(\varphi-\varphi_\mathrm{G}),
}
so that the 3D Dirac is normalized to unity
\eqn{
\int_{V} \delta(\vect{r}-\vect{r_\mathrm{G}})\ dV = \int_{V} \delta(\vect{r}-\vect{r_\mathrm{G}})\ r^2 dr \ \sin{\theta} d\theta \ d\varphi = 1.
}
}

With this density, \refeqn{torque} becomes
\eqn{
\vect{\Gamma_\mathrm{G/c}} = - \frac{G\ m_\mathrm{c}\ m_\mathrm{G}}{2a} \sum_{n=1}^{\infty} \sqrt{\frac{2n+1}{4\pi}} C_n^0 \left(\frac{a}{r_\mathrm{G}}\right)^{n+1} \left\{P_n^1[\cos{\theta_\mathrm{G}}] - P_n^1[\cos{(\pi-\theta_\mathrm{G})}] \right\} \vectu{\varphi}_\mathrm{G}.
}
One immediately has $P_n^1[\cos{(\pi-\theta_\mathrm{G})}] = P_n^1[-\cos{(\theta_\mathrm{G})}]$ and, knowing that
\eqn{
P_n^1(-x) = (-1)^n P_n^1(x),
}
one can see that the even terms cancel out because of the symmetry of the galaxy. Finally, one gets
\eqn{
\vect{\Gamma_\mathrm{G/c}} = - \frac{G m_\mathrm{c} m_\mathrm{G}}{a} \sum_{n=0}^{\infty} \sqrt{\frac{4n+3}{4\pi}} C_{2n+1}^0 \left(\frac{a}{r_\mathrm{G}}\right)^{2n+2} P_{2n+1}^1(\cos{\theta_\mathrm{G}})\ \vectu{\varphi}_\mathrm{G}.
}

%%%%%%%%%%%%%%%%%%%%%%%%%
\subsect[general_galaxy]{General case for the galaxy}

It is also possible to represent the density of the galaxy with a new set of spherical harmonics, evaluated on a shell, without loss of generality:
\eqn{
\rho_\mathrm{G}(\vect{r}) = \frac{m_\mathrm{G}\ \delta(r-r_\mathrm{G})}{r^2_\mathrm{G}} \sum_{\ell=0}^{\infty}\sum_{m=-\ell}^\ell D_{\ell}^m \Psi_\ell^m(\theta,\varphi),
}
so that \refeqn{torque} becomes
\eqn[secondcase]{
\vect{\Gamma_\mathrm{G/c}} = & - \frac{G m_\mathrm{c} m_\mathrm{G}}{a \ r^2_\mathrm{G}} \sum_{n=1}^{\infty}\sum_{\ell=0}^{\infty}\sum_{m=-\ell}^\ell \sqrt{\frac{2n+1}{4\pi}} C_n^0 D_{\ell}^m \nonumber \\
& \int_V \left(\frac{a}{r}\right)^{n+1} \delta(r-r_\mathrm{G}) \Psi_\ell^m(\theta,\varphi) P_n^1(\cos{\theta})\ \vectu{\varphi} \ dV.
}
One can write
\eqn{
\Psi_\ell^m(\theta,\varphi) = K_\ell^m P_\ell^{|m|}(\cos{\theta}) e^{\imath m \varphi},
}
with
\eqn[klm]{
K_\ell^m = \sqrt{\frac{2\ell +1}{4\pi}\ \frac{(\ell - |m|)!}{(\ell + |m|)!}}\times \left\{ \begin{array}{ll}(-1)^m & \textrm{ for } m \geq 0 \\ 1 & \textrm{ for } m < 0 \end{array}\right.,
}
and thus
\eqn{
\int_V \left(\frac{a}{r}\right)^{n+1} \delta(r-r_\mathrm{G}) \Psi_\ell^m(\theta,\varphi) P_n^1(\cos{\theta})\ \vectu{\varphi} \ dV = & K_\ell^m \ \int_{0}^{\infty} \left(\frac{a}{r}\right)^{n+1} \delta(r-r_\mathrm{G}) r^2\ dr \nonumber \\ &\int_0^{\pi} P_n^1(\cos{\theta}) P_\ell^{|m|}(\cos{\theta}) \sin{\theta}\ d\theta \nonumber \\
& \int_0^{2\pi} e^{\imath m \varphi} \vectu{\varphi}\ d\varphi.
}
The integral on $r$ reads
\eqn{
\int_{0}^{\infty} \left(\frac{a}{r}\right)^{n+1} \delta(r-r_\mathrm{G}) r^2\ dr = r^2_\mathrm{G} \left(\frac{a}{r_\mathrm{G}}\right)^{n+1},
}
and the one on $\theta$ can be re-writen
\eqn[ortholegendre]{
\int_0^{\pi} P_n^1(\cos{\theta}) P_\ell^{|m|}(\cos{\theta}) \sin{\theta}\ d\theta = \int_{-1}^{1} P_n^1(u) P_\ell^{|m|}(u)\ du.
}
The last integral is best expressed in the $\varphi$-independent cartesian base:
\eqn{
\int_0^{2\pi} \left[\cos{(m\varphi)} + \imath \sin{(m\varphi)}\right] \left[\cos{\varphi} \vectu{y} - \sin{\varphi} \vectu{x}\right] d\varphi = & \ \vectu{y} \int_0^{2\pi} \cos{(m\varphi)} \cos{\varphi}\ d\varphi \nonumber \\
& - \vectu{x} \int_0^{2\pi} \cos{(m\varphi)} \sin{\varphi}\ d\varphi \nonumber \\
& +\imath \vectu{y} \int_0^{2\pi} \sin{(m\varphi)} \cos{\varphi}\ d\varphi \nonumber \\
& -\imath \vectu{x} \int_0^{2\pi} \sin{(m\varphi)} \sin{\varphi}\ d\varphi.
}
We note that only the first and the last integrals are non-zero, giving
\eqn{
\int_0^{2\pi} \left[\cos{(m\varphi)} + \imath \sin{(m\varphi)}\right] \left[\cos{\varphi} \vectu{y} - \sin{\varphi} \vectu{x}\right] d\varphi = \pi \left[ \left(\vectu{y} + \imath \vectu{x}\right) \delta_m^{-1} + \left(\vectu{y} - \imath \vectu{x}\right) \delta_m^{1} \right].
}
Therefore, $|m|=1$ which forbids $\ell = 0$, so that the monopole term of the \emph{galaxy} does not create a torque on the cluster, in the reference frame centered on the cluster\footnote{... as oposed to the usual reference taken at the galactic barycenter.}. Furthermore, the orthogonality condition on the associated Legendre functions in \refeqn{ortholegendre} gives
\eqn{
\int_0^{\pi} P_n^1(\cos{\theta}) P_\ell^{|m|}(\cos{\theta}) \sin{\theta}\ d\theta & = \frac{2}{2n+1}\frac{(n+|m|)!}{(n-|m|)!}\delta_n^\ell \nonumber \\
& = \frac{2}{2n+1} n (n+1) \delta_n^\ell.
}
Back to \refeqn{secondcase}, one obtains
\eqn{
\vect{\Gamma_\mathrm{G/c}} = & -\frac{G m_\mathrm{c} m_\mathrm{G}}{a} \sum_{n=1}^{\infty}\sum_{m=-n}^n \sqrt{\frac{\pi}{2n+1}}\ n (n+1) \left(\frac{a}{r_\mathrm{G}}\right)^{n+1} \ C_n^0 \nonumber \\
& K_n^m D_n^m \left[ \left(\vectu{y} + \imath \vectu{x}\right) \delta_m^{-1} + \left(\vectu{y} - \imath \vectu{x}\right) \delta_m^{1} \right].
}
Furthermore, we recall from \refeqn{klm} that
\eqn{
K_n^{-1} = - K_n^1 = \sqrt{\frac{2n +1}{4\pi}\ \frac{1}{n(n+1)}}.
}
To derive the coefficient $D_n^m$, we note that
\eqn{
\int_S \Psi_n^{q*}(\theta,\varphi) \rho(\vect{r_\mathrm{G}})\ ds & = \frac{m_\mathrm{G}\ \delta(r-r_\mathrm{G})}{r^2_\mathrm{G}} \sum_{\ell=0}^{\infty}\sum_{m=-\ell}^\ell D_{\ell}^m \int_S \Psi_\ell^m(\theta,\varphi) \Psi_n^{q*}(\theta,\varphi)\ ds \nonumber \\
& = \frac{m_\mathrm{G}\ \delta(r-r_\mathrm{G})}{r^2_\mathrm{G}} D_n^{q}.
}
By applying this relation backward to $D_n^{-q*}$, one gets
\eqn{
\frac{m_\mathrm{G}\ \delta(r-r_\mathrm{G})}{r^2_\mathrm{G}} D_n^{-q*} & = \int_S \Psi_n^{-q}(\theta,\varphi) \rho(\vect{r_\mathrm{G}})\ ds \nonumber \\
& = (-1)^q \int_S \Psi_n^{q*}(\theta,\varphi) \rho(\vect{r_\mathrm{G}})\ ds \nonumber \\
& = (-1)^q \ \frac{m_\mathrm{G}\ \delta(r-r_\mathrm{G})}{r^2_\mathrm{G}} D_n^{q} \nonumber \\
D_n^{-q*} & = (-1)^q D_n^{q}.
}
In particular, one has
\eqn{
D_n^{-1} = - D_n^{1*},
}
so that
\eqn{
D_n^{-1}K_n^{-1} = D_n^{1*}K_n^{1}.
}
Thus, one finally has
\eqn{
\vect{\Gamma_\mathrm{G/c}} = & \frac{G m_\mathrm{c} m_\mathrm{G}}{a} \sum_{n=1}^{\infty} \sqrt{n(n+1)} \left(\frac{a}{r_\mathrm{G}}\right)^{n+1} \ C_n^0 \left[\mathfrak{R}(D_n^1) \vectu{y} + \mathfrak{I}(D_n^1) \vectu{x} \right],
}
where $\mathfrak{R}(z)$ and $\mathfrak{I}(z)$ denote the real and imaginary parts of the complex $z$. It is also possible to define an angle $\beta$ so that $\tan{\beta} = -\mathfrak{I}(D_n^1)/\mathfrak{R}(D_n^1)$, and write
\eqn{
\vect{\Gamma_\mathrm{G/c}} = & \frac{G m_\mathrm{c} m_\mathrm{G}}{a} \sum_{n=1}^{\infty} \sqrt{n(n+1)} \left(\frac{a}{r_\mathrm{G}}\right)^{n+1} \ C_n^0\ |D_n^1| \left(\cos{\beta}\ \vectu{y} + \sin{\beta}\ \vectu{x} \right) \nonumber \\
= & \frac{G m_\mathrm{c} m_\mathrm{G}}{a} \sum_{n=1}^{\infty} \sqrt{n(n+1)} \left(\frac{a}{r_\mathrm{G}}\right)^{n+1} \ C_n^0\ |D_n^1|\ \vectu{\beta}',
}
where $|D_n^1|$ is the modulus of the complex number $D_n^1$.

%%%%%%%%%%%%%%%%%%%%%%%%%%%%%%%%%%%%%%%%%%%%%%%%%%
\sect[spher_harm_general_case]{General case}

Up to now, we have considered the cluster to be cylindrically symmetric. If now we go to the general case, the gradient takes a more complicated form. By analogy with \refeqn{phi_cluster_cylindrical}, the potential of the cluster evaluated on the sphere of radius $a < r$ is now
\eqn{
\phi_\mathrm{c}(\vect{r}) = & \frac{G m_\mathrm{c}}{a} \sum_{n=0}^{\infty}\sum_{q=-n}^{n} \left(\frac{a}{r}\right)^{n+1} C_n^q\ \Psi_n^{q}(\theta,\varphi).
}
We still have
\eqn{
\vect{r} \times \vect{\nabla}\phi_\mathrm{c}(\vect{r}) & = -\frac{1}{\sin{\theta}}\frac{\partial \phi_\mathrm{c}(\vect{r})}{\partial \varphi}\ \vectu{\theta} + \frac{\partial \phi_\mathrm{c}(\vect{r})}{\partial \theta}\ \vectu{\varphi},
}
but now, both terms are non-zero:
\eqn{
\frac{\partial \Psi_n^{q}(\theta,\varphi)}{\partial \theta} & = \frac{Gm_\mathrm{c}}{a}\sum_{n=0}^{\infty}\sum_{q=-n}^{n} \left(\frac{a}{r}\right)^{n+1}C_n^q K_n^q\ e^{\imath q\varphi} \frac{d P_n^{|q|}(\cos{\theta})}{d \theta} \\
\frac{\partial \Psi_n^{q}(\theta,\varphi)}{\partial \varphi} & = \frac{Gm_\mathrm{c}}{a}\sum_{n=0}^{\infty}\sum_{q=-n}^{n} \left(\frac{a}{r}\right)^{n+1} (\imath q) C_n^q\ \Psi_n^{q}(\theta,\varphi).
}
One has
\eqn{
\frac{d P_n^{|q|}(\cos{\theta})}{d \theta} & = \frac{d}{d \theta}\left(\sin^{|q|}{(\theta)}\ \frac{d^{|q|} P_n^{0}(\cos{\theta})}{d \cos^{|q|}{(\theta)}} \right) \nonumber \\
& = |q| \sin^{|q|-1}{(\theta)} \cos{\theta}\ \frac{d^{|q|} P_n^{0}(\cos{\theta})}{d \cos^{|q|}{(\theta)}} + \sin^{|q|}{(\theta)} \frac{d\cos{\theta}}{d\theta}\frac{d}{d\cos{\theta}}\left(\frac{d^{|q|} P_n^{0}(\cos{\theta})}{d \cos^{|q|}{(\theta)}}\right) \nonumber \\
& = |q| \frac{\cos{\theta}}{\sin{\theta}}\ P_n^{|q|}(\cos{\theta}) - P_n^{|q|+1}(\cos{\theta}).
}
Note that this expression is not defined for $q = \pm n$ because of the second associated Legendre function. However, for these values, one can demonstrate that
\eqn{
\frac{d P_n^{|q|}(\cos{\theta})}{d \theta} = |q| \frac{\cos{\theta}}{\sin{\theta}}\ P_n^{|q|}(\cos{\theta}), 
}
i.e. same as previous but without the second term. Finally,
\eqn[general_solution_spherical_harmonics]{
\vect{\Gamma_\mathrm{G/c}} = & \frac{Gm_\mathrm{c}}{a}  \int_V \rho_\mathrm{G}(\vect{r})  \sum_{n=0}^{\infty} \left\{\sum_{q=-n}^{n} C_n^q\left(\frac{a}{r}\right)^{n+1} \left[-\frac{\imath q}{\sin{\theta}}\ \Psi_n^{q}(\theta,\varphi)\ \vectu{\theta} + |q| \frac{\cos{\theta}}{\sin{\theta}}\ \Psi_n^{q}(\theta,\varphi) \vectu{\varphi} \right] \right. \nonumber \\
& \left. - \sum_{q=-n+1}^{n-1} C_n^q \left(\frac{a}{r}\right)^{n+1} \frac{K_n^q}{K_n^{|q|+1}}\ e^{\imath\varphi (q-|q|-1)} \Psi_n^{|q|+1}(\theta,\varphi)\ \vectu{\varphi} \right\} dV.
}
From \refeqn{klm}, one recalls that
\eqn{
\frac{K_n^q}{K_n^{|q|+1}} = \sqrt{n^2+n-q^2-|q|} \times \left\{ \begin{array}{ll}-1 & \textrm{ for } q \geq 0 \\ (-1)^{|q|+1} & \textrm{ for } q < 0 \end{array}\right.,
}
and one retrieves the result of \refeqn{torque} for $q = 0$ (i.e. cylindrical symmetry).

It is now possible to replace the density of the galaxy with its decomposition in spherical harmonics (as in \refsec{general_galaxy}). Though technically possible, such an analytical derivation requires many terms to give a precise description of a real galaxy. Therefore, the use of $N$-body simulations would be prefered to a purely analytical work. In this case, the volume integral can be replaced by a discrete sum over all the particles of the galaxy:
\eqn[density_discrete]{
\vect{\Gamma_\mathrm{G/c}} = & \frac{Gm_\mathrm{c}}{a}  \sum_{j=1}^N m_j  \sum_{n=0}^{\infty} \left\{\sum_{q=-n}^{n} C_n^q\left(\frac{a}{r_j}\right)^{n+1} \left[-\frac{\imath q}{\sin{\theta_j}}\ \Psi_n^{q}(\theta_j,\varphi_j)\ \vectu{\theta}_j + |q| \frac{\cos{\theta_j}}{\sin{\theta_j}}\ \Psi_n^{q}(\theta_j,\varphi_j) \vectu{\varphi}_j \right] \right. \nonumber \\
& \left. - \sum_{q=-n+1}^{n-1} C_n^q \left(\frac{a}{r_j}\right)^{n+1} \frac{K_n^q}{K_n^{|q|+1}}\ e^{\imath\varphi_j (q-|q|-1)} \Psi_n^{|q|+1}(\theta_j,\varphi_j)\ \vectu{\varphi}_j \right\},
}
where $m_j$ is the mass of the $j$-th particle of the galaxy, located at coordinates $(r_j, \theta_j, \varphi_j)$.

\sect{Opposite torque}

The derivations presented above use the potential of the cluster and the density of the galaxy to compute the torque. In some cases, it is easier to consider the opposite. One could be tempted to simply switch the two subscripts and get minus the torque ($\vect{\Gamma_\mathrm{c/G}} = -\vect{\Gamma_\mathrm{G/c}}$). If this is true in principle, one should not forget that the way the potential is evaluated from the Anderson's sphere depends on whether one stands \emph{inside} or \emph{outside} this sphere. \refeqn{Anderson} gives
\eqn{
\phi_\mathrm{c}(\vect{r}) = \frac{1}{4\pi}\sum_{n=0}^{\infty} (2n+1) \left(\frac{a}{r}\right)^{n+1} \int_S P_n\left(\frac{\vect{s}\cdot\vect{r}}{r}\right) \phi_\mathrm{c}(a\vect{s})\ ds,
}
for $r > a$, which is true when the potential of the cluster is needed at the position of the galaxy. But when doing the opposite, the potential of the galaxy is required at the positions of the stars of the cluster, i.e. inside the sphere ($r < a$). In this case, one has to write
\eqn{
\phi_\mathrm{G}(\vect{r}) = \frac{1}{4\pi}\sum_{n=0}^{\infty} (2n+1) \left(\frac{r}{a}\right)^n \int_S P_n\left(\frac{\vect{s}\cdot\vect{r}}{r}\right) \phi_\mathrm{G}(a\vect{s})\ ds.
}
Because of the cross product with $\vect{r}$, the gradient of $\phi$ along $r$ is never computed and thus, this change does not affect the final results more than transforming
\eqn{
\left(\frac{a}{r}\right)^{n+1} \qquad\mathrm{into}\qquad -\left(\frac{r}{a}\right)^n
}
when switching the subscripts $G$ and $c$.

%%%%%%%%%%%%%%%%%%%%%%%%%%%%%%%%%%%%%%%%%%%%%%%%%%

\cha{virial}{Virial theorem}
\chaphead{In this Appendix, we derive the virial theorem in the general case, and then in the context of an isotropic tidal field. We also present the concept of virial radius.}

%%%%%%%%%%%%%%%%%%%%%%%%%%%%%%%%%%%%%%%%%%%%%%%%%%
\sect{Derivation of the scalar virial theorem}

%%%%%%%%%%%%%%%%%%%%%%
\subsect{General case}
Let's consider a system of $N$ particles. Its moment of inertia is given by
\eqn{
I = \sum_{i=1}^N m_i \ r_i^2.
}
By deriving it, we obtain
\eqn[virialinertia]{
\frac{dI}{dt} & = 2 \sum_{i=1}^N m_i \ v_i \ r_i \nonumber \\
\frac{1}{2} \frac{d^2I}{dt^2} & = \sum_{i=1}^N m_i \ a_i \ r_i +  \sum_{i=1}^N m_i \ v_i^2 \nonumber \\
& = \sum_{i=1}^N F_i \ r_i +  \sum_{i=1}^N m_i \ v_i^2,
}
where $a_i$ is the acceleration of the $i$-th particle. The forces and their potential energies can be writen as
\eqn{
E = \kappa \ r^\xi \qquad \mathrm{and} \qquad F = -\nabla E = -\kappa \ \xi \ r^{\xi-1},
}
where $\kappa$ is a constant and $\xi$ an integer. By replacing the term $\left( F_i\ r_i \right)$ in \refeqn{virialinertia}, one gets
\eqn{
\frac{1}{2} \frac{d^2I}{dt^2} = - \sum_\mathrm{potentials} \xi \ E  +  2 K,
}
where $K$ is the kinetic energy. This equation is called general virial theorem but is rarely used. In the case of a stationnary system, the first derivative of the moment of inertia is null, and \emph{a fortiori} is the second. Therefore we have a simpler form for the theorem:
\eqn[virialgeneral]{
2K - \sum_\mathrm{potentials} \xi \ E = 0.
}

%%%%%%%%%%%%%%%%%%%%%%
\subsect{Gravitational case}
Usual assuptions in stellar dynamics set that the only interaction in the system is the gravitation ($\xi = -1$), whose potential energy is $\Omega$. Therefore, the virial theorem for a stationary system (\refeqn{virialgeneral}) with gravitational interaction only becomes
\eqn[virial]{
2K + \Omega = 0.
}

%%%%%%%%%%%%%%%%%%%%%%
\subsect[virial_tides]{In a tidal field}
Let's consider a star cluster embedded in an isotropic tidal field represented by the tensor
\eqn{
T = \left( \begin{array}{ccc} \lambda & 0 & 0 \\ 0 & \lambda & 0 \\ 0 & 0 & \lambda \end{array} \right).
}
The differencial force between an element $i$ of mass $m_i$ and the barycenter of the cluster is
\eqn{
\delta \vect{F}_{\mathrm{tides},i} = m_i \lambda \ \vect{r}_i.
}
We note that in this case $\xi = 2$ and thus
\eqn{
E_{\mathrm{tides},i} = -\frac{1}{2}m_i \lambda r_i^2.
}
When summing on all the mass elements, if we assume that the tidal field is uniform over the entire cluster, we have
\eqn[virialtides]{
\sum_{i=1}^N F_{\mathrm{tides},i} \ r_i & = \int_0^M \lambda r^2 \ dm \nonumber \\
& = \lambda \alpha \ M R_\mathrm{t}^2,
}
where $M$ is the total mass of the cluster, $R_\mathrm{t}$ is the radius where the density drops to zero, and
\eqn[alpha]{
\alpha \equiv \frac{1}{M\ R_\mathrm{t}^2} \int_0^M r^2 \ dm
}
(see \refsec[virial]{alpha} for various examples). Finally, we can write for the entire cluster
\eqn[tidal_energy]{
E_\mathrm{tides} = -\frac{1}{2}\lambda \alpha \ M \ R_\mathrm{t}^2
}
and, from \refeqn{virialtides}, the new virial theorem becomes
\eqn[virial_with_tides]{
2K + \Omega + \lambda \alpha \ M \ R_\mathrm{t}^2 = 0.
}

%%%%%%%%%%%%%%%%%%%%%%%%%%%%%%%%%%%%%%%%%%%%%%%%%%
\sect[virialradius]{Velocity dispersion and virial radius}

In the relation above, global quantities have been derived by summing over all the mass elements of the cluster. It is also possible to use quantities that globaly characterize the cluster. For example, the velocity dispersion $\sigma$ can be used in the expression of the kinetic energy:
\eqn{
K = \frac{1}{2}\sum_{i=1}^N m_i \ v_i^2 = \frac{1}{2} M \sigma^2
}
and the same can be done with the gravitational energy
\eqn{
\Omega = -\frac{GM^2}{2r_\mathrm{v}},
}
where $r_\mathrm{v}$ is called virial radius\footnote{This definition is based on \mycitet{Meylan1997}, but \mycitet{Binney1987} set another expression.}. Therefore, the total energy of a system in virial equilibrium (without tidal field: $K = -\Omega /2$) is
\eqn[energy_virial]{
E = K + \Omega = \frac{\Omega}{2} = -\frac{GM^2}{4r_\mathrm{v}}.
}

%%%%%%%%%%%%%%%%%%%%%%%%%%%%%%%%%%%%%%%%%%%%%%%%%%
\sect[alpha]{$\alpha$ parameter}

As defined in \refeqn{alpha}, the $\alpha$ parameter describes the mass distribution of a spatially extended object. Using usual mass profiles (like, e.g. those of \refapp{triplets}), it is possible to calculate $\alpha$ via the density $\rho$ and the relation $dm = 4\pi \rho(r) r^2 \ dr$:
\eqn{
\alpha = \frac{4\pi}{M R_\mathrm{t}^2} \int_0^\infty \rho(r) r^4 dr.
}
Note that the upper limit of the integral can now be set to infinity because the density will drop to zero where the integrated mass reaches $M$.

%%%%%%%%%%%%%
\subsect{Homogeneous sphere}
Let's first consider a homogeneous sphere of density $\rho_0$ and radius $R_\mathrm{t}$
\eqn{
\rho(r) = \left\{ \begin{array}{ll} 
\rho_0 & \textrm{if }r < R_\mathrm{t} \\
0 & \textrm{else}.
\end{array} \right.
}
Immediately, one gets
\eqn{
\alpha = \frac{4\pi \rho_0}{5 M}  R_\mathrm{t}^3
}
and, with
\eqn{
\rho_0 = \frac{3 M}{4\pi R_\mathrm{t}^3},
}
one has
\eqn{
\alpha = \frac{3}{5}.
}

%%%%%%%%%%%%%
\subsect{Power law profile}
Let's now consider a density profile defined by
\eqn{
\rho(r) = \left\{ \begin{array}{ll} 
\rho_0 \left(\frac{r}{r_0}\right)^{-\gamma} & \textrm{if }r < R_\mathrm{t} \\
0 & \textrm{else}.
\end{array} \right.
}
We have
\eqn{
\alpha & = \frac{4\pi \rho_0 r_0^\gamma}{M R_\mathrm{t}^2} \int_0^{R_\mathrm{t}} r^{4-\gamma} \ dr \nonumber \\
& = \frac{4\pi \rho_0 r_0^\gamma \ R_\mathrm{t}^{3-\gamma}}{M (5-\gamma)}.
}
We also know that
\eqn{
M & = 4\pi \rho_0 r_0^\gamma \int_0^{R_\mathrm{t}} r^{2-\gamma} \ dr \nonumber \\
& = \frac{4\pi \rho_0 r_0^\gamma \ R_t^{3-\gamma}}{3-\gamma}
}
so,
\eqn{
\alpha = \frac{3-\gamma}{5-\gamma}.
}
In the case $\gamma = 0$, we retrieve the homogeneous sphere.

%%%%%%%%%%%%%
\subsect{Plummer sphere}
For a Plummer sphere, we use the density given in \refeqnp{plummerdensity} and change the variable thanks to $\tan{\theta} = r/r_0$, to get
\eqn{
\alpha & = \frac{3r_0^2}{R_\mathrm{t}^2} \int_0^{\infty}\frac{r^4 \ dr}{\left(r_0^2+r^2\right)^{5/2}} \nonumber \\
& = \frac{3r_0^2}{R_\mathrm{t}^2} \int_0^{\hat{\theta}} \frac{\tan^4{\theta} \ d\theta}{\cos^2{\theta} \left(\frac{1}{\cos^2{\theta}}\right)^{5/2}} \nonumber \\
& = \frac{3r_0^2}{R_\mathrm{t}^2} \int_0^{\hat{\theta}} \frac{\sin^4{\theta} \ d\theta}{\cos{\theta}} \nonumber \\
& = \frac{3r_0^2}{R_\mathrm{t}^2} \int_0^{\hat{\theta}} \frac{\sin^4{\theta}}{1 - \sin^2{\theta}} \cos{\theta} \ d\theta.
}
Then, we set $y = \sin{\theta}$ and get
\eqn{
\alpha & = \frac{3r_0^2}{R_\mathrm{t}^2} \int_0^{\hat{y}} \frac{y^4 \ dy}{1-y^2} \nonumber \\
& = \frac{3r_0^2}{R_\mathrm{t}^2} \int_0^{\hat{y}} \left( \frac{1}{1 - y^2} - y^2 -1 \right) dy \nonumber \\
& = \frac{3r_0^2}{R_\mathrm{t}^2} \left(\arg\tanh{\hat{y}} - \frac{\hat{y}^3}{3} - \hat{y} \right).
}
However, if the Plummer model yields a finite mass, it does not have a truncation radius. Therefore, we have to set an artificial one. Knowing the integrated mass (see \refeqnt{plummerintegratedmass}), we can see that $99\%$ of the total mass $M$ is enclosed in the sphere of radius $R_\mathrm{t} \approx 12 r_0$.

\mybox{
To determine the value of $\hat{y}$:
\eqn{
\hat{y} = \sin{\hat{\theta}} = \sin{\left[\arctan{\left(\frac{R_\mathrm{t}}{r_0}\right)}\right]},
}
let's use the complex notation ($\imath^2 = -1$ and $\Im(z)$ is the imaginary part of the complex $z$). We know that
\eqn{
\sin{z} = \Im \left[\exp{\left(\imath z\right)}\right] \qquad \mathrm{and} \qquad \arctan{z} = \frac{1}{2\imath} \ln{\left(\frac{1+\imath z}{1-\imath z}\right)}.
}
\follow}\mybox[nocnt]{
Therefore, 
\eqn{
\sin{\left(\arctan{z}\right)} = \Im \left(  \sqrt{\frac{1+\imath z}{1-\imath z}}  \right) = \frac{z}{\sqrt{1+z^2}}
}
which is obviously true for real numbers too. 
}

With our arbitrary truncation radius, we get
\eqn{
\hat{y} = \frac{12}{\sqrt{145}}
}
which gives
\eqn{
\alpha \approx 3.86 \times 10^{-2}.
}

A similar development can be used for the Hernquist profile, this time by setting the arbitrary truncation radius at $R_\mathrm{t} = 200 r_0$. In this case, one obtains $\alpha = 9.33 \times 10^{-3}$.

%%%%%%%%%%%%%%%%%%%%%%%%%%%%%%%%%%%%%%%%%%%%%%%%%%
\sect[sfe_hills]{Star formation efficiency}

%%%%%%%%%%%%%
\subsect[sfe_hills_isolation]{In isolation}

The total energy of young cluster in isolation is simply the sum of its potential and kinetic energies:
\eqn{
E_0 & = K + \Omega \nonumber \\
& = \frac{1}{2} M \sigma^2 -\frac{GM^2}{2r_\mathrm{v}}.
}
This cluster formed with a \acr{}{SFE}{star formation efficiency} $\epsilon$ defined as the ratio between the stellar mass to the total (stars + gas) mass:
\eqn{
\epsilon \equiv \frac{M_\star}{M}.
}
Now, let's assume that this cluster expels its gaseous content, via stellar feedback. If the expulsion is rapid enough (impulsive mass loss, see also \mycitealt{Hills1980}; \mycitealt{Boily2003a}), one can consider that neither the velocity dispersion nor the virial radius have changed during the process. Therefore, the energy becomes
\eqn{
E_1 & = \frac{1}{2} M_\star \sigma^2 -\frac{GM_\star^2}{2r_\mathrm{v}} \nonumber \\
& = \frac{1}{2} M \epsilon \ \sigma^2 -\frac{GM^2 \ \epsilon^2}{2r_\mathrm{v}}.
}
Assuming that the cluster was initially virialized, one has
\eqn{
M \sigma^2 = \frac{GM^2}{2r_\mathrm{v}}, 
}
therefore
\eqn{
E_1 & = \frac{GM^2 \ \epsilon}{4r_\mathrm{v}} -\frac{GM^2 \ \epsilon^2}{2r_\mathrm{v}} \nonumber \\
& = \frac{GM^2 \ \epsilon}{2r_\mathrm{v}} \left( \frac{1}{2}-\epsilon \right).
}
The stars of such a cluster remain bound if the energy is negative, i.e. if $\epsilon > 50\%$. In other words, when a molecular cloud collapses, its material is gravitationally bound in a potential well created by a mass $M$. When the first stars blow the remaining gas away, this mass changes, which flattens the well and allows the less bound stars to escape. If the fraction of the remaining gas is less than $1/2$, the mass loss is not important enough to allows this unbinding process, and the cluster survives

%%%%%%%%%%%%%
\subsect{In a tidal field}

When the star cluster is not isolated but embedded in a tidal field, one must take the tidal energy into account:
\eqn{
E_0 & = K + \Omega + E_\mathrm{tides} \nonumber \\
& = \frac{1}{2} M \sigma^2 -\frac{GM^2}{2r_\mathrm{v}} -\frac{1}{2}\lambda \alpha \ M \ R_\mathrm{t}^2.
}

%%%%%%%%%%%%%
\subsubsection{Impulsive approximation}

Once again, after the instantaneous expulsion of the gas (which leaves $\sigma$, $r_\mathrm{v}$ and $R_\mathrm{t}$ unchanged), one can write
\eqn{
E_1 & = \frac{1}{2} M_\star \sigma^2 -\frac{GM_\star^2}{2r_\mathrm{v}} -\frac{1}{2}\lambda \alpha \ M_\star \ R_\mathrm{t}^2 \nonumber  \\
& = \frac{\epsilon}{2} \left( M \sigma^2 -\frac{GM^2 \ \epsilon}{r_\mathrm{v}} - \lambda \alpha \ M\ R_\mathrm{t}^2 \right).
}
Using the virial theorem (\refeqn{virial_with_tides}) at the initial stage, one gets
\eqn{
E_1 = \frac{G \ \epsilon \ M^2}{2r_\mathrm{v}} \left(\frac{1}{2} - \epsilon \right)  - \epsilon \ \lambda \alpha \ M\ R_\mathrm{t}^2.
}

\fig{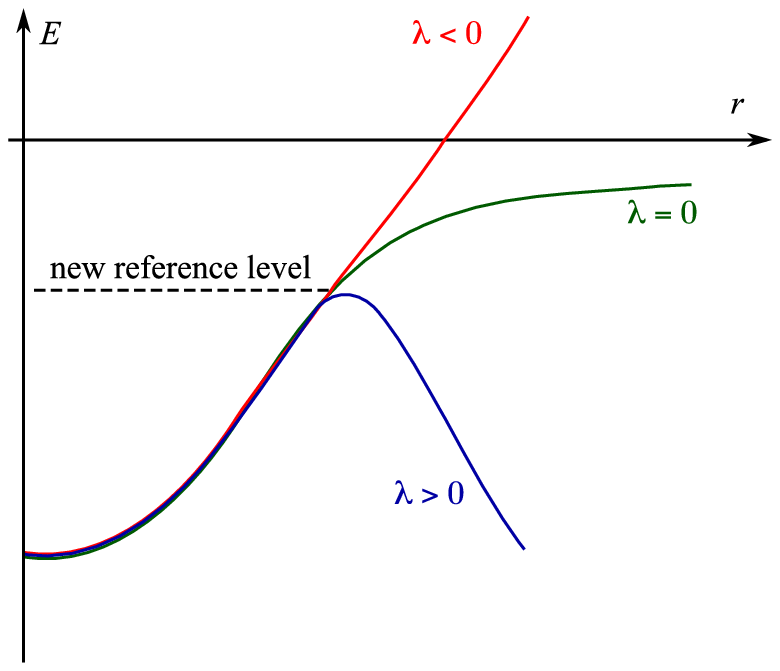}{Energy of a cluster}
{Schematic view of the energy profile of a cluster. With no tidal field (green), the criterion for a bound system is $E<0$. When adding an extensive field (blue), this level is shifted so that even a cluster with $E<0$ can be unbound. This situation does not occurs with a compressive tidal field (red).}

Because of the tidal term, the potential energy is now the sum of the intrinsic potential of the cluster, and the external potential (via the tides) seen as an harmonic potential (\reffig{bound_cluster}). This shifts the reference level of energy and thus, a bound system does not necessarily correspond to $E<0$. Instead, we compare the energy $E_1$ of the cluster after the gas expulsion to those of a (virtual) cluster of the same mass but virialized, and still embedded in the same tidal field:
\eqn{
E' & = \frac{1}{2} M_\star \sigma^2 -\frac{GM_\star^2}{2r'_\mathrm{v}} -\frac{1}{2}\lambda \alpha' \ M_\star \ {R'}_\mathrm{t}^2 \nonumber \\
& = \frac{\epsilon}{2} M \sigma^2 -\frac{G \ \epsilon^2 \ M^2}{2r'_\mathrm{v}} -\frac{\epsilon}{2}\lambda \alpha' \ M \ {R'}_\mathrm{t}^2,
}
which becomes, thanks to the virial theorem,
\eqn[virial_post_gas]{
E' = -\frac{G \ \epsilon^2 \ M^2}{4r'_\mathrm{v}} - \epsilon \ \lambda \alpha' \ M\ {R'}_\mathrm{t}^2.
}
Hence, our cluster remains bound if its energy after expulsion $E_1$ is lower than the reference level $E'$ of a virialiazed system of same mass, i.e.
\eqn{
E_1 \le E'
}
\eqn{
\frac{G \ \epsilon \ M^2}{4r_\mathrm{v}} \left(1 - 2\epsilon + \epsilon\  \frac{r_\mathrm{v}}{r'_\mathrm{v}} \right) - \epsilon \ \lambda \ M \left( \alpha \ R_\mathrm{t}^2 - \alpha' \ {R'}_\mathrm{t}^2 \right) \le 0.
}
We set
\eqn[cste_a]{
A \equiv \epsilon \ \lambda \alpha \ M \ R^2_\mathrm{t} \left(\frac{G \ \epsilon \ M^2}{r_\mathrm{v}}\right)^{-1} = \frac{\lambda \alpha \ R^2_\mathrm{t}\ r_\mathrm{v}}{G M},
}
which leads to
\eqn[bound_impuls]{
1 - 2\epsilon + \epsilon \left(\frac{r'_\mathrm{v}}{r_\mathrm{v}}\right)^{-1} - 4 A \left[ 1 - \frac{\alpha'}{\alpha} \left( \frac{{R'}_\mathrm{t}}{R_\mathrm{t}} \right)^2 \right] \le 0.
}

%%%%%%%%%%%%%
\subsubsection{Adiabatic evolution}
If one does not consider the expulsion of gas as an instantaneous process but rather as a slow evolution, the energy of the gas-free cluster takes into account the modified shape parameters ($\alpha'$, $r'_\mathrm{v}$ and ${R'}_\mathrm{t}$):
\eqn{
E_1 = \frac{1}{2} M_\star \sigma^2 -\frac{GM_\star^2}{2r'_\mathrm{v}} -\frac{1}{2}\lambda \alpha' \ M_\star \ {R'}_\mathrm{t}^2.
}
Again, we use the virial theorem from the initial state:
\eqn{
E_1 = \frac{G \ \epsilon \ M^2}{2r_\mathrm{v}} \left(\frac{1}{2} - \epsilon\ \frac{r_\mathrm{v}}{r'_\mathrm{v}} \right)  - \frac{1}{2} \epsilon \ \lambda \ M \left(\alpha\ R_\mathrm{t}^2 + \alpha'\ {R'}_\mathrm{t}^2\right).
}
This energy has to be compared to the enegy of the same system if it was virialized (recall \refeqn{virial_post_gas}):
\eqn{
\frac{G \ \epsilon \ M^2}{4r_\mathrm{v}} \left(1 - \epsilon\ \frac{r_\mathrm{v}}{r'_\mathrm{v}} \right) - \frac{1}{2} \epsilon \ \lambda \ M \left(\alpha\ R_\mathrm{t}^2 - \alpha' \ {R'}_\mathrm{t}^2 \right) \le 0.
}
With \refeqn{cste_a}, we obtain
\eqn[bound_adia]{
1 - \epsilon \left(\frac{r'_\mathrm{v}}{r_\mathrm{v}}\right)^{-1} - 2A \left[ 1 - \frac{\alpha'}{\alpha} \left( \frac{{R'}_\mathrm{t}}{R_\mathrm{t}} \right)^2 \right] \le 0.
}

%%%%%%%%%%%%%
\subsubsection{Solution}
In both cases (impulsive or adiabatic), one is able to get the properties (mainly the radii) of the bound cluster after the gas expulsion, depending on the strenght of the tidal field (which is encapsulated in the term $A$). However, the analytical solving of \refeqn{bound_impuls} or \refeqn{bound_adia} is only feasible with a simplifying assumption of the ratios $\alpha'/\alpha$, $r'_\mathrm{v}/r_\mathrm{v}$ and ${R'}_\mathrm{t}/R_\mathrm{t}$.

The minimal hypothesis which consists in assuming a homologous transformation (i.e. $r'_\mathrm{v}/r_\mathrm{v} = {R'}_\mathrm{t}/R_\mathrm{t}$) does not recover the different responses of the core and the external part of the cluster to the mass-loss, as noted by \mycitet{Boily2003b}. However, an analytical solution does not exist in the general case, because the relations giving the evolution of the radii and mass profile are not known. An example of derivation in the frame of the (non-realistic) homologous transformation is given in the \refsecp[formation]{sfe_massloss}.

%%%%%%%%%%%%%%%%%%%%%%%%%%

\cha{}{List of publications}
\chaphead{This Appendix lists the publications (in preparation, submitted, in press, refereed and conference proceedings) I collaborated to. A short teaser (instead of an abstract) is available for the major papers.}

%%%%%%%%%%%%%%%%%%%%%%%%%%%%%%%%%%%%
\sect{In prep., submitted...}

\begin{enumerate}[{$\bullet\phantom{[1]}$}] % deals with margin properly (instead of itemize)  The phantom is used to align with [1] below
\item	\textbf{Renaud F.}, Boily C.M. \& Theis C.\\ 
	\emph{Dark matter halos profiles and tidal fields}\\
	MNRAS Letters, submitted\\
	In this short contribution (5 pages, 3 figures), we study analytically the tidal field of \acr{}{DM}{dark matter} halos profiles, some derived from cosmological simulations (\acr{}{NFW}{NFW}, Einasto). Our results show that the superposition of some potentials (Dehnen's $\gamma$-family, Einasto), as in configurations close to galaxy-galaxy interactions, yields large areas of compressive tides, while \acr{}{NFW}{NFW} do not. Because the star formation is very likely enhanced when embedded in a compressive tidal field, we claim that the precise shape of the \acr{}{DM}{dark matter} halo should be chosen with care, when modeling star formation in galaxy mergers.

\item	\textbf{Renaud F.}, Appleton P. \& Xu K.\\ 
	\emph{$N$-body simulation of the Stephan's Quintet}\\
	ApJ, submitted\\
	In this paper (12 pages, 13 figures), we propose a possible scenario for the formation of the Stephan's Quintet (a compact group of five galaxies) thanks to collisionless $N$-body simulations. With an innovative serial approach, we cover a large range of parameters and rule out several scenarios. We discuss the possible future of the group, in the more general frame of the hierarchical formation and evolution scenarios for compact groups of galaxies.
\end{enumerate}

%%%%%%%%%%%%%%%%%%%%%%%%%%%%%%%%%%%%
\sect{Refereed papers}
\begin{enumerate}[{[}1{]$\phantom{\bullet}$}]
\item	Karl S., Naab T., Johansson P., Kotarba H., Boily C.M., \textbf{Renaud F.} \& Theis C.\\ 
	\emph{One moment in time - Modeling star formation in the Antennae}\\
	ApJ Letters, 715, L88 \linkads{http://adsabs.harvard.edu/abs/2010ApJ...715L..88K}\\
	In this Letter (6 pages, 4 figures), we present a \acr{}{SPH}{smoothed particle hydrodynamics} simulation of the Antennae galaxies. The hydrodynamical aspects of star formation, feedback and radiative cooling are included to measure the \acr{}{SFR}{star formation rate} along time, and the distributions of star clusters in the central region. An enhancement of the \acr{}{SFR}{star formation rate} is noted shortly after the two passages. This new model complements our purely gravitational approach by confirming the role of the large-scale interactions on the demographics of the star clusters.
	
\item	\textbf{Renaud F.}, Boily C.M., Naab T. \& Theis C.\\ 
	\emph{Fully compressive tides in galaxy mergers}\\
	2009, ApJ, 706, 67 \linkads{http://adsabs.harvard.edu/abs/2009ApJ...706...67R}\\
	This paper (16 pages, 16 figures) is the direct continuation of the Letter Renaud et al. (2008) published one year before in MNRAS. The concept of compressive tides is introduced analytically and  explored through a series of $N$-body simulations. The Antennae galaxies are taken as an observational reference. Then, a survey on the parameters of interacting galaxies is presented, to broaden the conclusions from the Antennae. The paper closes with a discussion on the link between compressive tidal field and star formation, in term of energy, enhancement of the \acr{}{SFR}{star formation rate} and multiple stellar populations.

\item	\textbf{Renaud F.}, Boily C.M., Fleck J.-J., Naab T. \& Theis C.\\ 
	\emph{Star cluster survival and compressive tides in Antennae-like mergers}\\
	2008, MNRAS Letters, 391, L98 \linkads{http://adsabs.harvard.edu/abs/2008MNRAS.391L..98R}\\
	In this Letter (5 pages, 3 figures), we briefly introduce the concept of compressive tides and concentrate on an $N$-body model of the Antennae galaxies. Details of the spatial distribution of the compressive regions, as well as their characteristic times are presented and confronted to recent observational data on young clusters in this merger.
\setcounter{saveenumi}{\theenumi} % save counter value
\end{enumerate}

\newpage

%%%%%%%%%%%%%%%%%%%%%%%%%%%
\sect{Conference proceedings}

\begin{enumerate}[{[}1{]$\phantom{\bullet}$}]
\setcounter{enumi}{\thesaveenumi} % restore counter
	
\item	\textbf{Renaud F.}, Theis C., Naab T. \& Boily C.M.\\ 
	\emph{Substructures formation induced by gravitational tides?}\\
	Galaxy Wars - Johnson City, Tn, USA (19-22 July 2009)\\
	2010, ASP Conference Series, 423, 191 \linkads{http://adsabs.harvard.edu/abs/2010ASPC..423..191R}

\item	Hwang J.-S., Struck C., \textbf{Renaud F.} \& Appleton P.\\ 
	\emph{Models of the intergalactic gas in Stephan's Quintet}\\
	Galaxy Wars - Johnson City, Tn, USA (19-22 July 2009)\\
	2010, ASP Conference Series, 423, 232 \linkads{http://adsabs.harvard.edu/abs/2010ASPC..423..232H}

\item	Boily, C.M., Fleck J.-J., Lan\c{c}on A. \& \textbf{Renaud F.}\\ 
	\emph{The mass-to-light ratio of rich star clusters}\\
	Young massive star clusters - Granada, Spain (11-14 September 2007)\\
	2009, Astrophysics and Space Science, 324, 265 \linkads{http://adsabs.harvard.edu/abs/2009Ap&SS.324..265B}

\item	\textbf{Renaud F.}, Boily C.M. \& Theis C.\\ 
	\emph{Starburst triggered by compressive tides in galaxy mergers}\\
	Galactic and Stellar Dynamics 2008 - Strasbourg, France (16-20 March 2008)\\
	2008, Astronomische Nachrichten, 329, 1050 \linkads{http://adsabs.harvard.edu/abs/2008AN....329.1050R}
\end{enumerate}

\cdp

%%%%%
\begin{thegeneralbibliography}{}
\bibitem[Aarseth(1963)]{Aarseth1963}\linkup{autobib:Aarseth1963} Aarseth, S.~J.\ 1963, \mnras, 126, 223\linkads{http://adsabs.harvard.edu/abs/1963MNRAS.126..223A}
\bibitem[Aarseth(1999)]{Aarseth1999}\linkup{autobib:Aarseth1999} Aarseth, S.~J.\ 1999, \pasp, 111, 1333\linkads{http://adsabs.harvard.edu/abs/1999PASP..111.1333A}
\bibitem[Anders et al.(2007)]{Anders2007}\linkup{autobib:Anders2007} Anders, P., Bissantz, N., Boysen, L., de Grijs, R., \& Fritze-v.~Alvensleben, U.\ 2007, \mnras, 377, 91\linkads{http://adsabs.harvard.edu/abs/2007MNRAS.377...91A}
\bibitem[Anderson(1992)]{Anderson1992}\linkup{autobib:Andersson1992} Anderson, C.\ 1992, SIAM Journal on Scientific and Statistical Computing, 13, 923
\bibitem[Athanassoula et al.(1998)]{Athanassoula1998}\linkup{autobib:Athanassoula1998} Athanassoula, E., Bosma, A., Lambert, J.-C., \& Makino, J.\ 1998, \mnras, 293, 369\linkads{http://adsabs.harvard.edu/abs/1998MNRAS.293..369A}
\bibitem[Baldi et al.(2006)]{Baldi2006}\linkup{autobib:Baldi2006} Baldi, A., Raymond, J.~C., Fabbiano, G., Zezas, A., Rots, A.~H., Schweizer, F., King, A.~R., \& Ponman, T.~J.\ 2006, \apj, 636, 158\linkads{http://adsabs.harvard.edu/abs/2006ApJ...636..158B}
\bibitem[Barnes \& Hut(1986)]{Barnes1986}\linkup{autobib:Barnes1986} Barnes, J., \& Hut, P.\ 1986, \nat, 324, 446\linkads{http://adsabs.harvard.edu/abs/1986Natur.324..446B}
\bibitem[Barnes(1988)]{Barnes1988a}\linkup{autobib:Barnes1988a} Barnes, J.~E.\ 1988, \apj, 331, 699\linkads{http://adsabs.harvard.edu/abs/1988ApJ...331..699B}
\bibitem[Barnes et al.(1988)]{Barnes1988b}\linkup{autobib:Barnes1988b} Barnes, J., Hernquist, L.~E., Hut, P., \& Teuben, P.\ 1988, \baas, 20, 706\linkads{http://adsabs.harvard.edu/abs/1988BAAS...20..706B}
\bibitem[Barnes(2004)]{Barnes2004}\linkup{autobib:Barnes2004} Barnes, J.~E.\ 2004, \mnras, 350, 798\linkads{http://adsabs.harvard.edu/abs/2004MNRAS.350..798B}
\bibitem[Barnes \& Hibbard(2009)]{Barnes2009}\linkup{autobib:Barnes2009} Barnes, J.~E., \& Hibbard, J.~E.\ 2009, \aj, 137, 3071\linkads{http://adsabs.harvard.edu/abs/2009AJ....137.3071B}
\bibitem[Bastian et al.(2009)]{Bastian2009}\linkup{autobib:Bastian2009} Bastian, N., Trancho, G., Konstantopoulos, I.~S., \& Miller, B.~W.\ 2009, \apj, 701, 607\linkads{http://adsabs.harvard.edu/abs/2009ApJ...701..607B}
\bibitem[Bate, Bonnell, \& Bromm(2003)]{Bate2003}\linkup{autobib:Bate2003} Bate, M.~R., Bonnell, I.~A., \& Bromm, V.\ 2003, \mnras, 339, 577\linkads{http://adsabs.harvard.edu/abs/2003MNRAS.339..577B}
\bibitem[Baumgardt \& Makino(2003)]{Baumgardt2003}\linkup{autobib:Baumgardt2003} Baumgardt, H., \& Makino, J.\ 2003, \mnras, 340, 227\linkads{http://adsabs.harvard.edu/abs/2003MNRAS.340..227B}
\bibitem[Baumgardt \& Kroupa(2007)]{Baumgardt2007}\linkup{autobib:Baumgardt2007} Baumgardt, H., \& Kroupa, P.\ 2007, \mnras, 380, 1589\linkads{http://adsabs.harvard.edu/abs/2007MNRAS.380.1589B}
\bibitem[Baumgardt et al.(2010)]{Baumgardt2010}\linkup{autobib:Baumgardt2010} Baumgardt, H., Parmentier, G., Gieles, M., \& Vesperini, E.\ 2010, \mnras, 401, 1832\linkads{http://adsabs.harvard.edu/abs/2010MNRAS.401.1832B}
\bibitem[Bedin et al.(2004)]{Bedin2004}\linkup{autobib:Bedin2004} Bedin, L.~R., Piotto, G., Anderson, J., Cassisi, S., King, I.~R., Momany, Y., \& Carraro, G.\ 2004, \apjl, 605, L125\linkads{http://adsabs.harvard.edu/abs/2004ApJ...605L.125B}
\bibitem[Bekenstein \& Milgrom(1984)]{Bekenstein1984}\linkup{autobib:Bekenstein1984} Bekenstein, J., \& Milgrom, M.\ 1984, \apj, 286, 7\linkads{http://adsabs.harvard.edu/abs/1984ApJ...286....7B}
\bibitem[Bekki \& Mackey(2009)]{Bekki2009}\linkup{autobib:Bekki2009} Bekki, K., \& Mackey, A.~D.\ 2009, \mnras, 394, 124\linkads{http://adsabs.harvard.edu/abs/2009MNRAS.394..124B}
\bibitem[Binney \& Tremaine(1987)]{Binney1987}\linkup{autobib:Binney1987} Binney, J., \& Tremaine, S.\ 1987, Galactic Dynamics, First Edition, Princeton Univ. Press\linkads{http://adsabs.harvard.edu/abs/1987gady.book.....B}
\bibitem[Binney \& Tremaine(2008)]{Binney2008}\linkup{autobib:Binney2008} Binney, J., \& Tremaine, S.\ 2008, Galactic Dynamics, Second Edition, Princeton Univ. Press\linkads{http://adsabs.harvard.edu/abs/2008gady.book.....B}
\bibitem[Boily, Kroupa, \& Pe{\~n}arrubia-Garrido(2001)]{Boily2001}\linkup{autobib:Boily2001} Boily, C.~M., Kroupa, P., \& Pe{\~n}arrubia-Garrido, J.\ 2001, New Astronomy, 6, 27\linkads{http://adsabs.harvard.edu/abs/2001NewA....6...27B}
\bibitem[Boily \& Kroupa(2003a)]{Boily2003a}\linkup{autobib:Boily2003a} Boily, C.~M., \& Kroupa, P.\ 2003a, \mnras, 338, 665\linkads{http://adsabs.harvard.edu/abs/2003MNRAS.338..665B}
\bibitem[Boily \& Kroupa(2003b)]{Boily2003b}\linkup{autobib:Boily2003b} Boily, C.~M., \& Kroupa, P.\ 2003b, \mnras, 338, 673\linkads{http://adsabs.harvard.edu/abs/2003MNRAS.338..673B}
\bibitem[Boily \& Athanassoula(2006)]{Boily2006}\linkup{autobib:Boily2006} Boily, C.~M., \& Athanassoula, E.\ 2006, \mnras, 369, 608\linkads{http://adsabs.harvard.edu/abs/2006MNRAS.369..608B}
\bibitem[Bournaud, Jog, \& Combes(2005)]{Bournaud2005}\linkup{autobib:Bournaud2005} Bournaud, F., Jog, C.~J., \& Combes, F.\ 2005, \aap, 437, 69\linkads{http://adsabs.harvard.edu/abs/2005A\%26A...437...69B}
\bibitem[Boutloukos \& Lamers(2003)]{Boutloukos2003}\linkup{autobib:Boutloukos2003} Boutloukos, S.~G., \& Lamers, H.~J.~G.~L.~M.\ 2003, \mnras, 338, 717\linkads{http://adsabs.harvard.edu/abs/2003MNRAS.338..717B}
\bibitem[Boylan-Kolchin et al.(2009)]{Boylan2009}\linkup{autobib:Boylan2009} Boylan-Kolchin, M., Springel, V., White, S.~D.~M., Jenkins, A., \& Lemson, G.\ 2009, \mnras, 398, 1150\linkads{http://adsabs.harvard.edu/abs/2009MNRAS.398.1150B}
\bibitem[Brandl et al.(2005)]{Brandl2005}\linkup{autobib:Brandl2005} Brandl, B.~R., et al.\ 2005, \apj, 635, 280\linkads{http://adsabs.harvard.edu/abs/2005ApJ...635..280B}
\bibitem[Brodie \& Strader(2006)]{Brodie2006}\linkup{autobib:Brodie2006} Brodie, J.~P., \& Strader, J.\ 2006, \araa, 44, 193\linkads{http://adsabs.harvard.edu/abs/2006ARA\%26A..44..193B}
\bibitem[Bruzual \& Charlot(2003)]{Bruzual2003}\linkup{autobib:Bruzual2003} Bruzual, G., \& Charlot, S.\ 2003, \mnras, 344, 1000\linkads{http://adsabs.harvard.edu/abs/2003MNRAS.344.1000B}
\bibitem[Bushouse, Telesco, \& Werner(1998)]{Bushouse1998}\linkup{autobib:Bushouse1998} Bushouse, H.~A., Telesco, C.~M., \& Werner, M.~W.\ 1998, \aj, 115, 938\linkads{http://adsabs.harvard.edu/abs/1998AJ....115..938B}
\bibitem[Cappa \& Benaglia(1998)]{Cappa1998}\linkup{autobib:Cappa1998} Cappa, C.~E., \& Benaglia, P.\ 1998, \aj, 116, 1906\linkads{http://adsabs.harvard.edu/abs/1998AJ....116.1906C}
\bibitem[Cardone, Piedipalumbo, \& Tortora(2005)]{Cardone2005}\linkup{autobib:Cardone2005} Cardone, V.~F., Piedipalumbo, E., \& Tortora, C.\ 2005, \mnras, 358, 1325\linkads{http://adsabs.harvard.edu/abs/2005MNRAS.358.1325C}
\bibitem[Carr(1994)]{Carr1994}\linkup{autobib:Carr1994} Carr, B.\ 1994, \araa, 32, 531\linkads{http://adsabs.harvard.edu/abs/1994ARA\%26A..32..531C}
\bibitem[Castor, McCray, \& Weaver(1975)]{Castor1975}\linkup{autobib:Castor1975} Castor, J., McCray, R., \& Weaver, R.\ 1975, \apjl, 200, L107\linkads{http://adsabs.harvard.edu/abs/1975ApJ...200L.107C}
\bibitem[Chandar, Fall, \& Whitmore(2006)]{Chandar2006}\linkup{autobib:Chandar2006} Chandar, R., Fall, S.~M., \& Whitmore, B.~C.\ 2006, \apjl, 650, L111\linkads{http://adsabs.harvard.edu/abs/2006ApJ...650L.111C}
\bibitem[Chandrasekhar(1943)]{Chandrasekhar1943}\linkup{autobib:Chandrasekhar1943} Chandrasekhar, S.\ 1943, \apj, 97, 255\linkads{http://adsabs.harvard.edu/abs/1943ApJ....97..255C}
\bibitem[D'Ercole et al.(2008)]{DErcole2008}\linkup{autobib:DErcole2008} D'Ercole, A., Vesperini, E., D'Antona, F., McMillan, S.~L.~W., \& Recchi, S.\ 2008, \mnras, 391, 825\linkads{http://adsabs.harvard.edu/abs/2008MNRAS.391..825D}
\bibitem[Davis et al.(1985)]{Davis1985}\linkup{autobib:Davis1985} Davis, M., Efstathiou, G., Frenk, C.~S., \& White, S.~D.~M.\ 1985, \apj, 292, 371\linkads{http://adsabs.harvard.edu/abs/1985ApJ...292..371D}
\bibitem[de Blok(2010)]{deBlok2010}\linkup{autobib:deBlok2010} de Blok, W.~J.~G.\ 2010, Advances in Astronomy, 2010, 5\linkads{http://adsabs.harvard.edu/abs/2010AdAst2010....5D}
\bibitem[Decressin et al.(2007)]{Decressin2007}\linkup{autobib:Decressin2007} Decressin, T., Meynet, G., Charbonnel, C., Prantzos, N., \& Ekstr{\"o}m, S.\ 2007, \aap, 464, 1029\linkads{http://adsabs.harvard.edu/abs/2007A\%26A...464.1029D}
\bibitem[Decressin, Baumgardt, \& Kroupa(2008)]{Decressin2008}\linkup{autobib:Decressin2008} Decressin, T., Baumgardt, H., \& Kroupa, P.\ 2008, \aap, 492, 101\linkads{http://adsabs.harvard.edu/abs/2008A\%26A...492..101D}
\bibitem[Dehnen(1993)]{Dehnen1993}\linkup{autobib:Dehnen1993} Dehnen, W.\ 1993, \mnras, 265, 250\linkads{http://adsabs.harvard.edu/abs/1993MNRAS.265..250D}
\bibitem[Dehnen(2000)]{Dehnen2000}\linkup{autobib:Dehnen2000} Dehnen, W.\ 2000, \apjl, 536, L39\linkads{http://adsabs.harvard.edu/abs/2000ApJ...536L..39D}
\bibitem[Dehnen(2001)]{Dehnen2001}\linkup{autobib:Dehnen2001} Dehnen, W.\ 2001, \mnras, 324, 273\linkads{http://adsabs.harvard.edu/abs/2001MNRAS.324..273D}
\bibitem[Dekel, Devor, \& Hetzroni(2003)]{Dekel2003}\linkup{autobib:Dekel2003} Dekel, A., Devor, J., \& Hetzroni, G.\ 2003, \mnras, 341, 326\linkads{http://adsabs.harvard.edu/abs/2003MNRAS.341..326D}
\bibitem[di Matteo et al.(2007)]{diMatteo2007}\linkup{autobib:diMatteo2007} di Matteo, P., Combes, F., Melchior, A.-L., \& Semelin, B.\ 2007, \aap, 468, 61\linkads{http://adsabs.harvard.edu/abs/2007A\%26A...468...61d}
\bibitem[di Matteo et al.(2008)]{diMatteo2008}\linkup{autobib:diMatteo2008} di Matteo, P., Bournaud, F., Martig, M., Combes, F., Melchior, A.-L., \& Semelin, B.\ 2008, \aap, 492, 31\linkads{http://adsabs.harvard.edu/abs/2008A\%26A...492...31d}
\bibitem[Dieball, M{\"u}ller, \& Grebel(2002)]{Dieball2002}\linkup{autobib:Dieball2002} Dieball, A., M{\"u}ller, H., \& Grebel, E.~K.\ 2002, \aap, 391, 547\linkads{http://adsabs.harvard.edu/abs/2002A\%26A...391..547D}
\bibitem[Dobbs et al.(2010)]{Dobbs2010}\linkup{autobib:Dobbs2010} Dobbs, C.~L., Theis, C., Pringle, J.~E., \& Bate, M.~R.\ 2010, \mnras, 403, 625\linkads{http://adsabs.harvard.edu/abs/2010MNRAS.403..625D}
\bibitem[Draine \& Salpeter(1979)]{Draine1979}\linkup{autobib:Draine1979} Draine, B.~T., \& Salpeter, E.~E.\ 1979, \apj, 231, 438\linkads{http://adsabs.harvard.edu/abs/1979ApJ...231..438D}
\bibitem[Drake et al.(2007)]{Drake2007}\linkup{autobib:Drake2007} Drake, A.~J., Djorgovski, S.~G., Williams, R., Mahabal, A., Graham, M.~J., Christensen, E., Beshore, E.~C., \& Larson, S.~M.\ 2007, CBET, 1172, 1\linkads{http://adsabs.harvard.edu/abs/2007CBET.1172....1D}
\bibitem[Dubinski(2008)]{Dubinski2008}\linkup{autobib:Dubinski2008} Dubinski, J.\ 2008, New Journal of Physics, 10, 125002\linkads{http://adsabs.harvard.edu/abs/2008NJPh...10l5002D}
\bibitem[Duc \& Mirabel(1999)]{Duc1999}\linkup{autobib:Duc1999} Duc, P.-A., \& Mirabel, I.~F.\ 1999, Galaxy Interactions at Low and High Redshift, 186, 61\linkads{http://adsabs.harvard.edu/abs/1999IAUS..186...61D}
\bibitem[Duncan(1923)]{Duncan1923}\linkup{autobib:Duncan1923} Duncan, J.~C.\ 1923, \apj, 57, 137\linkads{http://adsabs.harvard.edu/abs/1923ApJ....57..137D}
\bibitem[Einasto(1965)]{Einasto1965}\linkup{autobib:Einasto1965} Einasto, J.\ 1965, Trudy Inst. Astroz. Alma-Ata, 51, 87
\bibitem[Elmegreen \& Efremov(1997)]{Elmegreen1997}\linkup{autobib:Elmegreen1997} Elmegreen, B.~G., \& Efremov, Y.~N.\ 1997, \apj, 480, 235\linkads{http://adsabs.harvard.edu/abs/1997ApJ...480..235E}
\bibitem[Elmegreen(2000)]{Elmegreen2000}\linkup{autobib:Elmegreen2000} Elmegreen, B.~G.\ 2000, \apj, 530, 277\linkads{http://adsabs.harvard.edu/abs/2000ApJ...530..277E}
\bibitem[Eneev, Kozlov, \& Sunyaev(1973)]{Eneev1973}\linkup{autobib:Eneev1973} Eneev, T.~M., Kozlov, N.~N., \& Sunyaev, R.~A.\ 1973, \aap, 22, 41\linkads{http://adsabs.harvard.edu/abs/1973A&A....22...41E}
\bibitem[Evans(1993)]{Evans1993}\linkup{autobib:Evans1993} Evans, N.~W.\ 1993, \mnras, 260, 191\linkads{http://adsabs.harvard.edu/abs/1993MNRAS.260..191E}
\bibitem[Fabbiano, Zezas, \& Murray(2001)]{Fabbiano2001}\linkup{autobib:Fabbiano2001} Fabbiano, G., Zezas, A., \& Murray, S.~S.\ 2001, \apj, 554, 1035\linkads{http://adsabs.harvard.edu/abs/2001ApJ...554.1035F}
\bibitem[Fall \& Rees(1977)]{Fall1977}\linkup{autobib:Fall1977} Fall, S.~M., \& Rees, M.~J.\ 1977, \mnras, 181, 37P\linkads{http://adsabs.harvard.edu/abs/1977MNRAS.181...37F}
\bibitem[Fall \& Zhang(2001)]{Fall2001}\linkup{autobib:Fall2001} Fall, S.~M., \& Zhang, Q.\ 2001, \apj, 561, 751\linkads{http://adsabs.harvard.edu/abs/2001ApJ...561..751F}
\bibitem[Fall, Chandar, \& Whitmore(2005)]{Fall2005}\linkup{autobib:Fall2005} Fall, S.~M., Chandar, R., \& Whitmore, B.~C.\ 2005, \apjl, 631, L133\linkads{http://adsabs.harvard.edu/abs/2005ApJ...631L.133F}
\bibitem[Fleck et al.(2006)]{Fleck2006}\linkup{autobib:Fleck2006} Fleck, J.-J., Boily, C.~M., Lan{\c c}on, A., \& Deiters, S.\ 2006, \mnras, 369, 1392\linkads{http://adsabs.harvard.edu/abs/2006MNRAS.369.1392F}
\bibitem[Fleck(2007)]{Fleck2007}\linkup{autobib:Fleck2007} Fleck, J.-J.\ 2007, unpublished thesis\linkads[http://tel.archives-ouvertes.fr/tel-00184822]{http://tel.archives-ouvertes.fr/tel-00184822}
\bibitem[Fortin(2006)]{Fortin2006}\linkup{autobib:Fortin2006} Fortin, P.\ 2006, unpublished thesis\linkads[http://tel.archives-ouvertes.fr/tel-00135843]{http://tel.archives-ouvertes.fr/tel-00135843}
\bibitem[Freeman(1970)]{Freeman1970}\linkup{autobib:Freeman1970} Freeman, K.~C.\ 1970, \apj, 160, 811\linkads{http://adsabs.harvard.edu/abs/1970ApJ...160..811F}
\bibitem[Geyer \& Burkert(2001)]{Geyer2001}\linkup{autobib:Geyer2001} Geyer, M.~P., \& Burkert, A.\ 2001, \mnras, 323, 988\linkads{http://adsabs.harvard.edu/abs/2001MNRAS.323..988G}
\bibitem[Gieles, Athanassoula, \& Portegies Zwart(2007)]{Gieles2007a}\linkup{autobib:Gieles2007a} Gieles, M., Athanassoula, E., \& Portegies Zwart, S.~F.\ 2007, \mnras, 376, 809\linkads{http://adsabs.harvard.edu/abs/2007MNRAS.376..809G}
\bibitem[Gieles, Lamers, \& Portegies Zwart(2007)]{Gieles2007b}\linkup{autobib:Gieles2007b} Gieles, M., Lamers, H.~J.~G.~L.~M., \& Portegies Zwart, S.~F.\ 2007, \apj, 668, 268\linkads{http://adsabs.harvard.edu/abs/2007ApJ...668..268G}
\bibitem[Gieles \& Baumgardt(2008)]{Gieles2008}\linkup{autobib:Gieles2008} Gieles, M., \& Baumgardt, H.\ 2008, \mnras, 389, L28\linkads{http://adsabs.harvard.edu/abs/2008MNRAS.389L..28G}
\bibitem[Gieles(2009)]{Gieles2009}\linkup{autobib:Gieles2009} Gieles, M.\ 2009, \mnras, 394, 2113\linkads{http://adsabs.harvard.edu/abs/2009MNRAS.394.2113G}
\bibitem[Gieles, Sana, \& Portegies Zwart(2010)]{Gieles2010}\linkup{autobib:Gieles2010} Gieles, M., Sana, H., \& Portegies Zwart, S.~F.\ 2010, \mnras, 402, 1750\linkads{http://adsabs.harvard.edu/abs/2010MNRAS.402.1750G}
\bibitem[Gingold \& Monaghan(1977)]{Gingold1977}\linkup{autobib:Gingold1977} Gingold, R.~A., \& Monaghan, J.~J.\ 1977, \mnras, 181, 375\linkads{http://adsabs.harvard.edu/abs/1977MNRAS.181..375G}
\bibitem[Glatt et al.(2008)]{Glatt2008}\linkup{autobib:Glatt2008} Glatt, K., et al.\ 2008, \aj, 136, 1703\linkads{http://adsabs.harvard.edu/abs/2008AJ....136.1703G}
\bibitem[Gnedin, Hernquist, \& Ostriker(1999)]{Gnedin1999}\linkup{autobib:Gnedin1999} Gnedin, O.~Y., Hernquist, L., \& Ostriker, J.~P.\ 1999, \apj, 514, 109\linkads{http://adsabs.harvard.edu/abs/1999ApJ...514..109G}
\bibitem[Goodwin(1997)]{Goodwin1997}\linkup{autobib:Goodwin1997} Goodwin, S.~P.\ 1997, \mnras, 284, 785\linkads{http://adsabs.harvard.edu/abs/1997MNRAS.284..785G}
\bibitem[Guillard(2010)]{Guillard2010}\linkup{autobib:Guillard2010} Guillard, P.\ 2010, PhD thesis, Univ. Paris-Sud XI, arXiv:1001.3613\linkads{http://adsabs.harvard.edu/abs/2010arXiv1001.3613G}
\bibitem[Hansen \& Moore(2006)]{Hansen2006}\linkup{autobib:Hansen2006} Hansen, S.~H., \& Moore, B.\ 2006, New Astronomy, 11, 333\linkads{http://adsabs.harvard.edu/abs/2006NewA...11..333H}
\bibitem[Harris(1991)]{Harris1991}\linkup{autobib:Harris1991} Harris, W.~E.\ 1991, \araa, 29, 543\linkads{http://adsabs.harvard.edu/abs/1991ARA\%26A..29..543H}
\bibitem[Hayashi \& Navarro(2006)]{Hayashi2006}\linkup{autobib:Hayashi2006} Hayashi, E., \& Navarro, J.~F.\ 2006, \mnras, 373, 1117\linkads{http://adsabs.harvard.edu/abs/2006MNRAS.373.1117H}
\bibitem[Hayashi, Navarro, \& Springel(2007)]{Hayashi2007}\linkup{autobib:Hayashi2007} Hayashi, E., Navarro, J.~F., \& Springel, V.\ 2007, \mnras, 377, 50\linkads{http://adsabs.harvard.edu/abs/2007MNRAS.377...50H}
\bibitem[Healy et al.(2003)]{Healy2003}\linkup{autobib:Healy2003} Healy D., Rockmore D., Kostelec P. \& Moore S.\ 2003, Journal of Fourier Analysis and Applications, 9:4, 341
\bibitem[Heggie \& Mathieu(1986)]{Heggie1986}\linkup{autobib:Heggie1986} Heggie, D.~C., \& Mathieu, R.~D.\ 1986, The Use of Supercomputers in Stellar Dynamics, Eds. P.~Hut and S.~McMillan, 267, 233\linkads{http://adsabs.harvard.edu/abs/1986LNP...267..233H}
\bibitem[Hemsendorf, Sigurdsson, \& Spurzem(2002)]{Hemsendorf2002}\linkup{autobib:Hemsendorf2002} Hemsendorf, M., Sigurdsson, S., \& Spurzem, R.\ 2002, \apj, 581, 1256\linkads{http://adsabs.harvard.edu/abs/2002ApJ...581.1256H}
\bibitem[H{\'e}non(1961)]{Henon1961}\linkup{autobib:Henon1961} H{\'e}non, M.\ 1961, Annales d'Astrophysique, 24, 369\linkads{http://adsabs.harvard.edu/abs/1961AnAp...24..369H}
\bibitem[Hernquist(1990)]{Hernquist1990}\linkup{autobib:Hernquist1990} Hernquist, L.\ 1990, \apj, 356, 359\linkads{http://adsabs.harvard.edu/abs/1990ApJ...356..359H}
\bibitem[Hernquist(1993)]{Hernquist1993}\linkup{autobib:Hernquist1993} Hernquist, L.\ 1993, \apjs, 86, 389\linkads{http://adsabs.harvard.edu/abs/1993ApJS...86..389H}
\bibitem[Hibbard \& van Gorkom(1996)]{Hibbard1996}\linkup{autobib:Hibbard1996} Hibbard, J.~E., \& van Gorkom, J.~H.\ 1996, \aj, 111, 655\linkads{http://adsabs.harvard.edu/abs/1996AJ....111..655H}
\bibitem[Hibbard et al.(2001)]{Hibbard2001}\linkup{autobib:Hibbard2001} Hibbard, J.~E., van der Hulst, J.~M., Barnes, J.~E., \& Rich, R.~M.\ 2001, \aj, 122, 2969\linkads{http://adsabs.harvard.edu/abs/2001AJ....122.2969H}
\bibitem[Hibbard et al.(2005)]{Hibbard2005}\linkup{autobib:Hibbard2005} Hibbard, J.~E., et al.\ 2005, \apjl, 619, L87\linkads{http://adsabs.harvard.edu/abs/2005ApJ...619L..87H}
\bibitem[Hills(1980)]{Hills1980}\linkup{autobib:Hills1980} Hills, J.~G.\ 1980, \apj, 235, 986\linkads{http://adsabs.harvard.edu/abs/1980ApJ...235..986H}
\bibitem[Holmberg(1941)]{Holmberg1941}\linkup{autobib:Holmberg1941} Holmberg, E.\ 1941, \apj, 94, 385\linkads{http://adsabs.harvard.edu/abs/1941ApJ....94..385H}
\bibitem[Holtzman et al.(1992)]{Holtzman1992}\linkup{autobib:Holtzman1992} Holtzman, J.~A., et al.\ 1992, \aj, 103, 691\linkads{http://adsabs.harvard.edu/abs/1992AJ....103..691H}
\bibitem[Hubble(1929)]{Hubble1929}\linkup{autobib:Hubble1929} Hubble, E.~P.\ 1929, \apj, 69, 103\linkads{http://adsabs.harvard.edu/abs/1929ApJ....69..103H}
\bibitem[Hubble(1936)]{Hubble1936}\linkup{autobib:Hubble1936} Hubble, E.~P.\ 1936, Realm of the Nebulae, Oxford Univ. Press\linkads{http://adsabs.harvard.edu/abs/1936rene.book.....H}
\bibitem[Hummel \& van der Hulst(1986)]{Hummel1986}\linkup{autobib:Hummel1986} Hummel, E., \& van der Hulst, J.~M.\ 1986, \aap, 155, 151\linkads{http://adsabs.harvard.edu/abs/1986A\%26A...155..151H}
\bibitem[Hurley, Pols, \& Tout(2000)]{Hurley2000}\linkup{autobib:Hurley2000} Hurley, J.~R., Pols, O.~R., \& Tout, C.~A.\ 2000, \mnras, 315, 543\linkads{http://adsabs.harvard.edu/abs/2000MNRAS.315..543H}
\bibitem[Hwang et al.(2010)]{Hwang2010}\linkup{autobib:Hwang2010} Hwang, J.-S., Struck, C., Renaud, F., \& Appleton, P.~N.\ 2010, ASPC, 423, 232\linkads{http://adsabs.harvard.edu/abs/2010ASPC..423..232H}
\bibitem[Jaffe(1983)]{Jaffe1983}\linkup{autobib:Jaffe1983} Jaffe, W.\ 1983, \mnras, 202, 995\linkads{http://adsabs.harvard.edu/abs/1983MNRAS.202..995J}
\bibitem[Jeans(1915)]{Jeans1915}\linkup{autobib:Jeans1915} Jeans, J.~H.\ 1915, \mnras, 76, 70\linkads{http://adsabs.harvard.edu/abs/1915MNRAS..76...70J}
\bibitem[Jeon, Kim, \& Ann(2009)]{Jeon2009}\linkup{autobib:Jeon2009} Jeon, M., Kim, S.~S., \& Ann, H.~B.\ 2009, \apj, 696, 1899\linkads{http://adsabs.harvard.edu/abs/2009ApJ...696.1899J}
\bibitem[Johansson, Naab, \& Burkert(2009)]{Johansson2009}\linkup{autobib:Johansson2009} Johansson, P.~H., Naab, T., \& Burkert, A.\ 2009, \apj, 690, 802\linkads{http://adsabs.harvard.edu/abs/2009ApJ...690..802J}
\bibitem[Joseph \& Wright(1985)]{Joseph1985}\linkup{autobib:Joseph1985} Joseph, R.~D., \& Wright, G.~S.\ 1985, \mnras, 214, 87\linkads{http://adsabs.harvard.edu/abs/1985MNRAS.214...87J}
\bibitem[Kapferer et al.(2009)]{Kapferer2009}\linkup{autobib:Kapferer2009} Kapferer, W., Sluka, C., Schindler, S., Ferrari, C., \& Ziegler, B.\ 2009, \aap, 499, 87\linkads{http://adsabs.harvard.edu/abs/2009A\%26A...499...87K}
\bibitem[Karl et al.(2008)]{Karl2008}\linkup{autobib:Karl2008} Karl, S.~J., Naab, T., Johansson, P.~H., Theis, C., \& Boily, C.~M.\ 2008, Astronomische Nachrichten, 329, 1042\linkads{http://adsabs.harvard.edu/abs/2008AN....329.1042K}
\bibitem[Karl et al.(2010)]{Karl2010}\linkup{autobib:Karl2010} Karl, S.~J., Naab, T., Johansson, P.~H., Kotarba, H., Boily, C., Renaud, F., \& Theis, C.\ 2010, \apjl, 715, L88\linkads{http://adsabs.harvard.edu/abs/2010ApJ...715L..88K}
\bibitem[Katz(1992)]{Katz1992}\linkup{autobib:Katz1992} Katz, N.\ 1992, \apj, 391, 502\linkads{http://adsabs.harvard.edu/abs/1992ApJ...391..502K}
\bibitem[Kawai \& Makino(1998)]{Kawai1998}\linkup{autobib:Kawai1998} Kawai, A., \& Makino, J.\ 1998, arXiv:9812431\linkads{http://adsabs.harvard.edu/abs/1998astro.ph.12431K}
\bibitem[Kennicutt(1998)]{Kennicutt1998}\linkup{autobib:Kennicutt1998} Kennicutt, R.~C., Jr.\ 1998, \apj, 498, 541\linkads{http://adsabs.harvard.edu/abs/1998ApJ...498..541K}
\bibitem[Kim, Wise, \& Abel(2009)]{Kim2009}\linkup{autobib:Kim2009} Kim, J.-h., Wise, J.~H., \& Abel, T.\ 2009, \apjl, 694, L123\linkads{http://adsabs.harvard.edu/abs/2009ApJ...694L.123K}
\bibitem[King(1962)]{King1962}\linkup{autobib:King1962} King, I.\ 1962, \aj, 67, 471\linkads{http://adsabs.harvard.edu/abs/1962AJ.....67..471K}
\bibitem[King(1966)]{King1966}\linkup{autobib:King1966} King, I.\ 1966, \aj, 71, 64\linkads{http://adsabs.harvard.edu/abs/1966AJ.....71...64K}
\bibitem[Klessen, Burkert, \& Bate(1998)]{Klessen1998}\linkup{autobib:Klessen1998} Klessen, R.~S., Burkert, A., \& Bate, M.~R.\ 1998, \apjl, 501, L205\linkads{http://adsabs.harvard.edu/abs/1998ApJ...501L.205K}
\bibitem[Klessen \& Burkert(2001)]{Klessen2001}\linkup{autobib:Klessen2001} Klessen, R.~S., \& Burkert, A.\ 2001, \apj, 549, 386\linkads{http://adsabs.harvard.edu/abs/2001ApJ...549..386K}
\bibitem[Kopp(2008)]{Kopp2008}\linkup{autobib:Kopp2008} Kopp, J.\ 2008, Int. J. Mod. Phys., C19, 523\linkads{http://adsabs.harvard.edu/abs/2008IJMPC..19..523K}
\bibitem[Kouwenhoven \& de Grijs(2008)]{Kouwenhoven2008}\linkup{autobib:Kouwenhoven2008} Kouwenhoven, M.~B.~N., \& de Grijs, R.\ 2008, \aap, 480, 103\linkads{http://adsabs.harvard.edu/abs/2008A\%26A...480..103K}
\bibitem[Kroupa(2001)]{Kroupa2001}\linkup{autobib:Kroupa2001} Kroupa, P.\ 2001, \mnras, 322, 231\linkads{http://adsabs.harvard.edu/abs/2001MNRAS.322..231K}
\bibitem[Kuijken \& Dubinski(1994)]{Kuijken1994}\linkup{autobib:Kuijken1994} Kuijken, K., \& Dubinski, J.\ 1994, \mnras, 269, 13\linkads{http://adsabs.harvard.edu/abs/1994MNRAS.269...13K}
\bibitem[Kuijken \& Dubinski(1995)]{Kuijken1995}\linkup{autobib:Kuijken1995} Kuijken, K., \& Dubinski, J.\ 1995, \mnras, 277, 1341\linkads{http://adsabs.harvard.edu/abs/1995MNRAS.277.1341K}
\bibitem[Kundic \& Ostriker(1995)]{Kundic1995}\linkup{autobib:Kundic1995} Kundic, T., \& Ostriker, J.~P.\ 1995, \apj, 438, 702\linkads{http://adsabs.harvard.edu/abs/1995ApJ...438..702K}
\bibitem[Lada \& Lada(2003)]{Lada2003}\linkup{autobib:Lada2003} Lada, C.~J., \& Lada, E.~A.\ 2003, \araa, 41, 57\linkads{http://adsabs.harvard.edu/abs/2003ARA\%26A..41...57L}
\bibitem[Laine et al.(2003)]{Laine2003}\linkup{autobib:Laine2003} Laine, S., van der Marel, R.~P., Rossa, J., Hibbard, J.~E., Mihos, J.~C., B{\"o}ker, T., \& Zabludoff, A.~I.\ 2003, \aj, 126, 2717\linkads{http://adsabs.harvard.edu/abs/2003AJ....126.2717L}
\bibitem[Lamers et al.(2005)]{Lamers2005}\linkup{autobib:Lamers2005} Lamers, H.~J.~G.~L.~M., Gieles, M., Bastian, N., Baumgardt, H., Kharchenko, N.~V., \& Portegies Zwart, S.\ 2005, \aap, 441, 117\linkads{http://adsabs.harvard.edu/abs/2005A\%26A...441..117L}
\bibitem[Larsen(2004)]{Larsen2004}\linkup{autobib:Larsen2004} Larsen, S.~S.\ 2004, \aap, 416, 53\linkads{http://adsabs.harvard.edu/abs/2004A\%26A...416...53L}
\bibitem[Larsen(2009)]{Larsen2009}\linkup{autobib:Larsen2009} Larsen, S.~S.\ 2009, \aap, 494, 539\linkads{http://adsabs.harvard.edu/abs/2009A\%26A...494..539L}
\bibitem[Lee et al.(1999)]{Lee1999}\linkup{autobib:Lee1999} Lee, Y.-W., Joo, J.-M., Sohn, Y.-J., Rey, S.-C., Lee, H.-C., \& Walker, A.~R.\ 1999, \nat, 402, 55\linkads{http://adsabs.harvard.edu/abs/1999Natur.402...55L}
\bibitem[Lindzen \& Chapman(1969)]{Lindzen1969}\linkup{autobib:Lindzen1969} Lindzen, R.~S., \& Chapman, S.\ 1969, Space Science Reviews, 10, 3\linkads{http://adsabs.harvard.edu/abs/1969SSRv...10....3L}
\bibitem[Mac Low \& Klessen(2004)]{MacLow2004}\linkup{autobib:MacLow2004} Mac Low, M.-M., \& Klessen, R.~S.\ 2004, Reviews of Modern Physics, 76, 125\linkads{http://adsabs.harvard.edu/abs/2004RvMP...76..125M}
\bibitem[Ma{\'{\i}}z-Apell{\'a}niz(2001)]{MaizApellaniz2001}\linkup{autobib:MaizApellaniz2001} Ma{\'{\i}}z-Apell{\'a}niz, J.\ 2001, \apj, 563, 151\linkads{http://adsabs.harvard.edu/abs/2001ApJ...563..151M}
\bibitem[Martin et al.(2005)]{Martin2005}\linkup{autobib:Martin2005} Martin, D.~C., et al.\ 2005, \apjl, 619, L1\linkads{http://adsabs.harvard.edu/abs/2005ApJ...619L...1M}
\bibitem[Mart{\'{\i}}n, Cappa, \& Benaglia(2008)]{Martin2008}\linkup{autobib:Martin2008} Mart{\'{\i}}n, M.~C., Cappa, C.~E., \& Benaglia, P.\ 2008, RevMexAA Conference Series, 33, 164\linkads{http://adsabs.harvard.edu/abs/2008RMxAC..33..164M}
\bibitem[McCrady, Gilbert, \& Graham(2003)]{McCrady2003}\linkup{autobib:McCrady2003} McCrady, N., Gilbert, A.~M., \& Graham, J.~R.\ 2003, \apj, 596, 240\linkads{http://adsabs.harvard.edu/abs/2003ApJ...596..240M}
\bibitem[McMillan \& Dehnen(2007)]{McMillan2007}\linkup{autobib:McMillan2007} McMillan, P.~J., \& Dehnen, W.\ 2007, \mnras, 378, 541\linkads{http://adsabs.harvard.edu/abs/2007MNRAS.378..541M}
\bibitem[Mengel et al.(2005)]{Mengel2005}\linkup{autobib:Mengel2005} Mengel, S., Lehnert, M.~D., Thatte, N., \& Genzel, R.\ 2005, \aap, 443, 41\linkads{http://adsabs.harvard.edu/abs/2005A\%26A...443...41M}
\bibitem[Merritt(1996)]{Merritt1996}\linkup{autobib:Merritt1996} Merritt, D.\ 1996, \aj, 111, 2462\linkads{http://adsabs.harvard.edu/abs/1996AJ....111.2462M}
\bibitem[Merritt \& Fridman(1996)]{Merritt1996}\linkup{autobib:Merritt1996} Merritt, D., \& Fridman, T.\ 1996, \apj, 460, 136\linkads{http://adsabs.harvard.edu/abs/1996ApJ...460..136M}
\bibitem[Merritt et al.(2005)]{Merritt2005}\linkup{autobib:Merritt2005} Merritt, D., Navarro, J.~F., Ludlow, A., \& Jenkins, A.\ 2005, \apjl, 624, L85\linkads{http://adsabs.harvard.edu/abs/2005ApJ...624L..85M}
\bibitem[Meurer(1995)]{Meurer1995}\linkup{autobib:Meurer1995} Meurer, G.~R.\ 1995, \nat, 375, 742\linkads{http://adsabs.harvard.edu/abs/1995Natur.375..742M}
\bibitem[Meylan \& Heggie(1997)]{Meylan1997}\linkup{autobib:Meylan1997} Meylan, G., \& Heggie, D.~C.\ 1997, \aapr, 8, 1\linkads{http://adsabs.harvard.edu/abs/1997A\%26ARv...8....1M}
\bibitem[Michie(1963)]{Michie1963}\linkup{autobib:Michie1963} Michie, R.~W.\ 1963, \mnras, 125, 127\linkads{http://adsabs.harvard.edu/abs/1963MNRAS.125..127M}
\bibitem[Mihos, Bothun, \& Richstone(1993)]{Mihos1993}\linkup{autobib:Mihos1993} Mihos, J.~C., Bothun, G.~D., \& Richstone, D.~O.\ 1993, \apj, 418, 82\linkads{http://adsabs.harvard.edu/abs/1993ApJ...418...82M}
\bibitem[Milgrom(1983)]{Milgrom1983}\linkup{autobib:Milgrom1983} Milgrom, M.\ 1983, \apj, 270, 365\linkads{http://adsabs.harvard.edu/abs/1983ApJ...270..365M}
\bibitem[Milone et al.(2008)]{Milone2008}\linkup{autobib:Milone2008} Milone, A.~P., et al.\ 2008, \apj, 673, 241\linkads{http://adsabs.harvard.edu/abs/2008ApJ...673..241M}
\bibitem[Milone et al.(2009a)]{Milone2009a}\linkup{autobib:Milone2009a} Milone, A.~P., Bedin, L.~R., Piotto, G., \& Anderson, J.\ 2009a, \aap, 497, 755\linkads{http://adsabs.harvard.edu/abs/2009A\%26A...497..755M}
\bibitem[Milone et al.(2009b)]{Milone2009b}\linkup{autobib:Milone2009b} Milone, A.~P., Stetson, P.~B., Piotto, G., Bedin, L.~R., Anderson, J., Cassisi, S., \& Salaris, M.\ 2009b, \aap, 503, 755\linkads{http://adsabs.harvard.edu/abs/2009A\%26A...503..755M}
\bibitem[Mirabel et al.(1992)]{Mirabel1992}\linkup{autobib:Mirabel1992} Mirabel, I.~F., Dottori, H., \& Lutz, D.\ 1992, \aap, 256, L19\linkads{http://adsabs.harvard.edu/abs/1992A\%26A...256L..19M}
\bibitem[Moore et al.(1999)]{Moore1999}\linkup{autobib:Moore1999} Moore, B., Quinn, T., Governato, F., Stadel, J., \& Lake, G.\ 1999, \mnras, 310, 1147\linkads{http://adsabs.harvard.edu/abs/1999MNRAS.310.1147M}
\bibitem[Naab \& Burkert(2003)]{Naab2003}\linkup{autobib:Naab2003} Naab, T., \& Burkert, A.\ 2003, \apj, 597, 893\linkads{http://adsabs.harvard.edu/abs/2003ApJ...597..893N}
\bibitem[Naab, Jesseit, \& Burkert(2006)]{Naab2006}\linkup{autobib:Naab2006} Naab, T., Jesseit, R., \& Burkert, A.\ 2006, \mnras, 372, 839\linkads{http://adsabs.harvard.edu/abs/2006MNRAS.372..839N}
\bibitem[Navarro \& White(1993)]{Navarro1993}\linkup{autobib:Navarro1993} Navarro, J.~F., \& White, S.~D.~M.\ 1993, \mnras, 265, 271\linkads{http://adsabs.harvard.edu/abs/1993MNRAS.265..271N}
\bibitem[Navarro, Frenk, \& White(1997)]{Navarro1997}\linkup{autobib:Navarro1997} Navarro, J.~F., Frenk, C.~S., \& White, S.~D.~M.\ 1997, \apj, 490, 493\linkads{http://adsabs.harvard.edu/abs/1997ApJ...490..493N}
\bibitem[Navarro et al.(2004)]{Navarro2004}\linkup{autobib:Navarro2004} Navarro, J.~F., et al.\ 2004, \mnras, 349, 1039\linkads{http://adsabs.harvard.edu/abs/2004MNRAS.349.1039N}
\bibitem[Navarro et al.(2010)]{Navarro2010}\linkup{autobib:Navarro2010} Navarro, J.~F., et al.\ 2010, \mnras, 402, 21\linkads{http://adsabs.harvard.edu/abs/2010MNRAS.402...21N}
\bibitem[O'Connell, Gallagher, \& Hunter(1994)]{OConnell1994}\linkup{autobib:OConnell1994} O'Connell, R.~W., Gallagher, J.~S., III, \& Hunter, D.~A.\ 1994, \apj, 433, 65\linkads{http://adsabs.harvard.edu/abs/1994ApJ...433...65O}
\bibitem[Ocvirk et al.(2006)]{Ocvirk2006}\linkup{autobib:Ocvirk2006} Ocvirk, P., Pichon, C., Lan{\c c}on, A., \& Thi{\'e}baut, E.\ 2006, \mnras, 365, 46\linkads{http://adsabs.harvard.edu/abs/2006MNRAS.365...46O}
\bibitem[Oort(1932)]{Oort1932}\linkup{autobib:Oort1932} Oort, J.~H.\ 1932, \bain, 6, 249\linkads{http://adsabs.harvard.edu/abs/1932BAN.....6..249O}
\bibitem[Ostriker, Spitzer, \& Chevalier(1972)]{Ostriker1972}\linkup{autobib:Ostriker1972} Ostriker, J.~P., Spitzer, L., Jr., \& Chevalier, R.~A.\ 1972, \apjl, 176, L51\linkads{http://adsabs.harvard.edu/abs/1972ApJ...176L..51O}
\bibitem[Peebles(1982)]{Peebles1982}\linkup{autobib:Peebles1982} Peebles, P.~J.~E.\ 1982, \apjl, 263, L1\linkads{http://adsabs.harvard.edu/abs/1982ApJ...263L...1P}
\bibitem[Pflamm-Altenburg \& Kroupa(2009)]{PflammAltenburg2009}\linkup{autobib:PflammAltenburg2009} Pflamm-Altenburg, J., \& Kroupa, P.\ 2009, \mnras, 397, 488\linkads{http://adsabs.harvard.edu/abs/2009MNRAS.397..488P}
\bibitem[Pfleiderer \& Siedentopf(1961)]{Pfleiderer1961}\linkup{autobib:Pfleiderer1961} Pfleiderer, J., \& Siedentopf, H.\ 1961, Zeitschrift f\"ur Astrophysik, 51, 201\linkads{http://adsabs.harvard.edu/abs/1961ZA.....51..201P}
\bibitem[Piotto et al.(2005)]{Piotto2005}\linkup{autobib:Piotto2005} Piotto, G., et al.\ 2005, \apj, 621, 77\linkads{http://adsabs.harvard.edu/abs/2005ApJ...621...77P}
\bibitem[Piotto et al.(2007)]{Piotto2007}\linkup{autobib:Piotto2007} Piotto, G., et al.\ 2007, \apjl, 661, L53\linkads{http://adsabs.harvard.edu/abs/2007ApJ...661L..53P}
\bibitem[Piotto(2009)]{Piotto2009}\linkup{autobib:Piotto2009} Piotto, G.\ 2009, arXiv:0902.1422\linkads{http://adsabs.harvard.edu/abs/2009arXiv0902.1422P}
\bibitem[Plummer(1911)]{Plummer1911}\linkup{autobib:Plummer1911} Plummer, H.~C.\ 1911, \mnras, 71, 460\linkads{http://adsabs.harvard.edu/abs/1911MNRAS..71..460P}
\bibitem[Raddick et al.(2007)]{Raddick2007}\linkup{autobib:Raddick2007} Raddick, J., et al.\ 2007, BAAS, 38, 892\linkads{http://adsabs.harvard.edu/abs/2007AAS...211.9403R}
\bibitem[Renaud et al.(2008)]{Renaud2008}\linkup{autobib:Renaud2008} Renaud, F., Boily, C.~M., Fleck, J.-J., Naab, T., \& Theis, C.\ 2008, \mnras, 391, L98\linkads{http://adsabs.harvard.edu/abs/2008MNRAS.391L..98R}
\bibitem[Renaud et al.(2009)]{Renaud2009}\linkup{autobib:Renaud2009} Renaud, F., Boily, C.~M., Naab, T., \& Theis, C.\ 2009, \apj, 706, 67\linkads{http://adsabs.harvard.edu/abs/2009ApJ...706...67R}
\bibitem[Renaud, Boily, \& Theis(2010)]{Renaud2010a}\linkup{autobib:Renaud2010a} Renaud, F., Boily, C.~M, \& Theis, C.\ 2010, \mnras\ submitted
\bibitem[Renaud, Appleton, \& Xu(2010)]{Renaud2010b}\linkup{autobib:Renaud2010b} Renaud, F., Appleton, P.~N, \& Xu, K.~C.\ 2010, \apj\ submitted
\bibitem[Rownd \& Young(1999)]{Rownd1999}\linkup{autobib:Rownd1999} Rownd, B.~K., \& Young, J.~S.\ 1999, \aj, 118, 670\linkads{http://adsabs.harvard.edu/abs/1999AJ....118..670R}
\bibitem[Rubin, Ford, \& D'Odorico(1970)]{Rubin1970}\linkup{autobib:Rubin1970} Rubin, V.~C., Ford, W.~K.~J., \& D'Odorico, S.\ 1970, \apj, 160, 801\linkads{http://adsabs.harvard.edu/abs/1970ApJ...160..801R}
\bibitem[Salpeter(1955)]{Salpeter1955}\linkup{autobib:Salpeter1955} Salpeter, E.~E.\ 1955, \apj, 121, 161\linkads{http://adsabs.harvard.edu/abs/1955ApJ...121..161S}
\bibitem[Sanders \& McGaugh(2002)]{Sanders2002}\linkup{autobib:Sanders2002} Sanders, R.~H., \& McGaugh, S.~S.\ 2002, \araa, 40, 263\linkads{http://adsabs.harvard.edu/abs/2002ARA\%26A..40..263S}
\bibitem[Sarajedini \& Mancone(2007)]{Sarajedini2007}\linkup{autobib:Sarajedini2007} Sarajedini, A., \& Mancone, C.~L.\ 2007, \aj, 134, 447\linkads{http://adsabs.harvard.edu/abs/2007AJ....134..447S}
\bibitem[Saviane, Hibbard, \& Rich(2004)]{Saviane2004}\linkup{autobib:Saviane2004} Saviane, I., Hibbard, J.~E., \& Rich, R.~M.\ 2004, \aj, 127, 660\linkads{http://adsabs.harvard.edu/abs/2004AJ....127..660S}
\bibitem[Saviane et al.(2008)]{Saviane2008}\linkup{autobib:Saviane2008} Saviane, I., Momany, Y., da Costa, G.~S., Rich, R.~M., \& Hibbard, J.~E.\ 2008, \apj, 678, 179\linkads{http://adsabs.harvard.edu/abs/2008ApJ...678..179S}
\bibitem[Scalo(1998)]{Scalo1998}\linkup{autobib:Scalo1998} Scalo, J.\ 1998, The Stellar Initial Mass Function, 142, 201\linkads{http://adsabs.harvard.edu/abs/1998ASPC..142..201S}
\bibitem[Schechter(1976)]{Schechter1976}\linkup{autobib:Schechter1976} Schechter, P.\ 1976, \apj, 203, 29\linkads{http://adsabs.harvard.edu/abs/1976ApJ...203...29S}
\bibitem[Schmidt(1959)]{Schmidt1959}\linkup{autobib:Schmidt1959} Schmidt, M.\ 1959, \apj, 129, 243\linkads{http://adsabs.harvard.edu/abs/1959ApJ...129..243S}
\bibitem[Schmidt(1975)]{Schmidt1975}\linkup{autobib:Schmidt1975} Schmidt, K.-H.\ 1975, \apss, 34, 23\linkads{http://adsabs.harvard.edu/abs/1975Ap\%26SS..34...23S}
\bibitem[Schweizer(1978)]{Schweizer1978}\linkup{autobib:Schweizer1978} Schweizer, F.\ 1978, Structure and Properties of Nearby Galaxies, 77, 279\linkads{http://adsabs.harvard.edu/abs/1978IAUS...77..279S}
\bibitem[Schweizer et al.(2008)]{Schweizer2008}\linkup{autobib:Schweizer2008} Schweizer, F., et al.\ 2008, \aj, 136, 1482\linkads{http://adsabs.harvard.edu/abs/2008AJ....136.1482S}
\bibitem[S\'ersic(1968)]{Sersic1968}\linkup{autobib:Sersic1968} S\'ersic, J.~L.\ 1968, Atlas de Galaxies Australes, Univ. Nac. C\'ordoba\linkads{http://adsabs.harvard.edu/abs/1968adga.book.....S}
\bibitem[Shu(1969)]{Shu1969}\linkup{autobib:Shu1969} Shu, F.~H.\ 1969, \apj, 158, 505\linkads{http://adsabs.harvard.edu/abs/1969ApJ...158..505S}
\bibitem[Shu, Adams, \& Lizano(1987)]{Shu1987}\linkup{autobib:Shu1987} Shu, F.~H., Adams, F.~C., \& Lizano, S.\ 1987, \araa, 25, 23\linkads{http://adsabs.harvard.edu/abs/1987ARA\%26A..25...23S}
\bibitem[Spitzer(1940)]{Spitzer1940}\linkup{autobib:Spitzer1940} Spitzer, L.~J.\ 1940, \mnras, 100, 396\linkads{http://adsabs.harvard.edu/abs/1940MNRAS.100..396S}
\bibitem[Spitzer(1958)]{Spitzer1958}\linkup{autobib:Spitzer1958} Spitzer, L.~J.\ 1958, \apj, 127, 17\linkads{http://adsabs.harvard.edu/abs/1958ApJ...127...17S}
\bibitem[Spitzer(1987)]{Spitzer1987}\linkup{autobib:Spitzer1987} Spitzer, L.\ 1987, Princeton Univ. Press\linkads{http://adsabs.harvard.edu/abs/1987degc.book.....S}
\bibitem[Springel(2000)]{Springel2000}\linkup{autobib:Springel2000} Springel, V.\ 2000, \mnras, 312, 859\linkads{http://adsabs.harvard.edu/abs/2000MNRAS.312..859S}
\bibitem[Springel \& Hernquist(2003)]{Springel2003}\linkup{autobib:Springel2003} Springel, V., \& Hernquist, L.\ 2003, \mnras, 339, 289\linkads{http://adsabs.harvard.edu/abs/2003MNRAS.339..289S}
\bibitem[Springel(2005)]{Springel2005}\linkup{autobib:Springel2005a} Springel, V.\ 2005, \mnras, 364, 1105\linkads{http://adsabs.harvard.edu/abs/2005MNRAS.364.1105S}
\bibitem[Springel et al.(2005)]{Springel2005}\linkup{autobib:Springel2005b} Springel, V., et al.\ 2005, \nat, 435, 629\linkads{http://adsabs.harvard.edu/abs/2005Natur.435..629S}
\bibitem[Spurzem, Baumgardt \& Ibold(2003)]{Spurzem2003}\linkup{autobib:Spurzem2003} Spurzem, R., Baumgardt, H. \& Ibold, N., 2003,\\\linkads[ftp://ftp.ari.uni-heidelberg.de/pub/staff/spurzem/nb6mpi]{ftp://ftp.ari.uni-heidelberg.de/pub/staff/spurzem/nb6mpi}
\bibitem[Stadel et al.(2009)]{Stadel2009}\linkup{autobib:Stadel2009} Stadel, J., Potter, D., Moore, B., Diemand, J., Madau, P., Zemp, M., Kuhlen, M., \& Quilis, V.\ 2009, \mnras, 398, L21\linkads{http://adsabs.harvard.edu/abs/2009MNRAS.398L..21S}
\bibitem[Stewart et al.(2008)]{Stewart2008}\linkup{autobib:Stewart2008} Stewart, K.~R., Bullock, J.~S., Wechsler, R.~H., Maller, A.~H., \& Zentner, A.~R.\ 2008, \apj, 683, 597\linkads{http://adsabs.harvard.edu/abs/2008ApJ...683..597S}
\bibitem[Stewart(2009)]{Stewart2009}\linkup{autobib:Stewart2009} Stewart, K.~R.\ 2009, ASPC, 419, 243\linkads{http://adsabs.harvard.edu/abs/2009ASPC..419..243S} 
\bibitem[Sutherland \& Dopita(1993)]{Sutherland1993}\linkup{autobib:Sutherland1993} Sutherland, R.~S., \& Dopita, M.~A.\ 1993, \apjs, 88, 253\linkads{http://adsabs.harvard.edu/abs/1993ApJS...88..253S}
\bibitem[Tegmark et al.(2004)]{Tegmark2004}\linkup{autobib:Tegmark2004} Tegmark, M., et al.\ 2004, \apj, 606, 702\linkads{http://adsabs.harvard.edu/abs/2004ApJ...606..702T}
\bibitem[Teuben(1995)]{Teuben1995}\linkup{autobib:Teuben1995} Teuben, P.\ 1995, Astronomical Data Analysis Software and Systems IV, 77, 398\linkads{http://adsabs.harvard.edu/abs/1995ASPC...77..398T}
\bibitem[Teyssier(2002)]{Teyssier2002}\linkup{autobib:Teyssier2002} Teyssier, R.\ 2002, \aap, 385, 337\linkads{http://adsabs.harvard.edu/abs/2002A\%26A...385..337T}
\bibitem[Theis \& Kohle(2001)]{Theis2001}\linkup{autobib:Theis2001} Theis, C., \& Kohle, S.\ 2001, \aap, 370, 365\linkads{http://adsabs.harvard.edu/abs/2001A\%26A...370..365T}
\bibitem[Tiret \& Combes(2007)]{Tiret2007}\linkup{autobib:Tiret2007} Tiret, O., \& Combes, F.\ 2007, arXiv:0712.1459\linkads{http://adsabs.harvard.edu/abs/2007arXiv0712.1459T}
\bibitem[Tonry et al.(2000)]{Tonry2000}\linkup{autobib:Tonry2000} Tonry, J.~L., Blakeslee, J.~P., Ajhar, E.~A., \& Dressler, A.\ 2000, \apj, 530, 625\linkads{http://adsabs.harvard.edu/abs/2000ApJ...530..625T}
\bibitem[Toomre \& Toomre(1972)]{Toomre1972}\linkup{autobib:Toomre1972} Toomre, A., \& Toomre, J.\ 1972, \apj, 178, 623\linkads{http://adsabs.harvard.edu/abs/1972ApJ...178..623T}
\bibitem[Toomre(1977)]{Toomre1977}\linkup{autobib:Toomre1977} Toomre, A.\ 1977, Evolution of Galaxies and Stellar Populations, Yale Univ. Obs, New Haven, 401\linkads{http://adsabs.harvard.edu/abs/1977egsp.conf..401T}
\bibitem[Tremaine et al.(1994)]{Tremaine1994}\linkup{autobib:Tremaine1994} Tremaine, S., Richstone, D.~O., Byun, Y.-I., Dressler, A., Faber, S.~M., Grillmair, C., Kormendy, J., \& Lauer, T.~R.\ 1994, \aj, 107, 634\linkads{http://adsabs.harvard.edu/abs/1994AJ....107..634T}
\bibitem[Trimble(1987)]{Trimble1987}\linkup{autobib:Trimble1987} Trimble, V.\ 1987, \araa, 25, 425\linkads{http://adsabs.harvard.edu/abs/1987ARA\%26A..25..425T}
\bibitem[Valluri(1993)]{Valluri1993}\linkup{autobib:Valluri1993} Valluri, M.\ 1993, \apj, 408, 57\linkads{http://adsabs.harvard.edu/abs/1993ApJ...408...57V}
\bibitem[Verschueren(1990)]{Verschueren1990}\linkup{autobib:Verschueren1990} Verschueren, W.\ 1990, \aap, 234, 156\linkads{http://adsabs.harvard.edu/abs/1990A\%26A...234..156V}
\bibitem[Vesperini(1998)]{Vesperini1998}\linkup{autobib:Vesperini1998} Vesperini, E.\ 1998, \mnras, 299, 1019\linkads{http://adsabs.harvard.edu/abs/1998MNRAS.299.1019V}
\bibitem[von Hoerner(1960)]{vonHoerner1960}\linkup{autobib:vonHoerner1960} von Hoerner, S.\ 1960, Zeitschrift f\"ur Astrophysik, 50, 184\linkads{http://adsabs.harvard.edu/abs/1960ZA.....50..184V}
\bibitem[Wang et al.(2004)]{Wang2004}\linkup{autobib:Wang2004} Wang, Z., et al.\ 2004, \apjs, 154, 193\linkads{http://adsabs.harvard.edu/abs/2004ApJS..154..193W}
\bibitem[Watson et al.(1996)]{Watson1996}\linkup{autobib:Watson1996} Watson, A.~M., et al.\ 1996, \aj, 112, 534\linkads{http://adsabs.harvard.edu/abs/1996AJ....112..534W}
\bibitem[Weidner, Kroupa, \& Larsen(2004)]{Weidner2004}\linkup{autobib:Weidner2004} Weidner, C., Kroupa, P., \& Larsen, S.~S.\ 2004, \mnras, 350, 1503\linkads{http://adsabs.harvard.edu/abs/2004MNRAS.350.1503W}
\bibitem[White(1977)]{White1977}\linkup{autobib:White1977} White, S.~D.~M.\ 1977, \mnras, 179, 33\linkads{http://adsabs.harvard.edu/abs/1977MNRAS.179...33W}
\bibitem[Whitmore et al.(1993)]{Whitmore1993}\linkup{autobib:Whitmore1993} Whitmore, B.~C., Schweizer, F., Leitherer, C., Borne, K., \& Robert, C.\ 1993, \aj, 106, 1354\linkads{http://adsabs.harvard.edu/abs/1993AJ....106.1354W}
\bibitem[Whitmore \& Schweizer(1995)]{Whitmore1995}\linkup{autobib:Whitmore1995} Whitmore, B.~C., \& Schweizer, F.\ 1995, \aj, 109, 960\linkads{http://adsabs.harvard.edu/abs/1995AJ....109..960W}
\bibitem[Whitmore et al.(1999)]{Whitmore1999}\linkup{autobib:Whitmore1999} Whitmore, B.~C., Zhang, Q., Leitherer, C., Fall, S.~M., Schweizer, F., \& Miller, B.~W.\ 1999, \aj, 118, 1551\linkads{http://adsabs.harvard.edu/abs/1999AJ....118.1551W}
\bibitem[Whitmore, Chandar, \& Fall(2007)]{Whitmore2007}\linkup{autobib:Whitmore2007} Whitmore, B.~C., Chandar, R., \& Fall, S.~M.\ 2007, \aj, 133, 1067\linkads{http://adsabs.harvard.edu/abs/2007AJ....133.1067W}
\bibitem[Whitmore et al.(2010)]{Whitmore2010}\linkup{autobib:Whitmore2010} Whitmore, B.~C., et al.\ 2010, \aj, 140, 75\linkads{http://adsabs.harvard.edu/abs/2010AJ....140...75W}
\bibitem[Zezas \& Fabbiano(2002)]{Zezas2002}\linkup{autobib:Zezas2002} Zezas, A., \& Fabbiano, G.\ 2002, \apj, 577, 726\linkads{http://adsabs.harvard.edu/abs/2002ApJ...577..726Z}
\bibitem[Zhang \& Fall(1999)]{Zhang1999}\linkup{autobib:Zhang1999} Zhang, Q., \& Fall, S.~M.\ 1999, \apjl, 527, L81\linkads{http://adsabs.harvard.edu/abs/1999ApJ...527L..81Z}
\bibitem[Zwicky(1933)]{Zwicky1933}\linkup{autobib:Zwicky1933} Zwicky, F.\ 1933, Helvetica Physica Acta, 6, 110\linkads{http://adsabs.harvard.edu/abs/1933AcHPh...6..110Z}

\end{thegeneralbibliography}
\thispagestyle{frontmatter}
\ifthenelse{\isodd{\thepage}}{}{\newpage\thispagestyle{empty}\mbox{}}


\begin{thebibliography}{}

\bibitem[Aarseth(1963)]{Aarseth1963} Aarseth, S.~J.\ 1963, \mnras, 126, 223
\bibitem[Barnes(2004)]{Barnes2004} Barnes, J.~E.\ 2004, \mnras, 350, 798
\bibitem[Baumgardt \& Makino(2003)]{Baumgardt2003} Baumgardt, H., \& Makino, J.\ 2003, \mnras, 340, 227
\bibitem[Baumgardt \& Kroupa(2007)]{Baumgardt2007} Baumgardt, H., \& Kroupa, P.\ 2007, \mnras, 380, 1589
\bibitem[Baumgardt et al.(2010)]{Baumgardt2010} Baumgardt, H., Parmentier, G., Gieles, M., \& Vesperini, E.\ 2010, \mnras, 401, 1832
\bibitem[Boily \& Kroupa(2003a)]{Boily2003a} Boily, C.~M., \& Kroupa, P.\ 2003a, \mnras, 338, 665
\bibitem[Boily \& Kroupa(2003b)]{Boily2003b} Boily, C.~M., \& Kroupa, P.\ 2003b, \mnras, 338, 673
\bibitem[Boutloukos \& Lamers(2003)]{Boutloukos2003} Boutloukos, S.~G., \& Lamers, H.~J.~G.~L.~M.\ 2003, \mnras, 338, 717
\bibitem[Brodie \& Strader(2006)]{Brodie2006} Brodie, J.~P., \& Strader, J.\ 2006, \araa, 44, 193
\bibitem[Castor, McCray, \& Weaver(1975)]{Castor1975} Castor, J., McCray, R., \& Weaver, R.\ 1975, \apjl, 200, L107
\bibitem[Chandar, Fall, \& Whitmore(2006)]{Chandar2006} Chandar, R., Fall, S.~M., \& Whitmore, B.~C.\ 2006, \apjl, 650, L111
\bibitem[Dekel, Devor, \& Hetzroni(2003)]{Dekel2003} Dekel, A., Devor, J., \& Hetzroni, G.\ 2003, \mnras, 341, 326
\bibitem[di Matteo et al.(2007)]{diMatteo2007} di Matteo, P., Combes, F., Melchior, A.-L., \& Semelin, B.\ 2007, \aap, 468, 61
\bibitem[Draine \& Salpeter(1979)]{Draine1979} Draine, B.~T., \& Salpeter, E.~E.\ 1979, \apj, 231, 438
\bibitem[Elmegreen \& Efremov(1997)]{Elmegreen1997} Elmegreen, B.~G., \& Efremov, Y.~N.\ 1997, \apj, 480, 235 
\bibitem[Elmegreen(2000)]{Elmegreen2000} Elmegreen, B.~G.\ 2000, \apj, 530, 277
\bibitem[Eneev, Kozlov, \& Sunyaev(1973)]{Eneev1973} Eneev, T.~M., Kozlov, N.~N., \& Sunyaev, R.~A.\ 1973, \aap, 22, 41 
\bibitem[Fall, Chandar, \& Whitmore(2005)]{Fall2005} Fall, S.~M., Chandar, R., \& Whitmore, B.~C.\ 2005, \apjl, 631, L133
\bibitem[Geyer \& Burkert(2001)]{Geyer2001} Geyer, M.~P., \& Burkert, A.\ 2001, \mnras, 323, 988
\bibitem[Gieles, Athanassoula, \& Portegies Zwart(2007)]{Gieles2007a} Gieles, M., Athanassoula, E., \& Portegies Zwart, S.~F.\ 2007, \mnras, 376, 809
\bibitem[Gieles, Lamers, \& Portegies Zwart(2007)]{Gieles2007b} Gieles, M., Lamers, H.~J.~G.~L.~M., \& Portegies Zwart, S.~F.\ 2007, \apj, 668, 268
\bibitem[Gieles \& Baumgardt(2008)]{Gieles2008} Gieles, M., \& Baumgardt, H.\ 2008, \mnras, 389, L28
\bibitem[Gieles(2009)]{Gieles2009} Gieles, M.\ 2009, \mnras, 394, 2113
\bibitem[Goodwin(1997)]{Goodwin1997} Goodwin, S.~P.\ 1997, \mnras, 284, 785
\bibitem[Harris(1991)]{Harris1991} Harris, W.~E.\ 1991, \araa, 29, 543
\bibitem[H{\'e}non(1961)]{Henon1961} H{\'e}non, M.\ 1961, Annales d'Astrophysique, 24, 369
\bibitem[Hibbard \& van Gorkom(1996)]{Hibbard1996} Hibbard, J.~E., \& van Gorkom, J.~H.\ 1996, \aj, 111, 655
\bibitem[Hills(1980)]{Hills1980} Hills, J.~G.\ 1980, \apj, 235, 986
\bibitem[Holmberg(1941)]{Holmberg1941} Holmberg, E.\ 1941, \apj, 94, 385
\bibitem[Holtzman et al.(1992)]{Holtzman1992} Holtzman, J.~A., et al.\ 1992, \aj, 103, 691
\bibitem[Hubble(1929)]{Hubble1929} Hubble, E.~P.\ 1929, \apj, 69, 103
\bibitem[Hubble(1936)]{Hubble1936} Hubble, E.~P.\ 1936, Realm of the Nebulae, Oxford Univ. Press
\bibitem[Hurley, Pols, \& Tout(2000)]{Hurley2000} Hurley, J.~R., Pols, O.~R., \& Tout, C.~A.\ 2000, \mnras, 315, 543
\bibitem[Jeans(1915)]{Jeans1915} Jeans, J.~H.\ 1915, \mnras, 76, 70
\bibitem[Joseph \& Wright(1985)]{Joseph1985} Joseph, R.~D., \& Wright, G.~S.\ 1985, \mnras, 214, 87
\bibitem[Kapferer et al.(2009)]{Kapferer2009} Kapferer, W., Sluka, C., Schindler, S., Ferrari, C., \& Ziegler, B.\ 2009, \aap, 499, 87 
\bibitem[Kennicutt(1998)]{Kennicutt1998} Kennicutt, R.~C., Jr.\ 1998, \apj, 498, 541
\bibitem[King(1962)]{King1962} King, I.\ 1962, \aj, 67, 471
\bibitem[King(1966)]{King1966} King, I.\ 1966, \aj, 71, 64
\bibitem[Lada \& Lada(2003)]{Lada2003} Lada, C.~J., \& Lada, E.~A.\ 2003, \araa, 41, 57
\bibitem[Laine et al.(2003)]{Laine2003} Laine, S., van der Marel, R.~P., Rossa, J., Hibbard, J.~E., Mihos, J.~C., B{\"o}ker, T., \& Zabludoff, A.~I.\ 2003, \aj, 126, 2717
\bibitem[Lamers et al.(2005)]{Lamers2005} Lamers, H.~J.~G.~L.~M., Gieles, M., Bastian, N., Baumgardt, H., Kharchenko, N.~V., \& Portegies Zwart, S.\ 2005, \aap, 441, 117
\bibitem[Larsen(2004)]{Larsen2004} Larsen, S.~S.\ 2004, \aap, 416, 53
\bibitem[Larsen(2009)]{Larsen2009} Larsen, S.~S.\ 2009, \aap, 494, 539
\bibitem[Lindzen \& Chapman(1969)]{Lindzen1969} Lindzen, R.~S., \& Chapman, S.\ 1969, Space Science Reviews, 10, 3
\bibitem[Mac Low \& Klessen(2004)]{MacLow2004} Mac Low, M.-M., \& Klessen, R.~S.\ 2004, Reviews of Modern Physics, 76, 125
\bibitem[Ma{\'{\i}}z-Apell{\'a}niz(2001)]{MaizApellaniz2001} Ma{\'{\i}}z-Apell{\'a}niz, J.\ 2001, \apj, 563, 151
\bibitem[McCrady, Gilbert, \& Graham(2003)]{McCrady2003} McCrady, N., Gilbert, A.~M., \& Graham, J.~R.\ 2003, \apj, 596, 240
\bibitem[Meylan \& Heggie(1997)]{Meylan1997} Meylan, G., \& Heggie, D.~C.\ 1997, \aapr, 8, 1
\bibitem[Michie(1963)]{Michie1963} Michie, R.~W.\ 1963, \mnras, 125, 127
\bibitem[Mihos, Bothun, \& Richstone(1993)]{Mihos1993} Mihos, J.~C., Bothun, G.~D., \& Richstone, D.~O.\ 1993, \apj, 418, 82
\bibitem[O'Connell, Gallagher, \& Hunter(1994)]{OConnell1994} O'Connell, R.~W., Gallagher, J.~S., III, \& Hunter, D.~A.\ 1994, \apj, 433, 65
\bibitem[Ocvirk et al.(2006)]{Ocvirk2006} Ocvirk, P., Pichon, C., Lan{\c c}on, A., \& Thi{\'e}baut, E.\ 2006, \mnras, 365, 46
\bibitem[Ostriker, Spitzer, \& Chevalier(1972)]{Ostriker1972} Ostriker, J.~P., Spitzer, L., Jr., \& Chevalier, R.~A.\ 1972, \apjl, 176, L51 
\bibitem[Pfleiderer \& Siedentopf(1961)]{Pfleiderer1961} Pfleiderer, J., \& Siedentopf, H.\ 1961, Zeitschrift f\"ur Astrophysik, 51, 201
\bibitem[Plummer(1911)]{Plummer1911} Plummer, H.~C.\ 1911, \mnras, 71, 460
\bibitem[Raddick et al.(2007)]{Raddick2007} Raddick, J., et al.\ 2007, BAAS, 38, 892
\bibitem[Rownd \& Young(1999)]{Rownd1999} Rownd, B.~K., \& Young, J.~S.\ 1999, \aj, 118, 670
\bibitem[Sarajedini \& Mancone(2007)]{Sarajedini2007} Sarajedini, A., \& Mancone, C.~L.\ 2007, \aj, 134, 447
\bibitem[Schechter(1976)]{Schechter1976} Schechter, P.\ 1976, \apj, 203, 29
\bibitem[Schmidt(1959)]{Schmidt1959} Schmidt, M.\ 1959, \apj, 129, 243
\bibitem[Spitzer(1940)]{Spitzer1940} Spitzer, L.~J.\ 1940, \mnras, 100, 396
\bibitem[Spitzer(1958)]{Spitzer1958} Spitzer, L.~J.\ 1958, \apj, 127, 17
\bibitem[Toomre \& Toomre(1972)]{Toomre1972} Toomre, A., \& Toomre, J.\ 1972, \apj, 178, 623
\bibitem[Toomre(1977)]{Toomre1977} Toomre, A.\ 1977, Evolution of Galaxies and Stellar Populations, Yale Univ. Obs, New Haven, 401
\bibitem[Valluri(1993)]{Valluri1993} Valluri, M.\ 1993, \apj, 408, 57
\bibitem[Verschueren(1990)]{Verschueren1990} Verschueren, W.\ 1990, \aap, 234, 156
\bibitem[von Hoerner(1960)]{vonHoerner1960} von Hoerner, S.\ 1960, Zeitschrift f\"ur Astrophysik, 50, 184
\bibitem[Watson et al.(1996)]{Watson1996} Watson, A.~M., et al.\ 1996, \aj, 112, 534
\bibitem[Weidner, Kroupa, \& Larsen(2004)]{Weidner2004} Weidner, C., Kroupa, P., \& Larsen, S.~S.\ 2004, \mnras, 350, 1503
\bibitem[White(1977)]{White1977} White, S.~D.~M.\ 1977, \mnras, 179, 33
\bibitem[Whitmore et al.(1993)]{Whitmore1993} Whitmore, B.~C., Schweizer, F., Leitherer, C., Borne, K., \& Robert, C.\ 1993, \aj, 106, 1354
\bibitem[Zhang \& Fall(1999)]{Zhang1999} Zhang, Q., \& Fall, S.~M.\ 1999, \apjl, 527, L81

\end{thebibliography}

\begin{thebibliography}{}

\bibitem[Aarseth(1963)]{Aarseth1963} Aarseth, S.~J.\ 1963, \mnras, 126, 223
\bibitem[Athanassoula et al.(1998)]{Athanassoula1998} Athanassoula, E., Bosma, A., Lambert, J.-C., \& Makino, J.\ 1998, \mnras, 293, 369 
\bibitem[Barnes \& Hut(1986)]{Barnes1986} Barnes, J., \& Hut, P.\ 1986, \nat, 324, 446
\bibitem[Barnes et al.(1988)]{Barnes1988b} Barnes, J., Hernquist, L.~E., Hut, P., \& Teuben, P.\ 1988, \baas, 20, 706
\bibitem[Binney \& Tremaine(1987)]{Binney1987} Binney, J., \& Tremaine, S.\ 1987, Galactic Dynamics, First Edition, Princeton Univ. Press
\bibitem[Boylan-Kolchin et al.(2009)]{Boylan2009} Boylan-Kolchin, M., Springel, V., White, S.~D.~M., Jenkins, A., \& Lemson, G.\ 2009, \mnras, 398, 1150
\bibitem[Dehnen(2000)]{Dehnen2000} Dehnen, W.\ 2000, \apjl, 536, L39
\bibitem[Dehnen(2001)]{Dehnen2001} Dehnen, W.\ 2001, \mnras, 324, 273
\bibitem[Fortin(2006)]{Fortin2006} Fortin, P.\ 2006, unpublished thesis, http://tel.archives-ouvertes.fr/tel-00135843
\bibitem[Gingold \& Monaghan(1977)]{Gingold1977} Gingold, R.~A., \& Monaghan, J.~J.\ 1977, \mnras, 181, 375
\bibitem[Heggie \& Mathieu(1986)]{Heggie1986} Heggie, D.~C., \& Mathieu, R.~D.\ 1986, The Use of Supercomputers in Stellar Dynamics, Eds. P.~Hut and S.~McMillan, 267, 233
\bibitem[Kopp(2008)]{Kopp2008} Kopp, J.\ 2008, Int. J. Mod. Phys., C19, 523
\bibitem[Merritt(1996)]{Merritt1996} Merritt, D.\ 1996, \aj, 111, 2462
\bibitem[Teuben(1995)]{Teuben1995} Teuben, P.\ 1995, Astronomical Data Analysis Software and Systems IV, 77, 398
\bibitem[White(1977)]{White1977} White, S.~D.~M.\ 1977, \mnras, 179, 33

\end{thebibliography}

\begin{thebibliography}{}

\bibitem[Anders et al.(2007)]{Anders2007} Anders, P., Bissantz, N., Boysen, L., de Grijs, R., \& Fritze-v.~Alvensleben, U.\ 2007, \mnras, 377, 91
\bibitem[Baldi et al.(2006)]{Baldi2006} Baldi, A., Raymond, J.~C., Fabbiano, G., Zezas, A., Rots, A.~H., Schweizer, F., King, A.~R., \& Ponman, T.~J.\ 2006, \apj, 636, 158
\bibitem[Barnes(1988)]{Barnes1988a} Barnes, J.~E.\ 1988, \apj, 331, 699
\bibitem[Barnes(2004)]{Barnes2004} Barnes, J.~E.\ 2004, \mnras, 350, 798
\bibitem[Barnes \& Hibbard(2009)]{Barnes2009} Barnes, J.~E., \& Hibbard, J.~E.\ 2009, \aj, 137, 3071 
\bibitem[Bastian et al.(2009)]{Bastian2009} Bastian, N., Trancho, G., Konstantopoulos, I.~S., \& Miller, B.~W.\ 2009, \apj, 701, 607
\bibitem[Boily, Kroupa, \& Pe{\~n}arrubia-Garrido(2001)]{Boily2001} Boily, C.~M., Kroupa, P., \& Pe{\~n}arrubia-Garrido, J.\ 2001, New Astronomy, 6, 27
\bibitem[Brandl et al.(2005)]{Brandl2005} Brandl, B.~R., et al.\ 2005, \apj, 635, 280
\bibitem[Bruzual \& Charlot(2003)]{Bruzual2003} Bruzual, G., \& Charlot, S.\ 2003, \mnras, 344, 1000
\bibitem[Bushouse, Telesco, \& Werner(1998)]{Bushouse1998} Bushouse, H.~A., Telesco, C.~M., \& Werner, M.~W.\ 1998, \aj, 115, 938
\bibitem[Chandar, Fall, \& Whitmore(2006)]{Chandar2006} Chandar, R., Fall, S.~M., \& Whitmore, B.~C.\ 2006, \apjl, 650, L111
\bibitem[Drake et al.(2007)]{Drake2007} Drake, A.~J., Djorgovski, S.~G., Williams, R., Mahabal, A., Graham, M.~J., Christensen, E., Beshore, E.~C., \& Larson, S.~M.\ 2007, CBET, 1172, 1
\bibitem[Dubinski(2008)]{Dubinski2008} Dubinski, J.\ 2008, New Journal of Physics, 10, 125002
\bibitem[Duc \& Mirabel(1999)]{Duc1999} Duc, P.-A., \& Mirabel, I.~F.\ 1999, Galaxy Interactions at Low and High Redshift, 186, 61
\bibitem[Duncan(1923)]{Duncan1923} Duncan, J.~C.\ 1923, \apj, 57, 137
\bibitem[Fabbiano, Zezas, \& Murray(2001)]{Fabbiano2001} Fabbiano, G., Zezas, A., \& Murray, S.~S.\ 2001, \apj, 554, 1035
\bibitem[Fall \& Rees(1977)]{Fall1977} Fall, S.~M., \& Rees, M.~J.\ 1977, \mnras, 181, 37P
\bibitem[Fall \& Zhang(2001)]{Fall2001} Fall, S.~M., \& Zhang, Q.\ 2001, \apj, 561, 751
\bibitem[Fall, Chandar, \& Whitmore(2005)]{Fall2005} Fall, S.~M., Chandar, R., \& Whitmore, B.~C.\ 2005, \apjl, 631, L133
\bibitem[Harris(1991)]{Harris1991} Harris, W.~E.\ 1991, \araa, 29, 543
\bibitem[Hemsendorf, Sigurdsson, \& Spurzem(2002)]{Hemsendorf2002} Hemsendorf, M., Sigurdsson, S., \& Spurzem, R.\ 2002, \apj, 581, 1256
\bibitem[Hernquist(1990)]{Hernquist1990} Hernquist, L.\ 1990, \apj, 356, 359
\bibitem[Hernquist(1993)]{Hernquist1993} Hernquist, L.\ 1993, \apjs, 86, 389
\bibitem[Hibbard et al.(2001)]{Hibbard2001} Hibbard, J.~E., van der Hulst, J.~M., Barnes, J.~E., \& Rich, R.~M.\ 2001, \aj, 122, 2969
\bibitem[Hibbard et al.(2005)]{Hibbard2005} Hibbard, J.~E., et al.\ 2005, \apjl, 619, L87
\bibitem[Hummel \& van der Hulst(1986)]{Hummel1986} Hummel, E., \& van der Hulst, J.~M.\ 1986, \aap, 155, 151
\bibitem[Karl et al.(2008)]{Karl2008} Karl, S.~J., Naab, T., Johansson, P.~H., Theis, C., \& Boily, C.~M.\ 2008, Astronomische Nachrichten, 329, 1042
\bibitem[Karl et al.(2010)]{Karl2010} Karl, S.~J., Naab, T., Johansson, P.~H., Kotarba, H., Boily, C., Renaud, F., \& Theis, C.\ 2010, \apjl, 715, L88
\bibitem[Katz(1992)]{Katz1992} Katz, N.\ 1992, \apj, 391, 502
\bibitem[Kennicutt(1998)]{Kennicutt1998} Kennicutt, R.~C., Jr.\ 1998, \apj, 498, 541
\bibitem[Lada \& Lada(2003)]{Lada2003} Lada, C.~J., \& Lada, E.~A.\ 2003, \araa, 41, 57
\bibitem[Martin et al.(2005)]{Martin2005} Martin, D.~C., et al.\ 2005, \apjl, 619, L1
\bibitem[Mengel et al.(2005)]{Mengel2005} Mengel, S., Lehnert, M.~D., Thatte, N., \& Genzel, R.\ 2005, \aap, 443, 41
\bibitem[Meurer(1995)]{Meurer1995} Meurer, G.~R.\ 1995, \nat, 375, 742
\bibitem[Mihos, Bothun, \& Richstone(1993)]{Mihos1993} Mihos, J.~C., Bothun, G.~D., \& Richstone, D.~O.\ 1993, \apj, 418, 82
\bibitem[Mirabel et al.(1992)]{Mirabel1992} Mirabel, I.~F., Dottori, H., \& Lutz, D.\ 1992, \aap, 256, L19
\bibitem[Renaud et al.(2008)]{Renaud2008} Renaud, F., Boily, C.~M., Fleck, J.-J., Naab, T., \& Theis, C.\ 2008, \mnras, 391, L98
\bibitem[Renaud et al.(2009)]{Renaud2009} Renaud, F., Boily, C.~M., Naab, T., \& Theis, C.\ 2009, \apj, 706, 67
\bibitem[Rubin, Ford, \& D'Odorico(1970)]{Rubin1970} Rubin, V.~C., Ford, W.~K.~J., \& D'Odorico, S.\ 1970, \apj, 160, 801
\bibitem[Saviane, Hibbard, \& Rich(2004)]{Saviane2004} Saviane, I., Hibbard, J.~E., \& Rich, R.~M.\ 2004, \aj, 127, 660
\bibitem[Saviane et al.(2008)]{Saviane2008} Saviane, I., Momany, Y., da Costa, G.~S., Rich, R.~M., \& Hibbard, J.~E.\ 2008, \apj, 678, 179
\bibitem[Schmidt(1959)]{Schmidt1959} Schmidt, M.\ 1959, \apj, 129, 243
\bibitem[Schmidt(1975)]{Schmidt1975} Schmidt, K.-H.\ 1975, \apss, 34, 23
\bibitem[Schweizer(1978)]{Schweizer1978} Schweizer, F.\ 1978, Structure and Properties of Nearby Galaxies, 77, 279
\bibitem[Schweizer et al.(2008)]{Schweizer2008} Schweizer, F., et al.\ 2008, \aj, 136, 1482
\bibitem[Springel(2000)]{Springel2000} Springel, V.\ 2000, \mnras, 312, 859
\bibitem[Theis \& Kohle(2001)]{Theis2001} Theis, C., \& Kohle, S.\ 2001, \aap, 370, 365
\bibitem[Tonry et al.(2000)]{Tonry2000} Tonry, J.~L., Blakeslee, J.~P., Ajhar, E.~A., \& Dressler, A.\ 2000, \apj, 530, 625
\bibitem[Toomre \& Toomre(1972)]{Toomre1972} Toomre, A., \& Toomre, J.\ 1972, \apj, 178, 623
\bibitem[Vesperini(1998)]{Vesperini1998} Vesperini, E.\ 1998, \mnras, 299, 1019
\bibitem[Wang et al.(2004)]{Wang2004} Wang, Z., et al.\ 2004, \apjs, 154, 193
\bibitem[Whitmore \& Schweizer(1995)]{Whitmore1995} Whitmore, B.~C., \& Schweizer, F.\ 1995, \aj, 109, 960
\bibitem[Whitmore et al.(1999)]{Whitmore1999} Whitmore, B.~C., Zhang, Q., Leitherer, C., Fall, S.~M., Schweizer, F., \& Miller, B.~W.\ 1999, \aj, 118, 1551 
\bibitem[Whitmore, Chandar, \& Fall(2007)]{Whitmore2007} Whitmore, B.~C., Chandar, R., \& Fall, S.~M.\ 2007, \aj, 133, 1067
\bibitem[Whitmore et al.(2010)]{Whitmore2010} Whitmore, B.~C., et al.\ 2010, \aj, 140, 75
\bibitem[Zezas \& Fabbiano(2002)]{Zezas2002} Zezas, A., \& Fabbiano, G.\ 2002, \apj, 577, 726
\bibitem[Zhang \& Fall(1999)]{Zhang1999} Zhang, Q., \& Fall, S.~M.\ 1999, \apjl, 527, L81

\end{thebibliography}

\begin{thebibliography}{}

\bibitem[Aarseth(1999)]{Aarseth1999} Aarseth, S.~J.\ 1999, \pasp, 111, 1333
\bibitem[Anderson(1992)]{Anderson1992} Anderson, C.\ 1992, SIAM Journal on Scientific and Statistical Computing, 13, 923
\bibitem[Barnes(2004)]{Barnes2004} Barnes, J.~E.\ 2004, \mnras, 350, 798
\bibitem[Bate, Bonnell, \& Bromm(2003)]{Bate2003} Bate, M.~R., Bonnell, I.~A., \& Bromm, V.\ 2003, \mnras, 339, 577
\bibitem[Baumgardt \& Makino(2003)]{Baumgardt2003} Baumgardt, H., \& Makino, J.\ 2003, \mnras, 340, 227
\bibitem[Bedin et al.(2004)]{Bedin2004} Bedin, L.~R., Piotto, G., Anderson, J., Cassisi, S., King, I.~R., Momany, Y., \& Carraro, G.\ 2004, \apjl, 605, L125
\bibitem[Bekki \& Mackey(2009)]{Bekki2009} Bekki, K., \& Mackey, A.~D.\ 2009, \mnras, 394, 124
\bibitem[Binney \& Tremaine(1987)]{Binney1987} Binney, J., \& Tremaine, S.\ 1987, Galactic Dynamics, First Edition, Princeton Univ. Press
\bibitem[Boily \& Kroupa(2003a)]{Boily2003a} Boily, C.~M., \& Kroupa, P.\ 2003a, \mnras, 338, 665
\bibitem[Cappa \& Benaglia(1998)]{Cappa1998} Cappa, C.~E., \& Benaglia, P.\ 1998, \aj, 116, 1906
\bibitem[D'Ercole et al.(2008)]{DErcole2008} D'Ercole, A., Vesperini, E., D'Antona, F., McMillan, S.~L.~W., \& Recchi, S.\ 2008, \mnras, 391, 825
\bibitem[Decressin et al.(2007)]{Decressin2007} Decressin, T., Meynet, G., Charbonnel, C., Prantzos, N., \& Ekstr{\"o}m, S.\ 2007, \aap, 464, 1029
\bibitem[Decressin, Baumgardt, \& Kroupa(2008)]{Decressin2008} Decressin, T., Baumgardt, H., \& Kroupa, P.\ 2008, \aap, 492, 101
\bibitem[Dieball, M{\"u}ller, \& Grebel(2002)]{Dieball2002} Dieball, A., M{\"u}ller, H., \& Grebel, E.~K.\ 2002, \aap, 391, 547
\bibitem[di Matteo et al.(2007)]{diMatteo2007} di Matteo, P., Combes, F., Melchior, A.-L., \& Semelin, B.\ 2007, \aap, 468, 61
\bibitem[di Matteo et al.(2008)]{diMatteo2008} di Matteo, P., Bournaud, F., Martig, M., Combes, F., Melchior, A.-L., \& Semelin, B.\ 2008, \aap, 492, 31 
\bibitem[Fall, Chandar, \& Whitmore(2005)]{Fall2005} Fall, S.~M., Chandar, R., \& Whitmore, B.~C.\ 2005, \apjl, 631, L133
\bibitem[Fleck et al.(2006)]{Fleck2006} Fleck, J.-J., Boily, C.~M., Lan{\c c}on, A., \& Deiters, S.\ 2006, \mnras, 369, 1392
\bibitem[Fleck(2007)]{Fleck2007} Fleck, J.-J.\ 2007, unpublished thesis, http://tel.archives-ouvertes.fr/tel-00184822
\bibitem[Gieles, Athanassoula, \& Portegies Zwart(2007)]{Gieles2007a} Gieles, M., Athanassoula, E., \& Portegies Zwart, S.~F.\ 2007, \mnras, 376, 809
\bibitem[Gieles, Sana, \& Portegies Zwart(2010)]{Gieles2010} Gieles, M., Sana, H., \& Portegies Zwart, S.~F.\ 2010, \mnras, 402, 1750
\bibitem[Glatt et al.(2008)]{Glatt2008} Glatt, K., et al.\ 2008, \aj, 136, 1703
\bibitem[Gnedin, Hernquist, \& Ostriker(1999)]{Gnedin1999} Gnedin, O.~Y., Hernquist, L., \& Ostriker, J.~P.\ 1999, \apj, 514, 109
\bibitem[Healy et al.(2003)]{Healy2003} Healy D., Rockmore D., Kostelec P. \& Moore S.\ 2003, Journal of Fourier Analysis and Applications, 9:4, 341
\bibitem[Hibbard et al.(2001)]{Hibbard2001} Hibbard, J.~E., van der Hulst, J.~M., Barnes, J.~E., \& Rich, R.~M.\ 2001, \aj, 122, 2969
\bibitem[Hills(1980)]{Hills1980} Hills, J.~G.\ 1980, \apj, 235, 986
\bibitem[Hurley, Pols, \& Tout(2000)]{Hurley2000} Hurley, J.~R., Pols, O.~R., \& Tout, C.~A.\ 2000, \mnras, 315, 543
\bibitem[Karl et al.(2010)]{Karl2010} Karl, S.~J., Naab, T., Johansson, P.~H., Kotarba, H., Boily, C., Renaud, F., \& Theis, C.\ 2010, \apjl, 715, L88
\bibitem[Kawai \& Makino(1998)]{Kawai1998} Kawai, A., \& Makino, J.\ 1998, arXiv:9812431
\bibitem[Kim, Wise, \& Abel(2009)]{Kim2009} Kim, J.-h., Wise, J.~H., \& Abel, T.\ 2009, \apjl, 694, L123
\bibitem[Klessen, Burkert, \& Bate(1998)]{Klessen1998} Klessen, R.~S., Burkert, A., \& Bate, M.~R.\ 1998, \apjl, 501, L205
\bibitem[Klessen \& Burkert(2001)]{Klessen2001} Klessen, R.~S., \& Burkert, A.\ 2001, \apj, 549, 386
\bibitem[Kouwenhoven \& de Grijs(2008)]{Kouwenhoven2008} Kouwenhoven, M.~B.~N., \& de Grijs, R.\ 2008, \aap, 480, 103
\bibitem[Kroupa(2001)]{Kroupa2001} Kroupa, P.\ 2001, \mnras, 322, 231
\bibitem[Kundic \& Ostriker(1995)]{Kundic1995} Kundic, T., \& Ostriker, J.~P.\ 1995, \apj, 438, 702
\bibitem[Lee et al.(1999)]{Lee1999} Lee, Y.-W., Joo, J.-M., Sohn, Y.-J., Rey, S.-C., Lee, H.-C., \& Walker, A.~R.\ 1999, \nat, 402, 55
\bibitem[Mac Low \& Klessen(2004)]{MacLow2004} Mac Low, M.-M., \& Klessen, R.~S.\ 2004, Reviews of Modern Physics, 76, 125
\bibitem[Mart{\'{\i}}n, Cappa, \& Benaglia(2008)]{Martin2008} Mart{\'{\i}}n, M.~C., Cappa, C.~E., \& Benaglia, P.\ 2008, RevMexAA Conference Series, 33, 164
\bibitem[Meylan \& Heggie(1997)]{Meylan1997} Meylan, G., \& Heggie, D.~C.\ 1997, \aapr, 8, 1
\bibitem[Milone et al.(2008)]{Milone2008} Milone, A.~P., et al.\ 2008, \apj, 673, 241
\bibitem[Milone et al.(2009a)]{Milone2009a} Milone, A.~P., Bedin, L.~R., Piotto, G., \& Anderson, J.\ 2009a, \aap, 497, 755
\bibitem[Milone et al.(2009b)]{Milone2009b} Milone, A.~P., Stetson, P.~B., Piotto, G., Bedin, L.~R., Anderson, J., Cassisi, S., \& Salaris, M.\ 2009b, \aap, 503, 755
\bibitem[Navarro \& White(1993)]{Navarro1993} Navarro, J.~F., \& White, S.~D.~M.\ 1993, \mnras, 265, 271
\bibitem[Pflamm-Altenburg \& Kroupa(2009)]{PflammAltenburg2009} Pflamm-Altenburg, J., \& Kroupa, P.\ 2009, \mnras, 397, 488
\bibitem[Piotto et al.(2005)]{Piotto2005} Piotto, G., et al.\ 2005, \apj, 621, 77
\bibitem[Piotto et al.(2007)]{Piotto2007} Piotto, G., et al.\ 2007, \apjl, 661, L53 
%%% ARXIV
\bibitem[Piotto(2009)]{Piotto2009} Piotto, G.\ 2009, arXiv:0902.1422
\bibitem[Renaud et al.(2008)]{Renaud2008} Renaud, F., Boily, C.~M., Fleck, J.-J., Naab, T., \& Theis, C.\ 2008, \mnras, 391, L98
\bibitem[Renaud et al.(2009)]{Renaud2009} Renaud, F., Boily, C.~M., Naab, T., \& Theis, C.\ 2009, \apj, 706, 67
\bibitem[Salpeter(1955)]{Salpeter1955} Salpeter, E.~E.\ 1955, \apj, 121, 161
\bibitem[Scalo(1998)]{Scalo1998} Scalo, J.\ 1998, The Stellar Initial Mass Function, 142, 201
\bibitem[Shu, Adams, \& Lizano(1987)]{Shu1987} Shu, F.~H., Adams, F.~C., \& Lizano, S.\ 1987, \araa, 25, 23
\bibitem[Spitzer(1958)]{Spitzer1958} Spitzer, L., Jr.\ 1958, \apj, 127, 17
\bibitem[Spitzer(1987)]{Spitzer1987} Spitzer, L.\ 1987, Princeton Univ. Press
\bibitem[Springel(2000)]{Springel2000} Springel, V.\ 2000, \mnras, 312, 859
\bibitem[Springel \& Hernquist(2003)]{Springel2003} Springel, V., \& Hernquist, L.\ 2003, \mnras, 339, 289
\bibitem[Springel(2005)]{Springel2005a} Springel, V.\ 2005, \mnras, 364, 1105
\bibitem[Spurzem, Baumgardt \& Ibold(2003)]{Spurzem2003} Spurzem, R., Baumgardt, H. \& Ibold, N., 2003,\\ ftp://ftp.ari.uni-heidelberg.de/pub/staff/spurzem/nb6mpi
\bibitem[Sutherland \& Dopita(1993)]{Sutherland1993} Sutherland, R.~S., \& Dopita, M.~A.\ 1993, \apjs, 88, 253
\bibitem[Teyssier(2002)]{Teyssier2002} Teyssier, R.\ 2002, \aap, 385, 337
\bibitem[White(1977)]{White1977} White, S.~D.~M.\ 1977, \mnras, 179, 33

\end{thebibliography}

\begin{thebibliography}{}

\bibitem[Barnes(1988)]{Barnes1988a} Barnes, J.~E.\ 1988, \apj, 331, 699
\bibitem[Bournaud, Jog, \& Combes(2005)]{Bournaud2005} Bournaud, F., Jog, C.~J., \& Combes, F.\ 2005, \aap, 437, 69 
\bibitem[di Matteo et al.(2007)]{diMatteo2007} di Matteo, P., Combes, F., Melchior, A.-L., \& Semelin, B.\ 2007, \aap, 468, 61
\bibitem[Evans(1993)]{Evans1993} Evans, N.~W.\ 1993, \mnras, 260, 191
\bibitem[Hernquist(1990)]{Hernquist1990} Hernquist, L.\ 1990, \apj, 356, 359
\bibitem[Johansson, Naab, \& Burkert(2009)]{Johansson2009} Johansson, P.~H., Naab, T., \& Burkert, A.\ 2009, \apj, 690, 802 
\bibitem[King(1962)]{King1962} King, I.\ 1962, \aj, 67, 471
\bibitem[Kuijken \& Dubinski(1994)]{Kuijken1994} Kuijken, K., \& Dubinski, J.\ 1994, \mnras, 269, 13
\bibitem[Kuijken \& Dubinski(1995)]{Kuijken1995} Kuijken, K., \& Dubinski, J.\ 1995, \mnras, 277, 1341
\bibitem[McMillan \& Dehnen(2007)]{McMillan2007} McMillan, P.~J., \& Dehnen, W.\ 2007, \mnras, 378, 541 
\bibitem[Naab \& Burkert(2003)]{Naab2003} Naab, T., \& Burkert, A.\ 2003, \apj, 597, 893
\bibitem[Naab, Jesseit, \& Burkert(2006)]{Naab2006} Naab, T., Jesseit, R., \& Burkert, A.\ 2006, \mnras, 372, 839
\bibitem[Navarro, Frenk, \& White(1997)]{Navarro1997} Navarro, J.~F., Frenk, C.~S., \& White, S.~D.~M.\ 1997, \apj, 490, 493
\bibitem[Peebles(1982)]{Peebles1982} Peebles, P.~J.~E.\ 1982, \apjl, 263, L1
\bibitem[Renaud et al.(2009)]{Renaud2009} Renaud, F., Boily, C.~M., Naab, T., \& Theis, C.\ 2009, \apj, 706, 67
\bibitem[Shu(1969)]{Shu1969} Shu, F.~H.\ 1969, \apj, 158, 505
\bibitem[Springel et al.(2005)]{Springel2005b} Springel, V., et al.\ 2005, \nat, 435, 629
\bibitem[Stewart et al.(2008)]{Stewart2008} Stewart, K.~R., Bullock, J.~S., Wechsler, R.~H., Maller, A.~H., \& Zentner, A.~R.\ 2008, \apj, 683, 597
\bibitem[Stewart(2009)]{Stewart2009} Stewart, K.~R.\ 2009, ASPC, 419, 243
\bibitem[Toomre \& Toomre(1972)]{Toomre1972} Toomre, A., \& Toomre, J.\ 1972, \apj, 178, 623

\end{thebibliography}

\begin{thebibliography}{}

\bibitem[Bekenstein \& Milgrom(1984)]{Bekenstein1984} Bekenstein, J., \& Milgrom, M.\ 1984, \apj, 286, 7
\bibitem[Binney \& Tremaine(2008)]{Binney2008} Binney, J., \& Tremaine, S.\ 2008, Galactic Dynamics, Second Edition, Princeton Univ. Press
\bibitem[Boily \& Athanassoula(2006)]{Boily2006} Boily, C.~M., \& Athanassoula, E.\ 2006, \mnras, 369, 608
\bibitem[Cardone, Piedipalumbo, \& Tortora(2005)]{Cardone2005} Cardone, V.~F., Piedipalumbo, E., \& Tortora, C.\ 2005, \mnras, 358, 1325
\bibitem[Carr(1994)]{Carr1994} Carr, B.\ 1994, \araa, 32, 531
\bibitem[Davis et al.(1985)]{Davis1985} Davis, M., Efstathiou, G., Frenk, C.~S., \& White, S.~D.~M.\ 1985, \apj, 292, 371
\bibitem[de Blok(2010)]{deBlok2010} de Blok, W.~J.~G.\ 2010, Advances in Astronomy, 2010, 5
\bibitem[Dehnen(1993)]{Dehnen1993} Dehnen, W.\ 1993, \mnras, 265, 250
\bibitem[Einasto(1965)]{Einasto1965} Einasto, J.\ 1965, Trudy Inst. Astroz. Alma-Ata, 51, 87
\bibitem[Freeman(1970)]{Freeman1970} Freeman, K.~C.\ 1970, \apj, 160, 811
\bibitem[Hansen \& Moore(2006)]{Hansen2006} Hansen, S.~H., \& Moore, B.\ 2006, New Astronomy, 11, 333
\bibitem[Hayashi \& Navarro(2006)]{Hayashi2006} Hayashi, E., \& Navarro, J.~F.\ 2006, \mnras, 373, 1117 
\bibitem[Hayashi, Navarro, \& Springel(2007)]{Hayashi2007} Hayashi, E., Navarro, J.~F., \& Springel, V.\ 2007, \mnras, 377, 50
\bibitem[Hernquist(1990)]{Hernquist1990} Hernquist, L.\ 1990, \apj, 356, 359
\bibitem[Hibbard et al.(2001)]{Hibbard2001} Hibbard, J.~E., van der Hulst, J.~M., Barnes, J.~E., \& Rich, R.~M.\ 2001, \aj, 122, 2969
\bibitem[Jaffe(1983)]{Jaffe1983} Jaffe, W.\ 1983, \mnras, 202, 995
\bibitem[Jeon, Kim, \& Ann(2009)]{Jeon2009} Jeon, M., Kim, S.~S., \& Ann, H.~B.\ 2009, \apj, 696, 1899
\bibitem[Merritt \& Fridman(1996)]{Merritt1996} Merritt, D., \& Fridman, T.\ 1996, \apj, 460, 136
\bibitem[Merritt et al.(2005)]{Merritt2005} Merritt, D., Navarro, J.~F., Ludlow, A., \& Jenkins, A.\ 2005, \apjl, 624, L85 
\bibitem[Milgrom(1983)]{Milgrom1983} Milgrom, M.\ 1983, \apj, 270, 365
\bibitem[Moore et al.(1999)]{Moore1999} Moore, B., Quinn, T., Governato, F., Stadel, J., \& Lake, G.\ 1999, \mnras, 310, 1147
\bibitem[Navarro, Frenk, \& White(1997)]{Navarro1997} Navarro, J.~F., Frenk, C.~S., \& White, S.~D.~M.\ 1997, \apj, 490, 493
\bibitem[Navarro et al.(2004)]{Navarro2004} Navarro, J.~F., et al.\ 2004, \mnras, 349, 1039
\bibitem[Navarro et al.(2010)]{Navarro2010} Navarro, J.~F., et al.\ 2010, \mnras, 402, 21
\bibitem[Oort(1932)]{Oort1932} Oort, J.~H.\ 1932, \bain, 6, 249
%%%% in prep
\bibitem[Renaud, Boily, \& Theis(2010)]{Renaud2010a} Renaud, F., Boily, C.~M, \& Theis, C.\ 2010, \mnras\, submitted
\bibitem[Sanders \& McGaugh(2002)]{Sanders2002} Sanders, R.~H., \& McGaugh, S.~S.\ 2002, \araa, 40, 263
\bibitem[S\'ersic(1968)]{Sersic1968} S\'ersic, J.~L.\ 1968, Atlas de Galaxies Australes, Univ. Nac. C\'ordoba
\bibitem[Springel et al.(2005)]{Springel2005b} Springel, V., et al.\ 2005, \nat, 435, 629
\bibitem[Stadel et al.(2009)]{Stadel2009} Stadel, J., Potter, D., Moore, B., Diemand, J., Madau, P., Zemp, M., Kuhlen, M., \& Quilis, V.\ 2009, \mnras, 398, L21 
\bibitem[Tegmark et al.(2004)]{Tegmark2004} Tegmark, M., et al.\ 2004, \apj, 606, 702
%%%% arXiv
\bibitem[Tiret \& Combes(2007)]{Tiret2007} Tiret, O., \& Combes, F.\ 2007, arXiv:0712.1459 
\bibitem[Tremaine et al.(1994)]{Tremaine1994} Tremaine, S., Richstone, D.~O., Byun, Y.-I., Dressler, A., Faber, S.~M., Grillmair, C., Kormendy, J., \& Lauer, T.~R.\ 1994, \aj, 107, 634
\bibitem[Trimble(1987)]{Trimble1987} Trimble, V.\ 1987, \araa, 25, 425
\bibitem[Zwicky(1933)]{Zwicky1933} Zwicky, F.\ 1933, Helvetica Physica Acta, 6, 110

\end{thebibliography}

\begin{thebibliography}{}

\bibitem[Dobbs et al.(2010)]{Dobbs2010} Dobbs, C.~L., Theis, C., Pringle, J.~E., \& Bate, M.~R.\ 2010, \mnras, 403, 625
\bibitem[Guillard(2010)]{Guillard2010} Guillard, P.\ 2010, PhD thesis, Univ. Paris-Sud XI, arXiv:1001.3613
%%%% arXiv
\bibitem[Hwang et al.(2010)]{Hwang2010} Hwang, J.-S., Struck, C., Renaud, F., \& Appleton, P.~N.\ 2010, ASPC, 423, 232
%%% in prep.
\bibitem[Renaud, Appleton, \& Xu(2010)]{Renaud2010b} Renaud, F., Appleton, P.~N, \& Xu, K.~C.\ 2010, \apj\ submitted

\end{thebibliography}

\begin{thebibliography}{}

\bibitem[Chandrasekhar(1943)]{Chandrasekhar1943} Chandrasekhar, S.\ 1943, \apj, 97, 255 
\bibitem[Plummer(1911)]{Plummer1911} Plummer, H.~C.\ 1911, \mnras, 71, 460

\end{thebibliography}

\begin{thebibliography}{}

\bibitem[Cardone, Piedipalumbo, \& Tortora(2005)]{Cardone2005} Cardone, V.~F., Piedipalumbo, E., \& Tortora, C.\ 2005, \mnras, 358, 1325
\bibitem[Dehnen(1993)]{Dehnen1993} Dehnen, W.\ 1993, \mnras, 265, 250
\bibitem[Dehnen(2000)]{Dehnen2000} Dehnen, W.\ 2000, \apjl, 536, L39
\bibitem[Einasto(1965)]{Einasto1965} Einasto, J.\ 1965, Trudy Inst. Astroz. Alma-Ata, 51, 87
\bibitem[Hernquist(1990)]{Hernquist1990} Hernquist, L.\ 1990, \apj, 356, 359
\bibitem[Jaffe(1983)]{Jaffe1983} Jaffe, W.\ 1983, \mnras, 202, 995
\bibitem[Navarro, Frenk, \& White(1997)]{Navarro1997} Navarro, J.~F., Frenk, C.~S., \& White, S.~D.~M.\ 1997, \apj, 490, 493
\bibitem[Navarro et al.(2010)]{Navarro2010} Navarro, J.~F., et al.\ 2010, \mnras, 402, 21
\bibitem[Plummer(1911)]{Plummer1911} Plummer, H.~C.\ 1911, \mnras, 71, 460
\bibitem[Tremaine et al.(1994)]{Tremaine1994} Tremaine, S., Richstone, D.~O., Byun, Y.-I., Dressler, A., Faber, S.~M., Grillmair, C., Kormendy, J., \& Lauer, T.~R.\ 1994, \aj, 107, 

\end{thebibliography}

\begin{thebibliography}{}

\bibitem[Anderson(1992)]{Anderson1992} Anderson, C.\ 1992, SIAM Journal on Scientific and Statistical Computing, 13, 923
\bibitem[Binney \& Tremaine(1987)]{Binney1987} Binney, J., \& Tremaine, S.\ 1987, Galactic Dynamics, First Edition, Princeton Univ. Press
\bibitem[Kawai \& Makino(1998)]{Kawai1998} Kawai, A., \& Makino, J.\ 1998, arXiv:9812431

\end{thebibliography}

\begin{thebibliography}{}

\bibitem[Binney \& Tremaine(1987)]{Binney1987} Binney, J., \& Tremaine, S.\ 1987, Galactic Dynamics, First Edition, Princeton Univ. Press
\bibitem[Boily \& Kroupa(2003a)]{Boily2003a} Boily, C.~M., \& Kroupa, P.\ 2003a, \mnras, 338, 665
\bibitem[Boily \& Kroupa(2003b)]{Boily2003b} Boily, C.~M., \& Kroupa, P.\ 2003b, \mnras, 338, 673
\bibitem[Hills(1980)]{Hills1980} Hills, J.~G.\ 1980, \apj, 235, 986
\bibitem[Meylan \& Heggie(1997)]{Meylan1997} Meylan, G., \& Heggie, D.~C.\ 1997, \aapr, 8, 1

\end{thebibliography}
\end{document}